\documentclass[
11pt, % The default document font size, options: 10pt, 11pt, 12pt
english, % ngerman for German
singlespacing, % Single line spacing, alternatives: onehalfspacing or doublespacing
headsepline, % Uncomment to get a line under the header
]{MastersDoctoralThesis} % The class file specifying the document structure

\usepackage[utf8]{inputenc} % Required for inputting international characters
\usepackage[T1]{fontenc} % Output font encoding for international characters
\usepackage{graphics}
\usepackage{mathpazo} % Use the Palatino font by default

\usepackage{amsmath}	% Advanced maths commands

\usepackage{amssymb}	% Extra maths symbols

\usepackage[backend=bibtex,style=authoryear,natbib=true]{biblatex} % Use the bibtex backend with the authoryear citation style (which resembles APA)

\usepackage[autostyle=true]{csquotes} % Required to generate language-dependent quotes in the bibliography

\usepackage{acronym} % Acronyms

\usepackage{units} % Units

\usepackage{multirow}

\usepackage{lscape}

\usepackage{amsbsy}

\usepackage{rotating}
\usepackage{geometry}

\usepackage{tablefootnote}
\usepackage{threeparttable}

\thesistitle{Population synthesis of Galactic pulsars with machine learning} % Your thesis title, this is used in the title and abstract, print it elsewhere with \ttitle
\supervisor{Prof. Nanda Rea \\ Dr. Vanessa Graber} % Your supervisor's name, this is used in the title page, print it elsewhere with \supname
\examiner{} % Your examiner's name, this is not currently used anywhere in the template, print it elsewhere with \examname
\degree{Doctor of Philosophy in Physics} % Your degree name, this is used in the title page and abstract, print it elsewhere with \degreename
\author{Michele Ronchi} % Your name, this is used in the title page and abstract, print it elsewhere with \authorname
\addresses{} % Your address, this is not currently used anywhere in the template, print it elsewhere with \addressname

\subject{Physics and astrophysics} % Your subject area, this is not currently used anywhere in the template, print it elsewhere with \subjectname
\keywords{} % Keywords for your thesis, this is not currently used anywhere in the template, print it elsewhere with \keywordnames
\university{UNIVERSITAT AUTÒNOMA DE BARCELONA \\ DEPARTAMENT DE FÍSICA} % Your university's name and URL, this is used in the title page and abstract, print it elsewhere with \univname
\department{\href{}{Institute of Space Sciences (ICE-CSIC/IEEC)}} % Your department's name and URL, this is used in the title page and abstract, print it elsewhere with \deptname
\AtBeginDocument{
\hypersetup{pdftitle=\ttitle} % Set the PDF's title to your title
\hypersetup{pdfauthor=\authorname} % Set the PDF's author to your name
\hypersetup{pdfkeywords=\keywordnames} % Set the PDF's keywords to your keywords
}

\begin{document}

\frontmatter % Use roman page numbering style (i, ii, iii, iv...) for the pre-content pages

\pagestyle{plain} % Default to the plain heading style until the thesis style is called for the body content

%----------------------------------------------------------------------------------------
%	TITLE PAGE
%----------------------------------------------------------------------------------------

\begin{titlepage}
\begin{center}

\vspace*{.06\textheight}
{\scshape\large \univname\par}\vspace{1cm} % University name

\includegraphics[scale=0.16]{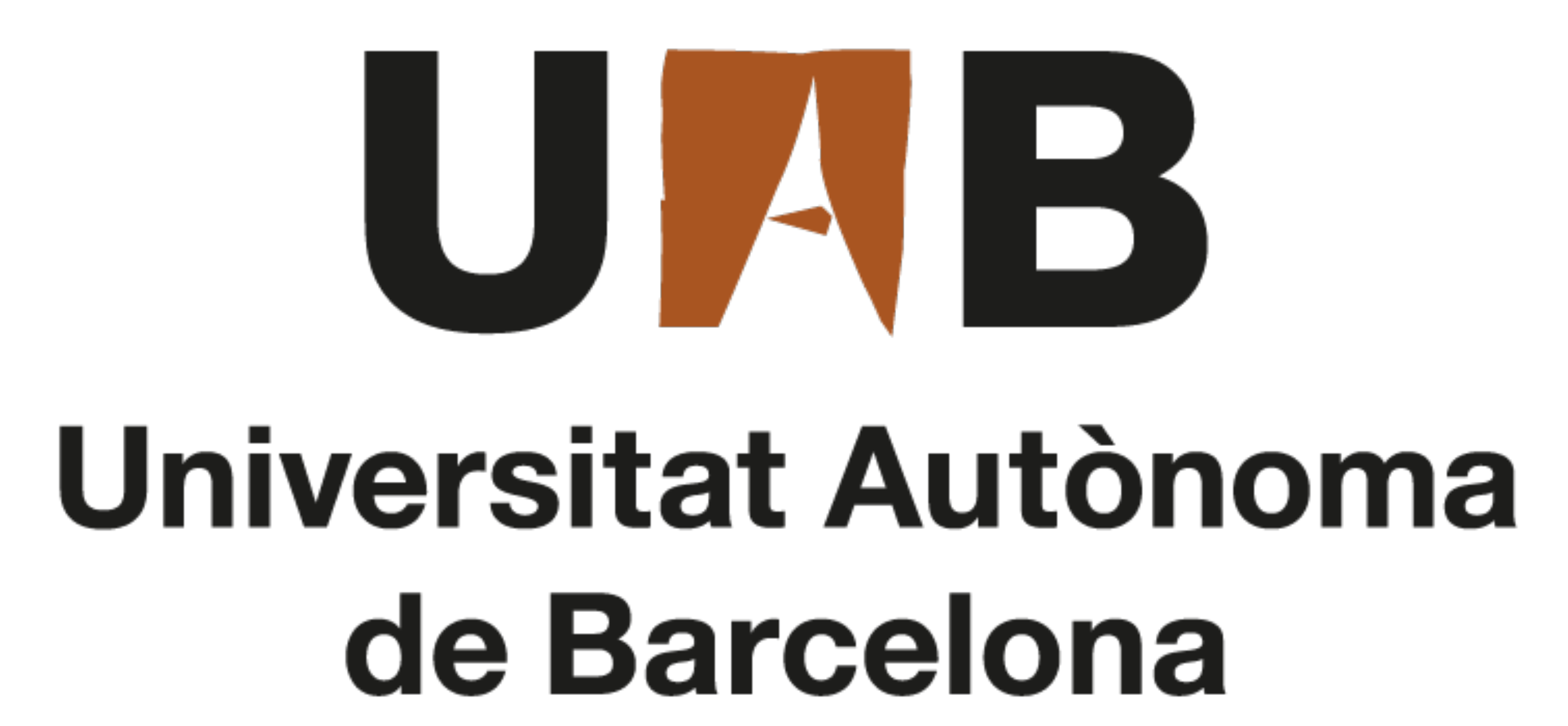}\vspace{0.5cm} % University/department logo - uncomment to place it

\textsc{\Large Doctoral Thesis}\\[0.5cm] % Thesis type

\HRule \\[0.4cm] % Horizontal line
{\huge \bfseries \ttitle\par}\vspace{0.4cm} % Thesis title
\HRule \\[1.5cm] % Horizontal line
 
\begin{minipage}[t]{0.4\textwidth}
\begin{flushleft} \large
\emph{Author:}\\
\href{}{\authorname} % Author name - remove the \href bracket to remove the link
\end{flushleft}
\end{minipage}
\begin{minipage}[t]{0.4\textwidth}
\begin{flushright} \large
\emph{Advisors:} \\
\href{}{\supname} % Supervisor name - remove the \href bracket to remove the link  
\end{flushright}
\end{minipage}\\[1cm]
\begin{minipage}[t]{0.4\textwidth}
\begin{flushleft} \large
% Just an empty minipage to create a blank space on the left
\end{flushleft}
\end{minipage}
\begin{minipage}[t]{0.4\textwidth}
\begin{flushright} \large
\emph{Tutor:} \\
\href{}{Prof. Lluís Font Guiteras} % Tutor name - remove the \href bracket to remove the link  
\end{flushright}
\end{minipage}\\[3cm]
 
\vfill

\large A thesis submitted for the degree of \\ \textit{\degreename}\\[0.3cm] % University requirement text

%\groupname\\
\deptname\\[2cm] % Research group name and department name
 
\vfill

%{\large \today}\\[4cm] % Date
{\large Barcelona, December 2023}\\[4cm]

\vfill
\end{center}
\end{titlepage}

%----------------------------------------------------------------------------------------
%	DECLARATION PAGE
%----------------------------------------------------------------------------------------
\vspace*{\fill}

\noindent The research reported in this thesis was carried out at the Institute of Space Sciences (ICE -- CSIC), within the framework of the doctoral program in Physics at the Autonomous University of Barcelona (UAB). It was supported by the ERC via the Consolidator Grant ``MAGNESIA'' (No. 817661, PI: Nanda Rea), by the program Unidad de Excelencia Mar\'ia de Maeztu CEX2020-001058-M and partially by the grant SGR2021-01269 (PI: Vanessa Graber). 

\vspace{1cm}

\noindent Cover image \textcopyright\ by Olena Shmahalo %Danielle Futselaar

\begin{abstract}
\addchaptertocentry{\abstractname} % Add the abstract to the table of contents
%The Thesis Abstract is written here (and usually kept to just this page). The page is kept centered vertically so can expand into the blank space above the title too\ldots

The recent advancements in data availability and computational power have allowed the development of new machine-learning algorithms and data-analysis techniques that have opened up new avenues to study the Galactic neutron-star population.

This thesis work represents the first efforts to combine population synthesis studies of the Galactic isolated neutron stars with deep-learning techniques with the aim of better understanding neutron-star birth properties and evolution.

In particular, we develop a flexible population-synthesis framework to model the dynamical and magneto-rotational evolution of neutron stars, their emission in radio and their detection with radio telescopes.
We first study the feasibility of using deep neural networks to infer the dynamical properties at birth of a simulated population of neutron stars from density maps storing the information on their sky position and proper motion. 
We then explore a simulation-based inference approach to constrain our physical models from the observed radio pulsar population. 
We employ a neural density estimator to predict the posterior distribution of the parameters describing the birth magnetic-field and spin-period distributions and the late-time magnetic-field decay from density maps containing information on the spin period and spin-period derivative. 
Our results for the initial magneto-rotational properties agree with the findings of previous works while we constrain the late-time evolution of the magnetic field in neutron stars for the first time. 

Besides the development of the population-synthesis framework, this thesis also studies possible scenarios to explain the puzzling nature of recently discovered periodic radio sources with very long periods of the order of thousands of seconds. 

In particular, by assuming a neutron-star origin, we study the spin-period evolution of a newborn neutron star interacting with a supernova fallback disk and find that the combination of strong, magnetar-like magnetic fields and moderate accretion rates can lead to very large spin periods on timescales of ten thousands of years.
Moreover, we perform population synthesis studies to assess the possibility for these sources to be either neutron stars or magnetic white dwarfs emitting coherently through magnetic dipolar losses. 
These discoveries have opened up a new perspective on the neutron-star population and have started to question our current understanding of how coherent radio emission is produced in pulsar magnetospheres.

Overall this thesis work represents the first step towards developing a simulation framework able to model the different neutron star classes in a unified scenario and constraining the properties of the neutron-star population as a whole from multi-wavelength observations through machine learning.  

\subsection*{This thesis is based on the following publications:} 

\begin{itemize}
	
\item \citet*{Ronchi2021} \\
\textit{“Analyzing the Galactic Pulsar Distribution with Machine Learning”} \\
Published in: ApJ 916.2, 100, p. 100

\item \citet*{Ronchi2022} \\
\textit{“Long-period Pulsars as Possible Outcomes of Supernova Fallback Accretion”} \\
Published in: ApJ 934.2, 184, p. 184

\item \citet*{Rea2022} \\
\textit{“Constraining the Nature of the 18 min Periodic Radio Transient GLEAM-X J162759.5-523504.3 via Multiwavelength Observations and Magneto-thermal Simulations”} \\ 
Published in: ApJ 940.1, 72, p. 72

\item \citet*{Rea2023} \\
\textit{“A long-period radio transient active for three decades: population study in the neutron star and white dwarf rotating dipole scenarios”} \\ 
Accepted for publication in ApJ

\item \citet*{Graber2023} \\
\textit{“Simulation-based inference for pulsar population synthesis”} \\
Submitted to ApJ

\end{itemize}

\end{abstract}

%----------------------------------------------------------------------------------------
%	ACKNOWLEDGEMENTS
%----------------------------------------------------------------------------------------

\begin{acknowledgements}
\addchaptertocentry{\acknowledgementname} % Add the acknowledgements to the table of contents
These past four years, besides being challenging, have been very enjoyable, stimulating and rewarding in a lot of aspects and I am very grateful to all the people that I met and who shared this part of my life with me.

First of all I would like to deeply thank my supervisors Nanda and Vanessa. Your guidance, exceptional expertise, enthusiasm, support and understanding have been fundamental during these years. I learnt a lot from you both from a scientific and a human point of view. Thanks to you I became more aware of both the positive aspects and drawbacks of academia and your advices are going to be incredibly valuable for choosing my future path.

People who have previously been and are currently part of the MAGNESIA group are all fantastic. The atmosphere has always been very friendly and supportive. Thanks to the entire Italian crew (Alice, Alessio, Francesco, Stefano, Daniele, Arianna and the new-comers Davide, Thomas and Anna) who always made me feel like at home and to all the people coming from Spain and other parts of the world (Albert I and Albert II, Celsa, Clara, Abu, Emilie, Kostas, Christine and Raj) who opened up my horizons. I feel enriched of having met so many interesting people with their diverse perspectives, vast knowledge and experience. I learned a lot from all of you. 
Special thanks to my PhD mates, Abu for your kindness and wisdom in every situation, Clara for your incredible determination and Celsa, also team-mate (and Spanish teacher), for your contagious enthusiasm and authenticity in everything you do. Together with Vanessa, the harmony and synergy in our population-synthesis group was fantastic and this thesis is only possible thanks to such an amazing team-work.

I extend my gratitude to all the people I met at ICE, a very international and stimulating environment with a lot of PhD students sharing the same journey and people at PIC, for all the helpful conversations on the computing side.
I would further like to thank my examiners, Ben Stappers, Alessandro Patruno and Helena Dominguez Sanchez, who took their time to read this thesis and give constructive feedback that helped me to improve this work.

Lastly, I would like to thank my family, my mum, dad, brother and my grandmother and all my friends back home for their support, encouragement and understanding in all these years.

\end{acknowledgements}

%----------------------------------------------------------------------------------------
%	LIST OF CONTENTS/FIGURES/TABLES PAGES
%----------------------------------------------------------------------------------------

\tableofcontents % Prints the main table of contents

\listoffigures % Prints the list of figures

\listoftables % Prints the list of tables

%----------------------------------------------------------------------------------------
%	ABBREVIATIONS
%----------------------------------------------------------------------------------------

\newcommand\aap{A\&A}                % Astronomy and Astrophysics
\let\astap=\aap                          % alternative shortcut
\newcommand\aapr{A\&ARv}             % Astronomy and Astrophysics Review (the)
\newcommand\aaps{A\&AS}              % Astronomy and Astrophysics Supplement Series
\newcommand\actaa{Acta Astron.}      % Acta Astronomica
\newcommand\afz{Afz}                 % Astrofizika
\newcommand\aj{AJ}                   % Astronomical Journal (the)
\newcommand\ao{Appl. Opt.}           % Applied Optics
\let\applopt=\ao                         % alternative shortcut
\newcommand\aplett{Astrophys.~Lett.} % Astrophysics Letters
\newcommand\apj{ApJ}                 % Astrophysical Journal
\newcommand\apjl{ApJ}                % Astrophysical Journal, Letters
\let\apjlett=\apjl                       % alternative shortcut
\newcommand\apjs{ApJS}               % Astrophysical Journal, Supplement
\let\apjsupp=\apjs                       % alternative shortcut
% The following journal does not appear to exist! Disabled.
%\newcommand\apspr{Astrophys.~Space~Phys.~Res.} % Astrophysics Space Physics Research
\newcommand\apss{Ap\&SS}             % Astrophysics and Space Science
\newcommand\araa{ARA\&A}             % Annual Review of Astronomy and Astrophysics
\newcommand\arep{Astron. Rep.}       % Astronomy Reports
\newcommand\aspc{ASP Conf. Ser.}     % ASP Conference Series
\newcommand\azh{Azh}                 % Astronomicheskii Zhurnal
\newcommand\baas{BAAS}               % Bulletin of the American Astronomical Society
\newcommand\bac{Bull. Astron. Inst. Czechoslovakia} % Bulletin of the Astronomical Institutes of Czechoslovakia 
\newcommand\bain{Bull. Astron. Inst. Netherlands} % Bulletin Astronomical Institute of the Netherlands
\newcommand\caa{Chinese Astron. Astrophys.} % Chinese Astronomy and Astrophysics
\newcommand\cjaa{Chinese J.~Astron. Astrophys.} % Chinese Journal of Astronomy and Astrophysics
\newcommand\fcp{Fundamentals Cosmic Phys.}  % Fundamentals of Cosmic Physics
\newcommand\gca{Geochimica Cosmochimica Acta}   % Geochimica Cosmochimica Acta
\newcommand\grl{Geophys. Res. Lett.} % Geophysics Research Letters
\newcommand\iaucirc{IAU~Circ.}       % IAU Cirulars
\newcommand\icarus{Icarus}           % Icarus
\newcommand\japa{J.~Astrophys. Astron.} % Journal of Astrophysics and Astronomy
\newcommand\jcap{J.~Cosmology Astropart. Phys.} % Journal of Cosmology and Astroparticle Physics
\newcommand\jcp{J.~Chem.~Phys.}      % Journal of Chemical Physics
\newcommand\jgr{J.~Geophys.~Res.}    % Journal of Geophysics Research
\newcommand\jqsrt{J.~Quant. Spectrosc. Radiative Transfer} % Journal of Quantitiative Spectroscopy and Radiative Transfer
\newcommand\jrasc{J.~R.~Astron. Soc. Canada} % Journal of the RAS of Canada
\newcommand\memras{Mem.~RAS}         % Memoirs of the RAS
\newcommand\memsai{Mem. Soc. Astron. Italiana} % Memoire della Societa Astronomica Italiana
\newcommand\mnassa{MNASSA}           % Monthly Notes of the Astronomical Society of Southern Africa
\newcommand\mnras{MNRAS}             % Monthly Notices of the Royal Astronomical Society
\newcommand\na{New~Astron.}          % New Astronomy
\newcommand\nar{New~Astron.~Rev.}    % New Astronomy Review
\newcommand\nat{Nature}              % Nature
\newcommand\nphysa{Nuclear Phys.~A}  % Nuclear Physics A
\newcommand\pra{Phys. Rev.~A}        % Physical Review A: General Physics
\newcommand\prb{Phys. Rev.~B}        % Physical Review B: Solid State
\newcommand\prc{Phys. Rev.~C}        % Physical Review C
\newcommand\prd{Phys. Rev.~D}        % Physical Review D
\newcommand\pre{Phys. Rev.~E}        % Physical Review E
\newcommand\prl{Phys. Rev.~Lett.}    % Physical Review Letters
\newcommand\pasa{Publ. Astron. Soc. Australia}  % Publications of the Astronomical Society of Australia
\newcommand\pasp{PASP}               % Publications of the Astronomical Society of the Pacific
\newcommand\pasj{PASJ}               % Publications of the Astronomical Society of Japan
\newcommand\physrep{Phys.~Rep.}      % Physics Reports
\newcommand\physscr{Phys.~Scr.}      % Physica Scripta
\newcommand\planss{Planet. Space~Sci.} % Planetary Space Science
\newcommand\procspie{Proc.~SPIE}     % Proceedings of the Society of Photo-Optical Instrumentation Engineers
\newcommand\rmxaa{Rev. Mex. Astron. Astrofis.} % Revista Mexicana de Astronomia y Astrofisica
\newcommand\qjras{QJRAS}             % Quarterly Journal of the RAS
\newcommand\sci{Science}             % Science
\newcommand\skytel{Sky \& Telesc.}   % Sky and Telescope
\newcommand\solphys{Sol.~Phys.}      % Solar Physics
\newcommand\sovast{Soviet~Ast.}      % Soviet Astronomy (aka Astronomy Reports)
\newcommand\ssr{Space Sci. Rev.}     % Space Science Reviews
\newcommand\zap{Z.~Astrophys.}       % Zeitschrift fuer Astrophysik

\newcommand{\bt}{\boldsymbol{\theta}}
\newcommand{\bp}{\boldsymbol{\psi}}
\newcommand{\bx}{\boldsymbol{x}}
\newcommand{\pr}{\mathcal{P}}

\def\xmm {\emph{XMM--Newton}}
\def\cxo {\emph{Chandra}}
\def\nustar {\emph{NuSTAR}}
\def\rst {\emph{ROSAT}}
\def\swift {\emph{Swift}}
\def\nicer {\emph{NICER}}
\def\hxmt {\emph{Insight}-HXMT}
\def\pks {Parkes}

\def\gleamfirst {\mbox{GLEAM-X\,J162759.5-523504.3}}
\def\gleam {\mbox{GLEAM-X\,J1627}}
\def\mtp{\mbox{PSR J0901-4046}}
\def\gpm {\mbox{GPM J1839–10}}
\def\rcw{\mbox{1E\,161348-5055}}
\def\lowbsgr{\mbox{SGR\,0418+5729}}
\def\sgrfrb{SGR\,1935+2154}
\def\xte{XTE\,J1810$-$197}

\def\msun{$M_{\odot}$\,}

\acrodef{ATNF}[ATNF]{Australia Telescope National Facility}
\acrodef{PHL}[PHL]{Parkes High-Latitude}
\acrodef{PMB}[PMB]{Parkes Multi-Beam}
\acrodef{SMB}[SMB]{Swinburne Intermediate-latitude}
\acrodef{MLP}[MLP]{multi-layer perceptron}
\acrodef{CNN}[CNN]{convolutional neural network}
\acrodef{MDN}[MDN]{mixture density network}
\acrodef{MAF}[MAF]{masked autoregressive flow}
\acrodef{NPE}[NPE]{neural posterior estimation}
\acrodef{SBC}[SBC]{simulation-based calibration}
\acrodef{DM}[$DM$]{dispersion measure}
\acrodef{SNR}[SNR]{Supernova Remnant}
\acrodef{GRBs}[GRBs]{Gamma-Ray Bursts}
\acrodef{FRBs}[FRBs]{Fast Radio Bursts}
\acrodef{ANNs}[ANNs]{artificial neural networks}
\acrodef{Adam}[Adam]{Adaptive Moment Estimation}
\acrodef{RMSE}[RMSE]{root-mean-square error}
\acrodef{MRE}[MRE]{mean relative error}
\acrodef{SKA}[SKA]{Square Kilometer Array}
\acrodef{ML}[ML]{machine learning}
\acrodef{RPPs}[RPPs]{rotation-powered pulsars}
\acrodef{SGRs}[SGRs]{soft gamma-ray repeaters}
\acrodef{AXPs}[AXPs]{anomalous X-ray pulsars}
\acrodef{RRATs}[RRATs]{rotating radio transients}
\acrodef{CCOs}[CCOs]{central compact objects}
\acrodef{XDINSs}[XDINSs]{X-ray dim isolated neutron stars}
\acrodef{CCSN}[CCSN]{core-collapse supernova}
\acrodef{ReLU}[ReLU]{rectified linear unit}
\acrodef{MWA}[MWA]{Murchison Widefield Array}
\acrodef{ATCA}[ATCA]{Australia Telescope Compact Array}
\acrodef{ASKAP}[ASKAP]{Australian Square Kilometre Array Pathfinder}
\acrodef{VLA}[VLA]{Very Large Array}
\acrodef{VLITE}[VLITE]{Low Band Ionospheric and Transient Experiment} \acrodef{GMRT}[GMRT]{Giant Metrewave Radio Telescope}
\acrodef{PMPS}[PMPS]{Parkes Multibeam Pulsar Survey}
\acrodef{SMPS}[SMPS]{Swinburne Intermediate-latitude Pulsar Survey}
\acrodef{HTRU}[HTRU]{High Time Resolution Universe}
\acrodef{ISM}[ISM]{interstellar medium}
\acrodef{CNN}[CNN]{convolutional neural network}
\acrodef{MDN}[MDN]{mixture density network}
\acrodef{GMM}[GMM]{Gaussian-mixture model}
\acrodef{ReLU}[ReLU]{rectified linear unit}
\acrodef{SBI}[SBI]{simulation-based inference}
\acrodef{MCMC}[MCMC]{Markov chain Monte Carlo}
\acrodef{NPE}[NPE]{neural posterior estimation}
\acrodef{NLE}[NLE]{neural likelihood estimation}
\acrodef{NRE}[NRE]{neural ratio estimation}
\acrodef{CI}[CI]{credible interval}
\acrodef{HDR}[HDR]{highest-density region}

%\begin{abbreviations}{ll} % Include a list of abbreviations (a table of two columns)

%\textbf{LAH} & \textbf{L}ist \textbf{A}bbreviations \textbf{H}ere\\
%\textbf{WSF} & \textbf{W}hat (it) \textbf{S}tands \textbf{F}or\\

%\end{abbreviations}

%----------------------------------------------------------------------------------------
%	PHYSICAL CONSTANTS/OTHER DEFINITIONS
%----------------------------------------------------------------------------------------

\begin{constants}{lr@{${}={}$}l} % The list of physical constants is a three column table

% The \SI{}{} command is provided by the siunitx package, see its documentation for instructions on how to use it

Speed of light in a vacuum & $c$ & $\unit[2.99792458 \times 10^{10}]{cm \, s^{-1}}$ \\
Gravitational constant & $G$ & $\unit[6.67259 \times 10^{-8}]{cm^3 \, g^{-1} \, s^{-2}}$ \\
Planck constant & $h$ & $\unit[6.6260755 \times 10^{-27}]{erg \, s}$ \\
Electron charge & $e$ & $\unit[4.8032068 \times 10^{-10}]{esu}$ \\
Electron mass & $m_e$ & $\unit[9.1093897 \times 10^{-28}]{g}$ \\
Proton mass & $m_p$ & $\unit[1.6726231 \times 10^{-24}]{g}$ \\
Neutron mass & $m_n$ & $\unit[1.6749286 \times 10^{-24}]{g}$ \\
Boltzmann constant & $k_{\rm B}$ & $\unit[1.380658 \times 10^{-16}]{erg \, K^{-1}}$ \\
Thomson cross-section & $\sigma_{\rm T}$ & $\unit[6.65245 \times 10^{-25}]{cm^{-2}}$  \vspace{5mm} \\

Solar mass & $M_{\odot}$ & $\unit[1.989 \times 10^{33}]{g}$ \\
Parsec & pc & $\unit[3.0856 \times 10^{18}]{cm}$ \\
Jansky & Jy & $\unit[1.0 \times 10^{-23}]{erg \, cm^{-2} \, s^{-1} \, Hz^{-1}}$ \\
Year & yr & $\unit[3.1557 \times 10^{7}]{s}$ \\
%Speed of Light & $c_{0}$ & \SI{2.99792458e8}{\meter\per\second} (exact)\\
%Constant Name & $Symbol$ & $Constant Value$ with units\\

\end{constants}

%----------------------------------------------------------------------------------------
%	SYMBOLS
%----------------------------------------------------------------------------------------

%\begin{symbols}{lll} % Include a list of Symbols (a three column table)

%$a$ & distance & \si{\meter} \\
%$P$ & power & \si{\watt} (\si{\joule\per\second}) \\
%Symbol & Name & Unit \\

%\addlinespace % Gap to separate the Roman symbols from the Greek

%$\omega$ & angular frequency & \si{\radian} \\

%\end{symbols}

%----------------------------------------------------------------------------------------
%	DEDICATION
%----------------------------------------------------------------------------------------

%\dedicatory{For/Dedicated to/To my\ldots} 

%----------------------------------------------------------------------------------------
%	THESIS CONTENT - CHAPTERS
%----------------------------------------------------------------------------------------

\mainmatter % Begin numeric (1,2,3...) page numbering

\pagestyle{thesis} % Return the page headers back to the "thesis" style

% Include the chapters of the thesis as separate files from the Chapters folder
% Uncomment the lines as you write the chapters

\chapter{Neutron stars} % Main chapter title

\label{Chapter1} % For referencing the chapter elsewhere, use \ref{Chapter1} 

%----------------------------------------------------------------------------------------

% Define some commands to keep the formatting separated from the content 
\newcommand{\keyword}[1]{\textbf{#1}}
\newcommand{\tabhead}[1]{\textbf{#1}}
\newcommand{\code}[1]{\texttt{#1}}
\newcommand{\file}[1]{\texttt{\bfseries#1}}
\newcommand{\option}[1]{\texttt{\itshape#1}}

%----------------------------------------------------------------------------------------

\section{Introduction}

The idea of the possible existence of neutron stars was first proposed in 1934, two years after the discovery of the neutron itself \citep{Chadwick1932}, when \citet{Baade1934} wondered if a celestial body made entirely of neutrons might remain after a supernova explosion. 
It was not until the 1960s that Baade and Zwicky’s hypothetical neutron stars were finally detected. While working as a PhD student at the University of Cambridge, Jocelyn Bell Burnell noticed a strange, regularly pulsed radio signal in the data collected using the Mullard Radio Astronomy Observatory. These signals were initially interpreted as the result of radial pulsations in either a white dwarf or a neutron star \citep{Hewish1968}.
This groundbreaking discovery opened up new avenues for exploring the intriguing relationship between neutron stars and their potential to emit highly focused beams of electromagnetic radiation. Soon after it was suggested by \citet{Pacini1968} and \citet{Gold1968} that coherent emission could occur in rotating neutron stars with strong magnetic fields where particles could be accelerated up to relativistic speeds and emit beams of electromagnetic radiation. This can be detected as pulsed signals as the neutron star rotates like a lighthouse. Hence the name "pulsar" was coined.
By that time about 20 similar objects had been identified including a source in the Crab nebula \citep{Staelin1968, Comella1969} and one in the Vela supernova remnant \citep{Large1968}. The identification of these pulsars with rotating neutron stars was in the end confirmed by the measurement of a spin-down in the period of the Crab pulsar \citep{Richards1969} as a spin-down is more easily achieved in rotating objects rather than in pulsating ones. Furthermore, the identification of some pulsars with supernova remnants further supported the initial idea of Baade and Zwicky.

With subsequent technological advancements in observational astronomy, the population of known neutron stars has grown exponentially. Since the first detection, around 3000 pulsars have been discovered, not only in the radio band \citep{Manchester2005}\footnote{see the ATNF pulsar catalog \url{https://www.atnf.csiro.au/research/pulsar/psrcat/}}. Pulsars can emit across a wider wavelength range from radio to gamma-rays and even be radio-quiet. However, an enduring challenge lies in accurately characterising their birth properties as well as deciphering the mechanisms that regulate their formation and evolution. Untangling the contributing factors, such as the initial magnetic field strength and spin period distributions, holds paramount importance in understanding the diverse outcomes of stellar evolution and the underlying physical processes within core-collapse supernovae. By combining observational constraints with data analysis and sophisticated numerical simulations, researchers can refine existing theoretical models and shed light on the complex physical processes involved in the formation and evolution of these dense remnants.
  
In this chapter we will explore the basics of neutron star physics and our current understanding of how they evolve, emit electromagnetic radiation, with particular focus on the radio emission, and how we can detect them (Sections from~\ref{sec:ch1_formation} to ~\ref{sec:ch1_radiometer}).
Additionally, we will provide a summary of the challenges involved in modelling the Galactic population of neutron stars (Sections~\ref{sec:ch1_ns_zoo} and~\ref{sec:ch1_birthrate}) and explore how employing a population synthesis framework can help in determining their overall properties (Section~\ref{sec:ch1_popsyn}).

%----------------------------------------------------------------------------------------

\section{Formation} \label{sec:ch1_formation}

The progenitors of most neutron stars are massive stars of spectral types O and B, characterised by initial masses between around 8 and 25 \msun \citep{Shapiro1983}. 
In general, depending on its initial mass, a star spend from a few million years up to more than ten billion years on the main sequence, burning hydrogen into helium \citep{Phillips1994, Prialnik2000}. In this stage, the thermal pressure released by the thermonuclear fusion reactions in its core supports the star against gravitational collapse. However, the more massive a star is, the faster it consumes its nuclear fuel and the faster it evolves towards its death. After exhausting all the hydrogen in the core the thermal pressure support coming from hydrogen burning vanishes. The inner regions of the star, now mainly composed of helium "ashes", start to contract due to their own gravity. As a consequence they heat up while the outer envelope expands and cools down. The star becomes a red giant. As the temperature in the helium core grows upon contraction, new nuclear fusion reactions are triggered. Helium starts to burn and to be converted into carbon, oxygen and neon. The released energy allows to restore again the equilibrium with gravity and the star enters a new stable phase that usually lasts one-tenth of the main sequence phase. However, once the helium fuel is exhausted, the core (now composed mainly of oxygen and carbon) starts to contract again. 

In stars with an initial mass of less than about 8 \msun, the dense core reaches equilibrium in a new state of matter called a degenerate electron gas. In this state, the stellar core is supported against further gravitational contraction thanks to the pressure exerted by the electrons due to the Pauli exclusion principle, even in the absence of nuclear reactions. In the meanwhile, the outer envelopes of the star expand further and are gradually dispersed. What is left is the exposed core of the star with a mass of $\sim 0.5 - 1$ \msun and a radius of $\sim \unit[10^8]{cm}$, made primarily by helium, carbon and oxygen. A white dwarf has formed.

For stars with an initial mass greater than about 8 \msun the electron degeneracy pressure is not enough to counteract gravity. Therefore they continue the sequence of core contraction and ignition of new nuclear reactions. As time goes by the inner core reaches temperatures and densities favourable for the synthesis of progressively heavier elements. After hydrogen and helium, these massive stars burn carbon, neon, oxygen, and silicon on progressively faster timescales. In this way, they undergo all the stages of nuclear burning up to the production of elements in the "iron group" with atomic number $A = 56$. The iron group elements are the most tightly bound nuclei and their synthesis into heavier elements consumes, rather than releases, energy. Since no further energy can be released from the nuclear fusion reactions as the iron core grows it becomes unstable and is not further supported against gravitational collapse.
As the iron core approaches the Chandrasekhar mass limit of about $1.4$ \msun \citep{Chandrasekhar1931}, the central temperature and density reaches values of about $\sim \unit[10^9]{K}$ and $\sim \unit[10^9]{g \, cm^{-3}}$ respectively. The temperature is high enough that nuclear photo-disintegration is triggered: the abundant energetic photons trapped in the hot dense core are partly consumed to unbind nuclei into lighter elements in endothermic reactions. In such a way the photon radiation pressure is lost. 
Furthermore, at such densities, reactions like electron capture $e^- + p \longrightarrow n + \nu_e$ become energetically favourable. The electrons are absorbed by the nuclei which become very neutron-rich in what is called neutronisation. This process depletes the core of electrons and their supporting degeneracy pressure and produces many escaping neutrinos which carry away a big amount of energy. This leads to an almost unrestrained collapse of the stellar core on a free-fall timescale of a few seconds. 

As the collapse proceeds, the core central density increases to the point that neutrons start to drip out of the nuclei and form a degenerate gas. At densities $\gtrsim \unit[10^{14}]{g \, cm^{-3}}$, the nuclei are completely dissolved into homogeneous nuclear matter mostly composed of neutrons. If the mass of the collapsing core is not too high ($\lesssim 3$ \msun), the degeneracy pressure exerted by the neutrons is finally able to halt the collapse. This triggers a shock wave, that propagates outwards through the collapsing stellar layers and blows them up in what is observed as a supernova explosion.
What is left is a very hot ($T \sim \unit[10^{8}]{K}$) and compact remnant of mass $M \sim 1 - 2$ \msun and radius $R \sim \unit[10^6]{cm}$: a neutron star.

What described above is the evolution of an isolated massive star. Another channel that leads to the formation of a neutron star is through accretion induced collapse. In this scenario a white dwarf in a binary system can collapse upon surpassing the Chandrasekhar mass limit due to mass accretion from a companion star \citep{Fryer1999}.

%---------------------------------------------------------------------------------------

\section{Internal structure and the mass-radius relation} \label{sec:ch1_internal_structure_MR_relation}

%----------------------------------------------------
\begin{figure}
\centering
\includegraphics[width=0.7\textwidth]{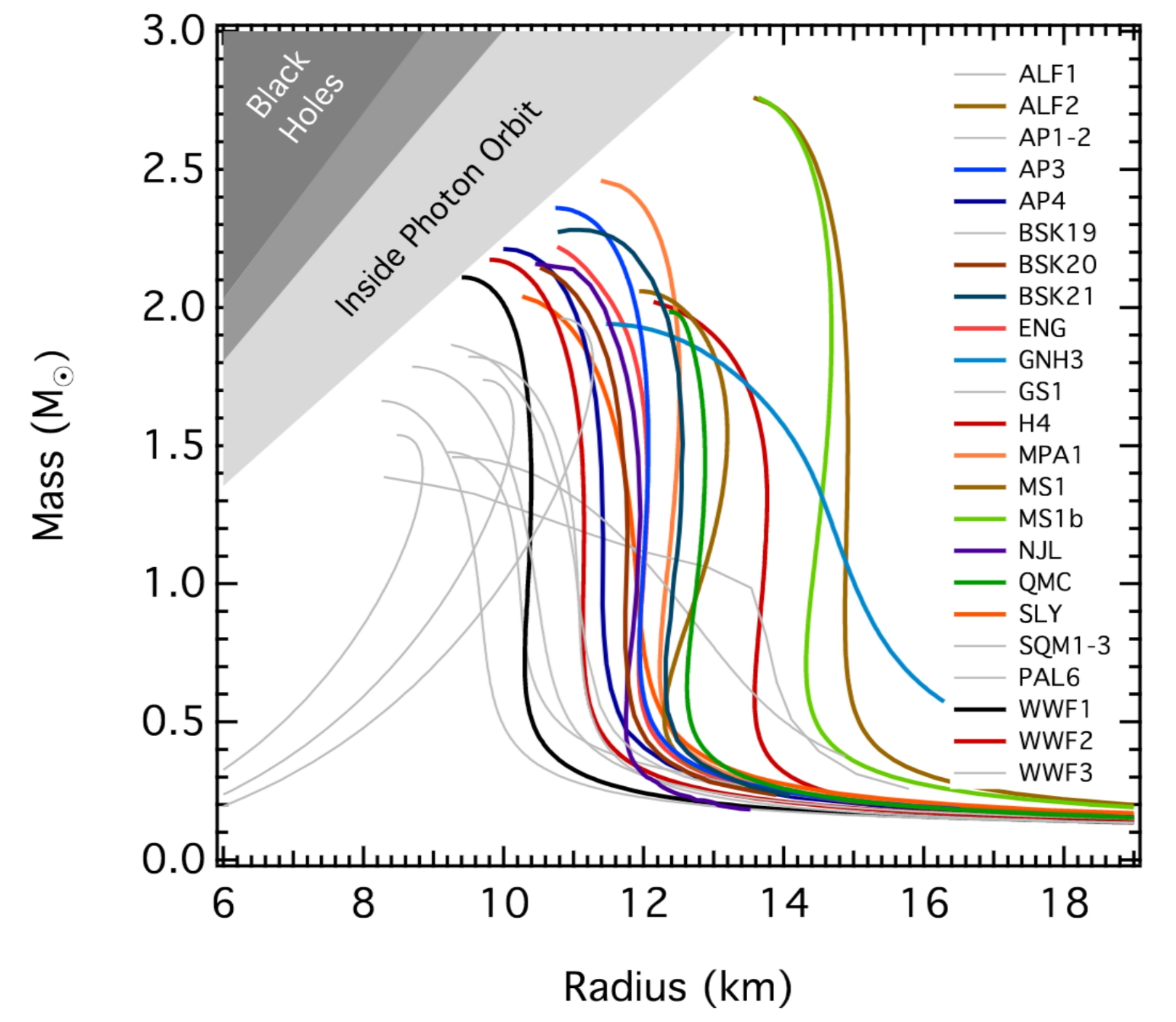}
\caption[Neutron star mass-radius relation.]{Neutron star mass-radius curves corresponding to different equations of state calculated under different physical assumptions and using a range of computational approaches \citep[see][for more details]{Ozel2016} \citep[Figure taken from][]{Ozel2016}.}
\label{fig:ch1_mass_radius}
\end{figure}
%----------------------------------------------------

To our knowledge, neutron stars are the most compact objects in the Universe where matter can still reach a stable configuration under the effect of extreme gravity. Their compactness, defined by the ratio between their Schwarzschild radius $R_{\rm S} = \frac{2GM_{\rm NS}}{c^2}$ and their radius $R_{\rm NS}$, is of the order:
%--------------------------------------------------------------------
\begin{align}
    \frac{R_{\rm S}}{R_{\rm NS}} \simeq 0.376 \left( \frac{M_{\rm NS}}{1.4 M_{\odot}} \right) \left( \frac{R_{\rm NS}}{\unit[11]{km}} \right)^{-1}.
\end{align}
%--------------------------------------------------------------------
This is close to the most extreme value given by a Schwarzschild black hole where the compactness is 1. 
The density grows from values less than $\unit[10^4]{g \, cm^{-3}}$ in the outermost envelope to nuclear density values $\unit[10^{14} - 10^{15}]{g \, cm^{-3}}$ in the neutron star core. As you move radially into the star from the surface, the pressure increases to support the increasing weight of the material above any given point. Hence the structure and state of matter change as we move inside \citep[see][for a review]{Chamel2008}. 
The outer layers ($\sim \unit[100]{m}$) are a liquid ocean of heavy nuclei of the iron group and degenerate free electrons. As one moves into the crust ($\sim \unit[1]{km}$), the nuclei are more packed together and the Coulomb interactions between them begin to dominate over the thermal energy. The nuclei rearrange themselves to form a solid elastic lattice. On the other end, the degenerate free electrons become relativistic.  
At the basis of the crust and the interface with the core, the heavy nuclei become more and more neutron-rich and deformed into cylindrical and planar shapes. This is the so-called "nuclear pasta phase".
Moving into the inner core the nuclei are completely dissolved into an ocean of free neutrons togeteher with protons, ultra-relativistic electrons and muons. The neutron degenerate pressure supports the neutron star against collapse. However, at these extreme densities, the exact composition and the properties of matter in the inner core are not clear as exotic particles may be present. Therefore the exact equation of state, i.e. the relation linking pressure and density, for neutron stars is still unknown. 
\citet{Oppenheimer1939, Tolman1939}  were the first ones to propose an equation of state of a relativistic gas of degenerate neutrons and a model for the hydrostatic equilibrium of general relativistic bodies also known as Oppenheimer-Volkoff-Tolmann equation:
%--------------------------------------------------------------------
\begin{align}
    \frac{{\rm d}p(r)}{{\rm d}r} = \frac{G M(r) \rho(r)}{r^2} \left[ 1 + \frac{p(r)}{\rho(r) c^2} \right] \left[ 1 + \frac{4\pi r^3 p(r)}{ M(r) c^2} \right] \left[1 - \frac{2GM(r)}{c^2 r} \right]^{-1},
\end{align}
%--------------------------------------------------------------------
where $p(r)$ and $\rho(r)$ are the pressure and density, respectively, and $M(r)$ is the mass within the radius $r$.
In particular, the solution of this equation maps a given equation of state to the corresponding mass-radius relation, linking in this way the microscopic to the macroscopic properties of the star. Since then the knowledge of the behaviour of matter at nuclear densities has progressed and several models for the equation of state of neutron stars have been proposed. 
The uncertainty is still big. For a given mass, these equations of state allow a range of radii between $\sim \unit[7]{km}$ up to $\sim \unit[16]{km}$ (see Figure~\ref{fig:ch1_mass_radius}).
Therefore, to constrain the equation of state for the neutron star matter, it is crucial to obtain independent measurements of the mass and radius \citep[see][for a review]{Ozel2016}.   

The currently available measurements of neutron star masses come primarily from the study of neutron stars in binary systems. Up to now the minimum mass measured for a neutron star is $1.174 \pm 0.004$ \msun for PSR J0453+1559 \citep{Martinez2015} while the heaviest neutron star is PSR J0952--0607 which has an estimated mass of $2.35 \pm 0.17$ \msun \citep{Romani2022}. Finding the maximum mass for neutron stars is of particular relevance since it can rule out the equations of state that predict maximum masses below this value. 

Until recently, the measurement of neutron-star radii mainly relied on X-ray observations and spectral analysis of quiescent and accreting neutron stars \citep{Miller2019, Riley2019}. The typical estimates for the radii obtained with these methods are in the range $\unit[10 - 14]{km}$ \citep[see, e.g.,][]{Steiner2018}. However, these analyses are often affected by systematics that are difficult to quantify. After the serendipitous detection of the famous binary neutron-star merger GW170817 and its electromagnetic counterparts \citep{Abbott2017a, Abbott2017b, Abbott2017c}, the beginning of the multi-messenger observation era has opened a new window on this front. For example, by combining electromagnetic and gravitational wave data \citet{Margalit2017} managed to put constraints on the maximum neutron star mass. \citet{Capano2020, Raaijmakers2021} combined nuclear theory with the data from the gravitational waves and electromagnetic emission from this event using a Bayesian framework. They were able to constrain the radius of a neutron star with mass $M = 1.4$ \msun to values around $\unit[11 - 12]{km}$.

%--------------------------------------------------------------------------------------------

\section{Spin periods} \label{sec:ch1_spin_periods}

By identifying the periodicity of the arrival time of pulses of radiation with the neutron star spin periods implies that these celestial bodies rotate at very high speeds. From the observed population their spin periods range from a few milliseconds to a few tens of seconds (see Figure~\ref{fig:ch1_PPdot_diagram}). The majority of these measurements come from radio timing analysis which has also unveiled that their spin periods increase with time at rates between around $\unit[10^{-20}]{s \, s^{-1}}$ and $\unit[10^{-10}]{s \, s^{-1}}$. This suggests that generally neutron stars spin down over time and they should be born with even shorter rotation periods.

Several works involving analyses of young supernova remnants \citep[e.g.][]{Chevalier2005}, population synthesis studies \citep[e.g.][]{Popov2012, Faucher2006, Gullon2014} and supernova simulations \citep[e.g.][]{Ott2006, Janka2022}, have suggested that the birth spin periods of neutron stars should range between a few tens of milliseconds up to a few hundreds of milliseconds. However, the origin of such fast rotation is still unclear.

A first insight into how neutron stars could achieve these fast rotation can be gained by considering the angular momentum conservation during the collapse of the iron core. For an order of magnitude estimation we consider a core with mass $M_{\rm core} \sim 1.4$ \msun, radius comparable to our Sun $R_{\rm core} \sim \unit[10^{11}]{cm}$ and spin period $P \sim \unit[1]{day}$. From the conservation of angular momentum, we get that:
%------------------------------------------------------------------------
\begin{align}
    M_{\rm core} R_{\rm core}^2 \omega_{\rm core} \sim  M_{\rm NS} R_{\rm NS}^2 \omega_{\rm NS},
\end{align}
%------------------------------------------------------------------------
where $\omega = 2 \pi / P$ is the spin angular velocity and we assumed mass conservation during the core collapse, i.e., $M_{\rm core} \sim  M_{\rm NS}$. During the collapse, as the radius shrinks to $R_{\rm NS} \sim \unit[10^{6}]{cm}$ reducing the moment of inertia by a factor $\sim 10^{10}$, the angular velocity should increase by the same amount. As a consequence, newborn neutron stars should spin with periods of the order of milliseconds.

However angular momentum conservation is not the only ingredient to explain the neutron star's fast rotation. Various mechanisms acting during and after the supernova explosion could contribute to spinning up or spinning down the collapsing core and the newly formed neutron star. During the core collapse and the development of the supernova, hydrodynamical instabilities and turbulent motion of mass could cause asymmetric mass ejection in the supernova. This asymmetry imparts accelerating forces on the core over timescales from several seconds to hours. In this way, the newly formed neutron star is displaced from the centre of the collapse and receives a kick that can be of several hundreds of kilometers per second \citep[see, e.g., ][but also Chapter~\ref{Chapter5}]{Hobbs2005}. As the neutron star moves in the supernova ejecta it can accrete a fraction of the matter that has remained gravitationally bound. This anisotropic accretion of mass could transfer a relevant amount of angular momentum to the compact remnant after the development of the supernova \citet{Janka2022, Coleman2022}. This scenario suggests also a possible link between the origin of the fast neutron stars' proper motion and their spin period.

%--------------------------------------------------------------------------------------------

\section{Magnetic fields} \label{sec:ch1_magnetic_field}

To our knowledge neutron stars are also the strongest magnets in the Universe. A direct measurement of the magnetic field strength is not straightforward though. One possibility is through the detection of cyclotron lines, that generally appear as absorption features in the UV and X-ray spectra of systems containing highly magnetised neutron stars. Almost all the observations of cyclotron absorption features come from neutron stars accreting matter from a stellar companion, also called X-ray binary pulsars. These features are produced as the photons produced by the hot column of plasma accreted onto the neutron star's magnetic poles are resonantly scattered by charged particles gyrating around the magnetic field lines \citep[see for example][for a review]{Staubert2019}. 

However, generally, neutron star magnetic fields are indirectly estimated from their rotational properties. 
The commonly adopted explanation for this rotational energy loss is that the neutron star experiences an electromagnetic torque generated by a very powerful dipolar magnetic field (see Section~\ref{sec:ch1_dip_spindown_evol}).
With this assumption, from the measurements of the spin period and spin period derivative one can infer dipolar magnetic field values that range between $\unit[10^8]{G}$ to almost $\unit[10^{15}]{G}$.

How such strong magnetic fields are generated is still a matter of debate.
One idea is that the neutron star inherits the magnetic field of the parent OB star. In this fossil field hypothesis, during the core collapse the magnetic field gets amplified due to magnetic flux freezing \citep{Ferrario2006, Hu2009}. 
Ordinary OB stars have a magnetic field of the order of $B \sim \unit[10^2]{G}$ with some of them exceeding $\unit[10^3]{G}$. 
% See: https://ui.adsabs.harvard.edu/abs/2014A%26A...564L..10H/abstract
From the conservation of the magnetic flux, we have that:
%------------------------------------------------------------------------
\begin{align}
    B_{\rm core} R_{\rm core}^2 \sim  B_{\rm NS} R_{\rm NS}^2.
\end{align}
%------------------------------------------------------------------------
As the core radius contracts by a factor $\sim 10^5$, the magnetic field should be amplified by a factor $\sim 10^{10}$. This translates to a magnetic field of the order of $B_{\rm NS} \sim \unit[10^{12}]{G}$ for a newborn neutron star, in line with current estimates.
A second hypothesis is that a strong magnetic field originates dynamically from a dynamo process. During the formation of the proto-neutron star, the magnetic field could be amplified by electric currents generated from rapid rotation or turbulent convective motions of conductive material inside the proto-neutron star a few seconds after its formation \citep[see e.g.][]{Burrows1987, Burrows1988, Duncan1992, White2022}. Probably both effects play a role in the creation of such intense magnetic fields.
% See: https://ui.adsabs.harvard.edu/abs/2008A%26A...479..167N/abstract

%--------------------------------------------------------------------------------------------

\section{Magnetic field evolution}  \label{sec:ch1_B_evol}

%----------------------------------------------------
\begin{figure}
	\includegraphics[width=0.47\textwidth]{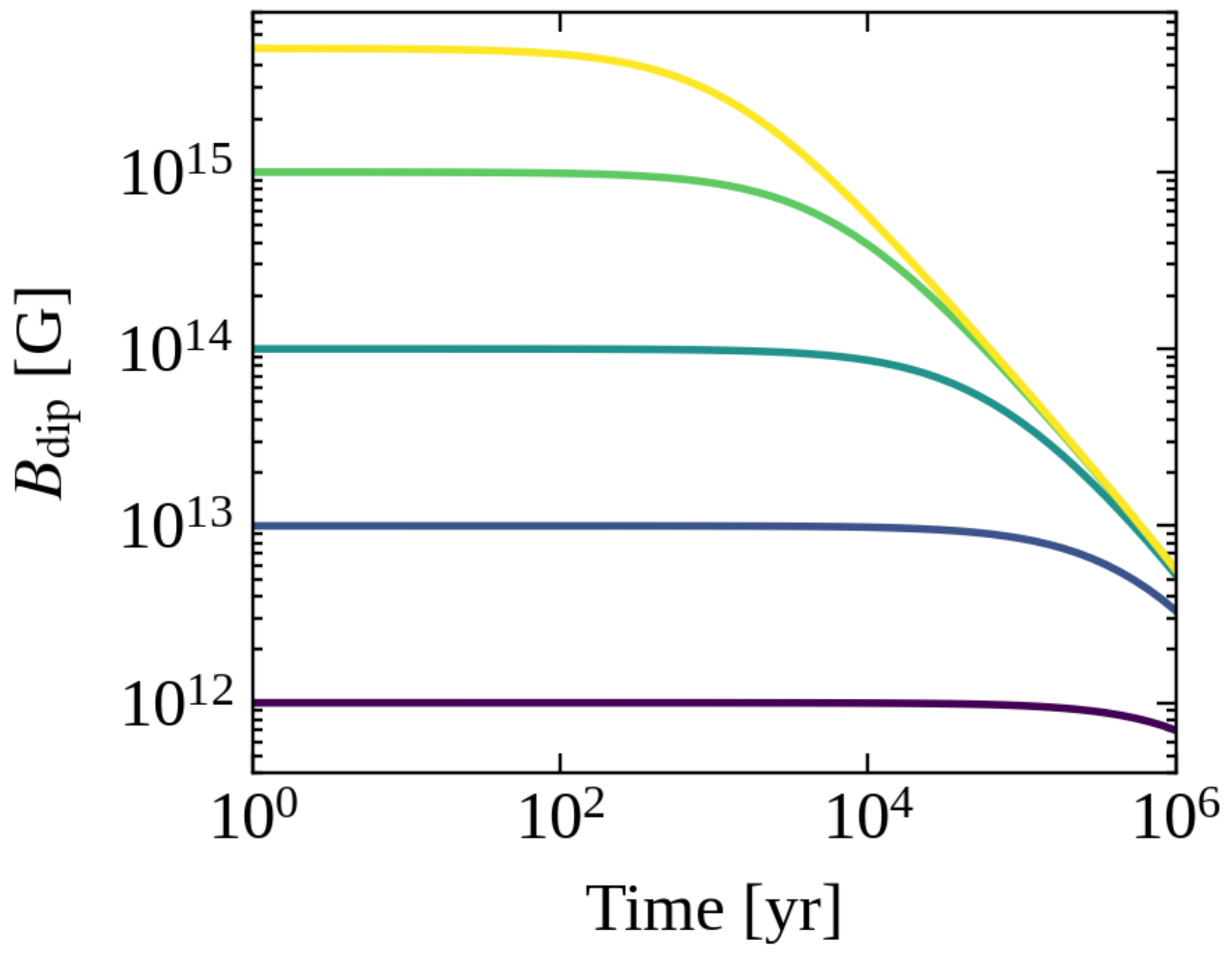}
	\includegraphics[width=0.47\textwidth]{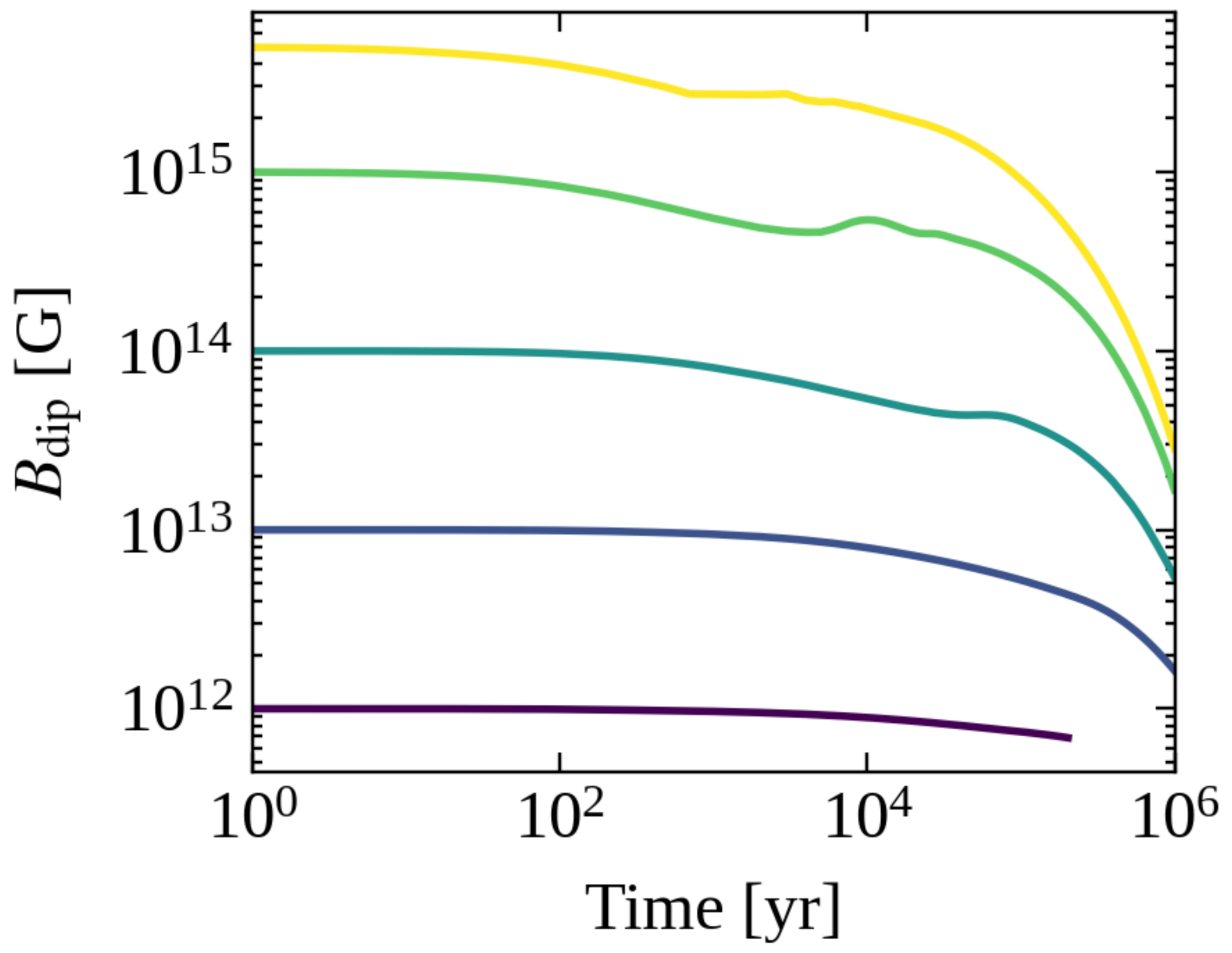}
	\caption[Dipolar magnetic field evolution curves]{Examples of dipolar magnetic field evolution curves. The analytical solution in Equation \eqref{eq:ch1_B_decay} (\textit{left panel}) is compared to the numerical solution from 2D magneto-thermal simulations \citep{Vigano2021} (\textit{right panel}). In the right panel, the purple curve corresponding to an initial magnetic field of $\unit[10^{12}]{G}$ appears truncated due to code constraints in modelling the micro-physical behaviour of the neutron-star crust at low temperatures (see also Appendix~\ref{app:B-field} for more details).}
	\label{fig:ch1_B_evolution}
\end{figure}
%----------------------------------------------------

Much effort has been put into modelling the magnetic field geometry and evolution inside a neutron star. 
Soon after birth, a solid crust of about $\unit[1]{km}$ thickness is formed.
At the densities of the crust ($\unit[10^{10} - 10^{14}]{g \, cm^{-3}}$), matter is pressure-ionised. Atoms are so densely packed together that electrons are not bound to a single nucleus but feel the electric fields of all the neighbouring nuclei. As a consequence, while the ions in the lattice or the pasta phase layer (see Section~\ref{sec:ch1_internal_structure_MR_relation}) have very restricted mobility, electrons are free to flow constituting the main source of conductivity. Therefore describing the magnetic field evolution in the crust is more straightforward as the electrons are the only charged particles affecting it and the microphysical properties of matter are relatively well known. On the other hand, describing the field geometry and evolution in the liquid core is more complex. Due to the extreme conditions prevalent within the core, the exact composition and the properties of matter are still uncertain. In the fluid core, besides electrons, also neutrons, protons, muons and possibly more exotic particles could influence the magnetic field evolution. In particular it is believed that neutron superfluidity and proton superconductivity can play a critical role for the internal dynamics \citep[see for example][]{Chamel2017, Wood2022}. In general, for the core, it is very challenging to accurately incorporate realistic microphysics and estimate the impact of these effects on the magnetic field's evolution \citep{Vigano2021}. Here we will focus instead on the evolution of the magnetic field which is confined in the neutron star crust.
In general, the equation governing the magnetic field evolution is Maxwell's induction equation:
%--------------------------------------------------------------
\begin{align} \label{eq:ch1_induction_eq_short}
\frac{\partial \boldsymbol{B}}{\partial t} = - c \left( \boldsymbol{\nabla} \times \boldsymbol{E} \right).
\end{align}
%--------------------------------------------------------------
For a conducting fluid of electrons, moving with velocity $\boldsymbol{v}_e$ and in the presence of a magnetic field $\boldsymbol{B}$, the electric field $\boldsymbol{E}$ is related to the current density $\boldsymbol{j}$ through the generalised Ohm's law:
%--------------------------------------------------------------
\begin{align} \label{eq:ch1_E_ohm_law}
\boldsymbol{E} = \frac{\boldsymbol{j}}{\sigma_e} - \frac{\boldsymbol{v_e}}{c} \times \boldsymbol{B},
\end{align}
%--------------------------------------------------------------
where $\sigma_e$ is the electrical conductivity that can be defined in terms of the mean free time between collisions or relaxation time $\tau_e$ \citep{Longair2011}:
%--------------------------------------------------------------
\begin{align}
\sigma_e = \frac{e^2 n_e \tau_e}{m_e}.
\end{align}
%--------------------------------------------------------------
The electron velocity is also related to the current and electron density $n_e$ trough:
%--------------------------------------------------------------
\begin{align} \label{eq:ch1_e_velocity}
\boldsymbol{v}_e = - \frac{\boldsymbol{j}}{e n_e}.
\end{align}
%--------------------------------------------------------------
We can also assume that the timescale for collisions inside the plasma is much shorter than the timescale of the variation of the electromagnetic field so that we can neglect the displacement current term in Ampere's law and express the current density as:
%--------------------------------------------------------------
\begin{align} \label{eq:ch1_current_density}
\boldsymbol{j} = - \frac{c}{4 \pi} \left( \boldsymbol{\nabla} \times \boldsymbol{B} \right).
\end{align}
%--------------------------------------------------------------
Therefore by combining the four equations above, the induction equation assumes the form \citep{Pons2007, Vigano2012}:
%--------------------------------------------------------------
\begin{align} \label{eq:ch1_induction_eq}
\frac{\partial \boldsymbol{B}}{\partial t} = - \boldsymbol{\nabla} \times \left( \frac{c^2}{4 \pi \sigma_e} \boldsymbol{\nabla} \times \boldsymbol{B} + \frac{c}{4 \pi e n_e} (\boldsymbol{\nabla} \times \boldsymbol{B}) \times \boldsymbol{B} \right).
\end{align}
%--------------------------------------------------------------

The first term in the parentheses describes the Ohmic dissipation. It is responsible for converting magnetic energy into heat according to $Q = j^2 / \sigma_e$ which is then deposited in the crust. In a neutron star crust, the conductivity $\sigma_e$ is dominated by the electronic transport and depends on the crustal temperature, the electron density, and the impurity concentration. To quantify the impurity content of the
lattice we define the impurity parameter as the mean quadratic deviation of the atomic number $Z$:
%--------------------------------------------------------------
\begin{align} \label{eq:ch1_impurity}
\mathcal{Q}_{\rm imp} = \sum_i Y_i \left( Z_i^2 - \langle Z^2 \rangle \right),
\end{align}
%--------------------------------------------------------------
where $Y_i$ is the relative abundance of the nuclide with $Z_i e$ charge, and $\langle Z^2 \rangle$ is the average squared atomic number in the lattice \citep[see][and reference therein]{Potekhin2015}.
In general, for higher temperatures, and for larger impurity concentration the conductivity is lower as there is more scattering between the electrons and the lattice.
As a result the magnetic field evolution is highly coupled to the temperature evolution. The higher the magnetic field, the greater the dissipation and the heat deposited into the crust. In this way the stellar surface is kept hot for longer times. This in turn affects the microphysical properties of the crust and the electrical conductivity so that the heat deposition itself influences how the magnetic field dissipates.
Furthermore, the electron density varies over about four orders of magnitude in the crust. As a consequence, the electric conductivity varies by many orders of magnitude across the neutron star crust as well. What drives the evolution is usually the resistivity of the region where currents are predominantly placed. For example, the dissipation will be greater if currents are mainly circulating in the pasta layer of the crust. Here the transport properties are uncertain but the irregularities in the nuclear matter shapes and charge distribution should enhance the resistivity and therefore the magnetic field dissipation \citep{Horowitz2015, Nandi2018}.

The second term in Equation~\ref{eq:ch1_induction_eq} is a non-linear term which strongly depends on the magnetic field strength and is related to the Hall drift.  It does not cause dissipation, but it is responsible for the drift and rearrangement of the electric currents inside the crust and the redistribution of the magnetic energy on smaller scales. By affecting the distribution and scale of the electric currents the Hall effect is responsible of enhancing the Ohmic dissipation. As this term depends only on the degenerate electron density it should be insensitive to temperature \citep[see also][]{Cumming2004}.

%The magnetic field geometry and evolution are also important to determine how and where the heat is deposited into the crust. Indeed the presence of the magnetic field makes the thermal conductivity anisotropic suppressing the heat transport across the magnetic field lines. This creates hot spots and inhomogeneities in the surface temperature distribution \citep{Gourgouliatos2018}.  

If one is concerned only with the evolution of the magnitude of the magnetic field in time, the induction equation can be rewritten in an approximated form as \citep[see][]{Aguilera2008b, Aguilera2008a}:
%-----------------------------------------------------------------
\begin{align}	\label{eq:ch1_Bdot}
\frac{{\rm d} B}{{\rm d} t} &\simeq - \frac{c^2 B}{4 \pi \sigma L^2} - \frac{c B^2}{4 \pi e n_e L^2} \\ \nonumber
&= - \frac{B}{\tau_{\rm Ohm}} - \frac{B^2}{B_0 \tau_{\rm Hall,0}},
\end{align}  
%-----------------------------------------------------------------      
where the spatial derivatives have been replaced by a factor $1/L$ where $L$ is a typical length scale over which the physical quantities $B$, $n_e$ and $\sigma_e$ change. We can also define two typical timescales, one for the Ohmic dissipation, and the other for the Hall drift:
%----------------------------------------------------
\begin{align} \nonumber
\tau_{\rm Hall,0} &= \frac{4 \pi e n_e L^2}{c B_0} \\ 
&\simeq \unit[6.4 \times 10^4]{yr} \left( \frac{n_e}{\unit[10^{35}]{cm^{-3}}} \right)
\left( \frac{L}{\unit[1]{km}} \right)^2 
\left( \frac{B_0}{\unit[10^{14}]{G}} \right)^{-1}, 
\label{eq:ch1_tau_hall} \\
\tau_{\rm Ohm} &= \frac{4 \pi \sigma L^2}{c^2} \nonumber \\ 
&\simeq \unit[4.4 \times 10^6]{yr} \left( \frac{\sigma}{\unit[10^{24}]{s^{-1}}} \right)
\left( \frac{L}{\unit[1]{km}} \right)^2, 
\label{eq:ch1_tau_ohm}
\end{align}
%----------------------------------------------------    
where the numerical values are taken from \citet{Cumming2004}. Note that the Hall timescale defined here is inversely proportional to the initial magnetic field magnitude $B_0$. 
By neglecting the influence of the temperature we find the following analytical solution for Equation~\eqref{eq:ch1_Bdot}:
%----------------------------------------------------
\begin{align}
B(t) = B_0 \frac{e^{-t/\tau_{\rm Ohm}}}{1 + \frac{\tau_{\rm Ohm}}{\tau_{\rm Hall,0}} \left[ 1 - e^{-t/\tau_{\rm Ohm}} \right]}. 
\label{eq:ch1_B_decay}
\end{align}
%----------------------------------------------------
Even if it is a very crude approximation, this equation captures a first stage that is characterised by rapid (non-exponential) decay regulated by the Hall timescale $\tau_{\rm Hall,0}$, and a second stage that is characterised by exponential decay due to Ohmic dissipation and regulated by the timescale $\tau_{\rm Ohm}$ (see left panel in Figure~\ref{fig:ch1_B_evolution}).

To model the magnetic field evolution and decay in neutron stars consistently, a complete magneto-thermal simulation is required that takes into account the coupled magnetic field and temperature evolution. Moreover inside the neutron star general relativistic effects become relevant. Therefore the induction equation \eqref{eq:ch1_induction_eq} has to be corrected to take into account the curvature of space-time. Generally the shwarzschild static metric is sufficient as any deformation induced by rotation and the magnetic fields are negligible. In recent years numerical codes able to simulate the magnetic field and temperature evolution of neutron stars have been developed. In particular large efforts have been put into 2D axis-symmetric simulations \citep[see for example][]{Aguilera2008a, Pons2009, Vigano2012, Vigano2021}. More recently the more complex but more realistic 3D case is being explored \citep[see][]{Gourgouliatos2016, DeGrandis2020, Dehman2023}. The right panel of Figure~\ref{fig:ch1_B_evolution} shows the magnetic field evolution curves predicted by 2D magneto-thermal simulations \citep{Vigano2021}.

%--------------------------------------------------------------------------------------------

\section{The magnetosphere} \label{sec:ch1_magnetosphere}

%----------------------------------------------------
\begin{figure}
	\centering
	\includegraphics[width=0.8\textwidth]{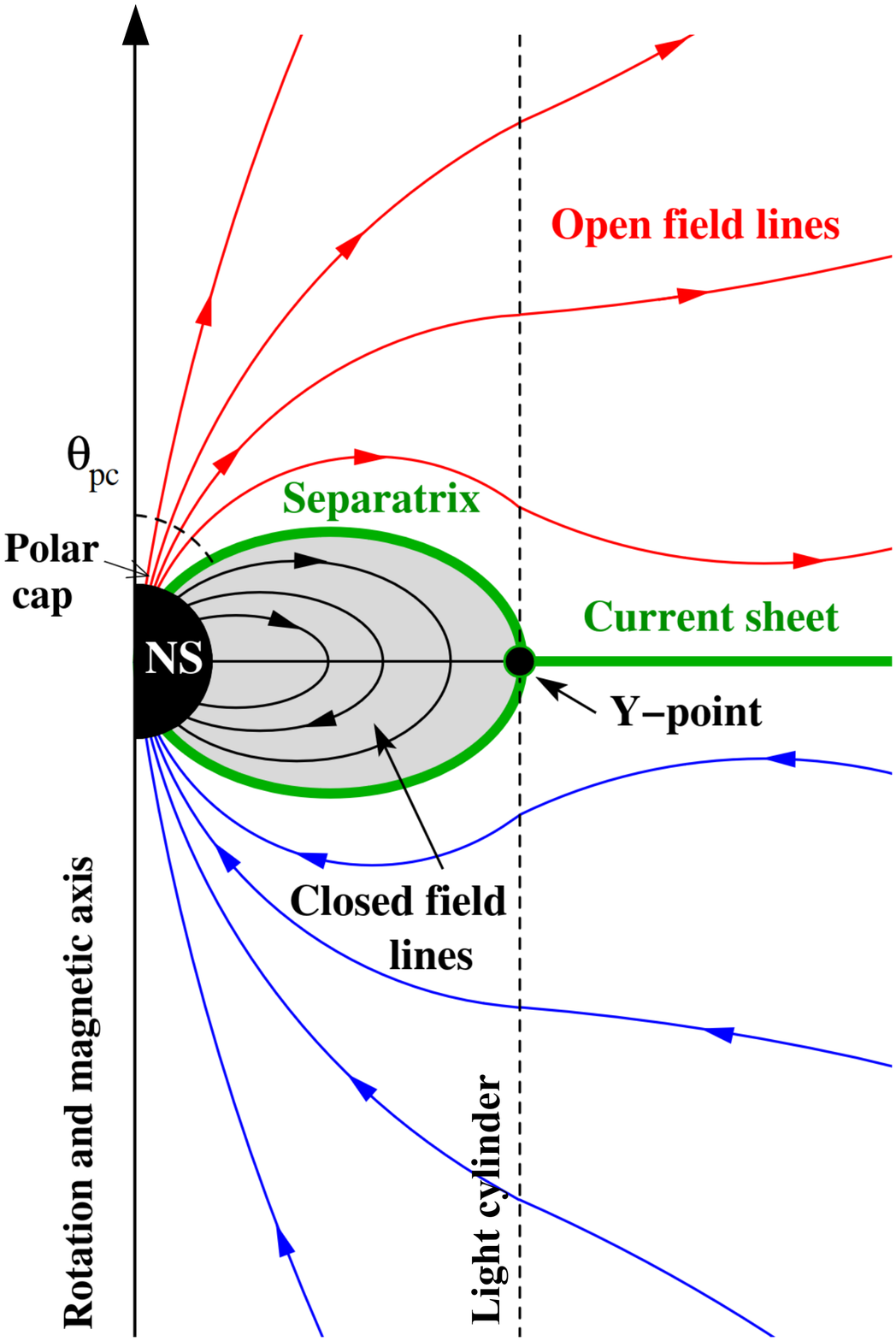}
	\caption[Sketch of the ideal force-free magnetosphere]{Sketch of the ideal force-free magnetosphere of the aligned pulsar. The main elements are: (i) The closed field line region (grey, and black field lines) lying between the star surface and the light cylinder. This zone is dead and does not participate to the pulsar activity. (ii) The open field line region (red and blue field lines) extending beyond the light cylinder. As particles are accelerated along the open field line region they emit electromagnetic radiation. (iii) The equatorial current sheet (green) between the opposite magnetic fluxes in the open field line region. It splits at the light cylinder into two separatrix current sheets that go around the closed zone, between the last open and the first closed field lines \citep[see][for further details]{Cerutti2017} \citep[Figure adapted from][]{Cerutti2017}.}
	\label{fig:ch1_sketch_magnetosphere}
\end{figure}
%----------------------------------------------------

%----------------------------------------------------
\begin{figure}
	\centering
	\includegraphics[width=0.8\textwidth]{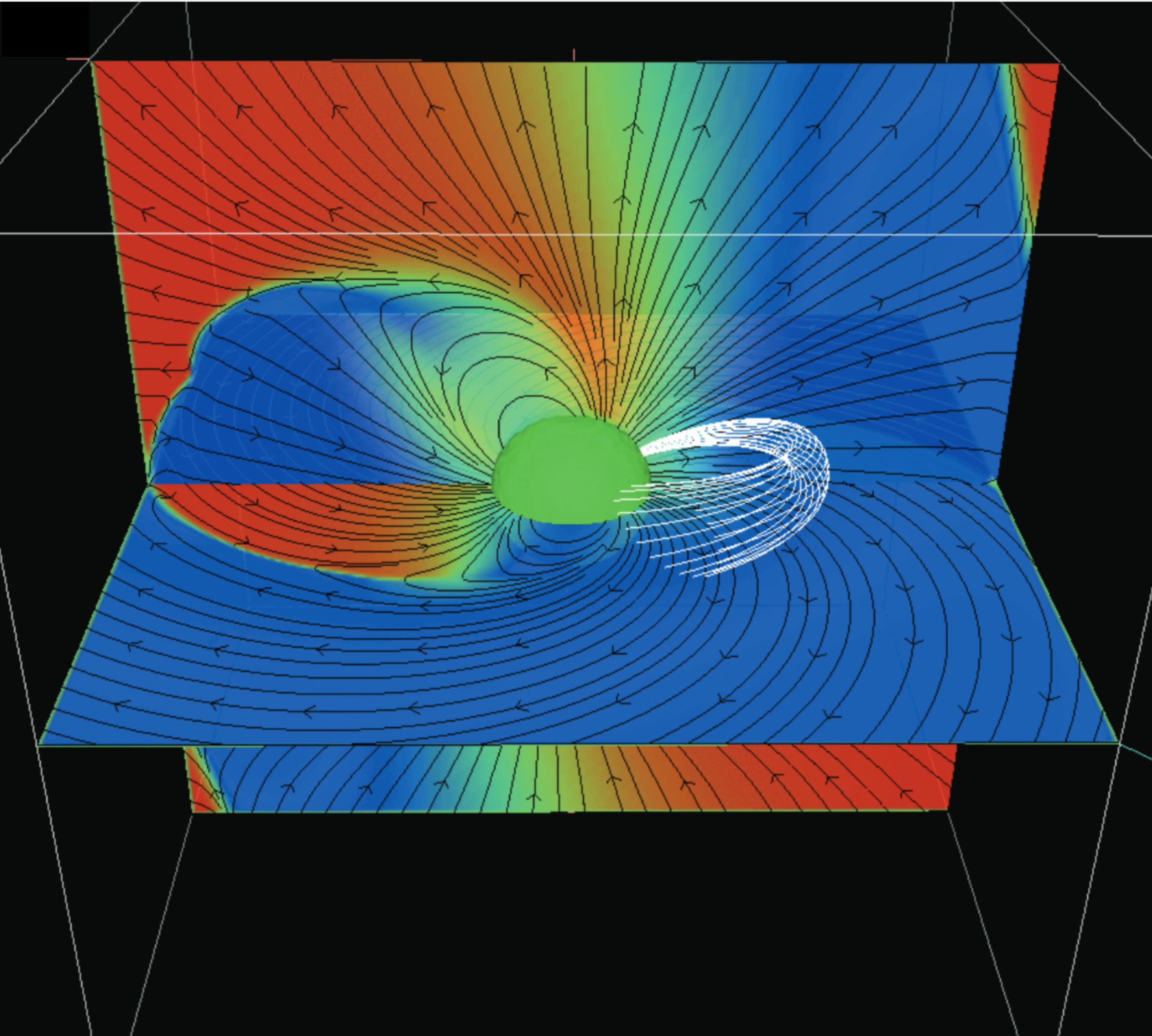}
	\caption[Simulated 3D force-free magnetosphere]{Slices through the simulated force-free magnetosphere of a neutron star with inclination angle of $60^{\circ}$ between the magnetic dipole moment and the rotation axis. The field lines in the horizontal and vertical slices are shown; colour represents the magnetic field component perpendicular to the slices. A sample three-dimensional magnetic flux tube is traced in white \citep[Figure taken from][]{Spitkovsky2006}.}
	\label{fig:ch1_3D_magnetosphere}
\end{figure}
%----------------------------------------------------

The magnetosphere of a neutron star is a complex and dynamic region filled with magnetised plasma where strong magnetic fields interact with the surrounding environment. It envelopes the entire star extending from the stellar surface up to several thousands of kilometers. Unlike Earth's magnetosphere, which is primarily influenced by the interaction with the solar wind, neutron star magnetospheres can encounter various external influences, including the interstellar medium, accretion disks, and even companion stars in binary systems.
\citet{Deutsch1955} was the first to study analytically the structure of the electromagnetic field surrounding an isolated rotating magnetised star with a misaligned magnetic dipole moment in a vacuum. Later, \citet{Goldreich1969} described the pulsar electrodynamics in the simplest case of a rotating magnetic dipole, aligned with the rotational axis, surrounded by a charge-separated plasma.
In general analytical solutions for the magnetospheric structure are possible only in very simplified and ideal cases. To fully study the overall geometry of the electromagnetic fields and how the surrounding plasma interacts with and influences them, numerical simulations are required. \citet{Contopoulos1999} were the first to present a numerical, realistic configuration for an aligned rotator in the force-free approximation (see below). His work was the first step towards more realistic 3D simulations that were able to quantitatively consider the effects of the misalignment between rotation and magnetic axes \citep[see Figure~\ref{fig:ch1_3D_magnetosphere} and][]{Spitkovsky2006, Philippov2014}. 
Understanding the structure and dynamics of neutron star magnetospheres is crucial for comprehending a wide range of phenomena associated with these celestial objects. In particular, the intricate interplay between the magnetic field and the highly energetic particles in the surrounding plasma leads to the generation of electromagnetic radiation across different wavelengths from radio to gamma rays. For more detailed reviews about pulsar magnetosphere see for example \citet{Petri2016, Petri2020}.

To gain an insight into the magnetosphere's structure we consider a neutron star initially surrounded by vacuum spinning with angular frequency $\omega = 2 \pi / P$. The neutron star is endowed with a strong magnetic field. However to define this magnetic field we need to make a distinction between two reference frames: an inertial reference frame of an external observer and a reference frame co-rotating with the star. For simplicity we consider the latter to be sufficiently small compared to the star's size and not so far from the axis of rotation as to be considered locally inertial. The position of the co-rotating frame can be described by a distance vector, $\boldsymbol{r}$, originating $\boldsymbol{r}$ at the centre of the neutron star. With these considerations, the electromagnetic fields as measured by the two observers can be linked by the Lorentz transformations \citep{Jackson1998}:
%--------------------------------------------------------------
\begin{align} \label{eq:ch1_lorentz_transf_B}
\boldsymbol{B} = \gamma_{\rm rot} \left(\boldsymbol{B}' + \boldsymbol{\beta}_{\rm rot} \times \boldsymbol{E}' \right) - \frac{\gamma_{\rm rot}^2}{\gamma_{\rm rot} + 1} \boldsymbol{\beta}_{\rm rot} (\boldsymbol{\beta}_{\rm rot} \cdot \boldsymbol{B}'),
\end{align}
%--------------------------------------------------------------
\begin{align} \label{eq:ch1_lorentz_transf_E}
\boldsymbol{E} = \gamma_{\rm rot} \left(\boldsymbol{E}' - \boldsymbol{\beta}_{\rm rot} \times \boldsymbol{B}' \right) - \frac{\gamma_{\rm rot}^2}{\gamma_{\rm rot} + 1} \boldsymbol{\beta}_{\rm rot} (\boldsymbol{\beta}_{\rm rot} \cdot \boldsymbol{E}'),
\end{align}
%--------------------------------------------------------------
where $\boldsymbol{E}$ and $\boldsymbol{B}$ are the fields measured by an external observer, while $\boldsymbol{E}'$ and $\boldsymbol{B}'$ are the fields measured in the co-rotating frame, $\boldsymbol{\beta}_{\rm rot} = (\boldsymbol{\omega} \times \boldsymbol{r}) / c$ is the ratio of the rotational velocity to the speed of light $c$ and $\gamma_{\rm rot} = 1 / (1 - \beta_{\rm rot}^2)^{1/2}$ is the corresponding Lorentz factor. 
If we consider the neutron star interior or we stay in the proximity of its surface, for typical observed spin periods we have that $\beta_{\rm rot} \equiv |\boldsymbol{\beta}_{\rm rot}| \ll 1$. Therefore the transformations above assume the simplified form:
%--------------------------------------------------------------
\begin{align} \label{eq:ch1_lorentz_transf_B_1}
\boldsymbol{B} \sim \left(\boldsymbol{B}' + \boldsymbol{\beta}_{\rm rot} \times \boldsymbol{E}' \right),
\end{align}
%--------------------------------------------------------------
\begin{align} \label{eq:ch1_lorentz_transf_E_1}
\boldsymbol{E} \sim \left(\boldsymbol{E}' - \boldsymbol{\beta}_{\rm rot} \times \boldsymbol{B}' \right),
\end{align}
%--------------------------------------------------------------
where we neglected second order terms in $\beta_{\rm rot}$ and considered $\gamma_{\rm rot} \sim 1$.

Little is known about the magnetic field geometry inside the neutron star. In the exterior, the magnetic field is typically described by a combination of multipolar components. As the higher-order multipoles decay faster with distance from the stellar surface, the dipolar component probably dominates the magnetic field geometry at a large scale. For the sake of simplicity we consider that in the neutron star's co-rotating frame the magnetic field is purely dipolar.
In this case the formula that relates the magnetic moment $\boldsymbol{\mu}$ with the strength of the magnetic field $\boldsymbol{B}'$ is given by:
%-----------------------------------------------------------------
\begin{align}	\label{eq:ch1_B_dipole}
\boldsymbol{B}'(\boldsymbol{r}) = \frac{3 (\boldsymbol{r} \cdot \boldsymbol{\mu} ) \boldsymbol{r}}{r^5} - \frac{ \boldsymbol{\mu} }{r^3}. 
\end{align}  
%-----------------------------------------------------------------
We further simplify by assuming a configuration with the magnetic moment $\boldsymbol{\mu}$ aligned with the rotation axis of the star. We will call this an aligned rotator.
In spherical coordinates $(r, \theta, \phi)$ with $\theta$ being the angle computed from the magnetic moment axis (or spin axis in the aligned case) and $\phi$ the azimuthal angle, the dipolar magnetic field can be expressed as:
%--------------------------------------------------------------
\begin{align} \label{eq:ch1_B_dipole_sph_coord}
\boldsymbol{B}'(r, \theta) = \frac{\mu}{r^3} \begin{bmatrix}
2 \cos{\theta} \\
\sin{\theta} \\
0
\end{bmatrix}.
\end{align}
%--------------------------------------------------------------

Due to the abundance of free charges in its interior, as a first approximation a neutron star can be considered as a perfect conductor. Therefore in the co-rotating frame, the internal electric field has to vanish so that in the neutron star interior the following equations are valid:
%--------------------------------------------------------------
\begin{align} \label{eq:ch1_lorentz_transf_B_2}
\boldsymbol{B} \sim \boldsymbol{B}',
\end{align}
%--------------------------------------------------------------
\begin{align} \label{eq:ch1_lorentz_transf_E_2}
\boldsymbol{E} \sim - \boldsymbol{\beta}_{\rm rot} \times \boldsymbol{B}' = - \frac{\boldsymbol{\omega} \times \boldsymbol{r}}{c} \times \boldsymbol{B}'.
\end{align}
%--------------------------------------------------------------
In other words, a perfect conductor rotating in a magnetic field behaves as a unipolar inductor converting the rotational energy into electrical energy. As seen by an inertial external observer the rotation of the star induces an electric field inside the star and the free charges inside rearrange themselves under its action. In particular, on the surface of the star particles of opposite charges are driven to the polar and equatorial regions respectively according to their signs. This creates a quadrupolar electric potential, $V$, on the surface that can be derived from $\boldsymbol{E} = - \boldsymbol{\nabla} V$. 

In particular, for an aligned rotator, from Equation~\eqref{eq:ch1_lorentz_transf_E_2}, the $\theta$ component of the electric field computed at the surface is given by:
%--------------------------------------------------------------
\begin{align} \label{eq:ch1_E_theta_surf}
E_{\theta} = - \frac{\omega R_{\rm NS} \sin{\theta} B_r}{c}.
\end{align}
%--------------------------------------------------------------
This determines the electrostatic potential drop with latitude on the stellar surface that can be computed considering that $E_{\theta} = 1/r (\partial V / \partial \theta) $ and integrating over the coordinate $\theta$. 
By using the radial component of the surface dipole magnetic field in Equation~\eqref{eq:ch1_B_dipole_sph_coord}, $B_r = 2 \mu \cos{\theta} / R_{\rm NS}^3$, we find \citep{Backus1956, Goldreich1969, Cerutti2017}:
%--------------------------------------------------------------
\begin{align} \label{eq:ch1_unipolar_potential}
V(R_{\rm NS}, \theta) - V(R_{\rm NS}, 0) = \int_0^\theta E_{\theta} R_{\rm NS} {\rm d}\theta = \frac{\omega \mu}{2 R_{\rm NS} c} \sin^2{\theta},
\end{align}
%--------------------------------------------------------------
where $V(R_{\rm NS}, 0) \equiv V_{\rm p}$ is the value of the electric potential at the pole.
This charge redistribution induces an external quadrupolar electric field with a strong component parallel to the magnetic field lines. As shown by \citet{Goldreich1969} this component of the electric field is strong enough to overcome gravity and pull out charged particles from the neutron star surface unless strong cohesive forces prevent them from escaping. This happens primarily for the electrons and protons as, in the presence of a strong magnetic field, heavier ions in the surface layer can rearrange themselves to form a dense and strongly bound molecular lattice \citep[see][and reference therein]{Ruderman1975}. Even if considered as a simplified and ideal case, the result from \citet{Goldreich1969} implies that a vacuum region around a neutron star cannot be maintained and a magnetosphere filled with conducting and co-rotating plasma has to surround the star.

As charges are extracted from the surface they are accelerated up to relativistic speeds along the magnetic field lines by the induced electric field. By moving along curved field lines these charges produce gamma-ray photons by curvature radiation \citep{Ruderman1975} or the inverse-Compton process by up-scattering lower energy photons \citep{Daugherty1986}. In the presence of strong magnetic fields gamma-ray photons can transform into electron-positron pairs, if the photon energy exceeds the threshold $E_{\gamma} \geq 2 m_e c^2$ where $m_e$ is the electron mass. This generation of new particles then produces new photons and pairs leading to a pair cascade that populates the magnetosphere with co-rotating highly conducting plasma. Assuming a continuous supply of charges, as the magnetosphere is replenished with perfectly conducting plasma (predominantly formed by $e^{-} \, e^{+}$ pairs) the electric field component along the magnetic field lines gets screened so that in a steady state one expects the condition $\boldsymbol{E} \cdot \boldsymbol{B} = 0$ to be satisfied. This prevents any further acceleration of particles along the field lines so that these can be considered equipotential lines. In this ideal case, called force-free limit, the magnetosphere is sufficiently populated with plasma for the conductivity in the medium to become infinite. Moreover, the electromagnetic field dominates the dynamics of the magnetosphere by several orders of magnitude with respect to pressure, gravity and inertia.
The unipolar induction mechanism on the other hand induces an electric field perpendicular to the magnetic field lines that leads to charge separation also in the magnetosphere. 

However, the flow of plasma in the magnetosphere has a retroactive effect on the electromagnetic field structure around the neutron star.  
As we move further away from the neutron star surface, the approximations and the electromagnetic field configuration outlined above break down. As the rotational velocity approaches the speed of light plasma co-rotation becomes impossible and a co-rotating frame cannot be defined anymore. We define an imaginary cylindrical surface called light cylinder, with the axis parallel to the rotation axis of the star. The radius of the light cylinder corresponds to the distance where the co-rotating speed is equal to the speed of light:
%--------------------------------------------------------------
\begin{align} \label{eq:ch1_light_cylinder}
r_{\rm lc} = \frac{c}{\omega} \simeq \unit[4.77 \times 10^9]{cm} \left( \frac{P}{\unit[1]{s}} \right).
\end{align}
%--------------------------------------------------------------

Near the light cylinder plasma inertia due to the approaching of the speed of light becomes relevant and the plasma is not able to stay in co-rotation. In particular, the plasma drags the magnetic field causing the field lines to sweep back in the azimuthal direction and spiral around the star (see Figure~\ref{fig:ch1_3D_magnetosphere}). This produces a toroidal magnetic field component. Furthermore, the co-rotating charges and these inertial effects create currents which modify the magnetic field configuration. These magnetic perturbations tend to repel field lines and as a result, the field lines near the light cylinder inflate to infinity and open up. In the vicinity of the light cylinder, the force-free condition is violated and dissipation effects and particle acceleration create electromagnetic emission detectable on Earth. 
The magnetosphere can then be divided into two regions with different properties (see Figure~\ref{fig:ch1_sketch_magnetosphere}). The magnetic field lines that do not cross the light cylinder form the so-called closed magnetosphere (grey region in Figure~\ref{fig:ch1_sketch_magnetosphere}). Here the plasma is trapped and the condition $\boldsymbol{E} \cdot \boldsymbol{B} = 0$ is valid everywhere. This region should therefore be electromagnetically inactive as particle acceleration is suppressed.
The field lines that cross the light cylinder and are inflated to infinity form the open field line region (red and blue lines in Figure~\ref{fig:ch1_sketch_magnetosphere}). Here charges are free to stream along the magnetic field lines and escape to infinity. Therefore here the condition $\boldsymbol{E} \cdot \boldsymbol{B} = 0$ can be violated and gaps with low-density plasma can be created where charged particles are accelerated. The open field line region is therefore active and it is the region where electromagnetic emission from pulsars is believed to be produced. 

%=================================================================

\section{Dipolar spin-down evolution} \label{sec:ch1_dip_spindown_evol}

The observation of the secular increase of spin periods for the detected neutron stars implies that they are losing rotational energy over time. Assuming that neutron stars are solid bodies the loss of rotational energy can be computed as:
%--------------------------------------------------------------
\begin{align} \label{eq:ch1_rotational_energy_loss}
\frac{{\rm d} E_{\rm rot}}{{\rm d} t} &= \frac{{\rm d}}{{\rm d}t} \left( \frac{1}{2} I_{\rm NS} \omega^2 \right) = I_{\rm NS} \omega \frac{{\rm d} \omega}{{\rm d}t} = - 4 \pi^2 I_{\rm NS} \frac{{\rm d}P}{{\rm d}t} \frac{1}{P^3} \nonumber \\
& \simeq - \unit[5 \times 10^{31}]{erg \, s^{-1}} \left( \frac{\dot P}{\unit[10^{-15}]{s \, s^{-1}}} \right) \left( \frac{P}{\unit[1]{s}} \right)^{-3},
\end{align}
%--------------------------------------------------------------
where we define the spin-down rate $\dot{P} \equiv {\rm d}P / {\rm d}t$ and assume a neutron star to be a perfect sphere of mass $M_{\rm NS} = 1.4$ \msun and radius $R_{\rm NS} = \unit[11]{km}$ and with moment of inertia $I_{\rm NS} \simeq 2/5 M_{\rm NS} R_{\rm NS}^2 \simeq \unit[1.4 \times 10^{45}]{g \, cm^2}$ which is constant in time. 

If a neutron star can be simply pictured as a rotating magnetic dipole with moment vector $\boldsymbol{\mu}$ misaligned by an angle $\chi$ with respect to the rotation axis, the spin-down can be explained using classical electrodynamics arguments \citep[see for example][]{Shapiro1983}.
In classical electrodynamics a rotating misaligned magnetic dipole emits electromagnetic energy \citep[see][]{Jackson1998}. In analogy with the Larmor formula for an accelerated charge, a time-varying magnetic dipole radiates away a total power integrated over the solid angle equal to: 
%-----------------------------------------------------------------
\begin{align}	\label{eq:ch1_mag_dipole_radiation}
\frac{{\rm d} E}{{\rm d} t} = \frac{2}{3 c^3} \left| \frac{ {\rm d}^2 \boldsymbol{\mu}}{{\rm d}t^2} \right|^2.
\end{align}  
%-----------------------------------------------------------------
For a misaligned magnetic dipole with an inclination angle $\chi$, rotating at an angular frequency $\omega$, we can write the magnetic dipole moment components as a function of time $t$ in Cartesian coordinates. We assume rotation about the $z$ axis:
%-----------------------------------------------------------------
\begin{align}	\label{eq:ch1_mag_moment_cart}
\boldsymbol{\mu} = \mu \begin{bmatrix}
\sin{\chi} \cos{\omega t} \\
\sin{\chi} \sin{\omega t} \\
\cos{\chi}
\end{bmatrix}.
\end{align}  
%-----------------------------------------------------------------
where $\mu = |\boldsymbol{\mu}|$.
If we compute the second time derivative for this vector we find:
%-----------------------------------------------------------------
\begin{align}	\label{eq:ch1_2der_mag_moment_cart}
\frac{ {\rm d}^2 \boldsymbol{\mu}}{{\rm d}t^2} = \mu \begin{bmatrix}
- \omega^2 \sin{\chi} \cos{\omega t} \\
- \omega^2 \sin{\chi} \sin{\omega t} \\
0
\end{bmatrix}.
\end{align}  
%-----------------------------------------------------------------
Here we have assumed that the spin frequency, the magnetic moment and the inclination angle change on timescales much longer than a rotation period so that they can be assumed as constants in computing these derivatives, i.e., they do not explicitly depend on the time $t$.
By computing the squared module of the vector above and substituting it into Equation~\eqref{eq:ch1_mag_dipole_radiation} we get an emitted power:
%-----------------------------------------------------------------
\begin{align}	\label{eq:ch1_mag_dipole_radiation_2}
\frac{{\rm d} {E}}{{\rm d} t} = \frac{2 \mu^2}{3 c^3} \omega^4 \sin^2{\chi}. 
\end{align}  
%-----------------------------------------------------------------
By assuming that this electromagnetic energy is emitted at the expense of the rotational energy we can equate Equation~\eqref{eq:ch1_rotational_energy_loss} with Equation~\eqref{eq:ch1_mag_dipole_radiation_2} to obtain an equation that describes how the spin frequency changes in time:
%--------------------------------------------------------------
\begin{align} \label{eq:ch1_omegadot_vacuum}
\frac{{\rm d} \omega}{{\rm d}t} = -\frac{2 \mu^2}{3 c^3 I_{\rm NS}} \omega^3 \sin^2{\chi}. 
\end{align}
%--------------------------------------------------------------

We find an analogous result by considering a more sophisticated calculation. By picturing the neutron star as a spherical conductor rotating in an electromagnetic field in vacuum it is possible to compute the net torque acting on it. In particular the electric and magnetic fields generate forces on the charges in the conductor which in turn generate a net torque acting on the conductor surface. This torque, $\boldsymbol{M}$, can be estimated by the following integral computed over the neutron star surface $S$ \citep[see][]{Michel1970, Philippov2014}: 
%--------------------------------------------------------------
\begin{align} \label{eq:ch1_torque_em}
\boldsymbol{M} = \int_S \boldsymbol{n} \cdot \left( \boldsymbol{r} \times \boldsymbol{T} \right) {\rm d} S,  
\end{align}
%--------------------------------------------------------------
where $\boldsymbol{r}$ is the radial vector originating in the neutron star centre, $\boldsymbol{T}$ is the Maxwell stress tensor and $\boldsymbol{n}$ is the unit vector perpendicular to the surface. By considering the expressions for the electromagnetic fields near the stellar surface computed by \citet{Deutsch1955} for a rotating star characterised as a misaligned magnetic dipole in vacuum, \citet{Michel1970} computed the non-vanishing torque components. The first one is antiparallel to the spin frequency vector and causes the neutron star to spin down:
%--------------------------------------------------------------
\begin{align} \label{eq:ch1_torque_em_parallel}
M_{\parallel} =  - \frac{2 \mu^2}{3 c^3} \omega^3 \sin^2{\chi}. 
\end{align}
%--------------------------------------------------------------
The second one is perpendicular to the spin frequency vector and acts in the plane defined by the two vectors $\boldsymbol{\omega}$ and $\boldsymbol{\mu}$. While it does not change the spin frequency, it is responsible for the alignment of the magnetic axis with the rotational axis:
%--------------------------------------------------------------
\begin{align} \label{eq:ch1_torque_em_perpendicular}
M_{\perp} =  - \frac{2 \mu^2}{3 c^3} \omega^3 \sin{\chi} \cos{\chi}.
\end{align}
%--------------------------------------------------------------
A third component is perpendicular to the $\boldsymbol{\omega} - \boldsymbol{\mu}$ plane and causes precession of the spin frequency vector around the magnetic axis \citep[][]{Philippov2014}. In the following, we will not take into account this precession since it does not affect the evolution of the spin period and inclination angle. 
The evolution of the angular momentum vector, $\boldsymbol{L}$, of the neutron star is then given by:
%--------------------------------------------------------------
\begin{align} \label{eq:ch1_torque}
\frac{{\rm d} \boldsymbol{L}}{{\rm d} t} = I_{\rm NS} \frac{{\rm d} \boldsymbol{\omega}}{{\rm d} t} = \boldsymbol{M},  
\end{align}
%--------------------------------------------------------------
which gives the following coupled differential equations:
%--------------------------------------------------------------
\begin{align} \label{eq:ch1_rot_evol_vacuum}
\frac{{\rm d} \omega}{{\rm d} t} &= -\frac{2 \mu^2}{3 c^3 I_{\rm NS}} \omega^3 \sin^2{\chi}, \\
\frac{{\rm d} \chi}{{\rm d} t} &= - \frac{2 \mu^2}{3 c^3 I_{\rm NS}} \omega^2 \sin{\chi} \cos{\chi}.
\end{align}
%--------------------------------------------------------------
Note that the first equation is the same as Equation~\eqref{eq:ch1_omegadot_vacuum}.
In general, neutron stars evolves towards longer periods and a configuration where the magnetic dipole axis is aligned with the rotation axis. Therefore rotational evolution tends towards a state where the energy losses due to the torques are minimised as can be seen from Equation~\eqref{eq:ch1_mag_dipole_radiation_2}. However, these equations are valid in the case of a vacuum magnetosphere which as we have seen in Section~\ref{sec:ch1_magnetosphere} is not the most realistic scenario.

For plasma-filled magnetospheres, an analytical solution of the torques can not be derived and simulations are required due to the non-linearity of the problem. Through 3D simulations of force-free plasma-filled magnetospheres \citet{Spitkovsky2006} found that the emitted electromagnetic luminosity can be well described by the following relation:
%-----------------------------------------------------------------
\begin{align}	\label{eq:ch1_pulsar_lum_forcefree}
\left( \frac{{\rm d} {E}}{{\rm d} t} \right)_{\rm force-free} = \frac{\mu^2}{c^3} \omega^4 (\kappa_0 + \kappa_1 \sin^2{\chi}). 
\end{align}  
%-----------------------------------------------------------------
The rotational evolution of a neutron star can then be approximated by the two following equations:
%-----------------------------------------------------------------
\begin{align}	\label{eq:ch1_rot_evol_forcefree}
\frac{{\rm d} \omega}{{\rm d} t} &=  - \frac{\mu^2}{c^3 I_{\rm NS}} \omega^3 \left( \kappa_0 + \kappa_1 \sin^2 \chi \right), \\
\frac{{\rm d} \chi}{{\rm d} t} &=  - \frac{\mu^2}{c^3 I_{\rm NS}} \omega^2 \left( \kappa_2 \sin\chi \cos\chi \right),
\end{align}  
%-----------------------------------------------------------------  
and substituting $\omega = 2\pi/P$ we get: 
%--------------------------------------------------------------
\begin{align} \label{eq:ch1_rot_evol_forcefree_2}
\frac{{\rm d} P}{{\rm d}t} &= \frac{ 4 \pi^2 \mu^2}{ c^3 I_{\rm NS} P } \left( \kappa_0 + \kappa_1 \sin^2 \chi \right), \\
\frac{{\rm d} \chi}{{\rm d}t} &= - \frac{ 4 \pi^2 \mu^2}{ c^3 I_{\rm NS} P^2 } \left( \kappa_2 \sin\chi \cos\chi \right),
\end{align}
%--------------------------------------------------------------
The parameters $\kappa_0$, $\kappa_1$ and $\kappa_2$ are derived by fitting simulation results and for force-free magnetosphere they take the values $\kappa_0 \simeq \kappa_1 \simeq \kappa_2 \simeq 1$ \citep{Philippov2014}.
Compared to the vacuum solution (where $\kappa_0 = 0$, $\kappa_1 = 2/3$ and $\kappa_2 = 1$) the electromagnetic torque and the energy loss does not vanish when the magnetic dipole axis aligns with the rotation axis and the torque exerted on the star can be a factor $\sim 2$ stronger.

\subsection{Dipole magnetic field estimate}

By considering the magnetic dipole breaking model outlined above we can estimate the dipolar magnetic field of a neutron star from its timing properties $P$ and $\dot{P}$. 
If we consider the magnetic field strength at the magnetic pole of the neutron star at $r = R_{\rm NS}$, from Equation~\eqref{eq:ch1_B_dipole} we get the expression:
%-----------------------------------------------------------------
\begin{align}	\label{eq:ch1_mag_dipole_B_pole}
B_{\rm p} = \frac{2 \mu}{R_{\rm NS}^3} \; \Longrightarrow \; \mu = \frac{B_{\rm p} R_{\rm NS}^3}{2}.
\end{align}  
%-----------------------------------------------------------------
If we consider the magnetic field strength at the magnetic equator of the neutron star we obtain that it is half of the value at the magnetic pole:
%-----------------------------------------------------------------
\begin{align}	\label{eq:ch1_mag_dipole_B_equat}
B_{\rm e} = \frac{\mu}{R_{\rm NS}^3} \; \Longrightarrow \; \mu = B_{\rm e} R_{\rm NS}^3.
\end{align}  
%-----------------------------------------------------------------
We can substitute Equation~\eqref{eq:ch1_mag_dipole_B_pole} or Equation~\eqref{eq:ch1_mag_dipole_B_equat} into Equation~\eqref{eq:ch1_rot_evol_forcefree_2} to obtain a relation between the observed quantities $P$ and $\dot{P}$ and the magnetic field strength at the poles or at the equator.
In the literature, the vacuum magnetosphere model is often assumed even if less realistic. This usually implies an overestimation of the magnetic field strength by a factor of $\sim 2$ with respect to the force-free magnetosphere solution. By assuming for the sake of simplicity an inclination angle $\chi = \pi/2$ (since it is usually unknown) we obtain a formula to roughly estimate the strength of the magnetic field at the neutron star poles or equator as a function of the timing properties:
%--------------------------------------------------------------
\begin{align} \label{eq:ch1_Bpol_vacuum}
B_{\rm p} = \left( \frac{ 3 c^3 }{ 2 \pi^2 } \frac{  I_{\rm NS} P \dot{P} }{R_{\rm NS}^6} \right)^{1/2} \simeq \unit[2 \times 10^{12}]{G} \left( \frac{P}{\unit[1]{s}} \right)^{1/2} \left( \frac{\dot{P}}{\unit[10^{-15}]{s \, s^{-1}}} \right)^{1/2},
\end{align}
%--------------------------------------------------------------
where we consider a neutron star moment of inertia $I_{\rm NS} \simeq \unit[10^{45}]{g \, cm^2}$ and radius $R_{\rm NS} \simeq \unit[10^6]{cm}$.

\subsection{Characteristic age estimate}

By assuming that both the magnetic field and the inclination angle stay constant during the neutron star's life, one can integrate of Equation~\eqref{eq:ch1_rot_evol_forcefree} in time to obtain a characteristic age:
%--------------------------------------------------------------
\begin{align} \label{eq:ch1_characteristic_age}
\tau_{\rm c} = -\frac{\omega^3}{2\dot{\omega}} \left( \frac{1}{\omega^2} - \frac{1}{\omega_0^2} \right), 
\end{align}
%--------------------------------------------------------------
where $\omega_0$ is the initial spin frequency and $\omega$ and $\dot{\omega}$ are the current values of spin frequency and spin frequency derivative.
If we assume that $\omega_0 \gg \omega$, which means that the neutron star was rotating much faster at birth than at the current time, Equation~\eqref{eq:ch1_characteristic_age} becomes:
%--------------------------------------------------------------
\begin{align} \label{eq:ch1_characteristic_age_approx}
\tau_{\rm c} = -\frac{\omega}{2\dot{\omega}} = \frac{P}{2\dot{P}} \simeq \unit[1.6 \times 10^{7}]{yr} \left( \frac{P}{\unit[1]{s}} \right) \left( \frac{\dot{P}}{\unit[10^{-15}]{s \, s^{-1}}} \right)^{-1}.
\end{align}
%--------------------------------------------------------------

The characteristic age in general represents an upper limit on the true neutron star age. Indeed for very young pulsars the assumption that $\omega_0 \gg \omega$ could not hold as their spin period can be still very close to the value at birth. Furthermore, for older pulsars, if one takes into account that the magnetic field decays over time the characteristic age could exceed the real age of the neutron star by orders of magnitude. In these cases the characteristic age fails to give a good estimate for the actual age.
%--------------------------------------------------------------------------------------------

\section{Pulsar radio emission} \label{sec:ch1_radio_em_geometry}

As we have described in Section~\ref{sec:ch1_magnetosphere}, the magnetosphere surrounding a neutron star is divided into a closed field line and an open field line region. According to the standard theory \citep{Goldreich1969, Sturrock1971, Ruderman1975}, in the open field line region particles can freely stream along the field lines. If charges are not continuously extracted from the surface the outflow of particles eventually generates vacuum or low-density regions, so-called plasma gaps, above the magnetic poles. Inside these gaps the force-free condition $\boldsymbol{E} \cdot \boldsymbol{B} = 0$ can be violated as the electric field parallel to the magnetic field lines is not perfectly screened anymore. New particles can therefore be extracted from the stellar surface and accelerated up to relativistic speed leading to an avalanche of pair production. The mechanism producing the radio emission in pulsars is not well understood, but it is believed that in this process bunches of $e^+ \, e^-$ pairs produced in the gap oscillate coherently to produce radio emission at a certain distance from the polar cap see Section~\ref{sec:ch1_radio_em_deathlines}. 

\subsection{The polar cap model and radio emission geometry}

%----------------------------------------------------
\begin{figure}
	\centering
	\includegraphics[width=0.7\textwidth]{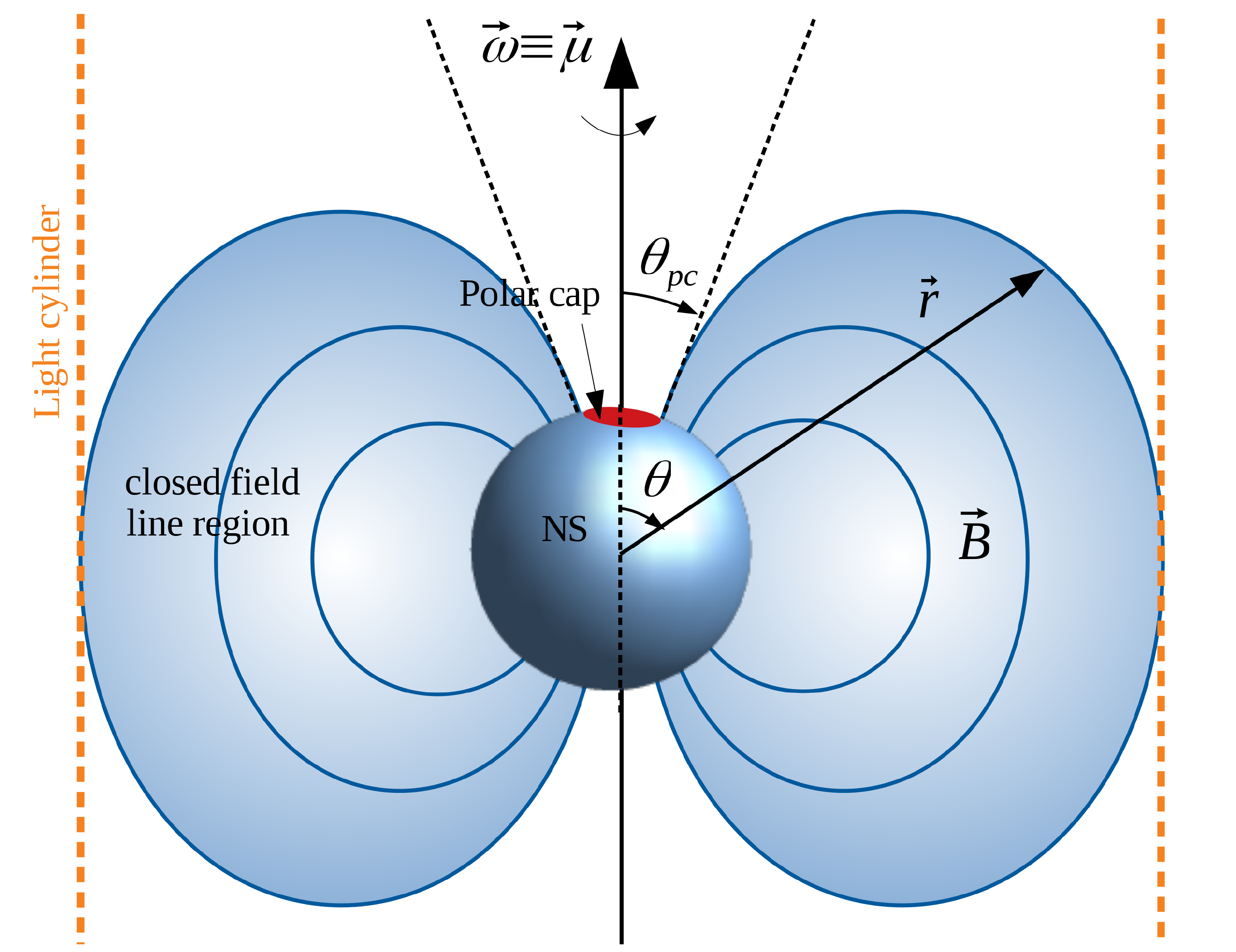}
	\caption[Geometry of the polar cap]{Sketch showing the geometry of the polar cap emission model for an aligned rotator. The polar cap region is delimited by the last closed field lines extending to the light cylinder. A magnetic field line can be parametrised as curve described by the coordinates $r$ and $\theta$ (see text for more details). }
	\label{fig:ch1_polar_cap}
\end{figure}
%----------------------------------------------------

The common picture is that this radio emission is generated in conical beams delimited by the open field lines around the magnetic poles. 
We can estimate the size of these conical beams by considering the size of the polar caps where the open field lines are anchored to the neutron star. The polar cap size depends on the amount of field lines that are opened which in turn depends on the size of the light cylinder (see Equation~\eqref{eq:ch1_light_cylinder}, i.e., on the spin period of the neutron star. To see this consider Figure~\ref{fig:ch1_polar_cap} showing a sketch of a neutron star with the magnetic dipole moment aligned to the rotation axis of the star. 
We parametrise a magnetic field line as a curve $\boldsymbol{r} = r(\theta) \hat{\boldsymbol{r}}$ in space, where $\theta$ is the angle between the rotation axis and the vector $\boldsymbol{r}$ whose origin is located in the star's centre and pointing to a random position on the field line. The magnetic field vector $\boldsymbol{B}$ is therefore proportional to the tangent vector of the curve given by $d\boldsymbol{r}/d\theta = (dr/d\theta) \hat{\boldsymbol{r}} + r \hat{\boldsymbol{\theta}}$, where $\hat{\boldsymbol{r}}$ and $\hat{\boldsymbol{\theta}}$ are the unit vectors in spherical coordinates.
By considering the expression of the magnetic dipole in spherical coordinates (Equation~\eqref{eq:ch1_B_dipole_sph_coord}) we obtain the following equations:
%--------------------------------------------------------------
\begin{align}
\frac{{\rm d} r}{{\rm d}\theta} &= \lambda(\theta) \frac{\mu}{r^3} 2 \cos(\theta), \\
r &= \lambda(\theta) \frac{\mu}{r^3} \sin(\theta), 
\end{align}
%--------------------------------------------------------------
where $\lambda(\theta)$ represents a generic proportionality coefficient.
If we consider the ratio between these two equations we end up with the following differential equation:
%--------------------------------------------------------------
\begin{align}
\frac{{\rm d} r}{r {\rm d}\theta} = \frac{2\cos{\theta}}{\sin{\theta}}. 
\end{align}
%--------------------------------------------------------------
We consider the last closed field line that delimits the polar cap region and is tangent to the light cylinder. The equation above can be integrated from a generic radius $r$ corresponding to a generic $\theta$ to the light cylinder radius $r = r_{\rm LC}$ corresponding to $\theta = \pi/2$:
%--------------------------------------------------------------
\begin{align}
\int_{r}^{r_{\rm LC}} \frac{{\rm d} r'}{r'} = \int_{\theta}^{\pi/2} \frac{2\cos{\theta'}}{\sin{\theta'}} d\theta' \; \Longrightarrow \;
\ln{\frac{r_{\rm LC}}{r}} = 2\ln{\frac{1}{\sin{\theta}}}. \\ \nonumber
\end{align}
%--------------------------------------------------------------
This implies:
%--------------------------------------------------------------
\begin{align} \label{eq:ch1_const_Bline}
\frac{\sin^2{\theta}}{r} = \frac{1}{r_{\rm LC}}.
\end{align}
%--------------------------------------------------------------
Generally the quantity $\sin^2{\theta}/r$ is constant along any dipolar field line.
With this relation, by setting $r = R_{\rm NS}$ we can find the angular radius of the polar cap on the neutron star surface:
%--------------------------------------------------------------
\begin{align} \label{eq:ch1_polar_cap_angular_radius}
\sin{\theta_{\rm pc}} &= \left( \frac{R_{\rm NS}}{r_{\rm LC}} \right)^{1/2} \\
\Longrightarrow \; \theta_{\rm pc} \sim \left( \frac{2 \pi R_{\rm NS}}{cP} \right)^{\frac{1}{2}} &= \unit[0.015]{rad} \, [\unit[0.87]{deg}] \left( \frac{R_{\rm NS}}{\unit[11]{km}} \right) \left( \frac{P}{\unit[1]{s}} \right)^{-1/2},
\end{align}
%--------------------------------------------------------------
where we assumed that the the polar cap angular aperture is small i.e., $\theta_{\rm pc} \lesssim 30^{\circ}$ \citep{Lorimer2012}.

We now consider an emission point on a field line at the edge of the polar cap, with coordinates $(r_{\rm em}, \, \theta_{\rm em})$. By using Equation~\eqref{eq:ch1_const_Bline} we can write:
%--------------------------------------------------------------
\begin{align}
\sin{\theta_{\rm em}} = \left( \frac{r_{\rm em}}{r_{\rm LC}} \right)^{1/2} \; \Longrightarrow \;
\theta_{\rm em} \sim \left( \frac{2 \pi r_{\rm em}}{cP} \right)^{1/2},
\end{align}
%--------------------------------------------------------------
where we have assumed again that the polar cap region is narrow and the emission point is close to the magnetic axis, i.e., $\theta_{\rm em} \lesssim 30^{\circ}$. As the field lines are curved, the opening angle $\rho_{\rm em}$ of the emission cone tangent to the field lines at the polar coordinates $(r_{\rm em}, \, \theta_{\rm em})$ is generally a little bigger than the polar angle $\theta_{\rm em}$. The relation between these two angles was found by \citet{Gangadhara2001}, and for small angles reduces to:
%--------------------------------------------------------------
\begin{align} \label{eq:ch1_beam_angular_aperture}
\rho_{\rm em} \sim \frac{3}{2} \theta_{\rm em} \sim \left( \frac{9 \pi r_{\rm em}}{2 cP} \right)^{\frac{1}{2}} = \unit[0.069]{rad} \, [\unit[3.95]{deg}] \left( \frac{r_{\rm em}}{\unit[100]{km}} \right)^{1/2} \left( \frac{P}{\unit[1]{s}} \right)^{-1/2}.
\end{align}
%--------------------------------------------------------------
This gives the half-opening angle of the radio emission cone at the two magnetic polar regions of the neutron star. The modifications arising by taking into account a misaligned magnetic dipole moment and general relativistic effects are small and in general can be neglected \citep{Kapoor1998}.

\subsection{Pulse width} \label{sec:ch1_pulse_width}

%-------------------------------------------------------------
\begin{figure}
	\centering
	\includegraphics[width = 0.8\textwidth]{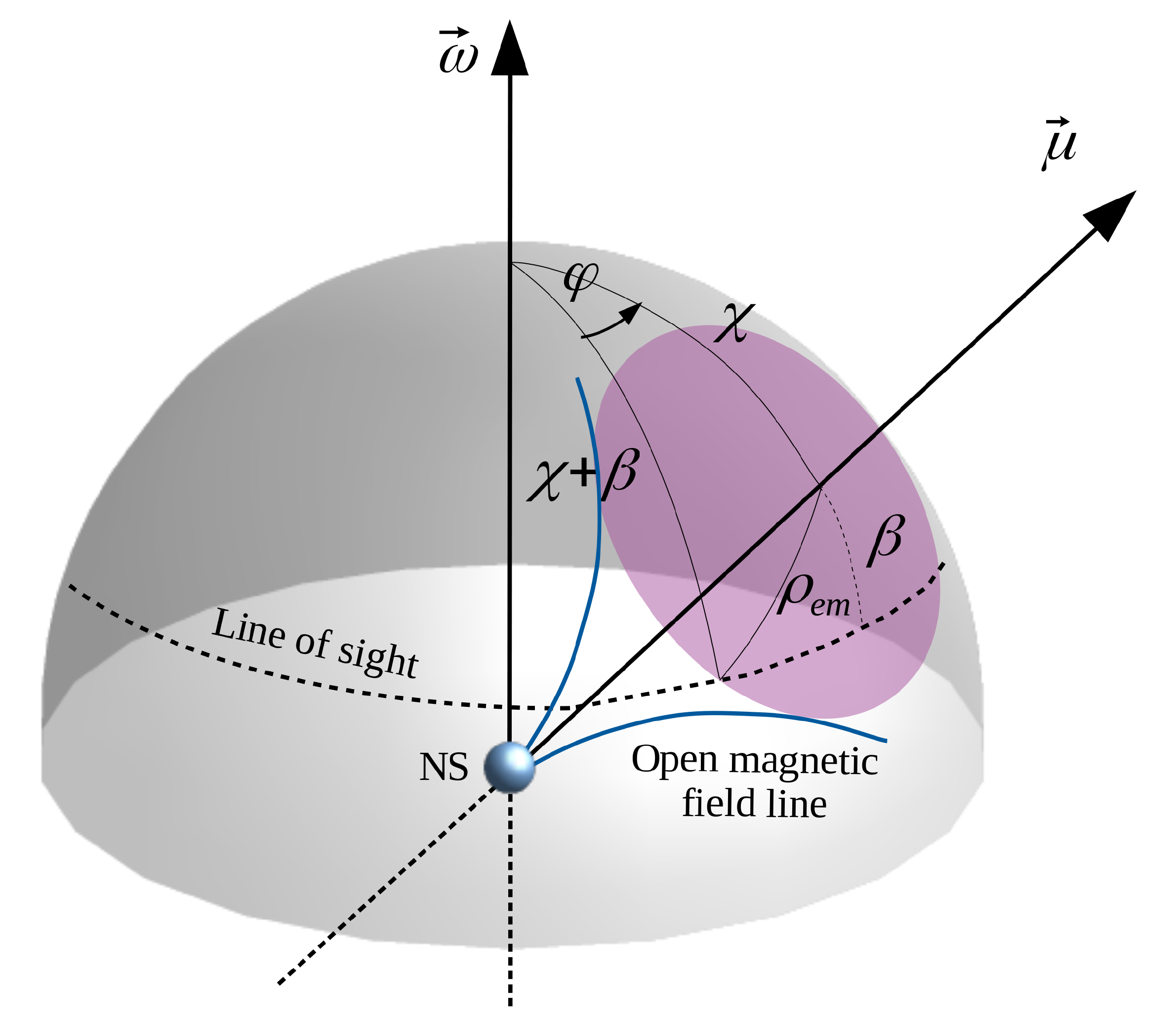}
	\caption[Radio beam geometry]{\label{fig:ch1_radio_beam_geometry}{Radio beam geometry. The magnetic dipole moment $\boldsymbol{\mu}$ is misaligned by an angle $\chi$ with respect to the rotation axis defined by $\boldsymbol{\omega}$. The line of sight cuts the emission cone at an angle $\beta$ relative to the centre of the beam. The angular pulse width is given by $\varphi$.}}
\end{figure}  
%-------------------------------------------------------------
%----------------------------------------------------
\begin{figure}
	\centering
	\includegraphics[width=0.8\textwidth]{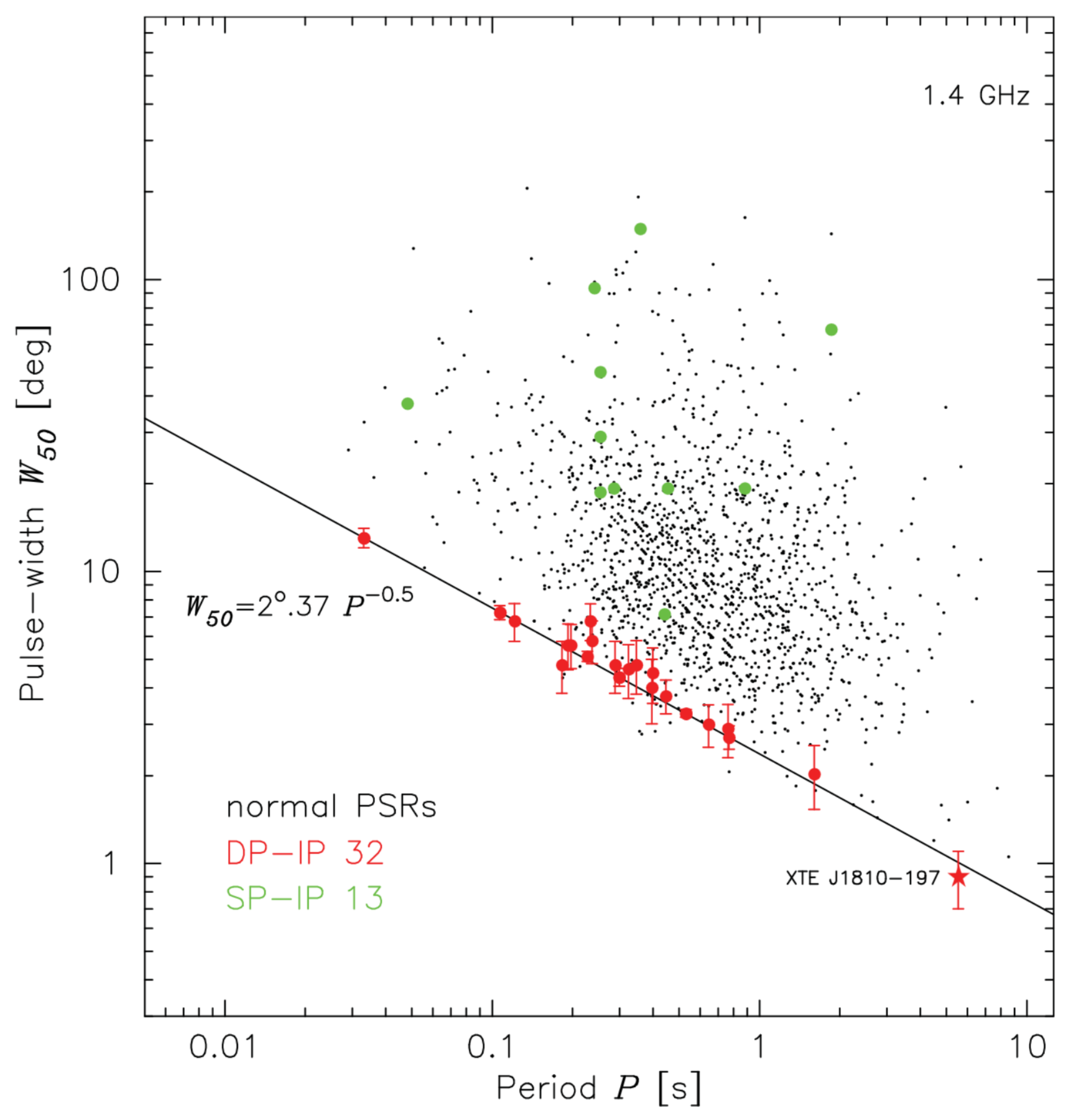}
	\caption[Pulse widths versus spin-period relation]{Plot of pulse-widths $w_{50}$ measured at the 50\% of the maximum pulse amplitude versus the spin period $P$ for 1450 normal pulsars detected at 1.4 GHz (excluding millisecond and other recycled pulsars) from the ATNF database \citep{Manchester2005}. The figure shows also 21 cases of pulsars being almost orthogonal rotators and manifesting inter-pulses due to our line of sight crossing both magnetic poles (red dots) and 11 cases of almost aligned pulsars where only one magnetic pole is visible but manifesting interpulses (green dots). The black solid line represents the fit of the orthogonal-rotator pulsars (red dots) \citep[see][for more details]{Maciesiak2011b, Maciesiak2012} \citep[Figure taken from][]{Maciesiak2012}.}
	\label{fig:ch1_w_P_relation}
\end{figure}
%----------------------------------------------------
As the neutron star rotates, a pulse of radiation can be detected if our line of sight cuts through at least one of the radio beams. Therefore the features of the observed pulses depend on the geometry of the radio beam and the relative position at which the line of sight intercepts it.
Having computed the angular aperture of the radio beam $\rho_{\rm em}$ \eqref{eq:ch1_beam_angular_aperture}, we can now set up a relation that links it to the intrinsic pulse half-width $\varphi$ (expressed in radians) that an observer sees. We provide Figure~\ref{fig:ch1_radio_beam_geometry}, valid for a generic inclined rotator with $\chi$ the inclination angle between the magnetic dipole moment and the rotation axis, as a reference. Let's assume that as the star rotates our the line of sight intercepts the beam at an angular distance $\beta$ from its centre. By assuming a sphere of unit radius, and considering the spherical triangle with sides $\chi + \beta$, $\chi$ and $\rho_{\rm em}$ we use the haversine law to write down the following relation:
%--------------------------------------------------------------
\begin{align}
{\rm hav}\left( \rho_{\rm em} \right) =  {\rm hav}\left( \chi + \beta - \chi \right) + \sin \left( \chi + \beta \right) \sin \left( \chi \right) {\rm hav}\left( \varphi\right),
\end{align}
%--------------------------------------------------------------
where in general given an angle $\alpha$ the haversine function is defined as ${\rm hav}\left( \alpha \right) = \sin^2 \left( \alpha/2 \right)$. 
Solving for the square sine of the half pulse width, we can write the following formula \citep[see][]{Maciesiak2011a}:
%--------------------------------------------------------------
\begin{align} \label{eq:ch1_radio_pulse_width}
\sin^2 \left( \frac{\varphi}{2} \right) &=  \frac{\sin^2\left( \frac{\rho_{\rm em}}{2} \right) - \sin^2\left( \frac{\beta}{2} \right)}{\sin \left( \chi+\beta \right) \sin\left( \chi \right)} \\ \Longrightarrow \;
\varphi &= 2 \arcsin{ \sqrt{ \frac{\sin^2\left( \frac{\rho_{\rm em}}{2} \right) - \sin^2\left( \frac{\beta}{2} \right)}{\sin \left( \chi+\beta \right) \sin\left( \chi \right)} }}.
\end{align} 
%--------------------------------------------------------------

To find the pulse width in time units we have to re-normalise for the spin period:
%--------------------------------------------------------------
\begin{align} \label{eq:ch1_pulse_width}
w &= 2 \varphi \frac{P}{2 \pi} = 2 \frac{P}{\pi} \arcsin{ \sqrt{ \frac{\sin^2\left( \frac{\rho_{\rm em}}{2} \right) - \sin^2\left( \frac{\beta}{2} \right)}{\sin \left( \chi+\beta \right) \sin\left( \chi \right)} }}.
\end{align} 
%--------------------------------------------------------------

A simplified formula for the pulse width can be found if one assumes that the line of sight intercepts the radio beam in its centre (i.e. with $\beta = 0$) and that $\rho_{\rm em}$ and $\varphi$ are small ($\lesssim 30^{\circ}$). In this case we can approximate the relation in Equation~\eqref{eq:ch1_radio_pulse_width} as:
%--------------------------------------------------------------
\begin{align}
\left( \frac{\varphi}{2} \right)^2 \sim  \frac{\left( \frac{\rho_{\rm em}}{2} \right)^2}{\sin^2 \left( \chi \right) }
\Longrightarrow \; \varphi \sim \frac{\rho_{\rm em}}{\sin \left( \chi \right) },
\end{align}
%--------------------------------------------------------------
which leads to:
%--------------------------------------------------------------
\begin{align} \label{eq:ch1_pulse_width_simpli}
w &= 2 \varphi \frac{P}{2 \pi} \sim 2 \frac{\rho_{\rm em}}{\sin{\chi}} \frac{P}{2 \pi} \sim \frac{2}{\sin{\chi}} \left( \frac{r_{\rm em} P}{2 \pi c} \right)^{\frac{1}{2}} \\ &\simeq \frac{\unit[0.015]{s}}{\sin{\chi}} \left( \frac{r_{\rm em}}{\unit[100]{km}} \right)^{1/2} \left( \frac{P}{\unit[1]{s}} \right)^{1/2}.
\end{align}
%--------------------------------------------------------------

Figure~\ref{fig:ch1_w_P_relation} shows the relation between the pulse-widths $w_{50}$ measured at the 50\% of the maximum pulse amplitude versus the spin period $P$ for 1450 normal pulsars detected at 1.4 GHz (excluding millisecond pulsars). The black line represents a fit of the pulsars showing interpulses and predicted to be orthogonal rotators, i.e. with $\chi \sim 90^{\circ}$.
Note that in Equation~\ref{eq:ch1_pulse_width_simpli}, the $\sin{\chi}$ term corrects for the inclination angle. For a given angular aperture, $\rho_{\rm em}$, of the emission cone, as the inclination angle decreases, the observed pulse width increases. This explain why in Figure~\ref{fig:ch1_w_P_relation} the pulsars that are likely orthogonal rotators (red dots) are found at the extreme bottom part of the diagram and more aligned rotators are scattered above depending on the inclination angle. For example, if we observe an aligned rotator ($\chi \sim 0^{\circ}$) we do not see pulsed emission since our line of sight is always inside the radio beam. In this case, according to the formula above the pulse width goes to $\infty$ but this simply indicates that we are seeing continuous emission from the polar cap. 
Furthermore, as $-\rho_{\rm em} < \beta < \rho_{\rm em}$, the angle $\chi + \beta$ could become negative. This occur only when $\rho_{\rm em} > \chi$, and in that case it happens that the line of sight is always inside the beam and therefore also in this case no pulsation is observed. Some periodic modulation in the observed pulse profile can be still observed if the intensity profile of the beam is not uniform. 

To obtain an estimate of the emission cone aperture $\rho_{\rm em}$ from the pulse width $w$ measurements one can in principle use Equation~\eqref{eq:ch1_pulse_width}. However, deriving an estimate of the intrinsic pulse width $w$ is not always trivial. The observed pulse width usually differ from the intrinsic pulse width due to pulse broadening as the radio waves propagates through the interstellar medium (see Section~\ref{sec:ch1_propagation_ism}). Furthermore Pulsars exhibit different pulse profiles and widths due to various intrinsic properties, including magnetospheric dynamics, emission processes, and rotational irregularities. These differences make it challenging to predict or measure the pulse width accurately, especially for pulsars with complex emission patterns \citep[see][]{Lorimer2012}. Moreover, we generally lack information on the relative angular position between the rotation axis, magnetic axis and line of sight.
For more details see also Chapter 3 in \citet{Lorimer2012}.

Overall even if this theoretical picture does not take into account all the complexity that might arise in reality, it predicts a relation $\rho_{\rm em} \sim P^{-1/2}$ which has been observed for samples of pulsars with reliable measurements of the pulse width \citep[see Figure~\ref{fig:ch1_w_P_relation} and][]{Rankin1993, Maciesiak2011b, Maciesiak2012}.
By using a sample of pulsars observed by the Parkes Murriyang radio telescope \citet{Johnston2019} found a shallower relation $\rho_{\rm em} \sim P^{-0.3}$. However the scatter of data in the pulse-width -- spin-period plane is large and is partly produced by the unknown magnetic inclination angle $\chi$ as can be seen from Equation~\eqref{eq:ch1_pulse_width_simpli}. As $\chi$ is unknown, its effect is difficult to take into account and affects the correlation between these quantities. This work also suggests that a narrow range of emission heights $r_{\rm em}$ between around $\unit[200]{km}$ and $\unit[400]{km}$ seems to be preferred. 

%========================================================================

\subsection{Radio emission mechanism and radio death valley} \label{sec:ch1_radio_em_deathlines}

%----------------------------------------------------
\begin{figure}
	\includegraphics[width=\textwidth]{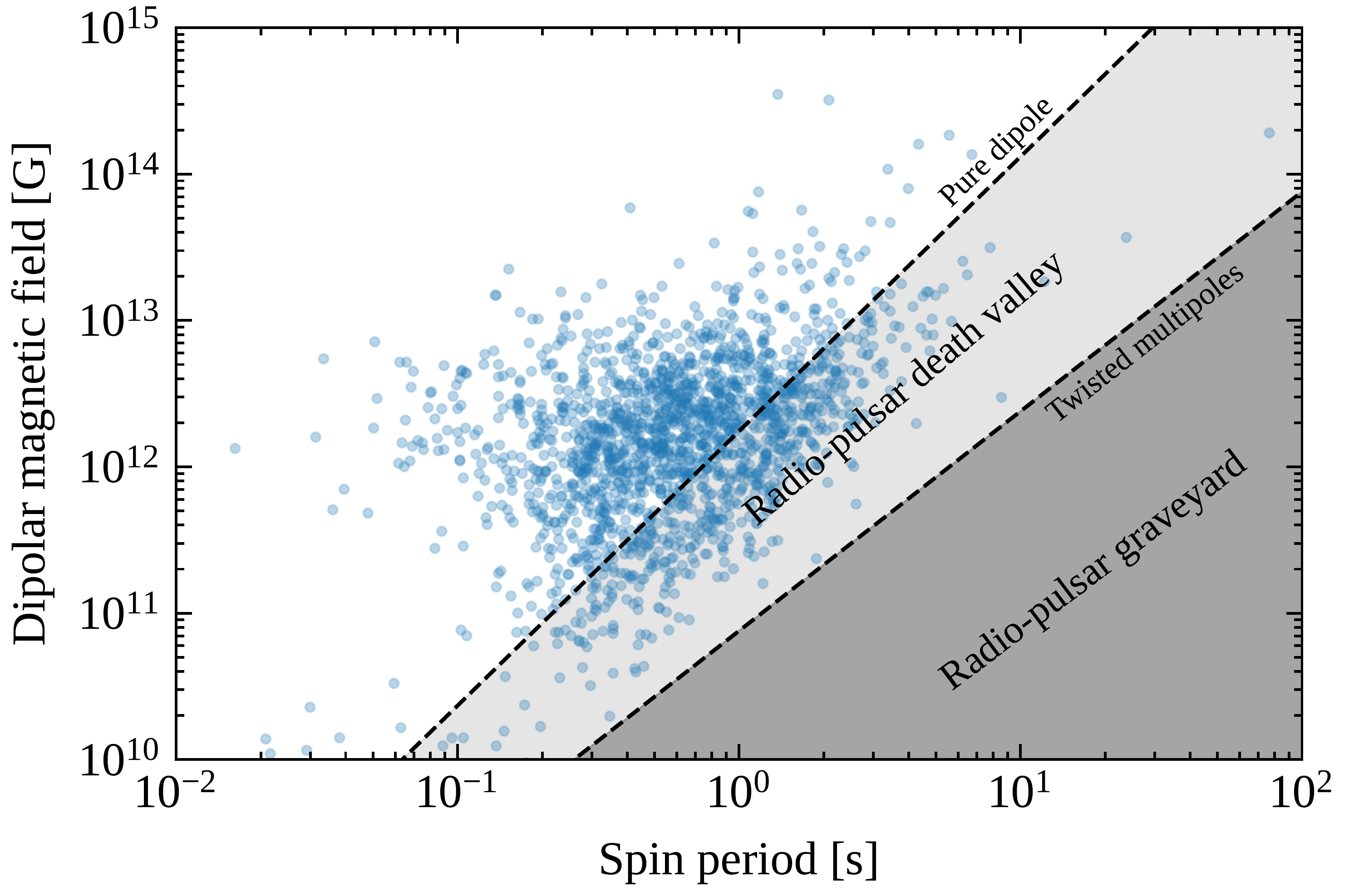}
	\caption[Radio emission death valley]{Radio emission death valley delimited by the two extreme death lines (\textit{dashed lines}) defined in Equations~\eqref{eq:ch1_death_line_1} for a pure dipole configuration, and~\eqref{eq:ch1_death_line_2} for a twisted multipole configuration, respectively. The dots represent the observed radio pulsars from the ATNF catalog \citep{Manchester2005}, where the dipolar magnetic field values are estimated from the timing properties ($P$ and $\dot{P}$) using Equation~\ref{eq:ch1_Bpol_vacuum}.}
	\label{fig:ch1_death_lines}
\end{figure}
%----------------------------------------------------

If a gap devoid of plasma is formed above the polar cap region an electric potential difference is generated along a magnetic field line. Following \citet{Ruderman1975} we can consider a narrow gap above the polar region extending from the neutron star surface up to a height $H$.
A given potential difference $\Delta V$ will accelerate an electron (or a positron) up to an energy $\gamma m_e c^2 = e \Delta V$ which implies a Lorentz factor of:
%--------------------------------------------------------------
\begin{align}
\gamma = \frac{e \Delta V} {m_e c^2}.
\end{align}
%--------------------------------------------------------------
By moving along the curved field lines the electron will emit curvature photons of energy \citep{Jackson1998}:
%--------------------------------------------------------------
\begin{align} \label{eq:ch1_curv_photon_energy}
h f = \frac{3}{4 \pi} \frac{h c}{r_{\rm c}} \gamma^3 = \frac{3}{4 \pi} \frac{h c}{r_{\rm c}} \left( \frac{e \Delta V}{m_e c^2} \right)^3,
\end{align}
%--------------------------------------------------------------
where $h$ is the Planck constant, $f$ is the photon frequency and $r_{\rm c}$ denots the curvature radius of the magnetic field line.
The condition that such a curvature photon with energy $h f > 2 m_e c^2$ moving through a magnetic field produces an electron-positron pair inside the gap is given by \citep[see][]{Ruderman1975, Chen1993}:
%--------------------------------------------------------------
\begin{align} \label{eq:ch1_pair_creation_lim}
\frac{h f}{2 m_e c^2} \frac{B_{\perp}}{B_{\rm QED}} \sim \frac{1}{15}.
\end{align}
%--------------------------------------------------------------
Here $B_{\rm QED} = 2 \pi m_e^2 c^3 / (e h)$ is the Schwinger magnetic field, defined as the magnetic field strength for which a cyclotron photon emitted by an electron gyrating around a field line will have an energy equal to $m_e c^2$.
$B_{\perp}$ is the component of the magnetic field perpendicular to the direction of propagation of the photon. Due to relativistic beaming the curvature photons are emitted along the direction of motion of the electron, i.e., almost parallel to the magnetic field lines, within an angle $\sim 1/\gamma$. Suppose that a curvature photon is emitted close to the stellar surface, where the magnetic field strength is $B_{\rm s}$. As the photon propagate inside the gap the component $B_{\perp}$ will grow up to an approximate maximum value given by $B_{\perp} \sim \frac{H}{r_{\rm c}} B_{\rm s}$. With all these considerations we can substitute Equation~\eqref{eq:ch1_curv_photon_energy} into Equation~\eqref{eq:ch1_pair_creation_lim} to find the relation:
%--------------------------------------------------------------
\begin{align} \label{eq:ch1_pair_creation_lim_2}
\frac{3}{8 \pi} \left( \frac{e \Delta V} {m_e c^2} \right)^3 \frac{h}{m_e c} \frac{H}{r_{\rm c}^2} \frac{B_{\rm s}}{B_{\rm QED}} \sim \frac{1}{15}.
\end{align}
%--------------------------------------------------------------
This equation defines the minimum potential difference that is required to trigger a pair cascade inside the gap. According to the standard theory if a neutron star is not able to generate such electric potential difference in its magnetosphere it cannot be visible as a radio pulsar. On the other hand, triggering the pair cascade will also limit any further growth of $\Delta V$ above the value established by Equation~\eqref{eq:ch1_pair_creation_lim_2} as the produced plasma will progressively screen the electric field.

Assuming an aligned dipolar configuration for the magnetic field, due to the unipolar induction mechanism (see Section~\ref{sec:ch1_magnetosphere}), the maximum potential difference that can be generated along an open field line in the magnetosphere is given by the difference between the electric potential on the stellar surface given by Equation~\eqref{eq:ch1_unipolar_potential}) and the electric potential of the interstellar medium to which the open field line is connected at infinity. The interstellar electric potential can be assumed to vanish. From Equation~\eqref{eq:ch1_unipolar_potential}, the maximum potential difference can be established on the last open field line at the edge of the polar cap, i.e. at a coordinate $\theta = \theta_{\rm pc}$. By assuming $V_{\rm p} = 0$ we have:
%--------------------------------------------------------------
\begin{align} \label{eq:ch1_electric_potential_max}
\Delta V_{\rm max} \simeq \frac{\omega \mu \sin^2{\theta_{\rm pc}}}{2 R_{\rm NS} c} = \frac{\omega B_{\rm p} R_{\rm NS}^2 \sin^2{\theta_{\rm pc}}}{2 c} = \frac{\omega^2 B_{\rm p} R_{\rm NS}^3}{2 c^2}.
\end{align}
%--------------------------------------------------------------
where we used Equation~\eqref{eq:ch1_mag_dipole_B_pole} and used the relation \eqref{eq:ch1_polar_cap_angular_radius}.
%This is indeed the potential difference between the polar magnetic field line and the last open field line at the edge of the polar cap.
This maximum potential difference depends solely on the dipole component of the field as it is the one determining the open field line configuration at large distance from the stellar surface, and is valid even if multipoles dominate near the stellar surface \citep{Ruderman1975}.
If the maximum potential difference $\Delta V_{\rm max}$ that can be produced in the magnetosphere is lower than the threshold defined by Equation~\eqref{eq:ch1_pair_creation_lim_2} the pair cascade can not be triggered and the neutron star is expected to be radio quiet. Therefore a necessary condition that has to be satisfied to turn on the radio emission is the following:
%--------------------------------------------------------------
\begin{align} \label{eq:ch1_death_line}
\frac{3}{8 \pi} \left( \frac{e \Delta V_{\rm max}} {m_e c^2} \right)^3 \frac{h}{m_e c} \frac{H}{r_{\rm c}^2} \frac{B_{\rm s}}{B_{\rm QED}} \gtrsim \frac{1}{15}
\end{align}
%--------------------------------------------------------------
Depending on the assumptions that one makes on the magnetic field configuration near the surface and the gap geometry, that is the values of $B_{\rm s}$, $r_{\rm c}$ and $H$, the condition \eqref{eq:ch1_death_line} defines a threshold for the turning on and off of the radio emission in a pulsar. As the suppression of the radio emission depends on the particular magnetospheric properties of each pulsar, relation \eqref{eq:ch1_death_line} defines a "death valley" for the radio pulsars \citep[see][]{Chen1993}.
We can consider two extreme cases. For an ideal case of a pure dipole configuration we consider $B_{\rm s} = B_{\rm p}$, the curvature radius for an open field line can be approximated as $r_{\rm c} \sim r_{\rm LC} \sin{\theta_{\rm pc}} = (R_{\rm NS} r_{\rm LC})^{1/2}$ and the gap height can be taken to be equal to the polar cap radius $H \sim R_{\rm NS} \sin{\theta_{\rm pc}} = R_{\rm NS} \left( R_{\rm NS}/r_{\rm LC} \right)^{1/2}$ \citep{Chen1993}.
With these order of magnitude assumptions the following condition for the radio emission turning on is derived from Equation~\eqref{eq:ch1_death_line}:
%--------------------------------------------------------------
\begin{align} \label{eq:ch1_death_line_1}
B_{\rm p} \gtrsim \unit[2.0 \times 10^{12}]{G} \left( \frac{R_{\rm NS}}{\unit[10^6]{cm}} \right)^{-19/8} \left( \frac{P}{\unit[1]{s}} \right)^{15/8}.
\end{align}
%--------------------------------------------------------------
If instead multipoles dominate the magnetic field configuration near the surface, the value $B_{\rm s}$ could exceed the one of a pure dipole. We can introduce the parametrisation $B_{\rm s} = b B_{\rm p}$ where $b$ is a numerical factor that takes into account deviations from the dipolar field value. Furthermore the curvature radius $r_{\rm c}$ in general scales inversely with the multiple order. We can consider an extreme case where the magnetic field near the surface is very twisted for the presence of strong multipole components. In this case for example we can assume $r_{\rm c} \sim H \sim R_{\rm NS}$. 
In this case, the condition for the radio emission becomes:
%--------------------------------------------------------------
\begin{align} \label{eq:ch1_death_line_2}
B_{\rm p} \gtrsim \unit[8.3 \times 10^{10}]{G} \, b^{-1/4} \left( \frac{R_{\rm NS}}{\unit[10^6]{cm}} \right)^{-2} \left( \frac{P}{\unit[1]{s}} \right)^{3/2}.
\end{align}
%--------------------------------------------------------------
Given a neutron star with a spin period $P$, Equation~\eqref{eq:ch1_death_line_1} and \eqref{eq:ch1_death_line_2} give the minimum polar magnetic field strength that the star should have to trigger the radio emission.
Neutron stars with a lower magnetic field are considered radio-dead (see bottom right corner in Figure~\ref{fig:ch1_death_lines}).

\subsection{Energetics} \label{sec:ch1_energetics}

As we have seen in Section~\ref{sec:ch1_magnetosphere} and Section~\ref{sec:ch1_dip_spindown_evol}, a neutron star behaves as a unipolar inductor and spins down due to magnetic dipole breaking. It converts its rotational kinetic energy into electrical potential energy that ultimately leads to particle acceleration and electromagnetic emission.
The rate of loss of rotational energy is given by Equation~\eqref{eq:ch1_rotational_energy_loss}. This represents the total budget of rotational power that can be converted into electromagnetic power by the unipolar inductor process. As the potential difference that is established inside the magnetosphere is created at the expense of the rotational energy we expect a relation between the maximum potential difference and the rotational energy loss rate.
By expressing the polar magnetic field strength in Equation~\eqref{eq:ch1_electric_potential_max} as a function of the timing properties $P$ and $\dot{P}$ through Equation~\eqref{eq:ch1_Bpol_vacuum} the maximum potential difference inside the magnetosphere can be expressed as:
%--------------------------------------------------------------
\begin{align} \label{eq:ch1_electric_potential_max_2}
\Delta V_{\rm max} \simeq \left( \frac{3}{2c} \frac{{\rm d} E_{\rm rot}}{{\rm d}t} \right)^{1/2}.
\end{align}
%--------------------------------------------------------------
We neglect the effect of the inclination angle which gives a correction of a factor 2 at most. 
The energy of the particles accelerated by this potential is given by $e \Delta V_{\rm max}$. This is the amount of energy that each particle can radiate away in the form of electromagnetic radiation.
Therefore the rotational energy is first converted into electrical potential energy which leads to particle acceleration and ultimately is radiated away in the form of electromagnetic radiation.
From this arguments it is reasonable to assume a link between the radio luminosity and the spin-down power. We consider a generic power-law dependence of the intrinsic radio luminosity $L$ of a pulsar on the rotational energy loss rate as:
%--------------------------------------------------------------
\begin{align} \label{eq:ch1_luminosity_Erot}
L = L_{0} \left( \frac{\dot{E}_{\rm rot}}{\dot{E}_{\rm rot,0}} \right)^{\epsilon} = \mathcal{L}_{0} \left( \frac{\dot{P}}{P^3} \right)^{\epsilon},
\end{align}
%--------------------------------------------------------------
where $L_0$ is the luminosity at the reference rotational power $\dot{E}_{\rm rot,0}$ and:
%--------------------------------------------------------------
\begin{align} 
\mathcal{L}_0 = L_0 \left( \frac{4 \pi^2 I_{\rm NS}}{\dot{E}_{\rm rot,0}} \right)^{\epsilon}.
\end{align}
%--------------------------------------------------------------
Considering a typical pulsar radio luminosity $L_0 \sim 10^{29}$ erg s$^{-1}$ corresponding to $\dot{E}_{\rm rot,0} = 10^{30}$ erg s$^{-1}$ \citep[see][]{Szary2014}, $\epsilon = 0.5$ and default values for the moment of inertia of a neutron star $I_{\rm NS} \sim 10^{45}$ g cm$^2$, we obtain $\mathcal{L}_0 \sim 3 \times 10^{35}$.
As the pulsar spins down it converts part of the rotational power in the form of radio waves. As the radio luminosity should not exceed the spin-down energy loss rate we can also express the equation above in terms of an efficiency $\eta$:
%--------------------------------------------------------------
\begin{align} \label{eq:ch1_luminosity_efficiency}
L & = \eta(\dot{E}_{\rm rot}) \dot{E}_{\rm rot}.
\end{align}
%--------------------------------------------------------------
The efficiency is a function of the rotational power itself and in principle should not exceed 1.
Equation~\eqref{eq:ch1_luminosity_Erot} represents the intrinsic total radio luminosity emitted by the two radiation beams. This intrinsic luminosity is however difficult to derive from observations and instead, a pseudo-luminosity is usually estimated as we will discuss in Section~\ref{sec:ch1_pseudo_luminosity}. 

%=======================================================================

\section{Pulse propagation in the interstellar medium} \label{sec:ch1_propagation_ism}

%----------------------------------------------------
\begin{figure}
	\centering
	\includegraphics[width=0.8\textwidth]{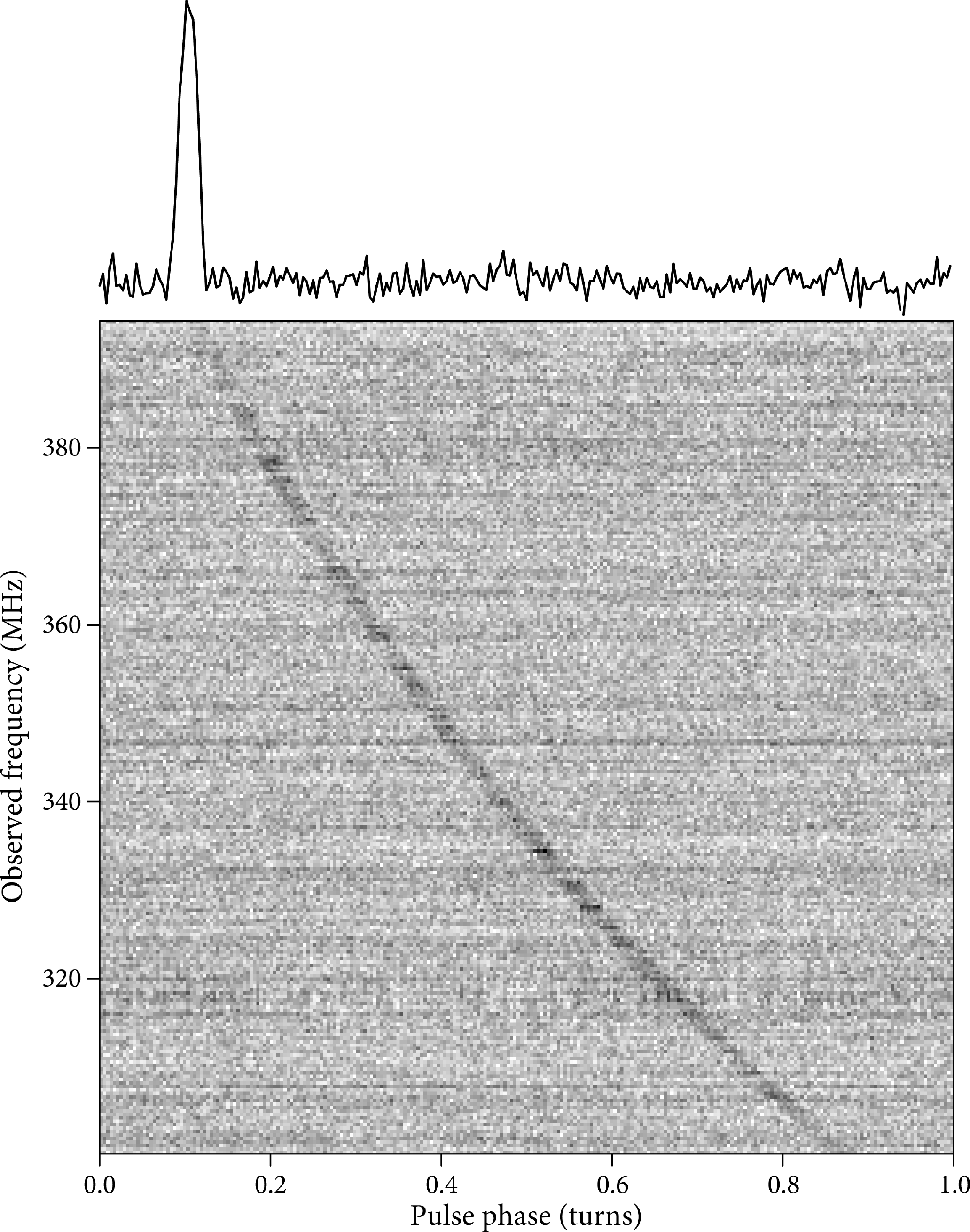}
	\caption[Pulse dispersion]{Pulse dispersion of the pulsar J1800+5034. The dark band in the frequency versus phase diagram shows the delay in arrival time of radio waves of different frequency. The "de-dispersed" band-integrated pulse profile is shown at the top. (Figure taken from \url{https://www.cv.nrao.edu/~sransom/web/Ch6.html}). }
	\label{fig:ch1_pulse_dispersion}
\end{figure}
%----------------------------------------------------
%----------------------------------------------------
\begin{figure}
	\centering
	\includegraphics[width=0.8\textwidth]{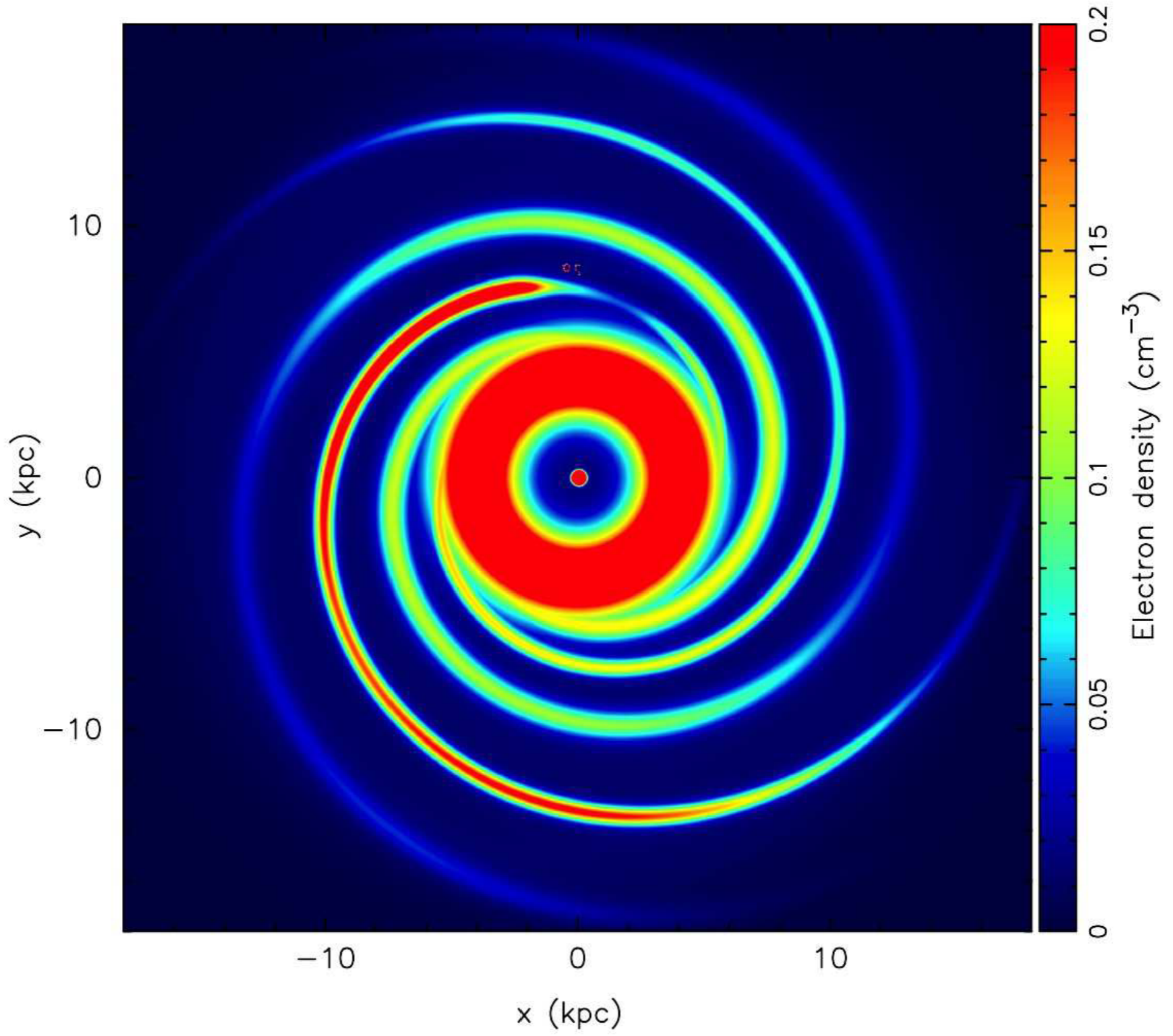}
	\caption[Galactic electron density for the YMW16 model]{Electron density in the Galactic plane ($z = 0$) for the YMW16 model \citep{Yao2017}. The Galactic Centre is at the origin and the Sun is at $x = 0$, $y = \unit[8.5]{kpc}$. The dense annulus of central radius 4 kpc represents the Galactic molecular ring. Spiral arms have a pre-determined logarithmic spiral form and generally decay exponentially at large Galactocentric radii.
	Exceptions are in the Carina and Sagittarius arms where there are over-dense and under-dense regions in Carina and Sagittarius
	respectively. The Gum Nebula and Local Bubble features are faintly visible near the position of the Sun \citep[see][for more details]{Yao2017} \citep[Figure taken from][]{Yao2017}.}
	\label{fig:ch1_e_density_model}
\end{figure}
%----------------------------------------------------

The intensity, shape and duration of a detected radio pulse are influenced by a variety of factors which depend on both the propagation in the interstellar medium and instrumental effects. In the following I will discuss these aspects individually. 

Suppose that a pulsar with spin period $P$ and emitting two beams with an intrinsic radio luminosity $L$ is located at a distance $d$ from Earth. If the radio signal from this pulsar were to travel in complete vacuum, the pulse of radiation would have a width $w$ given by Equation~\eqref{eq:ch1_radio_pulse_width} and the flux of radiation reaching the Earth would be:
%--------------------------------------------------------------
\begin{align} \label{eq:ch1_radio_flux}
S = \frac{L}{\Omega_{\rm em} d^2}.
\end{align}
%--------------------------------------------------------------
Here $\Omega_{\rm em}$ is the total solid angle covered by the two beams which are given by \citep[see][]{Lorimer2012}:
%--------------------------------------------------------------
\begin{align} \label{eq:ch1_solid_angle}
\nonumber \Omega_{\rm em}  & = 2 \times 2 \pi \int_{0}^{\rho_{\rm em}} \sin{\rho} {\rm d} \rho \\ \nonumber
& = 4 \pi \left( 1 - \cos{\rho_{\rm em}} \right) \\
& = 8 \pi \sin^2 \left( \frac{\rho_{\rm em}}{2} \right).
\end{align}
%--------------------------------------------------------------
The factor 2 comes from the two radio beams corresponding to each magnetic pole. In the last step we used the trigonometric identity $\left( 1 - \cos{\rho_{\rm em}} \right) = 2 \sin^2 \left( \rho_{\rm em}/2 \right)$.

Radio telescopes are sensitive to a limited range of frequencies around a central frequency $f_0$. 
The radio flux at a given frequency can be evaluated by considering the spectral shape of the signal.
Radio pulsars are usually characterised by a power-law spectral shape \citep[see][]{Jankowski2018}. 
Therefore to compute the flux density at a given frequency $f$ we can use the following relation:
%--------------------------------------------------------------
\begin{align}  \label{eq:ch1_flux_density}
S_{f}(f) = S_{f, 0} \left( \frac{f}{f_0} \right)^{\alpha} = \frac{S (\alpha + 1)}{f_{\rm max}^{\alpha+1} - f_{\rm min}^{\alpha+1}} f^{\alpha},  
\end{align}
%--------------------------------------------------------------
where we used the fact that $S$ is the flux integrated over the entire frequency range between $f_{\rm min}$ and $f_{\rm max}$:
%--------------------------------------------------------------
\begin{align} \label{eq:ch1_integral_flux}
S = \int_{f_{\rm min}}^{f_{\rm max}} S_{f}(f) {\rm d} f = S_{f,0} \frac{f_0^{-\alpha}}{\alpha + 1} \left( f_{\rm max}^{\alpha+1} - f_{\rm min}^{\alpha+1} \right).
\end{align}
%--------------------------------------------------------------

We can also define the fluence of this signal by integrating the flux over the interval of time defined by the spin period:
%--------------------------------------------------------------
\begin{align} \label{eq:ch1_fluence}
\mathcal{F}_{f}(f)  & = \int_{0}^{P} S_{f}(f, t) {\rm d} t = S_{f}(f)w,
\end{align}
%--------------------------------------------------------------
where for simplicity we considered a rectangular top-hat pulse shape with intrinsic width $w$.
The fluence has units of [erg cm$^{-2}$] and is a measure of the total energy carried by a radiation pulse passing through a unit surface area.

However, to reach the Earth the signal has to propagate through the interstellar medium which is characterised by clumpy regions with different densities where gas can be completely ionised.
In the presence of ionised gas, the electromagnetic interaction between the photons and the free electrons causes a delay in the propagation of the light, which is a function of photon frequency. More energetic photons tend to push past the free electrons with little effect on their propagation speed, whereas lower frequency photons (like radio waves) are more significantly delayed. 
We quantify these effects by introducing the refraction index $n$, which for a non-magnetised, non-relativistic and ionised plasma is given by:
%--------------------------------------------------------------
\begin{align} \label{eq:ch1_refraction_index}
n(f) & = \sqrt{ 1 - \left( \frac{f_{\rm p}}{f} \right)^2},
\end{align}
%--------------------------------------------------------------
where $f_{\rm p}$ is the so-called plasma frequency for a non-magnetised plasma:
%--------------------------------------------------------------
\begin{align}
f_{\rm p} = \left( \frac{e^2 n_e}{\pi m_e} \right)^{1/2} \simeq \unit[8.98]{kHz} \left( \frac{n_e}{\unit[1]{cm^{-3}}} \right)^{1/2}. 
\end{align}
%--------------------------------------------------------------
The delay in the arrival time of a wave of frequency $f$ with respect to the arrival time in vacuum can be estimated to be \citep[see][]{Lorimer2012}:
%--------------------------------------------------------------
\begin{align} \label{eq:ch1_delay_DM}
\Delta t(f) & = \left( \int_0^d \frac{{\rm d}l}{n(f) c} \right) - \frac{d}{c} \simeq \int_0^d \frac{{\rm d} l}{c} \frac{f_{\rm p}^2}{2 f^2} \nonumber \\ 
& = \frac{e^2}{2 \pi m_e c} f^{-2} \int_0^d {\rm d} l n_e(l) \nonumber \\ 
& \simeq \unit[0.41]{s} \left( \frac{f}{\unit[1]{GHz}} \right)^{-2} \frac{DM}{\unit[100]{pc \, cm^{-3}}},
\end{align}
%--------------------------------------------------------------
where we assumed $f_{\rm p} \ll f$. 
The integral
%--------------------------------------------------------------
\begin{align} \label{eq:DM}
DM & = \int_0^d {\rm d} l n_e(l),
\end{align}
%--------------------------------------------------------------
is performed along the line of sight $l$, over a distance $d$ and computes the electron column density along the line of sight which defines the dispersion measure $DM$.

To recover the original pulse shape, the detected signal has to be de-dispersed by re-aligning the observations of delayed pulses at different frequencies (see Figure~\ref{fig:ch1_pulse_dispersion}).
However, a radio detector has a finite frequency resolution. Usually, the total bandwidth of the instrument is divided into different channels, each one sensitive to photons in a particular range of frequencies $\Delta f_{\rm ch}$.
Since photons collected by a given spectral channel cannot be distinguished in frequency, the de-dispersion cannot be applied per channel. To obtain an estimate of the residual intra–channel dispersion smearing contribution to the observed pulse width we instead differentiate Equation~\eqref{eq:ch1_delay_DM} with respect to the frequency $f$ to find the amount of delay per unit frequency interval and multiply by the minimum frequency bin $\Delta f_{\rm ch}$ of the instrument:
%--------------------------------------------------------------
\begin{align} \label{eq:ch1_smearing_DM}
\tau_{DM} & = \left| \frac{{\rm d} \Delta t}{{\rm d} f} \right| \Delta f_{\rm ch}  \nonumber \\ 
& = \frac{e^2}{\pi m_e c} f^{-3} \Delta f_{\rm ch} DM \nonumber \\ 
& \simeq \unit[0.83]{ms} \left( \frac{f}{\unit[1]{GHz}} \right)^{-3} \frac{\Delta f_{\rm ch}}{\unit[1]{MHz}} \frac{DM}{\unit[100]{pc \, cm^{-3}}}.
\end{align}
%--------------------------------------------------------------

Furthermore, the electron density in the medium is not homogeneous and shows variations over a wide range of length scales. The distribution of scales of these variations is affected by the presence of turbulence in the medium, which introduces a relative motion between the source, the inhomogeneous
clumps and the observer. As the electromagnetic signal propagates, it encounters several such irregularities depending on the travelled distance $d$. At each encounter, it is delayed and deflected in a frequency-dependent way. The result is that a sharp pulse emitted by a point source is detected as a scattered-broadened pulse whose intensity decays exponentially with a characteristic time $\tau_{\rm scat}$ which depends on the signal
frequency and distance of the source \citep[see][]{Lorimer2012}. By assuming a Kolmogorov turbulence power spectrum for the inhomogeneity length scales \citep{Kolmogorov1941} the broadening timescale due to scattering is given by:
%--------------------------------------------------------------
\begin{align} \label{eq:ch1_scattering_timescale}
\tau_{\rm scat} & \propto d^{2.2} f^{-4.4}.
\end{align}
%--------------------------------------------------------------
Some studies show evidence of a relation between the scattering broadening timescale and the dispersion measure. For example, by using a sample of scattering measurements from 124 pulsars detected at $\unit[327]{MHz}$ by the Ooty Radio Telescope in Southern India, \citet{Krishnakumar2015} found the following empirical relation:
%--------------------------------------------------------------
\begin{align} \label{eq:ch1_tausc_DM_relation}
\tau_{\rm scat} = \unit[3.6 \times 10^{-9}]{s} \left(\frac{DM}{\unit[1]{pc \, cm^{-3}}} \right)^{2.2} \left[ 1.0 + 1.94 \times 10^{-3} \left(\frac{DM}{\unit[1]{pc \, cm^{-3}}} \right)^{2.0} \right].
\end{align}
%--------------------------------------------------------------
This empirical relation is useful to estimate the scattering timescale for sources which only have a measurement of the dispersion measure. 

Additionally, there are instrumental effects that contributes to the broadening of the pulse shape during the digital processing of the radio signal. The most relevant effect is given by the detector sampling time which limits the time resolution of the pulse and introduce an uncertainty in the pulse duration given by $\tau_{\rm samp}$.

As a result of the effects discussed here, the observed duration $w_{\rm obs}$ of a radio pulse is larger than the intrinsic duration $w$ and is given by the quadratic sum of the contributions described above as reported in \citet{Cordes2003}:
%--------------------------------------------------------------
\begin{align} \label{eq:ch1_eff_pulse_width}
w_{\rm obs} \simeq \sqrt {w^2 + \tau_{DM}^2 + \tau_{\rm sc}^2 + \tau_{\rm samp}^2},
\end{align}
%--------------------------------------------------------------

As the pulse propagates, gets dispersed and broadens, its fluence is defined by Equation~\eqref{eq:ch1_fluence} is conserved. Therefore the flux of the pulse detected on Earth can be found by:
%--------------------------------------------------------------
\begin{align} \label{eq:ch1_obs_radio_flux}
S_{f}(f) w  & = S_{f,\rm obs}(f) w_{\rm obs} \, \Longrightarrow \, S_{f,\rm obs}(f) = \frac{S_{f}(f) w}{w_{\rm obs}}.
\end{align}
%--------------------------------------------------------------

\section{Pulsar pseudo-luminosity} \label{sec:ch1_pseudo_luminosity}

By inverting the steps described in the last section, one should in principle be able to estimate the intrinsic luminosity of a pulsar from the observed quantities.
Usually in pulsar radio surveys the flux $S_{f, \rm mean}$ averaged over a period and at the observation frequency $f_0$ is reported. Assuming for simplicity a rectangular pulse shape with peak flux $S_{f, \rm obs}$ and observed width $w_{\rm obs}$ this is given by:
%--------------------------------------------------------------
\begin{align} \label{eq:ch1_mean_radio_flux}
S_{f, \rm mean}(f_0) & = \frac{\int_0^P S_{f, \rm obs}(f_0, t) {\rm d}t}{P} = \frac{S_{f, \rm obs}(f_0) w_{\rm obs}}{P}.
\end{align}
%--------------------------------------------------------------
To recover the intrinsic flux $S(f_0)$ we need to find the intrinsic pulse width. From Equation~\eqref{eq:ch1_eff_pulse_width} we have:
%--------------------------------------------------------------
\begin{align}
w \simeq \sqrt {w_{\rm obs}^2 - \tau_{DM}^2 - \tau_{\rm scat}^2 - \tau_{\rm samp}^2}.
\end{align}
%--------------------------------------------------------------
By combining Equation~\eqref{eq:ch1_obs_radio_flux} and Equation~\eqref{eq:ch1_mean_radio_flux} with the previous equation we then find:
%--------------------------------------------------------------
\begin{align} 
S_{f}(f_0) & \simeq S_{f, \rm mean}(f_0) \frac{P}{\sqrt {w_{\rm obs}^2 - \tau_{DM}^2 - \tau_{\rm scat}^2 - \tau_{\rm samp}^2}} = \frac{S_{f, \rm mean}(f_0)}{\delta},
\end{align}
%--------------------------------------------------------------
where here we define the intrinsic duty cycle $\delta \equiv w/P$.
Finally using Equation~\eqref{eq:ch1_radio_flux} and Equation~\eqref{eq:ch1_solid_angle} we obtain for the intrinsic luminosity density:
%--------------------------------------------------------------
\begin{align} 
L_{f}(f_0) \simeq \frac{8 \pi}{\delta} \sin^2\left( \frac{\rho_{\rm em}}{2} \right) d^2 S_{f, \rm mean}(f_0).
\end{align}
%--------------------------------------------------------------
And integrating over the frequencies (Equation~\eqref{eq:ch1_integral_flux}) we obtain the bolometric intrinsic radio  luminosity:
%--------------------------------------------------------------
\begin{align} 
L \simeq \frac{8 \pi}{\delta} \sin^2\left( \frac{\rho_{\rm em}}{2} \right) d^2 S_{\rm mean}(f_0) \frac{f_0^{-\alpha}}{\alpha + 1} \left( f_{\rm max}^{\alpha+1} - f_{\rm min}^{\alpha+1} \right).
\end{align}
%--------------------------------------------------------------

However, a precise estimate of the intrinsic radio luminosity is not trivial for the following reasons:
\begin{itemize}
	\item In most cases estimating the angular aperture $\rho_{\rm em}$ of the radio beam and the intrinsic pulse width $w$ can be challenging (see also Section~\ref{sec:ch1_pulse_width}). Pulsars often exhibit intricate emission patterns, making it highly complicated to deduce the emission geometry and accurately determine the pulse width. One possibility to estimate the beam angular aperture from the spin period is to use the relation given by Equation~\eqref{eq:ch1_beam_angular_aperture}, which has been confirmed by observation despite some uncertainties \citep{Rankin1993, Maciesiak2012}. Additionally, removing the broadening effects described above from the observed pulse width $w_{\rm obs}$ is not a straightforward task, as it can be difficult to differentiate these effects from the inherent properties of the pulse shape for instrumental limitations in the time and frequency resolution \citep{Lorimer2012}.
	
	\item Accurate modelling of radio pulsar spectra is not always possible. Radio spectra appear to be different for every pulsar. As shown in \citet{Jankowski2018} in many cases a single power-law model describes the observed spectrum well but many pulsars deviate from this simple behaviour. 
	
	\item The estimate of the distance $d$ is subject to many uncertainties. For radio pulsars, the main way to estimate the distance relies on models for the electron density distribution in the Galaxy. Three main models have been developed over the years from \citep[see][]{Taylor1993, Cordes2002, Yao2017}. By knowing the $DM$ value and the sky position of the pulsar, these models predict the amount of free electrons along the line of sight and give a rough estimate of the distance. However, due to the complexity of modelling the Galactic structures, these models suffer from a lot of uncertainties. The relative error on the distance estimated from the $DM$ values could indeed be around 40 \% \citep{Yao2017}. 
\end{itemize}

Due to these uncertainties fiducial values are usually adopted for the unknown quantities \citep[see][]{Lorimer2012}. For example by considering a fixed beam aperture of $\rho_{\rm em} \sim \unit[6]{deg}$, the ratio between the intrinsic pulse width and the spin period (the duty cycle) of $\delta = 0.04$,  a radio frequency range from $f_{\rm min} = \unit[10]{MHz}$ to $f_{\rm max} = \unit[100]{GHz}$ \citep[see for example][]{Izvekova1981, Morris1997}, and a spectral index $\alpha = -1.6$ \citep{Jankowski2018}, we have a pseudo luminosity of:
%--------------------------------------------------------------
\begin{align} \label{eq:ch1_pseudo-luminosity}
L_{f, \rm pseudo}(\unit[1.4]{GHz}) & \simeq \unit[7.4 \times 10^{27}]{erg \, s^{-1}} \left( \frac{S_{f, \rm mean}(\unit[1.4]{GHz})}{\unit[1]{mJy}} \right) \left( \frac{d}{\unit[1]{kpc}} \right)^2.
\end{align}
%--------------------------------------------------------------
However note that performing a population study and applying this expression to different pulsars without taking into account differences in their intrinsic duty cycles and beam geometries will mask eventual intrinsic correlations between the intrinsic luminosity and the spin-down properties like $P$, $\dot{P}$ and $\dot{E}_{\rm rot}$. This is because $\rho_{\rm em}$ and $w$ have a non-trivial dependence on the spin period (see Section~\ref{sec:ch1_radio_em_geometry}). For example \citet{Posselt2023} recently showed that the radio pseudo-luminosity has a shallow dependence on the spin-down energy loss $L_{f, \rm pseudo} \propto \dot{E}_{\rm rot}^{0.15}$ \citep[see also][]{Szary2014}. However, this might not be a proof of the absence of intrinsic correlations between the intrinsic radio luminosity and the rotational properties.

\section{Pulsar detection, the radiometer equation}
\label{sec:ch1_radiometer}

%----------------------------------------------------
\begin{figure}
	\centering
	\includegraphics[width=\textwidth]{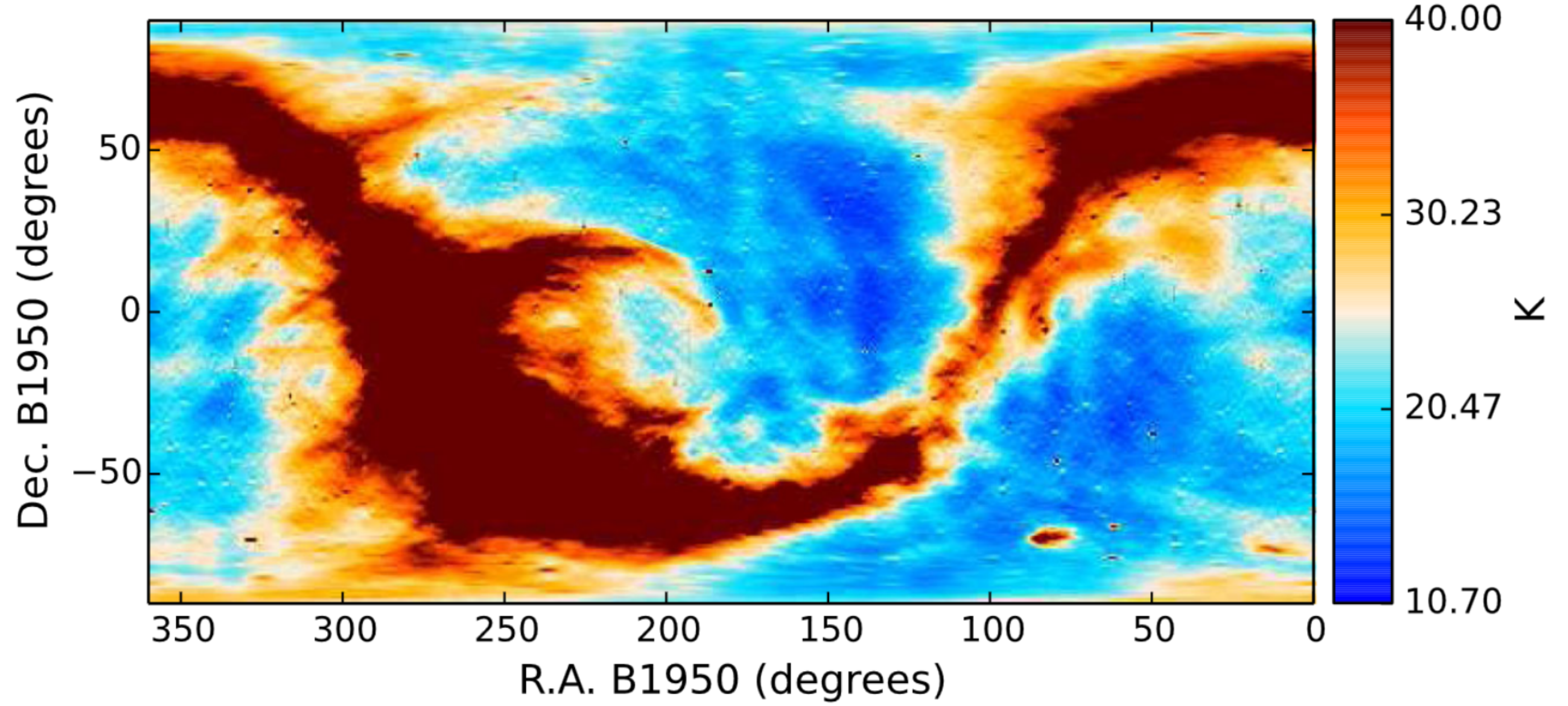}
	\caption[Sky temperature distribution]{Sky temperature distribution at 408 MHz from \citet{Haslam1982}, refined by \citet{Remazeilles2015} \citep[Figure taken from][]{Remazeilles2015}.}
	\label{fig:ch1_T_sky_distribution}
\end{figure}
%----------------------------------------------------

In this section we will describe how we can model the sensitivity of a radio telescope which is a crucial aspect to determine which pulsars are detectable. We follow \citet{Lorimer2012}.

Consider the observation of a signal with peak flux density $S_{f,\rm obs}(f)$ (in [erg s$^{-1}$ Hz$^{-1}$]), performed with an antenna telescope of effective area $A_{\rm eff}$. 
We consider an unpolarised signal where the electric field vector changes randomly in all possible directions perpendicular to the direction of wave propagation.
Since each of the two orthogonal polarisation directions contributes half the total flux, the power received from the source per unit frequency bandwidth and polarisation is $P_{f, \rm source} = S_{f, \rm obs}(f) A_{\rm eff} / 2$. Now consider modelling the antenna as a simple circuit consisting of a resistive load at temperature $T$. The thermal motions of the electrons inside the resistor induce a time variable voltage in this circuit which produces a power per unit frequency bandwidth that can be translated into a temperature by $P_{f, \rm circuit} = k_{\rm B} T$, where $k_{\rm B}$ is the Boltzmann constant. This is also known as Johnson–Nyquist noise \citep{Johnson1928, Nyquist1928}. Therefore an incident electromagnetic wave with a power per unit frequency $P_{f, \rm source}$ will produce a power in the circuit given by $P_{f, \rm circuit} = P_{f, \rm source}$. It follows that:
%--------------------------------------------------------------
\begin{align} \label{eq:ch1_flux_antenna_eq}
S_{f, \rm obs}(f) & = \frac{2 k_{\rm B} T_{\rm source}}{A_{\rm eff}} = \frac{T_{\rm source}}{G},
\end{align}
%--------------------------------------------------------------
where we define the antenna gain as $G \equiv A_{\rm eff} / ( 2 k_{\rm B} )$ which is a measure of the antenna sensitivity. In other words the source of flux density $S_{f, \rm obs}( f )$ have an equivalent temperature $T_{\rm source}$ given by Equation~\eqref{eq:ch1_flux_antenna_eq} in the telescope system.

However real observing systems have many sources of noise, that can be
modelled as a background system temperature $T_{\rm noise}$. This noise temperature accounts for the noise contributions from the receiver, transmission lines, and other components of the radio telescope system which are represented as a system temperature $T_{\rm sys}$. Furthermore, $T_{\rm noise}$ includes any spurious background signals coming from the sky with an equivalent temperature $T_{\rm sky}(l, b)$ which is a function of galactic sky coordinates $(l, b)$.
Figure~\ref{fig:ch1_T_sky_distribution} shows the Sky temperature distribution measured at $\unit[408]{MHz}$ as a function of equatorial coordinates \citep{Haslam1982, Remazeilles2015}. This temperature can be re-scaled to the central frequency of the survey assuming a frequency spectrum $T_{\rm sky} \propto f^{-2.6}$ \citep{Lawson1987, Johnston1992}.
Therefore in general we can write $T_{\rm noise} = T_{\rm sys} + T_{\rm sky}(l, b)$. These noise sources are usually much larger than the faint signals we receive from astronomical sources. For this reason, to establish if a source is detectable we rely on the concept of signal-to-noise ratio which measures deviations from the random noise oscillation of the receiver.   

The root mean square fluctuations $\sigma_T$ of the noise in the telescope for a given generic temperature $T$ decreases with the square root of the number of measured polarisations $n_{\rm pol}$, the receiver frequency bandwidth $\Delta f_{\rm bw}$ and integration time $t_{\rm obs}$ \citep{Dicke1946}:
%--------------------------------------------------------------
\begin{align} \label{eq:ch1_T_fluctuations}
\sigma_{T} & = \frac{T}{\sqrt{n_{\rm pol} \Delta f_{\rm bw} t_{\rm obs}}}.
\end{align}
%--------------------------------------------------------------
This equation is known as the radiometer equation and is used to perform estimation of the system sensitivity.

Consider an observation of a pulsar for a total integration time $t_{\rm obs}$. The pulsar has a top-hat pulse shape with width $w_{\rm obs}$, periodicity $P$ and an associated temperature $T_{\rm source}$. Over the whole integration time, the pulsar is on over a time $t_{\rm on} = t_{\rm obs} w_{\rm obs} / P$ and it is off for a time $t_{\rm off} = t_{\rm obs} (P - w_{\rm obs}) / P$.
For the pulsar to be detectable, its signal has to exceed the noise fluctuations in the system during the integration time. The ratio between $T_{\rm source}$ and the system noise fluctuations $\sigma_{T_{\rm noise}}$ define the signal to noise ratios $S/N$. To have a significative detection the $S/N$ has to surpass a threshold that is usually taken to be above 5 \citep[see][]{Vivekanand1982, Lorimer2012}. The root mean square noise fluctuations during the integration time can be computed as the quadrature sum:
%--------------------------------------------------------------
\begin{align} 
\sigma_{T_{\rm noise}} & = \sqrt{ \sigma_{T_{\rm on}}^2 + \sigma_{T_{\rm off}}^2} \nonumber \\
& = \sqrt{ \left( \frac{T_{\rm source} + T_{\rm noise}}{\sqrt{n_{\rm pol} \Delta f_{\rm bw} t_{\rm on}}} \right)^2  + \left( \frac{T_{\rm noise}}{\sqrt{n_{\rm pol} \Delta f_{\rm bw} t_{\rm off}}} \right)^2} \nonumber \\
& \simeq \frac{T_{\rm noise}}{\sqrt{n_{\rm pol} \Delta f_{\rm bw} t_{\rm obs}}} \frac{P}{\sqrt{w_{\rm obs} (P - w_{\rm obs})}},
\end{align}
%--------------------------------------------------------------
where we used Equation~\eqref{eq:ch1_T_fluctuations} and made the approximation that $T_{\rm source} \ll T_{\rm noise}$.
We can therefore define the signal-to-noise ratio as:
%--------------------------------------------------------------
\begin{align}
S/N &\equiv \frac{T_{\rm source}}{\sigma_{T_{\rm noise}}} = \frac{T_{\rm source} \sqrt{n_{\rm pol} \Delta f_{\rm bw} t_{\rm obs}}}{T_{\rm noise}} \frac{\sqrt{w_{\rm obs} (P - w_{\rm obs})}}{P} \nonumber \\
&= \frac{S_{\rm obs}(f) \sqrt{n_{\rm pol} \Delta f_{\rm bw} t_{\rm obs}}}{G [T_{\rm sys} + T_{\rm sky}(l, b)]} \frac{\sqrt{w_{\rm obs} (P - w_{\rm obs})}}{P},
\end{align}
%--------------------------------------------------------------
where in the last step we used Equation~\eqref{eq:ch1_flux_antenna_eq}.
To take into account system imperfections, for example due to the digitisation of the signal, a degradation factor $\beta \sim 1.5$ is commonly introduced so that the previous equation becomes: 
%--------------------------------------------------------------
\begin{align} \label{eq:ch1_StoN_radiometer_eq}
S/N = \frac{S_{f, \rm obs}(f) \sqrt{n_{\rm pol} \Delta f_{\rm bw} t_{\rm obs}}}{\beta G [T_{\rm sys} + T_{\rm sky}(l, b)]} \frac{\sqrt{w_{\rm obs} (P - w_{\rm obs})}}{P}.
\end{align}
%--------------------------------------------------------------
If we consider the period-averaged flux by substituting Equation~\eqref{eq:ch1_mean_radio_flux} into \eqref{eq:ch1_StoN_radiometer_eq} we find:
%--------------------------------------------------------------
\begin{align} \label{eq:ch1_StoN_radiometer_eq_Smean}
S/N = \frac{S_{f, \rm mean}(f) \sqrt{n_{\rm pol} \Delta f_{\rm bw} t_{\rm obs}}}{\beta G [T_{\rm sys} + T_{\rm sky}(l, b)]} \sqrt{ \frac{P - w_{\rm obs}}{w_{\rm obs}}}.
\end{align}
%--------------------------------------------------------------
This equation shows that the narrower the pulse, the bigger the signal-to-noise ratio, hence the easier it is to detect the pulsar.

%=========================================================================

\section{The neutron star zoo} \label{sec:ch1_ns_zoo}

%----------------------------------------------------
\begin{figure}
\centering
\includegraphics[width =\textwidth]{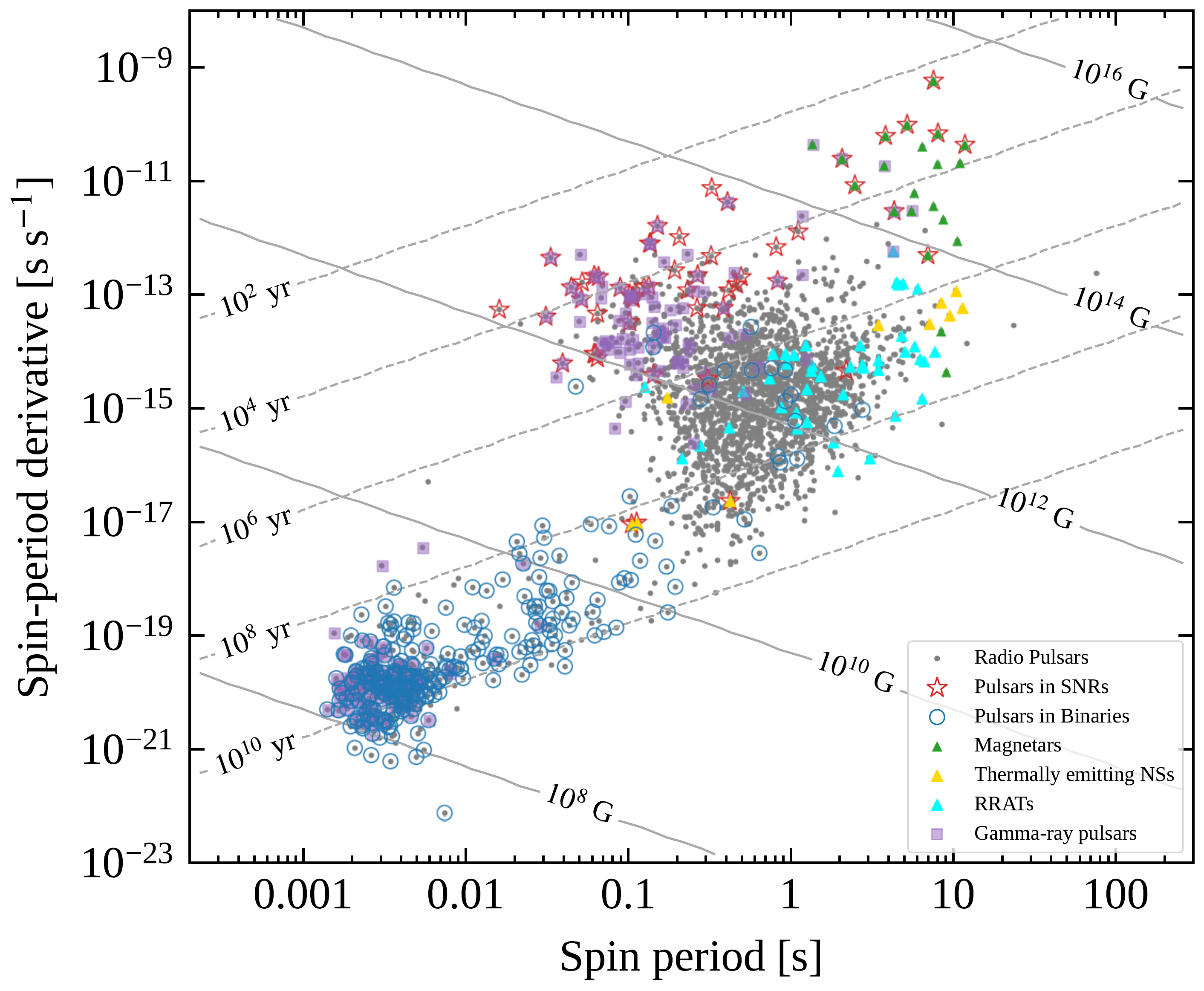}
\caption[Observed $P - \dot{P}$ diagram]{$P - \dot{P}$ diagram showing the spin period derivative {\it vs} spin period for the detected neutron stars \citep[Data from the ATNF Pulsar Catalog, \url{https://www.atnf.csiro.au/research/pulsar/psrcat/};][]{Manchester2005}.}
\label{fig:ch1_PPdot_diagram} 
\end{figure}  
%----------------------------------------------------

Figure~\ref{fig:ch1_PPdot_diagram} shows the observed population of non-accreting neutron stars in the $P-\dot{P}$ plane (where $P$ is the neutron star spin period and $\dot{P}$ is the spin period derivative with respect to time). Assuming a dipolar spin-down mechanism we also overlay the lines of constant polar magnetic field at the surface (\textit{solid lines}) (see Equation~\eqref{eq:ch1_Bpol_vacuum}) and characteristic age (\textit{dashed lines}) (see Equation~\eqref{eq:ch1_characteristic_age_approx}) as a reference.
Around $\sim 3000$ neutron stars are known to date and according to their observational properties we classify them as different types. This section will provide an overview of the different classes of neutron stars and their distinguishing characteristics \citep[see][for a more exhaustive review]{Kaspi2010, Harding2013}.

\paragraph{Rotation-powered pulsars.}

The bulk of the population is composed of \acf{RPPs} (\textit{grey dots}). As they spin down due to magnetospheric torques, the loss of rotational energy is converted into particle acceleration which in turn produces broadband electromagnetic emission, from radio to X-rays and gamma-rays. 
The origins of the radio and gamma-ray emission of pulsars are however not well understood. As described in Section~\ref{sec:ch1_radio_em_geometry}, the radio emission is thought to be generated close to the neutron star surface in a narrow beam centred on the magnetic axis within the magnetic pole region of the pulsar.
The high-energy gamma-ray emission is instead thought to be generated in the outer regions of the neutron star magnetosphere close to the light cylinder where the magnetic field lines open up (see also Section~\ref{sec:ch1_magnetosphere}).
Since the discovery of the first neutron star through its radio pulsations \citep{Hewish1968} around $\sim 2000$ radio pulsars have been detected. 
For some of them, X-ray, gamma-ray and optical pulsations were soon discovered. At present, there are over 100 \acs{RPPs} detected at X-ray energies and over 300 gamma-ray pulsars \citep{Abdo2013, Smith2023}. While most were found using the ephemerides of already known radio pulsars, many were discovered through their X-ray or gamma-ray pulsations and are radio-quiet. \\
Young neutron stars populate the top part of the diagram. They are often associated with Supernova Remnants (SNRs) (\textit{red stars}) and are often also detected as high-energy gamma-ray pulsars (\textit{purple squares}). As they spin down and their magnetic fields decay in time (see Section~\ref{sec:ch1_B_evol}, pulsars move towards the bottom right region of the diagram.
There are two main populations of RPPs. "Normal" pulsars with spin periods $P \gtrsim \unit[0.1]{s}$ and characteristic ages $\tau \lesssim \unit[100]{Myr}$, occupy the central part of the diagram whereas millisecond pulsars (MSPs), with spin periods down to a few milliseconds and generally higher characteristic ages $\tau \gtrsim \unit[100]{Myr}$ stretch towards the bottom left part of the diagram. MSPs are believed to be older neutron stars mostly in binary systems. Originally part of the X-ray binary population, they are recycled, i.e., spun up during a long phase of accretion from their binary companion \citep{Bhattacharya1991, Tauris1999, Tauris2012}.
There is another subpopulation of \acs{RPPs} known as \acf{RRATs} (\textit{light blue triangles}). They were discovered as powerful short-lasting radio bursts that seemed to recur randomly over timescales of minutes to hours \citep{McLaughlin2006}. By studying the arrival times of these bursts common periodicities have been discovered hinting to the spin periods of the emitting neutron stars. \acs{RRATs} have been interpreted as normal pulsars showing an irregular, sporadic emission of radio pulses (detected as radio bursts), instead of a regular emission associated with the rotation period. For these characteristics they are difficult to detect but they should be common in the neutron star population.

\paragraph{Magnetars.}

Neutron star emission can also be powered by magnetic energy. This is the case for the so-called magnetars, historically also known as \acf{SGRs} and \acf{AXPs} \citep{Woods2006}, which populate the upper-right corner of the $P-\dot{P}$ diagram (\textit{green triangles}). Magnetars have incredibly strong magnetic fields estimated to exceed $\unit[10^{14}]{G}$. This immense magnetic energy reservoirs often result in sporadic bursts of X-rays and gamma-rays which are thought to be caused by the magnetic field decay and significant magnetic field readjustments in the star's crust, leading to starquakes and intense releases of energy \citep[see][for reviews]{Kaspi2017, Esposito2021}.

\paragraph{Thermally emitting neutron stars.}

Some neutron stars emit thermal radiation, primarily in the form of X-rays, due to their high surface temperatures. These objects are collectively referred to as thermally emitting neutron stars (\textit{yellow triangles}). They include two main subclasses: \acf{CCOs} and the \acf{XDINSs}. \acs{CCOs} are found near the centres of supernova remnants and therefore should be young neutron stars with ages not exceeding a few tens of kyrs. However, they have low $\dot{P}$ values which translate into a low inferred magnetic fields in the dipole braking scenario. A possible explanation for this could be that their fields have been buried during an episode of fallback accretion of debris after the supernova explosion \citep{Shabaltas2012, Vigano2012b, TorresForne2016}. \acs{XDINSs} are also called the magnificent seven and are among the closest neutron stars that we know (few hundred parsecs from the Earth) \citep[see][for reviews]{vanKerkwijk2007, Turolla2009}. Only six of them have measured timing properties. Their thermal emission comes directly from the neutron star surface and is mainly powered by the residual heat stored in their interior and by Ohmic dissipation of the magnetic field in the neutron star crust. Therefore the study of thermally emitting neutron stars provides valuable insights into the cooling mechanisms and interior composition of neutron stars \citep{Vigano2013}.

\vspace{5mm}

The strict distinction between these different neutron star classes has started to be questioned after some recent discoveries.
For example, the dichotomy between rotational-powered pulsars and magnetars has been softened because magnetar-like X-ray activity was found in pulsars with large rotational energy loss rates, such as PSR J1846--0258 \citep{Gavriil2008} and PSR J1119--6127 \citep{Archibald2016}. Moreover pulsed radio emission has now been detected from several magnetars in outburst like XTE J1810--197 \citep{Camilo2006},  AXP J1810--197 \citep{Kramer2007} and J1818.0--1607 \citep{Esposito2020}. 

Furthermore recent radio surveys, in particular thanks to new radio interferometers such as the LOw Frequency ARray \citep[LOFAR;][]{VanHaarlem2013}, MeerKAT \citep{Jonas2009}, The Australian SKA Pathfinder \citep[ASKAP;][]{Hotan2021}, and the Murchison Widefield Array \citep[MWA;][]{Tingay2013, Wayth2018}, have started to uncover the existence of a new population of mysterious radio sources with very long period that challenge our understanding of the pulsar population and its evolution.

\subsection{Long-period pulsars}

Until recently, the only neutron star known with a very long spin period was the X-ray emitting neutron star \rcw\ at the centre of the $\unit[2]{kyr}$-old supernova remnant RCW103. This source manifests a measured modulation of $\sim \unit[6.67]{hr}$, and was responsible of a large magnetar-like X-ray outburst \citep{Rea2016, D'Ai2016b} which demonstrated the source's isolated magnetar nature despite its long period and young age. 

However, in the last couple of years a $\sim\unit[76]{s}$ radio pulsar \citep[\mtp,][]{Caleb2022} and two peculiar coherent radio sources with a periodicity of $\sim\unit[1091]{s}$ \citep[\gleamfirst;][]{Hurley-Walker2022a} and $\sim \unit[1318]{s}$ \citep[GPM J1839--10][]{Hurley-Walker2023} have been discovered. 
The nature of these two latter sources is still debated. Their emission properties such as the variable and irregular pulse profile, the high level of polarisation and the estimated luminosity exceeding their rotational power seem to point to a magnetar origin. Another proposed interpretation is that these long-period sources are magnetised white dwarfs with an active magnetosphere able to emit pulsar-like coherent radio emission.  

We will explore possible scenarios to produce long period pulsars, by considering spin-down induced on a neutron star by the interaction with supernova fallback matter (see Chapter~\ref{Chapter3}) and performing population synthesis studies of neutron stars and magnetic white dwarfs (see Chapter~\ref{Chapter4}).
These discoveries overall are showing that the boundaries between the different neutron star classes are blurring and motivate theoretical frameworks that consider the different neutron star types in an evolutionary scenario.

%----------------------------------------------------

\section{The birth rate and the need for an evolutionary scenario} \label{sec:ch1_birthrate}

As neutron stars are born from the core collapse of massive OB stars (see Section~\ref{sec:ch1_formation}), their birth rate should not exceed the expected \acf{CCSN} rate in the Milky Way.
The Galactic \acs{CCSN} rate can be estimated via different methods:
\begin{itemize}
    \item \citet{Reed2005} performed a census of the massive stars (with mass $M \gtrsim 10$ \msun) in the solar neighbourhood. From this sample, they extrapolated a Galactic \acs{CCSN} rate between 1 and 2 supernovae per century.
    \item \citet{Li2011} studied the \acs{CCSN}e in the local Universe and by assuming that the Milky Way has similar properties as other galaxies with the same morphological type they derived a \acs{CCSN} rate of $2.30 \pm 0.48$ per century.
    \item \citet{Diehl2006} modelled the gamma-ray emission from radioactive $^{26}$Al in the Milky Way by assuming that this emission traces the ongoing nucleosynthesis pollution by \acs{CCSN}e in the Milky Way. They estimated in this way a \acs{CCSN} rate of $1.9 \pm 1.1$ per century in our Galaxy. 
\end{itemize}

By combining these different ways of estimating the Galactic \acs{CCSN} rate, a recent study from \citet{Rozwadowska2021} found a best estimate of $\mathcal{R}_{\rm CCSN} = 1.79 \pm 0.55$ per century. The birth rate of Galactic neutron star should therefore be compatible with this estimate and not exceed this value.

In the past, several works have used the population of radio pulsars to make estimates of their birth rate.
Historically, pulsar population analyses used the fact that the typical timescales on which neutron stars are active as radio pulsars are much shorter than the age of the Galaxy. In this way, the population can be considered in a steady state and can be viewed as "flowing" in the $P-\dot{P}$ diagram as a "current" obeying a continuity equation \citep{Phinney1981, Vivekanand1981}. At a given spin period $P'$, this current, $J(P')$, equals the pulsar birth rate, $\mathcal{R}_{\rm birth}$, minus the pulsar death rate, $\mathcal{R}_{\rm death}$, in the period range from zero to the actual spin period $P'$, i.e.: 
%------------------------------------------------------------------------
\begin{align}
    J(P') = \mathcal{R}_{\rm birth}(0 < P < P') - \mathcal{R}_{\rm death}(0 < P < P').
\end{align}
%------------------------------------------------------------------------
As the known radio pulsar population is flux-limited (depending on the discovery survey), the maximum value of $J(P)$ provides a lower limit on the pulsar birth rate. For example, \citet{Lorimer2006}, using 1008 non-recycled pulsars detected by the Parkes Low-Latitude Survey estimated a birth rate of $1.38 \pm 0.21$ per century with this method. The advantage of this approach to compute birth rates is that it is agnostic to the details of how pulsars evolve in time.

A second approach to estimate pulsar birth rates is using population synthesis, discussed in detail below. Contrary to the previous approach, population synthesis relies on assuming physical models to describe pulsar properties and how the pulsar population evolves in time. 
For example, \citet{Faucher2006} use Monte Carlo simulations to model the birth properties of pulsars (velocity distributions, spin periods, magnetic fields) and the detectability in the Parkes and Parkes Swinburne Multibeam radio pulsar surveys. They evolve the initial population in time to obtain an observed synthetic pulsar sample to compare with observational data. In this way, they obtain a higher pulsar birth rate of $2.8 \pm 0.5$ per century compared to \citet{Lorimer2006}. 

We note that even without considering the existence of other neutron star classes, we are already at the limit of being compatible with the \acs{CCSN} rate. \citet{Keane2008} highlight that if one takes into account the birth rates estimated for other classes of neutron stars like \acs{RRATs}, \acs{XDINSs} and magnetars (see Section~\ref{sec:ch1_ns_zoo}), the total neutron star birth rate would exceed the Galactic \acs{CCSN} rate by a factor of $\sim 5 - 10$. Furthermore, the recent discovery of very long-period radio sources (see Chapters~\ref{Chapter4} and~\ref{Chapter5}) will further increase this discrepancy if their origin is associated with neutron stars. 
A possible solution is to consider an evolutionary scenario that introduce relationships between the different neutron star classes. Indeed by taking into account the coupled evolution of the magnetic field and the temperature in neutron stars, \citet{Vigano2013} explain the emission from magnetars, XDINs, CCOs and the X-ray emitting rotational powered pulsars as manifestations of neutron stars of different ages and with different magnetic-field strengths and geometries.

%----------------------------------------------------

\section{Population synthesis: current state of the art}
\label{sec:ch1_popsyn}

Population synthesis models aim to reproduce the observed population of neutron stars by simulating their formation, evolution, and observable properties. 
Typically the adopted approach relies on Monte Carlo simulations to model the different properties of neutron stars from some parametric distributions. These properties are then evolved in time through analytical and numerical models. Finally, models for the observational biases and filters are applied to assess which neutron stars can be detected to build a synthetic observed population.
The simulation outcomes are then compared to the empirical observed data to assess the validity of theoretical models, constrain model parameters and gain insights into the underlying physics.

Due to the abundance of data in the radio band, many works on population synthesis have mainly focused on reproducing the radio pulsar population, in particular trying to constrain the birth spin-period and magnetic-field distributions.
Some notable research papers in this area include \citet{Narayan1990, Faucher2006, Gonthier2007, Kiel2008, Kiel2009, Popov2010, Oslowski2011, Levin2013, Gullon2014, Bates2014, Cieslar2020}.
Overall these works adopt various prescriptions for the physical models and techniques to perform parameter optimisation. In Table~\ref{tab:ch1_pop_syn} I summarise the main prescriptions used by some reference works and compare them with the present work (see Chapters~\ref{Chapter5} and~\ref{Chapter6}). For example, \citet{Narayan1990} and \citet{Faucher2006} used the $\chi^2$ metric and Kolmogorov-Smirnov (KS) tests to compare the simulated distributions with the observed ones and guide a "by hand" parameter optimisation. These methods manifest some limitations as the computation of $\chi^2$-statistics relies on having a good amount of data while the KS-test is not particularly sensitive when applied to multi-dimensional distributions of data.
In recent years, improved modelling of neutron star magnetic field evolution and decay, refined cooling models, increased amounts of empirical data and the development of Bayesian techniques like Markov Chain Monte Carlo (MCMC) have all contributed to refining population synthesis codes. Current research efforts aim to combine the information coming from multi-wavelength observations, to better constrain the properties of the neutron star population as a whole and refine theoretical models.
A first attempt to include a more realistic magnetic field evolution based on magneto-thermal simulations of neutron stars and to consider both the radio and the X-ray pulsar population in a population synthesis study was performed by \citet{Popov2010, Gullon2014, Gullon2015}. In particular \citet{Gullon2014, Gullon2015} employed annealing methods for optimisation but found degeneracies between the simulation parameters were found and it was difficult to put meaningful constraints on the birth rate, especially for magnetars. \citet{Johnston2020} performed a population synthesis study combining both radio and gamma-ray pulsars and showed that highly energetic gamma-ray pulsars and young radio pulsars might come from the same underlying population. 
\citet{Cieslar2020} reproduce the observed radio pulsar population using MCMC, although using a simulation framework with simplified physical models.

This thesis work represents the first efforts to developing a population synthesis approach for isolated neutron stars in conjunction with modern deep-learning techniques. Indeed deep learning can be a powerful tool for extracting relevant features from multi-dimensional data coming from different surveys and performing parameter inference (Chapter~\ref{Chapter2}). In this work, we will focus on the radio pulsar population to benchmark our population synthesis approach and compare the results with previous ones in the literature. In particular, we will focus on predicting the birth properties, i.e., spatial and kick-velocity distributions (Chapter~\ref{Chapter5}) and initial spin-period and magnetic-field distributions and magnetic-field evolution (Chapter~\ref{Chapter6}). In Chapter~\ref{Chapter3} and~\ref{Chapter4} we will instead investigate the nature of recently discovered long-period radio sources assuming a neutron-star or white-dwarf origin scenarios. 

We first turn our attention to machine learning, introducing the key concepts crucial for the following chapters.

%-------------- TABLE -----------------------------------------------
\begin{landscape}

	\begin{table}
		\centering
		\caption[Comparison of population-synthesis works.]{Comparison between this work and several population-synthesis studies in the literature. We compare the following ingredients, which are given as individual table rows: the distributions of sources in the Galactic plane, $\mathcal{P}(r, \phi)$, and along Galactic heights, $\mathcal{P}(z)$, in cylindrical galactocentric coordinates; the inclusion of Galactic rotation and, if present, the corresponding rotation period, $T$; the distribution of neutron-star kick velocities, $\mathcal{P}(v_{\rm k})$; the distributions of initial dipolar magnetic-field strengths and initial periods, i.e., $\mathcal{P}(B_0)$ and $\mathcal{P}(P_0)$, as well as the prescriptions for their evolution (denoted as $B(t)$ and $P(t)$, respectively); the treatment of the misalignment-angle evolution, $\chi(t)$; the description of the radio beaming, where pulsars that intercept our line of sight are either determined with an \textit{empirical} relation between the beaming fraction and the period obtained from polarization data \citep{Tauris1998} or with a \textit{geometry-dependent} approach that considers the radio beam aperture and the inclination angle, $\chi$. We further provide information on the luminosity (distinguishing between \textit{pseudo} and \textit{intrinsic} luminosities), the respective surveys used for comparison and, finally, the method used to contrast simulated and observed populations (where KS denotes the Kolmogorov--Smirnov test).   \label{tab:ch1_pop_syn}}
		\tiny
		\begin{tabular}{c | c c c c c}
			%\multicolumn{5}{c}{} \\
			\toprule 
				& \tabhead{\citet{Faucher2006}} & \tabhead{\citet{Bates2014}} & \tabhead{\citet{Gullon2014, Gullon2015}} & \tabhead{\citet{Cieslar2020}} & \tabhead{This work} \\
			\midrule
			\multirow{2}{*}{$\boldsymbol{\mathcal{P}(r, \phi)}$} & spiral arms, & spiral arms, & spiral arms, & spiral arms, & $e$-density model \\
			& $\mathcal{P}(r)$ & $\mathcal{P}(r)$ & $\mathcal{P}(r)$ & $\mathcal{P}(r)$ & \citet{Yao2017} \\
			\midrule
			$\boldsymbol{\mathcal{P}(z)}$ & exponential & exponential & exponential & exponential & exponential \\
			\midrule
			\textbf{Galactic} & \multirow{2}{*}{-} & \multirow{2}{*}{-} & \multirow{2}{*}{-} & \multirow{2}{*}{-} & \multirow{2}{*}{$T \approx \unit[250]{Myr}$} \\
			\textbf{rotation} & & & & & \\
			\midrule
			$\boldsymbol{\mathcal{P}(v_{\rm k})}$ & exponential & exponential, normal & exponential & Maxwell & Maxwell \\
			\midrule
			$\boldsymbol{\mathcal{P}(B_0)}$ & log-normal & log-normal & log-normal & log-normal & log-normal \\
			\midrule
			$\boldsymbol{\mathcal{P}(P_0)}$ & normal & normal, log-normal & normal & normal & log-normal \\
			\midrule
			\multirow{3}{*}{$\boldsymbol{B(t)}$} & \multirow{3}{*}{-} & \multirow{3}{*}{-} & magneto-thermal models & exponential & magneto-thermal models \\
			& & & \citet{Vigano2013} & decay & \citet{Vigano2021}, \\
			& & & & & late-time power law \\
			\midrule
			$\boldsymbol{P(t)}$ & vacuum dipole & vacuum dipole & plasma-filled dipole & vacuum dipole & plasma-filled dipole \\
			\midrule
			$\boldsymbol{\chi(t)}$ & - & exponential & $P$-$\chi$ coupled & - & $P$-$\chi$ coupled \\
			\midrule
			\multirow{2}{*}{\textbf{Beaming}} & \multirow{2}{*}{empirical} & empirical, & \multirow{2}{*}{empirical} & \multirow{2}{*}{empirical} & \multirow{2}{*}{geometry-dependent} \\
			& & geometry-dependent & & & \\
			\midrule
			\multirow{2}{*}{\textbf{Luminosity}} & pseudo, & pseudo, & pseudo, & pseudo, & intrinsic, \\
			& $\propto \dot{E}_{\rm rot}^{\epsilon}$ & $\propto P^{\alpha} \dot{P}^{\beta}$ & $\propto \dot{E}_{\rm rot}^{\epsilon}$ & $\propto \dot{E}_{\rm rot}^{\epsilon}$ & $\propto \dot{E}_{\rm rot}^{\epsilon}$ \\
			\midrule
			\multirow{2}{*}{\textbf{Surveys}} & \multirow{2}{*}{PMPS, SMPS} & \multirow{2}{*}{PMPS, SMPS} & PMPS, SMPS & \multirow{2}{*}{PMPS} & \multirow{2}{*}{PMPS, SMPS, HTRU} \\
			& & & + X-ray pulsars (2015 study) & & \\
			\midrule
			\multirow{2}{*}{\textbf{Comparison}} & KS test, & \multirow{2}{*}{KS test} & annealing method, & \ac{MCMC} with & \multirow{2}{*}{\ac{SBI}} \\
			& by eye & & KS test & Gaussian likelihood & \\ 
			\bottomrule
		\end{tabular}
	\end{table}
\end{landscape}
%---------------------------------------------------------------

%--------------------------------------------------------------------------------------------

%=============================================================================

%--------------------------------------------------------------------------------------------
%\include{Chapters/Chapter_NS_theory}
% Chapter 2

\chapter{Machine learning} % Main chapter title

\label{Chapter2} % For referencing the chapter elsewhere, use \ref{Chapter2} 

%----------------------------------------------------------------------------------------

\section{Introduction}

In the past decade, the accumulation of extensive and heavy datasets has been almost ubiquitous in astronomy and astrophysics. In order to take full advantage of these data and perform data-driven science that complements ongoing theoretical modelling efforts, new techniques and analysis pipelines that can handle these large amounts of data (and do so in an automated way) are required. Machine learning (\acs{ML}) has played an important role in developing such new algorithms \citep{Ball2010, Allen2019, Baron2019, Fluke2020}.
As a subfield of artificial intelligence, machine learning involves the development of algorithms that allow computers to learn from and make predictions or decisions based on data without being explicitly programmed.

Depending on the task at hand, there are two different approaches to train machine-learning algorithms.
The first one is called \textit{supervised learning} and involves training a machine-learning model using labelled training data. In this approach, the dataset consists of input variables (features) and their corresponding output variables (labels). The goal is to learn a mapping function that predicts the labels for new, unseen data. 
In general to acquire good performance, a supervised learning algorithm requires a large amount of labelled data.
Depending on the task, these algorithms can be categorised into two main types:
\begin{itemize}
    \item \textbf{Classification}: This type of algorithm aims to predict discrete class labels for the given input data. An example is the prediction of the morphological type of a galaxy based on its image \citep[e.g.][]{Dieleman2015}.
    \item \textbf{Regression}: The goal of these algorithms is to predict continuous numerical values based on the input features. An example is the prediction of the star formation rate of a galaxy based on its redshift, luminosity and colour \citep[e.g.][]{Bonjean2019}.
\end{itemize}
Among the most common supervised learning algorithms, are decision trees \citep{Quinlan1986}, support vector machines \citep{Burges1998}, and artificial neural networks \citep{Bishop1995}.

The second approach is \textit{unsupervised learning}. In contrast to supervised learning, unsupervised learning involves training a model on unlabelled data. Here, the dataset only consists of input data without any corresponding output labels. The objective of unsupervised learning is to find patterns, structures, or relationships within the data. In principle unsupervised learning algorithms can potentially work with smaller amount of data.
They can be categorised into two main types:
\begin{itemize}
    \item \textbf{Clustering}: These algorithms aim to group similar data points based on their inherent similarities or distances in a given representation space. The goal is to discover hidden patterns or groupings within the dataset. For example \citet{Sanchez2013} used clustering to classify stellar spectra.
    \item \textbf{Dimensionality reduction}: These algorithms seek to reduce the number of input variables by transforming the data into a lower-dimensional representation. This helps in visualising and compressing complex datasets by focusing on their most relevant properties. Usually these techniques are used as a pre-processing step. For example they have been employed to extract physical parameters of stellar atmospheres from their spectra \citep[e.g.][]{ReFiorentin2007}. 
\end{itemize}
Examples of unsupervised learning algorithms include k-means clustering \citep{MacQueen1967}, principal component analysis \citep[PCA,][]{Hotelling1933}, and isomap embedding \citep{Tenenbaum2000}.

In the following, for the purpose of this thesis, I will focus on supervised machine-learning techniques using artificial neural networks to perform regression tasks.

%==========================================================================

\section{Deep learning} \label{sec:ch2_deep_learning}

Deep learning is a subset of machine learning that uses artificial neural networks formed of multiple layers of interconnected nodes \citep[see][for a review]{Goodfellow2016}. In analogy with biological brains, each node can be thought of as a neuron while the connections between nodes are the synapses. The importance of a connection is defined by a weight, $w$, represented by a real number. Each node receives weighted signals from other connected neurons, processes them and transmits the processed signal to other connected nodes. In this way, artificial neural networks are able to process a given input signal, $\boldsymbol{x}$, by decomposing it into more complex features and produce a final output, $\boldsymbol{y}$. Therefore, they can be regarded as a function $\boldsymbol{y} = F_{\boldsymbol{w}}(\boldsymbol{x})$ where the set of weights, $\boldsymbol{w}$, is adjusted to produce a desired output, $\boldsymbol{y}$, given an input, $\boldsymbol{x}$. The employment of deep neural networks enables the extraction of intricate features and representations from the input data, leading to refined predictions for classification or regression tasks.
In the following, I will describe several basic network architectures that are of concern for this work but that can also be used for a great variety of problems. I will specifically focus on the multi-layer perceptron and the convolutional neural network. In preparation for the following chapters, I will also describe how the training process works (Section~\ref{sec:ch2_training_process}).

\subsection{Multi-layer perceptron}
\label{sec:ch2_MLP}

%----------------------------------------------------
\begin{figure}
\centering
\includegraphics[width=0.5\textwidth]{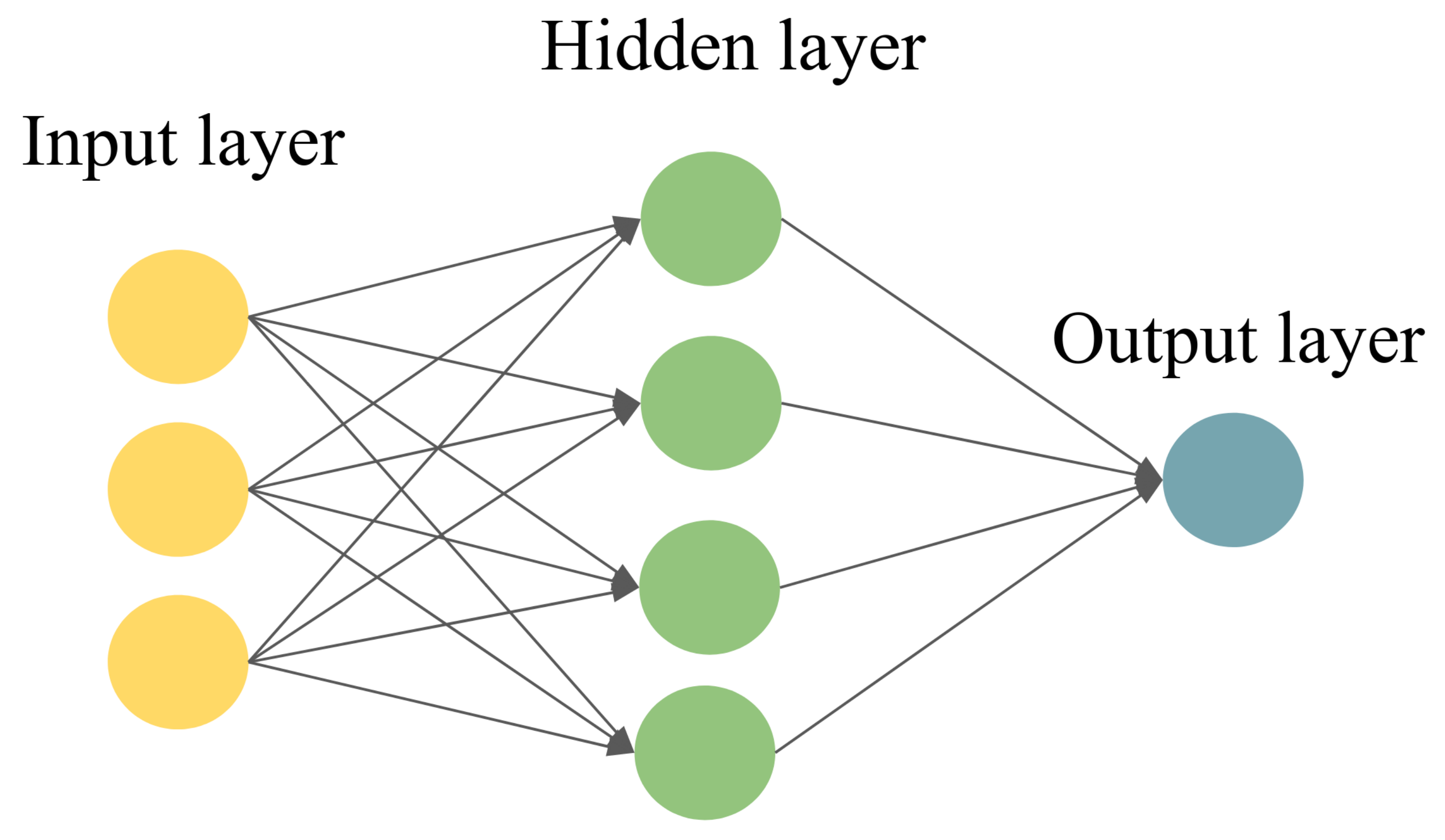}
\includegraphics[width=0.8\textwidth]{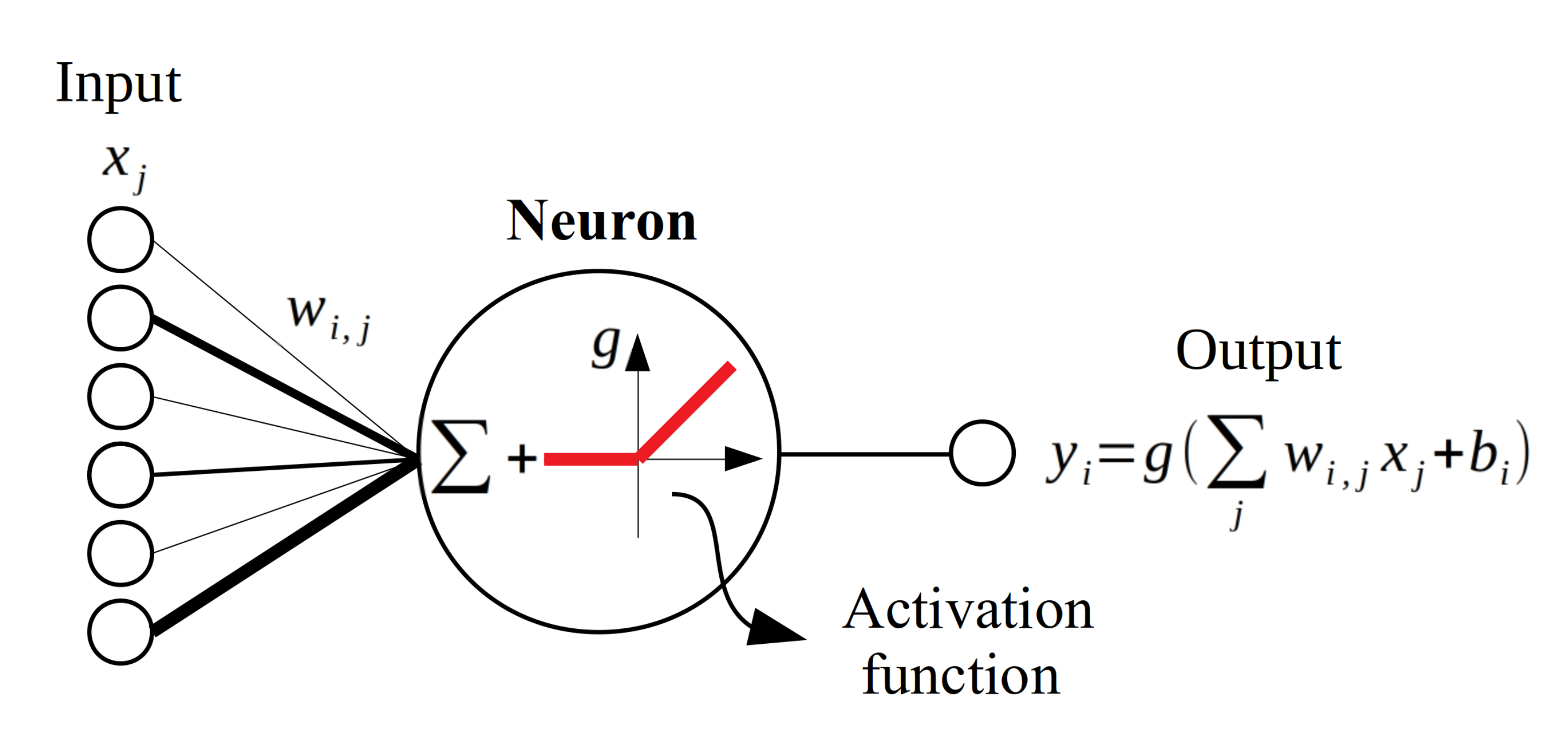}
\caption[Sketch of and an artificial neuron, the perceptron]{Sketch of a simple fully-connected feedforward neural network with one hidden layer (top panel) and of a neuron in a multi-layer perceptron (bottom panel). To produce an output $y_i$, each neuron first perform a linear combination of the input values $x_j$ and then applies a non linear activation function $g$.}
\label{fig:ch2_MLP}
\end{figure}
%----------------------------------------------------

The \acf{MLP} was originally conceived as a binary classifier \citep{Rosenblatt1958, Minsky1969} but the name today is also used to denote modern so-called fully-connected, deep, feedforward neural networks.
In this type of architecture, the neurons are organised in layers and the information flows unidirectionally from the first to the last layer (see top sketch in Figure~\ref{fig:ch2_MLP}). The first layer is denoted as the input layer, the last one as the output layer and the ones in the middle as hidden layers.
The nodes in a given layer are not connected and work in parallel, but the layers are fully connected to each other meaning that each neuron receives inputs from all nodes in the previous layer and passes its output to all nodes in the subsequent layer. The $i$-th neuron in a given layer performs a linear combination of the array of signals, $\boldsymbol{x}$, coming from the previous layer according to the weights, $\boldsymbol{w}_i$, associated with its connections and a bias term, $b_i$ (see bottom sketch in Figure~\ref{fig:ch2_MLP}): 
%-----------------------------------------------------------------------
\begin{align}   \label{eq:ch2_linear_layer}
    z_i = \boldsymbol{w}_i^{\rm T} \boldsymbol{x} + b_i = \sum_{j=1}^{n} w_{i,j} x_j + b_i,
\end{align}
%-----------------------------------------------------------------------
where $n$ is the number of neurons in the previous layer and the superscript "T" denotes the transposed vector.
The bias term is responsible of introducing an offset in the output value.
By adding an extra term to the input $\boldsymbol{x}$ and setting it to 1, we can write the linear combination above in a more compact form:
%-----------------------------------------------------------------------
\begin{align}
    z_i = \boldsymbol{w}_i^{\rm T} \boldsymbol{x} = \sum_{j=0}^{n} w_{i,j} x_j,
\end{align}
%-----------------------------------------------------------------------
where we adopt the convention that $x_0 = 1$, so that $b_i = w_{i,0} x_0$ becomes the bias term. In the following, I will also use the term weight to refer to the bias terms for simplicity.
Before passing the processed signal to all the neurons in the following layer a non-linear activation function, $g$, is usually applied to produce an output $y_i=g(z_i)$ (see Figure~\ref{fig:ch2_MLP}). Non-linear activation functions are used in neural networks to introduce non-linearity into the model (see Section~\ref{sec:ch2_activation_function}). This is crucial to enable the neural network to learn complex, non-linear patterns and relationships between input and output variables.
Therefore, in an \acs{MLP} with a total of $L$ layers, each layer represents a function $F^{(l)}$ with $l=1, \,...\, L$ that performs a linear combination of the input and applies a non-linear function $g$. Therefore by passing through the network the input signal $\boldsymbol{x}$ undergoes a composition of functions until the final output, $\boldsymbol{y}$ is computed:
%-----------------------------------------------------------------------
\begin{align}
\boldsymbol{y} = F^{(L)}(F^{(L-1)}...(F^{(2)}(F^{(1)}(\boldsymbol{x}))))
\end{align}
%-----------------------------------------------------------------------
The output layer will give back the prediction for either the regression or classification task, depending on the specific problem.

\subsection{Convolutional neural network}
\label{sec:ch2_CNN}

Convolutional neural networks (CNNs) \citep{LeCun1989} are deep-learning algorithms primarily used for computer vision tasks like image classification, object detection, and segmentation \citep{Krizhevsky2012}. They take inspiration from the visual cortex of living organisms and are designed to work with structured, grid-like data that are usually referred to as feature maps. 
%Indeed fully connected deep neural networks are not suitable to process multi-dimensional inputs, since the number of required neurons and connections will scale up exponentially with the input size.

\begin{figure}
\centering
\includegraphics[width =0.6\textwidth]{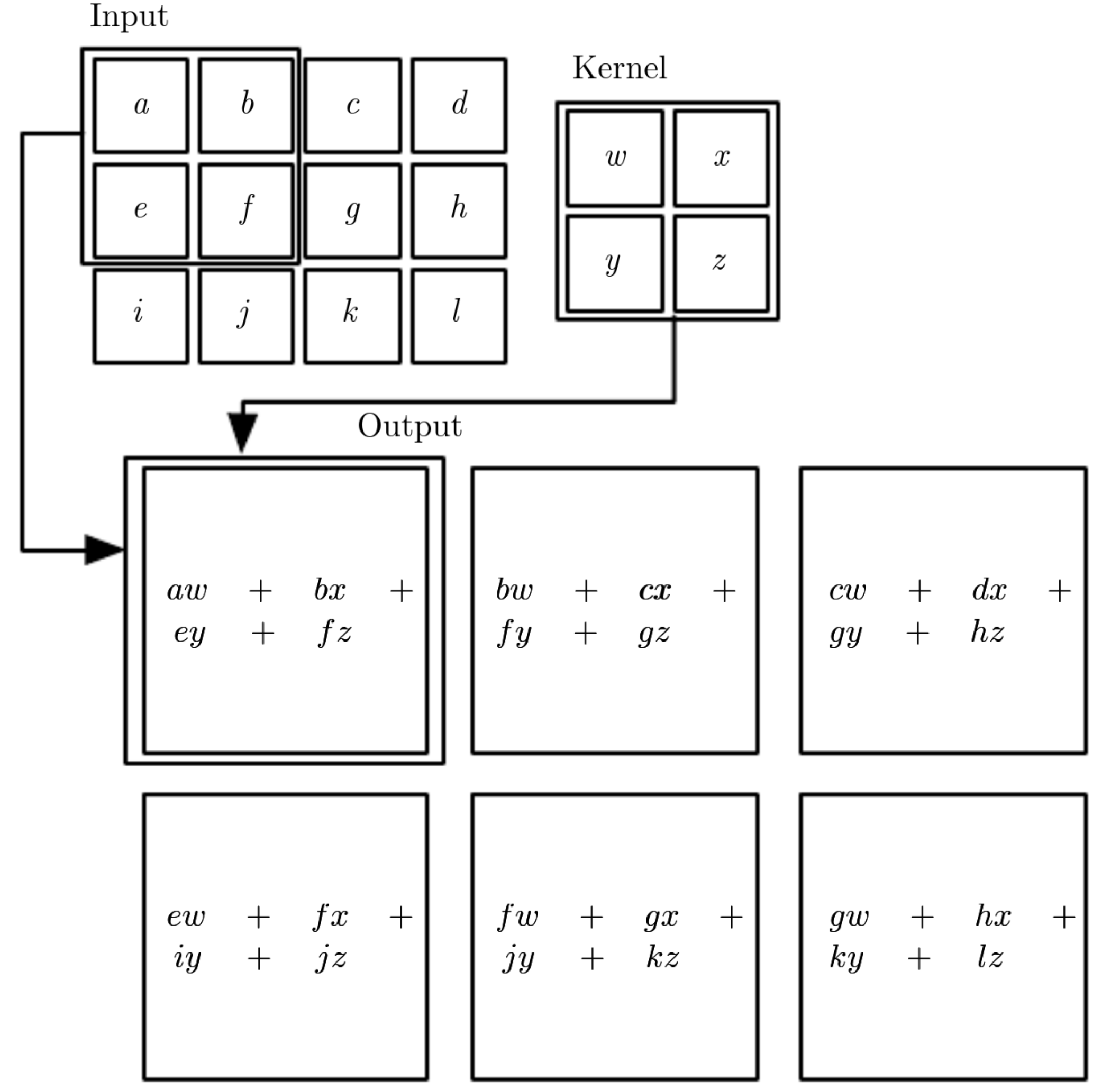}
\caption[Convolution operation]{An example showing how the convolution (cross-correlation in this case) operation works for a 2-D input. We draw boxes with arrows to indicate how the upper-left element of the output feature map is formed by applying the kernel to the corresponding upper-left region of the input feature map \citep[Figure taken from][]{Goodfellow2016}.}
\label{fig:ch2_convolution} 
\end{figure}

\acs{CNN}s are composed of multiple layers, including so-called \textit{convolutional layers}, pooling layers, and fully-connected layers. 
The key component is the convolutional layer, which performs the convolution operation on the input feature map using a set of filters or kernels composed of learnable weights. For example, Let us consider a two-dimensional feature map as input $\boldsymbol{x}$ and a two-dimensional kernel $\boldsymbol{w}$. We can define the convolution operation $\boldsymbol{w} * \boldsymbol{x}$ over a discretised grid as \citep[see Chapter 9 in][]{Goodfellow2016}:
%-----------------------------------------------------------------
 \begin{align}   \label{eq:ch2_convolution}
	z_{i,j} = (\boldsymbol{w} * \boldsymbol{x})_{i,j} = \sum_{m = -\infty}^{+\infty} \sum_{n = -\infty}^{+\infty} w_{m,n} x_{i-m, \, j-n}.
\end{align}
%----------------------------------------------------------------
The result of this operation will be non-zero only where the feature map and the filter overlap.
However, nearly all machine learning and deep learning libraries implement the cross-correlation operation $\boldsymbol{w} \star \boldsymbol{x}$ instead and improperly call it convolution. This is defined as:
%-----------------------------------------------------------------
 \begin{align}   \label{eq:ch2_cross_correlation}
	z_{i,j} = (\boldsymbol{w} \star \boldsymbol{x})_{i,j} = \sum_{m = -\infty}^{+\infty} \sum_{n = -\infty}^{+\infty} w_{m,n} x_{i+m, \, j+n}.
\end{align}
%----------------------------------------------------------------
Note that the difference between Equation~\eqref{eq:ch2_convolution} and \eqref{eq:ch2_cross_correlation} consists only in the "$\pm$" signs in the indexes of $\boldsymbol{x}$ which flips the way we access the entries of the input feature map. For simplicity, we will continue calling both operations convolution.
In practice, the convolution operation involves sliding the filter over the input feature map, computing element-wise multiplications, and summing them up to produce a new convolution feature map $\boldsymbol{z}$ (see Figure~\ref{fig:ch2_convolution}).
%----------------------------------------------------
%----------------------------------------------------
These feature maps capture local patterns or features present in the input map such as edges or corners in an image.
Generally, the convolutional filters have a much smaller size if compared with the input feature map.
By using small filters that slide over the input map each output value in the convolutional feature map depends only on a small subset of input values. This reduces the number of computations required compared to fully-connected layers where each input value affects all output values. Indeed fully-connected deep neural networks are not suitable to process multi-dimensional inputs, since the number of required neurons and connections will scale up exponentially with the input size.
Furthermore, the same filter (set of weights) is used across different spatial locations of the input feature map. This sharing of parameters significantly reduces the number of parameters in the network and enables the model to learn spatially invariant features. By sharing parameters, the network can detect the same feature regardless of its location and orientation in the input.
Since a single filter can recognise only a specific type of feature, usually many filters are stacked together to form a convolution layer. This allows one to perform many convolution operations on the input map and to extract different features in parallel.

To introduce non-linearity, an activation function is usually applied to the feature maps generated after the application of a convolutional layer.

%----------------------------------------------------
\begin{figure}
\centering
\includegraphics[width=0.8\textwidth]{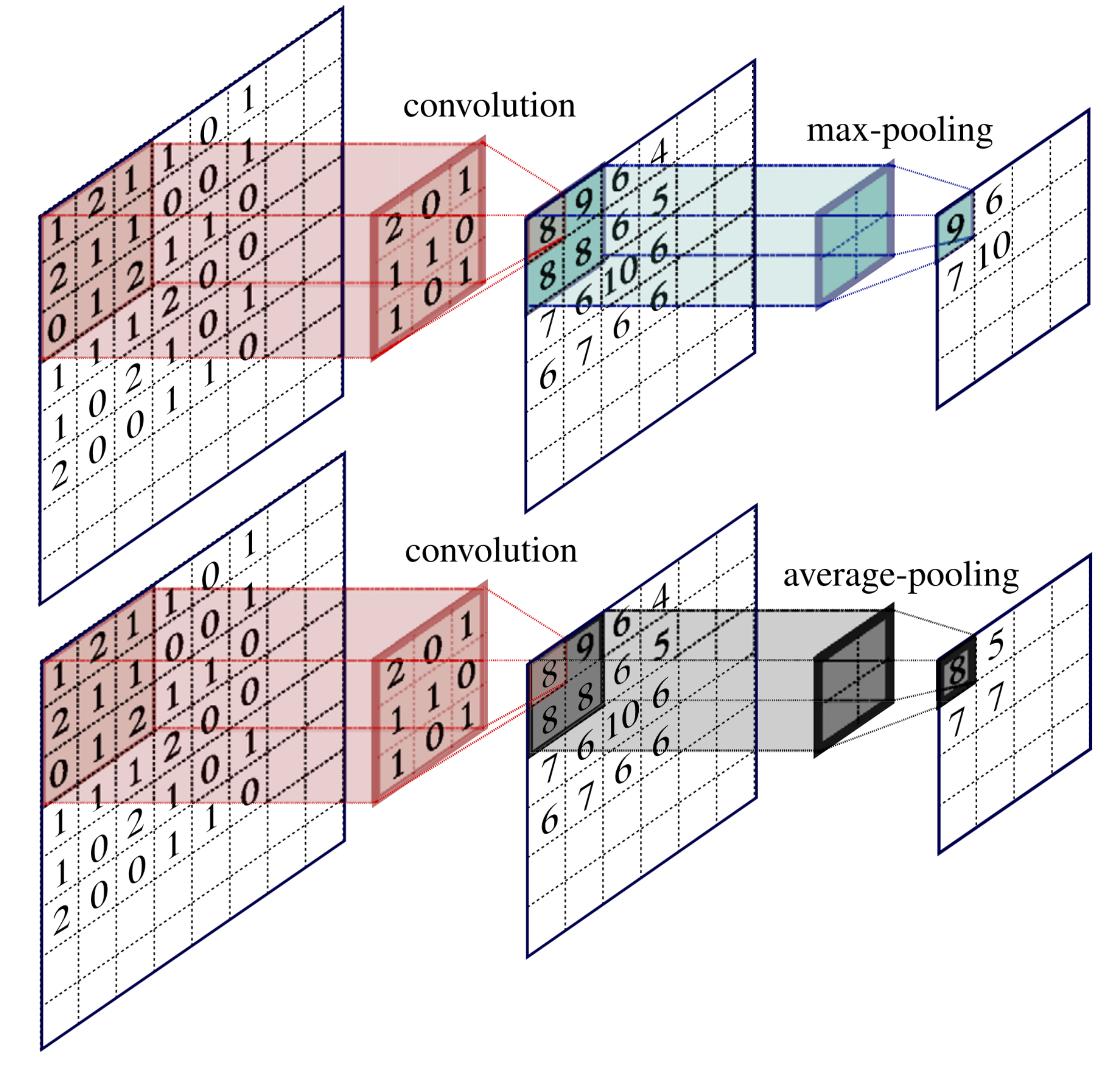}
\caption[Convolution and pooling operations]{Example showing how the convolution and pooling layers operate on the input data. First, the $8 \times 8$ input layer is reduced to a $6 \times 6$ output layer after going through a convolution layer consisting of a single filter with size $3 \times 3$ that moves along the input with a stride of 1. This output layer becomes the input layer with respect to either the max-pooling or average-pooling layer, each consisting of a single filter with size $2 \times 2$ that moves along the input with stride 2. The max-pooling layer picks the maximum number from the layer input within a selected window, while the average-pooling layer computes average values over the selected window. The final output layer is therefore a $4 \times 4$ array \citep[Figure taken from][]{Teo2021}.}
\label{fig:ch2_max_pool}
\end{figure}
%----------------------------------------------------
After the activation function has been applied to the feature maps, a common practice is to use \textit{pooling layers}.
The pooling operation performed by this kind of layers is specified, rather than learned. Two common functions used in the pooling operation are max-pooling and average-pooling.
Max-pooling and average-pooling layers consist of filters that slide on the feature map and select the maximum or compute the average value respectively inside the sub-region of the map that they overlap with (see Figure~\ref{fig:ch2_max_pool}). 
Both approaches are used to downsample the feature maps by reducing their spatial dimensions, thereby decreasing the computational complexity while preserving the most relevant information and the dominant features. In this way, these layers help reduce the influence of noisy or irrelevant information from the input.

Usually, deep \acs{CNN}s are formed by stacking together a series of convolutional layers composed of an increasing number of filters followed by pooling layers. This allows layers close to the input to learn low-level features (e.g. lines, edges, corners) and layers deeper in the model to learn high-order or more abstract features, like shapes or objects.

Finally, fully connected layers are added at the end of the \acs{CNN} to perform the tasks of classification or regression. They take the high-level features extracted by the previous layers and produce the final output.

\subsection{The activation function} \label{sec:ch2_activation_function}

%----------------------------------------------------
\begin{figure}
\centering
\includegraphics[width=0.6\textwidth]{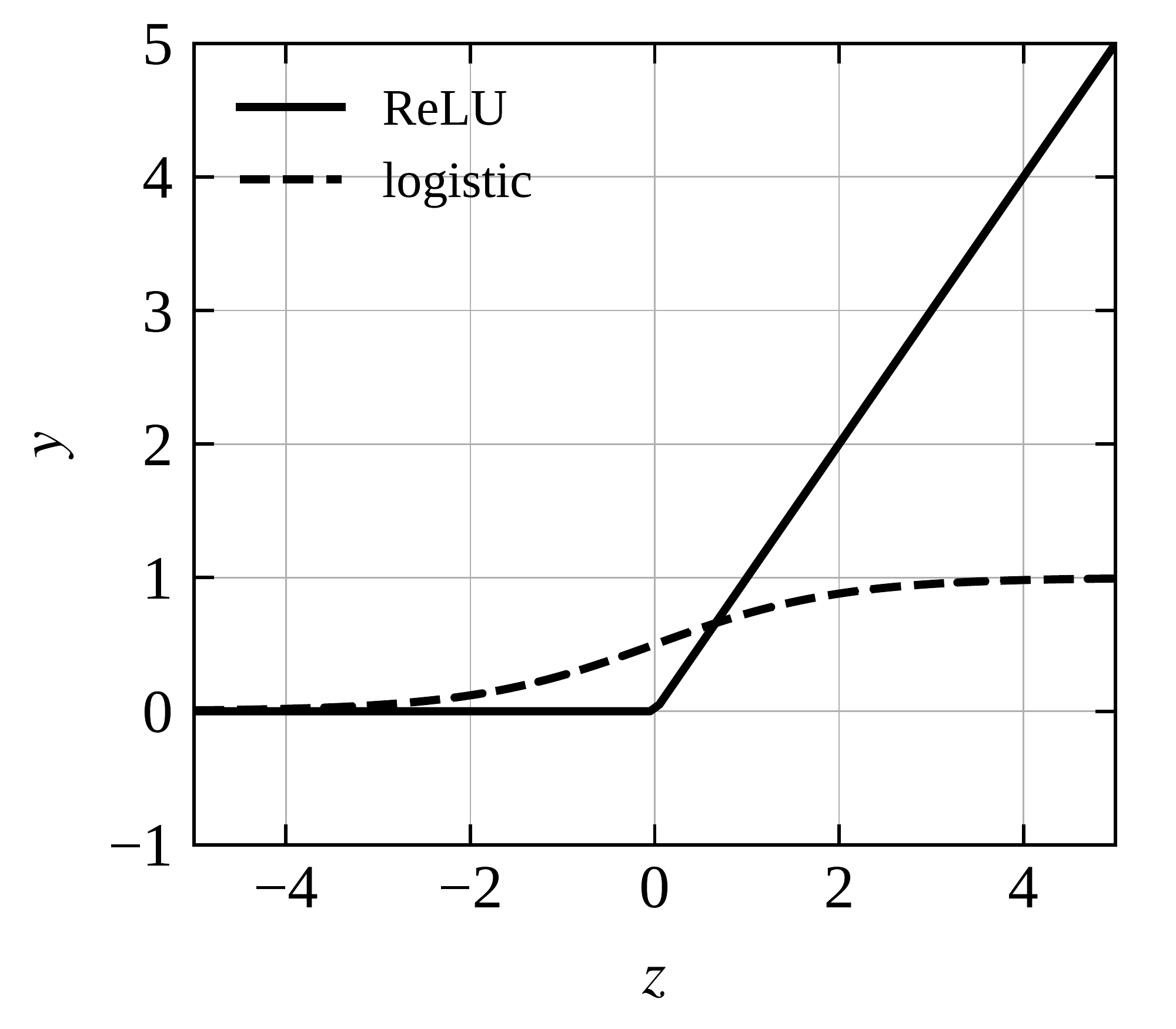}
\caption[Activation functions]{Comparison between the ReLU and the logistic (sigmoid) activation functions.}
\label{fig:ch2_activation_func}
\end{figure}
%----------------------------------------------------

Generally, the choice of the activation function depends on the problem and the type of data at hand. Besides introducing non-linearity the type of activation function is also important to implement a stable and effective learning process.
An example of activation function $g$ is the logistic (sigmoid) function:
%-----------------------------------------------------------------
 \begin{equation}	\label{eq:sigmoid}
	{\rm logistic}(z) = \frac{1}{1+e^{-z}}
\end{equation}
%----------------------------------------------------------------
This activation function is especially used in the output layer of neural networks employed as classifiers as it compresses the output to a value between 0 and 1 which can be interpreted as the probability to belong to a given class.
However as argued by \citet{Glorot2010, Glorot2011} using the sigmoid function in all the layers of deep neural networks can slow down the learning process as it is very sensitive to the input value $z$ especially when it is near 0. Moreover it saturates across most of its domain leading more easily to vanishing gradient issues (see Section~\ref{sec:ch2_training_process} and Figure~\ref{fig:ch2_activation_func}). 
Another common choice for the activation function is the \acf{ReLU} \citep{Glorot2010} which is defined as:
%-----------------------------------------------------------------
 \begin{align}	\label{eq:ch2_ReLU}
	{\rm ReLU}(z) = \max(0,z).
\end{align}
%----------------------------------------------------------------
One major benefit of using \acs{ReLU} is its simplicity (see Figure~\ref{fig:ch2_activation_func}). As we will see in the next section Section~\ref{sec:ch2_training_process}, the gradient calculation for ReLU is usually easier and more efficient compared to more complex activation functions (it is 0 for $z<0$ and always 1 for $z>0$), resulting in faster convergence during gradient-based optimisation. Indeed for this activation function, there is a reduced chance of its gradient to vanish contrary to other activation functions such as the sigmoid.
Furthermore by using the ReLU, some of the neurons are likely to be inactive (outputting zero) for negative input values. This sparse activation makes the network more efficient by reducing the number of active neurons and, in turn, the complexity and computational requirements \citep{Glorot2011}.

%==========================================================================

\section{The training process} \label{sec:ch2_training_process}

In supervised learning, a neural network is trained on a dataset containing input features $\boldsymbol{x}$ and the corresponding target labels $\hat{\boldsymbol{y}}$. By feeding the training data into the network, the training proceeds by adjusting the model parameters $\boldsymbol{w}$, i.e., the weights and biases, to minimise a loss function. The loss is a measure of the difference between the predicted outputs $\boldsymbol{y}$ and the expected target labels $\hat{\boldsymbol{y}}$. Its minimisation is achieved by computing the gradients of the loss function with respect to the weights and by performing gradient descent in a procedure called backpropagation (see Section~\ref{sec:ch2_backprop}).
Here I will describe the steps involved in the training process.

\subsection{Training, validation and test dataset}	\label{sec:ch2_datasets}

The labeled dataset that is available for training is usually divided into three separated subsets called training, validation, and test dataset, respectively. In machine learning, these are used for different purposes.

The training dataset is used to train a machine-learning model. It comprises a large portion (usually around 70\% to 80\%) of the original dataset. During the training phase, the model learns from this dataset by adjusting its parameters (i.e., the weights) to minimise the error between its predicted outputs and the actual target labels.

The validation dataset is used to evaluate the performance of the model during the training process. It usually constitutes a smaller portion (around 10\% to 20\%) of the original dataset. The validation dataset is not used to optimise the model parameters, but instead allows us to validate how well the model can generalise to unseen data. In general, it is used to fine-tune hyperparameters (like the learning rate, the size of input batches, etc., see below for an explanation), select the best model, and prevent overfitting, which refers to the model performing well on training data but poorly on unseen data.

Similar to the validation dataset, the test dataset is also a small fraction of the total dataset, usually around 10\%. However, contrary to the validation dataset, the test dataset is never seen by the model during the training process and it is not used during model development or hyperparameter tuning. Instead it is employed to assess the final performance and generalisation ability of the trained model on real-world data in an unbiased way. To ensure this, the test dataset is completely separate from the training and validation datasets and it should ideally have a similar distribution of data.

Before training, it is sometimes useful to apply some transformations to these datasets.
Some of these transformations include normalisation and standardisation of the input features $\boldsymbol{x}$ and in regression tasks also of the corresponding labels $\hat{\boldsymbol{y}}$. In general, the label $\hat{\boldsymbol{y}}$ can be a vector containing the different parameters to predict.
Normalisation is usually applied on every input feature separately. In particular, it performs the following operation:
%-------------------------------------------------------------
\begin{align}
    \boldsymbol{x}' = \frac{\boldsymbol{x} - \min{(\boldsymbol{x})}}{\max{(\boldsymbol{x})} - \min{(\boldsymbol{x})}}.
\end{align}
%-------------------------------------------------------------
where $\min(\boldsymbol{x})$ and $\max(\boldsymbol{x})$ are the minimum and maximum values of the input feature map $\boldsymbol{x}$ respectively.
This guarantees that the input values always fall in the range [0, 1]. 
The normalisation of the labels instead is performed on a dataset basis and for each specific label $\hat{y}$ in $\hat{\boldsymbol{y}}$ it consists of the following operation:
%-------------------------------------------------------------
\begin{align}
    \hat{y}' = \frac{\hat{y} - \min{(\hat{y})}}{\max{(\hat{y})} - \min{(\hat{y})}},
\end{align}
%-------------------------------------------------------------
where min and max are computed for each label parameter over the entire training dataset. In this way, all the labels are re-scaled to the range [0, 1] as well.
Standardisation instead consists of evaluating the average value $\langle \boldsymbol{x} \rangle$ and the standard deviation $\sigma(\boldsymbol{x})$ for each input feature $\boldsymbol{x}$ and then computing the z-scores by applying the following transformation: 
%-------------------------------------------------------------
\begin{align}
    \boldsymbol{x}' = \frac{\boldsymbol{x} - \langle \boldsymbol{x} \rangle}{\sigma(\boldsymbol{x})}.
\end{align}
%-------------------------------------------------------------
For the labels we standardise on a dataset basis and for each specific label $\hat{y}$ in $\hat{\boldsymbol{y}}$ we compute:
%-------------------------------------------------------------
\begin{align}
    y'_{\rm t} = \frac{\hat{y} - \langle \hat{y} \rangle} {\sigma(\hat{y})},
\end{align}
%-------------------------------------------------------------
where the mean and standard deviation are computed over the entire training dataset. Standardisation guarantees that the values of the input features and the labels are rescaled so that they have zero mean and unit variance.

Normalisation and standardisation aim to make the input features and targets have approximately the same range of values to improve the efficiency and effectiveness of the models' learning process. This is important because it prevents some variables from dominating others, which can lead to biased and ineffective learning. By normalising or standardising the input data, the model can therefore focus more accurately on the variations and patterns in the features rather than their absolute values.
These preprocessing techniques facilitate effective learning by reducing the impact of outliers, and by providing a consistent and balanced representation of the input data.

\subsection{Weights initialisation} \label{sec:ch2_w_initialization}

An important aspect of achieving good performance while training is the initialisation of the weights of the network. When defining the network architecture, the simplest approach is to initialise all the weights randomly in a given range of values. In this way at the beginning of the training process also the output of the network is random. However, if the weight values are too small, after a few layers the output values and the gradients will completely vanish. On the other hand, if weights are too big the output values and the gradients could explode. As a consequence, training will proceed very slowly or will stagnate.

To visualise this aspect let us consider an example with a fully-connected neural network. We focus on two neighboring layers in the network, $l-1$ and $l$, with layer $l-1$ having $n$ nodes. Suppose that the weights, $\boldsymbol{w}$, associated to the connections between these two layers, are initialised with values randomly sampled from a normal distribution with mean 0 and a generic variance ${\rm Var}(w)$ and the biases are initialised to be zero.

As shown in Equation~\eqref{eq:ch2_linear_layer}, the outputs of the nodes in layer $l$ are a linear combination of the input features $\boldsymbol{x}$ with weights $\boldsymbol{w}$, i.e., output $z$ is given by:
%-------------------------------------------------------------
\begin{align}
    z^{(l)} = \sum_{i=1}^{n} w_{i} x_i.
\end{align}
%-------------------------------------------------------------
The input features in turn are the result of the activation function applied to the output of the previous layer that is $x_i = y_i^{(l-1)} = g \left( z_i^{(l-1)} \right)$.
We can calculate the expectation value of $z^{(l)}$:
%-------------------------------------------------------------
\begin{align} \label{eq:ch2_Ez}
    {\rm E}(z^{(l)}) &= {\rm E} \left( \sum_{i=1}^{n} w_i x_i \right) \nonumber \\
                 &= n {\rm E}(w x) \nonumber \\
                 &= n {\rm E}(w) {\rm E}(x) = 0,
\end{align}
%-------------------------------------------------------------
where we treat the weights and the input features as two statistically independent random variables. Hence we dropped the index $i$ for clarity of notation and we used the fact that the average value of the weights ${\rm E}(w) = 0$.
Moreover for the variance of $z$ we find: 
%-------------------------------------------------------------
\begin{align} \label{eq:ch2_Varz}
    {\rm Var}(z^{(l)}) &= {\rm Var}\left(\sum_{i=1}^{n} w_{i} x_i \right) \nonumber \\
                   &= n {\rm Var}(w x) \nonumber \\
                   &= n [ {\rm Var}(w) {\rm Var}(x) + {\rm E}(w)^2 {\rm Var}(x) + {\rm Var}(w) {\rm E}(x)^2 ] \nonumber \\
                   &= n [ {\rm Var}(w) {\rm Var}(x) + {\rm Var}(w) {\rm E}(x)^2 ] \nonumber \\
                   &= n {\rm Var}(w) [{\rm Var}(x) + {\rm E}(x)^2 ] \nonumber \\
                   &= n {\rm Var}(w) {\rm E}(x^2),
\end{align}
%-------------------------------------------------------------
where in the third line we used the rule for the variance of a product of random variables \citep{Springer1979}. In the fourth line we took advantage of the fact that we define the weights to have zero mean and, in the last line we use the definition of the variance ${\rm Var}(x) = {\rm E}(x^2) - {\rm E}(x)^2$.
From this expression, we note that if the variance ${\rm Var}(w)$ and the expectation value ${\rm E}(x^2)$ are too big or too small, the variance ${\rm Var} \left( z^{(l)} \right)$ will explode or vanish after a few layers. 
Furthermore, it is worth noticing that as $x$ is the result of applying the activation function to $z^{(l-1)}$, the expectation value and variance of $x$ will, in general, differ from the ones of $z^{(l-1)}$ and they will depend on the type of activation function employed.

In the following we compute ${\rm E}(x^2)$ for the \acs{ReLU} activation function, that is considering $x = \max \left(0, z^{(l-1)} \right)$ assuming that the probability density functions for the variable $x$ and $z$ are $\mathcal{P}(x)$ and $\mathcal{P}(z)$ respectively:
%-------------------------------------------------------------
\begin{align}
    {\rm E}(x^2) &= \int_{-\infty}^{+\infty} x^2 \mathcal{P}(x) dx \nonumber \\
                   &= \int_{-\infty}^{+\infty} \max(0, z^{(l-1)})^2 \mathcal{P}(z) dz \nonumber \\
                   &= \int_{0}^{+\infty} \left( z^{(l-1)} \right)^2 \mathcal{P}(z) dz \nonumber \\
                   &= \frac{1}{2} \int_{-\infty}^{+\infty} \left( z^{(l-1)} \right)^2 \mathcal{P}(z) dz \nonumber \\
                   &= \frac{1}{2} {\rm Var}\left( z^{(l-1)} \right),
\end{align}
%-------------------------------------------------------------
where in the last step we recall that from Equation~\eqref{eq:ch2_Ez}, ${\rm E}(z) = 0$ in any layer.
Therefore, substituting in Equation~\eqref{eq:ch2_Varz} we find that:
%-------------------------------------------------------------
\begin{align}
    {\rm Var}\left( z^{(l)} \right) &= \frac{1}{2} n {\rm Var}(w) {\rm Var}\left( z^{(l-1)} \right).
\end{align}
%-------------------------------------------------------------
A good weight initialisation should ensure that ${\rm Var}\left( z^{(l)} \right) = {\rm Var} \left( z^{(l-1)} \right)$ which implies that the weights have a variance of:
%-------------------------------------------------------------
\begin{align}
    {\rm Var}(w) &= \frac{2}{n^{(l-1)}}.
\end{align}
%-------------------------------------------------------------
where we recall that $n^{(l-1)}$ is the number of nodes for the $l-1$ layer (i.e., the size of the input features for the $l$ layer). Notice that in the previous steps we did not choose a specific layer $l$. Thus, this expression holds for every layer of the network.
This is the initialisation procedure commonly adopted for models using the rectified activation function ReLU and is referred to as Kaiming initialisation \citep{Kaiming2015}. With this method, the initial weights between layers are sampled from a normal distribution with mean 0 and standard deviation $\sigma(w) = \sqrt{2 / n}$, where $n$ is the size of the input feature to each layer, while the biases are initially set to zero. This helps to reduce the problem of vanishing or exploding gradients and to make the training more stable and effective. In the following, we will adopt this initialisation procedure.

\subsection{Forward pass and loss calculation} 
\label{sec:ch2_forward}

During the forward pass, input data from the training dataset is passed through the network, and intermediate activations of neurons are computed using the current weights. This process starts from the input layer and progresses through the hidden layers until it reaches the output layer. The output of the network is then compared to the desired output. 
The distance between the network's output $\boldsymbol{y}$ and the expected target label $\hat{\boldsymbol{y}}$ is quantified using a loss function $\mathcal{L}(\boldsymbol{y}, \hat{\boldsymbol{y}})$, which measures the network's performance on a particular task. For regression tasks where one wants to predict some continuous real values, an example of a loss function is the mean squared error defined as:
%-----------------------------------------------------------------
 \begin{align}	\label{eq:ch2_MSE}
	MSE(\boldsymbol{y}, \hat{\boldsymbol{y}}) = \frac{1}{n^{\rm (o)}} \sum_{i=1}^{n^{\rm (o)}}(y_i - \hat{y}_i)^2,
\end{align}
%----------------------------------------------------------------
where $n^{\rm (o)}$ is the size of the output $\boldsymbol{y}$.

Usually, the training dataset is passed through the network in batches composed of more than one sample. By passing a single training sample at a time, the gradient estimation can be noisy and the weight updates can be more erratic. On the other hand, using a larger batch size usually allows more stable training and gradient estimation since the loss computation and its gradients are averaged over the batch of data. This leads to a better estimate of the overall direction in which the weights need to be updated (see Section~\ref{sec:ch2_backprop}). The typical batch sizes are usually powers of 2, such as 32, 64, or 128. However, the appropriate batch size can vary depending on the dataset size, available computational resources, and the specific problem being addressed, and might need to be determined through experimentation and performance evaluation. Using a large batch size has indeed the downside of using more memory resources.

Once a batch of samples from the training dataset has been passed through the network and the loss has been computed the next step is the backpropagation.

\subsection{Backward pass (backpropagation) and weight update} \label{sec:ch2_backprop}

%----------------------------------------------------
\begin{figure}
\centering
\includegraphics[width=0.8\textwidth]{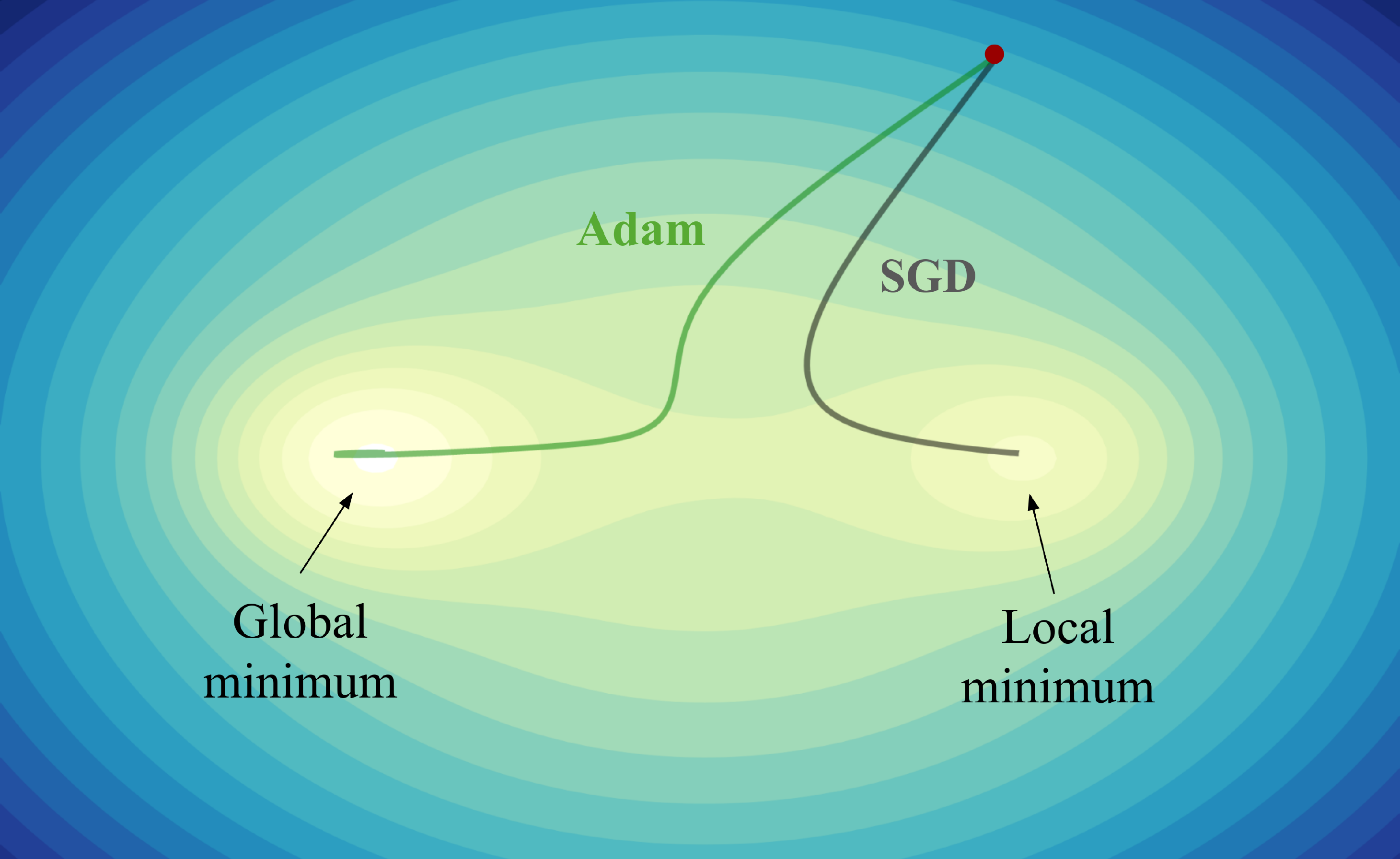}
\caption[Behaviour of the optimisation algorithms Adam and stochastic gradient descent]{Visualisation showing the comparison in behaviour of the optimisation algorithms Adam (\textit{green curve}) and stochastic gradient descent (SGD) (\textit{dark grey curve}). The loss landscape has two minima, the global minimum is on the left while a local minimum is found on the right.
Starting from the same point (\textit{red dot}), Adam is able to find the global minimum while SGD converge to the local one (Figure adapted from \url{https://emiliendupont.github.io/2018/01/24/optimization-visualization/}).}
\label{fig:ch2_grad_descent}
\end{figure}
%----------------------------------------------------

In this step, the loss is propagated backward through the network to compute the gradients with respect to the weights.
Indeed the value of the loss function $\mathcal{L}(\boldsymbol{y}, \hat{\boldsymbol{y}})$ depends on the output $\boldsymbol{y}$ which in turn depends on the values of the weights in the network. 
The chain rule of calculus is applied to efficiently calculate the gradients of the loss with respect to each weight in the network. For example let us assume for simplicity that $y = g(z) = g(\sum_{i=0}^{n} w_i x_i)$ where $g$ is again the activation function. We can now focus on a particular weight $w_i$. The gradient of the loss function with respect to this weight is computed as:
%------------------------------------------------------------------
\begin{align}	\label{eq:ch2_grad_loss}
	 \frac{\partial \mathcal{L}(y, \hat{y})}{\partial w_i} = \frac{\partial \mathcal{L}(y, \hat{y})}{\partial y} \frac{\partial y(z)}{\partial z} \frac{\partial z(\boldsymbol{w}, \boldsymbol{x})}{ \partial w_i}.
\end{align}
%------------------------------------------------------------------
We note here that the gradient of the loss function vanishes if either $\frac{\partial y}{\partial z}$ or $\frac{\partial z}{ \partial w_i}$ or both are zero as could happen for an unsuitable choice of activation functions or a bad weight initialisation (see Section~\ref{sec:ch2_activation_function} and Section~\ref{sec:ch2_w_initialization}).

The gradients calculated for the backpropagation are then used to update the weights of the network. This is usually performed through gradient descent. Let us again focus on a specific weight $w_i$ for simplicity.
%-----------------------------------------------------------
\begin{itemize}
\item If $\frac{\partial \mathcal{L}}{\partial w_i} > 0$, an increase in the value of $w_i$ increases the loss $\mathcal{L}$;

\item If $\frac{\partial \mathcal{L}}{\partial w_i} < 0$, an increase in the value of $w_i$ decreases $\mathcal{L}$. 
\end{itemize}
%-----------------------------------------------------------
Therefore to always decrease the value of the loss $\mathcal{L}$, the update $\delta w_i$ in the weight value has to be proportional to the opposite of the gradient computed with respect to that weight. This means that, in general, at the training iteration $t$, the weight values can be updated in the following way:
%------------------------------------------------------------------
\begin{align}	\label{eq:ch2_grad_descent}
	\boldsymbol{w}_{t} = \boldsymbol{w}_{t-1} + \boldsymbol{\delta w}_{t} = \boldsymbol{w}_{t-1} - \eta \boldsymbol{\nabla_w} \mathcal{L}(\boldsymbol{w}_{t-1}),
\end{align}
%------------------------------------------------------------------
where $\eta$ is called the learning rate defined as $\eta > 0$. The learning rate establishes the size of the displacement between the updated values and the old values of the weights at training step $t-1$. 

In the simplest optimisation algorithm called stochastic gradient descent (SGD), the learning rate is kept constant. More sophisticated optimisation algorithms introduce a more efficient way to update the weights. One of the most popular optimisers is Adam \citep{Kingma2014}. As in SGD, Adam computes the gradients of the loss function with respect to the model's parameters. However, it keeps track of the exponentially decaying averages of past gradients and square gradients (the so called first and second moment, respectively) according to \citep{Kingma2014}:
%------------------------------------------------------------------
\begin{align}	\label{eq:ch2_adam_moments}
	\boldsymbol{m}_{t} = \beta_1 \boldsymbol{m}_{t-1} + (1 - \beta_1) \boldsymbol{\nabla_w} \mathcal{L}(\boldsymbol{w}_{t-1}), \\
    \boldsymbol{v}_{t} = \beta_2 \boldsymbol{v}_{t-1} + (1 - \beta_2) \boldsymbol{\nabla_w} \mathcal{L}(\boldsymbol{w}_{t-1}).
\end{align}
%------------------------------------------------------------------
The first moment $\boldsymbol{m}_{t}$ gives an average estimate of the magnitude of the gradients.
The second moment $\boldsymbol{v}_{t}$ instead represents an uncentered variance of the gradients and is a measure of how much the gradients oscillate around a 0 mean gradient.
The rates $\beta_1$ and $\beta_2$ that regulate the exponential decay are usually set to around $0.9$.
When the training starts, these moments are initialised to zero, causing them to be biased towards zero in the early training stages.
To mitigate this bias, Adam applies bias corrections during the first few iterations. This involves adjusting the estimates of the first and second moments by dividing them by correction factors which are different for the mean and the variance:
%------------------------------------------------------------------
\begin{align}
	\hat{\boldsymbol{m}}_{t} = \frac{\boldsymbol{m}_{t}}{1 - \beta_1^{t}}, \\
    \hat{\boldsymbol{v}}_{t} = \frac{\boldsymbol{v}_{t}}{1 - \beta_2^{t}}.
\end{align}
%------------------------------------------------------------------
Note the training iteration $t$ at the exponent that makes the bias correction vanish as the training proceeds.
Adam then adapts the learning rate separately for each weight. This means that weights with a history of large gradients are assigned smaller learning rates, and weights with a history of smaller gradients are assigned larger learning rates. The weight update rule is then given by:
%------------------------------------------------------------------
\begin{align}	\label{eq:ch2_adam_grad_descent}
	\boldsymbol{w}_{t} = \boldsymbol{w}_{t-1} + \boldsymbol{\delta w}_{t} = \boldsymbol{w}_{t-1} - \eta \frac{\hat{\boldsymbol{m}}_{t}}{\sqrt{\hat{\boldsymbol{v}}_{t}} + \epsilon},
\end{align}
%------------------------------------------------------------------
where $\epsilon$ is a small factor to avoid dividing by zero.
This optimisation algorithm helps converge faster and prevents overshooting near minima of the loss landscape. Furthermore, it helps push the optimiser towards the right direction by reducing oscillations during updates.
However in general it is not guaranteed that the optimiser converges to a global minimum. If the loss landscape is complex and very irregular it can happen that the optimiser gets trapped in a local minimum (see Figure~\ref{fig:ch2_grad_descent}). 

To summarise, the update reduces the loss and moves the network's predictions closer to the expected output.

\subsection{Network optimisation and model evaluation}

%----------------------------------------------------
\begin{figure}
\centering
\includegraphics[width=0.8\textwidth]{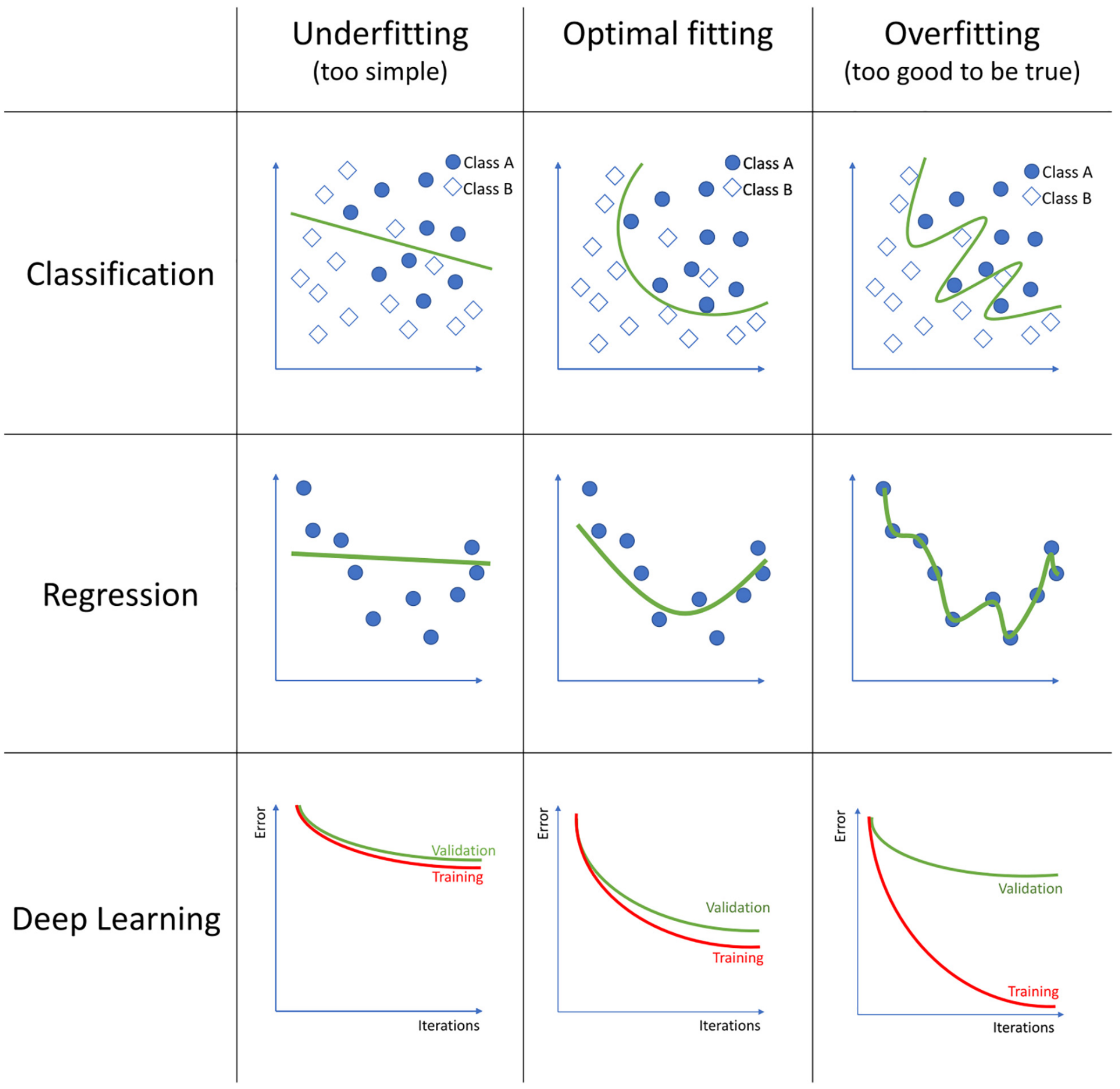}
\caption[Underfitting, overfitting in deep learning]{Examples of model underfitting, model fitting good, and a model overfitting to the data on classification, regression, and deep learning models. In classification, squares and circles represent two different classes. In deep learning, the figures represent the training and validation error over each iteration during the model training process. Figure taken from \citet{Solanes2022}.}
\label{fig:ch2_overfitting}
\end{figure}
%----------------------------------------------------

Once the first update of the weights has been performed, the network proceeds to learn by receiving another batch of samples and updating the weights again. The forward pass, backpropagation, and weights update steps are repeated in cycles until the entire training dataset has been passed in batches through the network. At this point, a single training epoch has been completed.
The validation dataset is then used to evaluate the performance of the network on data that it has not seen during the first round of optimisation. The accuracy of the network predictions over the validation dataset is monitored by computing the loss function on the validation dataset. This helps to assess how well the network performs on unseen data and indicates if the training data is overfitted. Signs of overfitting are observed by comparing the evolution of the training and validation losses. If the training and validation loss curves start to diverge during training, with the training loss reducing while the validation loss increases or remains high, this indicates overfitting (see Figure~\ref{fig:ch2_overfitting}). A way to mitigate overfitting is to introduce early-stopping during the training which consists of stopping the training process if the validation loss ceases to improve.

The steps outlined above are repeated for a fixed number of iterations (epochs) or until the network achieves a desired level of accuracy on the validation dataset. During each iteration, the forward and backward passes help to refine the network's weights, minimising the loss and improving the network's performance.

After the training is completed, the final weights are saved and the neural network's performance is evaluated on a separate test dataset to gauge its generalisation ability on completely unseen data. While the preparation of the training dataset and the training process itself can be time consuming, once the network has been trained the evaluation on a new set of data is very fast. This is an advantage for example in automatic detection and classification of transient events (see Section~\ref{sec:ch2_astro_application} for some example of applications in astrophysics).

%==========================================================================

\section{Simulation based (likelihood-free) inference} \label{sec:ch2_sbi}

In physics and astrophysics, we commonly employ stochastic simulators that aim to emulate some real-world observations or phenomena. Such a simulator incorporates variability and randomness by sampling some variables from probability distributions using Monte Carlo techniques. It relies on theoretical parametric models that depend on certain input parameters $\boldsymbol{\theta} = \{ \theta_1, \theta_2, ... \}$. Given this set of input parameters $\boldsymbol{\theta}$, the simulator generates a synthetic realisation of the observed data $\boldsymbol{x}$. In general, underlying the usage of simulation models is the assumption that these models are correct and hence provide an accurate emulation of the real-world processes that generate the observed data.
A key challenge in simulation based problems is constraining the free parameters in our simulation models with observational data. The goal is to find the region in this parameter space that is consistent with both the available empirical data and our prior knowledge. In general, we encode our prior knowledge of the phenomenon under study in a prior probability distribution, $\pi (\boldsymbol{\theta})$, over the model parameters $\boldsymbol{\theta}$. The stochastic simulator defines the likelihood probability distribution $\mathcal{P}(\boldsymbol{x}|\boldsymbol{\theta})$ of some data $\boldsymbol{x}$ (that is generated by the simulator or the product of real observations) given some set of model parameters $\boldsymbol{\theta}$. The aim is to compute a posterior probability distribution $\mathcal{P}(\boldsymbol{\theta}|\boldsymbol{x})$ of the model parameters $\boldsymbol{\theta}$ given the data $\boldsymbol{x}$. This can be achieved by using the Bayes theorem \citep{Stuart1994}:
%------------------------------------------------------
\begin{align} \label{eq:ch2_Bayes_theorem}
    \mathcal{P}(\boldsymbol{\theta}|\boldsymbol{x}) = \frac{\pi(\boldsymbol{\theta}) \mathcal{P}(\boldsymbol{x}|\boldsymbol{\theta})}{\mathcal{P}(\boldsymbol{x})}
\end{align}
%------------------------------------------------------
where $\pr(\bx |\bt)$ is the likelihood of our data, $\bx$, given the parameter, $\bt$, and
%------------------------------------------------------
\begin{equation}
\pr(\bx) \equiv \int \pr(\bx |\bt') \pi(\bt') \, {\rm d} \bt' ,
\label{eq:ch2_evidence}
\end{equation}
%------------------------------------------------------
denotes the evidence obtained by marginalizing over all $\bt$.
where $\mathcal{P}(\boldsymbol{x})$.
However, for complex simulators, the likelihood $\mathcal{P}(\boldsymbol{x}|\boldsymbol{\theta})$ is typically intractable as it is implicitly defined by the simulator. Even if it were tractable, for multidimensional problems the posterior computation can become very costly as Equation~\eqref{eq:ch2_evidence} involves an integral over $\bt$, which becomes challenging for simulators with high-dimensional parameter spaces. 
Simulation-based likelihood-free inference aims to bypass these problems and directly compute the posterior distribution $\mathcal{P}(\boldsymbol{\theta}|\boldsymbol{x})$ without having access to the likelihood probability density \citep[see][for a review]{Cranmer2020}.

%----------------------------------------------------
\begin{figure}
\includegraphics[width=\textwidth]{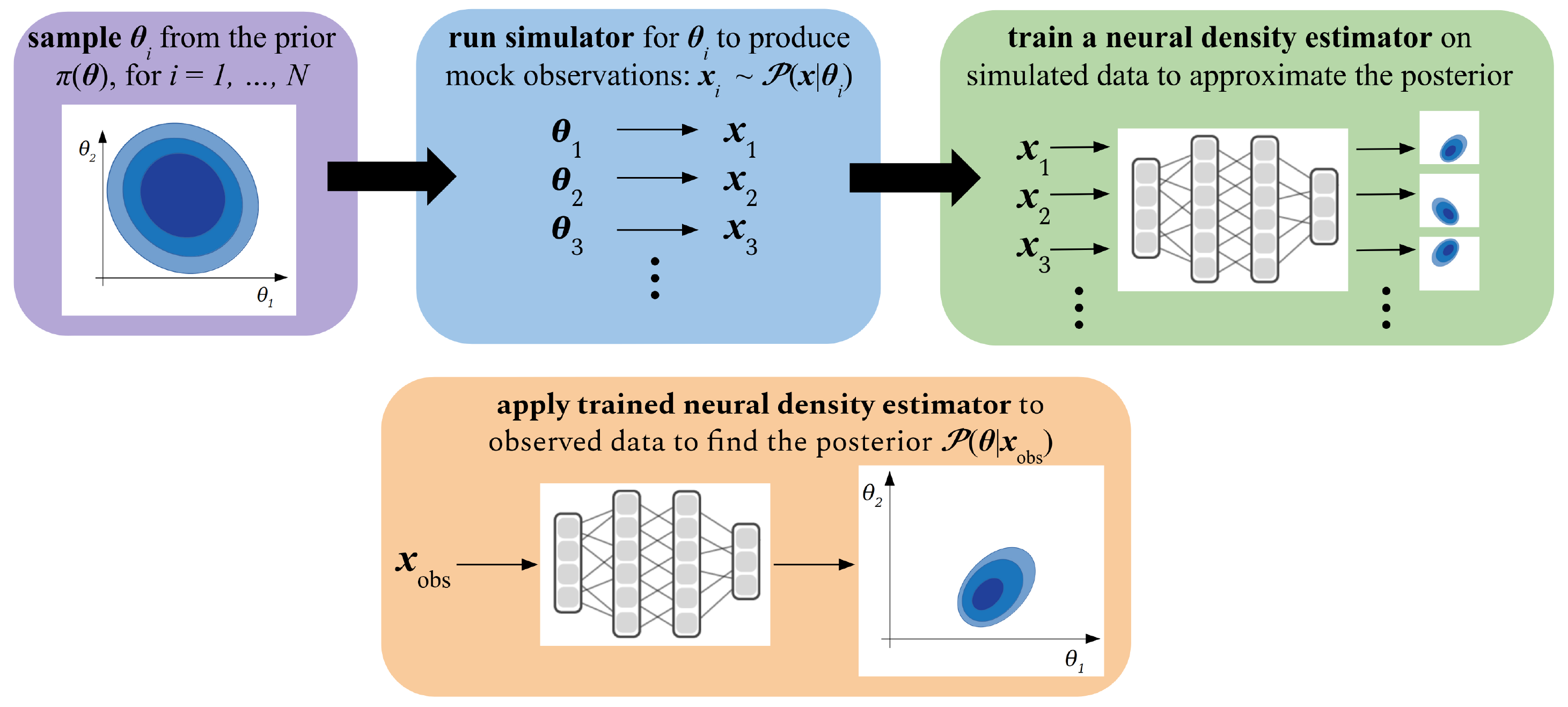}
\caption{Neural posterior estimation workflow}{Scheme of the workflow for the neural posterior estimation (NPE) algorithm (See text for more details).}
\label{fig:ch2_sbi_workflow}
\end{figure}
%----------------------------------------------------

Deep neural networks are particularly useful for this task because they can be used to learn a (probabilistic) association between data and the underlying model parameters, i.e., extract a posterior probability distribution from simulated data without the need to explicitly compute the likelihood. 
There are three main approaches to achieve this goal:
\begin{itemize}
	\item \textbf{Neural posterior estimation (NPE)}: The network learns to directly map the simulator output, $\bx$, onto the posterior distribution, $\pr(\bt |\bx)$, for the underlying parameters, $\bt$. This requires the use of a flexible neural density estimator such as a normalising flow or a \ac{MDN} \citep[e.g.][]{Papamakarios2016, Lueckmann2017, Greenberg2019, Mishra-Sharma2022, Vasist2023, Dax2021, Hahn2023}.
	\item \textbf{Neural likelihood estimation (NLE)}: The network emulates the simulator by learning an association between $\bt$ and $\bx$, thus providing direct access to an approximation of the likelihood, $\pr(\bx |\bt)$. Because the prior is known, the posterior can then be obtained by an additional \ac{MCMC} sampling step \citep[e.g.,][]{Papamakarios2018, Alsing2019}.
	\item \textbf{Neural ratio estimation (NRE)}: Here, the network learns the likelihood-to-evidence ratio, $r(\bt, \bx) \equiv \pr(\bx |\bt) / \pr (\bx)$, which is equivalent to $\pr(\bt  |\bx) / \pr (\bt)$ using Bayes' theorem~\eqref{eq:ch6_Bayes_theorem}. Once $r(\bt, \bx)$ is known, the posterior can be recovered through \ac{MCMC} by sampling the prior weighted by the ratio, $r(\bt, \bx)$ \citep[e.g.,][]{Hermans2019, Miller2021, Bhardwaj2023}.
\end{itemize}

In the following I will concentrate on NPE as it allows to directly derive an estimation of the posterior distribution without the extra step of a MCMC sampling. As we will adopt the NPE method in Chapter~\ref{Chapter6}, in this section I will discuss further details of this approach.

\subsection{Neural posterior estimation} \label{sec:ch2_npe}

Typically, in the \acs{NPE} approach, a neural network, $F$, is trained to map a given simulation outcome $\boldsymbol{x}$ onto the parameters $\boldsymbol{\psi}$ of a density estimator $q_{\boldsymbol{\psi}}$ which is used to approximate the posterior distribution, $\mathcal{P}(\boldsymbol{\theta}|\boldsymbol{x})$, of the underlying model parameters $\boldsymbol{\theta}$ \citep[][see also the schematic workflow in Figure~\ref{fig:ch2_sbi_workflow}]{Papamakarios2016, Lueckmann2017, Greenberg2019}.
To train the network, suppose that we have a dataset $\{ (\boldsymbol{\theta}_i, \boldsymbol{x}_i) \}$ with $i=1, ..., N$, where $\{ \boldsymbol{\theta}_i \}$ is a set of model parameters drawn from the prior distribution $\pi (\boldsymbol{\theta})$ and $\{\boldsymbol{x}_i \}$ is the set of corresponding simulation outcomes. For a given sample $(\boldsymbol{\theta}_i, \boldsymbol{x}_i)$, we define the following loss function:
%------------------------------------------------------
\begin{align} \label{eq:ch2_loss_MDN}
    \mathcal{L}(\boldsymbol{\theta}_i) = - \log q_{F(\boldsymbol{x}_i, \boldsymbol{w})}(\boldsymbol{\theta}_i),
\end{align}
%------------------------------------------------------
where we make explicit that the density estimator parameters depend on the input $\boldsymbol{x}_i$ and the network weights $\boldsymbol{w}$, i.e., $\boldsymbol{\psi} =F(\boldsymbol{x}_i, \boldsymbol{w})$ .
Minimising this loss function means adjusting the network weights so that the probability defined by the density estimator of the label parameters $\boldsymbol{\theta}_i$ is maximised.
As demonstrated in \citet{Papamakarios2016} for sufficiently complex networks and flexible density estimators, minimising the expectation value of the loss in Equation~\eqref{eq:ch2_loss_MDN} leads to $q_{F(\boldsymbol{x}, \boldsymbol{w})}(\boldsymbol{\theta}) \approx \mathcal{P}(\boldsymbol{\theta}|\boldsymbol{x})$ as $N \to \infty$. This can be seen by using the Kullback-Leibler divergence, $D_{\rm KL}(\pr_1 || \pr_2)$, which is a measure of the difference between two probability distributions, $\pr_1$ and $\pr_2$ \citep{Kullback1951}. First we note that $q_{F(\boldsymbol{x}, \boldsymbol{w})}(\boldsymbol{\theta}) \approx \mathcal{P}(\boldsymbol{\theta}|\boldsymbol{x})$ is equivalent to $\mathcal{P}(\boldsymbol{x} | \boldsymbol{\theta}) \pi(\boldsymbol{\theta}) \approx \mathcal{P}(\boldsymbol{x}) q_{F(\boldsymbol{x}, \boldsymbol{w})}(\boldsymbol{\theta})$ by using Bayes theorem (Equation~\eqref{eq:ch2_Bayes_theorem}), so that we can write the following \citep{Papamakarios2016}: 
%------------------------------------------------------
\begin{align}
    \nonumber
    D_{\rm KL}[\mathcal{P}(\boldsymbol{\theta}|\boldsymbol{x}) \, || \, q_{F(\boldsymbol{x}, \boldsymbol{w})}(\boldsymbol{\theta})] &= D_{\rm KL}[\mathcal{P}(\boldsymbol{x} | \boldsymbol{\theta}) \pi(\boldsymbol{\theta}) \, || \, \mathcal{P}(\boldsymbol{x}) q_{F(\boldsymbol{x}, \boldsymbol{w})}(\boldsymbol{\theta})] \\ \nonumber 
    &\equiv \int \mathcal{P}(\boldsymbol{x} | \boldsymbol{\theta}) \pi(\boldsymbol{\theta}) \log{\frac{\mathcal{P}(\boldsymbol{x} | \boldsymbol{\theta}) \pi(\boldsymbol{\theta})}{ \mathcal{P}(\boldsymbol{x}) q_{F(\boldsymbol{x}, \boldsymbol{w})}(\boldsymbol{\theta})}} {\rm d}\boldsymbol{\theta} {\rm d}\boldsymbol{x} \\ \nonumber
    &= \int \mathcal{P}(\boldsymbol{x} | \boldsymbol{\theta}) \pi(\boldsymbol{\theta}) \log{\frac{\mathcal{P}(\boldsymbol{\theta} | \boldsymbol{x})}{q_{F(\boldsymbol{x}, \boldsymbol{w})}(\boldsymbol{\theta})}} {\rm d}\boldsymbol{\theta} {\rm d}\boldsymbol{x} \\ \nonumber
    &= \int \mathcal{P}(\boldsymbol{x} | \boldsymbol{\theta}) \pi(\boldsymbol{\theta}) \log{\mathcal{P}(\boldsymbol{\theta} | \boldsymbol{x})} {\rm d}\boldsymbol{\theta} {\rm d}\boldsymbol{x} \\  \nonumber 
    &{} \qquad - \int \mathcal{P}(\boldsymbol{x} | \boldsymbol{\theta}) \pi(\boldsymbol{\theta}) \log{q_{F(\boldsymbol{x}, \boldsymbol{w})}(\boldsymbol{\theta})} {\rm d}\boldsymbol{\theta} {\rm d}\boldsymbol{x} \\ 
    &= {\rm E}_{\mathcal{P}(\boldsymbol{x} | \boldsymbol{\theta}) \pi(\boldsymbol{\theta})}[\log{\mathcal{P}(\boldsymbol{\theta} | \boldsymbol{x})}] - {\rm E}_{\mathcal{P}(\boldsymbol{x} | \boldsymbol{\theta}) \pi(\boldsymbol{\theta})}[\log q_{F(\boldsymbol{x}, \boldsymbol{w})}(\boldsymbol{\theta})].
\end{align}
%------------------------------------------------------
The expectation values are computed over the joint probability distribution $\mathcal{P}(\boldsymbol{x}, \boldsymbol{\theta}) = \mathcal{P}(\boldsymbol{x} | \boldsymbol{\theta}) \pi(\boldsymbol{\theta})$ that can be sampled using the simulator. 
The first expectation value does not depend on the network parameters. Therefore it is a constant term that can be neglected in the minimisation procedure. This last expression shows that minimising the Kullback-Leibler divergence is equivalent to minimising the expectation value of the loss function defined by Equation~\eqref{eq:ch2_loss_MDN}.

A commonly employed density estimator is \acf{MAF}. A \acs{MAF} is a composition of invertible and differentiable transformations that are applied to a simple distribution, such as a standard Gaussian distribution, to transform it into more complex distributions \citep{Jimenez2015, Papamakarios2017}. Each transformation is parametrised by neural networks which depend on the input data $\boldsymbol{x}$. In this way, \acs{MAF}s can be trained to approximate any target distribution. 

An alternative to \acs{MAF}s are mixture density networks (MDNs). I will focus on this type of architecture as we will employ it in Chapater~\ref{Chapter6}. 
\acs{MDN}s can model complex probability distributions by approximating them using a mixture of distributions. A mixture of distributions (also known as a mixture model) is a statistical model that represents the probability distribution of a random variable as a combination of two or more component distributions. A \acs{MDN} thus takes some features $\boldsymbol{x}$ as input and outputs the parameters $\boldsymbol{\psi}$ defining the mixture model. More precisely, these include the parameters that determine the shape and characteristics of each component distribution and the mixing coefficients that determine their weights in the mixture model. 
An example of a mixture model is a mixture of multivariate Gaussian distributions in the $D$-dimensional space that is defined as:
%-------------------------------------------------------------------
\begin{equation}	\label{eq:ch2_mixture_gaussians}
	q_{\boldsymbol{\psi}}(\boldsymbol{\theta}) = \sum_{c = 1}^{C} \alpha_c \mathcal{N}( \boldsymbol{\theta} | \boldsymbol{\mu}_c,\boldsymbol{\Sigma}_c ), 
\end{equation}
%-------------------------------------------------------------------
where $C$ is total the number of Gaussian components $\mathcal{N}$, $\alpha_c$ are the mixing coefficients, $\boldsymbol{\mu}_c$ is the vector of means and $\boldsymbol{\Sigma}_c$ is the covariance matrix of the $c$ Gaussian component defined as:
%-------------------------------------------------------------------
\begin{equation}
	\mathcal{N}( \boldsymbol{\theta} | \boldsymbol{\mu}_c,\boldsymbol{\Sigma}_c ) = \frac{1}{(2 \pi)^{n/2} \det(\boldsymbol{\Sigma_c})^{1/2}} \exp \left[ -\frac{1}{2} \left( \boldsymbol{\theta} - \boldsymbol{\mu_c} \right)^{\rm T} \boldsymbol{\Sigma_c}^{-1} \left( \boldsymbol{\theta} - \boldsymbol{\mu_c} \right) \right].
\end{equation}
%-------------------------------------------------------------------
In this case, the mixture density network outputs the parameters $\alpha_c$, $\boldsymbol{\mu}_c$ and the entries of the covariance matrix $\boldsymbol{\Sigma}_c$ for every Gaussian component.
To evaluate the loss for a given simulated data sample $(\boldsymbol{\theta}_i, \boldsymbol{x}_i)$, we need to compute:
%-------------------------------------------------------------------
\begin{equation}	\label{eq:ch2_loss_mixture_gaussians}
	\mathcal{L}( \boldsymbol{\theta}_i ) = -\log \sum_{c = 1}^{C} \alpha_c \mathcal{N}( \boldsymbol{\theta}_i | \boldsymbol{\mu}_c,\boldsymbol{\Sigma}_c ), 
\end{equation}
%-------------------------------------------------------------------
where the parameters $\alpha_c$, $\boldsymbol{\mu}_c$ and $\boldsymbol{\Sigma}_c$ depend on the input features $\boldsymbol{x}_i$ and the network weights $\boldsymbol{w}$.
Since we have an exponential function in the normal distribution, we might obtain very small numbers when evaluating the loss in this form. Such values might be too small to be represented with the correct precision by the computing system (numerical underflow). To avoid this issue, we take advantage of the fact that $x = \exp(\log x)$. The loss function can thus be rewritten in the following form:
%-------------------------------------------------------------------
\begin{align}	\label{eq:ch2_loss_logsumexp}
	\mathcal{L}( \boldsymbol{\theta} ) &= -\log \sum_{c = 1}^{C} \exp \left[ \log \left( \alpha_c \mathcal{N}( \boldsymbol{\theta} | \boldsymbol{\mu}_c,\boldsymbol{\Sigma}_c ) \right) \right] \nonumber \\
              &= -\log \sum_{c = 1}^{C} \exp \left[ \log \alpha_c - \frac{n}{2} \log(2 \pi) - \frac{1}{2} \log \det(\boldsymbol{\Sigma_c}) - \frac{1}{2} \left( \boldsymbol{\theta} - \boldsymbol{\mu_c} \right)^{\rm T} \boldsymbol{\Sigma_c}^{-1} \left( \boldsymbol{\theta} - \boldsymbol{\mu_c} \right) \right].
\end{align}
%-------------------------------------------------------------------
Implementing this log of the sum of exponentials numerically allows a stable calculation of the loss.

Once trained on a large collection of simulations, this type of network is able to generate a precise estimation of the posterior density distribution for any given input of empirical data. This allows a fast inference procedure which is also amortised, meaning that it does not rely on a single specific observation.

%================================================================
\section{Diagnostic checks}

After a density estimator has been trained on a dataset of simulations and to produce posteriors, its predictions should undergo several diagnostic tests, before being used for inference on actual observed data. A first qualitative check is the so-called posterior-predictive check. It consists of inspecting if the data generated from the simulator using the parameters sampled from the posterior looks similar to the actual input data used to estimate the posterior. Even if not fully systematic, this test gives already an indication if the predicted posterior is biased or inaccurate.
In the following, I report two more quantitative tests that should be performed to asses if the trained estimator is well calibrated (see Chapter~\ref{Chapter6} for an application of these tests).

\subsection{Simulation based calibration}

%----------------------------------------------------
\begin{figure}
\centering
\includegraphics[width=0.4\textwidth]{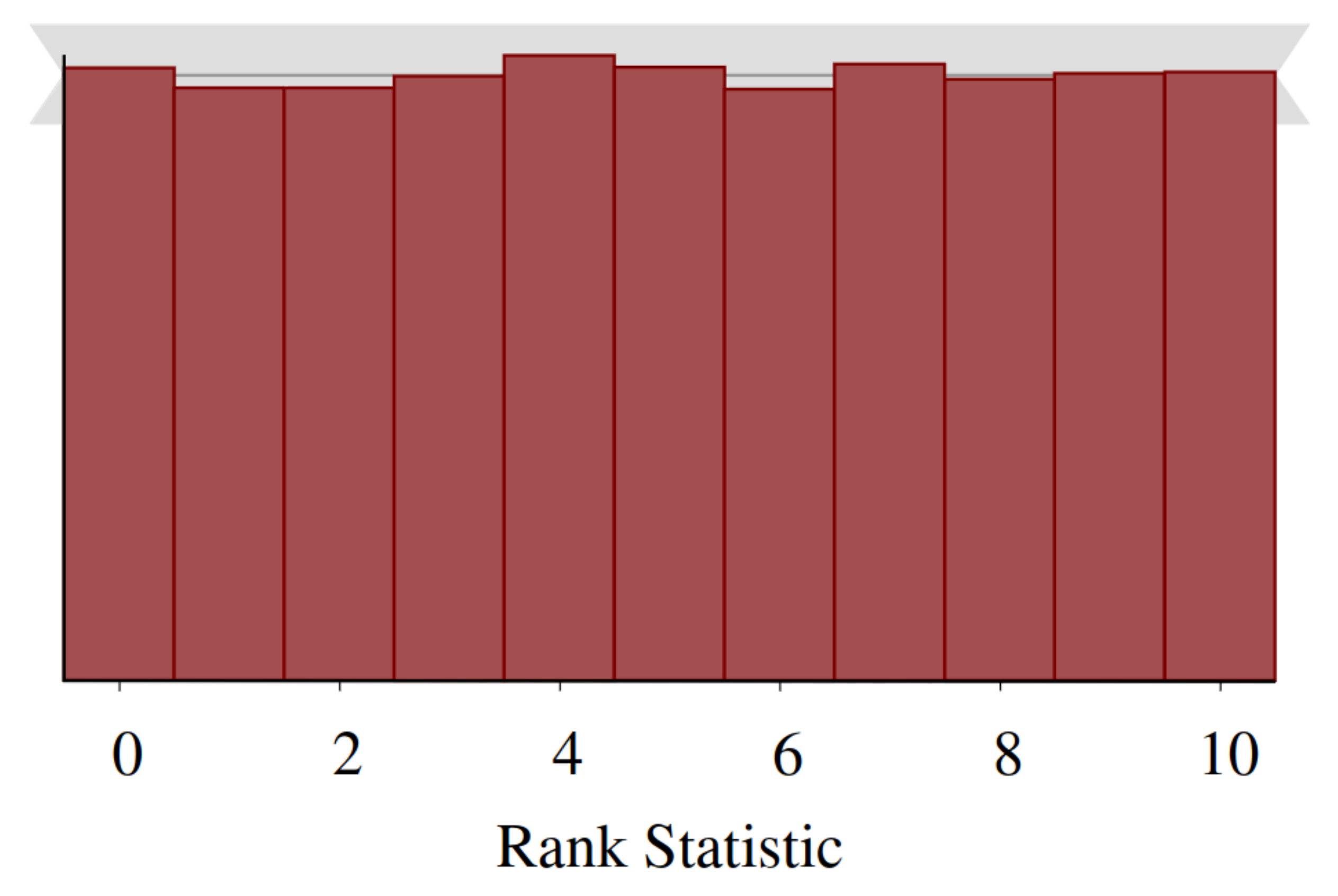}
\includegraphics[width=0.4\textwidth]{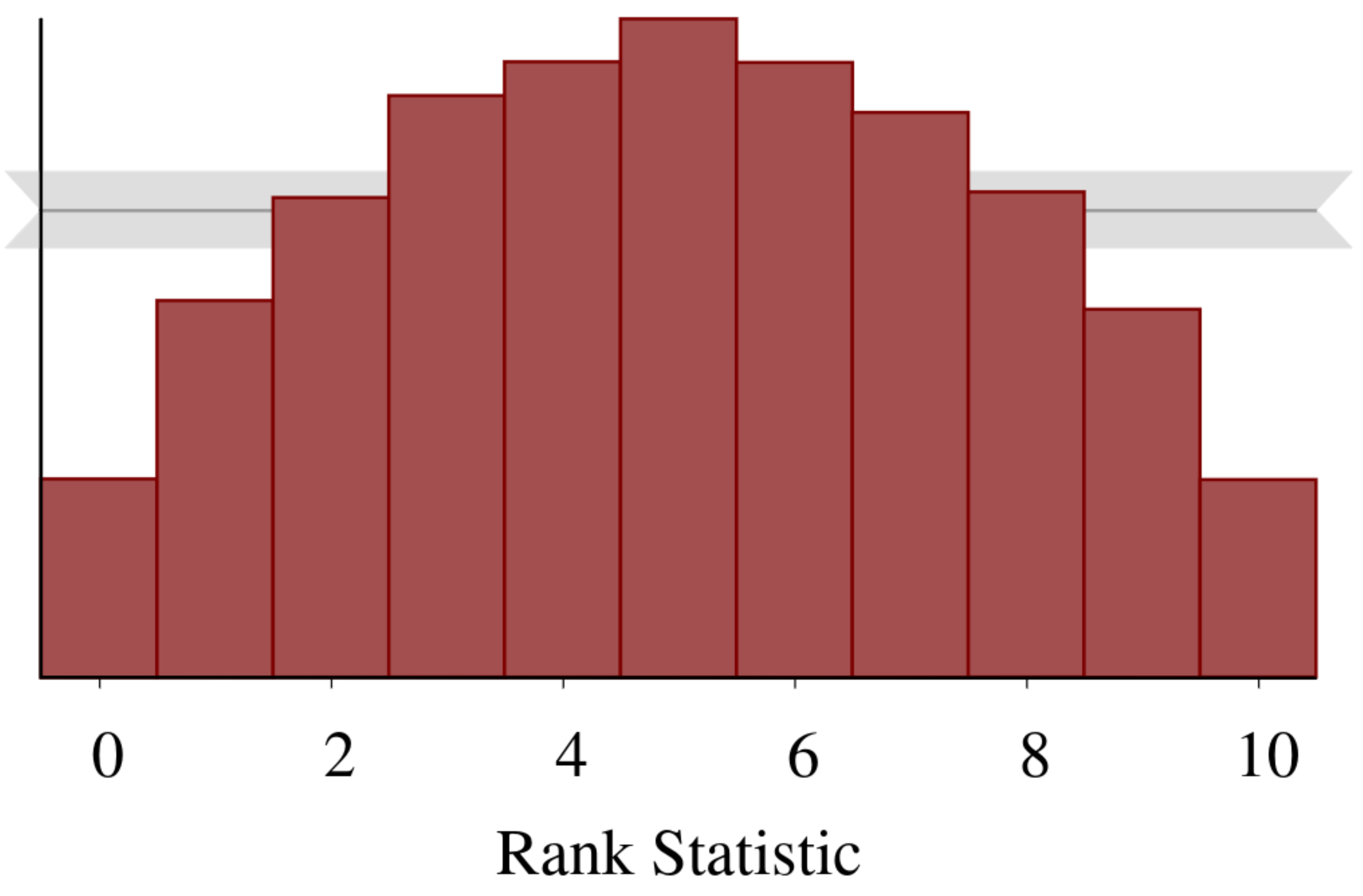}
\includegraphics[width=0.4\textwidth]{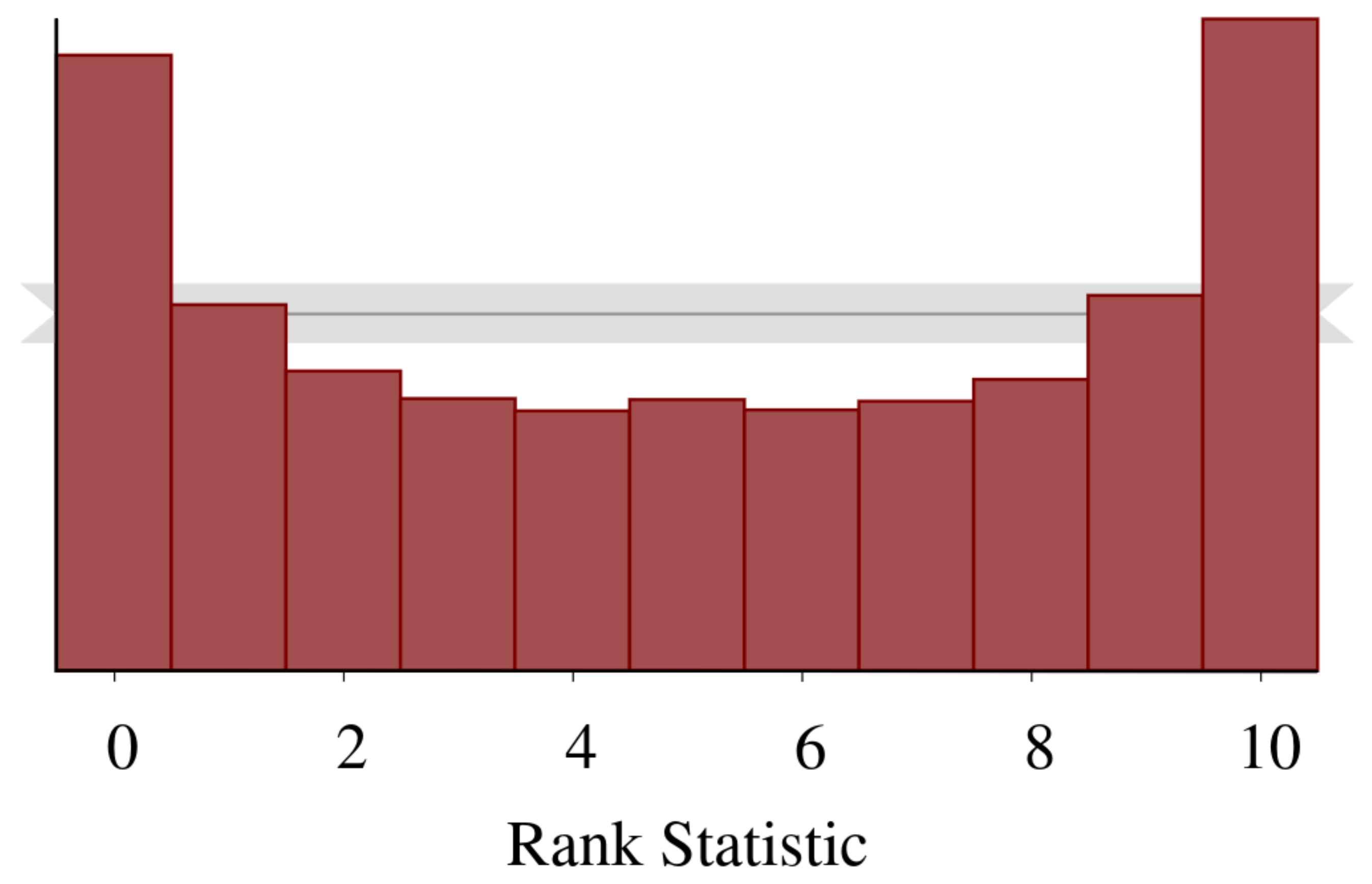}
\includegraphics[width=0.4\textwidth]{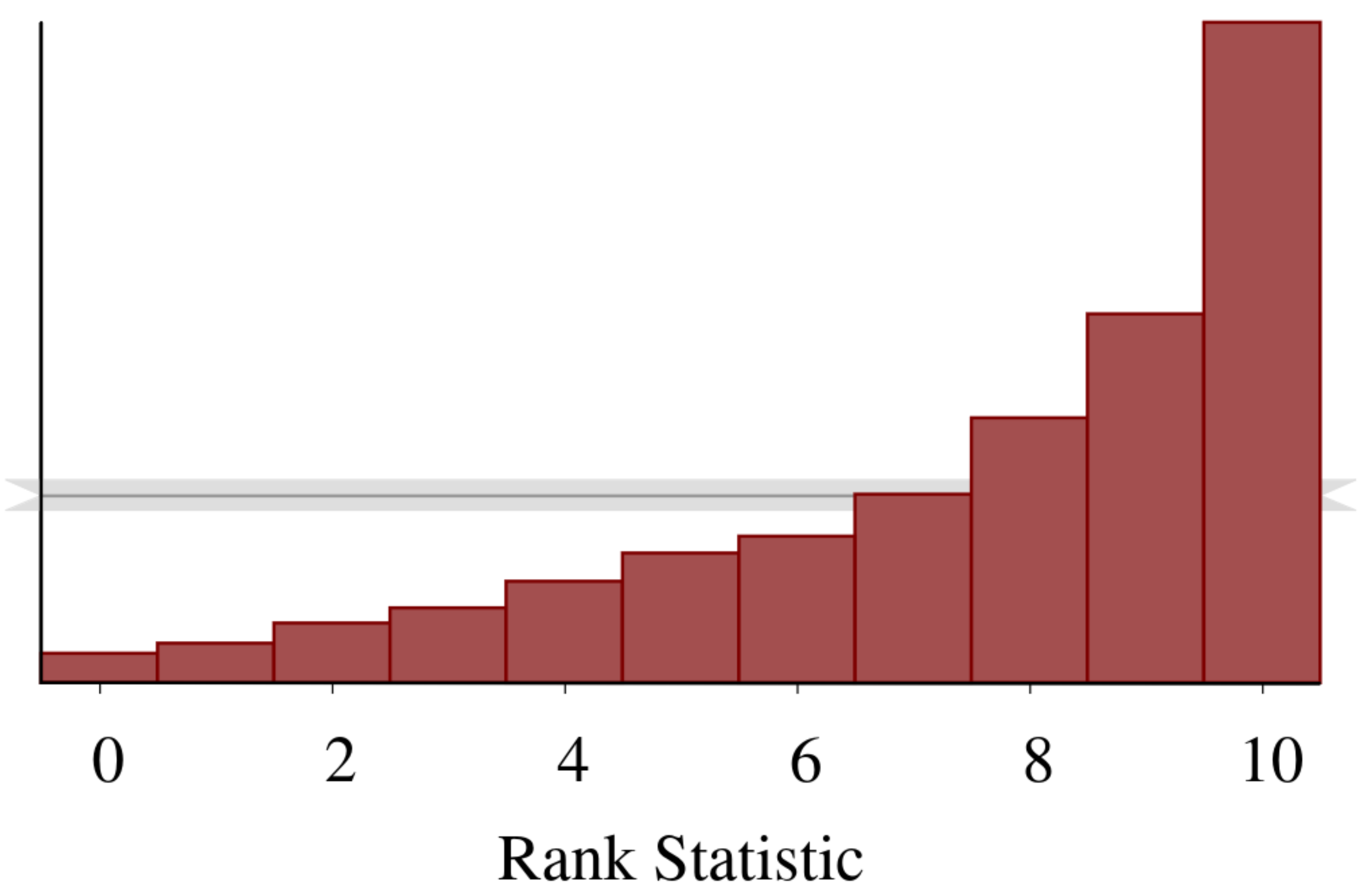}
\caption[Ranks statistics]{Example sketches showing the distribution of the ranks statistics in the case of a density estimator predicting well calibrated posteriors (\textit{top left plot}), posteriors that are too conservative (\textit{top right plot}), overconfident posteriors (\textit{bottom left plot}) and posteriors that are biased, i.e., that underestimate the parameter values in this specific case \citep[Figures taken from][]{Talts2018}.}
\label{fig:ch2_sbc_ranks}
\end{figure}
%----------------------------------------------------

\citet{Talts2018} proposed a method called \acf{SBC} that provides a qualitative view and a quantitive measure to check if posterior predictions are trustworthy.
\acs{SBC} can diagnose whether the estimated posterior predictions are biased, i.e., systematically overestimating or underestimating the true parameter values, and whether the posterior uncertainties are unbalanced, i.e., either over-confident or under-confident. Therefore, \acs{SBC} can be viewed as a necessary (but not sufficient) condition for a valid inference algorithm. If \acs{SBC} checks fail, the inference results should not be trusted. On the other hand, if SBC checks pass, the posterior estimation should be well calibrated but this is no guarantee for the method to be working properly.

\acs{SBC} is based on ranking parameters $\boldsymbol{\theta}$ sampled from the prior $\pi(\boldsymbol{\theta})$ and checking their rank statistics \citep[see also][]{Cook2006}. The procedure is as follows.
We consider a test dataset $\{ (\boldsymbol{\theta}_i, \boldsymbol{x}_i) \}$ (that has been used neither for training nor validating the network) and compute the posterior $\mathcal{P}(\boldsymbol{\theta}|\boldsymbol{x}_i)$ by processing each $\boldsymbol{x}_i$ through the trained network. 
From each posterior $\mathcal{P}(\boldsymbol{\theta}|\boldsymbol{x}_i)$, we sample $S$ values $\boldsymbol{\theta}_{\rm s}$ and count how many of these samples are smaller than the original $\boldsymbol{\theta}_i$. This count gives the rank associated with the parameter $\boldsymbol{\theta}_i$. Once this has been performed for the entire dataset $\{ (\boldsymbol{\theta}_i, \boldsymbol{x}_i) \}$, we analyze the obtained rank statistics.
For a well-calibrated posterior, the ranks follow a uniform distribution \citep{Talts2018}. If this is not the case, we obtain an indication that the estimated posteriors are not accurate on average (see Figure~\ref{fig:ch2_sbc_ranks}).
For example, if the rank distribution is skewed to the left, i.e., it is dominated by lower ranks, the predicted posteriors tend to overestimate the parameters $\boldsymbol{\theta}$. 
If the rank distribution is skewed to the right, i.e., it is dominated by high ranks, the predicted posteriors tend to underestimate the parameters $\boldsymbol{\theta}$.
Moreover, if the rank distribution has a "U" shape, and is dominated by low and high ranks, the predicted posteriors tend to be overconfident, that is they underestimate the uncertainties on the parameters $\boldsymbol{\theta}$. Finally, if the rank distribution has a "$\cap$" shape and is dominated by medium ranks, the predicted posteriors tend to be too conservative. This implies that they are overestimating the uncertainties on the parameters $\boldsymbol{\theta}$. 
As $\boldsymbol{\theta}$ is usually an array of parameters, the rank statistics are computed over the marginalised 1D posterior distributions of each these parameters. Therefore, this test might not be particularly sensitive to the shape of the posterior in the multi-dimensional space and the presence of correlations between parameters. While rank statistics are a useful metric to assessing the quality of a posterior distribution, we will primarily focus on another method outlined below in the remainder of this thesis.

\subsection{Coverage probability} \label{sec:ch2_coverage}

%----------------------------------------------------
\begin{figure}
\centering
\includegraphics[width=0.6\textwidth]{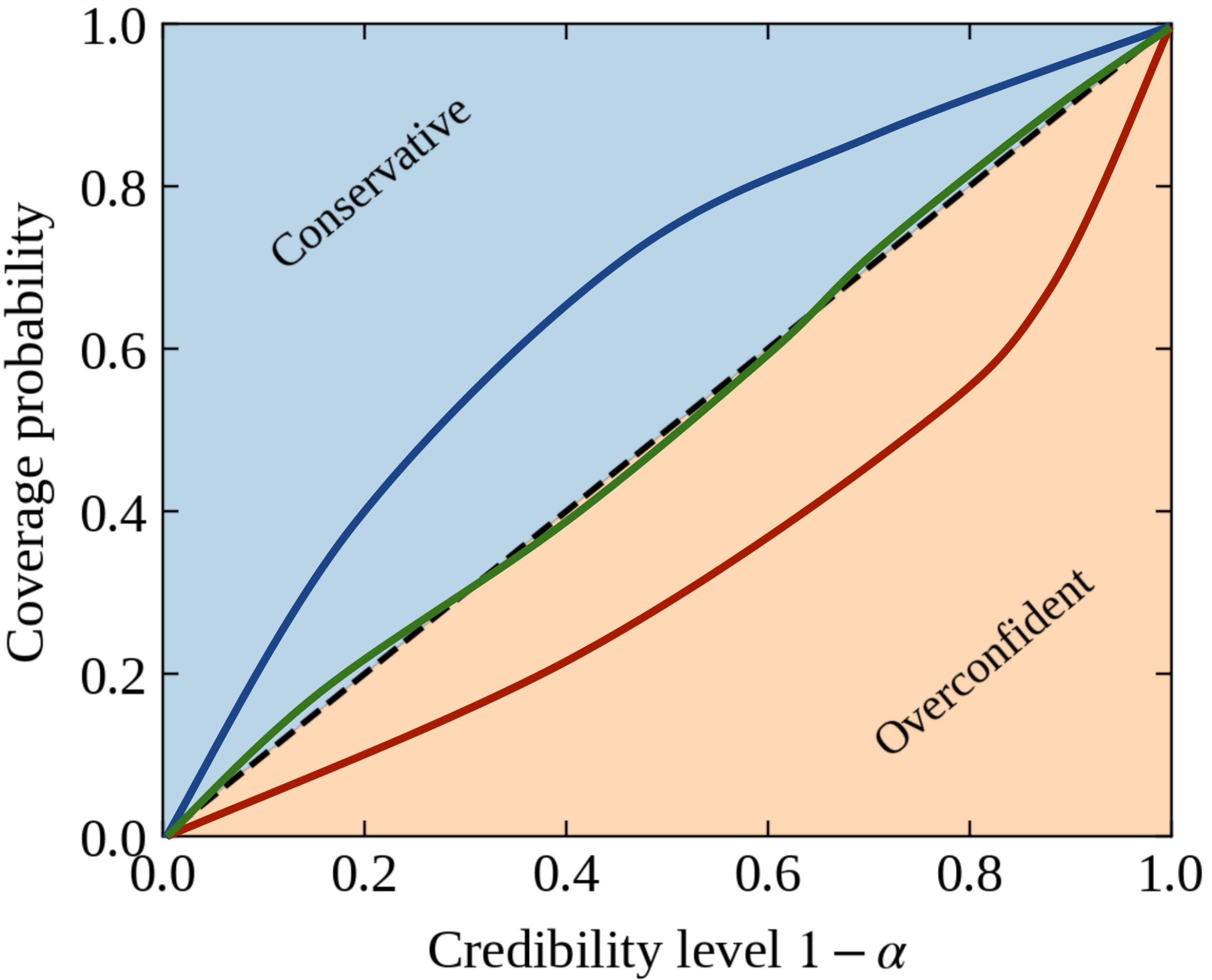}
\caption[Coverage probability]{Coverage probability plot showing the computed coverage probability versus the credibility level $1-\alpha$ for overconfident posterior distributions (\textit{red curve}), well-calibrated posteriors (\textit{green curve}) and overconfident posteriors (\textit{blue curve}).}
\label{fig:ch2_coverage}
\end{figure}
%----------------------------------------------------

Another test to asses if a posterior estimator is well calibrated is checking its coverage probability \citep{Hermans2021}. As in \acs{SBC}, we consider a test dataset $\{ (\boldsymbol{\theta}_i, \boldsymbol{x}_i) \}$ and compute the posterior $\mathcal{P}(\boldsymbol{\theta}|\boldsymbol{x}_i)$ for each $\boldsymbol{x}_i$. For each posterior, we then define a credible region $\Theta_i$ with the smallest volume in the multidimensional parameter space of $\boldsymbol{\theta}$ corresponding to a total probability $1 - \alpha$ with $\alpha \in [0, 1]$:
%-------------------------------------------------------------------
\begin{equation}	\label{eq:ch2_credibility_region}
	\int_{\Theta_i} \mathcal{P}(\boldsymbol{\theta}|\boldsymbol{x}_i) d \boldsymbol{\theta} = 1 - \alpha, 
\end{equation}
%-------------------------------------------------------------------
where $1 - \alpha$ defines the so-called credibility level.
Considering the region with the smallest volume guarantees that we are focusing on the region of the parameter space that encloses the parameter values $\boldsymbol{\theta}$ with the highest posterior probability density. In the literature, this is also called the highest posterior density region \citep{Hyndman1996}.

By counting how many $\boldsymbol{\theta}_i$ fall inside the corresponding credibility regions $\Theta_i$ we obtain a measure of how well the estimated posteriors are able to recover the label parameters $\boldsymbol{\theta}$. This count gives an estimate of the so-called coverage probability. If the posterior estimator is well calibrated, the coverage probability should be equal to the value $1 - \alpha$. If the coverage probability is higher than $1 - \alpha$, this is a symptom for a posterior estimator that tends to generate posteriors that are too conservative. On the other hand, if the coverage probability is lower than $1 - \alpha$, this indicates that the estimated posteriors are overconfident. We will use this test in Chapter~\ref{Chapter6} (see also Appendix~\ref{app:coverage} for more details and Figure~\ref{fig:ch2_coverage} for an illustrative representation).

\section{Applications in astrophysics}  \label{sec:ch2_astro_application}

In recent years there has been an increasing number of works that apply the machine learning techniques outlined above in the field of astronomy and astrophysics. In particular, for neutron-star and compact-object related science, \acs{ML} algorithms have for example been developed to classify new pulsars candidates \citep[see for example][]{Bethapudi2018, Balakrishnan2020, Lin2020} as well as transient radio events such as fast radio bursts (FRBs) \citep{Connor2018, Agarwal2020} or to visualize from another perspective the pulsar population through minimum spanning trees \citep{Garcia2022}. Other approaches have aimed to combine classification and regression tasks to forecast and analyse gravitational-wave signals in real-time and perform parameter estimation for the merger events \citep{George2018, Cabero2020, Wei2020, Skliris2020, Gerosa2020}. Some other studies focused on reconstructing the neutron-star equation of state from observed quantities, like their masses and radii \citep{Fujimoto2018, Morawski2020}.
In Chapter~\ref{Chapter5} we will study the feasibility of using deep neural networks (in particular CNNs) to infer the dynamical properties of the neutron star population.

In addition, the rapidly developing field of simulation-based inference is now offering new tools to perform parameter estimation with complex simulators \citep[see][for a review]{Cranmer2020}. For example, SBI has been employed for parameter estimation from gravitational-wave signals \citep{Green2020, Bhardwaj2023}. \citet{Vasist2023} inferred the physical properties of exoplanetary atmospheres from spectroscopic data. Moreover, \citet{Khullar2022} have been able to retrieve galaxy properties from their spectral energy distributions (SEDs). In the context of cosmology, \citet{Jeffrey2021} used an SBI framework to infer cosmological parameters from weak lensing maps.
In this work, we are going to employ for the first time an SBI framework developed by \citet{Tejero-Cantero2020} in conjunction with a population synthesis simulator to retrieve the birth properties of the Galactic population of isolated radio pulsars (see Chapter~\ref{Chapter6}). 
 
% Chapter 3

\chapter{Long-period pulsars as possible outcomes of supernova fallback accretion} % Main chapter title

\label{Chapter3} % For referencing the chapter elsewhere, use \ref{Chapter3} 

%----------------------------------------------------------------------------------------

\section{Introduction} 
\label{sec:ch3_intro}

The spin-period distribution of the pulsar population reflects intrinsic properties of neutron-star formation, early evolution, magnetic-field decay, and age (see sec. \ref{sec:ch1_ns_zoo}). Until a few years ago, the spin distribution of observed pulsars was ranging between $\sim$ 0.002--12~s. At the fastest extreme, we have recycled millisecond pulsars (mostly in binaries), while the slowest extreme is populated by magnetars (see Figure~\ref{fig:ch1_PPdot_diagram}). The historical lack of isolated pulsars with periods $\gtrsim \unit[12]{s}$ has been intriguing and interpreted in different ways, ranging from the presence of a death line below which radio emission is quenched \citep[][see also Section~\ref{sec:ch1_radio_em_deathlines}]{Ruderman1975, Bhattacharya1991, Chen1993}, to magnetic-field decay coupled with the presence of a highly resistive layer in the inner crust \citep[possibly due to the existence of a nuclear pasta phase][see also Section~\ref{sec:ch1_internal_structure_MR_relation}]{Pons2013}, as well as due to an observational bias caused by high band pass filters in radio searches (albeit not in X-ray searches). Although the main reason is uncertain, all of these effects likely contribute at some level to the observational paucity of long-period pulsars \citep{Wu2020}.

However, recent radio surveys, in particular thanks to new radio interferometers such as the LOw Frequency ARray \citep[LOFAR;][]{VanHaarlem2013}, MeerKAT \citep{Jonas2009}, Australian SKA Pathfinder \citep[ASKAP;][]{Hotan2021}, and the Murchison Widefield Array \citep[MWA;][]{Tingay2013, Wayth2018}, have started to uncover the existence of a new population of slowly rotating radio pulsars that challenge our understanding of the pulsar population and its evolution.

Two radio pulsars, PSR\,J1903$+$0433 \citep{Han2021} and PSR\,J0250$+$5854 \citep{Tan2018}, have been recently discovered with periods of $\unit[14]{s}$ and $\unit[23]{s}$, respectively. Moreover, a $\sim\unit[76]{s}$ radio pulsar with a magnetic field of $B \sim \unit[1.3\times10^{14}]{G}$ \citep[\mtp,][]{Caleb2022} and two peculiar radio transient with a periodicity of $\sim\unit[1091]{s}$  \citep[\gleamfirst;][]{Hurley-Walker2022a} and $\sim \unit[1318]{s}$ \citep[\gpm][]{Hurley-Walker2023} have been discovered. These latter sources are currently still uncertain in nature. In particular, \gleam\ has a very variable flux, showing periods of ``radio outburst'' lasting a few months, a 90\% linear polarisation, and a very spiky and variable pulse profile. \gpm\, is also characterised by a very irregular and variable pulse profile and has been active for at least 34 years. Several interpretations have been advanced to explain the mysterious nature of these sources. As argued in the discovery papers, their emission characteristics are typical of observed radio magnetars \citep{Kaspi2017, Esposito2020} (although their magnetic fields are still poorly constrained). Alternatively, they could be strongly magnetised white dwarfs having spun down to the observed long spin period due to their larger moment of inertia (see also \citet{Tong2022} and Chapter~\ref{Chapter4} for a discussion of the nature of these sources).
Furthermore, a few years ago the X-ray emitting neutron star \rcw\ at the centre of the $\unit[2]{kyr}$ old supernova remnant (SNR) RCW103, with a measured modulation of $\sim \unit[6.67]{hr}$, showed a large magnetar-like X-ray outburst \citep{Rea2016, D'Ai2016b}, demonstrating the source's isolated magnetar nature despite its long period and young age. 
 
In general, studying the possible mechanisms that could lead to the formation of long-period pulsars is of great interest in order to understand their origin, eventual links with other types of neutron stars (see Section~\ref{sec:ch1_ns_zoo}) and possible connections to periodic activity of transient events such as fast radio bursts \citep[see for example][]{Beniamini2020, Xu2021}. One possible avenue that could lead to enhanced spin-down, especially in magnetars, is the presence of mass-loaded charged particle winds and outflows that could be particularly active after giant-flare episodes and temporarily expand the open magnetic flux region of the star \citep[see for example][]{Thompson2000, Tong2013, Beniamini2020}. 
Another possibility is the interaction of the neutron star with a fallback disk formed after the supernova explosion. 
Soon after their formation, neutron stars will necessarily witness fallback accretion with different accretion rates depending on the progenitor properties and explosion dynamics \citep{Ugliano2012, Perna2014, Janka2022}. If the fallback matter possesses sufficient angular momentum, it could form a long-lasting accretion disk that will interact with the neutron star. For certain ranges of the initial spin period $P_0$, magnetic field $B_0$ and disk accretion rate $\dot{M}_{\rm d,0}$, fallback after the supernova explosion can substantially affect the pulsar spin evolution, in some cases slowing down the pulsar period significantly more than standard dipolar spin-down losses alone. 
In this context, the long period observed in \rcw\ has been interpreted as the spin period of a magnetar interacting with a fallback disk in a propeller state by numerous authors \citep{Li2007, Ho2017, Tong2016, Xu2019}. 
Moreover \citet{Chatterjee2000, Ertan2009, Benli2016} developed detailed numerical models of the interaction between a fallback disk and a neutron star to explain the emission of magnetars. 
Although appealing, this model struggles to explain several observed features of these types of sources, such as their burst and flaring activity \citep[see][for a review]{Mereghetti2013, Turolla2015, Kaspi2017, Esposito2021}. 
Overall, these studies on fallback disk accretion suggest that the presence or absence of a fallback disk around newly born neutron stars could be a determining factor for their evolution in the $P$-$\dot{P}$ diagram and (Figure~\ref{fig:ch1_PPdot_diagram}) as well as their emission properties and could further be relevant to explain the connections between different neutron star classes \citep[see for example][and Section~\ref{sec:ch1_ns_zoo}]{Alpar2001}.  

In this chapter based on the work published in \citet*{Ronchi2022}, we revisit the fallback scenario to specifically analyze the spin-period evolution of newly born pulsars that witness accretion from a fallback disk (Section~\ref{sec:ch3_fallback_scenario}), and determine the parameter ranges that allow pulsars to experience efficient spin-down in the presence of magnetic field decay. Furthermore, we study the characteristics of the newly discovered long-period radio sources \gleam\ (see Section~\ref{sec:ch3_gleam}), \gpm\ (see Section~\ref{sec:ch3_gpm}) and pulsar \mtp\ (see Section~\ref{sec:ch3_mtp0013}) in the context of the fallback scenario in order to constrain their nature and evolution (see Section~\ref{sec:ch3_discussion}). We provide a summary in Section~\ref{summary}. \footnote{Jupyter Notebooks to reproduce the plots and results in this chapter are publicly available at \url{https://github.com/MicheleRonchi/pulsar_fallback}}

%%%%%%%%%%%%%%%%%%%%%%%%%%%%%%%%%%%%%%%%%%%%%%%%%%%%%%
\section{Period evolution of pulsars slowing down via dipolar losses} 
\label{sec:ch3_dipolar_losses}

%----------------------------------------------------
\begin{figure}
\centering
\includegraphics[width =0.8\textwidth]{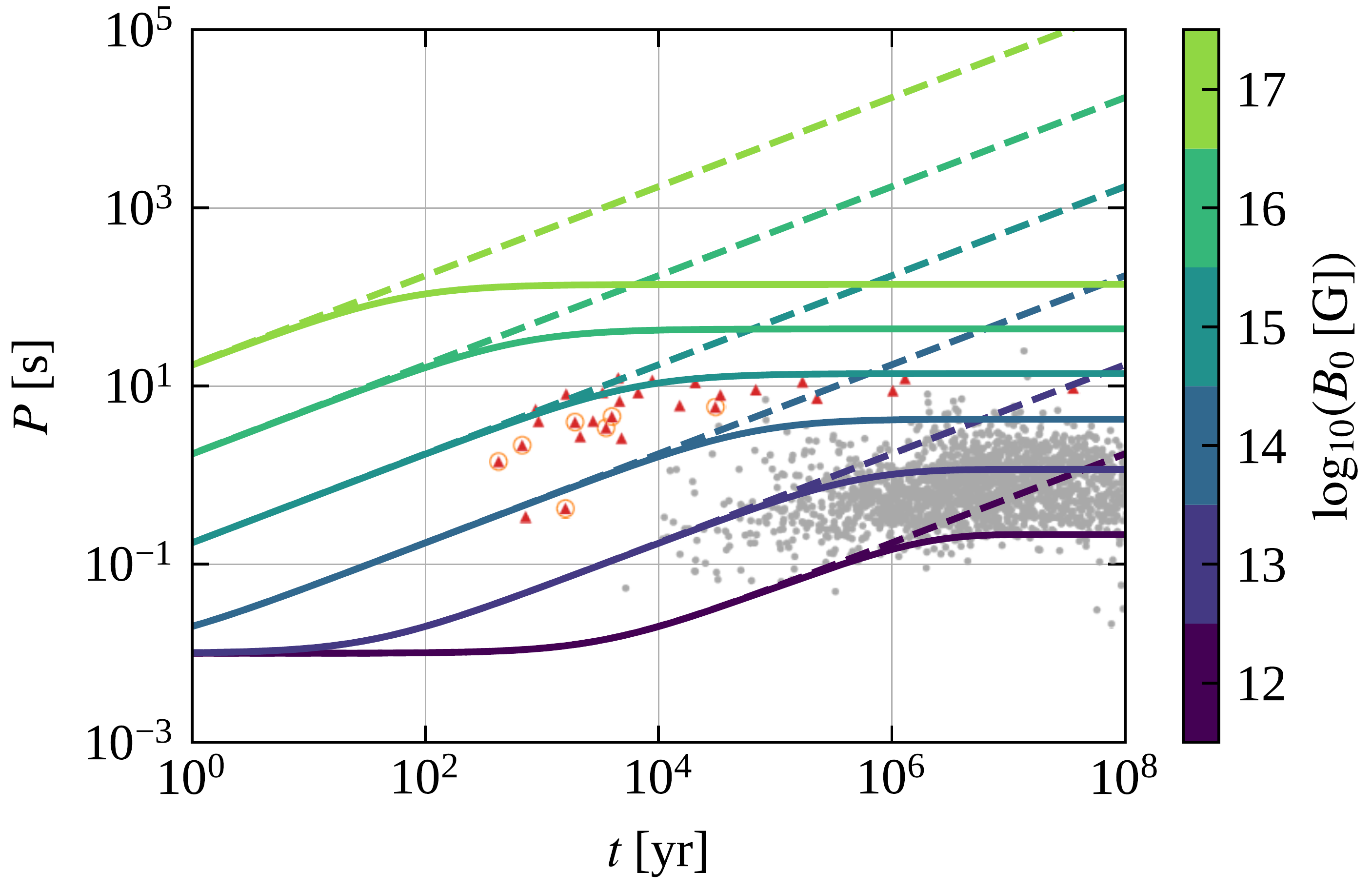}
\caption[Spin-period evolution for magnetic dipolar losses]{The evolution of the pulsar spin-period $P$ in time $t$ assuming dipolar spin-down for different values of the initial magnetic field $B_0$ (indicated by the colour code in the colour bar). The dashed lines represent the evolution curves for a constant magnetic field, while the solid lines represent the evolution tracks for a decaying magnetic field according to Equation~\eqref{eq:ch3_B_decay}. In the background, the gray points represent the observed radio pulsars \citep[data from the ATNF Pulsar Catalog, \url{https://www.atnf.csiro.au/research/pulsar/psrcat/};][]{Manchester2005} and red triangles represent the currently detected magnetars \citep[data from the Magnetar Outburst Online Catalogue, \url{http://magnetars.ice.csic.es/};][]{CotiZelati2018} highlighting those that exhibit radio emission (orange circles). We consider the characteristic age $\tau_{\rm c} = P/(2\dot{P})$ as a proxy for their real age.}
\label{fig:ch3_em_spin_down_P_evol} 
\end{figure}  
%----------------------------------------------------

Ordinary rotation-powered pulsars are expected to slow down via electromagnetic dipolar losses, with an additional component driven by their magnetic-field decay (see Section~\ref{sec:ch1_dip_spindown_evol}).
In this scenario, if $\omega = 2 \pi / P$ is the angular spin frequency of the pulsar, the electromagnetic torque causes the star to slow down according to (see Equation~\eqref{eq:ch1_rot_evol_forcefree}): 
%----------------------------------------------------
\begin{align}
    \frac{d\omega}{dt} = - \beta B(t)^2 \omega^3,
        \label{eq:ch3_domegadt}
\end{align}
%----------------------------------------------------
where for an aligned rotator $\beta \sim R_{\rm NS}^6/(4 c^3 I_{\rm NS}) \sim  \unit[1.5 \times 10^{-41}]{s \, G^{-2}}$ assuming a typical neutron-star radius $R_{\rm NS} \sim \unit[11]{km}$, and moment of inertia $I_{\rm NS} \sim \unit[1.5 \times 10^{45}]{g \, cm^2}$. As discussed in Section~\ref{sec:ch1_dip_spindown_evol} the inclination angle dependence and uncertainty on the neutron-star mass and radius introduce a correction of at most an order of magnitude to the value of $\beta$ in the equation above \citep[see also][]{Spitkovsky2006}.

If we consider a constant magnetic field $B(t) = B_0$, we can solve Equation~\eqref{eq:ch3_domegadt} to find the spin-period evolution in time:
%----------------------------------------------------
\begin{align}
    P(t) = P_0 \left(1 + \frac{t}{t_{\rm em}} \right)^{1/2},
        \label{eq:ch3_P_t_em_spindown}
\end{align}
%----------------------------------------------------
where we define an electromagnetic spin-down timescale $t_{\rm em} = 	P_0^2 / \left( 8 \pi^2 \beta B_0^2 \right)$.
Note that for $t \gg t_{\rm em}$ the period evolution loses memory of the initial spin period, $P_0$, and only depends on the magnetic field value:
%----------------------------------------------------
\begin{align} 
	P(t) = \left(8 \pi^2 \beta B_0^2 t \right)^{1/2}.
\end{align}
%----------------------------------------------------
This means that neutron stars born with different initial spin periods but same magnetic field strength will end up evolving identically.

However, several studies have shown that neutron star magnetic fields evolve over time and decay due to the combined action of Ohmic dissipation and the Hall effect in the star's crust \citep[][see also Section~\ref{sec:ch1_B_evol}]{Pons2007, Pons2009, Vigano2013, Pons2019, DeGrandis2020}. For simplicity, if we consider a crustal-confined magnetic field, we can use the phenomenological description of the field decay presented in \citet{Aguilera2008b, Aguilera2008a}, given by the analytic expression (see also Equation~\eqref{eq:ch1_B_decay}): 
%----------------------------------------------------
\begin{align}
    B(t) = B_0 \frac{e^{-t/\tau_{\rm Ohm}}}{1 + \frac{\tau_{\rm Ohm}}{\tau_{\rm Hall,0}} \left[ 1 - e^{-t/\tau_{\rm Ohm}} \right]}. 
        \label{eq:ch3_B_decay}
\end{align}
%----------------------------------------------------
This equation captures a first stage that is characterised by rapid (non-exponential) decay and regulated by the Hall timescale $\tau_{\rm Hall,0}$, and a second stage that is characterised by exponential decay due to Ohmic dissipation and regulated by the timescale $\tau_{\rm Ohm}$.
These two characteristic timescales were defined in eqs. \eqref{eq:ch1_tau_hall} and \eqref{eq:ch1_tau_ohm}. I report them here again as easy reference:
%----------------------------------------------------
\begin{align} \nonumber
    \tau_{\rm Hall,0} &= \frac{4 \pi e n_e L^2}{c B_0} \\ 
    &\simeq \unit[6.4 \times 10^4]{yr} \left( \frac{n_e}{\unit[10^{35}]{cm^{-3}}} \right)
    \left( \frac{L}{\unit[1]{km}} \right)^2 
    \left( \frac{B_0}{\unit[10^{14}]{G}} \right)^{-1}, 
        \label{eq:ch3_tau_hall} \\
    \tau_{\rm Ohm} &= \frac{4 \pi \sigma L^2}{c^2} \nonumber \\ 
    &\simeq \unit[4.4 \times 10^6]{yr} \left( \frac{\sigma}{\unit[10^{24}]{s^{-1}}} \right)
    \left( \frac{L}{\unit[1]{km}} \right)^2, 
        \label{eq:ch3_tau_ohm}
\end{align}
%----------------------------------------------------
where we remind that $\sigma$ is the dominant conductivity based on phonon or impurity scattering and $L$ is the typical lengthscale over which the relevant physical quantities (i.e., $n_e$, $B_0$ and $\sigma$) change inside the crust \citep[see][]{Cumming2004, Gourgouliatos2014a}. 
Note also that in Equation~\eqref{eq:ch3_B_decay}, $\tau_{\rm Hall,0}$ represents the Hall timescale for the initial magnetic field strength $B_0$ \citep[we refer to][for more details]{Aguilera2008b}. 
In general, the value of the Hall and the Ohmic timescales can vary by orders of magnitude within the crust and during the evolution, depending strongly on the density profile and magnetic field intensity and curvature. The Ohmic timescale additionally varies with temperature due to its dependence on the conductivity (see Section~\ref{sec:ch1_B_evol}). 
Although a rough simplification, Equation~\eqref{eq:ch3_B_decay} and the timescales defined above are consistent with the evolution history inferred for Galactic magnetars \citep{Colpi2000, Dall'Osso2012, Beniamini2019} and are able to qualitatively capture the magnetic field evolution obtained with more sophisticated magneto-thermal numerical simulations \citep[see for example][]{Vigano2013, DeGrandis2020, Vigano2021}.

Using these prescriptions for the magnetic field, we solve Equation~\eqref{eq:ch3_domegadt} and determine the evolution of the spin-period in time. In Figure~\ref{fig:ch3_em_spin_down_P_evol}, we present the corresponding behaviour for $B_0$ values in the range $\unit[10^{12-17}]{G}$. Since the long-term evolution for $t \gg t_{\rm em}$ is insensitive to the initial spin period value, we fix $P_0$ to a fiducial value of $\unit[10]{ms}$ which is compatible with the birth spin-period distributions inferred from population synthesis studies \citep[see for example][]{Faucher2006, Gullon2014, Cieslar2020}. After an initial phase of duration $t_{\rm em}$, where the spin period remains almost constant and equal to its initial value, $P$ starts to evolve $\propto t^{1/2}$. If the magnetic field remains constant, this evolution proceeds indefinitely (dashed lines). If $B$ decays over time according to Equation~\eqref{eq:ch3_B_decay}, the electromagnetic torque eventually becomes negligible and the spin period stops increasing and stabilises (solid lines).

From this plot, it is evident that, if one assumes a decaying magnetic field in the neutron star crust, explaining the existence of slowly rotating neutron stars with spin periods $P \gtrsim \unit[100]{s}$ becomes problematic since one would require rather extreme magnetic field values ($B \gtrsim \unit[10^{16}]{G}$). If instead, some mechanism prevents the crustal magnetic field from decaying in time, the star could reach longer spin periods more easily since a strong electromagnetic torque is maintained over longer times. After a few Hall timescales, the magnetic field in the crust reorganises towards smaller scales through the Hall cascade \citep[e.g.][]{Brandenburg2020} and approaches a quasi-equilibrium configuration during which the dissipation of the magnetic field is slowed down. In this phase, commonly named the ``Hall attractor'' \citep{Gourgouliatos2014b}, the electric currents are predominantly confined to the inner crust where subsequent dissipation occurs on the Ohmic timescale, which in turn depends on the local properties of the inner crust. As a result, magnetic field decay cannot be halted indefinitely but will proceed on timescales of a few Myr \citep[see however][on how the pasta phase could greatly shorten these timescales]{Pons2013}.
Another possibility for maintaining strong magnetic fields is that the electric currents could be predominantly present in the neutron star core allowing the magnetic field to stay stable and barely decay over time \citep[see for example][]{Vigano2021}. However it is still unclear if such conditions can be realised since little is known about magnetic field evolution in neutron star cores \citep[in particular the effects of superfluid and superconducting components on field evolution is uncertain;][]{Graber2015, Passamonti2017, Ofengeim2018, Gusakov2020}. In general, we cannot exclude the possibility that long-period pulsars are the result only of electromagnetic spin-down in the presence of strong and persistent magnetic fields, possibly supported by a Hall attractor or a core component. However, in light of our current knowledge of magnetic field evolution in neutron stars, we suggest that dipolar spin-down alone struggles to explain the existence of long-period pulsars. Instead, we argue below that these sources could be the result of a different scenario, whose ingredients are readily available in standard neutron-star formation models.   

%%%%%%%%%%%%%%%%%%%%%%%%%%%%%%%%%%%%%%%%%%%%%%%%%%%%%%
\section{Accretion from a fallback disk} 
\label{sec:ch3_fallback_scenario}

An alternative scenario that could explain the existence of strongly magnetised and slowly rotating pulsars involves the interaction between a newly born highly magnetic neutron star and fallback material from the supernova explosion. 

In the early stages after the supernova explosion, the proto-neutron star emits a powerful neutrino wind which exerts a pressure on the outer envelope of the progenitor star. The duration of this wind is believed to be of the order of $\sim \unit[10]{s}$ after core bounce, which corresponds to the neutrino-cooling timescale for a newborn neutron star \citep{Ugliano2012, Ertl2016}. On this timescale, the magnetosphere of the neutron star reaches an equilibrium configuration in the region swept by the neutrino wind. After the subsiding of this neutrino-driven wind, the gas in the inner envelopes of the exploding star decelerates due to the gravitational pull of the neutron star and collapses back. This leads to the onset of fallback accretion. According to simulations \citep{Ugliano2012, Ertl2016, Janka2022}, the total fallback mass can reach values up to $M_{\rm fb} \lesssim 0.1 M_{\odot}$, while the fallback mass rate can reach values of around $\sim \unit[10^{27-31}]{g \, s^{-1}}$ in the first $\sim \unit[10-100]{s}$ after bounce, afterwards decreasing according to a power law $\sim t^{-5/3}$, which is compatible with theoretical predictions for spherical supernova fallback \citep{Michel1988, Chevalier1989}. However, if part of the fallback matter possesses sufficient angular momentum, it will circularise to form an accretion disk. 

\citet{Mineshige1997} and \citet{Menou2001} have studied the formation and time evolution of fallback accretion disks around compact objects. In particular, by using the self-similar solution of \citet{Cannizzo1990}, \citet{Menou2001} found that the fallback material with excess angular momentum circularises to form a disk on a typical viscous timescale: 
%----------------------------------------------------
\begin{align} \label{eq:ch3_visc_timescale}
    t_{\rm v} \sim \unit[2.08]{s} \times 10^3 \left( \frac{T_{\rm c}}{\unit[10^6]{K}} \right)^{-1} \left( \frac{r_{\rm d}}{ \unit[10^8]{cm} } \right)^{1/2}, 
\end{align}
%----------------------------------------------------
where $T_{\rm c}$ is the disk's central temperature and $r_{\rm d}$ is the circularisation radius of the disk. We assume the circularisation radius $r_{\rm d}$ to be equal to the Keplerian radius corresponding to the initial angular momentum of the fallback matter. Consider a mass element orbiting with a velocity $v$ at the Keplerian radius $r_{\rm d}$ around a central neutron star of mass $M_{\rm NS}$. The angular momentum density (i.e., the angular momentum per unit mass) associated with this mass element is $j = v r_{\rm d}$. The Keplerian radius of the orbit is given by $r_{\rm d} = G M_{\rm NS}/v^2$. Therefore, by combining these two equations, we find that:
%----------------------------------------------------
\begin{align} \label{eq:ch3_circ_radius}
    j = \left( G M_{\rm NS} r_{\rm d} \right)^{1/2} \, \Longrightarrow \, r_{\rm d} = \frac{j^2}{G M_{\rm NS}}
\end{align}
%----------------------------------------------------
Supernova simulations have shown that typical values for the angular momentum density of the fallback matter are around $j_{\rm fb} \sim \unit[10^{16-17}]{cm^2 \, s^{-1}}$ \citep{Janka2022}, which corresponds to a circularisation radius of around $\unit[10^{6-8}]{cm}$. In the following, we will always assume the fiducial values $T_{\rm c} = \unit[10^6]{K}$ and $r_{\rm d} = \unit[10^8]{cm}$ \citep[see also][]{Hameury1998}. 

The viscous timescale defined in Equation~\eqref{eq:ch3_visc_timescale} determines the duration of an initial transient accretion phase characterised by a nearly constant accretion rate. Afterwards, as the supply of fallback matter decreases, the accretion rate into the disk itself declines as a power law. Furthermore, the disk starts to spread due to viscous effects. We follow these prescriptions and model the long-term time evolution of the accretion rate and outer radius of the disk as \citep[see][]{Menou2001, Ertan2009}:
%----------------------------------------------------
\begin{align} \label{eq:ch3_disk_rate_evolution}
    \dot{M}_{\rm d}(t) = \dot{M}_{\rm d, 0} \left( 1 + \frac{t}{t_{\rm v}} \right)^{-\alpha},
\end{align}
%----------------------------------------------------
%----------------------------------------------------
\begin{align} \label{eq:ch3_outer_radius_evolution}
    r_{\rm out}(t) = r_{\rm d} \left( 1 + \frac{t}{t_{\rm v}} \right)^{\gamma},
\end{align}
%----------------------------------------------------
\noindent where the coefficients $\alpha$ and $\gamma$ depend on the main mechanism determining the opacity of the disk. In particular, $\alpha = 19/16 \simeq 1.18$ and $\gamma = 3/8 \simeq 0.38$ if the disk opacity is dominated by electron scattering or $\alpha = 5/4 = 1.25$ and $\gamma = 1/2$ if the disk is dominated by Kramer's opacity \citep{Cannizzo1990, Menou2001, Ertan2009}. In the following, we adopt intermediate values of $\alpha = 1.2$ and $\gamma = 0.44$. 

In general, determining the fraction of fallback matter that eventually forms the disk is not trivial, since this depends on the progenitor's properties and on the supernova mechanism itself. Here, we consider a broad range of values for the disk initial accretion rate $\dot{M}_{\rm d, 0}$ between $\unit[10^{19-29}]{g \, s^{-1}}$. This is compatible with the disk accreting at a fraction of the overall supernova fallback rate as well as having a fraction of the total fallback mass $M_{\rm fb}$. 

From the accretion rate inside the disk we can compute an estimate of the accretion luminosity.
Consider a mass element ${\rm d}M$ in the accretion disk around the neutron star. To fall from a Keplerian orbit at radius $r + {\rm d}r$ to an orbit of radius $r$, the mass element must lose gravitational potential energy. The virial theorem \citep{Prialnik2000, Binney2008} dictates that half the potential energy is converted into additional kinetic energy. The remaining half is converted to heat due to viscous friction. The thermal energy released as the mass element moves inward will therefore be:
%----------------------------------------------------
\begin{align} 
    {\rm d}E_{\rm th} = \frac{1}{2} G M_{\rm NS} {\rm d}M \left( \frac{1}{r} - \frac{1}{r + dr} \right) \simeq \frac{1}{2} G M_{\rm NS} {\rm d}M \frac{{\rm d}r}{r^2},
\end{align}
%----------------------------------------------------
where we assume that ${\rm d}r \ll r$.

For a radiatively efficient disk we assume that the hot gas radiates its thermal energy as a black body at the same radius where the gravitational energy is liberated. The luminosity from an annulus in the disk characterised by the accretion rate $\dot{M}_{\rm d} = {\rm d}M / {\rm d}t$ is then given by:
%----------------------------------------------------
\begin{align} 
    {\rm d} L = \frac{{\rm d}E_{\rm th}}{{\rm d}t} =  \frac{1}{2} G M_{\rm NS} \dot{M}_{\rm d} \frac{{\rm d}r}{r^2}.
\end{align}
%----------------------------------------------------
The total luminosity of the accretion disk with inner and outer radii, $r_{\rm in}$ and $r_{\rm out}$ respectively, is found by integrating the luminosity over all annuli:
%----------------------------------------------------
\begin{align} \label{eq:ch3_accretion_L}
    L_{\rm acc} = \int_{r_{\rm in}}^{r_{\rm out}} \frac{1}{2} \frac{G M_{\rm NS} \dot{M}_{\rm d}}{r^2} {\rm d}r = \frac{1}{2} G M_{\rm NS} \dot{M}_{\rm d} \left( \frac{1}{r_{\rm in}} - \frac{1}{r_{\rm out}} \right) \simeq \frac{G M_{\rm NS} \dot{M}_{\rm d}}{2 r_{\rm in}},
\end{align}
%----------------------------------------------------
where we assumed that the mass accretion rate in the disk $\dot{M}_{\rm d}$ does not depend on the radius $r$ and that $r_{\rm in} \ll r_{\rm out}$.
If the flow of matter is constant throughout the disk up to the inner disk radius $r_{\rm in}$, fiducial values $M_{\rm NS} = 1.4 M_{\odot}$, $\dot{M}_{\rm d} = \unit[10^{28}]{g \, s^{-1}}$ and $r_{\rm in} = \unit[10^8]{cm}$ result in luminosities up to $L_{\rm acc} \simeq \unit[10^{46}]{erg \, s^{-1}}$. These luminosities by far exceed the Eddington limit given by \citep{Rybicki1986}:
%----------------------------------------------------
\begin{align} 
L_{\rm Edd} = \frac{4 \pi G M_{\rm NS} m_{\rm p} c}{\sigma_{\rm T}} \simeq \unit[1.8 \times 10^{38}]{erg \, s^{-1}} \left( \frac{M_{\rm NS}}{1.4 \, M_{\odot}} \right),
\end{align}
%----------------------------------------------------
where $m_{\rm p}$ is the proton mass and $\sigma_{\rm T}$ is the Thomson cross section and we naively assumed a disk composition of pure ionised hydrogen.
It is generally expected that such super-critical inflows of matter produce winds and outflows that reduce the accretion rate in the inner disk region to values below the Eddington limit \citep{poutanen2007}. 
We note however that in the presence of strong magnetic fields ($B \gtrsim \unit[10^{14}]{G}$), the Eddington limit could be increased due to a reduction in the electron-scattering opacity \citep{Canuto1971, Bachetti2014}. In our model, we neglect this effect as it is relevant only in the very early stages of the evolution when the inner disk radius closely approaches the neutron star surface but it does not significantly affect the long-term dynamics outlined below.
In a simplified scenario, we hence assume that the accretion rate at the inner disk radius $r_{\rm in}$ has to be limited by the Eddington accretion rate, given by $\dot{M}_{\rm Edd} \simeq 2 L_{\rm Edd} r_{\rm in} / (G M_{\rm NS}) \simeq \unit[10^{18} (r_{\rm in}/R_{\rm NS})]{g \, s^{-1}}$. 
Therefore, we model the accretion rate at $r_{\rm in}$ as:
%----------------------------------------------------
\begin{align} \label{eq:ch3_accretion_rate_evolution}
    \dot{M}_{\rm d, in}(t) = 
        \begin{cases}
          \dot{M}_{\rm Edd} & \text{if $\dot{M}_{\rm d} \geq \dot{M}_{\rm Edd}$}, \\
          \dot{M}_{\rm d}(t)   & \text{if $\dot{M}_{\rm d}(t) < \dot{M}_{\rm Edd}$}.
        \end{cases}
\end{align}
%----------------------------------------------------
In the following we discuss how the evolution of the disk and its influence on the spin-period evolution of the central neutron star depends on the complex interaction with the neutron star magnetosphere.

\section{Period evolution in the presence of a fallback disk} \label{sec:ch3_torque_fallback}

Hereafter, we will study in detail the case where a disk forms successfully and interacts with the central neutron star. In the following for simplicity, we assume that the star's magnetic dipole moment is aligned with the rotation axis and that the magnetic field is crust-dominated and decays in time according to Equation~\eqref{eq:ch3_B_decay}. We also assume that the accretion disk's rotation is prograde with respect to the spin of the neutron star. We note that in order to observe pulsed radio emission from sources of this kind, a misalignment between the magnetic axis and the spin axis is, in principle, required. We do not study the impact of this effect in our calculations but point out that for small misalignment angles the evolutionary scenario should not differ substantially from the one studied in this work.

We consider two different cases depending on if the disk is able to influence the closed magnetosphere of the neutron star or not.
If the accretion rate is sufficiently high, the in-falling matter is able to deform and penetrate the closed magnetosphere whose boundary can be roughly defined by the light cylinder radius $r_{\rm lc}$ (Equation~\eqref{eq:ch1_light_cylinder}).
In this case, we obtain the \textit{magnetospheric radius} $r_{\rm m}$ which represents the distance from the star where the magnetic pressure equals the ram pressure of the accreted flow \citep{Davidson1973, Elsner1977, Ghosh1979} through:
%----------------------------------------------------
\begin{align} \label{eq:ch3_B_ram_pressure_balance}
    \frac{B^2}{8 \pi} = \frac{1}{2} \rho v_{\rm ff},
\end{align}
%----------------------------------------------------
where $\rho$ is the mass density of the accreting matter and $v_{\rm ff} = (2 G M_{\rm NS} / r)^{1/2}$ is its free-fall velocity at radius $r$. Assuming that the disk interacts with the magnetosphere at the magnetic equator, we can express the magnetic field in terms of the magnetic dipole moment as $B = \mu / r^3$ (see Equation~\eqref{eq:ch1_mag_dipole_B_equat}). Furthermore, by considering for simplicity spherical accretion we can express the accretion rate as $\dot{M}_{\rm d, in} = 4 \pi r^2 \rho v_{\rm ff}$. By substituting this information into Equation~\eqref{eq:ch3_B_ram_pressure_balance} we can solve to find the magnetospheric radius:
%----------------------------------------------------
\begin{align} \label{eq:ch3_magnetospheric_radius}
    r_{\rm m} = \xi \left( \frac{\mu^4}{2 G M_{\rm NS} \dot{M}_{\rm d, in}^2} \right)^{1/7} ,
\end{align}
%----------------------------------------------------
where $\xi \sim 0.5$ is a corrective factor that takes into account that the accretion disk has a non-spherical geometry \citep{Long2005, Bessolaz2008, Zanni2013}.
To relate the stellar magnetic moment to the time-dependent magnetic field at the pole $B_{\rm p}(t)$ (assuming a dipolar structure for the magnetosphere), we can use Equation~\eqref{eq:ch1_mag_dipole_B_pole} to obtain $\mu = B_{\rm p}(t) R_{\rm NS}^3 / 2$. Note that the magnetospheric radius evolves in time due to the decay of both the disk accretion rate and the magnetic field. Inside the magnetospheric radius, the plasma dynamics are dominated by the magnetic field such that the accretion flow is forced to corotate with the star's closed magnetosphere. Under such conditions, the magnetospheric radius roughly determines the inner edge of the accretion disk so that we can assume $r_{\rm in} \sim r_{\rm m}$. 

If the accretion rate is sufficiently low, the magnetic pressure of the closed magnetosphere is able to keep the accretion flow outside the light cylinder. However beyond $r_{\rm lc}$, the dipole configuration breaks down and the magnetic field lines open up. As a consequence, they are no longer able to exert significant pressure on the accreted plasma. Under these conditions, we expect the inner radius $r_{\rm in}$ of the disk to be roughly equal to $r_{\rm lc}$. Thus, in general we adopt the same prescription as \citet{Yan2012} and assume $r_{\rm in} \simeq \min(r_{\rm m}, r_{\rm lc})$.

Note that a condition for the disk to form and remain active is that the disk's outer radius satisfies $r_{\rm out} > r_{\rm in}$. Otherwise, the stellar magnetic field would completely disrupt the disk, restoring a configuration where the neutron star's spin-down is determined by dipolar losses only.

Another critical lengthscale is the \textit{corotation radius}
%----------------------------------------------------
\begin{align} \label{eq:ch3_corotation_radius}
    r_{\rm cor} = \left( \frac{G M_{\rm NS}}{\omega^2} \right)^{1/3},
\end{align}
%----------------------------------------------------
which represents the distance at which the gravitational pull of the neutron star balances the centrifugal force for a test mass that is corotating with the star at the spin frequency $\omega$.

The position of the magnetospheric radius with respect to the light cylinder radius and the corotation radius determines the total torque $N_{\rm tot}$ exerted on the star, which in turn drives the time evolution of the spin period according to:
%----------------------------------------------------
\begin{align} \label{eq:ch3_omega_evolution}
I_{\rm NS} \dot{\omega} = N_{\rm tot}.
\end{align}
%----------------------------------------------------
Following \citet{Piro2011, Metzger2018} the total torque can be modelled by the following equation:
%----------------------------------------------------
\begin{align} \label{eq:ch3_torque}
    N_{\rm tot} &= N_{\rm acc} + N_{\rm dip} \nonumber \\ \nonumber
    			&= \dot{M}_{\rm d, in} r_{\rm in}^2 \left[ \Omega_{\rm K}(r_{\rm in}) - \omega \right] - I_{\rm NS} \left( \frac{r_{\rm lc}}{r_{\rm in}} \right)^2 \beta B^2 \omega^3 \\ 
    			&=
        \begin{cases}
          \dot{M}_{\rm d, in} r_{\rm lc}^2 \left[ \Omega_{\rm K}(r_{\rm lc}) - \omega \right] - I_{\rm NS} \beta B^2 \omega^3   & \text{if $r_{\rm m} > r_{\rm lc}$}, \\
          \dot{M}_{\rm d, in} r_{\rm m}^2 \left[ \Omega_{\rm K}(r_{\rm m}) - \omega \right] - I_{\rm NS} \left( \frac{r_{\rm lc}}{r_{\rm m}} \right)^2 \beta B^2 \omega^3 & \text{if $r_{\rm m} \leq r_{\rm lc}$}.
        \end{cases}
\end{align}
%----------------------------------------------------
where $N_{\rm dip}$ accounts for the electromagnetic torque of the magnetosphere while $N_{\rm acc}$ accounts for the torque exerted by the accretion process and $\Omega_{\rm K}(r) = \left( G M_{\rm NS} / r^3 \right)^{1/2}$ is the Keplerian orbital angular velocity at radius $r$.
%----------------------------------------------------
\begin{figure}
\centering
\includegraphics[width = 0.6\textwidth]{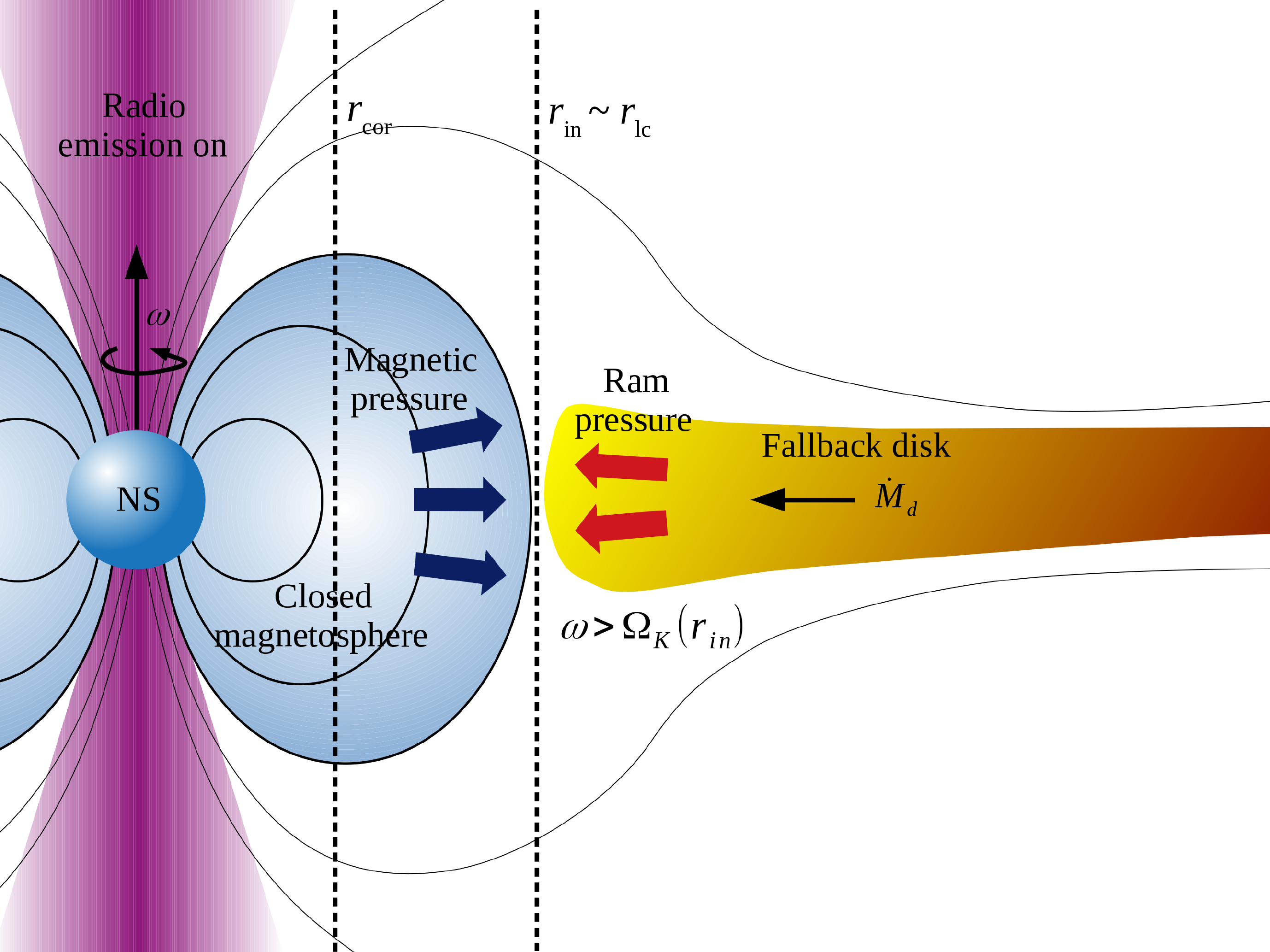}
\includegraphics[width = 0.6\textwidth]{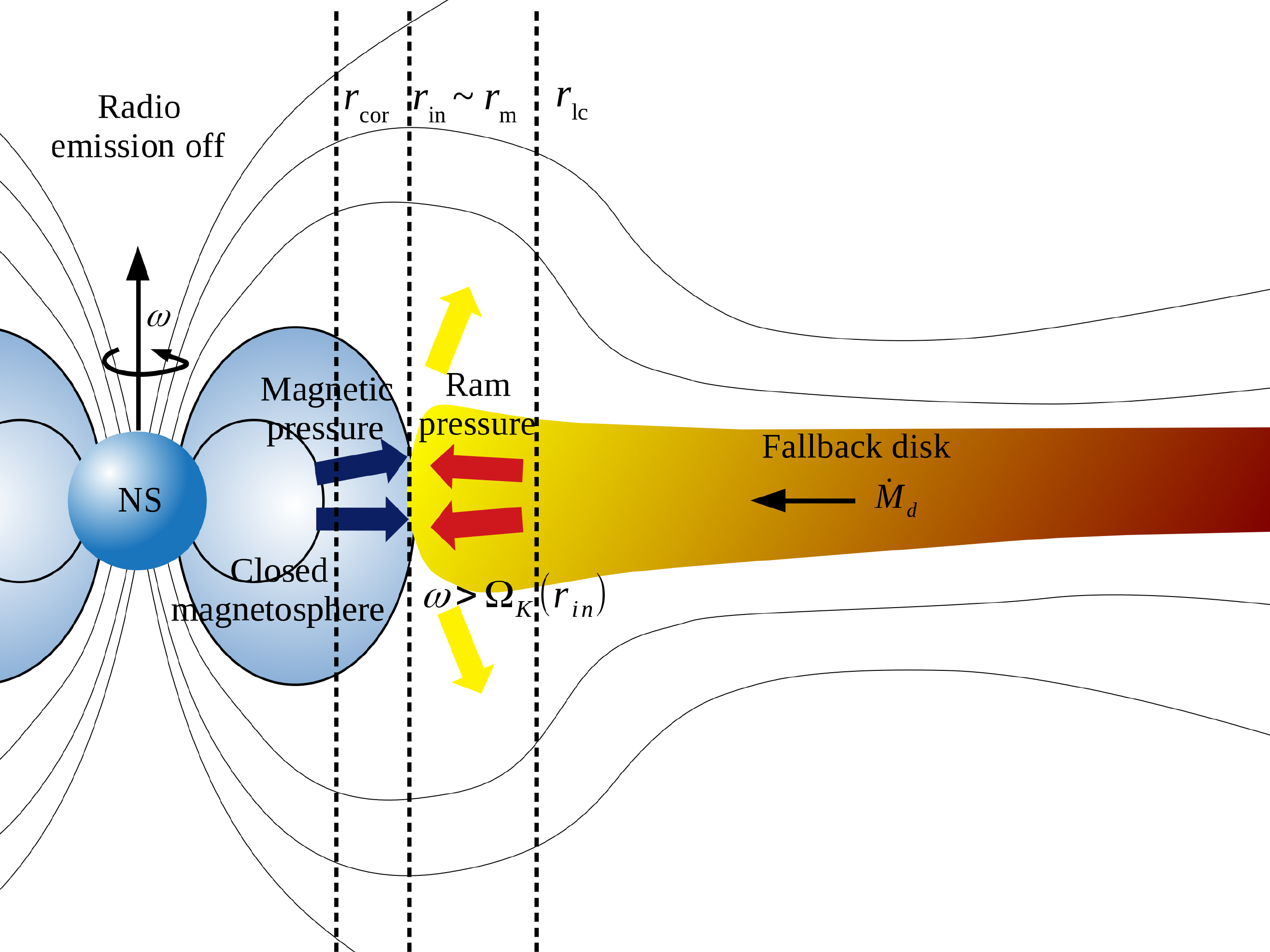}
\includegraphics[width = 0.6\textwidth]{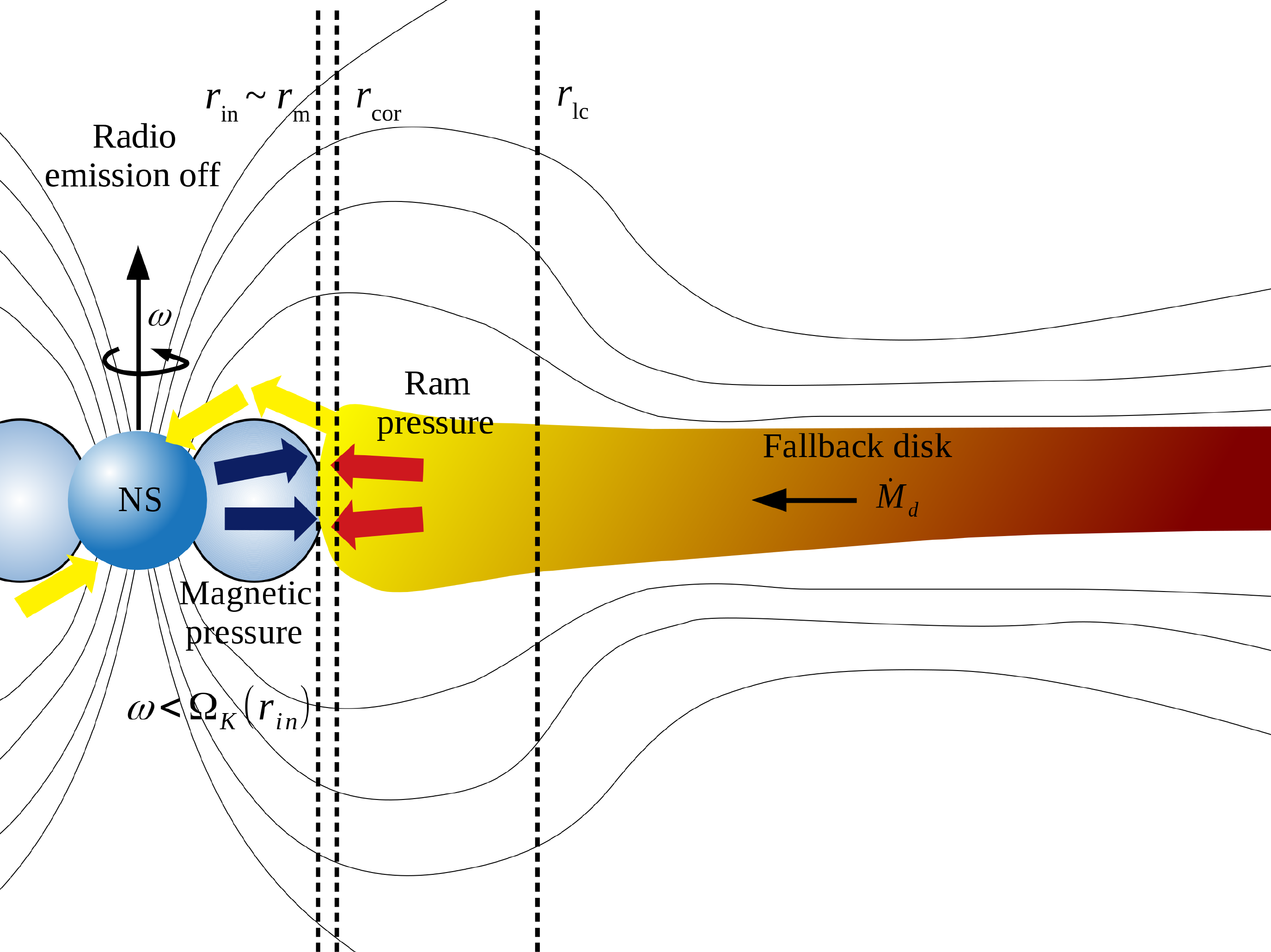}
\caption[Sketch of the three phases: ejector, propeller and direct accretion]{Sketch showing the three phases a neutron star with a fallback disk system can experience during its evolution. During the ejector phase (\textit{top panel}) the fallback disk is confined outside the light cylinder and the neutron star spin down mainly by dipolar electromagnetic losses. In the propeller phase (\textit{middle panel}) the fallback disk penetrate inside the closed magnetosphere but at the $r_{\rm in}$ it finds a centrifugal barrier. The neutron star spin down mainly by propeller torque. In the direct accretion phase (\textit{bottom panel}) the fallback disk penetrates inside the closed magnetosphere and it is super-Keplerian at $r_{\rm in}$. The neutron star spins up and accretes on the surface. } 
\label{fig:ch3_ejector_prop_sketch}
\end{figure}  
%----------------------------------------------------
%----------------------------------------------------
\begin{figure}
\centering
\includegraphics[width = 0.8\textwidth]{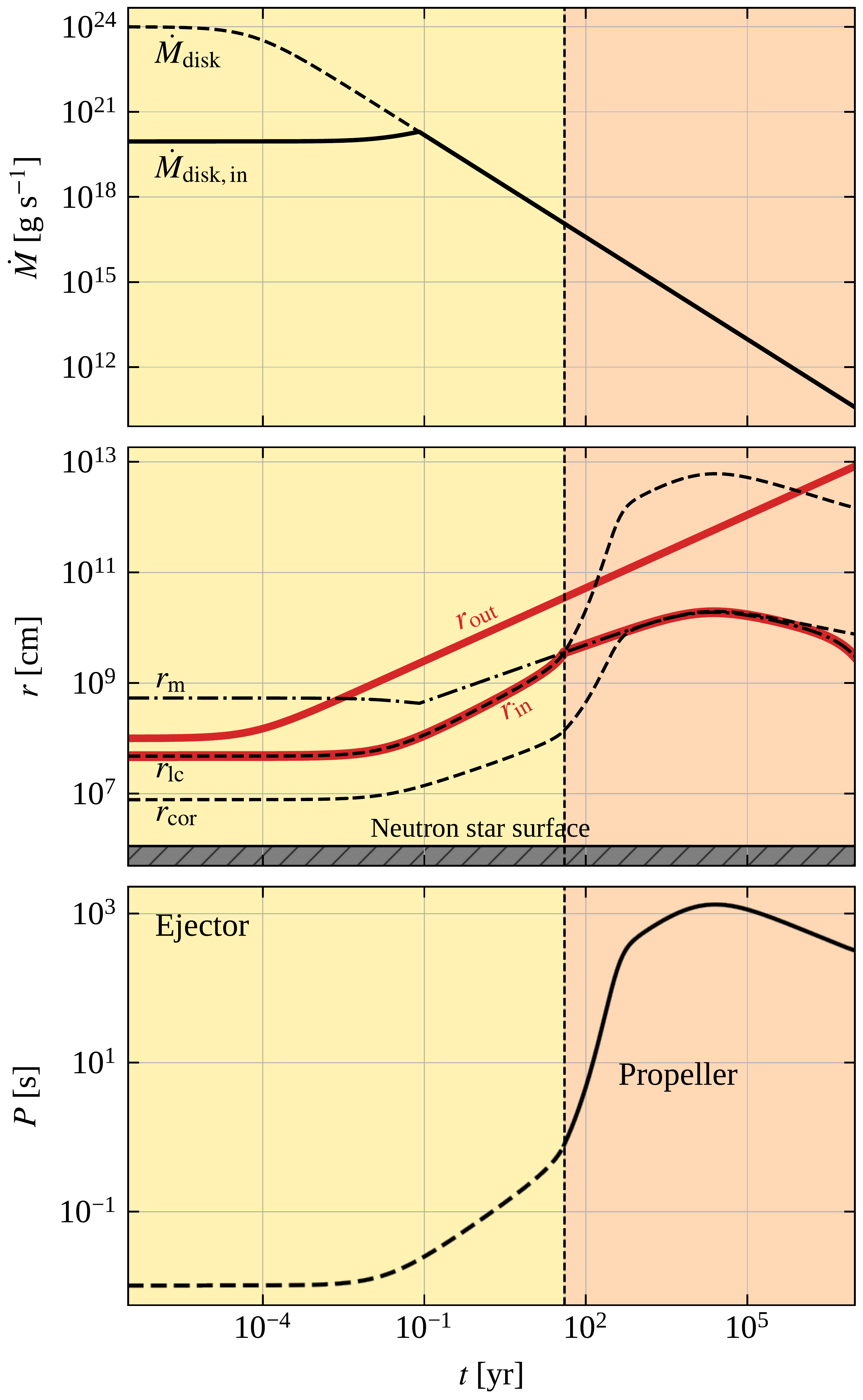}
\caption[Example of neutron star spin-down in the presence of a fallback disk]{Example of neutron star spin-down in the presence of a fallback disk on a timescale of $\unit[10^7]{yr}$. The top panel shows the total disk accretion $\dot{M}_{\rm d}$ and the Eddington-limited accretion rate at the inner radius $\dot{M}_{\rm d, in}$. The middle panel illustrates the evolution of the three critical radii $r_{\rm cor}$, $r_{\rm m}$, $r_{\rm lc}$. The neutron star radius is also indicated as a reference. The evolution of the disk's inner and outer radii is highlighted by the red lines. The bottom panel shows the resulting time evolution of the spin period. We assume an initial spin period $P_0 = \unit[10]{ms}$, initial magnetic field $B_0 = \unit[4 \times 10^{14}]{G}$ and an initial disk accretion rate $\dot{M}_{\rm d,0} = \unit[10^{24}]{g \, s^{-1}}$. We also highlight the duration of the ejector and propeller phases by shading the background in yellow and orange, respectively.} 
\label{fig:ch3_prop_evolution_ex}
\end{figure}  
%----------------------------------------------------

Figure~\ref{fig:ch3_prop_evolution_ex} shows an example of a solution of Equation~\eqref{eq:ch3_omega_evolution} for a pulsar interacting with a fallback disk. We consider an initial spin period $P_0 = \unit[10]{ms}$, initial magnetic field $B_0 = \unit[4 \times 10^{14}]{G}$ and an initial disk accretion rate $\dot{M}_{\rm d,0} = \unit[10^{24}]{g \, s^{-1}}$. Note that for this choice of initial parameters the disk can form and stay active since $r_{\rm out} > r_{\rm in}$ at all times. 
Furthermore, in the early phases, accretion at the inner radius of the disk is limited by the Eddington limit (see Section~\ref{sec:ch3_fallback_scenario}). 

Different regimes are present depending on the relative ordering of the three radii defined above. In particular, with reference to Figures~\ref{fig:ch3_ejector_prop_sketch} and~\ref{fig:ch3_prop_evolution_ex}, we distinguish the three following phases.

\paragraph{Ejector phase:} 

for $r_{\rm m} > r_{\rm lc} > r_{\rm cor}$, the accreted material remains at the boundary of the closed magnetosphere and does not influence the neutron star's internal dynamics, i.e., the star spins down mainly due to dipolar electromagnetic torques, that is $N_{\rm tot} \sim N_{\rm dip}$ (top panel in Figure~\ref{fig:ch3_ejector_prop_sketch}). After a phase of duration $t_{\rm em}$ where the spin period stays constant, $P$ starts to increase $\propto t^{1/2}$ (see Equation~\eqref{eq:ch3_P_t_em_spindown}). As a consequence, the characteristic radii $r_{\rm lc}$ and $r_{\rm cor}$ increase as $t^{1/2}$ and $t^{2/3}$, respectively. This phase is commonly referred to as the ejector phase (shaded yellow region in Figure~\ref{fig:ch3_prop_evolution_ex}). The mechanism responsible for radio emission can be active and the neutron star could be observed as a radio pulsar. Morover, as $\dot{M}_{\rm d}$ decreases and eventually becomes sub-Eddington, the magnetospheric radius grows slower than the other two critical radii ($\propto t^{2 \alpha/7} \simeq t^{0.34}$ for $\alpha=1.2$). Thus, $r_{\rm m}$ will eventually cross the light cylinder.

\paragraph{Propeller phase:} 

for $r_{\rm lc} > r_{\rm m} > r_{\rm cor}$, the accretion flow is able to penetrate inside the closed magnetosphere (middle panel in Figure~\ref{fig:ch3_ejector_prop_sketch}). As it reaches the magnetospheric radius, the plasma flow is forced to corotate with the magnetosphere at super-Keplerian speeds causing it to be ejected due to centrifugal forces. This introduces a viscous torque that spins down the star very efficiently, a phase commonly referred to as the propeller (shaded orange region in Figure~\ref{fig:ch3_prop_evolution_ex}). In this case, we have $\omega > \Omega_K(r_{\rm m})$.
The electromagnetic torque $N_{\rm dip}$, i.e., the second term in Equation~\eqref{eq:ch3_torque}, is also influenced by accretion. When the accretion flow penetrates inside the closed magnetosphere, the fraction of the magnetic field lines, which connect the star to the disk, is forced to open and as a consequence, the polar cap region containing open field lines expands. This enhances the spin-down torque caused by the magnetosphere \citep{Parfrey2016, Metzger2018}. In particular \citet{Parfrey2016} argue that for $r_{\rm m} < r_{\rm lc}$, the magnetic flux through the expanded polar cap increases by a factor $\sim (r_{\rm lc}/ r_{\rm m})$. As a consequence, since the dipole spin-down torque is proportional to the square of the magnetic flux through the open field-line region, $N_{\rm dip}$ is enhanced by a factor $\sim (r_{\rm lc}/ r_{\rm m})^2$.
Moreover, since the radio emission is associated with magnetospheric currents and pair production \citep[][see also Section~\ref{sec:ch1_magnetosphere} and Section~\ref{sec:ch1_radio_em_deathlines}]{Beloborodov2008, Philippov2020}, the mechanism generating the radio emission is likely perturbed or even stopped as the closed-magnetosphere geometry is disturbed by the accretion flow. This effectively switches off the radio-loud nature of these sources \citep{Li2006}. Therefore, we expect a neutron star in the propeller regime unlikely to be observable as a radio pulsar.
During this propeller phase, the spin frequency decreases over time and eventually $\omega \sim \Omega_K(r_{\rm m})$ (i.e., when $r_{\rm cor} \sim r_{\rm m}$), so that the net torque exerted on the neutron star vanishes and a spin equilibrium is reached. From this point onward, further evolution of the spin period will be regulated mainly by time variation of the accretion rate in the inner disk $\dot{M}_{\rm d, in}$ and the stellar magnetic field $B$. For example, the decay of the magnetic field strength causes the slow decrease of the magnetospheric radius at late times. This drives a progressive shift of the spin-equilibrium radius towards the neutron star surface, causing the neutron star to slightly spin up; in Figure~\ref{fig:ch3_prop_evolution_ex}, this happens when the magnetic field starts to decay after a Hall timescale, i.e., after $\sim \unit[10^4]{yr}$.

\paragraph{Direct accretion phase:} 

for $r_{\rm lc} > r_{\rm cor} > r_{\rm m}$, we have $\omega < \Omega_K(r_{\rm m})$ (bottom panel in Figure~\ref{fig:ch3_ejector_prop_sketch}). The accretion flow still manages to penetrate inside the closed magnetosphere but as it reaches the magnetospheric radius, the plasma is forced to corotate with the magnetosphere at sub-Keplerian speeds. In other words, at the boundary defined by the magnetospheric radius, the accreted plasma is orbiting faster than the neutron star magnetosphere. This introduces a viscous torque that transfers angular momentum from the plasma flow to the star and tends to spin up the star. Besides this, the enhanced electromagnetic spin-down torque described above is still acting, opposing the spin-up due to accretion. Therefore, the evolution in this regime is controlled by the relative strength of $N_{\rm acc}$ and $N_{\rm dip}$. In the example in Figure~\ref{fig:ch3_prop_evolution_ex}, this phase is experienced at very late times, i.e. $\sim \unit[10^6]{yr}$, when the magnetic field has decayed so much that the star exits the spin equilibrium and the magnetospheric radius becomes smaller than the corotation radius.

\vspace{5mm}

From these considerations, we observe that the parameters that mainly regulate the evolution of a pulsar surrounded by a fallback disk are the initial stellar magnetic field $B_0$ and the initial disk accretion rate $\dot{M}_{\rm d,0}$. The latter determines the time at which the accretion rate affecting the compact object becomes sub-Eddington and starts decreasing. In contrast, the value of the neutron star's initial spin period only affects the early evolution stages, while the long-term evolution of the neutron star is almost insensitive to the value of $P_0$. 

%----------------------------------------------------
\begin{figure}
\centering
\includegraphics[width = 1\textwidth]{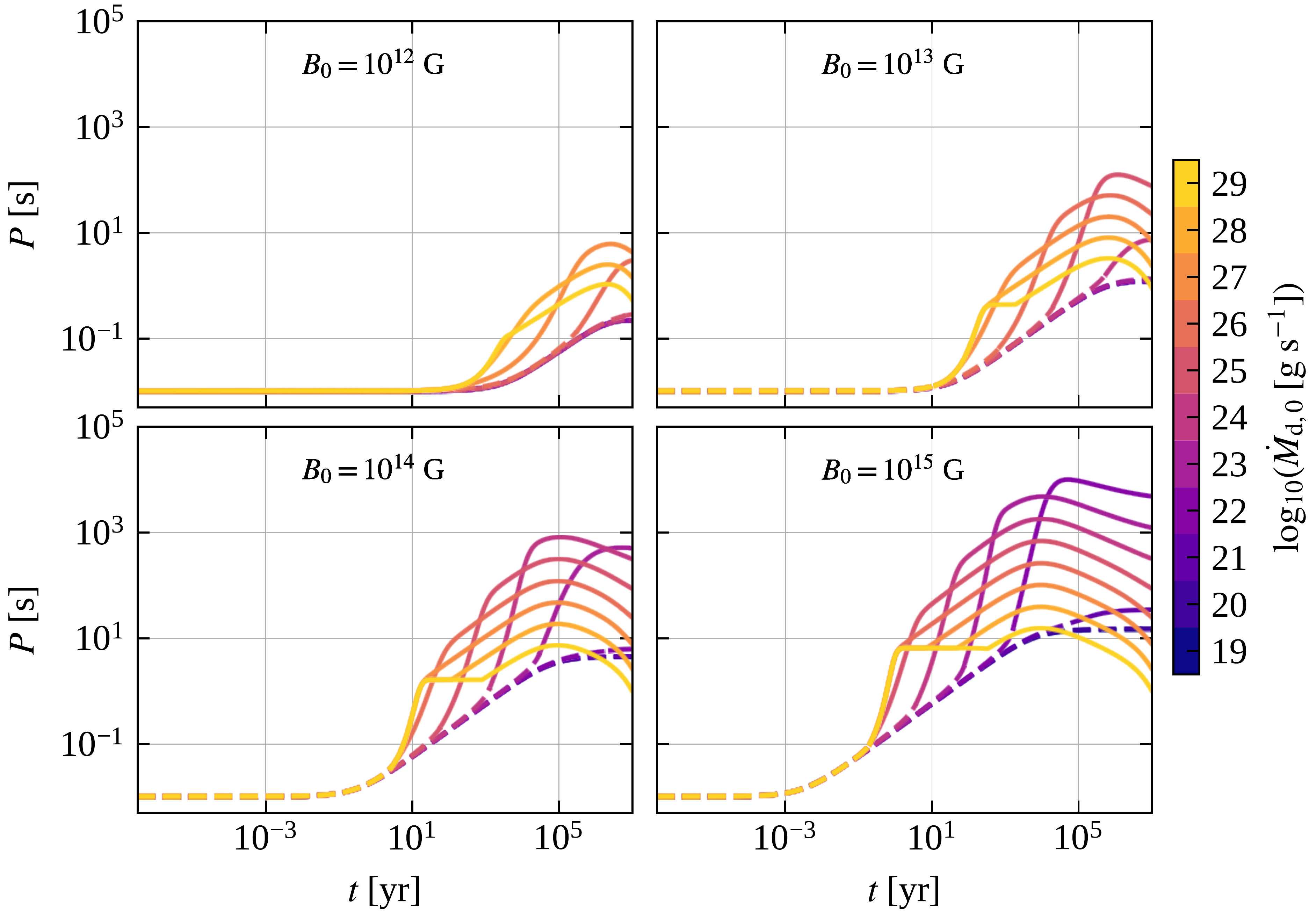}
\caption[Example of neutron star spin-down in the presence of a fallback disk for different accretion rates]{Example curves showing the time evolution of the spin period for a pulsar interacting with a fallback disk on a timescale of $\unit[10^7]{yr}$, for different assumptions on the initial $B_0$ field strength, and varying disk fallback rate $\dot{M}_{\rm d,0}$. The dashed portion of the curves indicates when the neutron star is in the radio-loud ejector phase, i.e., when $r_{\rm m} > r_{\rm lc}$, while the solid portion indicates when the neutron star is in the radio-quiet propeller stage, i.e., when $r_{\rm m} < r_{\rm lc}$.}
\label{fig:ch3_P_evolution_B}
\end{figure}  
%----------------------------------------------------

%----------------------------------------------------
\begin{figure}
\centering
\includegraphics[width = 1\textwidth]{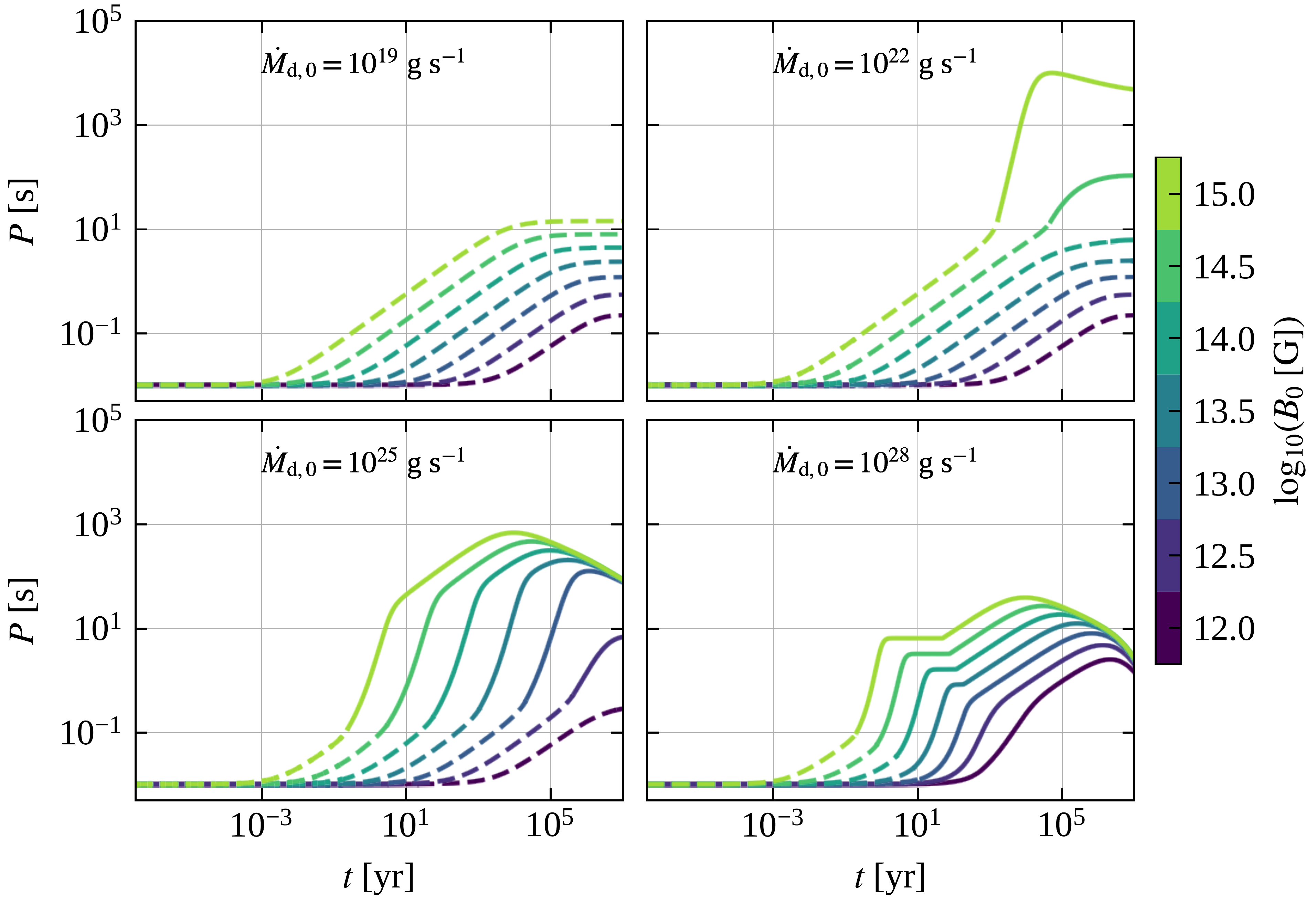}
\caption[Example of neutron star spin-down in the presence of a fallback disk for different initial magnetic fields]{Example curves showing the time evolution of the spin period for a pulsar interacting with a fallback disk on a timescale of $\unit[10^7]{yr}$, for different assumptions of the disk fallback rate $\dot{M}_{\rm d,0}$, and varying the initial magnetic field $B_0$. The dashed portion of the curves indicates when the neutron star is in the radio-loud ejector phase, i.e., when $r_{\rm m} > r_{\rm lc}$, while the solid portion indicates when the neutron star is in the radio-quiet propeller stage, i.e., when $r_{\rm m} < r_{\rm lc}$.}
\label{fig:ch3_P_evolution_Mdot}
\end{figure}  
%----------------------------------------------------

Figures~\ref{fig:ch3_P_evolution_B} and \ref{fig:ch3_P_evolution_Mdot} show several examples of evolutionary curves of the spin period. We always assume an initial spin period $P_0 = \unit[10]{ms}$. In particular, Figure~\ref{fig:ch3_P_evolution_B} shows the spin-period evolution for several values of the initial magnetic field $B_0$ and varying initial disk accretion rate $\dot{M}_{\rm d,0}$, while Figure~\ref{fig:ch3_P_evolution_Mdot} shows the evolution curves for several values of initial disk accretion rate and varying initial magnetic field. In the early phases, the accretion at the inner radius of the disk is limited by the Eddington limit. 
For each simulated evolutionary curve we check that the disk's outer radius is always greater than the inner radius guaranteeing that the disk can form and influence the rotational evolution of the neutron star. In general, it can be noted that for low values of the initial disk accretion rate ($< \unit[10^{22}]{g \, s^{-1}}$), the neutron star remains in the ejector phase for its entire evolution, independent of the value of its initial magnetic field. In contrast, higher accretion rates and higher magnetic fields allow the star to enter the propeller phase at earlier times. However, the propeller regime is most effective at spinning down the neutron star to periods $\gtrsim \unit[10]{s}$ only for relatively strong magnetic fields $\gtrsim \unit[10^{13}]{G}$ and intermediate initial disk accretion rates in the range $\unit[10^{22-27}]{g \, s^{-1}}$. 

We note that a neutron star that enters the propeller phase (and subsequently reaches spin equilibrium) will remain in this state until an abrupt change in the disk accretion rate occurs. For example, if the accretion rate in the disk suddenly drops, a neutron star can exit the propeller phase and enter the ejector phase again. In this case, the neutron star can transition from a faint X-ray source (due to thermal emission from material accreted onto the magnetosphere) to a standard rotation-powered radio pulsar or potentially radio-loud magnetar (see Section~\ref{sec:ch3_discussion} for a detailed explanation of how this transition can occur and a description of the expected X-ray and radio luminosity in the different fallback accretion states).

\section{Accretion from spherical fallback}

If the fallback matter does not possess sufficient angular momentum to form a disk, the fallback will proceed quasi-spherically. Even in this case, the neutron star could experience different accretion phases depending on the magnitude of the fallback rate. If the matter inflow is radiatively inefficient, accretion could proceed at super-Eddington rates, especially in the early stages. For such high fallback rates, the accretion flow is likely able to penetrate and squeeze the proto-neutron star magnetosphere, causing an initial phase of direct accretion onto the surface. This might also result in the burial of the magnetic field \citep[see for example][]{Taam1986, Li2021, Lin2021}, a scenario that (combined with the subsequent secular re-emergence of the magnetic field) has been invoked to explain the observed properties of Central Compact Objects (CCOs) \citep{Halpern2010, Fu2013, Ho2015, Zhong2021}; a class of young, generally weak-field neutron stars found close to the centres of supernova remnants (introduced in Section~\ref{sec:ch1_ns_zoo}). After this initial direct accretion phase, as the fallback rate decreases in time, the neutron star could, in principle, enter a propeller phase that causes the star to spin down as in the disk scenario. However, in the case of quasi-spherical accretion, the fallback episode is expected to last at most a free-fall timescale $t_{\rm ff} \sim (G \rho)^{-1/2}$, determined by the density $\rho$ of the infalling outer layers of the progenitor participating in the fallback. In general, $t_{\rm ff}$ could reach at most $\sim \unit[10^7]{s}$ for red supergiant progenitor stars, whose envelopes have typical densities $\rho \sim \unit[10^{-7}]{g \, cm^{-3}}$ \citep{Metzger2018}.
Therefore, it is unlikely that a propeller phase acting on such short timescales (of the order of $\sim \unit[1]{yr}$) could result in equally long spin periods as the disk scenario.

%%%%%%%%%%%%%%%%%%%%%%%%%%%%%%%%%%%%%%%%%%%%%%%%%%%%%%
\section{The 18-min periodic radio transient GLEAM-X\,J162759.5-523504.3}
\label{sec:ch3_gleam}

%----------------------------------------------------
\begin{figure}
\centering
\includegraphics[width=0.6\textwidth]{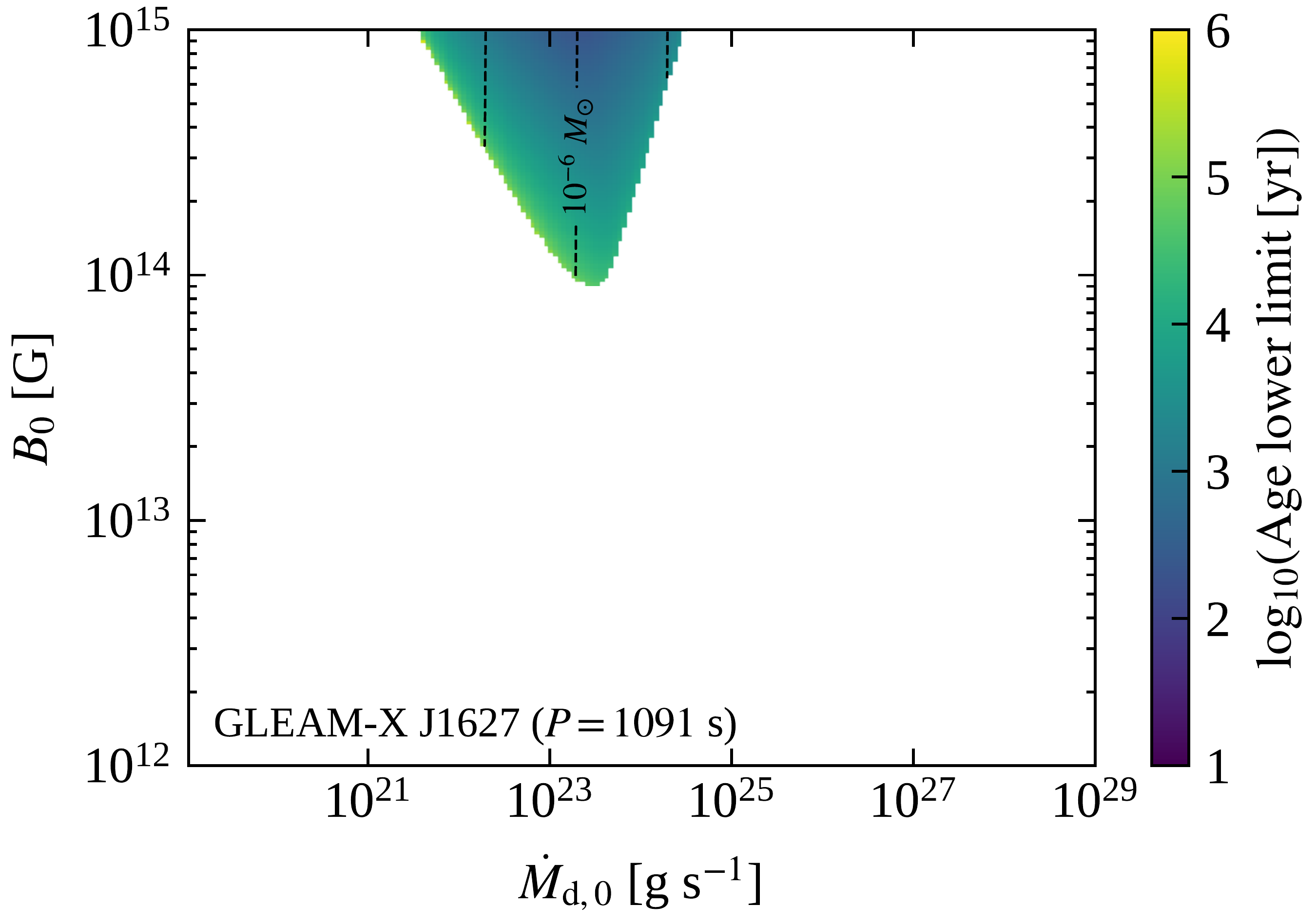}
\includegraphics[width=0.6\textwidth]{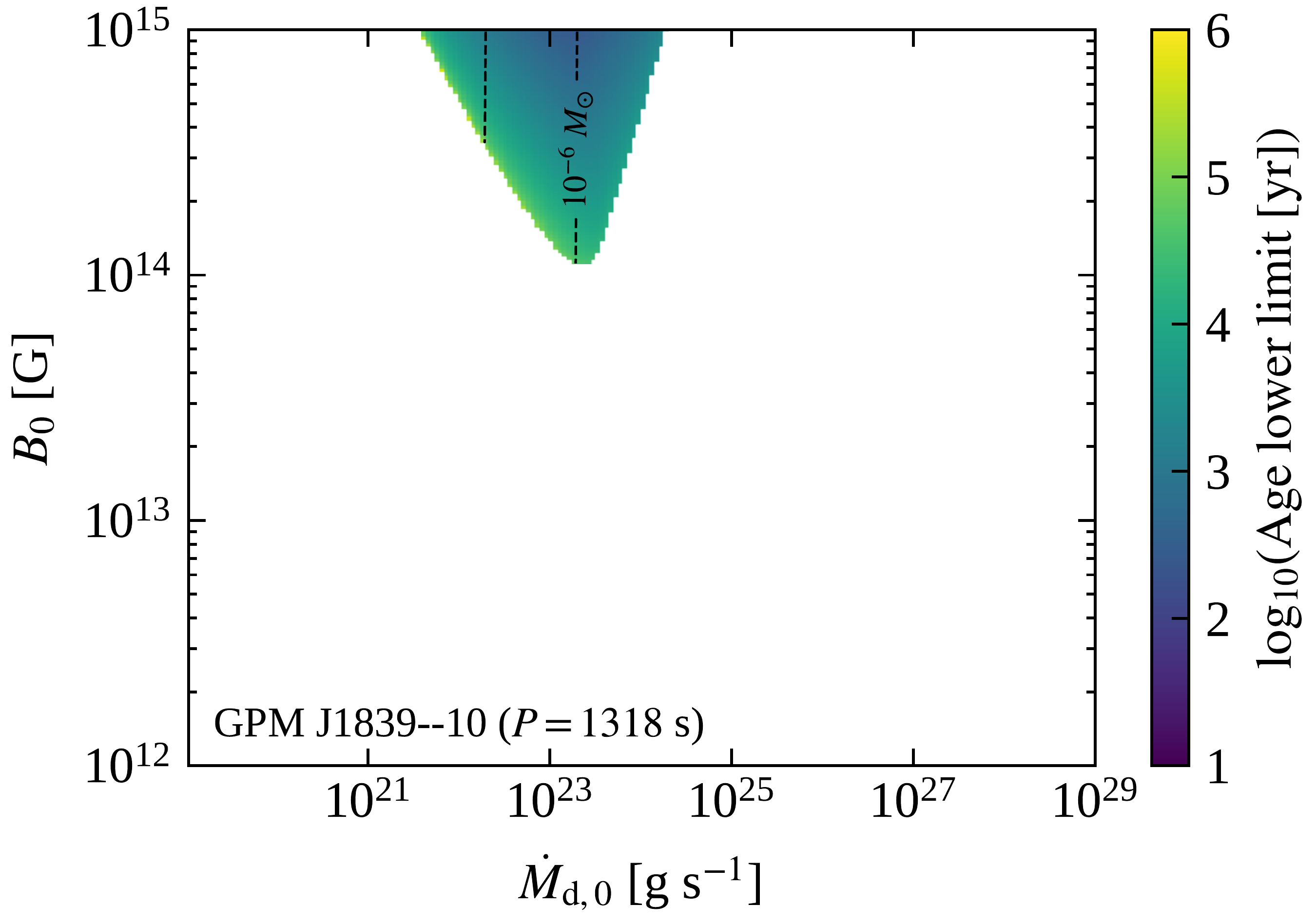}
\includegraphics[width=0.6\textwidth]{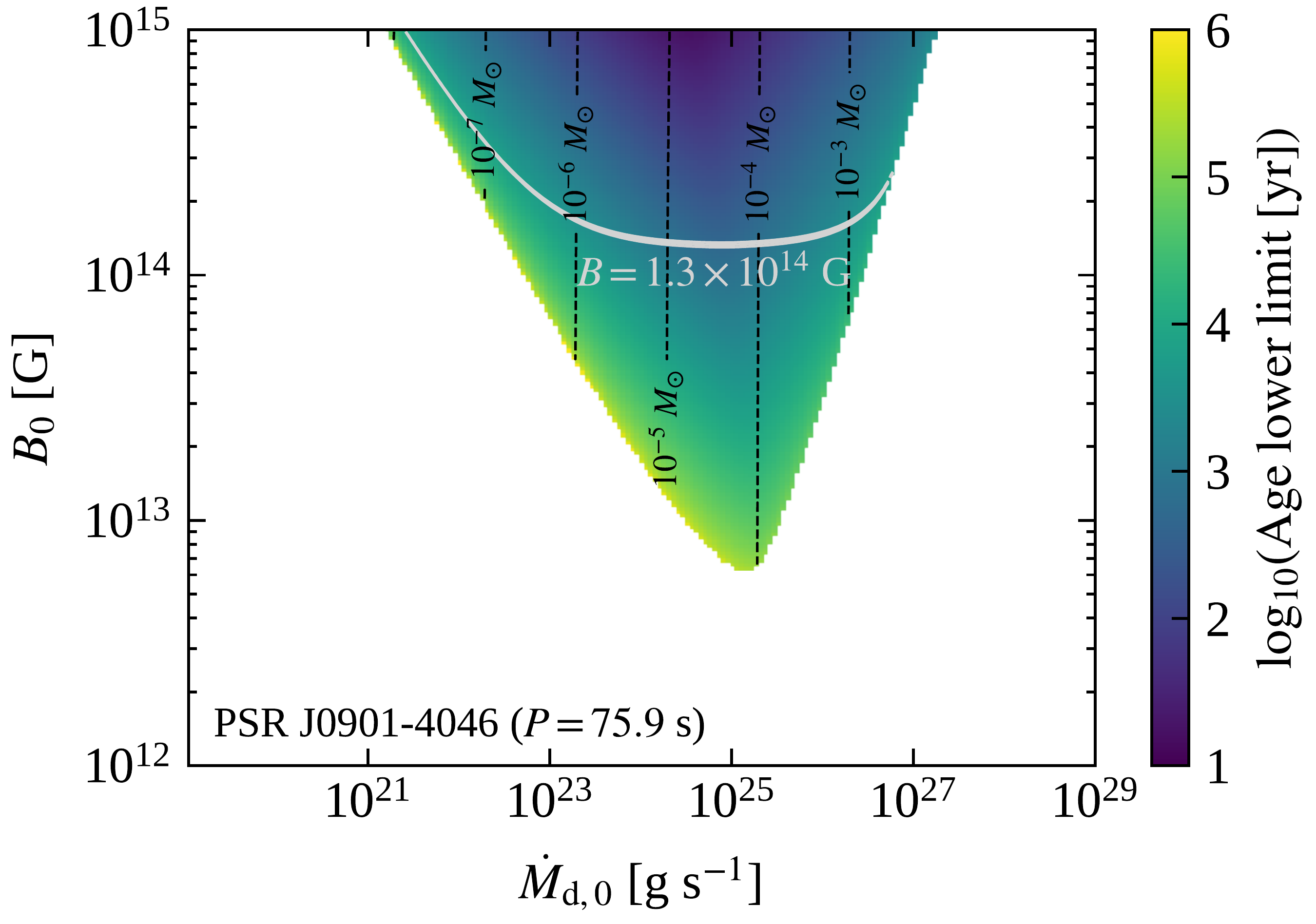}
\caption[Parameter search for long-period pulsars in the supernova fallback-accretion scenario]{Long-period pulsars in the supernova fallback-accretion picture. The coloured regions show the values of the initial magnetic field $B_0$ and disk fallback rate $\dot{M}_{\rm d,0}$ that allow the neutron star to reach a spin period of $\unit[1091]{s}$ for \gleam\ (top left panel), $\unit[1318]{s}$ for \gpm\ (top right panel) and $\unit[75.9]{s}$ for \mtp\ (bottom panel) in less than $\unit[10^7]{yr}$. The colour code indicates the time at which the neutron stars have reached their respective periods and can be interpreted as a lower limit on the source's current age. The contour lines indicate the total fallback mass that has been accreted by the disk in the same time interval. For \mtp, the gray contour line represents where the magnetic field value at $t=\rm Age$ is equal to the estimated magnetic field of \mtp\ from the spin-down formula.}
\label{fig:ch3_age_B_mdot}
\end{figure}  
%----------------------------------------------------

During the GaLactic and Extragalactic All-sky MWA eXtended survey (GLEAM-X) \citep{Hurley-Walker2022b} with the Murchison Widefield Array (MWA), a peculiar periodic radio transient has been discovered displaying a periodicity of $\unit[1091]{s}$ \citep{Hurley-Walker2022a}. The source was detected during two radio outburst periods in January and March 2018, displaying a very variable flux (going from undetected to values as high as $\unit[20-50]{Jy}$), $\sim 5\%$ duty cycle, 90\% linear polarisation, and a very spiky and variable pulse profile. 

From a detailed timing analysis, a dispersion measure of DM $=\unit[57\pm1]{pc \, cm^{-3}}$ was calculated, converting to a distance of $\unit[1.3]{kpc}$ according to the Galactic electron-density model of \citet{Yao2017}. The period derivative was loosely constrained to  $\dot{P}<\unit[1.5\times10^{-9}]{s \, s^{-1}}$. Assuming that the source is an isolated neutron star spinning down due to the classical electromagnetic dipole formula, this gives a relatively weak constraint on the dipolar magnetic field of $B_{\rm dip} <\unit[10^{17}]{G}$ through Equation~\eqref{eq:ch1_Bpol_vacuum}. 
We can also obtain an upper limit on the spin-down power $\dot{E}_{\rm rot} = I_{\rm NS} \omega \dot{\omega} < \unit[10^{28}]{erg \, s^{-1}}$. If we roughly estimate an isotropic radio luminosity from the flux detected during the outburst phase and the distance estimated from the DM, we obtain a value of $\sim \unit[10^{31}]{erg \, s^{-1}}$. From this, we deduce that the spin-down alone is insufficient to power these very bright radio flares.
The radio characteristics, such as the large flux variability, the radio outburst activity, and the high linear polarisation seem analogous to those observed in radio-active magnetars, whose activity is believed to be triggered and powered by the evolution and decay of their strong magnetic fields \citep[see also Section~\ref{sec:ch1_ns_zoo} and][]{Thompson1995, Duncan1996}. However, \gleam's exceptionally long spin period would make this source stand out among them.

As shown in Figure~\ref{fig:ch3_em_spin_down_P_evol}, assuming \gleam\, is indeed a magnetar that has spun down via dipolar losses alone, in the limiting case of a constant magnetic field, the magnetar would require an age $\gtrsim \unit[10^6 (10^8)] {yr}$ and a very strong magnetic field of $\unit[10^{16} (10^{15})]{G}$ to reach its spin period of $\unit[1091]{s}$. As argued in Section~\ref{sec:ch3_dipolar_losses}, sustaining such a strong magnetic field over this lifetime is difficult to reconcile with crustal magnetic-field evolution models that predict field decay on the Hall timescale; in this case $\sim \unit[10^{3}]{yr}$. For comparison, \lowbsgr\ is most likely the oldest magnetar detected so far with a characteristic age of $\sim \unit[10^7]{yr}$ and has an inferred magnetic field of $\sim \unit[10^{13}]{G}$ \citep[Magnetar Outburst Online Catalogue \url{http://magnetars.ice.csic.es/}][]{CotiZelati2018}. Moreover, if the star's magnetic energy reservoir has decayed in time, it becomes challenging to explain the current magnetar-like activity observed from this source. 

A more promising explanation for \gleam\ that requires less extreme conditions is a magnetar that has experienced accretion from a fallback disk soon after the supernova. As outlined in \S \ref{sec:ch3_fallback_scenario}, a magnetar surrounded by a fallback disk will pass through the propeller phase and spin down very efficiently on short timescales. To study this scenario for \gleam, we first fix the initial spin period to $P_0 = \unit[10]{ms}$ (remember that as long as $P \gg P_0$, $P_0$ has very little influence on the long-term evolution); for the neutron star mass and radius we adopt the fiducial values $M_{\rm NS} = 1.4 \, M_{\odot}$ and $R_{\rm NS} = \unit[11]{km}$. By varying the two parameters $B_0$ and $\dot{M}_{\rm d,0}$ and using Equation~\eqref{eq:ch3_torque} to determine the torque acting on the star in the different stages, we can numerically integrate Equation~\eqref{eq:ch3_omega_evolution} in time. This allows us to find those parameter combinations that lead to a spin-down evolution reaching a period of at least $\unit[1091]{s}$. In what follows, we consider a maximum time of $\unit[10^7]{yr}$ for the evolution. The motivation for this limit is two-fold. Firstly, it ensures that the spin-period evolution curves reach their maximum values during the propeller phase before magnetic field decay enters into play (see the discussion of Figures~\ref{fig:ch3_prop_evolution_ex}--\ref{fig:ch3_P_evolution_Mdot} in Section~\ref{sec:ch3_torque_fallback}). Secondly, a limit of $\unit[10^7]{yr}$ ensures that we encompass all age estimates of currently known magnetars (see the Magnetar Outburst Online Catalogue
\url{http://magnetars.ice.csic.es/} \citet{CotiZelati2018}).

The top panel in Figure~\ref{fig:ch3_age_B_mdot} shows the parameter space of $B_0$ and $\dot{M}_{\rm d,0}$, where we choose an initial magnetic field between $\unit[10^{12-15}]{G}$ and an initial disk accretion rate between $\unit[10^{19-29}]{g \, s^{-1}}$. The coloured region represents the combination of parameters that guarantees the neutron star to reach a spin period of $\unit[1091]{s}$. The colour code indicates the age at which the neutron star reaches the desired period for the first time. The vertical contour lines show the total fallback mass that has been accreted into the disk over the same time interval (calculated by integrating Equation~\eqref{eq:ch3_disk_rate_evolution} in time). 
In this scenario, we observe that (thanks to the propeller phase) a magnetar with a magnetic field of around $\unit[10^{14}]{G}$ can reach a spin period of $\unit[1091]{s}$ on a relatively short time scale of about $\unit[10^{3-5}]{yr}$ for an initial disk accretion rate of around $\unit[10^{23}]{g \, s^{-1}}$ and a total accreted mass of $\sim 10^{-6} M_{\odot}$. Note that since \gleam\ has been observed to emit periodic radio signals, disk accretion must have ceased to allow for the radio activity to be restored (see \S \ref{sec:ch3_discussion}). Assuming this is the case, the ages shown in Figure~\ref{fig:ch3_age_B_mdot} should be interpreted as the times at which the neutron star exited the propeller phase and shifted to standard dipolar spin-down; i.e., they represent lower limits on the real age of the observed source.

%%%%%%%%%%%%%%%%%%%%%%%%%%%%%%%%%%%%%%%%%%%%%%%%%%%%%%

\section{The 21-min periodic radio transient GPM J1839-10}
\label{sec:ch3_gpm}

During recent monitoring of the Galactic plane using the \acs{MWA} and subsequent follow-up observations with the \acf{ATCA}, Parkes Murriyang radio telescope, the \acf{ASKAP} and MeerKAT another mysterious periodic radio source with a periodicity of $\unit[1318]{s}$ has been discovered \citep{Hurley-Walker2023}. Further searches in archival data from the \acf{VLA}, the \acf{VLITE} and the \acf{GMRT} demonstrated that this source has been active for at least 33 years. As for \gleam\ the morphology and brightness of the observed pulses is very variable, going from undetected to maximum flux densities in the range of $\unit[0.1]{Jy} - \unit[10]{Jy}$ with linear polarisation varying between 10\% and 100\% and a duty cycles as large as 20\%. The timing analysis allowed an accurate estimation of the dispersion measure ${\rm DM} = \unit[273.5 \pm 2.5]{pc \, cm^{-3}}$ which corresponds to a distance of $\unit[5.7 \pm 2.9]{kpc}$ according to the Galactic electron density model of \citet{Yao2017}. The long-duration activity of this source allowed us to infer a stringent constraint on its period derivative $\dot{P} \lesssim \unit[3.6 \times 10^{-13}]{s \, s^{-1}}$. Assuming the source to be an isolated neutron star spinning down via dipole radiation, we find upper limits on the polar magnetic field $B_{\rm dip} \lesssim \unit[1.2 \times 10^{15}]{G}$ and on the rotational power $\dot{E}_{\rm rot} \lesssim \unit[8.4 \times 10^{24}]{erg \, s^{-1}}$. A rough estimate of the radio luminosity obtained from the observed flux densities and the estimated distance leads to $L \sim \unit[10^{28}]{erg \, s^{-1}}$. As for \gleam\ source, the estimated rotational energy loss is not enough to power the observed radio luminosity.
Furthermore assuming a magnetar is producing the emission of these bright pulses, extreme values of the magnetic field are necessary to reach such long spin periods (see Figure~\ref{fig:ch3_em_spin_down_P_evol}).

We extend the analysis of \citet{Ronchi2022} and we also study the period evolution in the fallback scenario for \gpm. We solve the spin evolution equation Equation~\eqref{eq:ch3_omega_evolution} for different combinations of the two parameters $B_0$ and $\dot{M}_{\rm d,0}$ and determine the parameter space that allows \gpm\ to reach a spin period of at least $P = \unit[1318]{s}$ within $< \unit[10^{7}]{yr}$. We show the results in the middle panel of Figure~\ref{fig:ch3_age_B_mdot}. We notice that the allowed region in the $B_0 - \dot{M}_{d,0}$ plane is very similar to that for \gleam. We again highlight that \gpm's disk accretion must have stopped in order to allow the re-activation of the radio-pulsar mechanism.

%%%%%%%%%%%%%%%%%%%%%%%%%%%%%%%%%%%%%%%%%%%%%%%%%%%%%%
\section{The 76 seconds magnetar: \mtp}
\label{sec:ch3_mtp0013}

The long-period pulsar \mtp\, was recently discovered by the South African radio telescope MeerKAT. The pulsar manifests a periodicity of $P = \unit[75.9]{s}$ with a period derivative of $\dot{P} = \unit[2.25 \times 10^{-13}]{s \, s^{-1}}$ \citep{Caleb2022}. From the timing analysis, a dispersion measure of ${\rm DM} = \unit[52 \pm 1]{pc\, cm^{-3}}$ has been derived which translates into a distance of $\sim \unit[0.4]{kpc}$.
Assuming that this neutron star is spinning down simply by dipolar losses, the strength of the dipolar magnetic field can be computed to $B_{\rm dip} \simeq \unit[1.3 \times 10^{14}]{G}$ and its spin-down power to $\dot{E}_{\rm rot} \simeq \unit[2 \times 10^{28}]{erg \, s^{-1}}$. Such a strong field is in the typical range of observed magnetars (see Figure~\ref{fig:ch1_PPdot_diagram}).

Using similar considerations as for \gleam\ and \gpm\ and assuming that \mtp\ is an isolated neutron star, which has spun down due to electromagnetic dipolar losses alone, we can reproduce the observed period, if its magnetic field remained almost constant over a lifetime of almost $\unit[10^{7}]{yr}$ (the source's characteristic age is $\tau_c = P/(2 \dot{P}) \simeq \unit[5.4]{Myr}$). 
If we instead consider the fallback disk scenario, we can relax these conditions. 
As before, we solve the spin evolution equation Equation~\eqref{eq:ch3_omega_evolution} for different combinations of the two parameters $B_0$ and $\dot{M}_{\rm d,0}$ and determine the parameter space that allows \mtp\ to reach a spin period of at least $P = \unit[75.9]{s}$ within $< \unit[10^{7}]{yr}$. The results are reported in the bottom panel of Figure~\ref{fig:ch3_age_B_mdot}. Also for \mtp, disk accretion must have stopped in order for the radio emission mechanism to be restored. From Figure~\ref{fig:ch3_age_B_mdot}, we deduce that observational characteristics can be explained within our model if this happened at an age of around $\unit[10^{3-4}]{yr}$, which we again interpret as a lower limit on the pulsar's true age. This constraint on the age is in line with other radio-emitting magnetars or high-$B$ radio pulsars that show similar radio emission. Furthermore in the case of \mtp, an initial disk accretion rate of around $\unit[10^{24}]{g \, s^{-1}}$, and a total accreted mass of $\sim 10^{-5} M_{\odot}$ are able to explain the observed period.

%%%%%%%%%%%%%%%%%%%%%%%%%%%%%%%%%%%%%%%%%%%%%%%%%%%%%%
\section{Discussion} 
\label{sec:ch3_discussion}

\subsection{Re-establishing the radio emission after the propeller}
\label{subsec:disk_instability}

One key piece that we have to discuss is how a neutron star, that enters the propeller phase to experience effective spin-down, can exit this phase again in order to be observed as a radio-loud object. This is a crucial requirement, given that we need to explain the radio detection of \gleam, \gpm\ and \mtp. As briefly mentioned in \S \ref{sec:ch3_fallback_scenario}, the transition from the propeller to the ejector phase can be naturally explained by an abrupt drop in the accretion rate. As a result, the magnetospheric radius would move beyond the light cylinder, providing the condition for the radio emission to be reactivated. 

Several factors could play a role in causing such a drop in the accretion flow. For example, since magnetars are very active young neutron stars that undergo outburst and flaring episodes \citep{CotiZelati2018, Esposito2021}, the energy released in such events could cause the disk to unbind or be completely disrupted. This would subsequently stop accretion and radio emission could be reinstated.

Another explanation could be simply that the fallback disk runs out of matter. During the propeller phase, the material that reaches the magnetospheric radius is ejected due to the centrifugal barrier. If the ejected material possesses a velocity superior to the escape velocity, it will become unbound from the system; otherwise, it will fall back and be reprocessed inside the accretion disk. As the matter feeding the disk from the supernova fallback is not replenished, we expect the accretion disk to eventually be completely consumed if the propeller is efficient at unbinding matter from the system \citep[see for example][]{Eksi2005, Romanova2005}. 

Another possibility is that the fallback disk, which itself evolves with time, undergoes a thermal ionisation instability, which has been outlined in detail in \citet{Mineshige1993, Menou2001, Ertan2009, Liu2015}. As the disk spreads and the accretion rate decays with time, energy dissipation decreases and the disk gradually cools down. In particular, \citet{Menou2001} argued that as the disk accretion rate falls below a critical value of around $\unit[10^{15}]{g \, s^{-1}}$ and the temperature in the outermost part of the disk drops below $\sim \unit[10^4]{K}$, the recombination of free electrons with heavy nuclei in the plasma is triggered. This transition alters the corresponding magnetic and viscous properties of the disk, reducing the efficiency of angular momentum transfer and eventually stopping accretion. As the disk evolves further in time, the recombination front propagates from the outermost regions inwards so that the active region of the disk shrinks. Eventually, accretion onto the neutron star will halt completely if the disk becomes totally neutralised. \citet{Menou2001} have argued that this transition occurs at around $\unit[10^{3-4}]{yr}$ when the outer disk radius is $\sim \unit[10^{10-11}]{cm}$. 
This timescale is comparable to the ones we require in our fallback accretion scenario for neutron stars to reach spin periods $>\unit[10]{s}$ (see Figure~\ref{fig:ch3_age_B_mdot}). However, it has also been suggested that the irradiation from the central source may prevent the disk from becoming completely neutral, allowing it to stay active for longer times at even lower temperatures \citep[see for example][]{Alpar2001, Inutsuka2005, Alpar2013}. 
In general, the evolution of a fallback disk is not trivial to model, since it also depends on the complex interaction with the central compact source. However, if the disk becomes inactive and the accretion flow stops due to any of the mechanisms outlined above, this could explain the transition from the propeller back to the ejector phase.

%----------------------------------------------------
\begin{sidewaysfigure}
\centering
\includegraphics[width = 0.8\textwidth]{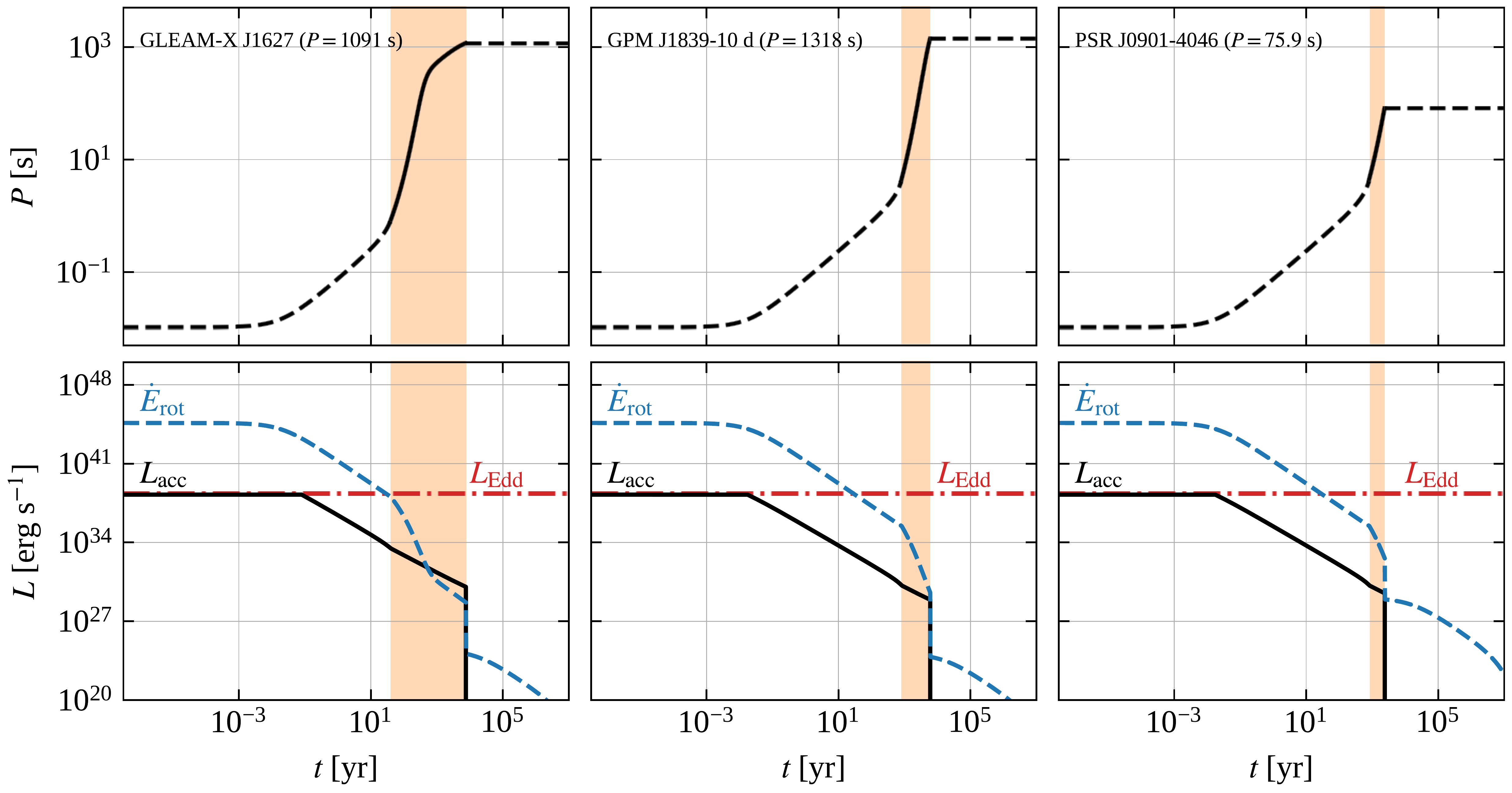}
\caption[Evolution example of the long-period radio sources]{Example of the time evolution of the spin period (top panels), disk luminosity and spin-down power (bottom panels) of \gleam\ (left panels), \gpm\ (central panels) and \mtp\ (right panels). For all sources we assume an initial magnetic field $B_0=\unit[4 \times 10^{14}]{G}$ and an initial disk accretion rate of $\dot{M}_{\rm d,0} = \unit[10^{24}]{g \, s^{-1}}$ for \gleam\ and $\dot{M}_{\rm d,0} = \unit[10^{23}]{g \, s^{-1}}$ for \gpm\ and \mtp, respectively. The orange-shaded region indicates the interval during which the three neutron stars spin down in the propeller phase. We assume that after the neutron stars have reached their observed periods of $\unit[1091]{s}$, $\unit[1318]{s}$ and $\unit[75.9]{s}$, respectively, an abrupt drop in accretion rate causes them to recover the ejector phase.}
\label{fig:ch3_GLEAM_GPM_MTP_evolution_L}
\end{sidewaysfigure}  
%----------------------------------------------------

In Figure~\ref{fig:ch3_GLEAM_GPM_MTP_evolution_L}, we show a possible evolutionary scenario for \gleam, \gpm\ and \mtp\ that incorporates this drop in accretion rate. In particular, the top panels show the spin-period evolution. For all three sources we choose an initial magnetic field $B_0=\unit[4 \times 10^{14}]{G}$ and initial disk accretion rates of $\dot{M}_{\rm d,0} = \unit[10^{24}]{g \, s^{-1}}$ for \gleam\ and $\dot{M}_{\rm d,0} = \unit[10^{23}]{g \, s^{-1}}$, for \gpm\ and \mtp\ respectively. These values fall within the allowed parameter spaces that guarantee the sources to reach their spin period in less than $\unit[10^7]{yr}$ (see Figure~\ref{fig:ch3_age_B_mdot}). Note that for \mtp, we fix the value of $B_0$ to a higher value than the estimated current magnetic field strength in order to take into account the effects of magnetic field decay.
To keep things simple, we set the disk accretion rate to zero as soon as the three neutron stars reach their observed spin periods of $\unit[1091]{s}$, $\unit[1318]{s}$ and $\unit[75.9]{s}$, respectively. This abrupt stop in the accretion process causes the sources to transition into the ejector phase and enables the restoration of the radio emission. From this point on, the spin periods remain effectively constant as the combination of their large values and magnetic field decay causes the dipolar losses to become negligible.

On the other hand, if accretion would persist, the neutron stars would continue to evolve in a spin-equilibrium, unable to exit the propeller phase. This is possibly the evolutionary phase currently witnessed for \rcw, the $\unit[2]{kyr}$-old magnetar associated with the SNR RCW103. This source has shown magnetar-like outbursts \citep{Rea2016, D'Ai2016b} and an observed period of $\unit[6.67]{hr}$ but is detected only in the X-ray band. Its period evolution has been modeled by \citet{Ho2017, Xu2019} using a similar model to our own. However, while \citet{Ho2017} assumed a constant accretion flow, in this work, we adopt the more realistic prescription of a time-varying accretion (similar to \citet{Xu2019}) as discussed in \S \ref{sec:ch3_fallback_scenario}. 
In this scenario, the long period together with the young age of \rcw\ could be explained with a field of $\gtrsim \unit[10^{15}]{G}$ \citep[in line with the results of][]{Ho2017, Xu2019} and a relatively low initial disk accretion rate of $\sim \unit[10^{21}]{g  \, s^{-1}}$ (see bottom right panel of Figure~\ref{fig:ch3_P_evolution_B} as a reference). As \rcw\ is currently not observed in radio, and thus still in the propeller phase, we infer that the accretion disk is still active, which is compatible with the source's age of $\unit[2]{kyr}$.

An association of long-period pulsars with supernova remnants could present additional proof for the fallback disk scenario. However, for \gleam, GPM J1839–10 and \mtp\ no clear evidence of SNR associations has been discovered so far. In general, SNRs are expected to have an observational lifetime of around $\unit[10^{4-5}]{yr}$ \citep{Braun1989, Leahy2020}, which is comparable with our inferred timescales for the fallback disk to be active and accretion spinning down the three sources to their current periods. It is therefore possible that the associated SNRs are too faint to be detectable at the present time. Moreover, only around one-third of known young pulsars have a detected SNR remnant \citep[see the ATNF Pulsar Catalog][]{Manchester2005}. The reasons for this are still a matter of debate and study, but the absence of detected SNRs for most neutron stars could simply indicate differences in their progenitors, supernova explosions and interstellar environments \citep[see for example][]{Gaensler1995, Cui2021}. We therefore do not consider the lack of SNR associations for \gleam, \gpm\ and \mtp\ an issue for the validity of a fallback disk scenario. 

\subsection{Prediction for the X-ray and radio luminosities of long-period pulsars}
\label{subsec:luminosity}

%----------------------------------------------------
\begin{figure}
	\centering
	\includegraphics[width = \textwidth]{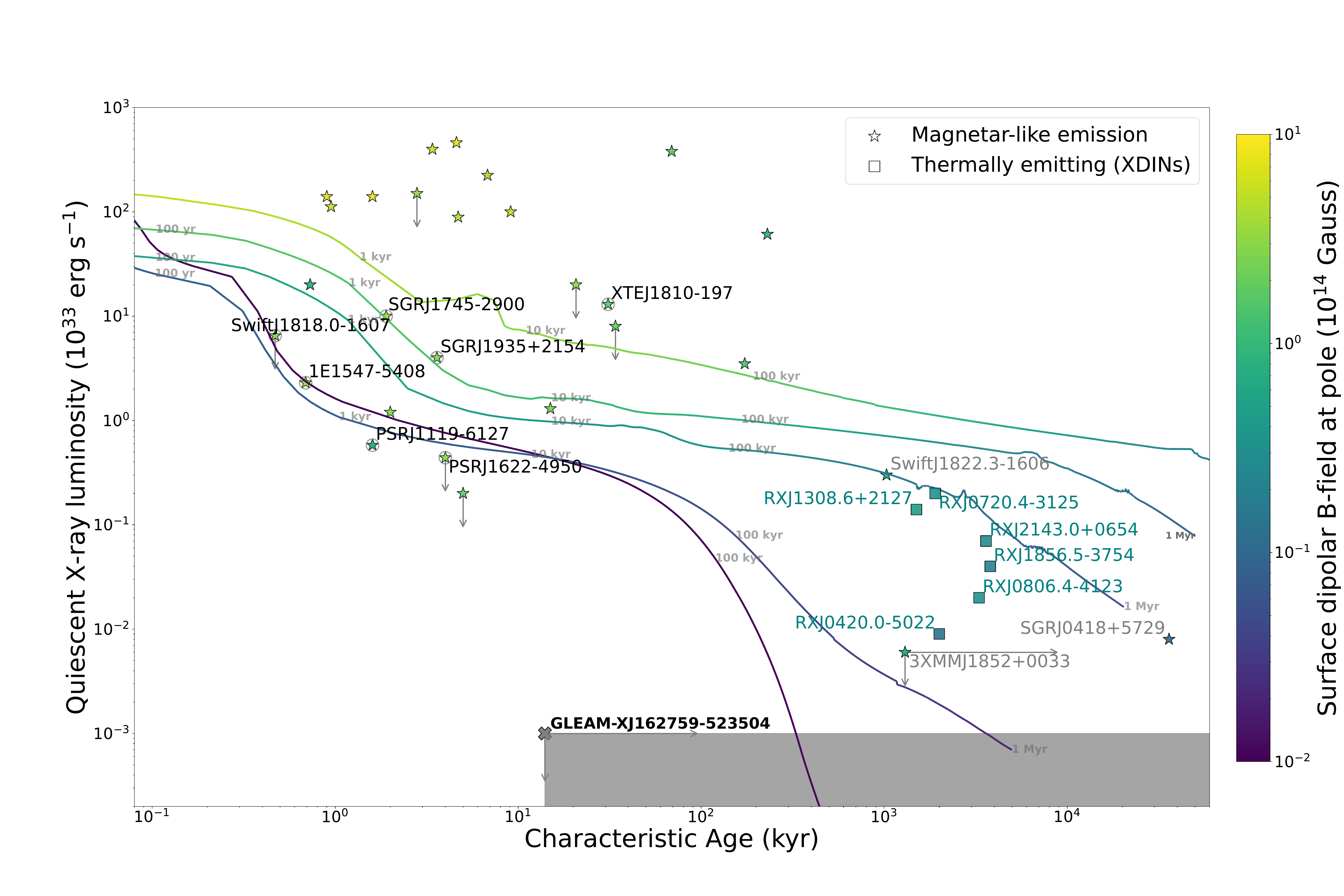}
	\includegraphics[width = \textwidth]{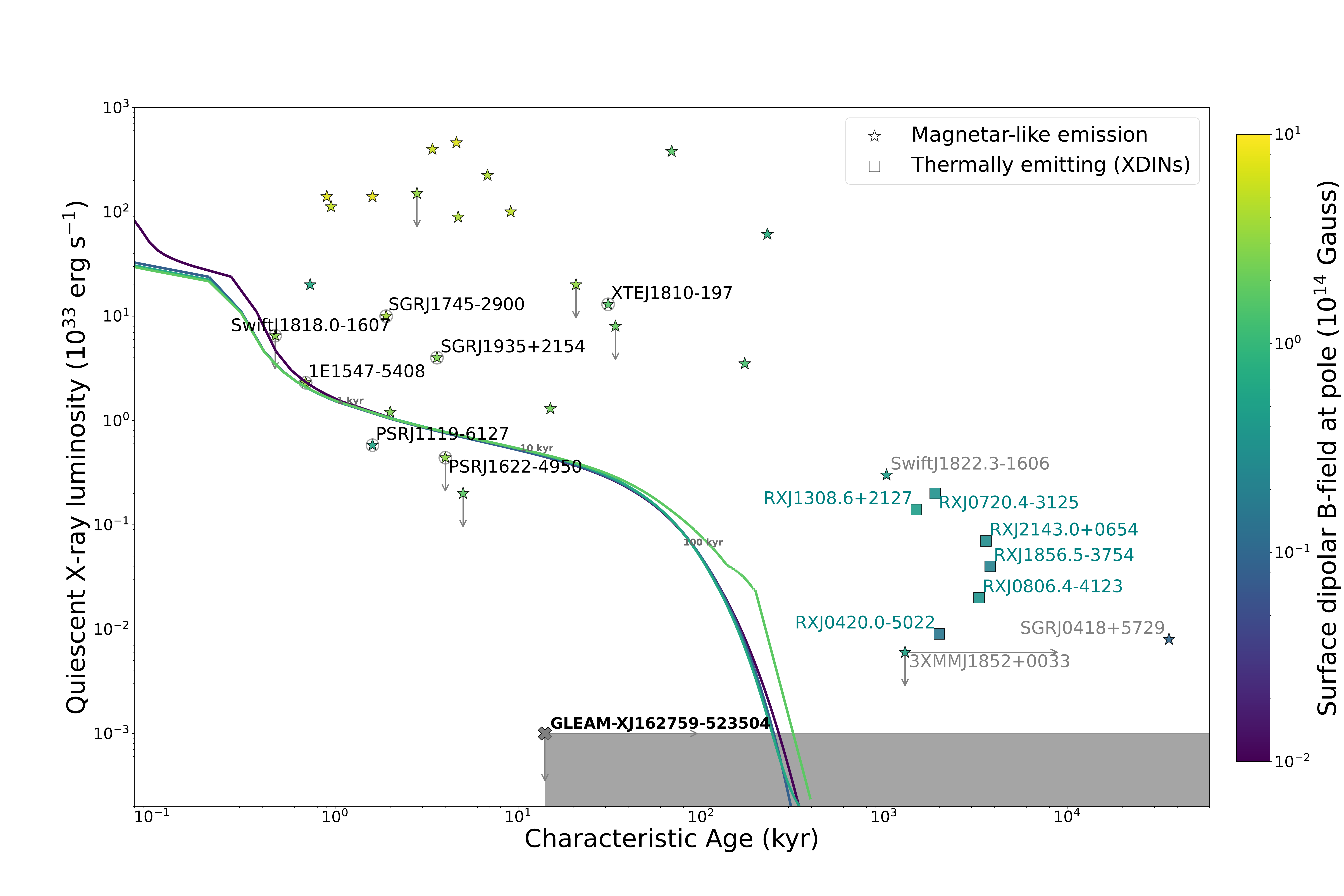}
	\caption[Magneto-thermal cooling models for \gleam]{Evolutionary tracks for crust-dominated (\textit{top panel}) and core-dominated (\textit{bottom panel}) $B$-field configurations as a function of the characteristic age. Labels showing the real age are also marked along the curves in \textit{grey}. Superimposed, are the X-ray luminosity of different pulsar classes, in particular radio-loud magnetars (\textit{grey circles}), XDINSs (\textit{light blue}) and low-field magnetars (\textit{gray names}). The shaded region corresponds to the limits on the X-ray luminosity and the characteristic age derived for \gleam\ \citep[Figure taken from][]{Rea2022}.}
	\label{fig:ch3_gleam_cooling}
\end{figure}  
%----------------------------------------------------

In the bottom panels of Figure~\ref{fig:ch3_GLEAM_GPM_MTP_evolution_L}, we show the evolution of the disk luminosity and spin-down power for \gleam, \gpm\ and \mtp. The accretion luminosity is computed as in Equation~\eqref{eq:ch3_accretion_L}, where we recall that $r_{\rm in} \simeq \min(r_{\rm m}, r_{\rm lc})$. At very early times, i.e. $\lesssim \unit[1]{yr}$ after the supernova, the accretion flow is limited by the Eddington limit, suggesting that the disk emits in X-rays at the Eddington luminosity. As the system evolves and the disk accretion rate starts to decrease as $t^{-\alpha}$, the inner disk radius increases following the evolution of the magnetospheric radius as $r_{\rm in} \sim r_{\rm m} \propto t^{2 \alpha / 7}$. As a consequence, the luminosity decreases roughly $\propto t^{-9 \alpha/7}$. Once the disk becomes inactive or is completely consumed, the accretion rate is expected to vanish and the disk becomes undetectable in the X-rays.
Therefore assuming the fallback scenario for long-period radio transients such as \gleam, \gpm\ and \mtp\ which require ceasing of the accretion flow, we predict that X-ray observations should not be able to detect emission from a residual disk if present. The cold debris of inactive disks could instead be detectable in the infrared \citep{Wang2006, Posselt2018}. However, if the central neutron stars are indeed young (around $\unit[10^{5}]{yr}$) and have strong magnetic fields, they could emit thermal X-rays from their surfaces with luminosities up to $\sim \unit[10^{31-35}]{erg \, s^{-1}}$ due to the dissipation of magnetic energy in the crust \citep{Vigano2013}.

However deep X-ray observations performed with the Chandra
X-ray Observatory and XMM-Newton satellite have imposed upper limits of $\sim \unit[10^{30}]{erg \, s^{-1}}$ \citep{Rea2022} and $\sim \unit[10^{33}]{erg \, s^{-1}}$ \citep{Hurley-Walker2023} on the X-ray luminosity of \gleam\ and \gpm, respectively. These limits challenge the magnetar interpretation especially for \gleam. Indeed comparing the X-ray luminosity of this source with cooling curves from magneto-thermal simulations \citep{Vigano2021}, the predicted age of \gleam\ should be $>\unit[1]{Myr}$ for any reasonable crustal-confined magnetic field ($B > \unit[10^{13}]{G}$) (see upper panel in Figure~\ref{fig:ch3_gleam_cooling}). This age constraint is two orders of magnitude
higher than that of typical radio-loud magnetars (which have
estimated ages $<\unit[20]{kyr}$). At this age, the bright radio bursts emitted by \gleam\ would be unusual for such an old magnetar. On the other hand if the magnetar has a core-dominated magnetic field or has witnessed unusual fast cooling (both effects have never been unambiguously observed in a pulsar or a magnetar), the observed X-ray upper limits would be compatible with a younger age (see bottom panel in Figure~\ref{fig:ch3_gleam_cooling}).

In Figure~\ref{fig:ch3_GLEAM_GPM_MTP_evolution_L} we also show the evolution of the spin-down power that is typically taken as the energy source for the radio emission of pulsars. In general, after these neutron stars have exited the propeller phase and recovered the ejector regime, we expect the spin-down to be caused by electromagnetic torques only so that $\dot{E}_{\rm rot} \propto B^2 / P^4$ (see Equation~\eqref{eq:ch1_mag_dipole_radiation_2} and~\eqref{eq:ch1_pulsar_lum_forcefree}). Therefore, if we consider an upper limit for the magnetic field of around $\unit[10^{15}]{G}$ and a lower limit for the spin periods after the propeller regime of around $\unit[10]{s}$, we expect that the spin-down power has decayed to values $\lesssim \unit[10^{33}]{erg \, s^{-1}}$ for long-period pulsar. This energy budget together with the magnetic energy stored in their strong fields is enough to power the observed radio emission and cause magnetar-like activity for neutron stars of this kind.

%%%%%%%%%%%%%%%%%%%%%%%%%%%%%%%%%%%%%%%%%%%%%%%%%%%%%%%%%%%%%%%%%%%%%

\section{Summary}
\label{summary}

We have studied the spin evolution of young, isolated neutron stars under the influence of fallback accretion. We specifically focused on the fallback-disk scenario as a promising origin of long-period pulsars, a class of objects that recent radio surveys are starting to unveil. 
By solving the torque balance equation for a disk accreting neutron star, we demonstrate that the evolution of such an object can differ significantly from the standard dipole spin-down. In particular, we find that for a combination of high (but not extreme) magnetic field strengths and moderate fallback disk accretion rates in agreement with current core-collapse supernova simulations, neutron stars can enter the propeller phase during their evolution. This leads to effective spin-down and allows neutron stars to reach spin periods $\gg \unit[10]{s}$ on time scales on the order of $\sim \unit[10^{3-5}]{yr}$. Magnetic dipolar losses alone have problems explaining long spin periods and would require extreme conditions like strong and long-lasting magnetic fields potentially supported either by a core field component or other mechanisms such as the Hall attractor.

We have then interpreted the recently discovered objects \gleam, GPM J1839–10 and \mtp, with rotation periods of $\unit[1091]{s}$, $\unit[1318]{s}$ and $\unit[75.9]{s}$, respectively, in light of this model, and showed that all three objects could be explained as highly magnetised neutron stars with a fallback disk accretion history. The possibility to reach such long spin periods in much less than $\unit[10^{7}]{yr}$ is crucial to maintain the magnetic field and thus an energy reservoir to power their radio or X-ray activity. This is particularly important for \gleam, which was observed in outburst similar to other young radio-loud magnetars. 

We showed that for newly born neutron stars with birth fields of $B_0 \sim \unit[10^{14-15}]{G}$, a phase of fallback disk accretion with moderate initial accretion rates of $\unit[10^{22-27}]{g \, s^{-1}}$, could explain their detection as long-period radio or X-ray pulsars at relatively young ages ($\sim \unit[10^{3-5}]{yr}$). On the other hand, in systems where the initial magnetic fields are lower ($\sim \unit[10^{12-13}]{G}$), fallback disk accretion (even if present) is expected to have a negligible effect on the spin-period evolution. The majority of neutron stars will therefore primarily undergo standard dipolar electromagnetic spin-down and recover rotation periods below $\sim \unit[12]{s}$. In our framework, we therefore naturally recover the pulsar population which is observed to spin in the range $\sim$ 0.002--12~s (see Figure~\ref{fig:ch1_PPdot_diagram}). Note that the recently discovered radio pulsars PSR\,J1903$+$0433 \citep{Han2021} and PSR\,J0250$+$5854 \citep{Tan2018} with periods of $\unit[14]{s}$ and $\unit[23]{s}$, respectively, could be easily accommodated within our fallback accretion scenario. However, both sources can in principle also be explained within the standard evolutionary scenario provided that crustal field decay is very weak \citep[essentially requiring the absence of a highly resistive pasta layer; see][]{Pons2013}.

We also mentioned the X-ray emitting magnetar at the centre of the $\unit[2]{kyr}$-old SNR RCW103, which requires (ongoing) fallback to explain its $\unit[6.67]{hr}$ period and radio-quiet nature. Finally note that classical magnetars with periods $\lesssim \unit[12]{s}$ would correspond to those systems where only strong fields are present, but fallback disk accretion does not take place or is inefficient because of a low $\dot{M}_{\rm d,0}$. 

In conclusion, fallback disk accretion after the supernova explosions of massive stars is expected to affect the evolution of newly born neutron stars. Depending on the relative intensities of the initial pulsar magnetic field and accretion rates, this scenario could represent an important ingredient to explain the connection between different neutron star classes and specifically shed light on the nature of the long-period radio sources recently discovered.
% Chapter 4

\chapter{Population study of long-period pulsars in the neutron star and white dwarf rotating dipole scenarios} % Main chapter title

\label{Chapter4} % For referencing the chapter elsewhere, use \ref{Chapter4} 

%----------------------------------------------------------------------------------------

\section{Introduction} \label{sec:ch4_intro}

As we have discussed in Chapter~\ref{Chapter3}, the discovery of ultra-long periodic coherent radio emitters challenges our current knowledge of neutron-star emission and evolution. Some of these new sources seem to be extreme neutron-star pulsars (e.g., the 76-s source \mtp; \citealt{Caleb2022}), while the interpretation of others is still uncertain (e.g., the 18-minute source \gleam\ \citealt{Hurley-Walker2022a}, and the 21-minute \gpm\ \citealt{Hurley-Walker2023}). The periodic radio emission features of \gleam\ and \gpm\ are similar to other radio magnetars, and their long periodicity can be explained through past supernova fallback accretion \citep[see Chapter~\ref{Chapter3}, e.g.,][]{Alpar2001, Chatterjee2000, Ertan2009, Tong2016, Ronchi2022}. 
X-ray observations with the Chandra X-ray Observatory and XMM-Newton satellite allowed to derive deep X-ray luminosity limits of $\lesssim \unit[10^{30}]{erg \, s^{-1}}$ and $\lesssim \unit[10^{33}]{erg \, s^{-1}}$ for \gleam\ and \gpm\,, respectively that challenge the magnetar interpretation \citep{Rea2022, Hurley-Walker2023}. By considering these upper limits, magneto-thermal evolution models \citep{Vigano2013, Vigano2021} predict cooling ages of these two new sources that exceed 1 Myr, far beyond any currently observed magnetar \citep{Rea2022}.

In contrast, slow spin periods are common in magnetic white dwarfs \citep{Ferrario2005, Ferrario2020}. Although isolated magnetic white dwarfs exhibit magnetic-field strengths between $10^6$ to $10^9$\,G \citep{Ferrario2015, Ferrario2020}, lower than neutron star $B$-fields spanning $10^8$ to $10^{15}$\,G (see also Figure~\ref{fig:ch1_PPdot_diagram}, magnetic white dwarfs have also been proposed to emit spin-down-driven radio emission similar to neutron stars \citep{Zhang2005}. To date, two radio-emitting white dwarfs have been detected, the binary systems AR\,Sco  ($P\sim1.95$\,min in a 3.5\,hr orbit; \citealt{Marsh2016}) and J1912$-$4410 ($P\sim 5.3$\,min in a 4\,hr orbit; \citealt{Pelisoli2023}). The radio emission of both systems is partly compatible with dipolar spin-down \citep{Geng2016, Buckley2017, duPlessis2019}, but also has a significant component resulting from the intrabinary shock with the wind of the companion star. The lack of an optical/infrared counterpart to \gleam\ at the estimated distance of 1.3\,kpc rules out a similar binary system for this source \citep{Rea2022}. However, it does not exclude lower mass companions or the possibility of a relatively cold isolated magnetic white dwarf.

In this chapter based on the work in \citet*{Rea2023}, we study \gleam\ and \gpm\, in the context of the radio emission expected from the spin-down of an isolated neutron star and white dwarf by means of death-line analyses (Section~\ref{sec:ch4_deathlines}) and population-synthesis simulations (Section~\ref{sec:ch4_popsyn}). In the following section I will review briefly our current knowledge on magnetic white dwarfs.

%------------------------------------------------------------------------------------------------------------

\section{Magnetic white dwarfs}

To date we have detected around 600 isolated white dwarfs exhibiting magnetic fields in the range $\unit[10^3 - 10^9]{G}$ and another 200 in binary systems \citep[see][for a review]{Ferrario2015, Ferrario2020}. 
As white-dwarf atmospheres are rich in hydrogen and helium, the Zeeman splitting of spectral lines provides a direct measure of the magnetic field strength on the surface of these stars. Circular polarisation in the spectrum, caused by the presence of a strong magnetic field, is another indicator of the magnetism in white dwarfs \citep{Ferrario2015, Ferrario2020, Caiazzo2021}. Studies of the local distribution of white dwarfs have revealed that between 10\% and 20\% of the white dwarf population exhibits relevant magnetic fields \citep[see, e.g.,][]{Holberg2016, Bagnulo2021}.

The origin of magnetism in white dwarfs is not well understood. Different processes can play a role in generating strong magnetic fields during the evolution of white dwarfs and their progenitors also depending on their physical properties. As for the neutron star case (see Section \ref{sec:ch1_magnetic_field}) one hypothesis is the fossil origin scenario. In this case the magnetic field is inherited from the progenitor star and is amplified due to flux conservation during the contraction to a white dwarf. This require a progenitor already having a strong magnetic field. It is commonly believed that the progenitors of magnetic white dwarfs are Ap and Bp-type stars which usually possess a strong magnetic field up to tens of kiloGauss \citep{Ferrario2005}. However, given the incidence of magnetism in the white-dwarf population \citet{Kawka2004} showed that the local density of Ap and Bp stars is not sufficient to account for the entire magnetic white dwarf population. Therefore, it has also been suggested that dynamo processes during binary interactions and stellar mergers can play a role in shaping the magnetism of these objects \citep{Tout2008}. An alternative theory is that magnetic fields in white dwarfs can be produced and amplified due to convective processes during the crystallisation of the liquid core at times of around 1 Gyr after the white dwarf formation \citep{Isern2017}. This hypothesis is supported also by observational evidence of an increasing occurrence of magnetism in older white dwarfs \citep{Bagnulo2021}.   
Observations have also shown that magnetic white dwarfs tend to be on average more massive than the non-magnetic ones with an excess mass of around 0.07 \msun \citep{Liebert1988, Kepler2013, Bagnulo2021}. This supports the theory that magnetic white dwarfs originate from massive A and B-type progenitors, close to the limit of around 8 \msun above which neutron stars are created.
Also, the merger scenario can be a viable explanation since after the merger a more massive outcome is expected \citep{Garcia-Berro2012}.

\begin{figure}
\centering
\includegraphics[width=0.7\textwidth]{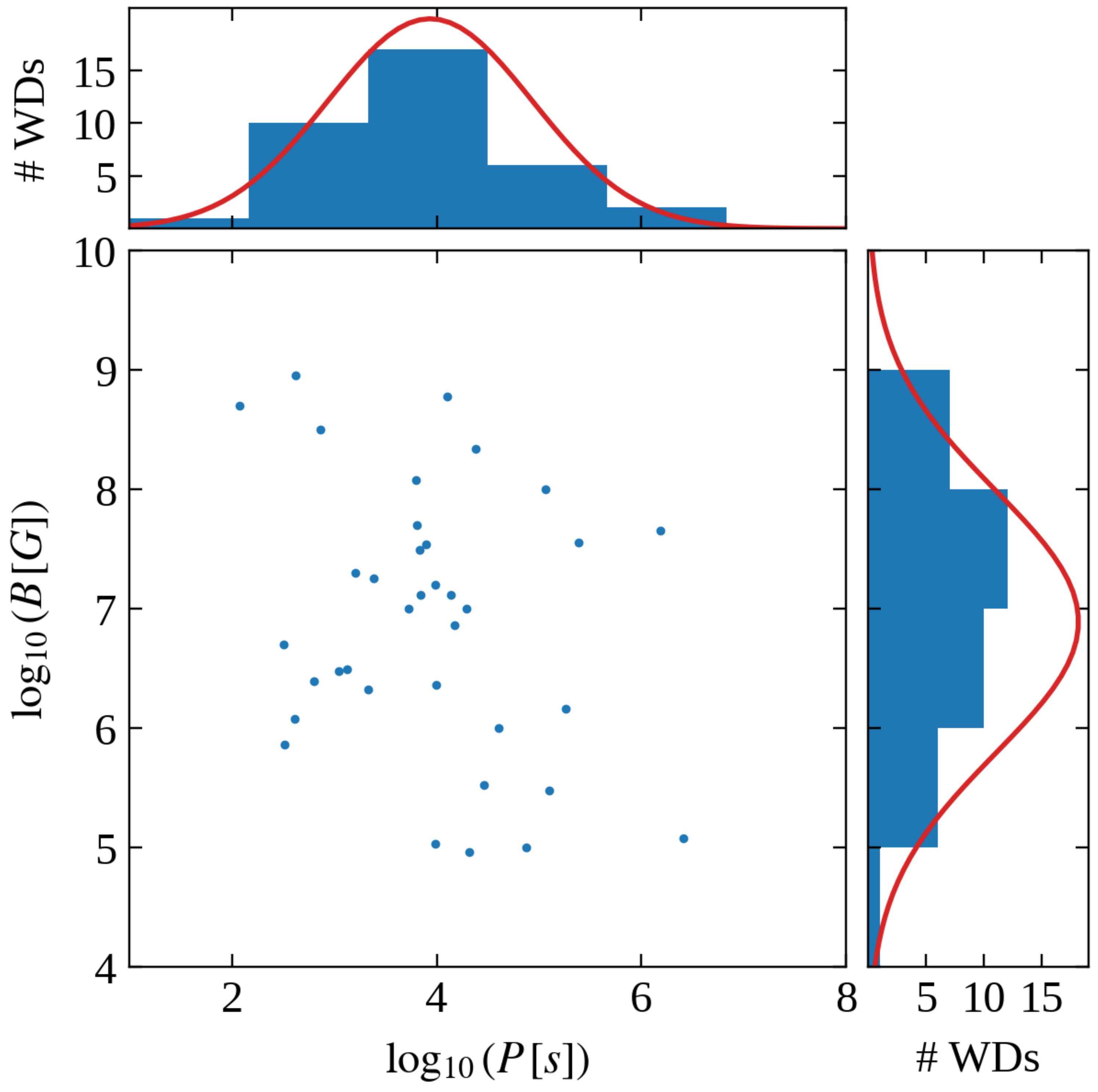}
\caption[Magnetic field strengths versus spin periods of the observed magnetic white dwarfs]{Relation between the logarithm of magnetic field strengths and spin periods of the currently known magnetic white dwarfs with a reliable measurement of these two quantities \citep[see][]{Ferrario2020}. On the top and right panels we report the corresponding histograms for the two quantities with the corresponding fit with Gaussian distributions (see text for more details).  }
\label{fig:ch4_MWD_B_P}
\end{figure}

In general, white dwarf spin periods are not easily determined. For magnetic white dwarfs, the magnetic activity and the presence of magnetic spots on the surface can induce spectroscopic, polarimetric and photometric variability that together with asteroseismology techniques can be used to estimate the spin period \citep[see][and references therein]{Wickramasinghe2000, Ferrario2015, Ferrario2020}. The measured spin periods are in the range of a few hundred seconds up to decades or even centuries and show that white dwarfs are slow rotators. Such a large range of spin periods can not be explained with angular momentum conservation during the contraction phase and is likely evidence for different physical processes playing a role during the formation and evolution of white dwarfs. For example, efficient angular momentum transfer from the core to the envelope and large-scale angular momentum loss during post-main-sequence evolution can contribute to slowing down the rotation. Moreover, magnetic fields are likely to lead to increased braking of the stellar core as it approaches the white dwarf state. Binary interaction and accretion phases during the progenitor evolution and mergers also play a role in shaping the final spin period \citep{Wickramasinghe2000}. 
Figure~\ref{fig:ch4_MWD_B_P} shows the magnetic field versus rotation period of all currently known magnetic white dwarfs. We note that there is no significant correlation between field strength and spin period.
%We do however caution that, despite the poor statistics, only strongly magnetic white dwarfs exhibit very long rotational periods of decades or centuries.
% -----------------------------------------------
% ----------------------------------------------

%------------------------------------------------------------------------------------------------------------
\section{Death valleys for neutron-star and white-dwarf radio pulsars}
\label{sec:ch4_deathlines}

% -----------------------------------------------
\begin{figure*}
\centering
\includegraphics[width=\textwidth]{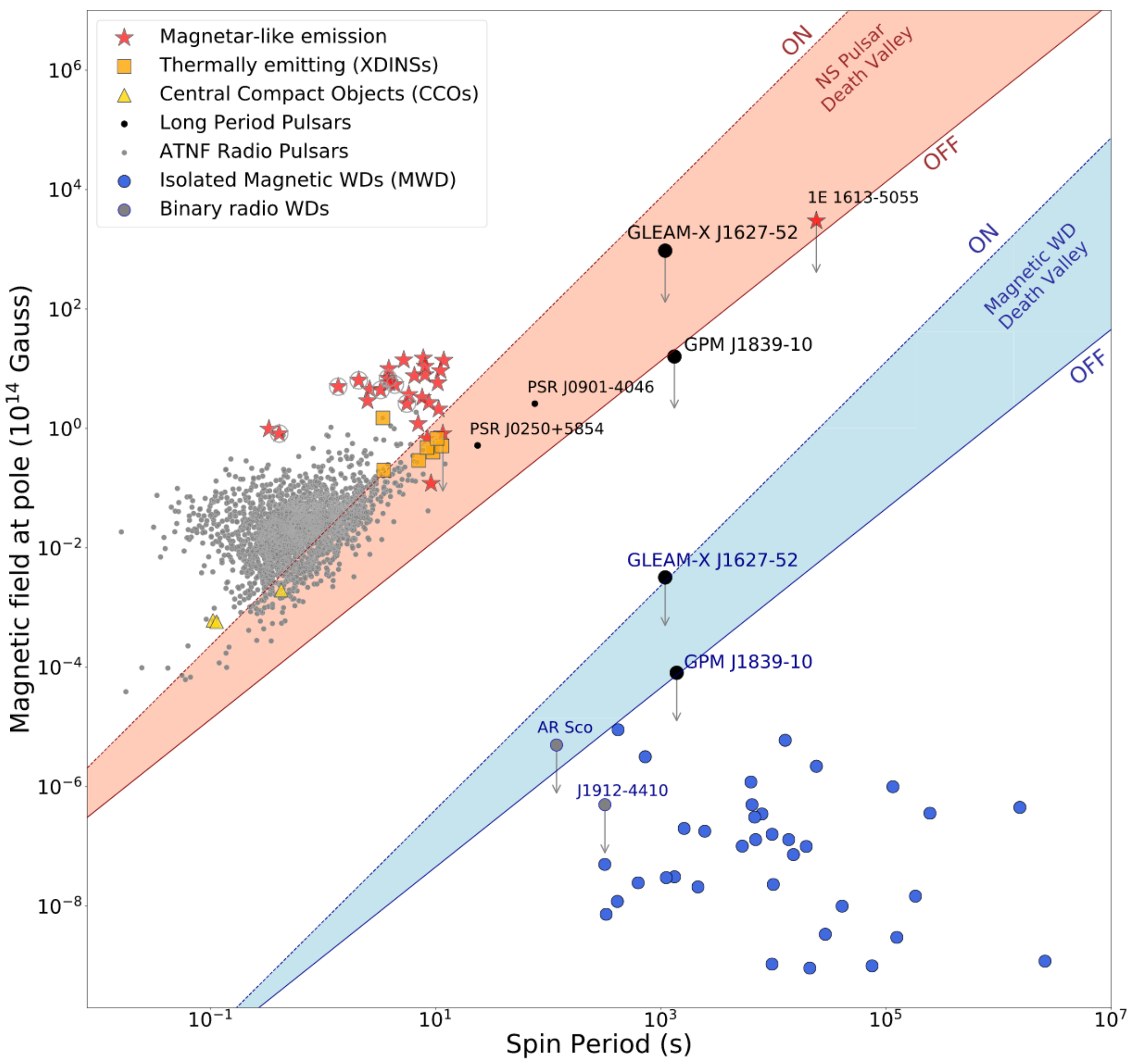}
\caption[Radio death lines in the neutron-star and magnetic white-dwarf scenarios]{Surface dipolar magnetic field, $B$, against spin period, $P$, for observed isolated neutron stars and magnetic white dwarfs. \gpm{} and \gleam{} are interpreted as isolated neutron stars or white dwarfs. Arrows represent upper $B$-field limits. We show isolated ATNF radio pulsars \citep{Manchester2005} (gray dots), pulsars with magnetar-like X-ray emission (red stars; gray circles highlight the radio magnetars), including the long-period magnetar 1E~161348-5055 \citep{DeLuca2006, Rea2016, D'Ai2016b}, X-ray Dim Isolated neutron stars (XDINSs; orange squares) and Central Compact Objects (CCOs; gold triangles) \citep{Olausen2014, CotiZelati2018}. Other long-period radio pulsars are reported as black circles \citep{Tan2018, Caleb2022}. Isolated magnetic white dwarfs are represented by blue dots \citep{Ferrario2020, Caiazzo2021, Buckley2017}. Gray dots show upper $B$-field limits for the binary white dwarfs AR\,Sco \citep{Buckley2017} and J1912-4410 \citep{Pelisoli2023, Pelisoli2024} computed from Equation~\ref{eq:ch4_B_upper_limit_WDbinary}. Dashed (solid) lines correspond to theoretical death lines for a pure dipole (extremely twisted multipole) configuration. Red and blue shaded regions indicate neutron star and white dwarf death valleys, respectively.}
\label{fig:ch4_p_B}
\end{figure*}
% ----------------------------------------------

Radio emission from rotating magnetospheres is usually explained as a result of pair production just above the polar caps \citep{Ruderman1975}. However, as we have seen in Section~\ref{sec:ch1_radio_em_deathlines}, for certain limiting periods, magnetic-field strengths and geometries, radio pulsars can no longer produce pairs, and radio emission ceases.
 
The parameter space in the $P-B$ plane (or equivalently $P-\dot{P}$ plane as $B$ and $\dot{P}$ are related via Equation \eqref{eq:ch1_Bpol_vacuum}) below which radio emission is quenched is called the ``death valley'' \citep[][see also Section~\ref{sec:ch1_radio_em_deathlines}]{Chen1993, Zhang2000}. This death valley encompasses a large variety of death lines depending on the magnetic-field configuration (e.g., dipolar, multi-polar, twisted), the nature of the seed gamma-ray photons for pair production (i.e., curvature or inverse Compton photons), the pulsar obliquity, the stellar radius and moment of inertia (see \citealt{Suvorov2023} for a death-valley discussion for long-period pulsars). To date, these death-line models have been applied exclusively to neutron star pulsars because, until very recently, no white dwarf pulsed radio emission had been detected. However, magnetic white dwarfs might not be unlike neutron star pulsars in generating coherent radio emission through magnetospheric spin-down losses \citep{Zhang2005}, albeit with different stellar radii, masses and magnetic fields.

Figure~\ref{fig:ch4_p_B} shows death valleys for neutron star and white dwarf pulsars as red and blue-shaded regions, respectively. Their boundaries are marked by two death-line extremes \citep{Chen1993}, representing the broadest range of $B$-field configurations (see Section~\ref{sec:ch1_radio_em_deathlines}). I report them here for convenience:
\begin{itemize}
\item a pure dipole configuration (see Equation~\eqref{eq:ch1_death_line_1}):
%--------------------------------------------------------------
\begin{align} \label{eq:ch4_death_line_1}
   B_{\rm p} \gtrsim \unit[2.0 \times 10^{12}]{G} \left( \frac{R}{\unit[10^6]{cm}} \right)^{-19/8} \left( \frac{P}{\unit[1]{s}} \right)^{15/8};
\end{align}
%--------------------------------------------------------------
\item an extremely twisted, multipolar magnetic field located in a small spot above the polar cap (see Equation~\eqref{eq:ch1_death_line_2}):
%--------------------------------------------------------------
\begin{align} \label{eq:ch4_death_line_2}
   B_{\rm p} \gtrsim \unit[8.3 \times 10^{10}]{G} \, b^{-1/4} \left( \frac{R}{\unit[10^6]{cm}} \right)^{-2} \left( \frac{P}{\unit[1]{s}} \right)^{3/2}.
\end{align}
%--------------------------------------------------------------
where $b$ is the ratio between the spot's $B$-field and the dipolar strength $B_{\rm p}$ (here we assume an extreme value of $b=10$ as in Section~\ref{sec:ch1_radio_em_deathlines}).
\end{itemize}

These death lines can be also represented in the $P-\dot{P}$ diagram by considering Equation~\eqref{eq:ch1_Bpol_vacuum} as an estimate of the magnetic field strength at the pole and making explicit the dependence on the mass and radius:
%--------------------------------------------------------------
\begin{align} \label{eq:ch4_Bpol_MR_dependence}
B_{\rm p} &= \left( \frac{3 c^3}{5 \pi^2} \right)^{1/2} M^{1/2} R^{-2} P^{1/2} \dot{P}^{1/2} \\ \nonumber
		  &\simeq \unit[5.7 \times 10^{19}]{G} \left( \frac{M}{1 M_{\odot}} \right)^{1/2} \left( \frac{R}{\unit[10^6]{cm}} \right)^{-2} P^{1/2} \dot{P}^{1/2},
\end{align}
%--------------------------------------------------------------
where we assumed a moment of inertia $I = 2/5 M R^2$.
By substituting Equation~\ref{eq:ch4_Bpol_MR_dependence} into Equations~\ref{eq:ch4_death_line_1} and ~\ref{eq:ch4_death_line_2} we obtain the following corresponding death lines:
%--------------------------------------------------------------
\begin{align} \label{eq:ch4_death_line_1_ppdot}
\dot{P} \gtrsim \unit[1.2 \times 10^{-15}]{s \, s^{-1}} \left( \frac{M}{1 M_{\odot}} \right)^{-1} \left( \frac{R}{\unit[10^6]{cm}} \right)^{-3/4} \left( \frac{P}{\unit[1]{s}} \right)^{11/4};
\end{align}
%--------------------------------------------------------------
%--------------------------------------------------------------
\begin{align} \label{eq:ch4_death_line_2_ppdot}
\dot{P} \gtrsim \unit[2.1 \times 10^{-18}]{s \, s^{-1}} b^{1/2} \left( \frac{M}{1 M_{\odot}} \right)^{-1} \left( \frac{P}{\unit[1]{s}} \right)^{2}.
\end{align}
%--------------------------------------------------------------
Note that the second death line turns out to be independent of the radius of the star.

Although these death lines rely on simplifications compared to more recent works \citep[e.g.,][]{Zhang2000}, our focus is on the extremes of the death valley. All newer models incorporating more detailed physics fall within the shaded region for any reasonable neutron star or white dwarf parameters.
For Figure~\ref{fig:ch4_p_B}, we use a fiducial neutron star radius of $R_\mathrm{NS} = 11$\,km, in line with recent measurements \citep[][see Section~\ref{sec:ch1_internal_structure_MR_relation}]{Riley2021, Raaijmakers2021}. 
For white dwarfs, we assume $R_\mathrm{WD} = 6000$\,km consistent with the Hamada-Salpeter relation \citep{Hamada1961} and measurements of isolated magnetic white dwarfs \citep{Ferrario2015, Ferrario2020}.

We also add observed neutron-star pulsars and magnetic white dwarfs to the $P-B$ plane in Figure~\ref{fig:ch4_p_B}. For observed neutron star pulsars, we derive surface magnetic fields at their poles using $P$ and $\dot{P}$ measurements, employing the dipolar loss formula (Equation~\eqref{eq:ch1_Bpol_vacuum}) assuming $M_{\rm NS} = 1.4M_{\odot}$ and $R_{\rm NS} = \unit[11]{km}$. $B$-fields at the poles of isolated white dwarfs are obtained from observations (i.e., Zeeman splitting of spectral lines; \citealt{Ferrario2015, Ferrario2020, Caiazzo2021}). For the radio pulsating white dwarfs AR\,Sco and J1912$-$4410, we estimate upper $B$-field limits assuming the emission to result from dipolar losses alone \citep[see][]{Buckley2017}. By combining Equations~\eqref{eq:ch1_mag_dipole_radiation_2} and \eqref{eq:ch1_mag_dipole_B_pole} and assuming an inclination angle $\chi = \pi/2$, we find the expression:
%--------------------------------------------------------------
\begin{align} \label{eq:ch4_B_upper_limit_WDbinary}
   B_{\rm p} &\sim \left( \frac{3 c^3}{8 \pi^4} \right)^{1/2} R_{\rm WD}^{-3} P^{2} L^{1/2} \nonumber \\
             &\sim \unit[4.7 \times 10^{8}]{G} \left( \frac{R_{\rm WD}}{\unit[6000]{km}} \right)^{-3} \left( \frac{P}{\unit[100]{s}} \right)^{2} \left( \frac{L}{\unit[10^{33}]{erg\,s^{-1}}}\right)^{1/2}.
\end{align}
%--------------------------------------------------------------
For AR\,Sco and J1912$-$4410-like systems, Equation~\eqref{eq:ch4_B_upper_limit_WDbinary} represents an upper limit on the $B$-field since, besides magnetic dipolar losses, the bulk of the spin-down power is likely lost by a combination of different effects: magnetohydrodynamic (MHD) interactions of the white dwarf’s magnetic field with the secondary star, outflows of relativistic charged particles from the magnetic white dwarf and winds from the companion star \citep{Marsh2016, Buckley2017}.

Finally, we also show the upper limits on the surface dipolar $B$-fields of the two long-period radio sources \gleam\ and \gpm\ in Figure~\ref{fig:ch4_p_B}.

% -----------------------------%

\section{Population synthesis for neutron-star and white-dwarf radio pulsars}
\label{sec:ch4_popsyn}

We simulate isolated neutron star and white dwarf populations using the framework of Chapter~\ref{Chapter6} (see also \citet{Ronchi2021} and Chapter~\ref{Chapter5}) with model parameters adjusted for each object type. Initially, we randomly sample the logarithm of the birth periods and magnetic fields from normal distributions, and the inclination angle between the magnetic and the rotational axis from a uniform distribution in spherical coordinates. Assuming that neutron stars and white dwarfs spin down due to magnetospheric torques, we then evolve their periods, $P$, and inclination angles, $\chi$, over time by solving the coupled differential equations \citep[][see also Section~\ref{sec:ch1_dip_spindown_evol}]{Spitkovsky2006, Philippov2014}:
%---------------------------------------------------------------
\begin{align}	\label{eq:ch4_magrot}
\dot{P} &= \frac{\pi^2}{c^3}\frac{B^2 R^6}{I P} \left( \kappa_0 + \kappa_1 \sin^2 \chi \right), \\
\dot{\chi} &= -\frac{\pi^2}{c^3}\frac{B^2 R^6}{I P^2} \left( \kappa_2 \sin\chi \cos\chi \right),
\end{align} 
%---------------------------------------------------------------
where we assume for simplicity $I = 2/5 (M R^2)$ and $\kappa_0 \simeq \kappa_1 \simeq \kappa_2 \simeq 1$ for pulsars surrounded by magnetospheres (see discussion in Section~\ref{sec:ch1_dip_spindown_evol}). Finally, we determine the number of stars that point towards the Earth by assuming a random direction for the line of sight and employing a prescription for the aperture of the radio beam as outlined below (see Chapter~\ref{Chapter6} for more details). 

To compare the impact of various initial model assumptions on the final spin-period distributions, we carry out the population simulations summarised in Table~\ref{tab:ch4_results} and Figures~\ref{fig:ch4_pop_syn_NS_1}--\ref{fig:ch4_pop_syn_WD}. Specifically, in Table~\ref{tab:ch4_results}, to help the reader, we count the objects falling within the period ranges $10-10^2$\,s, $10^2-10^3$\,s, and $10^3-10^5$\,s. We then distinguish objects intercepting our line of sight and those with $\dot{E} > 10^{27} \, {\rm erg} \, {\rm s}^{-1}$ (see Figures~\ref{fig:ch4_pop_syn_NS_1}--\ref{fig:ch4_pop_syn_WD} for the exact $\dot{E}$ and $P$ distributions). The latter limit has no intrinsic meaning, but was chosen as a reference to show how many sources would have sufficient rotational power to support \gpm's radio luminosity.

% ---------------------------------------%%

\subsection{Neutron star population synthesis}
\label{subsec:popsyn_NS}

% --------------- FIGURE --------------%

\begin{figure*}
\centering
\includegraphics[width=\textwidth]{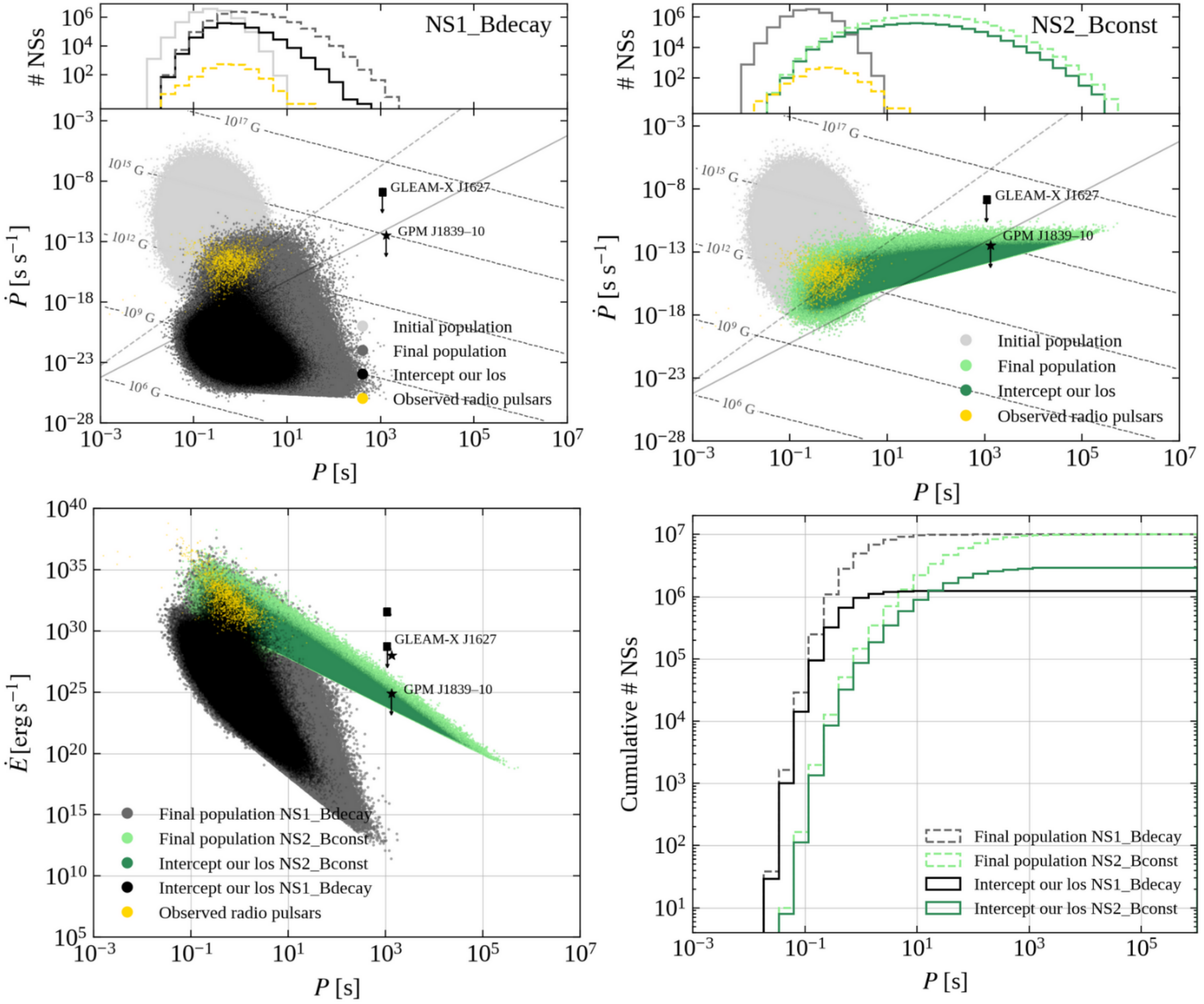}
\caption[Population-synthesis results in the neutron-star scenario]{Population-synthesis results for models {\tt NS1\_Bdecay} (black) and {\tt NS2\_Bconst} (green). Top panels show $P-\dot{P}$ diagrams for both simulations, respectively. Light grey dots represent initial neutron star populations, dark grey and light green final populations. The subsets of objects intercepting our line of sight (los) are shown in black and dark green. Lines of constant $B$-fields and the death lines defined in Equations~\ref{eq:ch4_death_line_1_ppdot} and~\ref{eq:ch4_death_line_2_ppdot} are indicated for reference. Histograms above the $P-\dot{P}$ diagrams represent the corresponding period distributions. The bottom left panel shows $\dot{E}$ versus spin period for the evolved populations, while the bottom right panel highlights the cumulative period distributions.  With the black square and black star we also report the two sources \gleam\ and \gpm. In the $\dot{E}$ versus spin period diagram we report both the upper limit on the $\dot{E}$ and the estimated radio luminosity for the two sources. Yellow dots highlight the observed isolated pulsar population given in the ATNF pulsar catalogue \citep{Manchester2005}. }
\label{fig:ch4_pop_syn_NS_1}
\end{figure*}

\begin{figure*}
\centering
\includegraphics[width=\textwidth]{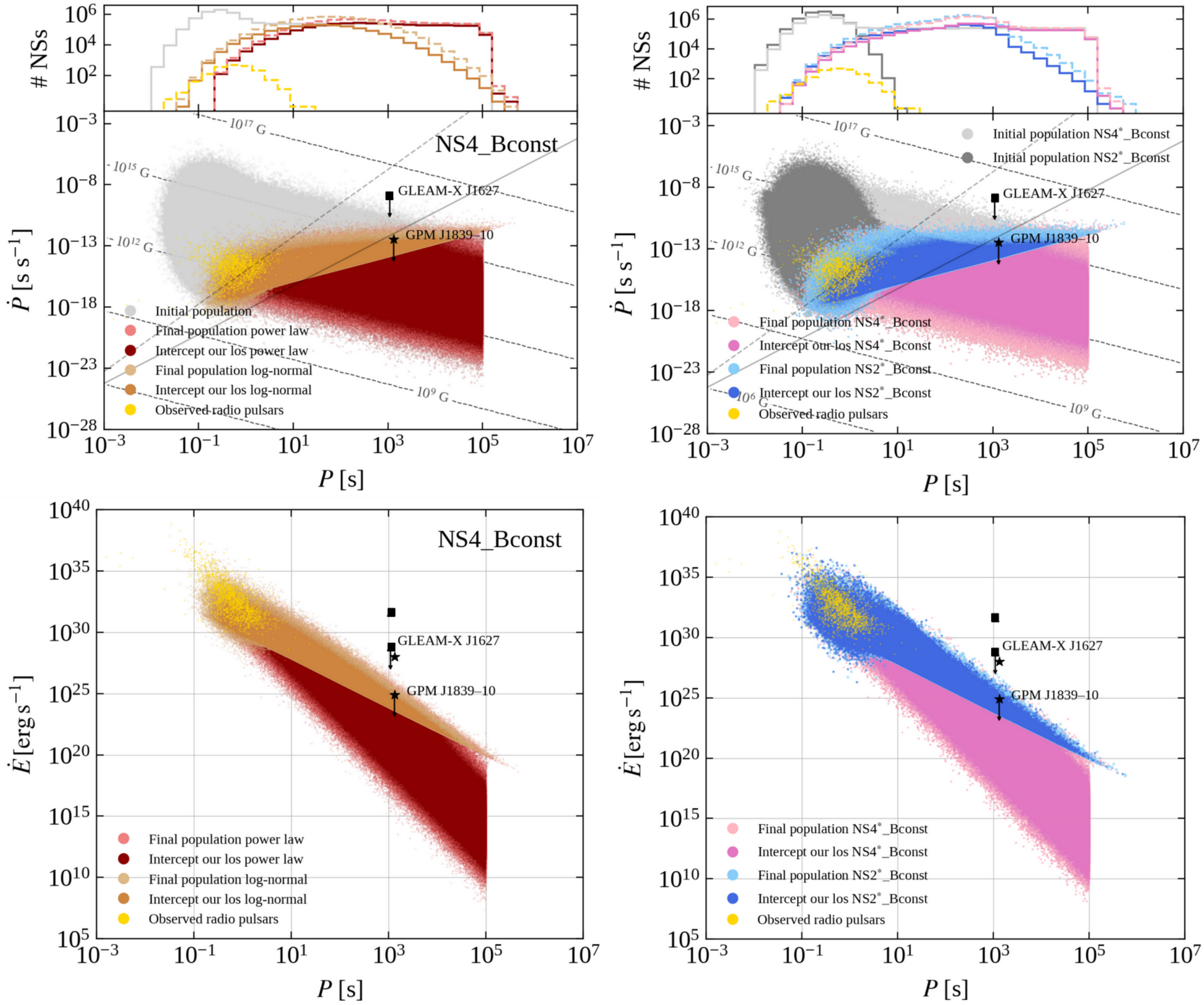}
\caption[Population-synthesis results in the neutron-star scenario with fallback accretion]{Population-synthesis results for models {\tt NS4\_Bconst} (orange/red), {\tt NS2$^*$\_Bconst} (blue), and {\tt NS4$^*$\_Bconst} (pink). Panels, lines and yellow dots are equivalent to those in Figure\,\ref{fig:ch4_pop_syn_NS_1}.
%Top panels display $P-\dot{P}$ diagrams. Gray dots represent initial NS populations. Lines of constant magnetic field and rotational spin-down power are indicated for reference. Histograms above the $P-\dot{P}$ diagrams represent the corresponding period distributions. Bottom panels illustrate the rotational spin-down power, $\dot{E}$, versus spin period, $P$. 
In both left panels, evolved objects sampled from the log-normal (power-law) contribution to the initial period distribution are shown in orange (red). Right panels show evolved populations of models {\tt NS4$^*$\_Bconst} and {\tt NS2$^*$\_Bconst} based on a bimodal $B$-field distribution. Across all panels, light shades depict evolved populations, while objects intercepting our lines of sight are shown in dark shades.}
\label{fig:ch4_pop_syn_NS_2}
\end{figure*}

% -----------------------------%

We simulate $10^7$ neutron stars with random ages sampled from a uniform distribution up to a maximum age of $10^9$ yrs. This translates to a birth rate of one neutron star per century, consistent with the Galactic core-collapse supernova rate \citep{Rozwadowska2021}. 
To assign each neutron star a birth field, we then sample the logarithm of the field (in Gauss) from a normal distribution with mean $\mu_{\log B}=13.25$ and a standard deviation of $\sigma_{\log B} = 0.75$ \citep[see, e.g.,][]{Gullon2014, Gullon2015, Cieslar2020}. Unless stated otherwise, we adopt $M_{\rm NS} = 1.4 \, M_{\odot}$ and $R_{\rm NS} = 11 \,$km. %Results of eleven simulation configurations are summarized in Tab.~\ref{tab:results}.
We sample the logarithm of the initial period from a normal distribution with mean $\mu_{\log P}$ = -0.6 (corresponding to $0.25$\,s) and standard deviation $\sigma_{\log P} = 0.3$ \citep{Popov2010, Gullon2014, Xu2023}. 

We further incorporate magnetic-field decay due to Ohmic dissipation and the Hall effect through magneto-thermal evolution curves from \citet{Vigano2013, Vigano2021} (see also Chapter~\ref{Chapter6} for more details) and assume a radio beam angular aperture $\propto P^{-1/2}$ \citep[see Equation~\eqref{eq:ch1_beam_angular_aperture} and ][]{Lorimer2012} where we assume an emission radius $r_{\rm em} = \unit[300]{km}$ \citep{Johnston2019}. Model {\tt NS1\_Bdecay} serves as a reference with standard population assumptions (Figure~\ref{fig:ch4_pop_syn_NS_1} top-left). These are typical initial parameters compatible with the current observed pulsar population (yellow dots in Figures \ref{fig:ch4_pop_syn_NS_1} and \ref{fig:ch4_pop_syn_NS_2}). However, they cannot predict any long-period pulsars.

We continue with investigating more extreme scenarios, focusing first on an evolution without field decay. Strong fields could be maintained over long timescales if electric currents are predominantly present in the neutron star core \citep[e.g.,][]{Vigano2021}. Consequently, neutron stars experience a more pronounced spin-down, reaching longer periods.
For model {\tt NS1\_Bconst}, we thus repeat the set-up of {\tt NS1\_Bdecay} but with constant $B$-field at the very limit of what is physically viable. However, adding the constant-$B$ assumption is insufficient to slow down the population substantially (see Table~\ref{tab:ch4_results}). For subsequent models, we continue with the extreme constant $B$-field case to explore the impact of other assumptions.

In model {\tt NS2\_Bconst}, we also relax the standard beaming assumption, adjusting the radio beam angular aperture to obtain a duty cycle of roughly 20\% independent from the spin period $P$, in line with observations of \gleam{} and \gpm{}. To do so, we consider an orthogonal rotator and compute the beam aperture that would result in a duty cycle of 20\% ($\delta = 0.2$) for this configuration. This implies that the angular aperture of the beam has to be $\rho_{\rm em} = \delta \pi = 36^{\circ}$ (see also Section~\ref{sec:ch1_radio_em_geometry}.
This adjustment results in an increase of the number of pulsars crossing our line of sight especially in the long period tale (see Figure\,\ref{fig:ch4_pop_syn_NS_1} top-right and bottom panels). For the remaining simulations, we thus maintain this prescription of the beaming unless stated otherwise.

Next, we explore different initial spin-period distributions, mimicking a possible interaction with initial fallback accretion \citep[see, e.g.,][]{Alpar2001, Ertan2009, Tong2016, Ronchi2022}. For models {\tt NS3\_Bconst} to {\tt NS6\_Bconst}, we add a power law with an arbitrary cut-off at a period of $\unit[10^5]{s}$ to the aforementioned log-normal distribution of the observed pulsar population. We specifically consider a power law, as the spin-down is likely determined by different fallback accretion rates. Note that the cut-off does not affect our final results, but reflects the maximum spin reachable by fallback accretion \citep[see Figures \ref{fig:ch3_P_evolution_B} and \ref{fig:ch3_P_evolution_Mdot} and][]{Ronchi2022}. We arbitrarily assume that both distributions are equally normalised, sampling 50\% of neutron stars from either distribution, maintaining a birth rate of 1 neutron star per century. This prescription is still consistent with the log-normal population resulting in the observed radio pulsars \citep[][see also yellow dots in Figures \ref{fig:ch4_pop_syn_NS_1} and \ref{fig:ch4_pop_syn_NS_2}]{Gullon2015}. For models {\tt NS3\_Bconst} and {\tt NS4\_Bconst} (see Figure~\ref{fig:ch4_pop_syn_NS_2} left panels), we assume a corresponding power-law index of -3 and -1, respectively. {\tt NS5\_Bconst} investigates a duty cycle of $10\%$ ($\delta = 0.1$), i.e., $\rho_{\rm em} = \delta \pi = 18^{\circ}$, while for {\tt NS6\_Bconst}, we explore the effect of the assumed mass, setting $M_{\rm NS} = 2 M_{\odot}$. 

Since stronger magnetic fields enhance the spin-down, we also investigate the effect of a bimodal $B$-field distribution (four models denoted with an asterisk ($*$) in Table~\ref{tab:ch4_results}). In particular, besides the log-normal distribution, we consider that 50\% of neutron stars are formed with a strong field uniformly distributed in $\log B \in [13.5, 14.5]$ following \citet{Gullon2015}. {\tt NS2$^*$\_Bdecay} and {\tt NS2$^*$\_Bconst} consider only the log-normal for the initial period distribution and a decaying and constant magnetic field, respectively, while for {\tt NS4$^*$\_Bdecay} and {\tt NS4$^*$\_Bconst}, we explore the log-normal plus power law for the initial period (see Figure~\ref{fig:ch4_pop_syn_NS_2} right panels).

%Using the same prescription as \citet{Gullon2014}, we have checked that all population models presented here are consistent with the observed pulsar population highlighted as the yellow dots in figs. \ref{fig:ch4_pop_syn_NS_1} and \ref{fig:ch4_pop_syn_NS_2}. This is mainly due to the low rotational power and long periods of the resulting hidden pulsars \textbf{say it differently?}.

%%%%%%%%%%%%%%%%%%%%%%%%%%%%%%%%%%%%%%%%%%%%%%%

\subsection{White dwarf population synthesis}
\label{subsec:popsyn_WD}

% ----------- FIGURE ----------------------------%%
\begin{figure*}
\centering
\includegraphics[width=\textwidth]{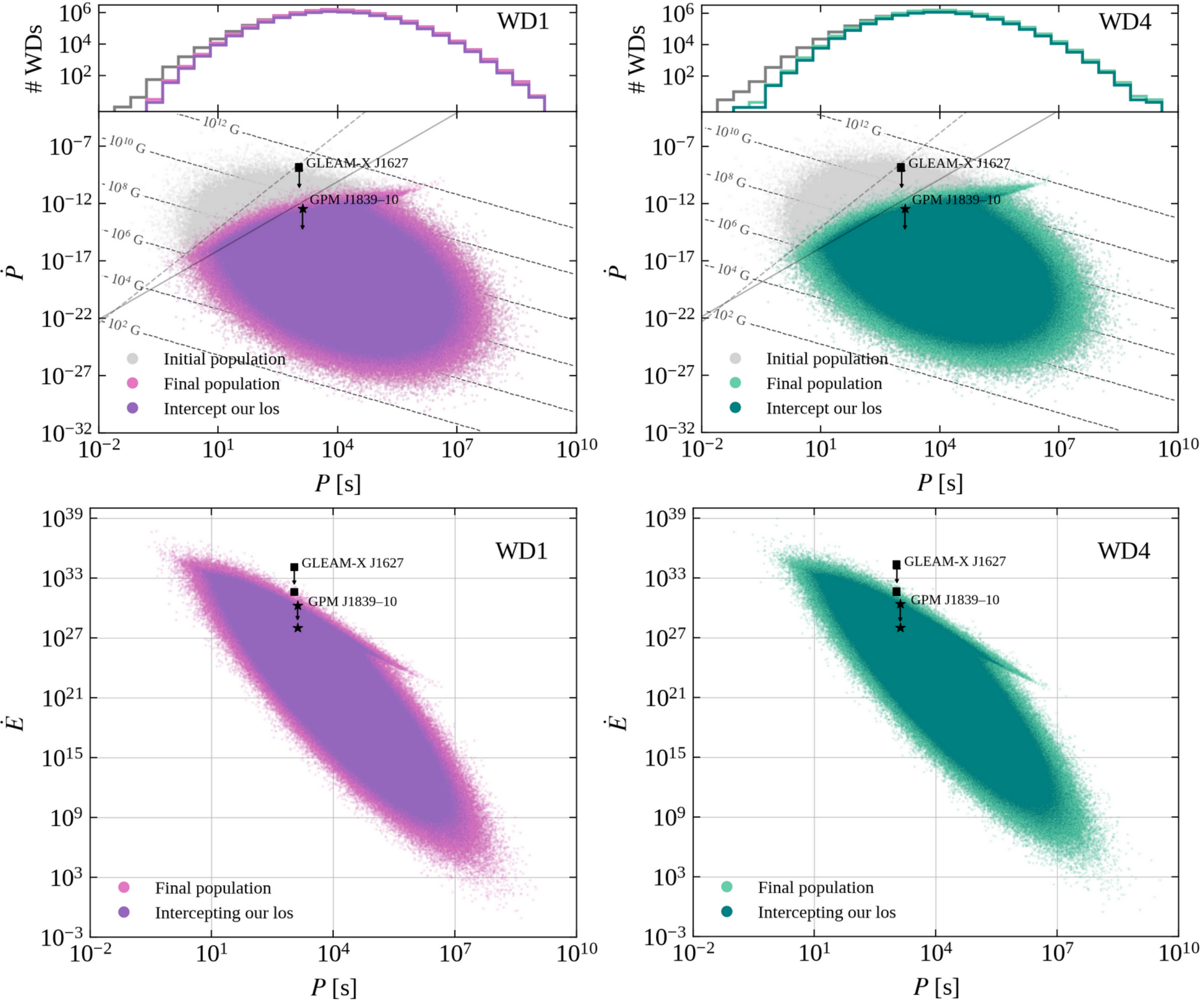}
\caption[Population-synthesis results in the white-dwarf scenario]{Population-synthesis results for models {\tt WD1} (purple) and {\tt WD4} (seagreen). Panels and lines are similar to Figure~\ref{fig:ch4_pop_syn_NS_1} and  Figure\,\ref{fig:ch4_pop_syn_NS_2}.
%Top panels show $P-\dot{P}$ diagrams for both simulations, respectively. 
Gray dots represent initial white dwarf populations, pink and light seagreen final populations. The subsets of objects intercepting our line of sight are shown in purple and dark seagreen. Note that the death valley and the upper limits for the two long period sources refer to the WD case (see Figure~\ref{fig:ch4_p_B}, Section~\ref{sec:ch4_deathlines}, and Section~\ref{subsec:popsyn_WD}). Note that the two cases have different masses and radii, which reflects in different B-field lines, death valley and $\dot{E}$ limits. }
%Lines of constant magnetic field and rotational spin-down power are indicated for reference. Histograms above the $P-\dot{P}$ diagrams represent the corresponding period distributions. The bottom panels illustrate the rotational spin-down power, $\dot{E}$, versus spin period, $P$.}
\label{fig:ch4_pop_syn_WD}
\end{figure*}
% -----------------------------%

Magnetic white dwarfs spin down slower than neutron stars due to larger moments of inertia, larger spin periods and lower $B$-fields (see Equation~\ref{eq:ch1_rot_evol_forcefree_2}). 
Moreover, magnetic fields of magnetic white dwarfs do not exhibit relevant magnetic-field decay due to longer Ohmic dissipation timescales \citep[e.g.,][]{Cumming2002} and can be taken as constant \citep{Ferrario2020}. Considering Equation~\eqref{eq:ch1_tau_ohm} we obtain a rough estimate of the Ohmic timescale in white dwarfs:
%---------------------------------------------------- 
\begin{align}
\tau_{\rm Ohm} \simeq \unit[4.4 \times 10^9]{yr} \left( \frac{\sigma}{\unit[10^{21}]{s^{-1}}} \right)
    \left( \frac{L}{\unit[1000]{km}} \right)^2, 
\end{align}
%---------------------------------------------------- 
where we consider typical values of the conductivity from \citet{Cumming2002}.
Consequently, current isolated white dwarf periods and magnetic-field strengths closely reflect those at birth.

To model these birth distributions, we consider a sample of 37 magnetic white dwarfs with reliable spin-period and magnetic-field measurements \citep{Ferrario2020}. We fit Gaussian functions to the distributions of the logarithm of the periods and $B$-fields, deriving a mean of $\mu_{\log P} = 3.94$ and standard deviation of $\sigma_{\log P} = 1.0$, and $\mu_{\log B} = 6.91$ and $\sigma_{\log B} = 1.09$, respectively (see Figure~\ref{fig:ch4_MWD_B_P}).
The white dwarf birth rate in the Milky Way has been predicted to be around 100 per century assuming a Galaxy radius of 20 kpc and a Galactic disk height of 0.5 kpc \citep{Holberg2016}. 
For our population synthesis, we then simulate $10^8$ magnetic white dwarfs with ages drawn from a uniform distribution up to a maximum of $10^9\,$yr, consistent with a birth rate of 10 per century assuming 10\% of the white dwarfs being magnetic (see, e.g., \citealt{Holberg2016} but also \citealt{Bagnulo2021} who recently found 22\%). We then assign initial $P$ and $B$ values from our fitted distributions. Results of four simulation configurations are summarised in Table~\ref{tab:ch4_results} and Figure\,\ref{fig:ch4_pop_syn_WD}.

We further assume a white dwarf radio beam angular aperture independent of $P$. For models {\tt WD1} and {\tt WD2}, we adjust our approach to obtain a 20\% and 10\% duty cycle, corresponding to beam apertures $\rho_{\rm em} = 18^{\circ}$ and $36^{\circ}$, respectively (see Section~\ref{subsec:popsyn_NS}). For {\tt WD3} and {\tt WD4}, we set the beaming as in {\tt WD1} but vary mass and radius. In particular, using the Hamada-Salpeter mass-radius relation for He white dwarfs \citep{Hamada1961}, we consider $M_{\rm WD} = 1.2 \, M_{\odot}$ and $R_{\rm WD} = 4000$\,km for a high-mass white dwarf in {\tt WD3} and $M_{\rm WD} = 0.6 \, M_{\odot}$ with $R_{\rm WD} = 9000$\,km for a low-mass white dwarf in {\tt WD4}.

%-------------- TABLE -----------------------------------------------
\begin{landscape}
\begin{table}
\centering
\caption[Population-synthesis results in the neutron-star and white-dwarf scenarios]{Population-synthesis results for isolated neutron stars and magnetic white dwarfs. All models were evolved for $10^{9}$yr assuming a birth rate of 1 neutron star and 10 magnetic white dwarf per century. Unless specified otherwise, neutron stars have 1.4$M_{\odot}$ and 11\,km radii, while magnetic white dwarfs have 1$M_{\odot}$ and 6000\,km radii. Numbers in parentheses denote the assumed power-law (PL) slope. Numbers in brackets are the sub-samples of simulated neutron stars that cross our line of sight [l.o.s.] for their respective beaming. Note that we normalise the neutron star and white dwarf numbers to $10^3$. \label{tab:ch4_results}}
\footnotesize
\begin{threeparttable}
\begin{tabular}{l | l l | l l l | l l l}
%\multicolumn{9}{c}{} \\
\toprule
& & & 
\multicolumn{3}{c}{Total N/$10^3$ [l.o.s.]}  &
\multicolumn{3}{|c}{Total N/$10^3$ with $\dot{E} > 10^{27}$ erg s$^{-1}$ [l.o.s.]}\\
\midrule
\tabhead{Model} &
\tabhead{Initial $\mathbold{P}$} &
\tabhead{Beaming} &
\tabhead{$\mathbold{P = 10^{1-2}}$\,s} &
\tabhead{$\mathbold{P = 10^{2-3}}$\,s} &
\tabhead{$\mathbold{P = 10^{3-5}}$\,s} &
\tabhead{$\mathbold{P = 10^{1-2}}$\,s} &
\tabhead{$\mathbold{P = 10^{2-3}}$\,s} &
\tabhead{$\mathbold{P = 10^{3-5}}$\,s} \\
\midrule
\multicolumn{9}{c}{NEUTRON STARS} \\
\midrule
{\tt NS1\_Bdecay}  & Log-N &  pulsars    & 629.51 [9.22] & 6.69 [0.02] & 0.01 [0.00] & 2.49 [0.05] & 0.01 [0.00] & 0.00 [0.00] \\
{\tt NS1\_Bconst} & Log-N  & pulsars    & 4560.95 [31.15] & 3357.42 [3.70] & 574.38 [0.09] & 2797.74 [24.68] & 34.62 [0.06] & 0.00 [0.00]\\
{\tt NS2\_Bconst} & Log-N  & 20\%   &  4564.26 [1360.90] & 3355.02 [799.39] & 573.61 [120.38] & 2799.49 [876.59] & 34.70 [8.75] & 0.00 [0.00]\\
{\tt NS3\_Bconst} & Log-N+PL(-3)  & 20\%    & 4568.02 [1417.86] & 3364.06 [818.03] & 573.68 [122.11] & 2799.46 [915.60] & 34.78 [9.09] & 0.00 [0.00]\\
{\tt NS4\_Bconst} & Log-N+PL(-1)  & 20\%    & 3425.36 [1302.58] &  3344.05 [1341.96] & 2291.68 [1475.93] & 1920.09 [706.18] & 26.38 [8.52] & 0.00 [0.00]\\
{\tt NS5\_Bconst} & Log-N+PL(-1)  & 10\%    & 3425.41 [603.07] & 3345.82 [646.41] & 2289.66 [838.73] &  1920.40 [320.11] & 26.44 [3.62] & 0.00 [0.00]\\
{\tt NS6\_Bconst}\tnote{o} & Log-N+PL(-1) & 20\%   & 3551.79 [1364.31] & 3158.14 [1298.75] & 2200.36 [1453.03] & 2269.94 [837.21] & 37.55 [12.22] & 0.00 [0.00]\\
{\tt NS2\_Bdecay}\tnote{*} & Log-N  & 20\%    & 827.50 [367.95] & 3.37 [0.93] & 0.00 [0.00] & 3.75 [1.74] & 0.00 [0.00] & 0.00 [0.00] \\
{\tt NS2\_Bconst}\tnote{*} & Log-N  & 20\%    & 2621.41 [765.23] & 5849.46 [1342.22] & 774.43 [168.48] & 1656.60 [504.92] & 77.12 [19.16] & 0.00 [0.00] \\
{\tt NS4\_Bdecay}\tnote{*} & Log-N+PL(-1)  & 20\%    & 1511.96 [977.49] & 895.09 [665.54] & 1786.98 [1330.37] & 3.92 [2.27] & 0.01 [0.00] & 0.00 [0.00] \\
{\tt NS4\_Bconst}\tnote{*} & Log-N+PL(-1)  & 20\%    & 1942.87 [717.60] & 5077.88 [1733.97] & 2508.41 [1565.64] & 1136.06 [406.37] & 58.33 [18.82] & 0.00 [0.00] \\  
\midrule
\multicolumn{9}{c}{WHITE DWARFS} \\
\midrule
{\tt WD1} & Log-N &  20\%  & 1929.55 [1406.56] & 14485.85 [10649.08] & 69050.05 [51234.73] & 1823.72 [1328.54] & 7956.49 [5803.80] & 3880.60 [2729.91] \\
{\tt WD2} &  Log-N  & 10\%  & 1928.76 [817.16] & 14484.90 [6225.08] & 69042.10 [30034.14] & 1822.90 [771.63] & 7955.59 [3386.88] & 3883.30 [1577.85] \\
{\tt WD3}\tnote{o} &  Log-N   & 20\%   & 2147.21 [1576.08] & 14705.23 [10871.34] & 68592.09 [51016.48] & 1894.94 [1389.56] & 5687.17 [4172.56] & 1664.71 [1176.52] \\
{\tt WD4}\tnote{o} &  Log-N & 20\%  & 1579.74 [1141.63] & 13837.27 [10076.93] & 70003.37 [51617.09] & 1542.90 [1114.49] & 9672.34 [6989.46] & 7872.76 [5454.37] \\
\bottomrule
\end{tabular}
\begin{tablenotes}
\item[*]{Bimodal magnetic field distribution (see text).}
\item[o]{{\tt NS6\_Bconst} assumes a 2$M_{\odot}$ neutron star mass, {\tt WD3} a mass of 1.2$M_{\odot}$ and 4000\,km radius, and {\tt WD4} a mass of 0.6$M_{\odot}$ and 9000\,km radius.}
\end{tablenotes}
\end{threeparttable}
\end{table}
\end{landscape}
%---------------------------------------------------------------

%%%%%%%%%%%%%%%%%%%%%%%%%%%%%%%%%%%%%%%%%%%%%%%
%%%%%%%%%%%%%%%%%%%%%%%%%%%%%%%%%%%%%%%%%%%%%%%

\section{Results and discussion}
\label{sec:ch4_conclusion}

% Wide-field radio interferometers have begun to revolutionize our understanding of the transient radio sky. Until recently, coherent, polarized and periodic radio emission was characteristic of neutron star pulsars with periods $\lesssim 20$\,s, a period range attributed to magnetic-field decay and a resistive crust \citep{Pons2013}. However, in the last year, two ultra-long period systems, \gleam\, and \gpm\, \citep{Hurley-Walker2022a, Hurley-Walker2023}, and the slow pulsar PSR J0901-4046 \citep{Caleb2022} were discovered. 

% Ongoing detectability studies (Hurley-Walker et al., in prep) suggest that these discoveries are just the tip of the iceberg and that these sources are very common in the Galaxy. However, due to the limited sample size, it is premature to draw definitive conclusions regarding their unique nature or their classification as distinct astrophysical objects.

Figure~\ref{fig:ch4_p_B}, provides an overview of long period sources in the classical scenario of neutron star pulsar radio emission (see section~\ref{sec:ch1_radio_em_deathlines}) based on magnetospheric pair production. While this scenario can in principle accommodate \gleam, it cannot account for \gpm\, as the source sits below even the most extreme death line. However, note that both objects have radio luminosities exceeding their $\dot{E}$s by 2-3 orders of magnitudes \citep[see][and Section~\ref{sec:ch3_gleam}, and~\ref{sec:ch3_gpm}]{Hurley-Walker2022a, Hurley-Walker2023} implying that dipolar spin-down alone cannot explain the observed luminosities. Hence, the emission scenario for long-period emitters is necessarily more complex than for normal radio pulsars, possibly resembling radio magnetars, where the electromagnetic emission can be powered by the energy stored in the strong magnetic field \citep[see also][and Section~\ref{sec:ch1_ns_zoo}]{Rea2022}.

Figure~\ref{fig:ch4_p_B} also highlights that a similar mechanism in magnetic white dwarfs could in principle contribute to the radio emission of \gleam, AR\,Sco and J1912$-$4410 in the white-dwarf scenario. However, \gpm's bright radio emission cannot be easily reconciled even in the isolated magnetic white dwarf case. For white dwarfs in binary systems such as AR\,Sco and J1912$-$4410, the interactions of the companion's wind with the white dwarf's magnetosphere can replenish the magnetosphere with plasma and enhance the radio emission \citep[][]{Geng2016}. Deep optical and IR observations ruled out main sequence stars for \gleam\ \citep{Rea2022}, but deeper observations are needed to exclude any binarity. On the other hand, \gpm's limits \citep{Hurley-Walker2023} cannot provide strong constraints given its larger distance. 

Our neutron star population synthesis models summarised in Table~\ref{tab:ch4_results} show that a large population of long-period radio emitters can not be easily explained as neutron star pulsars. Neither standard population assumptions nor the most extreme scenarios invoking no field decay (Figure~\ref{fig:ch4_pop_syn_NS_1}), initial slow-down via fallback accretion, 20\% duty cycles or stronger birth fields (Figure~\ref{fig:ch4_pop_syn_NS_2}) result in sufficiently energetic neutron star pulsars with periods $>1000$\,s pointing towards the Earth (irrespective of mass). A difference by a factor of a few in the neutron star birth rate does not alter this conclusion.

On the other hand, white dwarf population synthesis highlights that long-period magnetic white dwarfs are more common than neutron star pulsars. This is primarily because they are born with longer spin periods. This combined with their lower magnetic field and larger moment of inertia reduce the effectiveness of the dipolar spin-down process. Only white dwarfs with small spin periods and large magnetic field experience appreciable spin down (see Figure~\ref{fig:ch4_pop_syn_WD}). Therefore the abundance of slow rotators (Table~\ref{tab:ch4_results}) in the white-dwarf scenario mainly reflects their birth properties. 
However, Figure~\ref{fig:ch4_p_B} shows that the known sample of isolated magnetic white dwarfs are not expected to emit coherent radio emission via standard pair production, being all located below the most extreme death line. 

%=========================================================================

\section{Conclusions}

In this chapter, we studied long-period pulsars in the rotating neutron-star and white-dwarf dipole scenario, one of the most likely interpretations given their coherent and highly polarised emission. We also perform population synthesis simulations to try to asses the likelihood of detecting long-period sources originating from these scenarios.  

We find that the classical particle acceleration mechanism for rotating dipoles (see Section~\ref{sec:ch1_radio_em_deathlines}) fails to provide a satisfactory explanation for the radio emission of \gpm\ in either the neutron star or white dwarf scenario. 
In contrast, all observed isolated magnetic white dwarfs with measured $B$-fields fall below the most extreme death lines, possibly explaining their radio non-detection. The radio emission observed from the binary white dwarfs AR\,Sco and J1912$-$4410 might be enhanced by the presence of their companion star within the white dwarf pulsar light cylinder. However, for \gleam\ optical and IR observations could rule out main sequence companion stars \citep{Rea2022}. For \gpm\, a similar constrain was not possible given the larger distance \citep{Hurley-Walker2023}.

Moreover, in the neutron star scenario, we do not expect a large population of ultra-long period pulsars to possess enough energy to power the observed coherent emission under any (physically motivated or extreme) assumptions. While many more slow white dwarf pulsars can be expected possessing enough rotational energy, we however still lack a mechanism to explain the bright radio emission. Therefore, if \gleam\, and \gpm\, are confirmed as isolated neutron star or white dwarf pulsars, this would call for a revision of our understanding of radio emission from dipolar magnetospheres. Corroborating the neutron star scenario would further require a significant re-examination of our understanding of initial neutron star parameters (birth rates, magnetic-field distribution, etc.) and their evolution at the population level.

% Chapter 5

\chapter{Analyzing the Galactic pulsar dynamics with machine learning} % Main chapter title
%\chapter{Analyzing the Galactic pulsar distribution with machine learning} % Main chapter title

\label{Chapter5} % For referencing the chapter elsewhere, use \ref{Chapter2} 

%----------------------------------------------------------------------------------------

\section{Introduction}
\label{sec:ch5_intro}

Neutron stars have been observed to travel through the Galaxy with typical velocities of around several hundreds of kilometers per second, reaching more than a thousand kilometers per second in some extreme cases \citep{Chatterjee2005,Hobbs2005,Hui2006,Pavan2014}. Accurate information on neutron star positions and velocities in the Milky Way usually comes from radio timing and interferometric observations \citep[see Chapters 8 and 9 in][and references therein]{Lorimer2012, Liu2020} or high spatial resolution X-ray observations with, e.g., the  Chandra X-ray observatory \citep{Motch2009}. These observations provide measurements of the pulsars' angular positions in the sky and their proper motions projected onto the celestial sphere. In some cases, the radio pulse \acf{DM} or the X-ray absorption density ($N_{\rm H}$) together with Galactic free electron-density and hydrogen-density models \citep{Balucinska-Church1992, Taylor1993, Cordes2002, Yao2017} can also yield a rough distance estimate. Moreover, in a few cases, a parallax measurement \citep{Deller2009, Matthews2016, Wang2017, Deller2019} or the presence of a \acf{SNR} \citep{Yao2017} might provide better distance measurements.

Such high proper velocities of the neutron star population as a whole exceed those of their progenitors (typically massive OB stars) \citep[see][and references therein]{Hansen1997, Lai2001}, and cannot be explained by the neutron stars' motion in the Galactic gravitational potential alone. The mechanisms providing such high velocities are still unclear but are likely related to the underlying supernova explosion. One possibility is that the central core of an exploding star receives a kick due to an asymmetric ejection of material from the star's outer layers; a direct result of momentum conservation \citep{Shklovskii1970, Dewey1987, Mandel2020}. Additionally, the anisotropic emission of neutrinos has been suggested to impart kicks on compact remnants \citep{Bisnovatyi-Kogan1993, Fryer2006, Tamborra2014, Nagakura2019}.

However, constraining the neutron stars' natal kick-velocity distribution from current observational data is not straightforward. Most pulsars, especially those with very high velocities, have moved far away from their birth places, and their proper motions have been modified by the Galactic gravitational potential. Thus, the current velocity of a pulsar may differ substantially from its velocity at birth. Knowing the exact pulsar age and its current 3D spatial velocity, we are in principle able to recover the initial conditions by integrating the pulsar's orbit back in time. However, in general we lack information about the pulsar's line-of-sight velocity, and accurate knowledge about its age, since the characteristic age estimated from the pulsar period and its derivative can differ significantly from the true age \citep[see e.g.][and Section~\ref{sec:ch1_dip_spindown_evol}]{Kaspi2001, Vigano2013}. Furthermore, estimates of pulsar distances have typically large associated errors due to uncertainties in the underlying density models used to convert pulsar \acs{DM} or $N_{\rm H}$ into distance estimates \citep{Lorimer2006, He2013, Deller2019}.

Reconstruction of the three-dimensional initial position and velocity distribution of pulsars, and comparison with the observed Galactic neutron star population is therefore a complicated task that requires careful simulations as well as detailed estimates of the observational biases of multi-band surveys. Several studies have performed statistical and population synthesis analyses to recover the distributions of important neutron star parameters from the observed population \citep[see e.g.][]{Arzoumanian2002, Brisken2003, Hobbs2005, Faucher2006, Gullon2014, Verbunt2017, Cieslar2020}. While these models are broadly able to explain the observational data, high degrees of degeneracy between the different input parameters make it difficult to exactly pin down the distributions that control the pulsars' birth properties, such as their natal kick velocities. Nonetheless, disentangling the birth properties of the isolated neutron star population in our Galaxy is crucial as it has important implications for several lines of research, including formation mechanisms of these compact stars, the evolution of massive stars, as well as extreme events such as \acf{GRBs}, \acf{FRBs} and peculiar types of supernovae.

In this chapter, based on the work published in \citet*{Ronchi2021}, we focus on characterising initial pulsar properties using machine learning techniques (see Chapter~\ref{Chapter2}). I will describe the technical aspects related to these efforts and show that a machine-learning framework can be used to estimate parameters with high accuracy. For this feasibility study, we restrict ourselves to a simplified approach, where selection effects and observational biases are neglected and reduced physical models are sufficient. In particular, we focus on the dynamical properties of the pulsar population and explore the possibility of inferring the parameters that control a given Galactic pulsar kick-velocity and scale-height distribution at birth (the two quantities that largely control the spatial distribution of pulsars in the Milky Way) through neural networks. For this purpose, we implement a basic population synthesis code in \texttt{Python} and simulate the dynamical evolution of a synthetic population of isolated neutron stars for a variety of different birth-position and natal-kick distributions. These evolved mock populations are then fed into a suitably structured machine-learning pipeline with the aim of recovering the underlying parameters controlling the distributions. We show that this procedure is successful at estimating birth characteristics. Additionally, we link our framework to the observed sample of pulsars with measured proper motion in a phenomenological way and discuss implications for future pulsar survey efforts, e.g., with the Square Kilometer Array (SKA).

In Section~\ref{sec:ch5_popsynth}, I describe the methods used to simulate and evolve a mock neutron star population in time. Section~\ref{sec:ch5_MLsetup} contains a description of the machine-learning framework, including the generation of our datasets (Section~\ref{sec:ch5_dataset_creation}), the employed network architectures (Section~\ref{sec:ch5_net_architecture}) as well as details of the training process (Section~\ref{sec:ch5_train_process}). In Section~\ref{sec:ch5_experiments}, I present our experiments, which are discussed in detail and connected with observational data in Section~\ref{sec:ch5_discussion}. Finally, I provide a summary and outlook in Section~\ref{sec:ch5_summary}.

%%%%%%%%%%%%%%%%%%%%%%%%%%%%%%%%%%%%%%%%%%%%%%%%%%
%%%%%%%%%%%%%%%%%%%%%%%%%%%%%%%%%%%%%%%%%%%%%%%%%%

\section{Population synthesis: dynamics}
\label{sec:ch5_popsynth}

A widely used approach to investigate the properties of the observed neutron star population is through population synthesis \citep[see e.g.][and Section~\ref{sec:ch1_popsyn}]{Narayan1990, Faucher2006, Gonthier2007, Kiel2008, Kiel2009, Oslowski2011, Levin2013, Gullon2014, Bates2014, Cieslar2020}. These frameworks aim to simulate the evolution of a population of neutron stars from birth until today. The resulting mock population is then compared with the real observed population in order to constrain and validate the physical model assumptions that entered the simulation. In particular, the population synthesis approach relies on assumptions about the distributions of the birth properties of the mock neutron stars, and typically takes advantage of Monte--Carlo methods to construct the initial parameters of each simulated star. Starting from these initial conditions, the mock population is then evolved over time according to some evolutionary prescriptions, and eventually contrasted with real data. For the development of our population synthesis framework, we largely follow \citet{Faucher2006}, \citet{Gullon2014} and \citet{Cieslar2020}. In this chapter we will focus on the dynamical properties while in the next chapter we will extend the analysis to the magneto-rotational properties of neutron stars. The necessary ingredients are briefly summarised in the following.

%%%%%%%%%%%%%%%%%%%%%%%%%%%%%%%%%%%%%%%%%%%%%%%%%%

\subsection{Age}
\label{subsec:age}

The age, $t_{\rm age}$, of each neutron star is randomly drawn from a uniform probability distribution between $1$ and $\unit[10^7]{yr}$. By choosing a uniform distribution, we assume that the birth rate of neutron stars is constant in the chosen time range. For all simulations of the synthetic neutron star population, we choose an average neutron star birth rate of 1 neutron star per century, compatible with the core-collapse supernova rate in the Galaxy \citep[][see also Section~\ref{sec:ch1_birthrate}]{Rozwadowska2021}. This yields a total of $10^{5}$ simulated neutron stars for each synthetic population, whose evolution we can compute within reasonable timescales. 

%%%%%%%%%%%%%%%%%%%%%%%%%%%%%%%%%%%%%%%%%%%%%%%%%%

\subsection{Birth position}
\label{subsec:position}

To define the initial positions, we use both a Cartesian reference frame $(x,~y,~z)$ and a cylindrical reference frame $(r,~\phi,~z)$, whose origins are located at the centre of the Galaxy. Here, $r$ represents the distance in kiloparsec from the Galactic centre, $\phi$ is the azimuthal angle in radians and $z$ is the distance from the Galactic plane. The two coordinate systems are related by the transformation: 
%---------------------------------------------------------------
\begin{align}
    \begin{cases}
    x=r\cos \phi, \\
    y = r\sin \phi, \\
    z = z.
    \end{cases}
\end{align}
%---------------------------------------------------------------
We assume that the Sun is located at the coordinates $x= \unit[0]{kpc}$, $y=r_{\odot}$, $z=z_{\odot}$, where $r_{\odot}= \unit[8.3]{kpc}$ and $z_{\odot}= \unit[0.02]{kpc}$ \citep[see][and references therein]{Pichardo2012}. We calculate the initial position at birth of each neutron star in both cylindrical and Cartesian galactocentric reference frames. To do so, we execute the following steps:
\begin{itemize}
\item[\it{i)}] First, we draw a random distance $r$ from the Galactic centre for each neutron star ranging between $10^{-4}$ and \unit[20]{kpc} according to a pulsar radial density distribution $P(r)$. In particular, we follow the Milky Way's pulsar surface density $\rho(r)$ defined by Equation~(15) in \citet{Yusifov2004} to determine the probability density function for the radial distance, i.e.: 
%----------------------------------------------------------------------------------------------
\begin{align}
  P(r) = 2 \pi r \rho(r),
\end{align}
%----------------------------------------------------------------------------------------------
with
%----------------------------------------------------------------------------------------------
\begin{align}
  \rho(r) = A \left( \frac{r+r_1}{r_{\odot}'+r_1}\right)^a \exp \left[ -b\left(\frac{r-r_{\odot}'}{r_{\odot}'+r_1} \right) \right],
    \label{eq:radialPDF}
\end{align}
%----------------------------------------------------------------------------------------------
where $A = \unit[37.6]{kpc^{-2}}$, $a = 1.64$, $b = 4.0$, $r_1 = \unit[0.55]{kpc}$ and $r_{\odot}' = \unit[8.5]{kpc}$ is the Sun's distance from the Galactic centre projected on the Galactic disk. Although different from the $r_{\odot}$ value assumed above, we keep $r_{\odot}' = \unit[8.5]{kpc}$ in this parametrisation in order to be consistent with the results of \citet{Yusifov2004}. We note that this is the distribution for evolved pulsars rather than that of their progenitors and \citet{Yusifov2004} find small discrepancies between this distribution and that of OB stars. However, \citet{Faucher2006} show that the evolved pulsar population is well described by birth positions drawn from the density distribution in Equation~\eqref{eq:radialPDF} and argue that differences fall within the current uncertainties of pulsar distance measurements. Given the lack of a more realistic description, we therefore adopt the above prescription.

\vspace{0.3cm}

\item[{\it ii)}] Neutron stars are born mainly within the Galactic spiral arms, as these regions are rich in massive OB stars \citep{Chen2019}. We implement a model for the galactic spiral structure that includes four arms with a logarithmic shape function, which gives the azimuthal coordinate, $\phi$, as a function of the distance from the Galactic centre:
%----------------------------------------------------------------------------------------------
\begin{align}	\label{eq:arm_structure1}
	\phi(r) = k \ln \left( \frac{r}{r_0} \right) + \phi_0. 
\end{align}
%---------------------------------------------------------------------------------------------- 
Our values of the model parameters, i.e., the winding constant $k$, the inner radius $r_0$ and the inner angle $\phi_0$ are reported in Table~\ref{tab:ch5_spiral_arms_params_YMW17} and evaluated from Table 1 in \citet{Yao2017} in order to match the same functional form as defined in Equation~\eqref{eq:arm_structure1}. For our analysis, we follow \citet{Faucher2006} and do not include the Local arm, whose density is much smaller than that of the four major arms \citep{Cordes2002, Yao2017}. For each star, we then randomly select one of the four spiral arms with equal probability, and evaluate the angular coordinate, $\phi$, for its given $r$ according to Equation~\eqref{eq:arm_structure1}.

%-------------------------------------------------------------
\begin{table}
\centering
\caption[Parameters of the Milky Way spiral arm structure]{Parameters of the Milky Way spiral arm structure: the winding constant $k$, the inner radius $r_0$ and the inner angle $\phi_0$  (adapted from Table 1 in \citet{Yao2017}; see Section~\ref{subsec:position} for more details).
\label{tab:ch5_spiral_arms_params_YMW17}}
\begin{tabular}{l l l l l}
\toprule
\tabhead{\multirow{2}{*}{Arm number}} &
\tabhead{\multirow{2}{*}{Name}} &
\tabhead{$\mathbold{k}$} &
\tabhead{$\mathbold{r_0}$} &
\tabhead{$\mathbold{\phi_0}$} \\
& 
& 
\tabhead{[rad]} & 
\tabhead{[kpc]} & 
\tabhead{[rad]} \\
\midrule
1 & Norma & 4.95 & 3.35 & 0.77 \\
2 & Carina-Sagittarius & 5.46 & 3.56 & 3.82 \\
3 & Perseus & 5.77 & 3.71 & 2.09 \\
4 & Crux-Scutum & 5.37 & 3.67 & 5.76 \\
\bottomrule\\
\end{tabular}
\end{table}
%---------------------------------------------------------------

The spiral pattern of the Galaxy is not static and as a first approximation can be considered as a rigid structure which rotates with an approximated period $T = \unit[250]{Myr}$ \citep{Vallee2017, Skowron2019}. Knowing the age of an object and assuming a rotational angular velocity of $\Omega = 2\pi/T$ for the spiral structure, we can derive the angular position at birth of each neutron star. Note that the Galaxy rotates in the clockwise direction, i.e., toward decreasing $\phi$ angles.

After obtaining the corresponding angular coordinate for each neutron star birth position, we add noise to both coordinates $r$ and $\phi$ to smear out the distribution and avoid artificial features near the Galactic centre. For this purpose, we add a correction $\phi_{\rm corr} = \phi_{\rm rand} \exp{\left( -0.35 r/ {\rm kpc} \right)}$ to the $\phi$ coordinate, where $\phi_{\rm rand}$ is randomly drawn from a uniform distribution in the interval $\left[ 0, 2\pi \right)$, and to the $r$ coordinate a correction $r_{\rm corr}$ randomly drawn from a normal distribution centred at 0 with standard deviation $\sigma = 0.07r$. Although this prescription was introduced by \citet{Faucher2006} (see their Section 3.2.1) in a somewhat arbitrary manner, the resulting stellar distribution broadly agrees with that observed for very young high-mass stars as shown in \citet{Reid2019}.

Then the birth position in polar coordinates of each neutron star is given by $(r+r_{\rm corr},~\phi(r)+\Omega t_{\rm age}+\phi_{\rm corr})$ with units [kpc, rad].

\vspace{0.3cm}

\item[{\it iii)}] To determine the height $z$ in kiloparsec from the Galactic plane of each neutron star, we adopt an exponential disk model as given by \citet{Wainscoat1992}. It is shaped by the characteristic scale-height parameter $h_{\rm c}$:
%----------------------------------------------------------------------------------------------
\begin{align}
  P(z) = \frac{1}{h_{\rm c}} \exp\left(-\frac{ \lvert z \rvert}{ h_{\rm c} } \right). 
\end{align}
%----------------------------------------------------------------------------------------------
For our machine-learning experiments, we will vary the scale height in the range $[0.02,2]$ $\unit[]{kpc}$ to simulate neutron star populations with different spread in Galactic height. This range encompasses the value $h_{\rm c} = \unit[0.18]{kpc}$, which was adopted by \citet{Gullon2014} to match radio pulsar observations, and which is also compatible with the population of young massive stars in the Galactic disk \citep{Li2019}. We will consider $h_{\rm c} = \unit[0.18]{kpc}$ below, whenever a fiducial scale height is required for our synthetic pulsar population. The coordinate $z$ of each neutron star is then randomly drawn according to this height probability distribution in a range of $10^{-4}$ to $\unit[5]{kpc}$. We choose a maximal distance of $\unit[5]{kpc}$ from the Galactic plane to model a fixed galactic volume for all of our simulation runs, while also ensuring sufficient resolution for the objects in the galactic disc for those models with small scale heights. Subsequently, for each star we randomly choose if $z$ is positive or negative determining in this way a position above or below the Galactic plane.
\end{itemize}

%%%%%%%%%%%%%%%%%%%%%%%%%%%%%%%%%%%%%%%%%%%%%%%%%%

\subsection{Initial velocity}
\label{subsec:velocity}

We assume that the initial velocity of the neutron stars in the Galaxy is given by two contributions: the progenitor velocity in the Galactic gravitational potential and a kick speed imparted onto the neutron stars as a result of the supernova explosion (see also Section \ref{sec:ch1_spin_periods}).
We consider a progenitor circular orbital speed given by the following relation:
%-------------------------------------------------------------
\begin{align}
	v_{\rm orb} = \sqrt{ r \frac{\partial \Phi_{\rm MW} \left( r,z \right)}{\partial r} },
\end{align}
%-------------------------------------------------------------
where $\Phi_{\rm MW}$ is the Milky Way gravitational potential discussed below.
We assume that each neutron star has an initial kick velocity $\boldsymbol{v}_{\rm k}$, whose 3D magnitude $v_{\rm k}$ is randomly drawn from a Maxwell distribution, shaped by the dispersion parameter $\sigma_{\rm k}$:
%-------------------------------------------------------------
\begin{align}	\label{eq:pdf_maxwell_kick}
	P(v_{\rm k}) = \sqrt{ \frac{2}{\pi} }  \frac{v_{\rm k}^2}{\sigma_{\rm k}^3} \exp\left(-\frac{v_{\rm k}^2}{ \sigma_{\rm k}^2 } \right).
\end{align}
%-------------------------------------------------------------
For our machine-learning purposes, we will vary $\sigma_{\rm k}$ in the range $[1, 700]$ $\unit[]{km \, s^{-1}}$ and randomly draw 3D velocity magnitudes from the resulting distribution in the range $[0, 2500]$ $\unit[]{km \, s^{-1}}$. This spread allows us to easily accommodate the fastest observed neutron stars whose velocities have been estimated to surpass 1000 $\unit[]{km \, s^{-1}}$ \citep[see for example][]{Chatterjee2005,Hui2006,Pavan2014}. Based on pulsar timing measurements, \citet{Hobbs2005} have suggested that $\sigma_{\rm k} = \unit[265]{km \, s^{-1}}$ provides a viable explanation for the proper motions of observed neutron stars. We will use this as a fiducial value below. For a given kick-velocity magnitude, we then associate a random direction to this velocity in order to evaluate the three components ($v_{{\rm k}, r}$, $v_{{\rm k}, \phi}$, $v_{{\rm k}, z}$) in galactocentric cylindrical coordinates. Therefore, the three components of the total initial velocity of each neutron star in the galactocentric reference frame are computed as ($v_{{\rm k}, r},~v_{\rm orb} + v_{{\rm k}, \phi},~v_{{\rm k}, z})$.  

%-------------------------------------------------------------
\begin{table}
\centering
\caption[Parameters of the Milky Way gravitational potential]{Parameters of the Milky Way gravitational potential taken from Table 1 in \citet{Marchetti2019}; see also \citet{Bovy2015}.
\label{tab:ch5_MW_pot_params_M19}}
\begin{tabular}{l l}
\toprule
\tabhead{Component} &
\tabhead{Parameters} \\
\midrule
nucleus (n) & $M_{\rm n} = 1.71 \times 10^{9} M_{\odot}$ \\
 & $R_{\rm n} = 0.07$ kpc \\
bulge (b) & $M_{\rm b} = 5.0 \times 10^{9} M_{\odot}$ \\
 & $R_{\rm b} = 1.0$ kpc \\
disk (d) & $M_{\rm d} = 6.8 \times 10^{10} M_{\odot}$ \\
 & $a_{\rm d} = 3.00$ kpc \\
 & $b_{\rm d} = 0.28$ kpc \\
halo (h) & $M_{\rm h} = 5.4 \times 10^{11} M_{\odot}$ \\
 & $R_{\rm h} = 15.62$ kpc \\
\bottomrule\\
\end{tabular}
\end{table}
%---------------------------------------------------------------

%%%%%%%%%%%%%%%%%%%%%%%%%%%%%%%%%%%%%%%%%%%%%%%%%%

\subsection{Galactic potential}
\label{subsec:potential}

As typical for spiral galaxies like the Milky Way, we assume an axisymmetric Galactic potential $\Phi_{\rm MW}$ \citep{Carlberg1987, Bovy2015} that does not incorporate the impact of the spiral arms themselves. We specifically consider a four-component Galactic potential model consisting of a nucleus ($\Phi_{\rm n}$), a bulge ($\Phi_{\rm b}$), a disk ($\Phi_{\rm d}$) and a halo ($\Phi_{\rm h}$) as discussed in \citet{Marchetti2019}:
%-------------------------------------------------------------
\begin{align}	
	\Phi_{\rm MW} = \Phi_{\rm n} + \Phi_{\rm b} + \Phi_{\rm d} + \Phi_{\rm h}.
\end{align}
%-------------------------------------------------------------
The nucleus and the bulge are described by a spherical Hernquist potential \citep{Hernquist1990}:
%-------------------------------------------------------------
\begin{align}	
	\Phi_{\rm n} = -\frac{ G M_{\rm n}}{ R_{\rm n} + R },
\end{align}
%-------------------------------------------------------------
\begin{align}	
	\Phi_{\rm b} = -\frac{ G M_{\rm b}}{ R_{\rm b} + R }. 
\end{align}
%-------------------------------------------------------------
where the coordinate $R = \sqrt{r^2 + z^2}$ is the distance from the Galactic centre in spherical coordinates.
The disk has a Miyamoto-Nagai disk potential \citep{Miyamoto1975}:
%-------------------------------------------------------------
\begin{align}	
	\Phi_{\rm d} = -\frac{ G M_{\rm d}}{ \sqrt{ K^2 + r^2} },
\end{align}
%-------------------------------------------------------------
where $K = a_{\rm d} + \sqrt{z^2+b_{\rm d}^2}$ is the shape parameter with $a_{\rm d}$ as the scale length and $b_{\rm d}$ the scale height of the disk. 
\noindent
The halo has a Navarro-Frenk-White potential \citep{Navarro1996}:
%-------------------------------------------------------------
\begin{align}	
	\Phi_{\rm h} = -\frac{ G M_{\rm h}}{ R } \ln{ \left( 1 + \frac{R}{R_{\rm h}}\right) }.
\end{align}
%-------------------------------------------------------------
The parameters of this model are reported in Table~\ref{tab:ch5_MW_pot_params_M19} and were derived by \citet{Bovy2015} through a fit of the mass profile of the Milky Way. We assume that these contributions to the galactic potential are stationary in time, i.e., they do not evolve over the time span we consider for the dynamical evolution.

%%%%%%%%%%%%%%%%%%%%%%%%%%%%%%%%%%%%%%%%%%%%%%%%%%

\subsection{Dynamical evolution}
\label{subsec:evolution}

Given the initial conditions defined above, i.e., the initial position, initial velocity and the Galactic gravitational potential, we can solve the equations of motion to determine the neutron stars' dynamical evolution. The system of dynamical equations that requires solving to determine the orbits of the neutron stars in the galactic potential is given by the Newtonian equations of motion: $\ddot{\boldsymbol{r}} = -\boldsymbol{\nabla} \Phi_{\rm MW}$.
In cylindrical galactocentric coordinates the three components of this vector equation take the form: 
%-------------------------------------------------------------
\begin{align}
	\begin{cases}	
    \vspace{0.2cm}
    \displaystyle 
	\ddot{r} - r \dot{\phi}^2 = -\frac{\partial \Phi_{\rm MW} }{ \partial r }, \\
    \vspace{0.2cm}
    \displaystyle 
	2\dot{r}\dot{\phi} + r \ddot{\phi} = - \frac{1}{r} \frac{\partial \Phi_{\rm MW} }{ \partial \phi }, \\
    \displaystyle 
	\ddot{z} = - \frac{\partial \Phi_{\rm MW} }{ \partial z }. 
	\end{cases}
\end{align}
%-------------------------------------------------------------
For each neutron star, we numerically integrate the above equations in time from $t = \unit[0]{yr}$ to $t = t_{\rm age}$ using a discrete time step. 
We use the Python package \texttt{scipy.integrate.\\odeint} and to speed up the computational time also employ the module \texttt{jit} (``just in time") from the \texttt{Numba} package (\url{https://numba.pydata.org/}, \citealt{Lam2015}).\footnote{\texttt{Numba} translates Python functions into optimised machine code at run-time, which allows us to achieve a speed-up by about a factor of 6.} To asses the performance of our integration method, we check that both the total energy (i.e., the potential plus kinetic energy) and the $z$-component of the total angular momentum of all the stars in our simulation are conserved. For simplicity, we assume that all pulsars have the same mass and we find that both quantities are conserved with a relative error of $\lesssim 10^{-7}$. The output of the dynamical evolution consists of the position and velocity of each neutron star computed in both galactocentric (GC) and equatorial ICRS (International Celestial Reference System) frames. To transform between different coordinate systems, we employed the method \texttt{coordinates} from the Python library \texttt{Astropy} \citep{astropy2013, astropy2018}, where we adopted a galactocentric distance of $r_{\odot}= \unit[8.3]{kpc}$ and Galactic height of $z_{\odot}= \unit[0.02]{kpc}$ for the Sun.

%%%%%%%%%%%%%%%%%%%%%%%%%%%%%%%%%%%%%%%%%%%%%%%%%%
%%%%%%%%%%%%%%%%%%%%%%%%%%%%%%%%%%%%%%%%%%%%%%%%%%

%-----------------------------------------------------------------
\begin{figure}
\centering
\includegraphics[width = 0.47\textwidth]{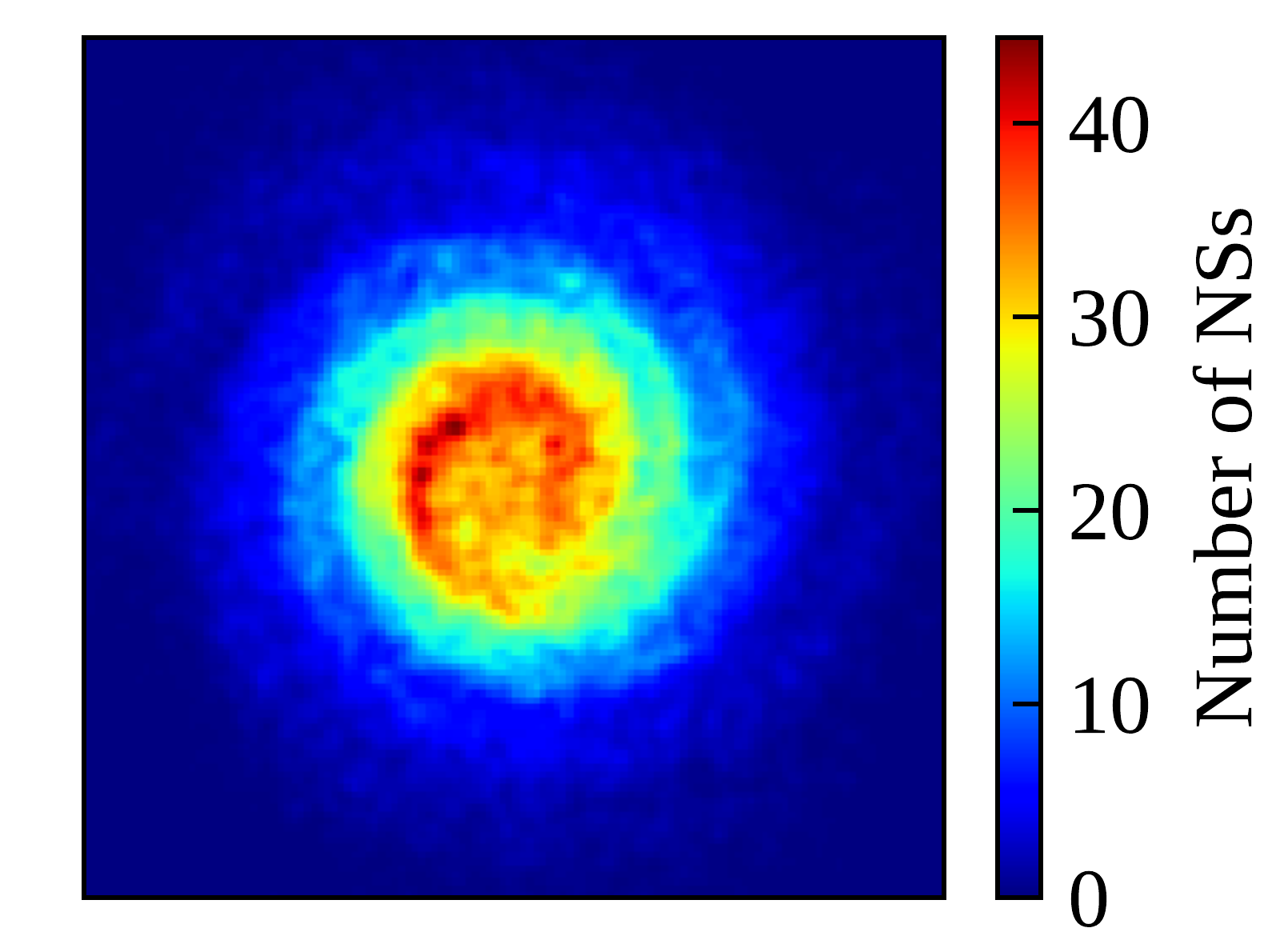}
\includegraphics[width = 0.47\textwidth]{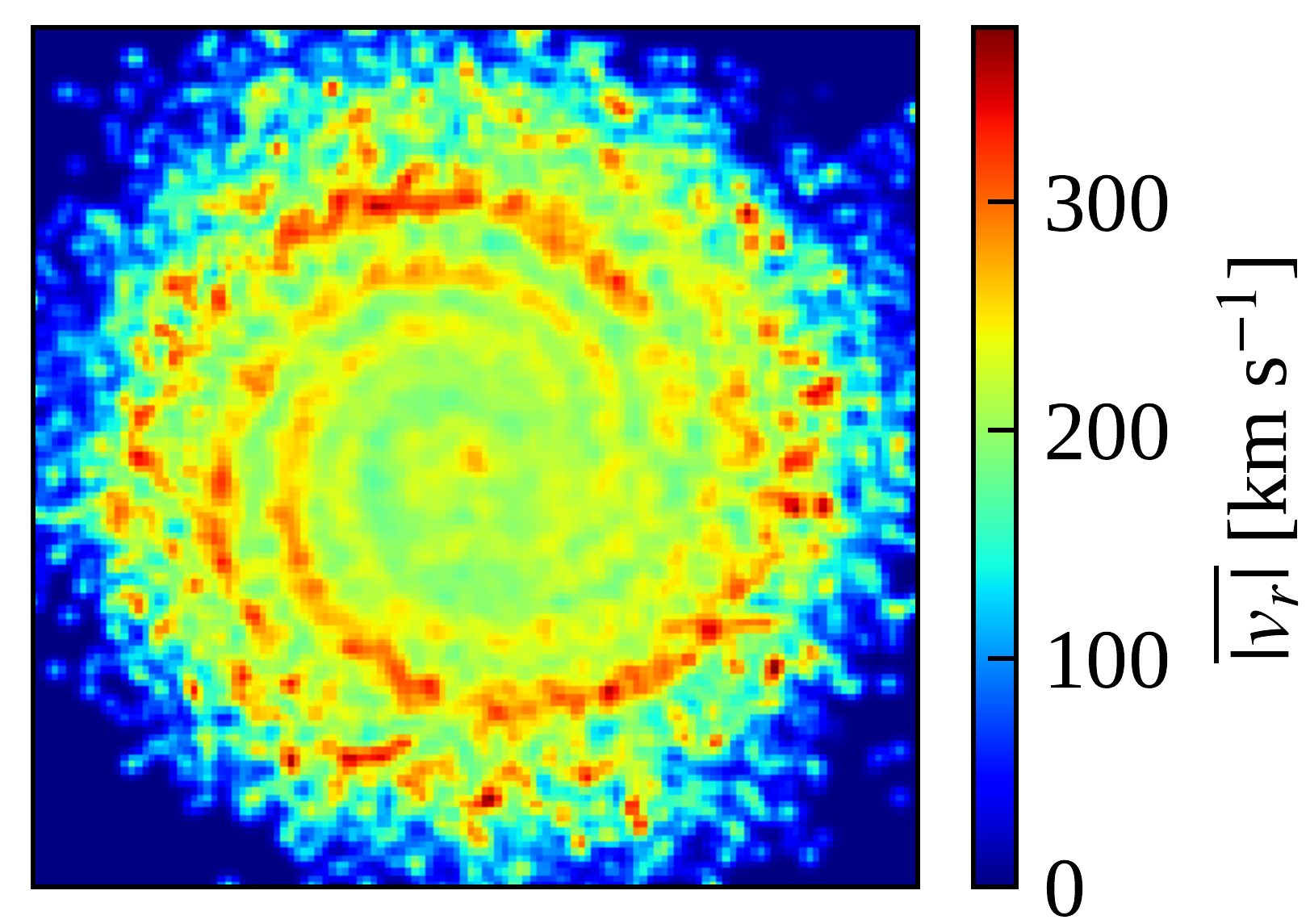}
\includegraphics[width = 0.47\textwidth]{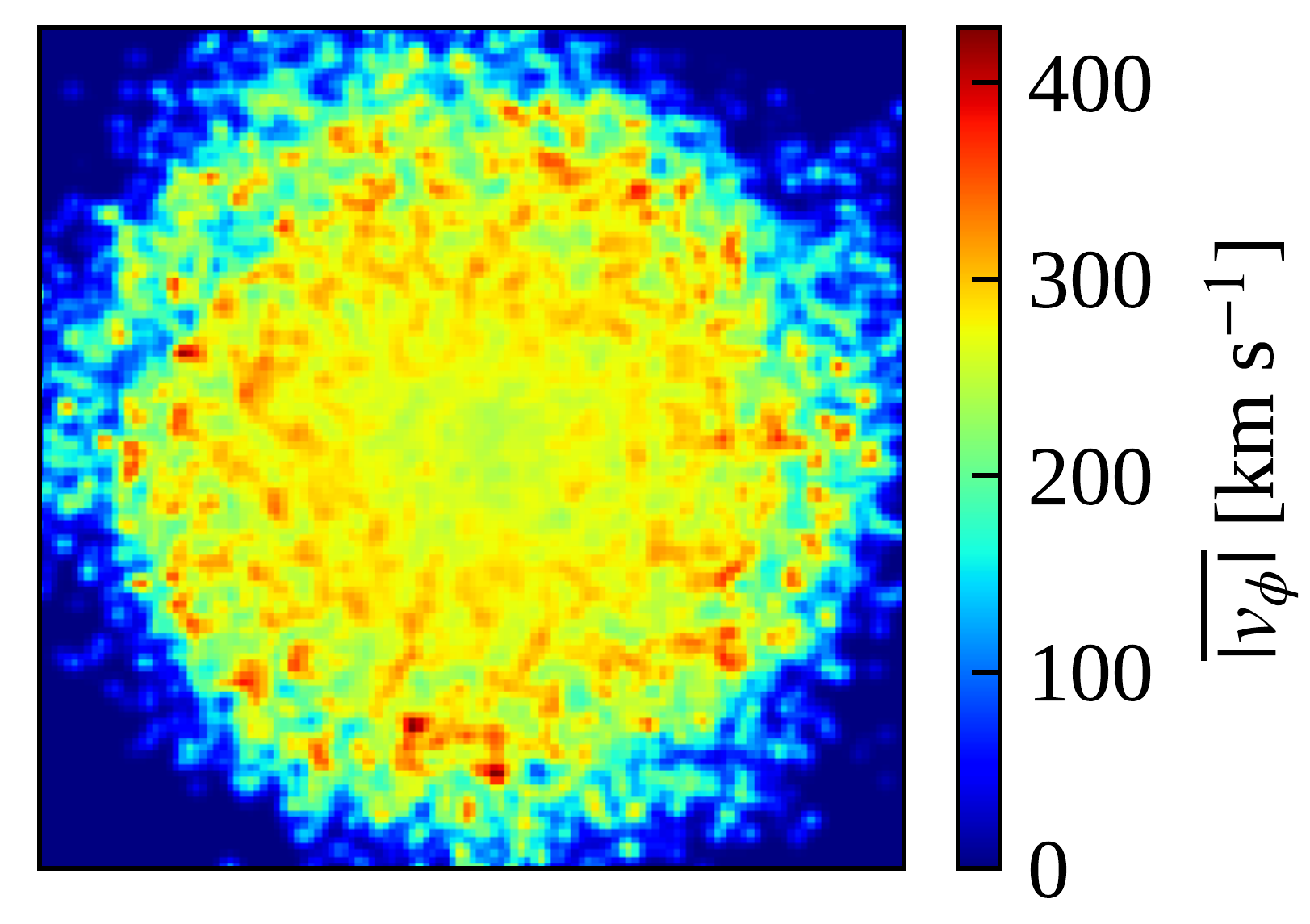}
\includegraphics[width = 0.47\textwidth]{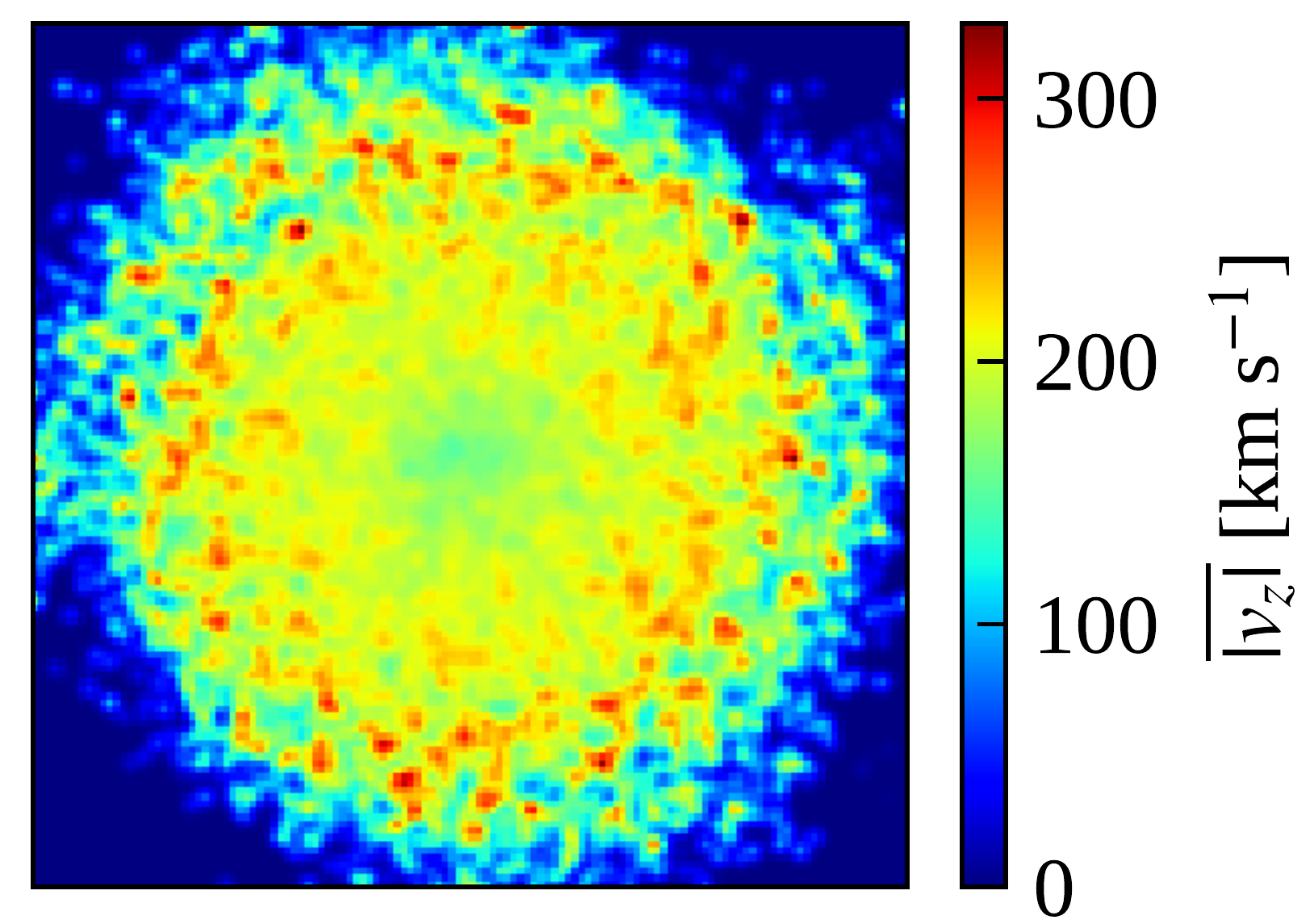}
\caption[Examples of $128 \times 128$ resolution maps in the galactocentric $xy$-plane]{\label{fig:ch5_gc_map}{Examples of $128 \times 128$ resolution maps in the galactocentric $xy$-plane extending from $-20$ to $\unit[20]{kpc}$ in both $x$ and $y$ direction and showing (in order from left to right) the density of simulated neutron stars, the distribution of average values of the $v_{r}$, $v_{\phi}$ and $v_{z}$ velocity components for a population of neutron stars simulated with $h_{\rm c} = \unit[0.18]{kpc}$ and $\sigma_{\rm k} = \unit[265]{km \, s^{-1}}$. For visualisation purposes, we represent the data using a colourmap to highlight the resulting structures; red regions are characterised by a higher density of stars or higher average magnitude of the velocity components, respectively, while blue areas correspond to lower densities and lower velocity magnitudes. We note that the spiral-arm pattern is still recognisable in the position-density map although high kick velocities tend to blur and disperse the stellar density distribution. In the $v_r$-velocity map, the inter-arm regions are visible as high-velocity areas, because during the dynamic evolution the space between the spiral arms is progressively filled with high velocity stars that have escaped from their birth places. The other two velocity components exhibit smoother behaviour because the spiral-arm structure is smeared out.}}
\end{figure}  
%-----------------------------------------------------------------
%-----------------------------------------------------------------
\begin{figure}
\centering
\includegraphics[width = 0.7\textwidth]{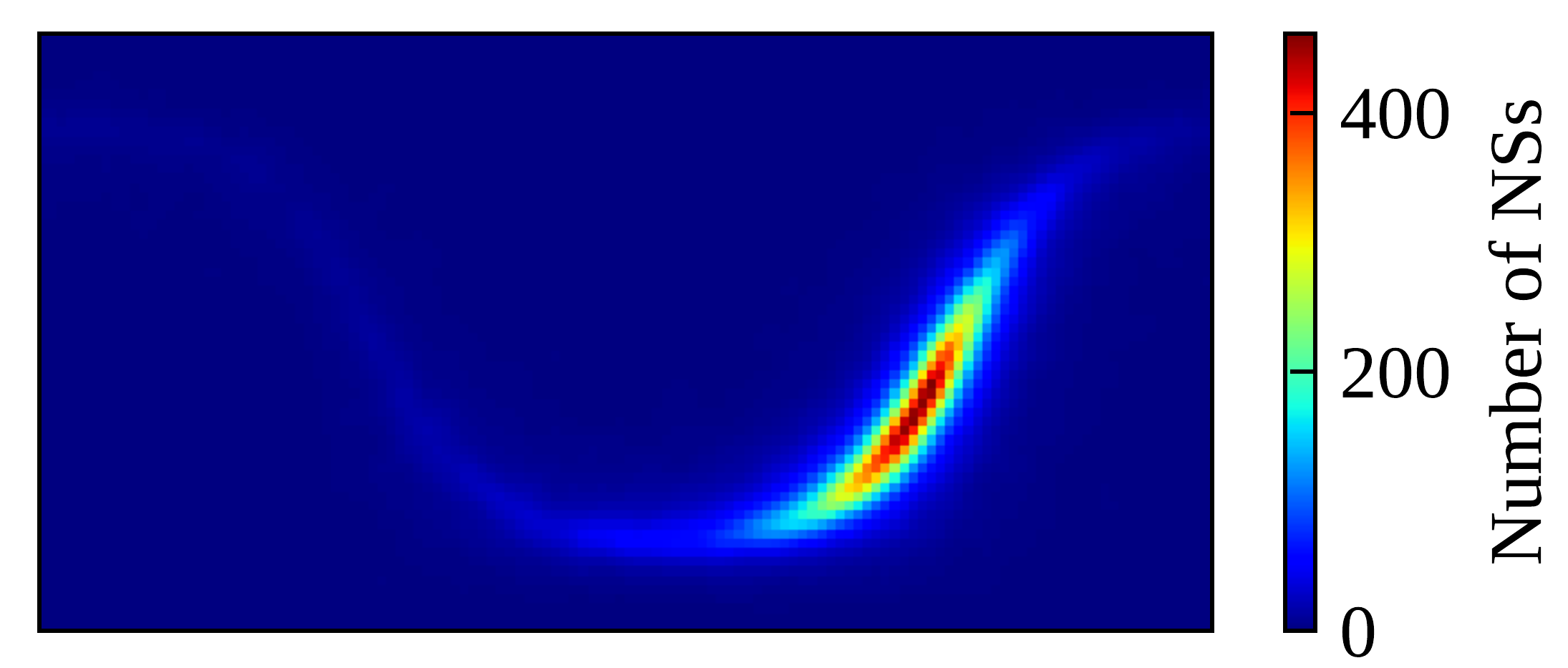}
\includegraphics[width = 0.7\textwidth]{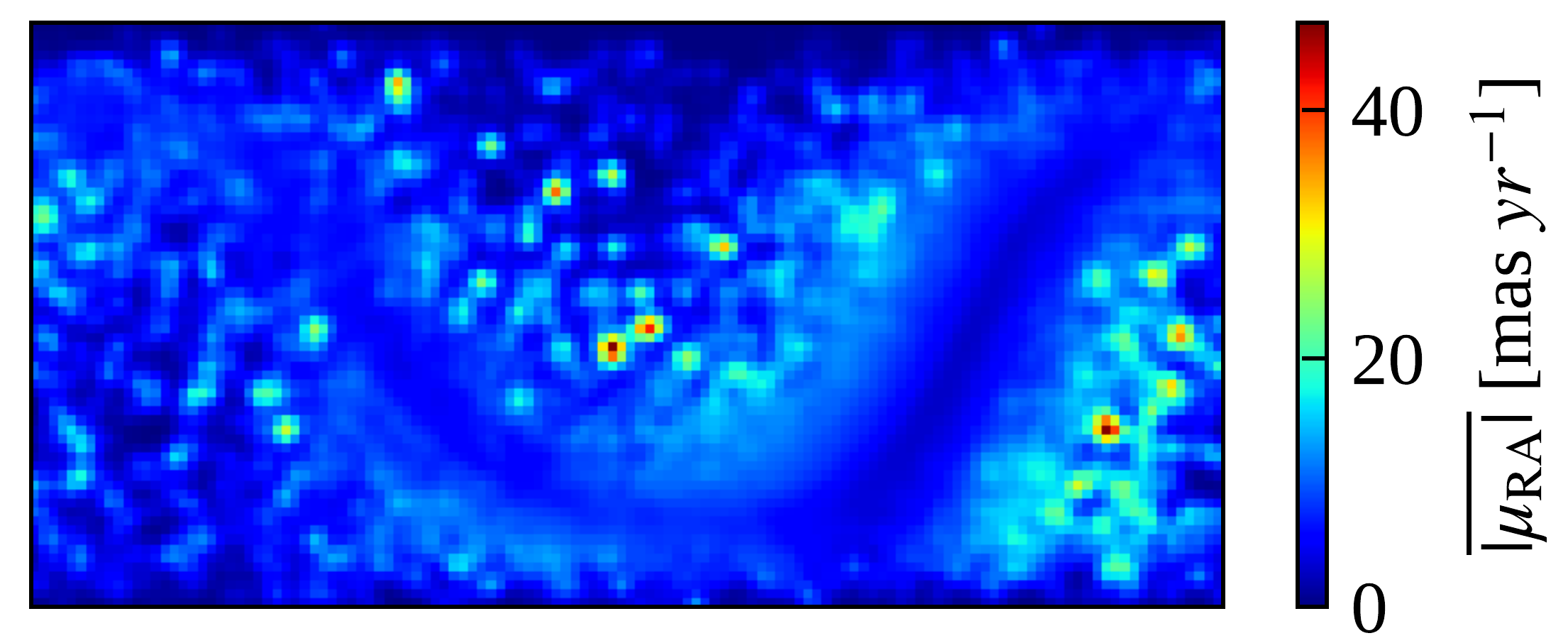}
\includegraphics[width = 0.7\textwidth]{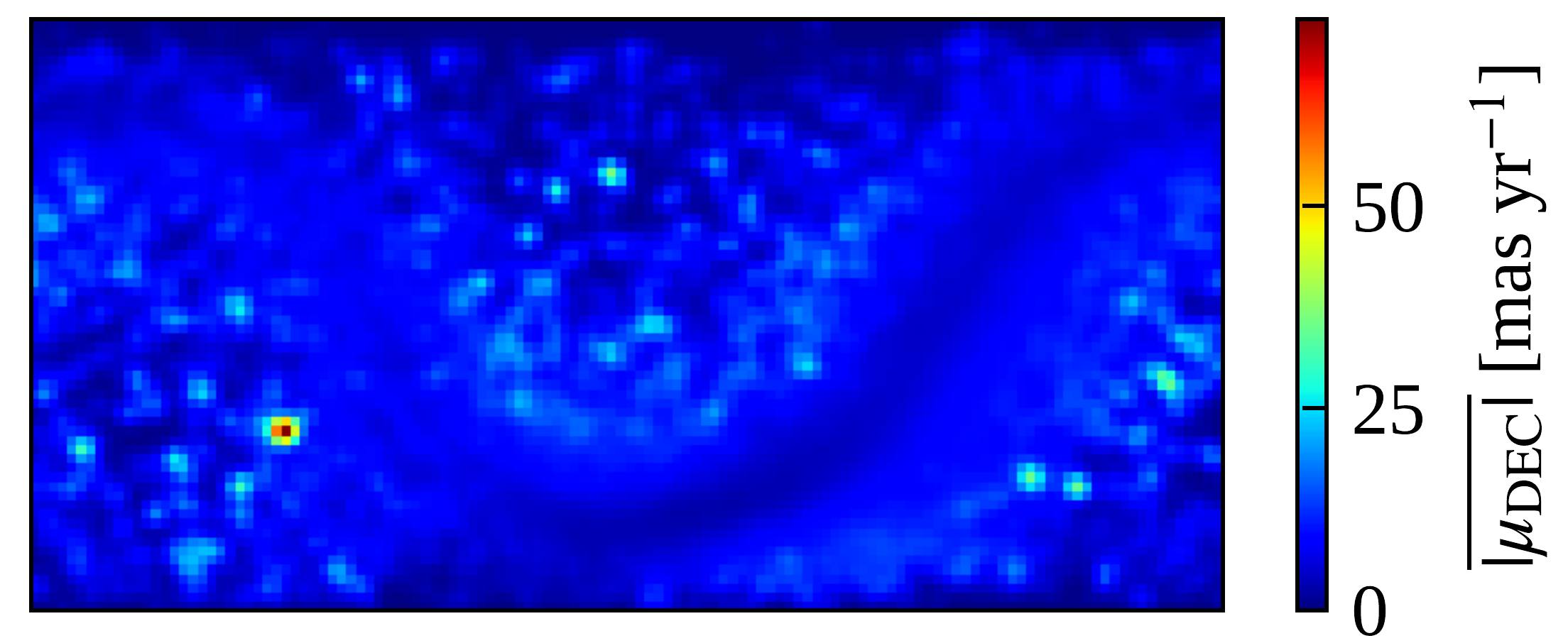}
\caption[Examples of $128 \times 64$ resolution maps in the equatorial ICRS frame]{\label{fig:ch5_icrs_map}{Examples of $128 \times 64$ resolution maps in the equatorial ICRS frame extending from $0^{\circ}$ to $360^{\circ}$ in RA and from $-90^{\circ}$ to $90^{\circ}$ in DEC and showing (in order from left to right) the density of simulated neutron stars, the distribution of average values for the $\mu_{\rm RA}$, $\mu_{\rm DEC}$ proper-motion components for a population of neutron stars simulated with $h_{\rm c} = \unit[0.18]{kpc}$ and $\sigma_{\rm k} = \unit[265]{km \, s^{-1}}$. For visualisation purposes, we represent the data using a colourmap to highlight the resulting structures}; red regions are characterised by a higher density of stars or a higher average magnitude of the velocity components, respectively, while blue areas correspond to lower densities and lower velocity magnitudes. Note that the Galactic silhouette is visible as a \textit{stream} in the position map with an enhanced stellar density close to the Galactic Centre. Due to low-number statistics, the regions outside the Galactic stream in the proper-motion maps are dominated by statistical fluctuations, i.e., the corresponding high-velocity regions are attributed to a small number of high proper-motion stars that have escaped the disk. As a result, the disk itself is dominated by stars with lower proper motion. }
\end{figure}  
%-----------------------------------------------------------------

\section{Machine-learning set-up}
\label{sec:ch5_MLsetup}

For our analysis, we will focus on \acf{ANNs}. \acs{ANNs} are algorithms inspired by the structure of biological brains that can be thought of as nets of interconnected neurons that exchange information from one to another (see Section~\ref{sec:ch2_deep_learning}). When the network receives an input, it is able to process it to produce an output, like a biological brain responds to external stimulation. In \acs{ANNs}, neurons are usually organised in a stack of layers. Each neuron in a layer receives input signals (typically real numbers) from the neurons in the previous layer and produces an output signal by applying a non-linear activation function to a linear combination of the input signals according to certain weights and a bias. The output is then passed to the neurons in the following layer, and so on, until the final layer is reached and the output is generated. In our particular case, we will focus on supervised learning, where training the neural network consists of making it produce a specific target output when a particular input is passed through it. This is achieved by (i) labeling the input samples in the training dataset with a label indicating the property that the network has to learn (the so-called ground truth) and (ii) iteratively adjusting the values of weights and biases, also called network \text{parameters}, in order to minimise a specific loss function which measures the distance between the network output prediction and the target ground truth (see Section~\ref{sec:ch2_backprop}). 

Among their numerous applications, \acs{ANNs} have been employed in regression problems where the network is trained to infer the values of continuous variables for the given input data. This is the kind of problem we are after since we want our network to infer certain parameter values given the evolved neutron star population. In the remainder of this section, we discuss (i) the simulation data we create for our \acs{ML} experiments (Section~\ref{sec:ch5_dataset_creation}), then (ii) focus on the specific network architecture employed (Section~\ref{sec:ch5_net_architecture}), and finally (iii) describe the details of our training process (Section~\ref{sec:ch5_train_process}).

%%%%%%%%%%%%%%%%%%%%%%%%%%%%%%%%%%%%%%%%%%%%%%%%%%

\subsection{Dataset creation and processing}
\label{sec:ch5_dataset_creation}

The goal of our \acs{ML} approach is to predict the parameters that control the dynamical properties of an evolved neutron star population. In particular, we focus on predicting the kick-velocity parameter $\sigma_{\rm k}$ and the scale-height parameter $h_{\rm c}$, which predominantly affect the distribution of pulsars in the Milky Way. To extract these from an evolved population, and follow a supervised learning approach, we first need to train a neural network on a series of simulated populations (created by exploring the ranges for $\sigma_{\rm k}$ and $h_{\rm c}$). Following the prescription described in Section~\ref{sec:ch5_popsynth}, we perform the following simulation runs:
\begin{itemize}
    \item[\textbf{S1}] We generate 10 datasets with an increasing number of samples (specifically 4, 8, 16, 32, 64, 128, 256, 512, 1024 and 20000 simulated populations) by uniformly varying the parameter $\sigma_{\rm k}$ of the kick velocity distribution in the range [1, 700] $\unit[]{km\,s^{-1}}$.\footnote{To generate our simulation data, we partially employ the package \texttt{Hydra} (\url{https://hydra.cc/}, \citealt{Yadan2019}), which allows us to easily sweep entire parameter ranges.} We also generate a test dataset with 1000 samples, each one simulated with $\sigma_{\rm k}$ randomly drawn from a uniform distribution in the same range of values. For these simulations, we keep the characteristic scale of the $z$-distribution fixed to its fiducial value $h_{\rm c} = \unit[0.18]{kpc}$.
    \item[\textbf{S2}] We fix the kick-velocity parameter to its fiducial value $\sigma_{\rm k} = \unit[265]{km\,s^{-1}}$ and generate a dataset of 20000 samples of simulated populations by uniformly varying the scale-height parameter $h_{\rm c}$ in the range [0.02, 2] $\unit[]{kpc}$. We also generate a test dataset with 1000 samples, each one simulated with $h_{\rm c}$ randomly drawn from a uniform distribution in the same range of values.
    \item[\textbf{S3}] We generate 6 datasets, where we uniformly vary the kick-velocity parameter $\sigma_{\rm k}$ in the range [1, 700] $\unit[]{km\,s^{-1}}$ \textit{as well as} the characteristic scale of the $z$-distribution $h_{\rm c}$ in the range [0.02, 2] $\unit[]{kpc}$. We choose the dataset sizes $16 = 4 \times 4$, $64 = 8 \times 8$, $256 = 16 \times 16$, $1024 = 32 \times 32$, $4096 = 64 \times 64$ and $16384 = 128 \times 128$ given by all the combinations of $\sigma_{\rm k}$ and $h_{\rm c}$ values. As an example: the 16 populations in the first set are obtained by combining each of the 4 values of the $\sigma_{\rm k}$ parameter with all 4 values of the $h_{\rm c}$ parameter. We also generate a test dataset with 1000 samples, each one simulated with both $\sigma_{\rm k}$ and $h_{\rm c}$ randomly drawn from uniform distributions in their respective parameter ranges specified above.
\end{itemize}
As addressed in detail in Section~\ref{sec:ch5_experiments}, the smaller simulation datasets will be used to explore the network behaviour. The largest datasets containing 20000 and 16384 samples, respectively, and the test datasets with 1000 samples will be used to perform the final training experiments and assess the actual network accuracy in generalisation scenarios.

After the runs have been performed, we transform the output of the simulation into a representation that can be interpreted by a \acs{ML} pipeline. Since \acs{ANNs} require the use of structured data, we represent the position and velocity output of the simulations in the form of 2D binned density and velocity maps both in the galactocentric and ICRS reference frames. The density maps give information about the density of neutron stars in the Galaxy by providing the number count of stars in each spatial bin. On the other hand, velocity maps contain information about the kinematic properties of the neutron stars by providing the average magnitude of the stellar velocity components in each spatial bin. In the galactocentric maps the Galaxy is represented face on and projected onto the $xy$-plane of the Cartesian galactocentric frame, extending from $\unit[-20]{kpc}$ to $\unit[20]{kpc}$ in $x$ and $y$ direction. The ICRS maps instead extend from $0^{\circ}$ to $360^{\circ}$ in right ascension (RA) and from $-90^{\circ}$ to $90^{\circ}$ in declination (DEC). To each map we apply a smoothing Gaussian filter (with radius $4 \sigma$ and $\sigma = 1$) in order to add some blurring and avoid sharp features. By doing so, we reduce noisy high-frequency features and thus make the training more stable, presumably resulting in better generalisation capabilities. Therefore, for each simulated population we have:
\begin{itemize}
    \item 1 density map in the galactocentric frame.
    \item 3 velocity maps, one for each component of the velocity in cylindrical galactocentric coordinates $v_{r}$, $v_{\phi}$ and $v_{z}$ in [$\unit[]{km \, s^{-1}}$].
    \item 1 density map in ICRS coordinates.
    \item 2 proper motion maps, one for each component of the angular proper motion projected on the celestial sphere $\mu_{\rm RA}$ and $\mu_{\rm DEC}$ in [$\unit[]{mas \, yr^{-1}}$].
\end{itemize}
This set of maps for each population is labeled with the corresponding values of the parameters $\sigma_{\rm k}$ and $h_{\rm c}$ used to simulate that specific population.

We will test the \acs{ML} performance on three different map resolutions. Thus, we generate each of the above datasets with resolutions $32 \times 32$, $128 \times 128$ and $512 \times 512$ square bins. Note that in the ICRS maps the DEC coordinate axis range is half that of the RA coordinate axis. Hence, these maps have half the bins along the DEC coordinate and their resolution is $32 \times 16$, $128 \times 64$ and $512 \times 256$ square bins. For brevity hereafter we will refer to the three resolutions as 32, 128 and 512 resolutions for both galactocentric and ICRS maps, respectively. An example of the maps with resolution 128 for a simulation with fiducial values $h_{\rm c} = \unit[0.18]{kpc}$ and $\sigma_{\rm k} = \unit[265]{km \, s^{-1}}$ is shown in Figures~\ref{fig:ch5_gc_map} and \ref{fig:ch5_icrs_map}. 

Before loading the maps into our \acs{ML} pipeline, they are normalised so that each bin contains a continuous value between 0 and 1. The same applies to the related labels so that their values range continuously between 0 and 1. The aim of normalisation is to speed up the training process and make convergence easier since all inputs will provide signals of similar magnitude to the loss-function minimisation (see Section~\ref{sec:ch2_datasets}). This is useful especially for multi-parameter and multi-channel training, that is when we train a network to predict more than one parameter or use channels that have different absolute magnitudes. In these cases, training without normalisation might lead to slower, worse or even no convergence at all. Apart from the blurring and normalisation described above, we do not apply any additional pre-processing steps to our input data.

%-----------------------------------------------------------------
\begin{figure*}
\centering
\includegraphics[width = \textwidth]{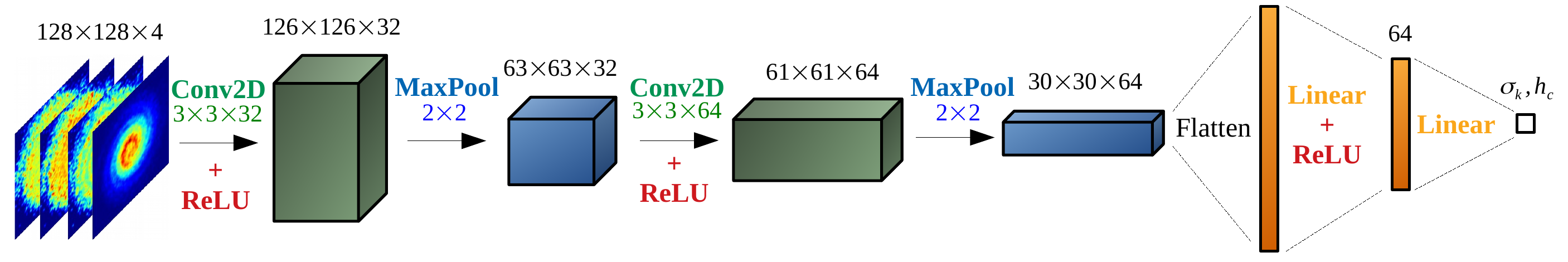}
\caption[Schematic representation of our \acs{CNN} architecture for an input of galactocentric maps]{\label{fig:ch5_CNN_architecture}{Schematic representation of our \acs{CNN} architecture for an input of galactocentric maps with resolution $128 \times 128$ and 4 channels (1 density map plus 3 velocity maps). A module formed by two blocks that each contain a convolution layer and max pooling layer is followed by a fully connected linear network with one hidden layer. The final network output is either a single parameter or two parameters depending on the experiment specifics.}}
\end{figure*}  
%-----------------------------------------------------------------

%%%%%%%%%%%%%%%%%%%%%%%%%%%%%%%%%%%%%%%%%%%%%%%%%%

\subsection{Network architecture}
\label{sec:ch5_net_architecture}

For the implementation of our \acs{ML} pipeline we use \texttt{PyTorch} \citep{Paszke2019}, an optimised tensor library for deep learning using GPUs and CPUs, that is written in \texttt{Python}. The simplest neural network that can be used for this task is a fully connected neural network also referred to as a \acf{MLP} (see Section~\ref{sec:ch2_MLP}) with only two layers of neurons, which are referred to as the input and output layer, respectively. As this is the starting point to develop more complicated and advanced architecture models, we first test how this simple configuration behaves. The number of neurons for the input layer is equal to the total number of input features, i.e., $C \times W \times H$. Here, $C$ is the number of input channels (corresponding to the total number of maps used), while $W$ and $H$ are the number of bins in width and height, respectively. The number of neurons in the output layer is equal to the number of regression parameters that we would like to predict, i.e., 1 or 2 in our experiments. For the activation function we use the Rectified Linear Unit (ReLU) defined as ${\rm ReLU}(x) = \max{(0,x)}$ (see Section~\ref{sec:ch2_activation_function}). To obtain the output values, the ReLU activation function is applied to a linear combination of the input features with weights and a bias.

A more sophisticated model architecture is represented by a \acf{CNN}. \acs{CNN}s are a particular type of deep neural network that have proven to be very successful in regression and classification tasks when applied to structured and matrix-like 2D inputs \citep[see Section~\ref{sec:ch2_CNN} and][for a review]{Rawat2017}. The basic structure of \acs{CNN}s consists of convolutional, pooling, and fully connected layers. Convolutional layers are multi-channel filters that slide along the 2D input maps and are able to extract feature maps. The role of the pooling layer is to down-sample the output of a convolutional layer. This inevitably causes a loss of information but in general helps to improve the training efficiency by increasing the size of the receptive field (i.e., the region of the input that produces the feature for each neuron) and reducing the number of trainable parameters. The fully connected layers collect all the output features from the convolution layers into a 1D input and return the final output prediction.

The detailed structure of the \acs{CNN} we built for our case study can be found in Table~\ref{tab:ch5_CNN_architecture}. A schematic representation of its structure for a 4-channel input with galactocentric maps is also shown in Figure~\ref{fig:ch5_CNN_architecture} as an example. It consists of the following layers:  
\begin{itemize}
    \item A 2D convolution layer with kernel size $3 \times 3$, $C$ input channels, 32 output channels, stride 1 and no padding.
    \item A 2D Max pooling layer of size $2 \times 2$ with stride 2 and no padding.
    \item A 2D convolution filter with kernel size $3 \times 3$, 32 input channels, 64 output channels, stride 1 and no padding.
    \item A 2D Max pooling layer of size $2 \times 2$ with stride 2 and no padding.
    \item A fully connected linear layer with flattened input from the convolutional modules' output and 64 output neurons.
    \item A fully connected linear layer with 64 input neurons and 1 or 2 output neurons (depending on the number of parameters we would like to predict).
\end{itemize}
For the convolutional and pooling layers the stride parameter regulates the amount of displacement in bins with which the filter moves over the map at each step. Padding adds one or more bins at the border of the 2D maps, so that the filters can move and cover the whole map without leaving any bins out. We use a padding of 0 because the borders of the maps do not contain relevant information. 

The choice of this architecture was found by trial and error experiments where we started from a very simple structure and progressively increased the complexity, adding more and more layers to acquire the desired accuracy in predicting the input parameters.

%-------------------------------------------------------------
\begin{table}
\centering
\caption[\acs{CNN} architecture layers]{\acs{CNN} architecture. The total number of input and output features is reported.  $C$ is the number of used channels, while $W$ and $H$ represent the number of bins (i.e. the resolution) in width and height of the density and velocity maps, respectively. Input and output feature numbers have been rounded down to the lower integer.
\label{tab:ch5_CNN_architecture}}
\begin{tabular}{l l l}
\toprule
\tabhead{Layer} &
\tabhead{Input} &
\tabhead{Output} \\
\midrule
Conv2d + ReLU & $C \times W \times H$ & $32 \times (W-2) \times (H-2)$ \\
MaxPool2d & $32 \times (W-2) \times (H-2)$ & $32 \times \left( \frac{W}{2}-1 \right) \times \left( \frac{H}{2}-1 \right)$ \\
Conv2d + ReLU & $32 \times \left( \frac{W}{2}-1 \right) \times \left( \frac{H}{2}-1 \right)$  & $64 \times \left( \frac{W}{2}-3 \right) \times \left( \frac{H}{2}-3 \right)$ \\
MaxPool2d & $64 \times \left( \frac{W}{2}-3 \right) \times \left( \frac{H}{2}-3 \right)$ & $ 64 \times \left( \frac{W}{4}-\frac{3}{2} \right) \times \left( \frac{H}{4}-\frac{3}{2} \right)$ \\
Linear + ReLU & $64 \times \left( \frac{W}{4}-\frac{3}{2} \right) \times \left( \frac{H}{4}-\frac{3}{2} \right)$ & 64 \\
Linear & 64 & 1(2) \\
\bottomrule \\
\end{tabular}

\end{table}
%---------------------------------------------------------------

%%%%%%%%%%%%%%%%%%%%%%%%%%%%%%%%%%%%%%%%%%%%%%%%%%

\subsection{Training process}
\label{sec:ch5_train_process}

For the training of the network, we use the \acf{RMSE} both for the loss function and validation metric, i.e., to compute the distance between the network predictions and the ground truths of the $h_{\rm c}$ and $\sigma_{\rm k}$ parameters. In general, validation occurs at the same time as training and consists of testing the network over a dataset different from the training set. This is needed to asses the ability of the network to generalise what it is learning to an unknown dataset (see Section~\ref{sec:ch2_training_process}). The minimisation of the loss function occurs through gradient descent and backpropagation \citep{Kelley1960, Ruder2017}, i.e., computation of the loss-function gradients with respect to all network parameters (weights and biases). These gradients taken with a negative sign indicate the directions towards which the network parameters should be updated so that the loss is reduced, and hence the network predictions move closer to the true, expected labels. In this regard, a crucial aspect to ensure the best performance of a neural network is to properly initialise the weights and biases (see Section~\ref{sec:ch2_w_initialization}). For this purpose we use the Kaiming initialisation method \citep{Kaiming2015} in order to avoid exploding or vanishing gradients during the training. 

The training process itself is regulated by several hyperparameters. The first one is the learning rate, which is a coefficient for the weight updates. In general, a larger learning rate results in updates of larger magnitude, which could in turn lead to faster convergence, but might also reduce the stability of the training process and thus increase the risk of overshooting the minima of the loss landscape. A second hyperparameter is the batch size, which defines the number of samples that are packed together and passed through the network before an optimisation step is performed. In general, for bigger training datasets, a larger batch size helps to increase the efficiency and stability of the training process, since the gradient-descent steps are averaged over many samples and noise is reduced (see Section~\ref{sec:ch2_forward}). For the gradient-descent optimiser, we use the \acf{Adam} \citep{Kingma2014}. As its name suggests, \acs{Adam} adds an adaptive momentum term to the gradient descent to automatically modify the learning rate and accelerate convergence (see Section~\ref{sec:ch2_backprop}). When using the \acs{Adam} optimiser, the chosen initial value of the learning rate represents only an upper limit. 

We fix the maximum number of learning epochs to 1024. Every epoch the network performs a series of optimisation steps by going through the whole training dataset once. Then epoch-averaged loss and validation metric values are computed. If the validation metric value has improved with respect to the previous epoch the current status of the optimised network is saved. We set an early stop of 128 epochs, so that if the validation metric does not improve over this epoch span, the training process automatically stops and the weights of the best epoch are stored. This prevents the network from overfitting the training samples, which would reduce its ability to generalise over unknown data.

%%%%%%%%%%%%%%%%%%%%%%%%%%%%%%%%%%%%%%%%%%%%%%%%%%
%%%%%%%%%%%%%%%%%%%%%%%%%%%%%%%%%%%%%%%%%%%%%%%%%%

%-----------------------------------------------------------------
\begin{figure}
\centering
\includegraphics[width = 0.47\textwidth]{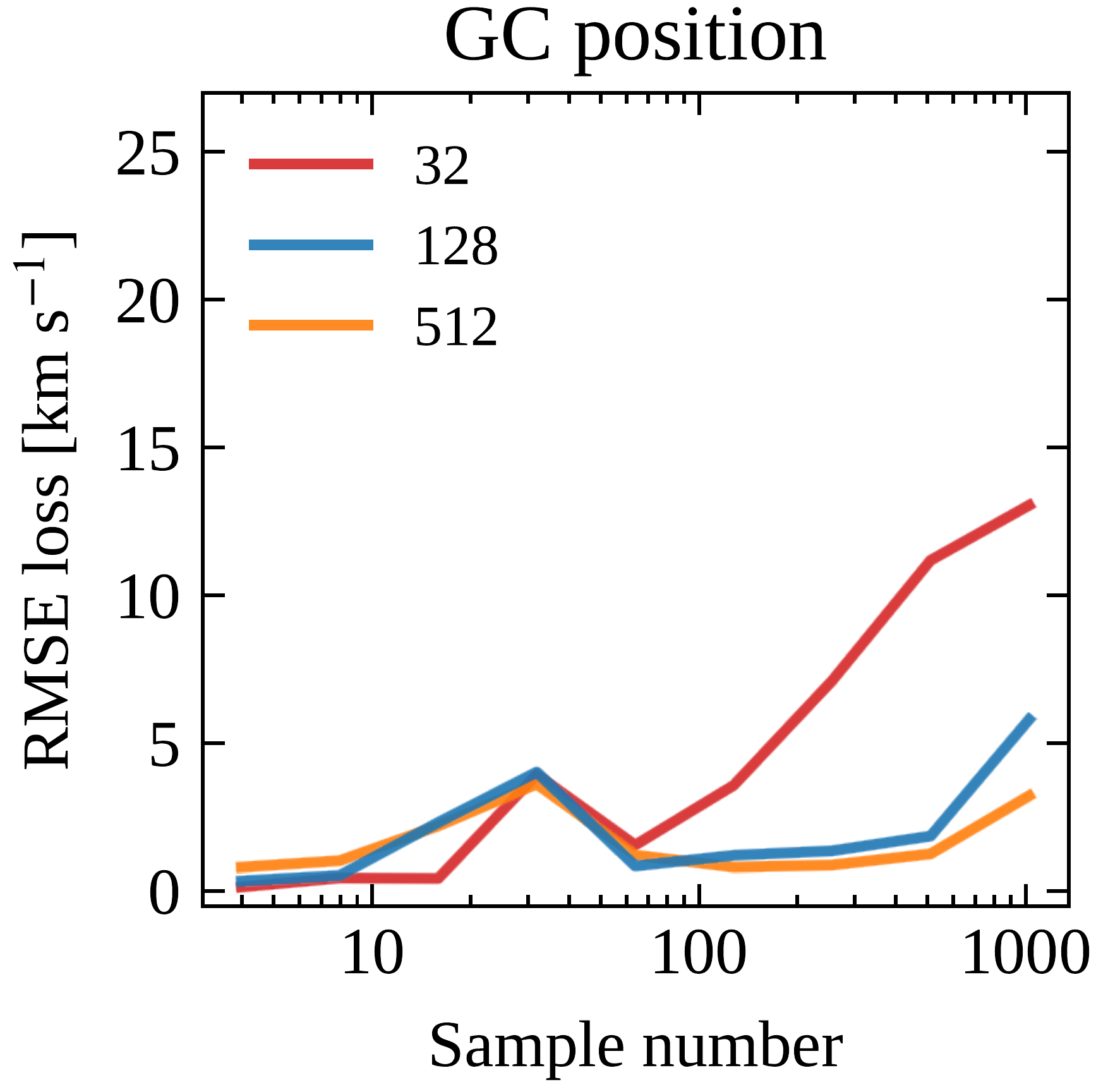}
\includegraphics[width = 0.47\textwidth]{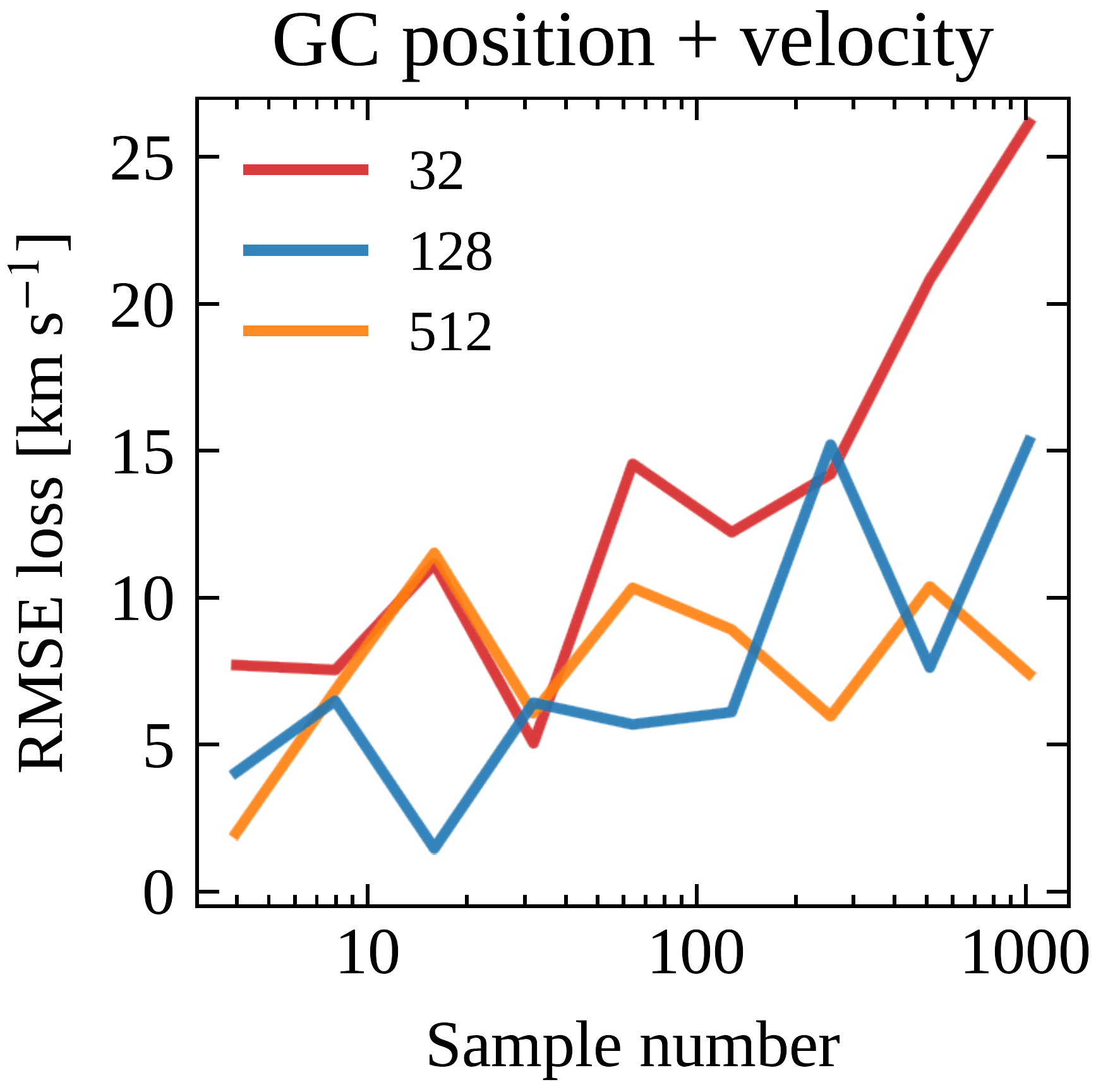}
\includegraphics[width = 0.47\textwidth]{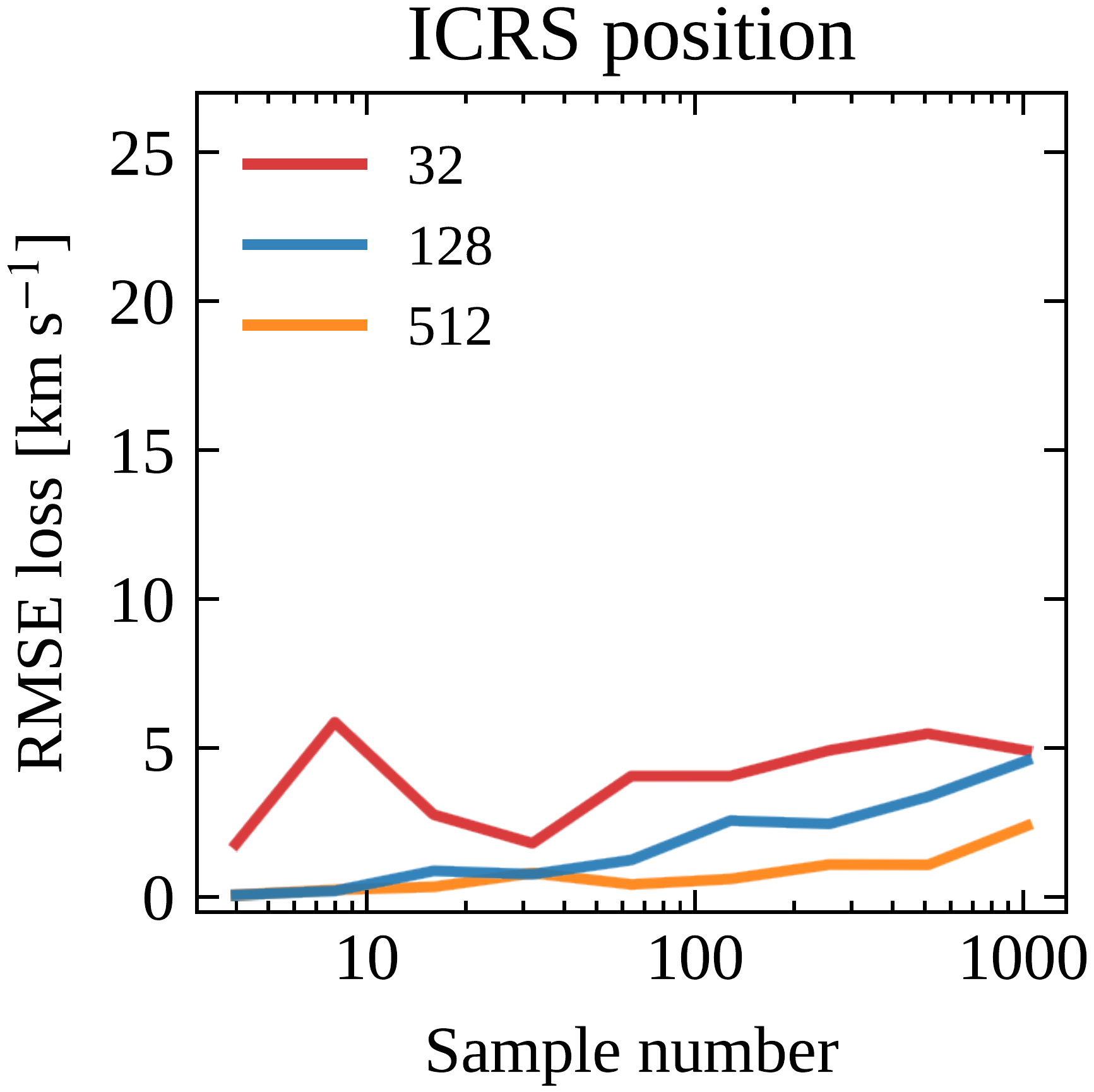}
\includegraphics[width = 0.47\textwidth]{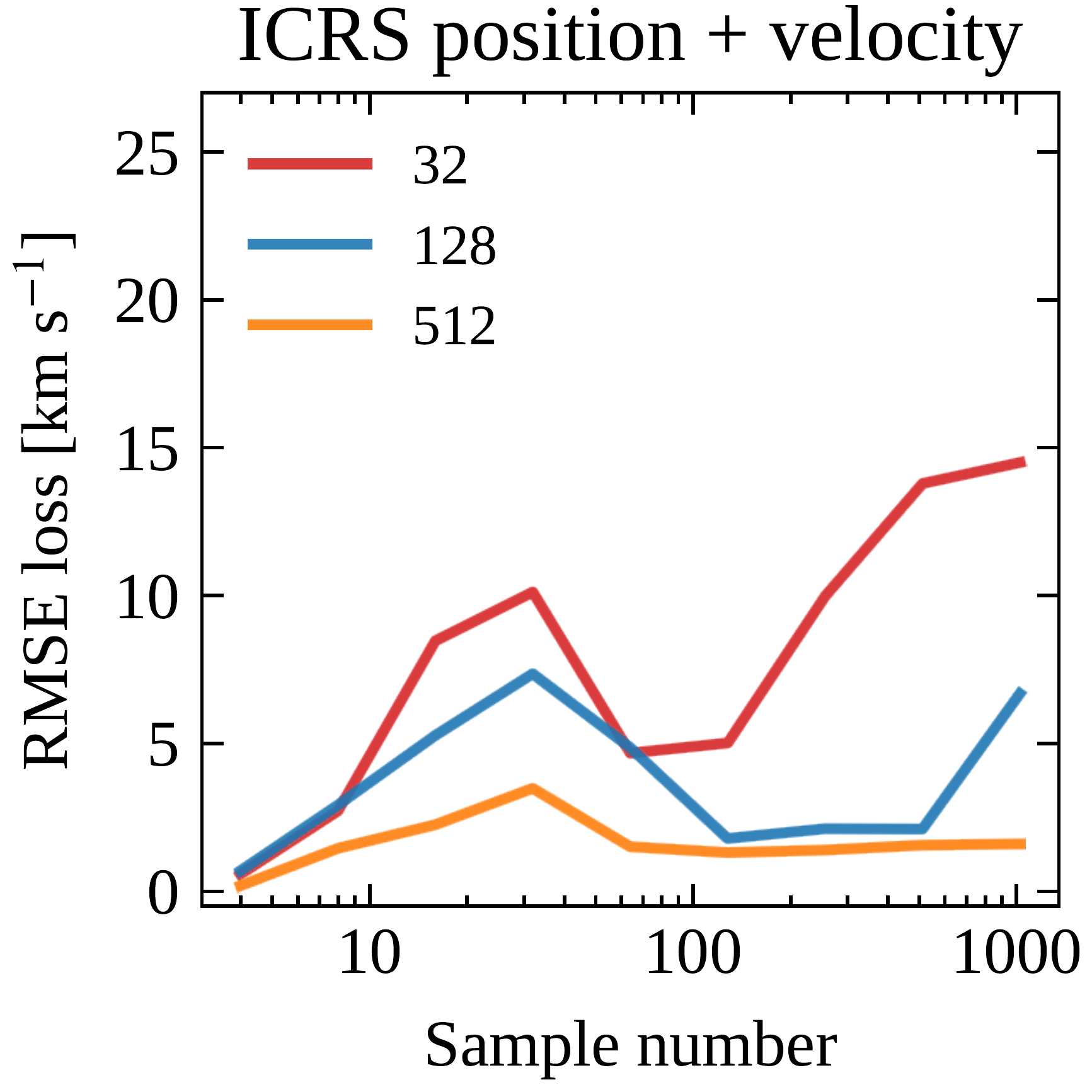}
\caption[Single parameter best training \acs{RMSE} values for the \acs{MLP}]{\label{fig:ch5_lnn_rmse_experiment}{\acs{MLP} best training \acs{RMSE} values for the training process on the single parameter $\sigma_{\rm k}$ of the Maxwell kick-velocity distribution, as a function of the training dataset size and the resolution ({\it red}, {\it blue}, and {\it orange} curves for 32, 128 and 512, respectively) using the four different input configurations T1 (GC position), T2 (GC position + velocity), T3 (ICRS position) and T4 (ICRS position + velocity). }}
\end{figure}  
%-----------------------------------------------------------------
%-----------------------------------------------------------------
\begin{figure}
\centering
\includegraphics[width = 0.47\textwidth]{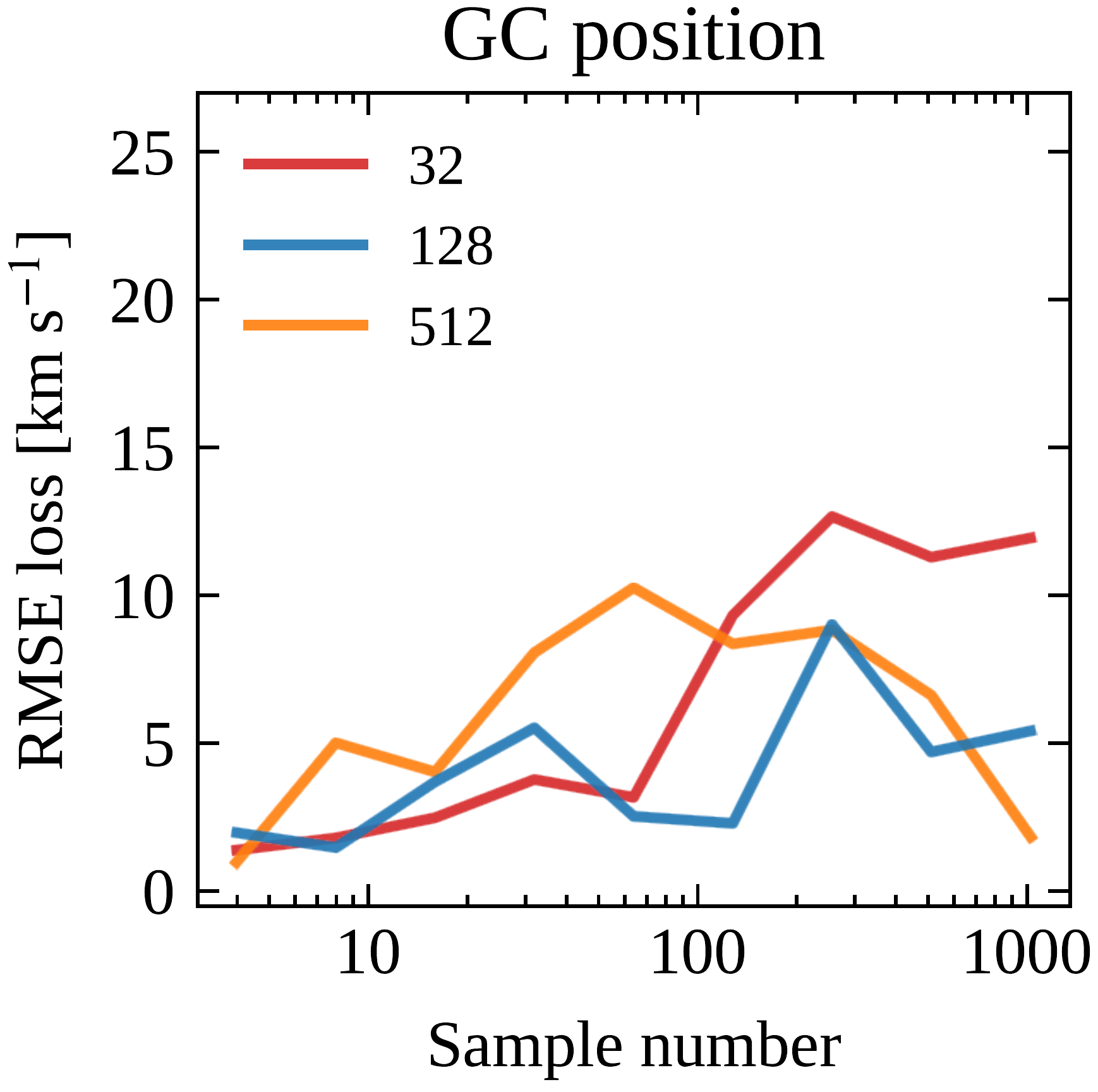}
\includegraphics[width = 0.47\textwidth]{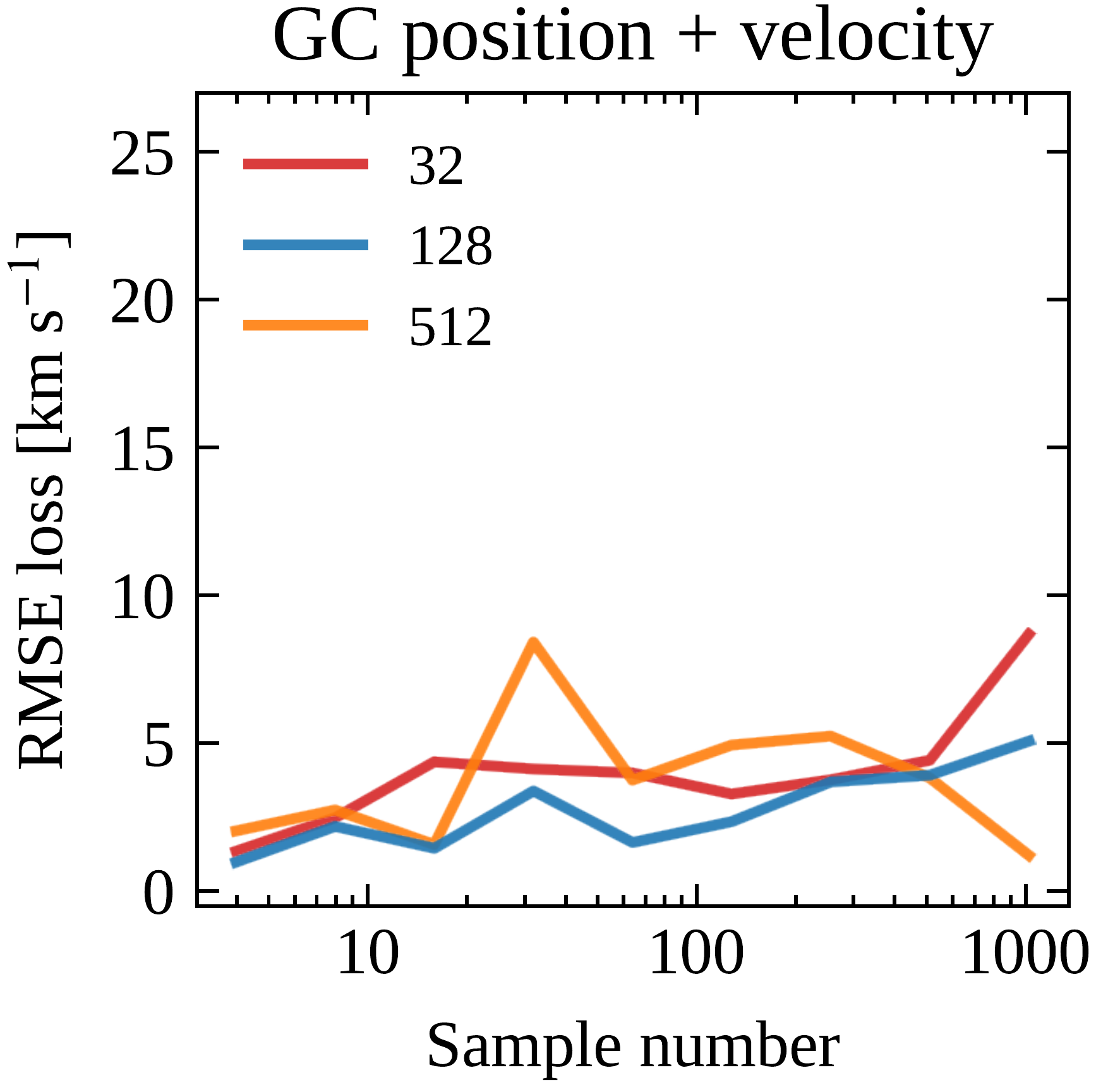}
\includegraphics[width = 0.47\textwidth]{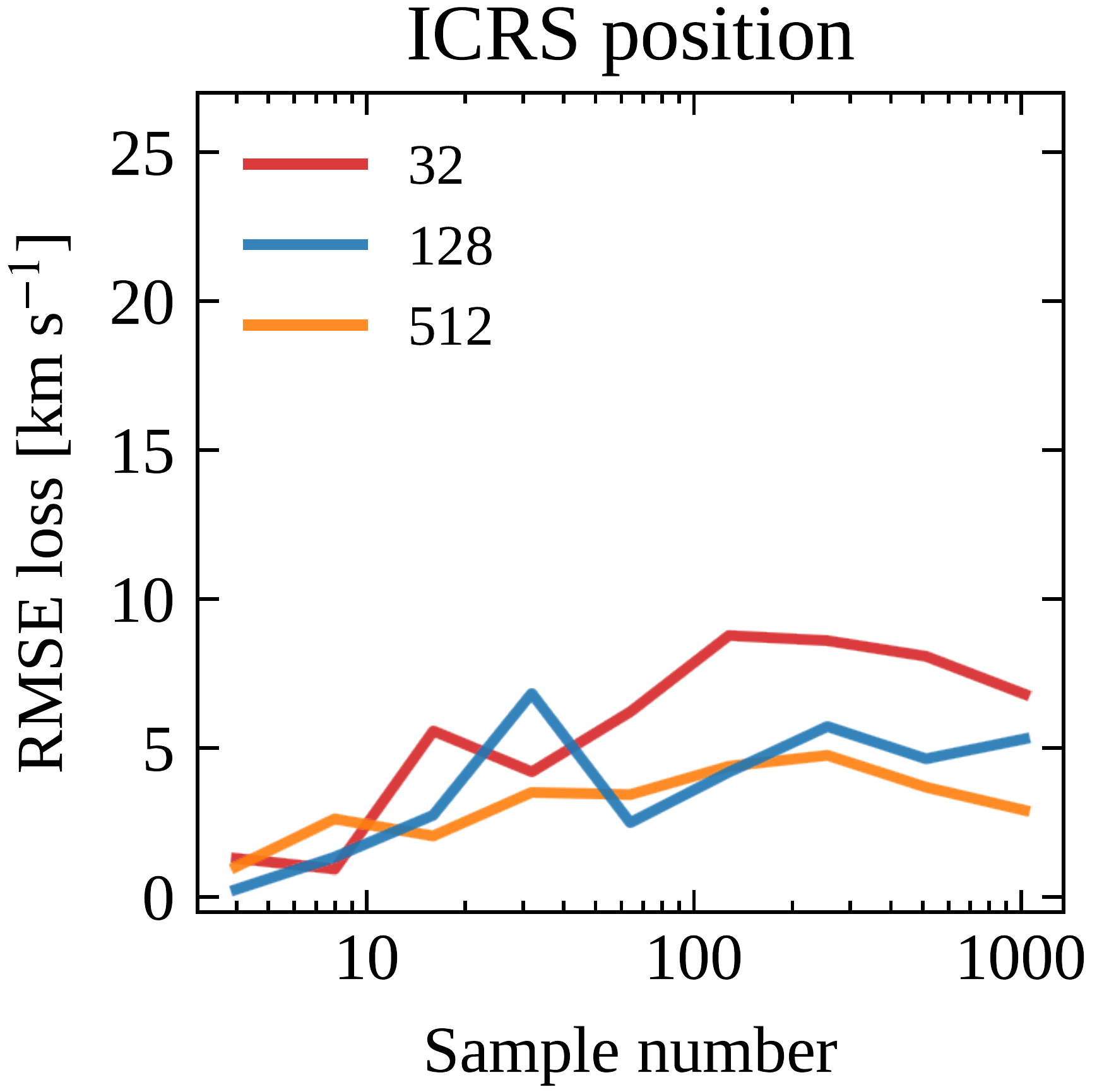}
\includegraphics[width = 0.47\textwidth]{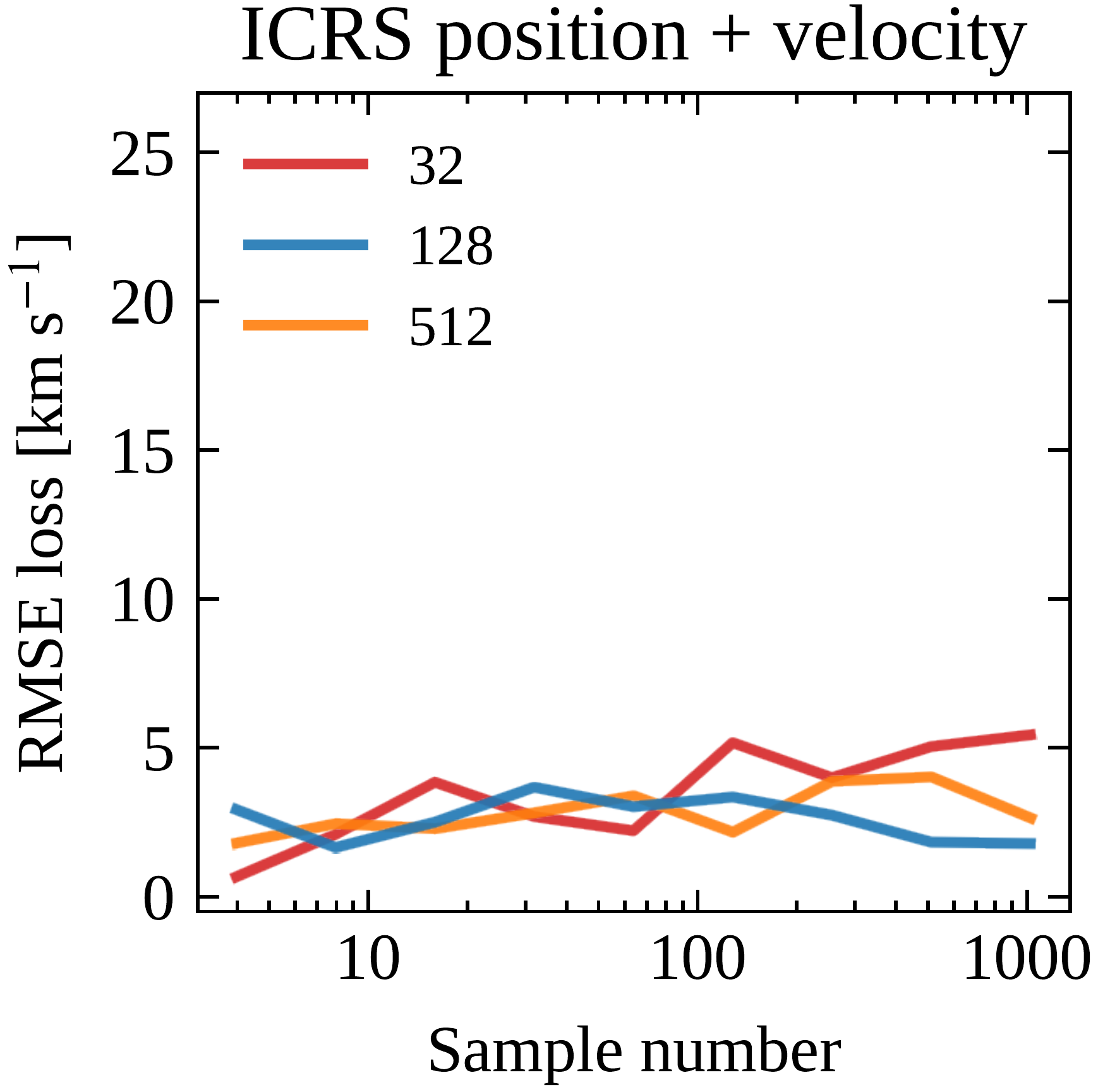}
\caption[Single parameter best training \acs{RMSE} values for the \acs{CNN}]{\label{fig:ch5_cnn_rmse_experiment}{\acs{CNN} best training \acs{RMSE} values for the training process on the single parameter $\sigma_{\rm k}$ of the Maxwell kick-velocity distribution, as a function of the training dataset size and the resolution ({\it red}, {\it blue}, and {\it orange} curves for 32, 128 and 512, respectively) using the four different input configurations T1 (GC position), T2 (GC position + velocity), T3 (ICRS position) and T4 (ICRS position + velocity). }}
\end{figure}  
%-----------------------------------------------------------------

\section{Experiments}
\label{sec:ch5_experiments}

For our machine-learning experiments we use the hardware and software specified in Appendix~\ref{app:hardware}. 
As a first step, we perform several tests to analyze which configuration of the input feature maps provides the best training experience and which proposed neural network architecture, either the \acs{MLP} or the \acs{CNN}, behaves better. Regarding the input configuration, we would like to understand (i) what the best type of maps is (galactocentric vs equatorial ICRS); (ii) how many input channels are needed to obtain good results (do density maps provide enough information alone or does the addition of velocity maps improve the results?); (iii) which resolution of the input maps provides the best result. To do so, we first compare the behaviour of the \acs{MLP} and the \acs{CNN} when trained to predict a single parameter. We explore different types of input signals by varying the resolution of the density and velocity maps as well as the number of input channels. Once we find the optimal configuration for the input maps and the best performing network, we proceed to test its generalisation power. A similar strategy is then followed for the two-parameter prediction.

\subsection{Single-parameter predictions}
\subsubsection{Data-representation and architecture comparison}
\label{sec:ch5_1par_tests}

We focus on predicting the parameter $\sigma_{\rm k}$ of the Maxwell kick velocity function and employ the density and velocity maps generated from simulation run S1. We keep aside the dataset with 20000 samples as it will be used to asses the generalisation power of the best performing network (see Section~\ref{sec:ch2_training_process}). Thus, we have training sets with increasing number of samples (from 4 to 1024) and increasing map resolution (32, 128, 512). For each of these training sets, we compare the performance on four different kinds of input information:
\begin{itemize}
    \item[\textbf{T1}] Galactocentric position information: 1 density map in galactocentric coordinates (1 channel). 
    \item[\textbf{T2}] Galactocentric position and velocity information: 1 density map plus 3 velocity maps in galactocentric coordinates (4 channels).
    \item[\textbf{T3}] ICRS position information: 1 density map in equatorial ICRS coordinates (1 channel).
    \item[\textbf{T4}] ICRS position and velocity information: 1 density map plus 2 proper motion maps in equatorial ICRS coordinates (3 channels).
\end{itemize}
We fix the network architectures and the training set-up as described in Section~\ref{sec:ch5_net_architecture} and Section~\ref{sec:ch5_train_process}. For these initial experiments, we do not incorporate validation but only focus on the training results for different types of datasets, as we are not yet interested in evaluating the generalisation power of the networks. To assess the convergence of the training runs, we set a threshold for the $\sigma_{\rm k}$ \acs{RMSE} training loss to $\unit[10]{km \, s^{-1}}$. If the final \acs{RMSE} value is higher than this threshold the training is repeated up to a maximum of 8 times. If convergence is not reached after 8 trials we take the trained model with the lowest final \acs{RMSE}. For each training experiment, we also monitor the computational time needed to perform a single optimisation step on a single data batch (see Appendix~\ref{app:timing_tests} for more details). 

Initially, we perform a search for the starting value of the learning rate that provides the best results. In particular, for the \acs{MLP} we find that to ensure a decaying \acs{RMSE} value during training, the initial learning rate needs to be decreased as the input-map resolution increases. Therefore, after several tests, we set the initial learning rate to $10^{-4}$, $10^{-5}$ and $10^{-6}$, respectively, for the 32, 128 and 512 resolution maps. The \acs{CNN} instead is more flexible and stable and all three initial learning rates are suitable for every resolution. In this case, we set it to $10^{-4}$ to obtain training convergence in the smallest number of epochs.

Next, we tune the batch size to make the learning process more stable and efficient as the number of training samples increases. In general, we keep the batch size to 1 for the dataset sizes 4, 8, 16, 32 and progressively increase it to 4, 8, 16, 32, 64 as the dataset size increases to 64, 128, 256, 512, 1024, respectively. Only when testing the performance of the \acs{MLP} on the T2 input configuration, we need to further fine tune the batch size in order to reach acceptable values of the \acs{RMSE}. In all other cases, the batch sizes mentioned above work well. 

%-----------------------------------------------------------------
\begin{sidewaysfigure}
\centering
\includegraphics[width = 0.22\textwidth]{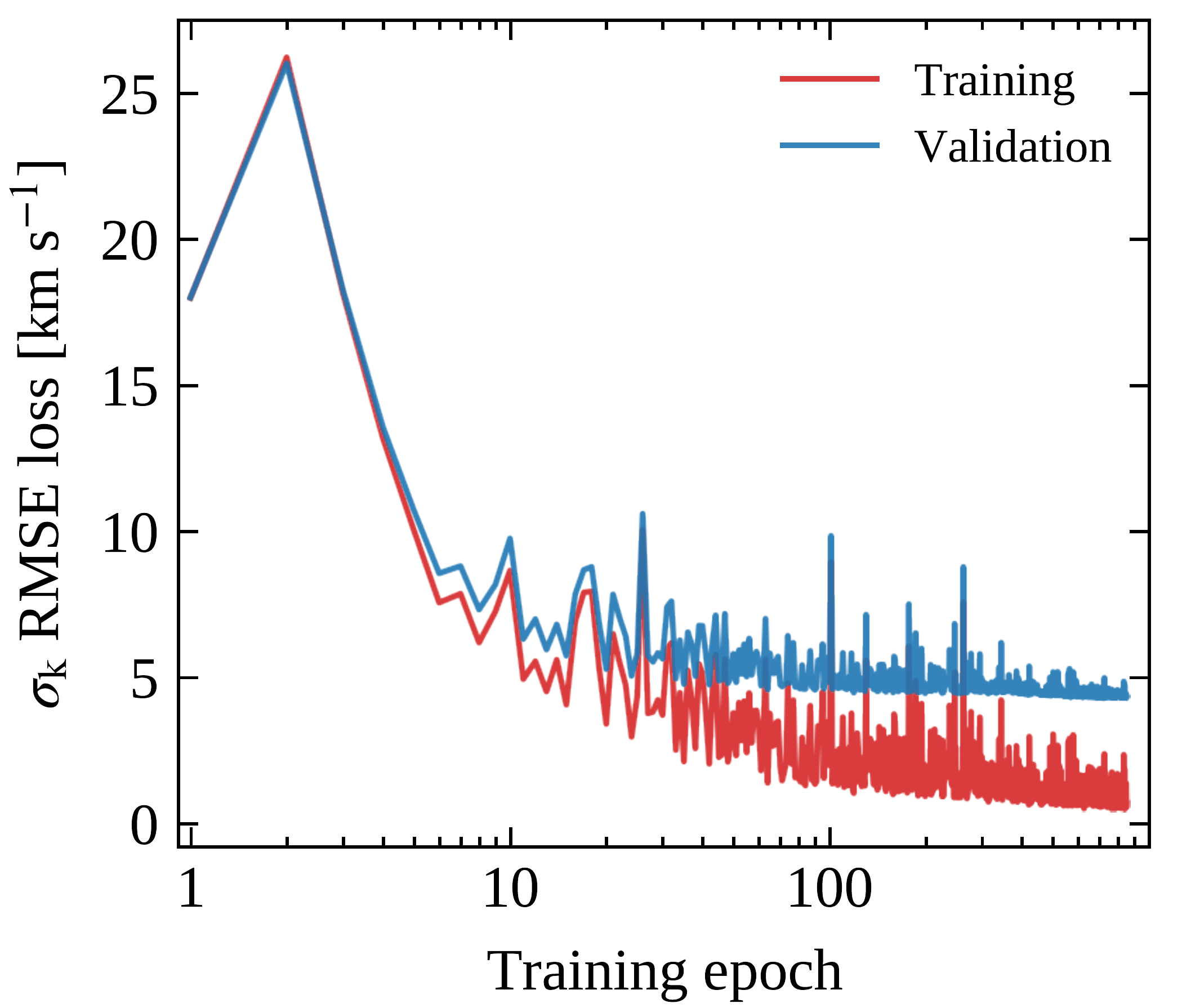}
\includegraphics[width = 0.22\textwidth]{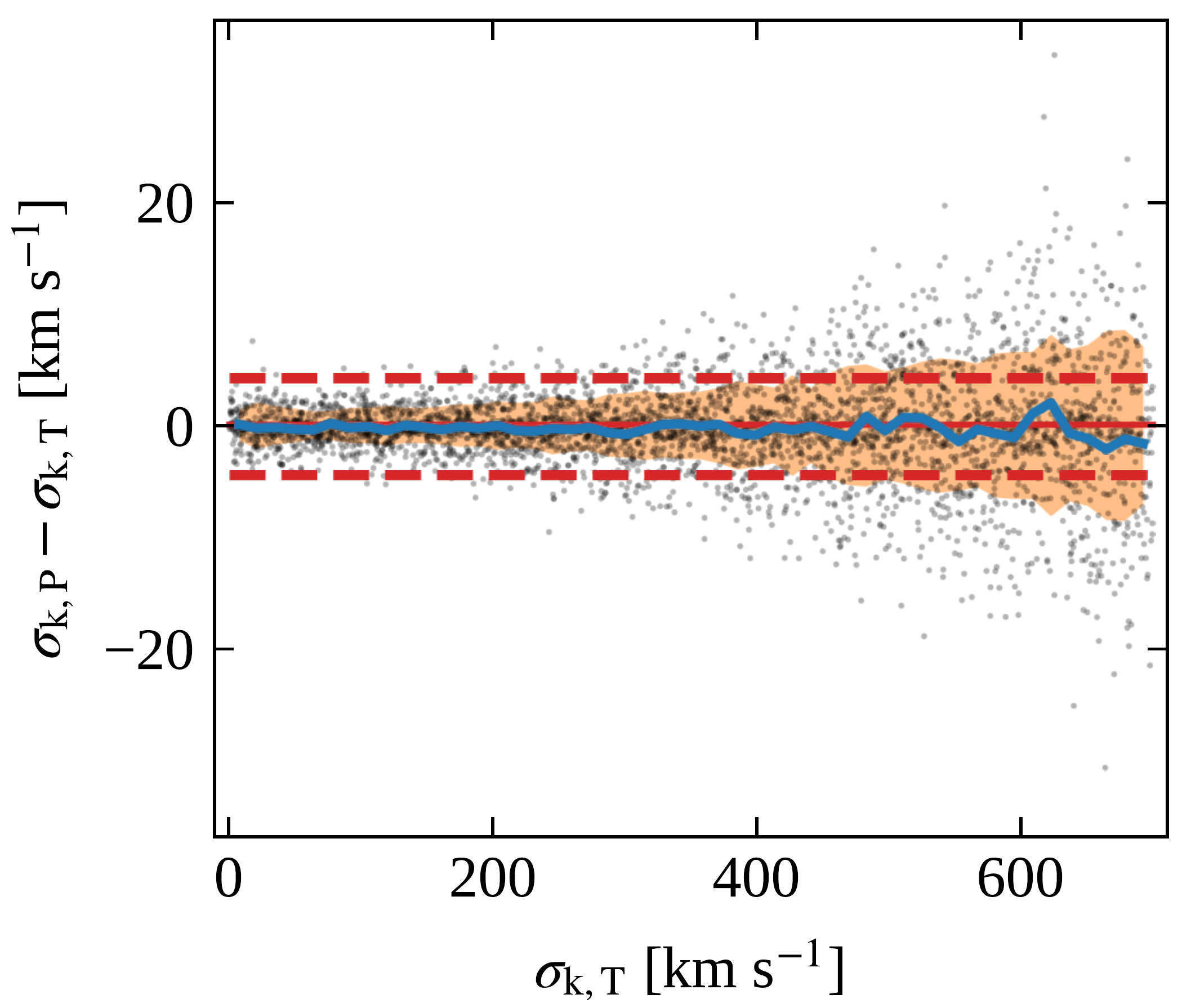}
\includegraphics[width = 0.22\textwidth]{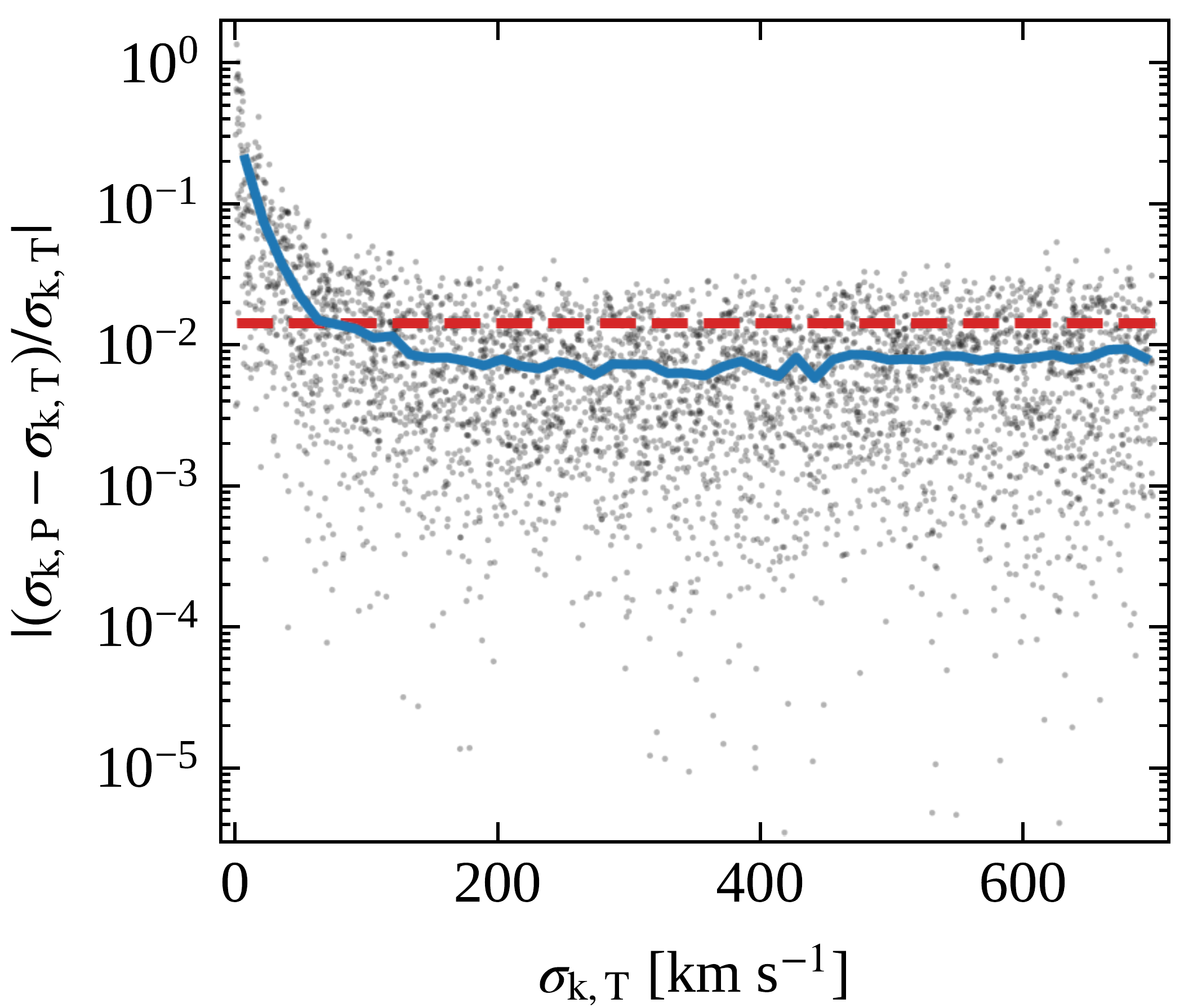}\\
\vspace{0.25cm}
\includegraphics[width = 0.22\textwidth]{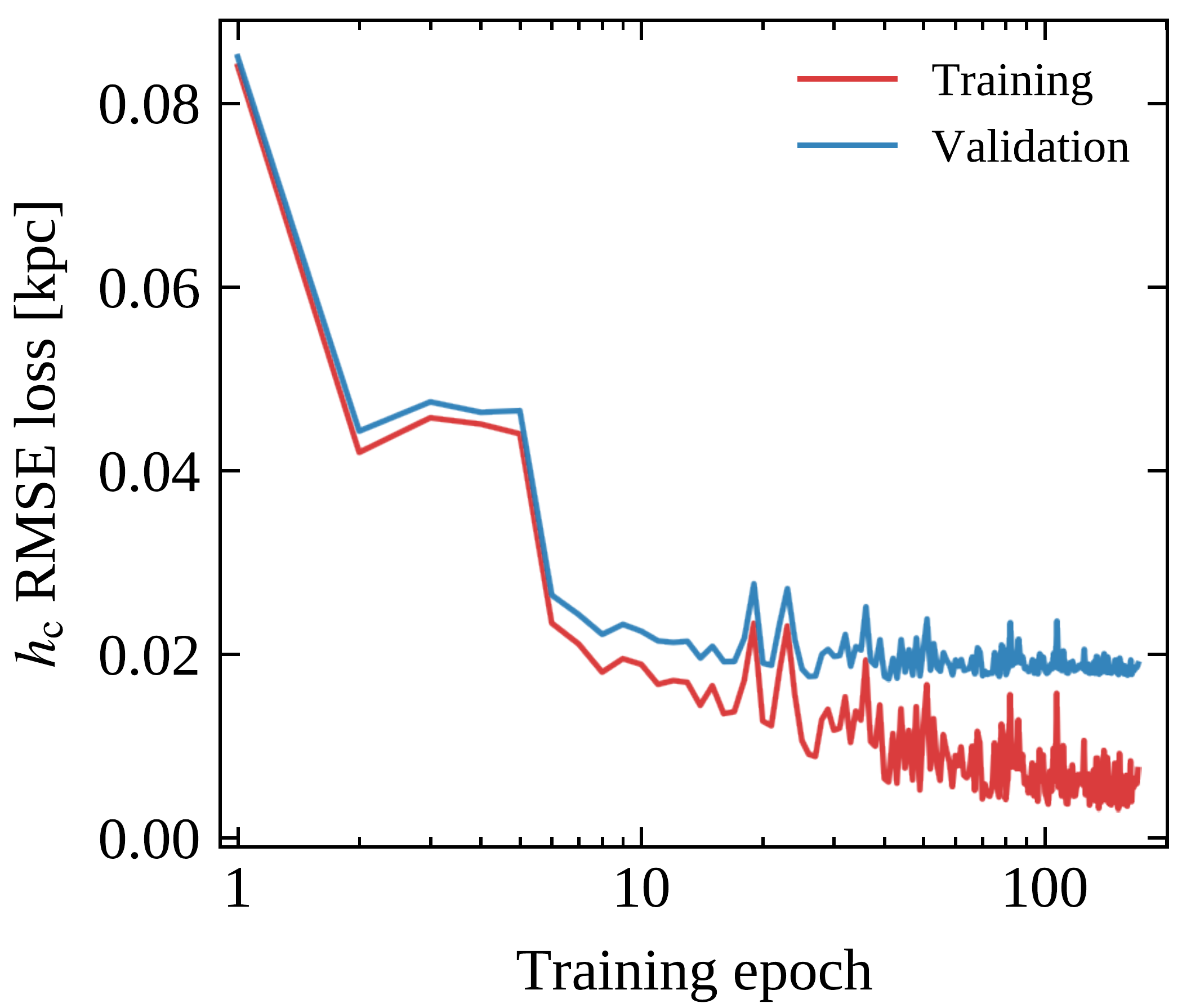}
\includegraphics[width = 0.22\textwidth]{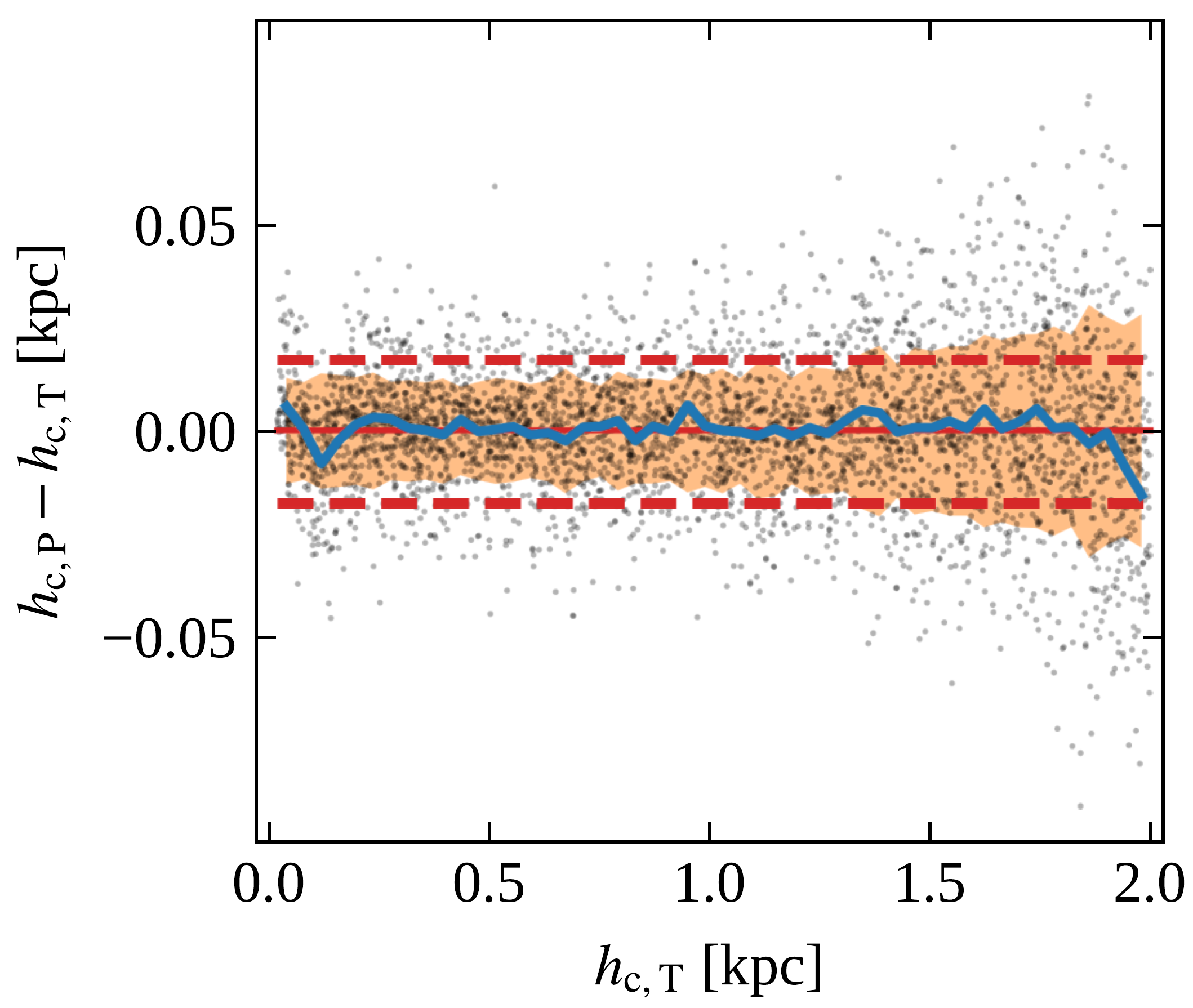}
\includegraphics[width = 0.22\textwidth]{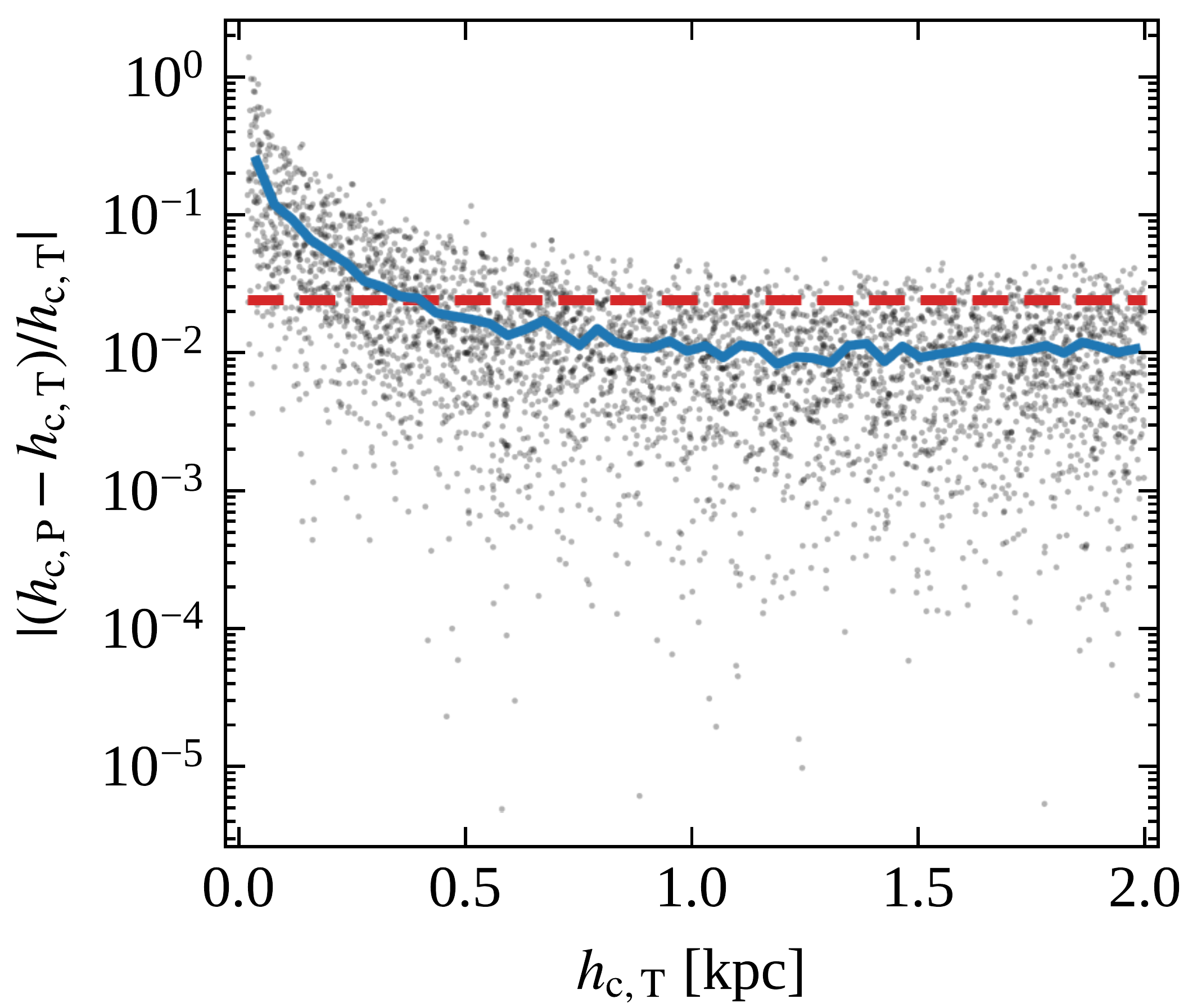}
\caption[Results of the \acs{CNN} single-parameter prediction for the kick-velocity parameter]{Results of the \acs{CNN} single-parameter prediction for the kick-velocity parameter, $\sigma_{\rm k}$, and scale height, $h_{\rm c}$, for the corresponding validation sets. {\it Top (bottom) left panel}: evolution of the \acs{RMSE} training ({\it red}) and validation ({\it blue}) losses as a function of the training/validation epoch for the $\sigma_{\rm k}$ ($h_{\rm c}$) parameters. {\it Top (bottom) central panel}: residuals of the prediction as a function of the target $\sigma_{\rm k}$ ($h_{\rm c}$) value; the subscripts P and T refer to the predicted and target values, respectively. The red dashed lines delimit the $68$\% uncertainty region corresponding to a \acs{RMSE} $=\unit[4.4]{km \, s^{-1}}$ ($\unit[0.017]{kpc}$) computed over the whole range $[1, 700]$ $\unit[]{km \, s^{-1}}$ ($[0.02, 2]$ $\unit[]{kpc}$). The {\it orange} region delimits the $68$\% uncertainty region computed as a running \acs{RMSE} for which we have divided the full $\sigma_{\rm k}$ ($h_{\rm c}$) range into $50$ bins of equal size. The {\it blue} line shows the trend of the average residuals which are well centred around the value 0. {\it Top (bottom) right panel}: relative error of the prediction as a function of the target $\sigma_{\rm k}$ ($h_{\rm c}$) value. The red dashed line corresponds to a \acs{MRE} $=0.014$ ($0.024$) computed over the whole range $[1, 700]$ $\unit[]{km \, s^{-1}}$ ($[0.02, 2]$ $\unit[]{kpc}$). The {\it blue} line shows the trend of the running \acs{MRE} computed over $50$ bins of equal size into which we have divided the full $\sigma_{\rm k}$ ($h_{\rm c}$) range.}
\label{fig:ch5_1par_cnn_inference_sigmak_hc}
\end{sidewaysfigure}  
%-----------------------------------------------------------------

The results of our experiments using the \acs{MLP} and the \acs{CNN} on the training datasets from S1 with configuration T1, T2, T3 and T4 are shown in Figures~\ref{fig:ch5_lnn_rmse_experiment} and \ref{fig:ch5_cnn_rmse_experiment}, respectively (see Appendix \ref{app:timing_tests} for the timing results), highlighting the best converged \acs{RMSE} values as a function of the training dataset size and the resolution.

First of all, we note that for both model architectures, ICRS maps allow us to obtain slightly better results in terms of the best \acs{RMSE} values. An explanation for this could be that only the ICRS maps contain 3D information of the Galaxy, i.e., they encode the stellar height distribution with respect to the Galactic disk which correlates with the kick-velocity magnitude. In fact, the higher the kick velocity of newborn neutron stars the more spread out their distribution in galactic height $z$ at the end of the dynamical evolution. On the other hand, galactocentric maps show the Galaxy represented face on and the information on the stars' height $z$ with respect to the Galactic plane is hidden. Therefore, it is likely easier for the networks to distinguish populations with different $\sigma_{\rm k}$ by processing the ICRS maps.

We also note that the addition of the velocity or proper motion information has opposite effects on the two model architectures. In particular, on one side it worsens the \acs{MLP} performance, as the best \acs{RMSE} almost doubles. On the other side, it helps the \acs{CNN} in reducing the overall \acs{RMSE}.
We interpret this as an indication that, as the complexity of the input data increases, a deeper (with more layers) and more sophisticated model architecture like the \acs{CNN} is more suitable to process the input data and extract meaningful features to perform the regression task. However, the improvement is not dramatic, which could suggest that the density maps already provide enough information to distinguish, with sufficient precision, populations simulated with different $\sigma_{\rm k}$ values.

Concerning the resolution, we observe that higher resolutions allow us to reach slightly lower \acs{RMSE} values with both the \acs{MLP} and the \acs{CNN} but at the expense of longer computation time (see Figures~\ref{fig:app_lnn_time_experiment} and \ref{fig:app_cnn_time_experiment} in Appendix~\ref{app:timing_tests}). We therefore consider the small differences between the best \acs{RMSE} values obtained with 128 and 512 resolutions not sufficient to justify the choice of the higher resolution. Hence, the use of the 128 resolution appears to be a good compromise to ensure good accuracy and reasonably fast training.

In light of these results for the single-parameter estimation, we conclude that the optimal representation to be used for training is composed of the ICRS density plus proper motion maps with 128 resolution. Moreover, as the \acs{CNN} obtains the best results and appears more stable and flexible when compared to the simple \acs{MLP} (especially for the multi-channel input features), we employ our \acs{CNN} architecture in the following experiments, which should also be less prone to overfitting.

%%%%%%%%%%%%%%%%%%%%%%%%%%%%%%%%%%%%%%%%%%%%%%%%%%

\subsubsection{Generalisation Results} 
\label{sec:ch5_1par_generalization_result}

As the next step, we separately train the \acs{CNN} to predict the $\sigma_{\rm k}$ and $h_{\rm c}$ parameters by using the two big datasets with 20000 simulations each (see S1 and S2). As input features, we use the 3-channel ICRS representation with one density map and two proper motion maps with 128 resolution that ensure the best results as suggested by our earlier experiments. We randomly split both datasets into training and validation subsets with a relative percentage of $80/20\%$, respectively. Therefore, training is performed over 16000 simulations, randomly sampled from the entire datasets, while validation is performed over the remaining 4000 simulations. We adopt an initial learning rate of $10^{-4}$, a batch size of 64, set the total number of learning epochs to 1024 and an early stop at 128 epochs to avoid overfitting. 
We set the convergence threshold for $\sigma_{\rm k}$ to $\unit[10]{km \, s^{-1}}$ and for $h_{\rm c}$ to $\unit[0.5]{kpc}$.
The evolution of the training and validation losses is shown in the left panels of Figure~\ref{fig:ch5_1par_cnn_inference_sigmak_hc}. 

The predictions of the trained network on the validation set for $\sigma_{\rm k}$ and $h_{\rm c}$ are summarised in the central panels of Figure~\ref{fig:ch5_1par_cnn_inference_sigmak_hc} and Table~\ref{tab:ch5_genralization_results}. In the first case, the network is able to predict the value of the kick-velocity parameter $\sigma_{\rm k}$ for the simulations in the validation dataset with a \acs{RMSE} uncertainty of $\unit[4.4]{km \, s^{-1}}$, computed over the whole range $[1, 700]$ $\unit[]{km \, s^{-1}}$. This is indicated by the red dashed lines in the residuals plot (see top central panel of Figure~\ref{fig:ch5_1par_cnn_inference_sigmak_hc}), which delimit the $68$\% uncertainty region. In the second case, the network is able to predict the value of the scale-height parameter $h_{\rm c}$ for the simulations in the validation dataset with a \acs{RMSE} uncertainty of $\unit[0.017]{kpc}$, computed over the whole range $[0.02, 2] $ $\unit[]{kpc}$ (see bottom central panel of Figure~\ref{fig:ch5_1par_cnn_inference_sigmak_hc}). Note that in both experiments, the residuals spread out as the target values increase. We visualise this by computing a running \acs{RMSE} with increasing target values. As is shown by the orange regions in the residuals plots, the running \acs{RMSE} increases from $1.4$ to $\unit[7.1]{km \, s^{-1}}$ for $\sigma_{\rm k}$ and from $0.013$ to $\unit[0.028]{kpc}$ for $h_{\rm c}$, respectively. However, the average residuals are consistent with $0$ over the entire target value ranges as marked by the blue line in the residual plots, showing no anomalous trends in the predicted values.

In the right panels of Figure~\ref{fig:ch5_1par_cnn_inference_sigmak_hc}, we also show the relative error to highlight the precision with which the network is able to predict the $\sigma_{\rm k}$ and $h_{\rm c}$ parameters for a given target value. We observe that the precision of the predictions improves with increasing $\sigma_{\rm k}$ and $h_{\rm c}$ and eventually stabilises to a relative error of around $0.01$. This is highlighted by the blue lines, which show the trend of the \acf{MRE} as a function of the target parameters. The red dashed lines correspond instead to the \acs{MRE} computed on the whole parameter ranges and are equal to $0.014$ and $0.024$ for $\sigma_{\rm k}$ and $h_{\rm c}$, respectively. The fact that the precision of the predictions decreases at the lower end of the target ranges suggests that (for our chosen network set-up) the input maps become harder to distinguish as the neutron stars' initial kick velocities and their galactic birth heights decrease in magnitude.

We then evaluate the generalisation capability of the two trained networks on the corresponding test-sets with 1000 samples each. We find RMSEs of $\unit[4.8]{km \, s^{-1}}$ and $\unit[0.019]{kpc}$ and MREs of $0.018$ and $0.029$ for $\sigma_{\rm k}$ and $h_{\rm c}$, respectively. To further assess the confidence intervals of both estimators, we evaluate the RMSE and the MRE of the parameter values predicted by the two networks over 1000 bootstrapped sets of the related test sets. We find that the RMSE variation is around $3$\%, while the MRE variation is around $11$\% for both predicted parameters. These results indicate that the trained networks are able to generalise well over unseen datasets.

%%%%%%%%%%%%%%%%%%%%%%%%%%%%%%%%%%%%%%%%%%%%%%%%%%

\subsection{Two-parameter predictions}
\subsubsection{Data-representation comparison}

To see how the \acs{CNN} behaves when two parameters, i.e., the kick-velocity parameter $\sigma_{\rm k}$ and the characteristic scale height $h_{\rm c}$, are inferred simultaneously, we use the dataset of maps generated from simulation run S3 (see Section~\ref{sec:ch5_dataset_creation}). In this case, we have training sets with increasing number of samples, 16 = 4x4, 64 = 8x8, 256 = 16x16, 1024 = 32x32, 4096 = 64x64. We leave aside the largest dataset with 16384 = 128x128 simulations for our final generalisation experiment. Given the results of the single-parameter training experiments, we choose the 128 resolution maps and compare the \acs{CNN}'s performance on the four kinds of input information T1, T2, T3 and T4 as we did for the single-parameter case (see Section~\ref{sec:ch5_1par_tests}). We fix the \acs{CNN} architecture and the training set-up as described in Sections~\ref{sec:ch5_net_architecture} and~\ref{sec:ch5_train_process}, respectively. As for the single-parameter comparison, we only focus on the training behaviour for this initial comparison. We keep track of the \acs{RMSE} training losses for both parameters separately, but training proceeds by minimising the total loss computed on both parameters. We set the convergence threshold for $\sigma_{\rm k}$ to $\unit[10]{km \, s^{-1}}$ and for $h_{\rm c}$ to $\unit[0.5]{kpc}$; if convergence is not reached for both parameters after 8 training trials we quote the experiment with the best performance. As before, we also keep track of the computational time for a single optimisation step (see Appendix \ref{app:timing_tests}). The initial learning rate is set to $10^{-4}$, while the batch size is changed according to the dataset size. In particular, we use a batch size of 1, 4, 16, 64, 128 for the dataset sizes 16, 64, 256, 1024 and 4096, respectively. 

The results of our experiments using the \acs{CNN} for the two-parameter prediction on the training datasets from S3, with configuration T1, T2, T3 and T4 are summarised in Figure~\ref{fig:ch5_2par_cnn_tests} (see Appendix \ref{app:timing_tests} for the timing results). As in the single-parameter case, we find that the best results are provided by the ICRS maps, which allow us to reach the lowest training \acs{RMSE} losses for both parameters. In particular, for the ICRS 3-channel input, the \acs{CNN} is able to reach a training \acs{RMSE} loss $\lesssim \unit[5]{km \, s^{-1}}$ for the $\sigma_{\rm k}$ parameter, comparable with the single-parameter case. For $h_{\rm c}$, the \acs{CNN} reaches a training \acs{RMSE} $\lesssim \unit[0.1]{kpc}$. However, we note a drop in accuracy for the $\sigma_{\rm k}$ parameter when only the ICRS density maps are used. As already mentioned in Section~\ref{sec:ch5_1par_tests}, information on the stars' $z$-coordinate is encoded in the ICRS maps. As we simultaneously vary the initial kick velocities and the galactic heights of the pulsars' birth places, the degeneracy between the effects of these two parameters becomes relevant. This makes a distinction of the impact of one parameter over the other more difficult for the network, when only ICRS density maps are provided (see also Section~\ref{sec:ch5_2par_generalization_result}). Adding two extra channels that contain information about the stars' proper motion thus helps to improve the accuracy on the kick-velocity parameter estimation. The results of these initial explorations are promising and indicate that our simple \acs{CNN} architecture has good predictive power for both parameters, if provided with all three input channels in the ICRS representation.

%-----------------------------------------------------------------
\begin{figure}
\centering
\includegraphics[width = 0.40\textwidth]{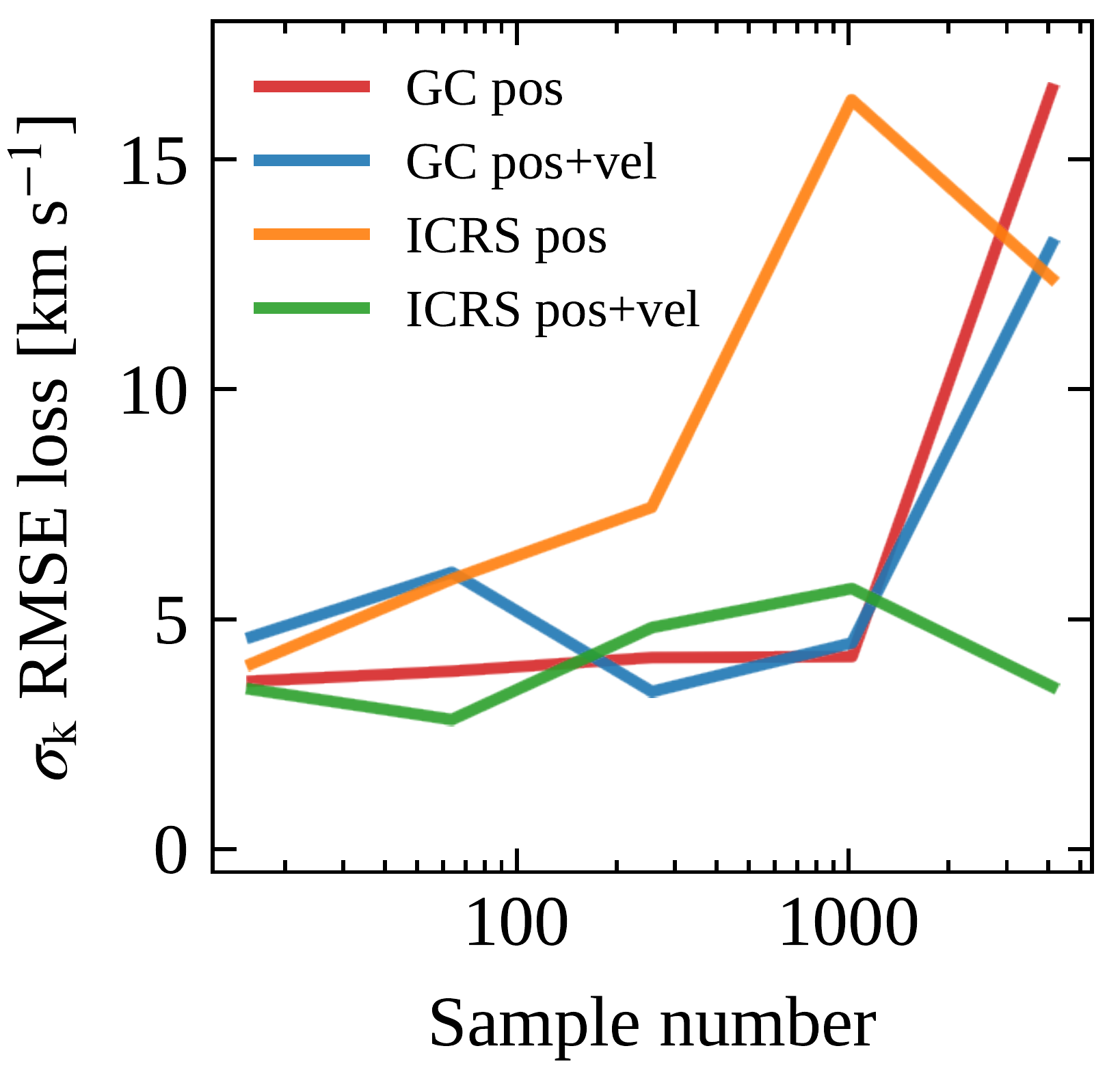}
\includegraphics[width = 0.40\textwidth]{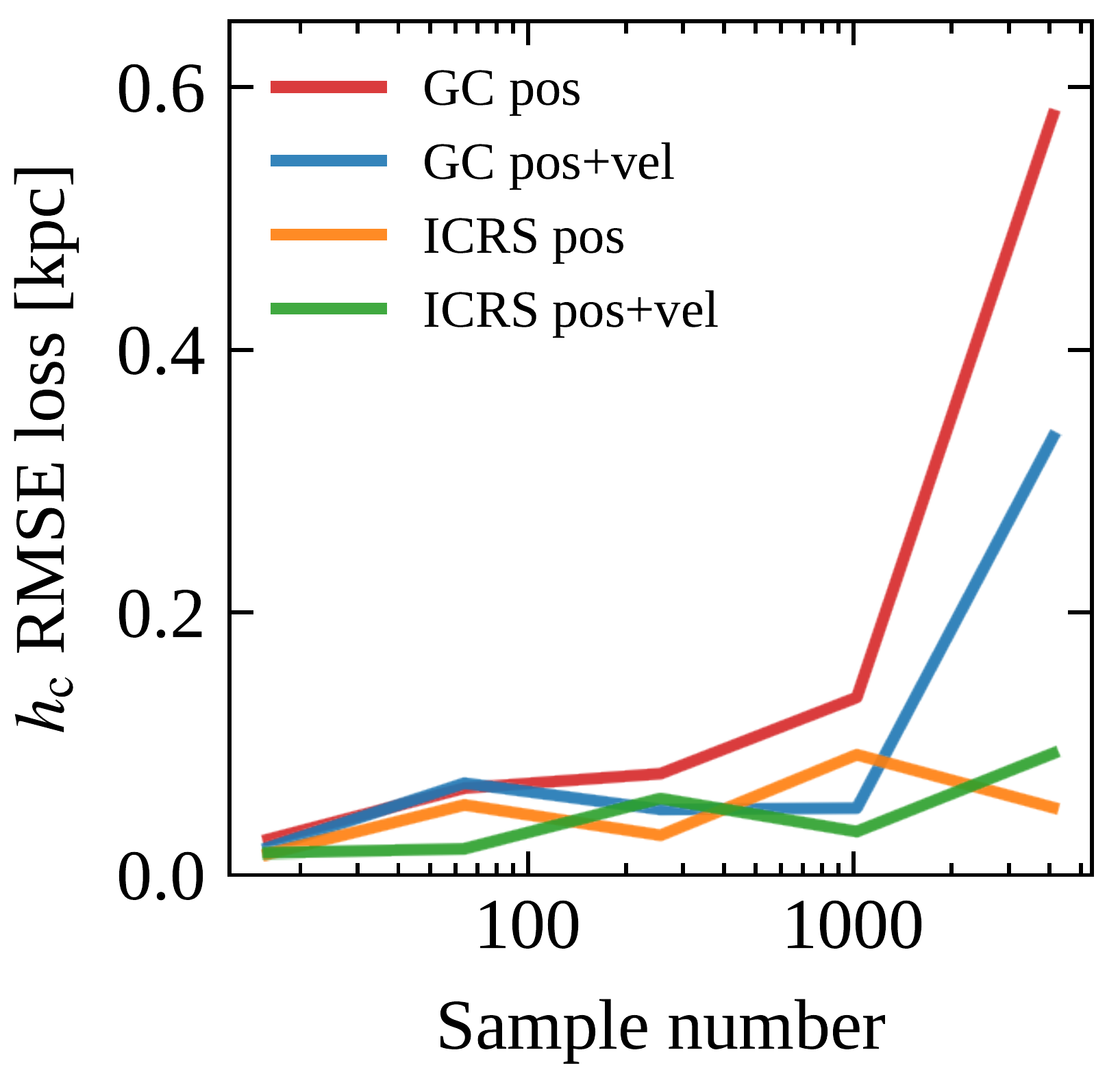}
\caption[Two-parameters best training \acs{RMSE} values for the \acs{CNN}]{\label{fig:ch5_2par_cnn_tests}{\acs{CNN} best training \acs{RMSE} values for the training process on the two parameters $\sigma_{\rm k}$ of the Maxwell kick-velocity distribution (\textit{left panel}) and characteristic scale height $h_{\rm c}$ (\textit{right panel}) as a function of the training dataset size for the four different input configurations T1 (GC position), T2 (GC position + velocity), T3 (ICRS position) and T4 (ICRS position+ velocity).}}
\end{figure}  
%-----------------------------------------------------------------

%-----------------------------------------------------------------
\begin{sidewaysfigure}
\centering
\includegraphics[width = 0.22\textwidth]{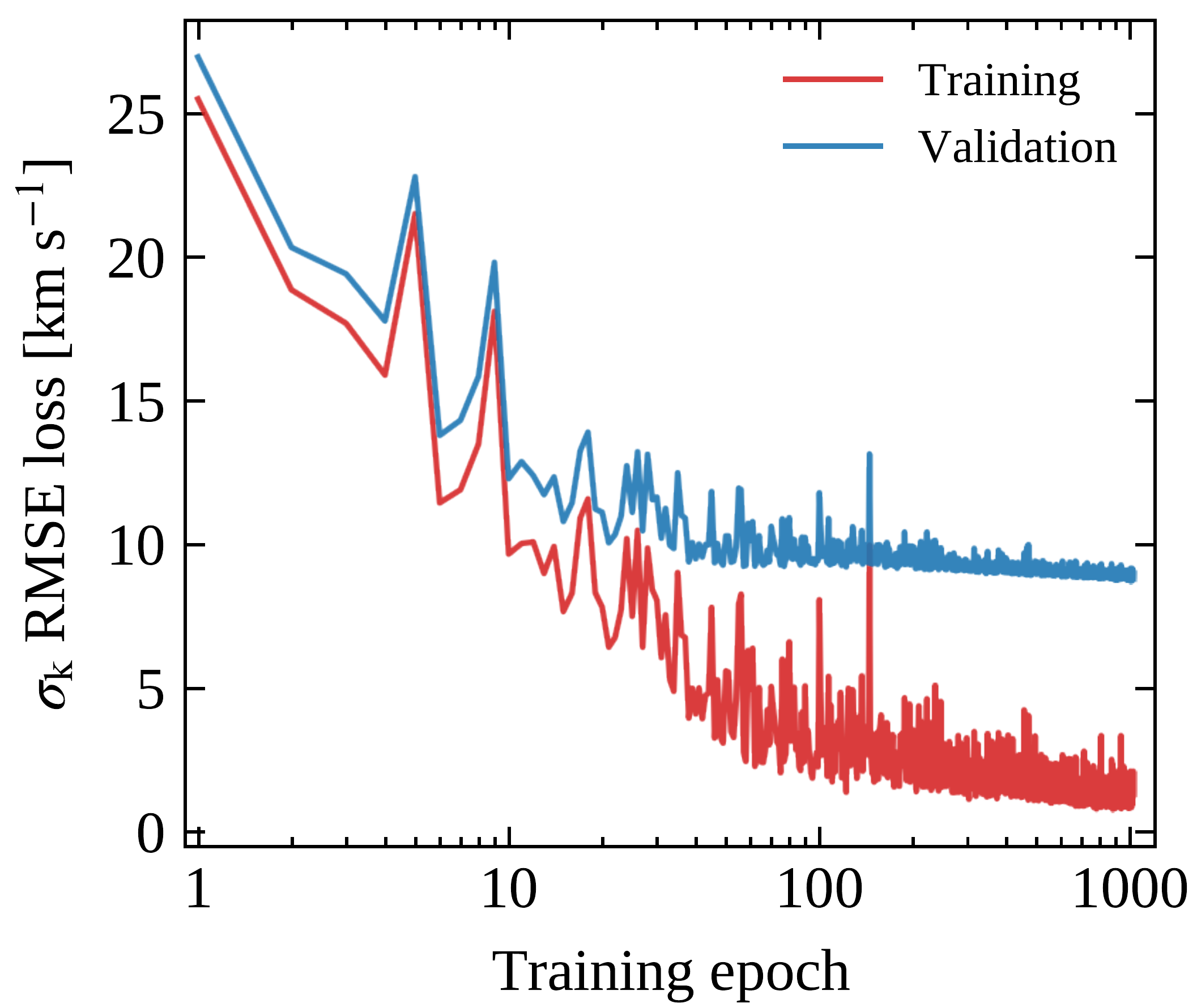}
\includegraphics[width = 0.22\textwidth]{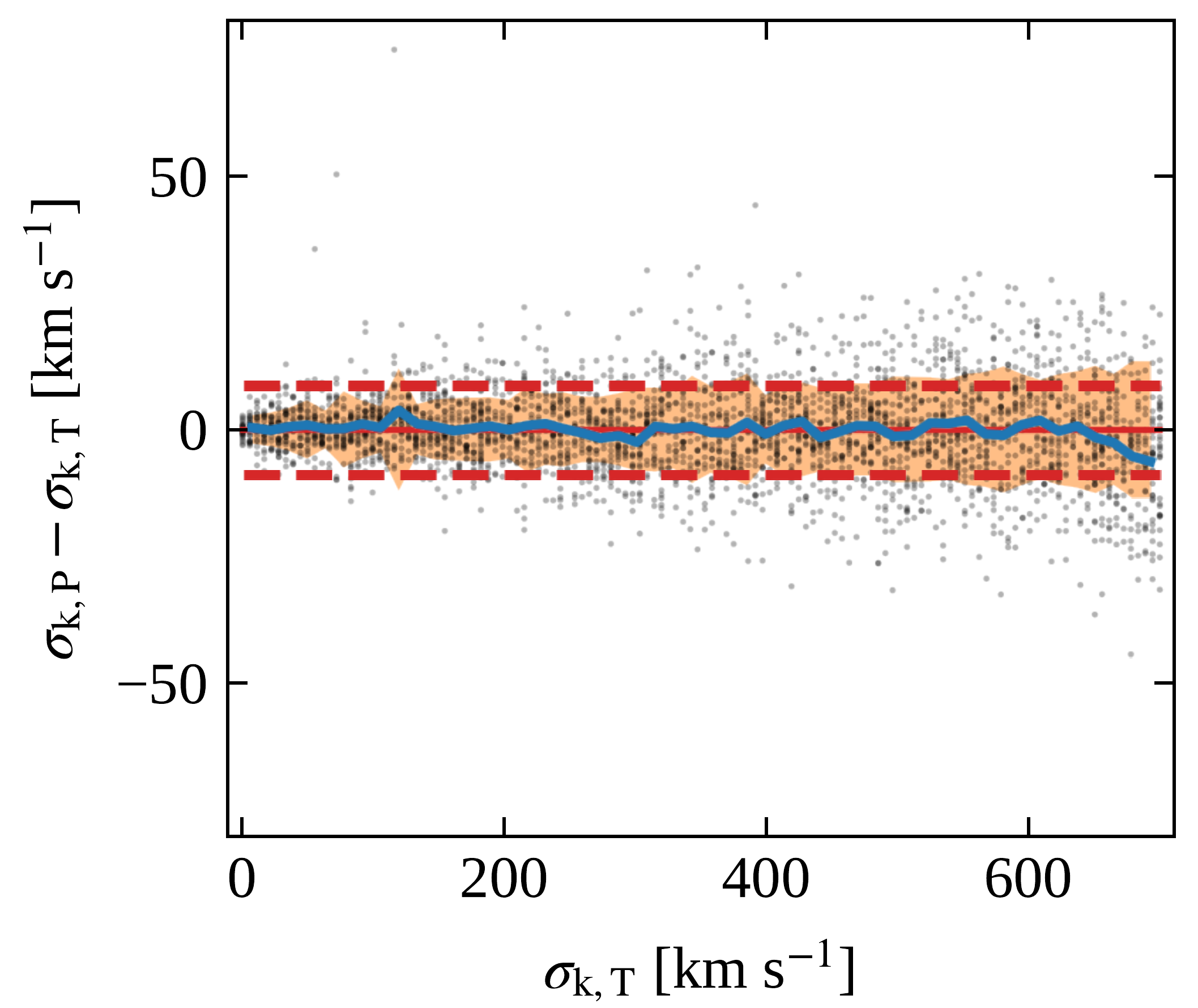}
\includegraphics[width = 0.22\textwidth]{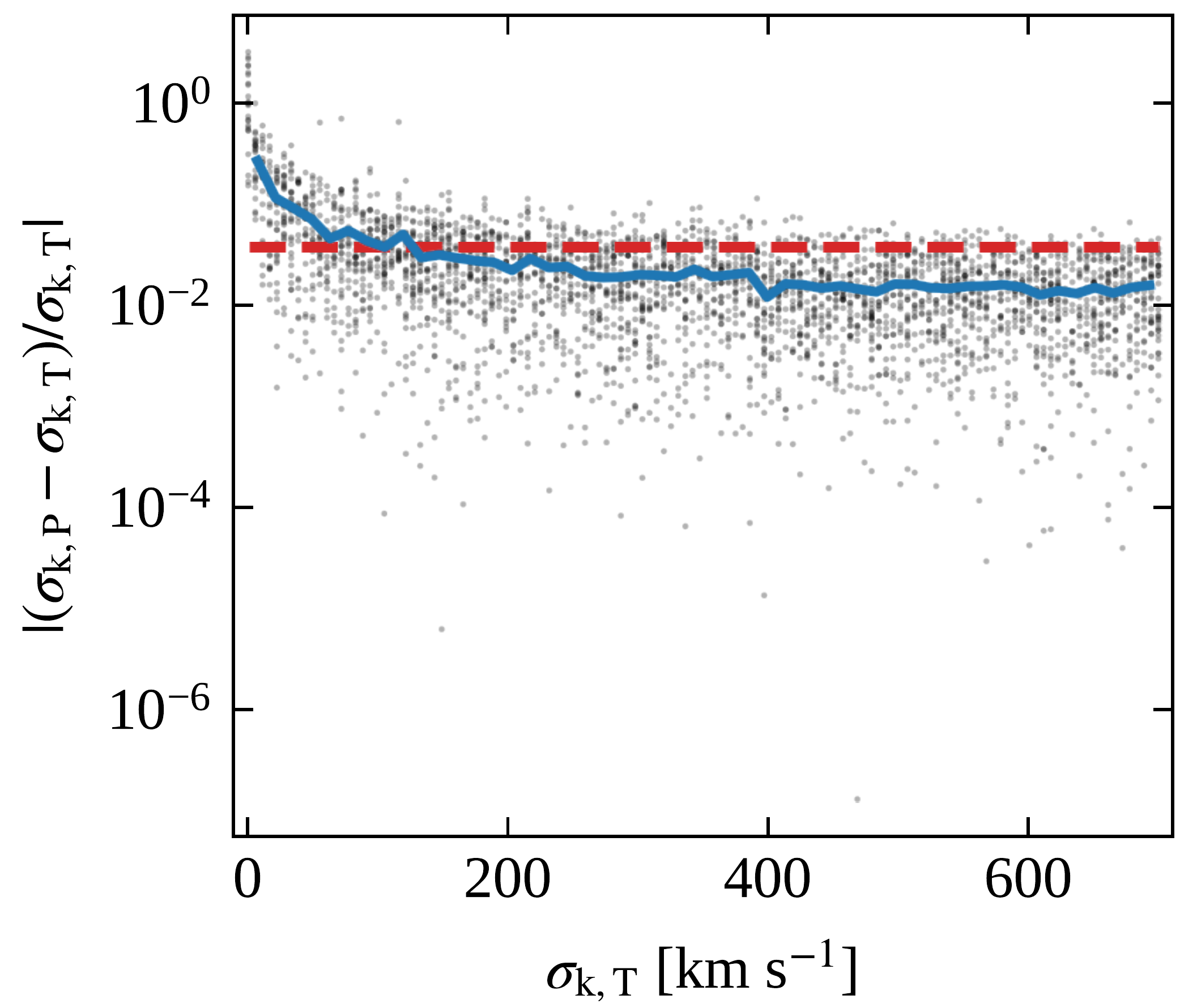}\\
\vspace{0.25cm}
\includegraphics[width = 0.22\textwidth]{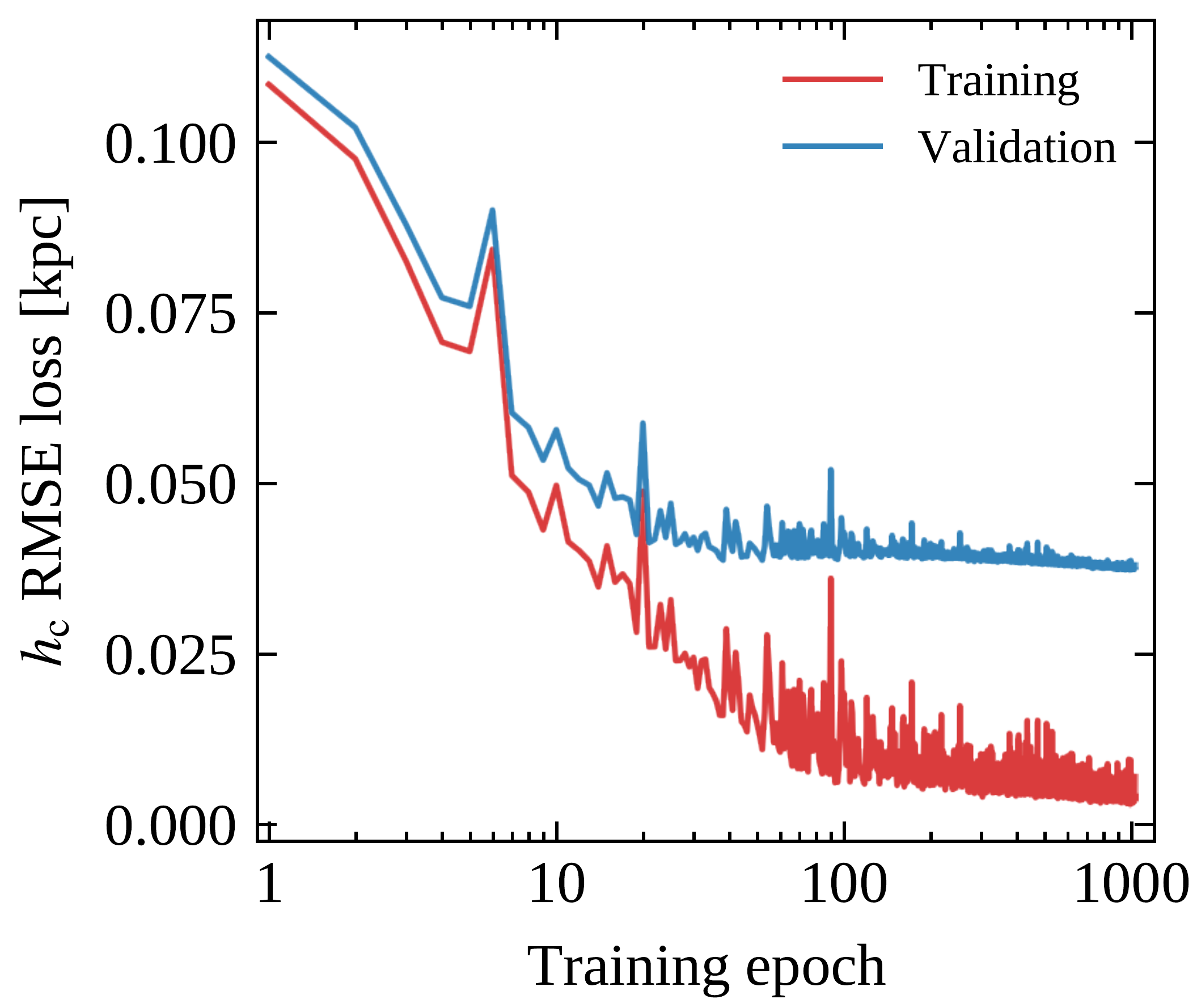}
\includegraphics[width = 0.22\textwidth]{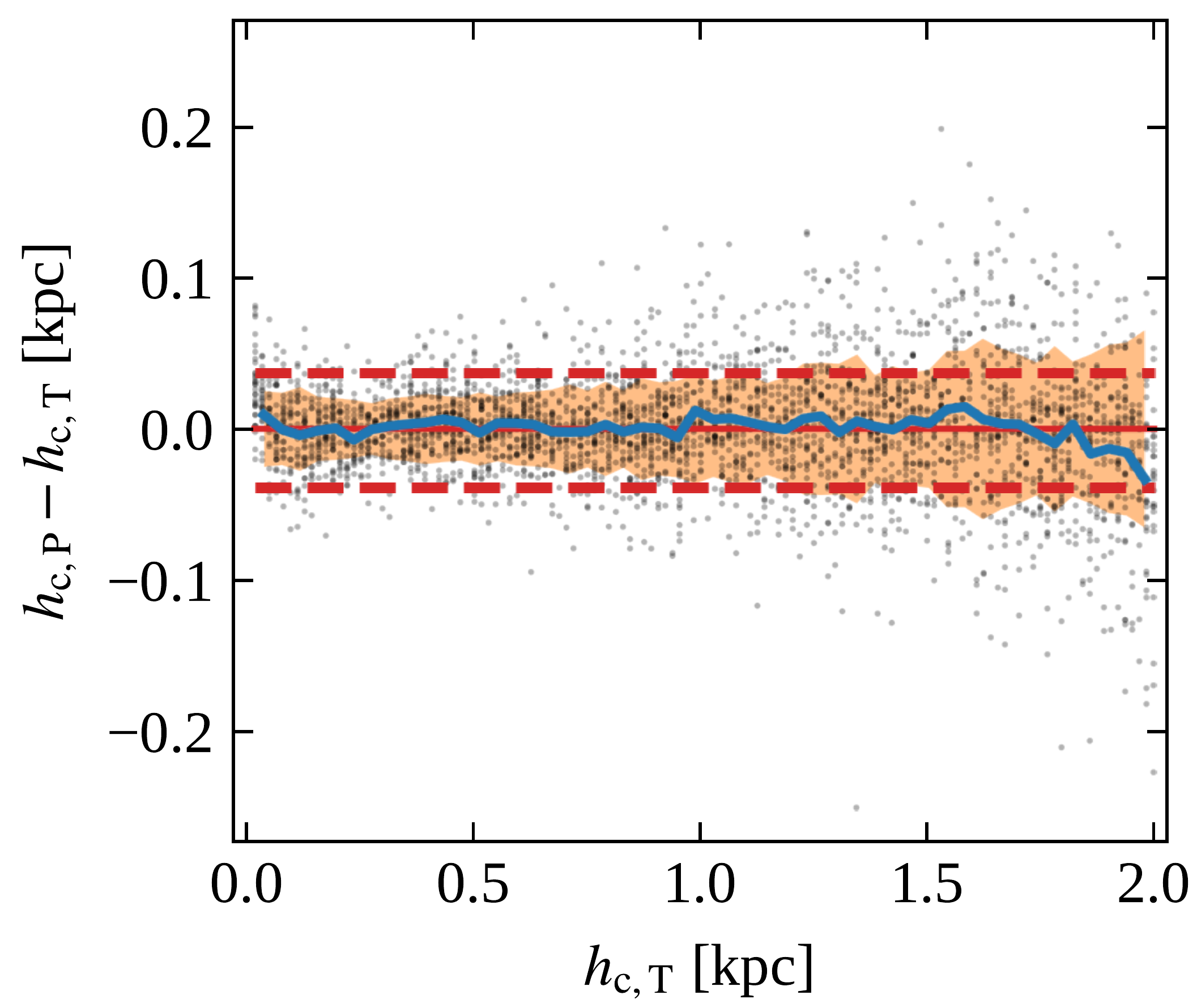}
\includegraphics[width = 0.22\textwidth]{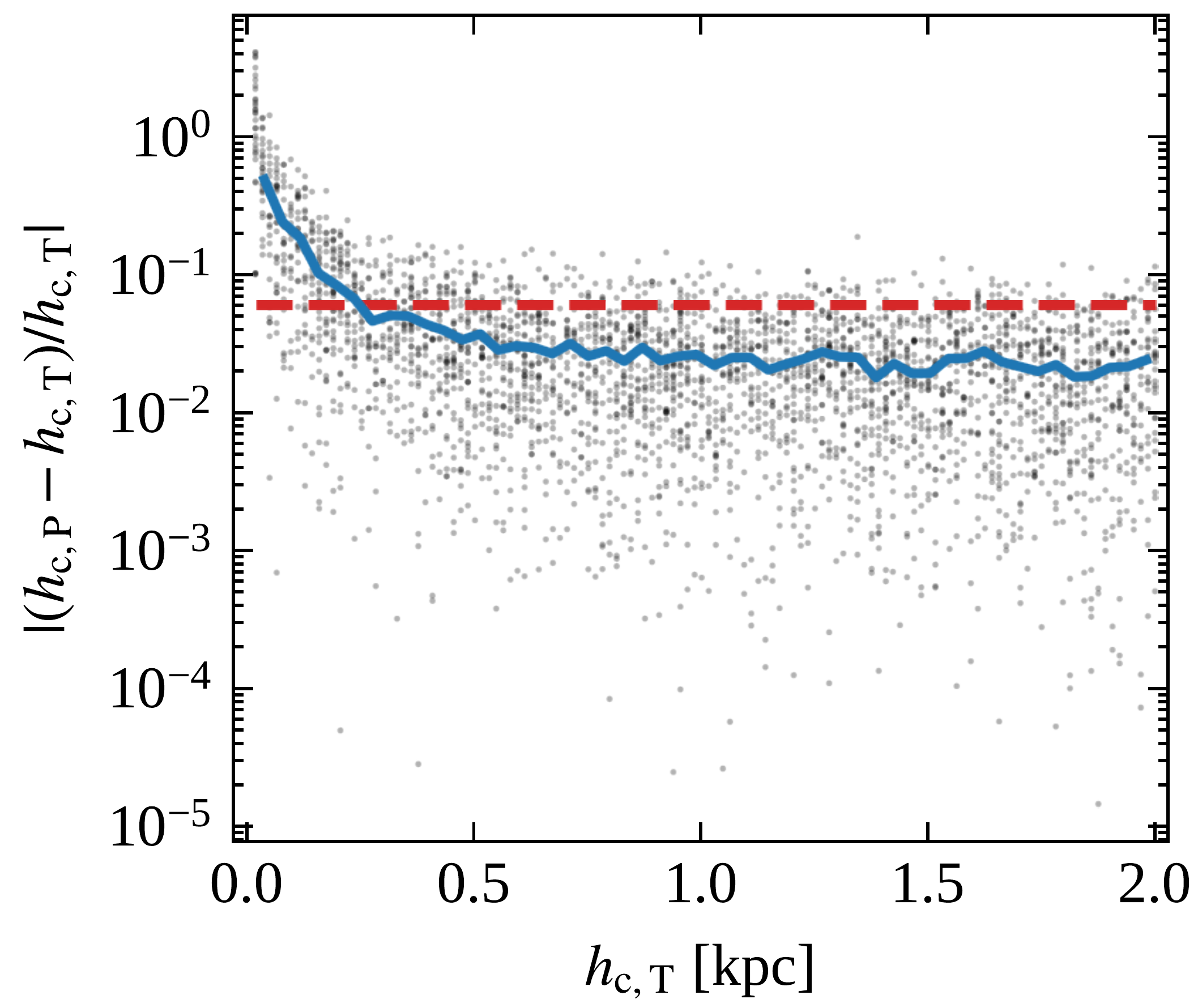}
\caption[Results of the \acs{CNN} two-parameter prediction for the kick-velocity parameter, $\sigma_{\rm k}$, and scale height, $h_{\rm c}$]{\label{fig:ch5_2par_cnn_inference_sigmak_hc}Results of the \acs{CNN} two-parameter prediction for the kick-velocity parameter, $\sigma_{\rm k}$, and scale height, $h_{\rm c}$, for the corresponding validation sets. {\it Top (bottom) left panel}: evolution of the \acs{RMSE} training ({\it red}) and validation  ({\it blue}) losses as a function of the training/validation epoch for the $\sigma_{\rm k}$ ($h_{\rm c}$) parameters. {\it Top (bottom) central panel}: residuals of the prediction as a function of the target $\sigma_{\rm k}$ ($h_{\rm c}$) value; the subscripts P and T refer to the predicted and target values, respectively. The red dashed lines delimit the $68$\% uncertainty region corresponding to a \acs{RMSE} $=\unit[8.8]{km \, s^{-1}}$ ($\unit[0.038]{kpc}$) computed over the whole range $[1, 700]$ $\unit[]{km \, s^{-1}}$ ($[0.02, 2]$ $\unit[]{kpc}$). The {\it orange} region delimits the $68$\% uncertainty region computed as a running \acs{RMSE} for which we have divided the full $\sigma_{\rm k}$ ($h_{\rm c}$) range into $50$ bins of equal size. The {\it blue} line shows the trend of the average residuals which are well centred around the value 0. {\it Top (bottom) right panel}: relative error of the prediction as a function of the target $\sigma_{\rm k}$ ($h_{\rm c}$) value. The red dashed line corresponds to a \acs{MRE} $=0.039$ ($0.061$) computed over the whole range $[1, 700]$ $\unit[]{km \, s^{-1}}$ ($[0.02, 2]$ $\unit[]{kpc}$). The {\it blue} line shows the trend of the running \acs{MRE} computed over $50$ bins of equal size into which we have divided the full $\sigma_{\rm k}$ ($h_{\rm c}$) range.}
\end{sidewaysfigure}  
%-----------------------------------------------------------------

%%%%%%%%%%%%%%%%%%%%%%%%%%%%%%%%%%%%%%%%%%%%%%%%%%

\subsubsection{Generalisation Results} \label{sec:ch5_2par_generalization_result}

As for the single-parameter analysis, we assess the actual performance of the network when simultaneously predicting $\sigma_{\rm k}$ and $h_{\rm c}$ by training the \acs{CNN} on the biggest dataset with $16384 = 128 \times 128$ simulations (see run S3). As discussed above, we use the 3-channel ICRS representation with one density map and two proper motion maps with 128 resolution as input features. We split the entire dataset into training and validation subsets with a relative percentage of $80/20$\%, respectively, leading to 13107 samples in the training and 3277 samples in the validation dataset and use the same configuration as in Section~\ref{sec:ch5_1par_generalization_result}. The evolution of the individual training and validation losses is shown in the left panels of Figure~\ref{fig:ch5_2par_cnn_inference_sigmak_hc}. 

%-------------------------------------------------------------
\begin{table}
\centering
\caption[Summary of the \acs{CNN} generalisation results]{Summary of the \acs{CNN} generalisation results on the validation and test (in parenthesis) datasets for the single-parameter and two-parameter cases.
\label{tab:ch5_genralization_results}}
\begin{tabular}{l l l l l}
\toprule
& \multicolumn{2}{c}{\tabhead{1-par. generalisation}} & \multicolumn{2}{c}{\tabhead{2-par. generalisation}} \\
\tabhead{Parameter} &
\tabhead{\acs{RMSE}} &
\tabhead{\acs{MRE}} &
\tabhead{\acs{RMSE}} &
\tabhead{\acs{MRE}} \\
\midrule \\
$\sigma_{\rm k}$ & $\unit[4.4 \,(4.8)]{km~s^{-1}}$ & 0.014 (0.018) & $\unit[8.8 \,(9.1)]{km~s^{-1}}$ & 0.039 (0.033) \\
$h_{\rm c}$ & $\unit[0.017 \,(0.019)]{kpc}$ & 0.024 (0.029) & $\unit[0.038 \,(0.041)]{kpc}$ & 0.061 (0.057) \\
\bottomrule \\
\end{tabular}

\end{table}
%---------------------------------------------------------------
The results for the trained network's prediction on the validation set for both parameters are shown in the central panels in Figure~\ref{fig:ch5_2par_cnn_inference_sigmak_hc} and summarised in Table~\ref{tab:ch5_genralization_results}. The network is able to predict $\sigma_{\rm k}$ and $h_{\rm c}$ with an average \acs{RMSE} uncertainty of $\unit[8.8]{km \, s^{-1}}$ and $\unit[0.038]{kpc}$, respectively, which is approximately doubled compared to the single-parameter experiment. These \acs{RMSE} values are computed over the full target ranges and represented by the red dashed lines in the residuals plots (see central panels of Figure~\ref{fig:ch5_2par_cnn_inference_sigmak_hc}). As in the single-parameter case, we observe that the \acs{RMSE} uncertainties increase with increasing target parameters as indicated by the orange regions in the residuals plots. The relative errors represented in the right panels of Figure~\ref{fig:ch5_2par_cnn_inference_sigmak_hc} show the same decreasing trend with the target value as for the single-parameter predictions, albeit with larger relative errors. When computing the \acs{MRE} over the whole range of the two target parameters, we obtain $0.039$ and $0.061$ for $\sigma_{\rm k}$ and $h_{\rm c}$, respectively. These values are highlighted by the red dashed lines in the right panels of Figure~\ref{fig:ch5_2par_cnn_inference_sigmak_hc}.

We then evaluate the generalisation capability of the trained network on the test set with 1000 samples. We find RMSEs of $\unit[9.1]{km \, s^{-1}}$ and $\unit[0.041]{kpc}$ and MREs of $0.041$ and $0.057$ for $\sigma_{\rm k}$ and $h_{\rm c}$, respectively. As before, we evaluate the confidence intervals of the two estimators over 1000 bootstrapped sets of the test set and find that the RMSE and MRE variations are around $3$\% and  $8$\%, respectively, for both predicted parameters. This indicates that also in the two-parameter prediction the trained network is stable and guarantees a good level of generalisation power. 

\begin{figure}
\centering
\includegraphics[height = 0.5\textwidth]{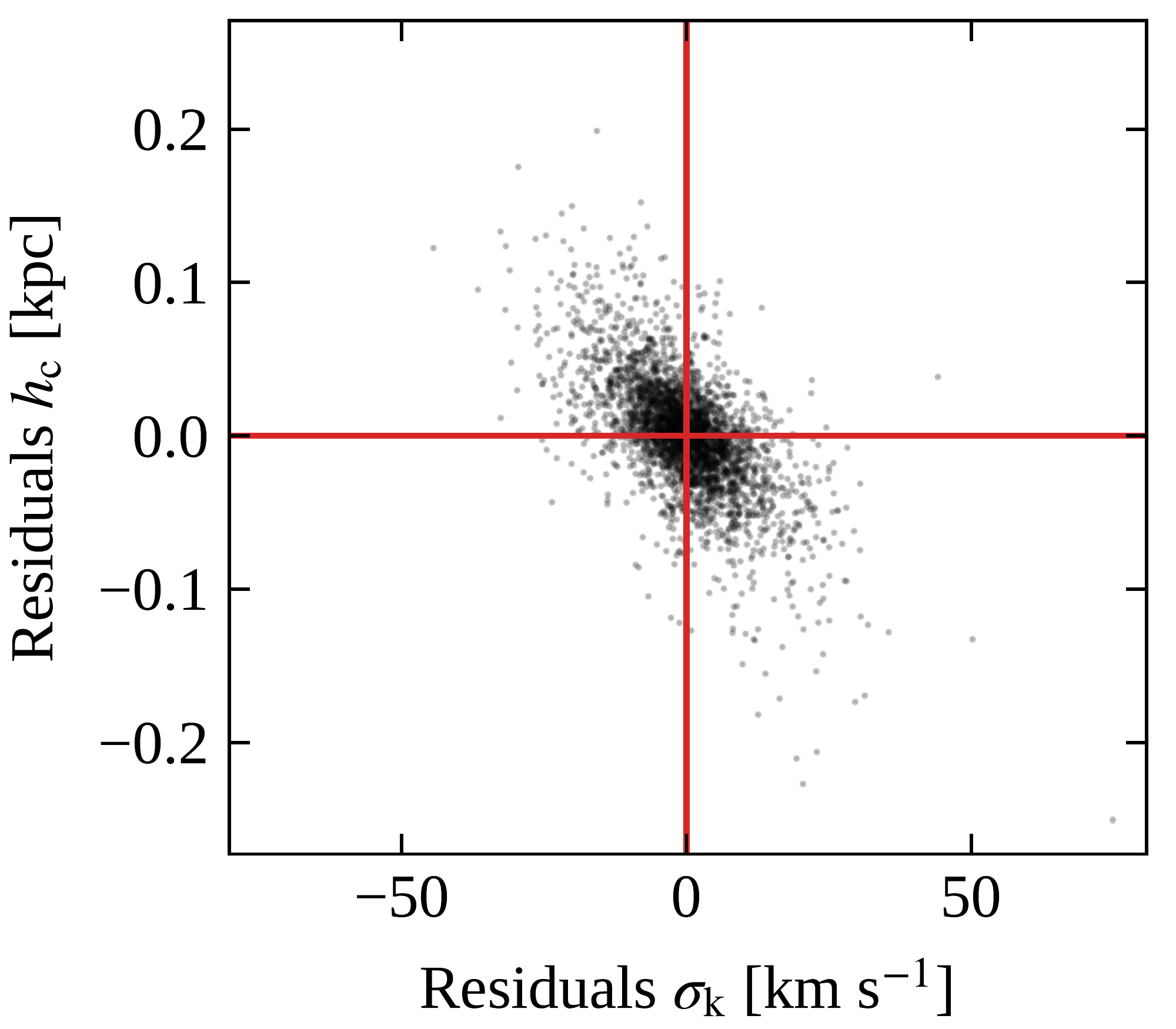}
\caption[residuals of the predicted scale-height parameter $h_{\rm c}$ versus residuals of the predicted kick-velocity parameter $\sigma_{\rm k}$]{\label{fig:ch5_2par_cnn_inference_res_corr}{Scatter plot of residuals of the predicted scale-height parameter $h_{\rm c}$ versus residuals of the predicted kick-velocity parameter $\sigma_{\rm k}$ for the validation dataset of our two-parameter generalisation experiment. An anticorrelation can be observed.}}
\end{figure}

However, we find that the \acs{CNN} trained to simultaneously predict $\sigma_{\rm k}$ and $h_{\rm c}$ is not able to reach the same level of accuracy as in the experiments where a single parameter was predicted at a time. This could be due to three distinct causes: (i) either our neural network is not sophisticated enough to discern between the effects of both parameters on the simulation outcomes represented in the ICRS maps, (ii) our choice of ICRS maps as input does not provide sufficient information for the network to distinguish between both parameters, or (iii) this is a physical (real) degeneracy, and there is a limit to what we can measure. To investigate this issue, we train the \acs{CNN} to predict \textit{only} the parameter $h_{\rm c}$ using the same two-parameter dataset used above where both $\sigma_{\rm k}$ and $h_{\rm c}$ are varied. After predicting on the validation dataset, we obtain a \acs{RMSE} accuracy of $\unit[0.038]{kpc}$ which is equal to the result obtained above for the two-parameter prediction. This suggests that the network complexity is suitable to predict either one or two parameters simultaneously. Limitations in performance are therefore either due to an inadequate input representation or a physical degeneracy that imposes a natural accuracy threshold. While we cannot distinguish these two with our current simulation and \acs{ML} pipeline, we can illustrate the underlying problem in the following way: Figure~\ref{fig:ch5_2par_cnn_inference_res_corr} shows the residuals of the scale-height parameter $h_{\rm c}$ versus the residuals of the kick-velocity parameter $\sigma_{\rm k}$ for the predictions over the validation set. The overall negative slope indicates that the network tends to overpredict $h_{\rm c}$ in those simulations where $\sigma_{\rm k}$ is underestimated and vice versa; i.e., large (small) $h_{\rm c}$ values have the same overall effect on the phenomenology of the pulsar population as large (small) $\sigma_{\rm k}$ values and the network struggles to distinguish these cases. This highlights the degeneracy between the two parameters already discussed above, which might be broken if the data itself were represented in a different way or additional input information about each neutron star (beyond position and velocity) were provided.

%-----------------------------------------------------------------
%-----------------------------------------------------------------

%%%%%%%%%%%%%%%%%%%%%%%%%%%%%%%%%%%%%%%%%%%%%%%%%%
%%%%%%%%%%%%%%%%%%%%%%%%%%%%%%%%%%%%%%%%%%%%%%%%%%

\section{Discussion}
\label{sec:ch5_discussion}

In this chapter, we have studied the potential of an artificial neural network to estimate with high accuracy the dynamical characteristics of a mock population of isolated pulsars. Implementing a simplified population-synthesis framework, we focused on the pulsar natal kick-velocity distribution and the distribution of birth distances from the Galactic plane. Taking into account the Galaxy's gravitational potential and evolving the pulsar motions in time, we generate a series of simulations that are used to train and validate a suitably structured convolutional neural network.

The generalised results presented in the previous sections are obtained in a very idealised and simplified scenario, implying that caution is required when the uncertainties for the prediction of the kick-velocity dispersion, $\sigma_{\rm k}$, and birth scale height, $h_{\rm c}$, are taken at face value and conclusions for the real pulsar population are drawn. 
%-----------------------------------------------------------------
\begin{figure*}
\centering
\includegraphics[width = \textwidth]{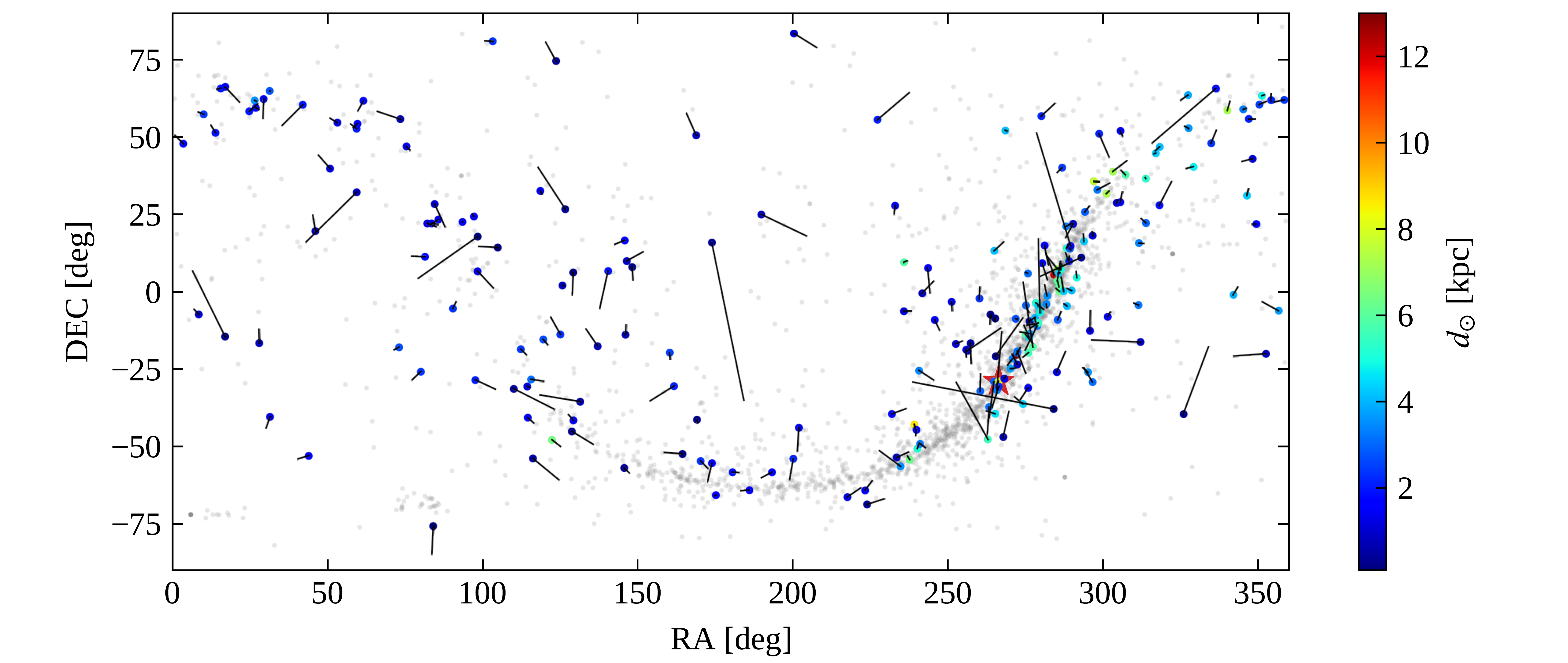}
\includegraphics[width = \textwidth]{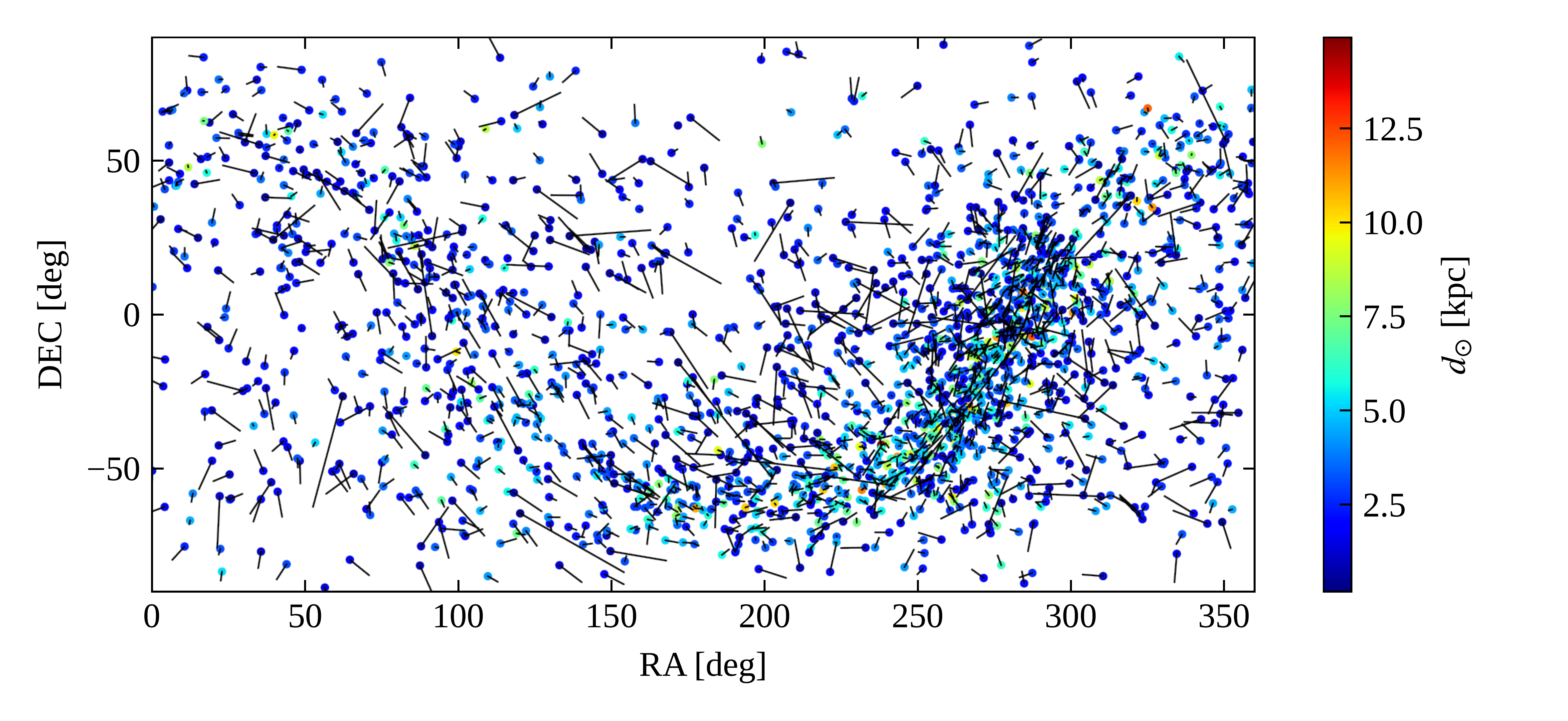}
\caption[Proper motions of observed 216 neutron stars compared with simulation in ICRS reference frame]{\label{fig:ch5_icrs_proper_motion}{Proper motion trajectories in the ICRS reference frame of the selected observed 216 neutron stars (\textit{upper panel}) and 2000 mock neutron stars simulated using the fiducial values for the kick velocity and scale height ($\sigma_{\rm k} = \unit[265]{km \, s^{-1}}$ and $h_{\rm c} = \unit[0.18]{kpc}$) and sampled using the wight function $f(d_\odot)$ (\textit{lower panel}). The current locations of the neutron stars are indicated by the coloured circles, whereas the tracks indicate their motion for the past $\unit[0.5]{Myr}$, assuming no radial velocity and neglecting the effects of the Galactic potential. The colour encodes the heliocentric distance, $d_\odot$, of the neutron stars. The corresponding data is provided in Table~\ref{tab:app_proper_motion}. In the background, we show in grey all non-binary pulsars in the ATNF catalogue (those in the Small and Large Magellanic Clouds as well as those in globular clusters are included). The red star highlights the position of the Galactic Centre.}}
\end{figure*}  
%-----------------------------------------------------------------
In particular, our simulations assume that the distribution of neutron-star progenitors in Galactic height is represented by the exponential thin-disk model, and that the kick-velocity magnitudes follow a Maxwellian distribution. While the choice of an exponentially thin disk is commonly adopted \citep{Wainscoat1992, Polido2013, Li2019} and can be justified theoretically as the outcome of a self-gravitating isothermal disk \citep{Spitzer1942}, the choice of a Maxwellian model for the kick-velocity distribution is difficult to motivate from a theoretical standpoint. The Maxwellian model has found empirical support as it has been shown to well describe the proper motions of observed pulsars \citep{Hobbs2005}. It is for this reason, and its rather simple mathematical form, that the Maxwell kick-velocity distribution is often adopted in population synthesis studies of compact stars \citep{Sartore2010, Cieslar2020}. However, the real functional form of the kick-velocity distribution is still unknown and debated. Several authors have studied the kick-velocity problem and concluded that other models explain observed proper motions of neutron stars equally well. For example, by using maximum-likelihood methods \citet{Arzoumanian2002} found that a bimodal Gaussian distribution with a low-velocity and a high-velocity component is the preferred model to describe the observed proper motion of a sample of 79 radio pulsars. \citet{Faucher2006} studied the velocity component along the Galactic longitude for a sample of 34 pulsars observed with interferometric techniques \citep{Brisken2002, Brisken2003}. After testing a two-component Gaussian model as well as a variety of single-component models, they opted for a single-component description with an exponential shape, although a two-component model could not be ruled out due to the poor statistics of their sample. More recently, \citet{Verbunt2017} and \citet{Igoshev2020} analyzed a sample of isolated young pulsars and found that a two-component Maxwellian model explained the observed sample best. 

In general, the presence of a low-velocity and a high-velocity component could indicate different progenitor properties as well as birth scenarios for the pulsar population. Numerical simulations of supernova explosions have, for example, suggested that neutron stars with lower kick velocities could be generated in the core-collapse supernovae of progenitors with small iron cores or in electron-capture supernovae \citep{Podsiadlowski2004, Mandel2020}. While also possible scenarios for isolated systems \citep{Janka2017}, such conditions might generally affect those neutron stars born in binaries \citep{Giacobbo2020}, where mass-loss episodes could strip their progenitors off their hydrogen envelopes. This might favour the formation of small iron cores or accretion-induced electron-capture supernovae, resulting in weaker natal kicks \citep{Schwab2010, Tauris2013b}. Only for the strongest kicks can the binaries be disrupted by the supernova and both companions expelled; otherwise the two stars remain gravitationally bound. Such effects are neglected in our simplified model but could in principle generate an imprint on the observed neutron-star population.

Up to this point, we have not considered any kind of selection effects or observational biases and effectively assumed that all the neutron stars in our simulation are detectable. While this provides direct insight into how various initial conditions affect the evolved population of neutron stars, a direct comparison with observational data in principle requires a careful treatment of biases. For example, due to beaming effects not all Galactic radio pulsars are visible from Earth \citep[][see also Section~\ref{sec:ch1_radio_em_geometry}]{Tauris1998, Melrose2017}, while survey sensitivity thresholds and instrumental limitations might hamper the detection of faint or far away sources \citep[][see also Section~\ref{sec:ch1_radiometer}]{Manchester2001, Johnston2008, Stovall2014, Coenen2014, Good2021}. Additionally, timing noise can significantly limit the sensitivity and precision in the detection of pulsar proper motions via timing analysis techniques \citep{Hobbs2004, Lentati2016, Parthasarathy2019}. 
With the aim of obtaining a rough idea on how selection effects and biases could potentially influence a future comparison with observational data, we perform the following experiment. We first collect those neutron stars that have measured proper motions. As the main resource, we use the ATNF pulsar catalogue\footnote{\url{https://www.atnf.csiro.au/research/pulsar/psrcat/}} \citep{Manchester2005}, but in some cases we provide proper motion results from more recent analyses (see Appendix~\ref{app:proper_motion} for details). We find a total of 417 neutron stars whose angular positions, proper motions in ICRS coordinates, spin periods, spin-period derivatives, \acs{DM} values and distance estimates are reported in Table~\ref{tab:app_proper_motion}. Out of these objects, we remove those stars that belong to the Magellanic Clouds, are associated with globular clusters or have a binary companion. We further select only those neutron stars with a spin-period derivative $\dot{P} > 10^{-17}$ to exclude those that have potentially been recycled. Finally, we consider only those for which an estimate of the heliocentric distance $d_{\odot}$ is available; in the case of radio pulsars we quote values that are derived from their respective \acs{DM}s using the free-electron density model of \citet{Yao2017} (YMW16 model hereafter). As some neutron stars have a $DM$ that exceeds the maximum Galactic $DM$ allowed by the YMW16 model, these cases are assigned a default distance of $\unit[25]{kpc}$. We exclude those cases unless an alternative distance measurement is available. Applying these filters we obtain a sample of 216 Galactic, likely isolated neutron stars, whose positions and proper motions are illustrated in the upper panel in Figure~\ref{fig:ch5_icrs_proper_motion}. 

%-----------------------------------------------------------------
\begin{figure*}
\centering
\includegraphics[height = 0.40\textwidth]{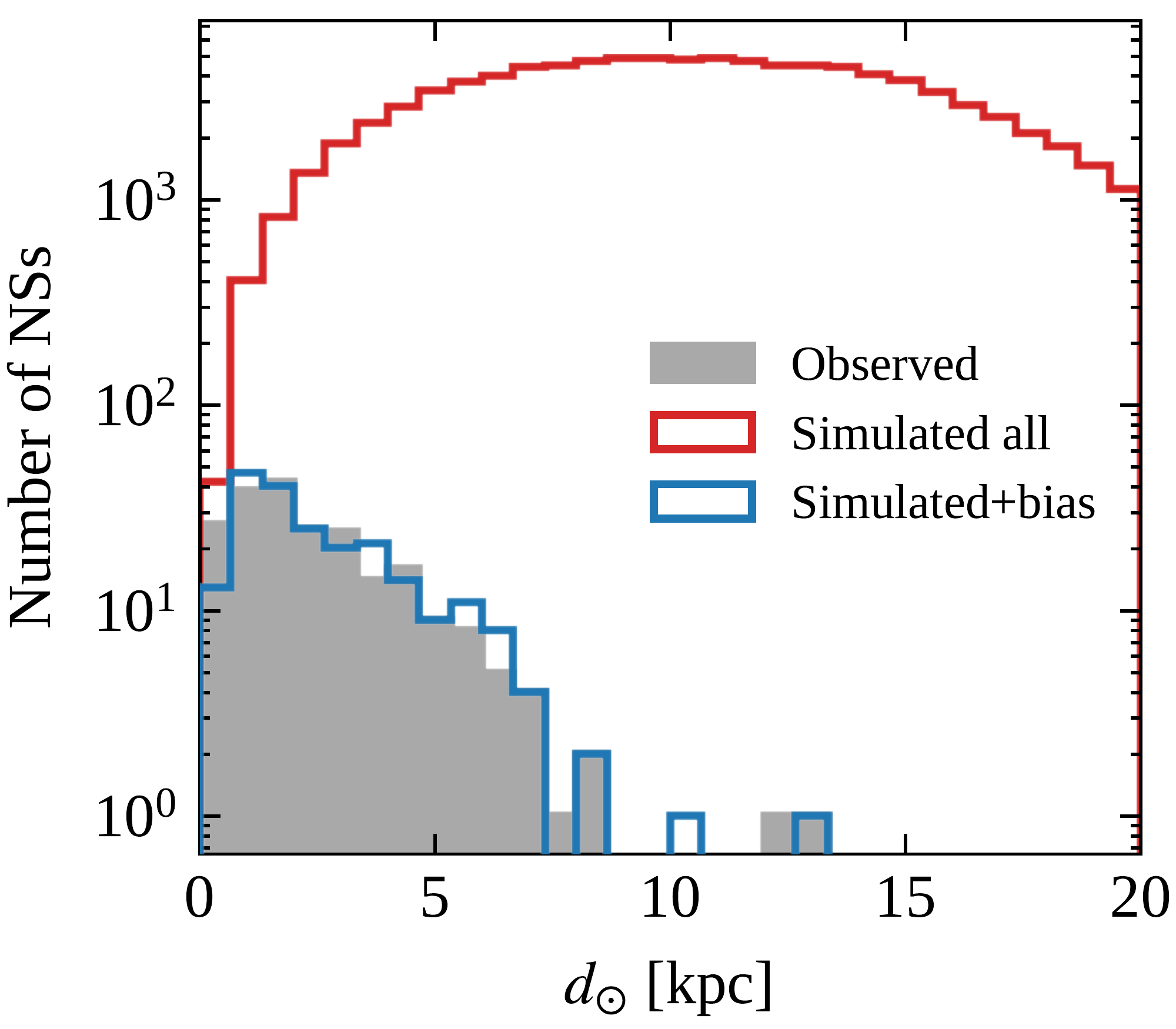}
%\hspace{0.5cm}
\includegraphics[height = 0.40\textwidth]{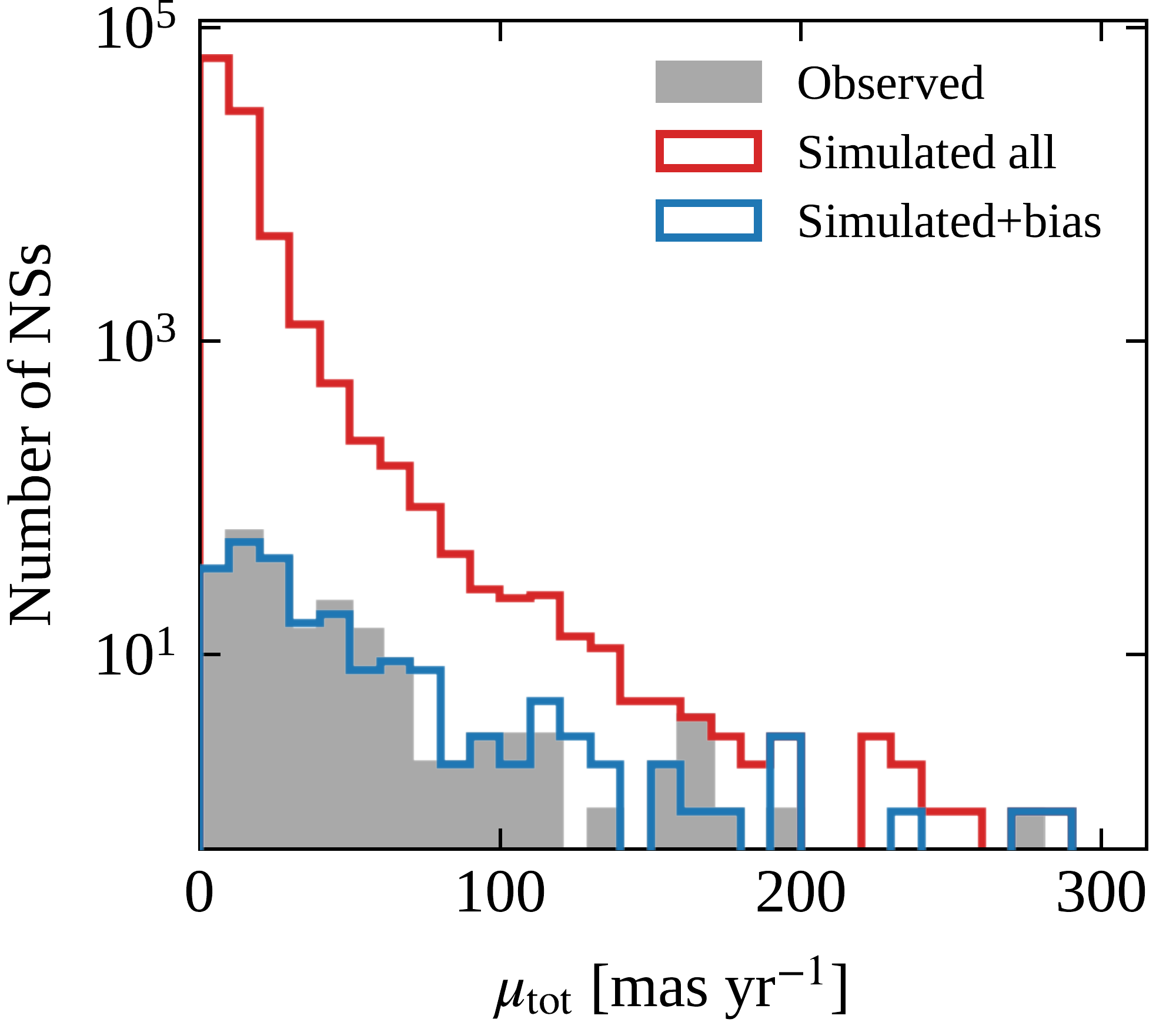}
\caption[Distribution of heliocentric distances $d_{\odot}$ and proper-motion magnitudes, $\mu_{\rm tot}$, for the 216 neutron stars with measured proper motion]{\label{fig:ch5_dist_pm_histo}{Distribution of heliocentric distances $d_{\odot}$ (\textit{left panel}) and proper-motion magnitudes, $\mu_{\rm tot}$, (\textit{right panel}) for the 216 neutron stars with measured proper motion ({\it grey histograms}). For comparison, we also show the distances and proper-motion magnitudes for 216 simulated neutron stars ({\it blue histograms}) randomly sampled from a mock population generated with the fiducial parameters $\sigma_{\rm k} = \unit[265]{km \, s^{-1}}$ and $h_{\rm c} = \unit[0.18]{kpc}$ ({\it red histograms}). For the weighted sampling, we use the weight function $f(d_{\odot}) = d_{\odot}^{-1} \exp (-0.5 d_{\odot})$ (see the text for more details).}}
\end{figure*}  
%-----------------------------------------------------------------

In the left panel of Figure~\ref{fig:ch5_dist_pm_histo}, the grey histogram shows the distribution of their heliocentric distances. Even if subject to some uncertainties due to imprecisions in the YMW16 model, the distance distribution peaks around $\unit[1]{kpc}$, followed by a sharp exponential decrease. For a realistic pulsar distribution and in the absence of selection effects, we would expect the number of neutron stars to increase with distance due to an increase in the explored volume, until reaching a maximum at a distance of about $\unit[10]{kpc}$, which comprises the region around the Galactic Centre (see the red histogram in the left panel of Figure~\ref{fig:ch5_dist_pm_histo}). Thus, the shape of the grey distribution in Figure~\ref{fig:ch5_dist_pm_histo}, as expected, indicates that our observed sample of neutron stars with measured proper motions is incomplete in distance and subject to selection biases. By looking at the right panel of Figure~\ref{fig:ch5_dist_pm_histo}, we also note that a selection bias on distance is also reflected in the distribution of the total proper motion magnitudes (grey histogram), computed as $\mu_{\rm tot} \equiv \sqrt{\mu_{\rm RA}^2 + \mu_{\rm DEC}^2}$. Indeed, since the nearest stars are also characterised by larger angular proper motions, there is a tendency to detect high proper-motion stars with higher probability.

In this first empirical approach, we follow a more agnostic approach to introduce a comparable selection effect in our simulated populations. Specifically, we use a weighted random-sampling routine to select $n$ pulsars from our mock populations, where each simulated star is assigned a weight according to a function $f(d_{\odot})$ of its heliocentric distance. This weight function has to assign larger weights to closer neutron stars in order to ensure their higher detection probabilities and has to be chosen such that we recover the observed distance distribution with sufficient accuracy. To find $f(d_{\odot})$, we simulate a mock population with the fiducial values of the kick velocity and scale height, that is  $\sigma_{\rm k} = \unit[265]{km \, s^{-1}}$ and $h_{\rm c} = \unit[0.18]{kpc}$, respectively (red histograms in Figure~\ref{fig:ch5_dist_pm_histo}). After using a given $f(d_{\odot})$ to weight the simulated neutron stars, we sample 216 mock stars and compare their distance and proper-motion distributions (shown as blue histograms in Figure~\ref{fig:ch5_dist_pm_histo}) with those of the observed sample by performing two-sample Kolmogorov-Smirnov (KS) tests. After testing various functional forms, we find that $f(d_{\odot}) = d_{\odot}^{-1} \exp (-0.5 d_{\odot})$ is able to reproduce the observed distributions with a good level of accuracy. More precisely, for this choice of $f(d_{\odot})$, the KS tests performed over $1000$ distinct comparisons give average p-values of $\sim 0.3$ and $\sim 0.6$ for the distance and proper-motion comparison, respectively. This means that at $95$\% confidence level, we cannot reject the null hypothesis that the simulated and observed samples are drawn from the same underline distribution.
We have also verified that for this weight function, the KS tests always provide p-values $>0.05$ when comparing the observed sample with the simulated populations for reasonable values of $\sigma_{\rm k}$ and $h_{\rm c}$. Only in the cases where $\sigma_{\rm k}$ and $h_{\rm c}$ assume extreme values (near the edges of their respective ranges) the p-values might drop below $0.05$. However, these cases are associated with simulations with extreme initial conditions that are unlikely to reproduce the observations. For our basic experiment, we further make the simplified assumptions that $f(d_{\odot})$ emulates all selection biases and that it is the same for every number $n$ of sampled neutron stars. We stress that for the purpose of this initial analysis, we do not aim to accurately constrain the selection function that affects the observed population of neutron stars. Instead, we study how the introduction of realistic selection biases will alter the predictive power of our machine learning framework. Although one might intuitively attribute the exponential factor in $f(d_{\odot})$ to scattering in the interstellar medium at large distances, the underlying nature and the precise form of the true selection function is certainly more complicated. We expect it to encompass a series of effects due to the physics of the interstellar medium and the pulsar emission itself, as well as selection effects of pulsar surveys and pulsar searches. We reserve a more accurate study disentangling the different effects that contribute to $f(d_{\odot})$ to future work.  

We then analyze how the predictive power of the \acs{CNN} evolves as a function of the number $n$ of neutron stars, sampled with the above weight function $f(d_{\odot})$. To do so, we vary $n$ from 200 to 2000 in steps of 200, and in each case re-sample the 16384 simulated populations from run S3, where both $\sigma_{\rm k}$ and $h_{\rm c}$ are varied. After applying a $80/20$\% training-validation split, we retrain the \acs{CNN} on each of the down-sampled simulations. As before, we use the 3-channel ICRS input maps (i.e., density plus proper motion information) but instead opt for a resolution of $32\times16$ bins to accommodate the smaller number of stars represented in our maps. We have verified that a higher resolution of $128 \times 64$ bins does not affect the training results significantly, but slows down the training process; we therefore choose the lower resolution. We use the same training hyperparameters as in Section~\ref{sec:ch5_1par_generalization_result}, that is an initial learning rate of $10^{-4}$, a batch size of 64 and an early stop at 128 epochs. Once trained for each $n$ value, we apply the \acs{CNN} to the validation sets and compute the \acf{RMSE} and \acf{MRE} for the $\sigma_{\rm k}$ and $h_{\rm c}$ predictions as a function of $n$.

\begin{figure*}
\centering
\includegraphics[height = 0.40\textwidth]{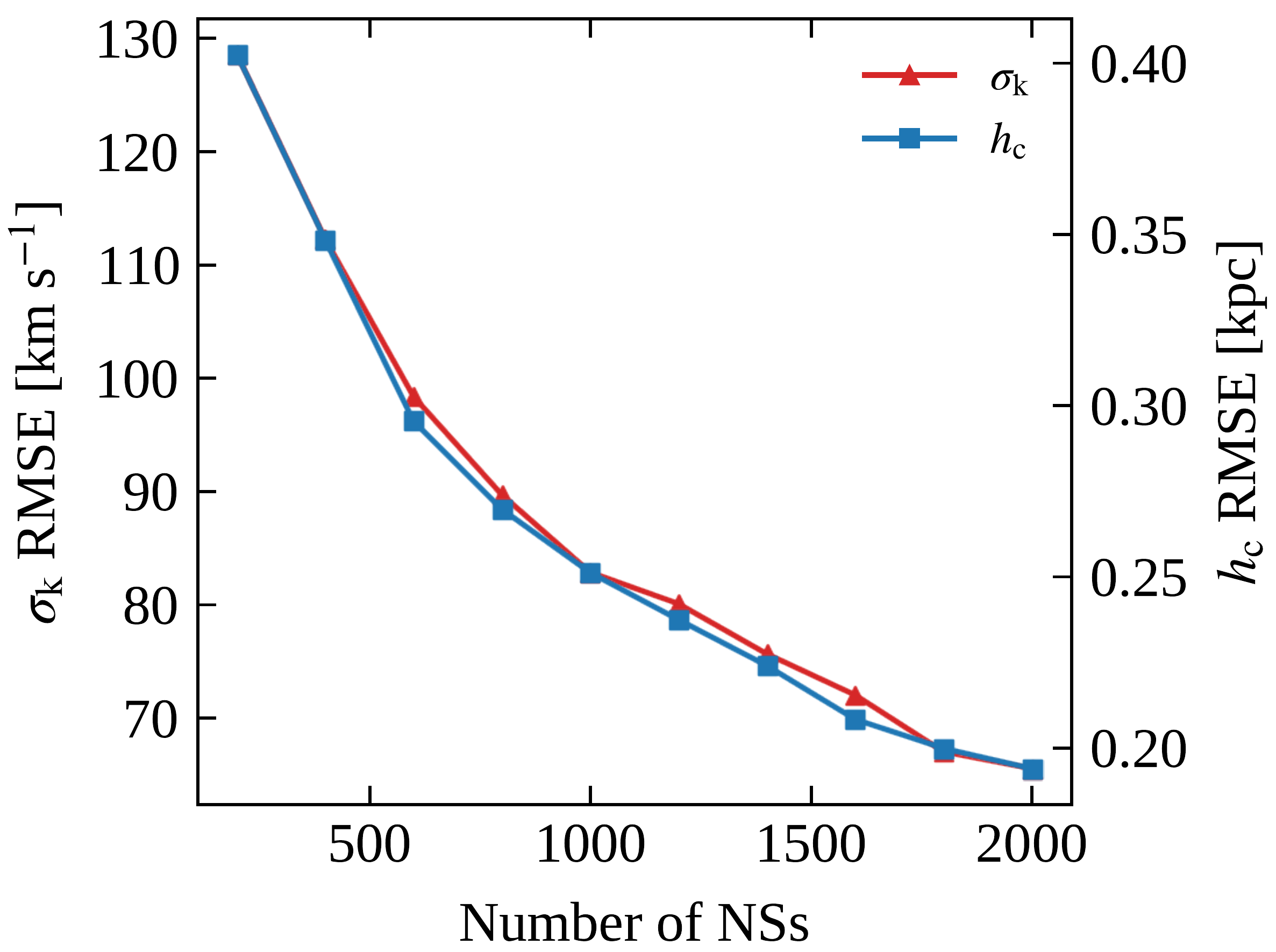}
%\hspace{0.2cm}
\includegraphics[height = 0.40\textwidth]{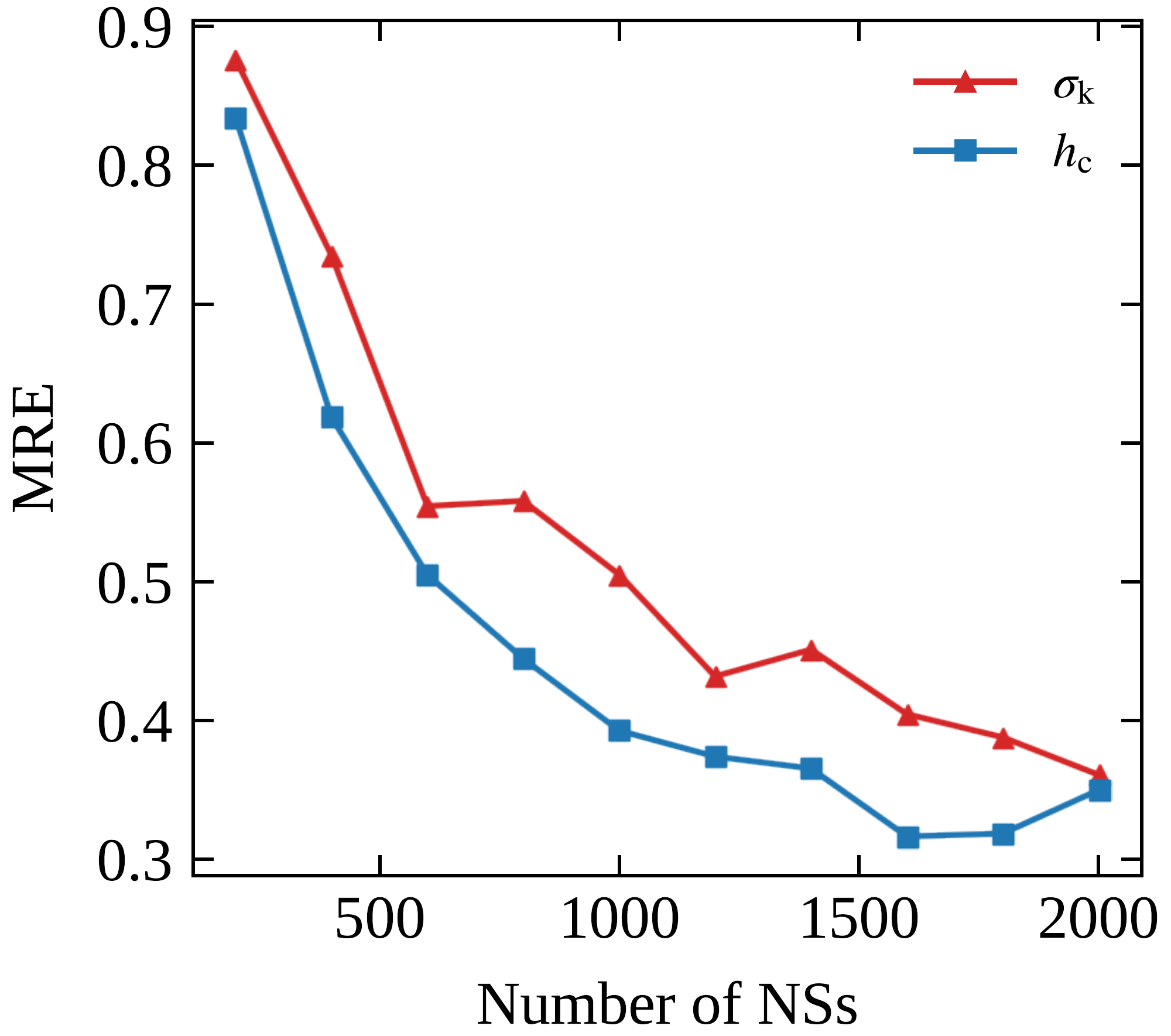}
\caption[Trend of the \acs{RMSE} and \acs{MRE} uncertainties on the two parameter prediction as a function of the number of neutron stars in the resampled simulations]{\label{fig:ch5_rmse_mre_nsnumber}{Trend of the \acs{RMSE} ({\it left panel}) and \acs{MRE} ({\it right panel}) uncertainties on the prediction of the two parameters $\sigma_{\rm k}$ ({\it blue}) and $h_{\rm c}$ ({\it red}) as a function of the number of neutron stars $n$ in the resampled simulations in the validation sets. To train and validate the \acs{CNN}, we use the 16384 simulated populations from simulation run S3 (with a $80/20$\% training/validation split), where both $\sigma_{\rm k}$ and $h_{\rm c}$ are varied, and which are resampled with increasing number of stars $n$ according to the weight function $f(d_{\odot}) = d_{\odot}^{-1} \exp (-0.5 d_{\odot})$.}}
\end{figure*}

In the left panel of Figure~\ref{fig:ch5_rmse_mre_nsnumber}, we show how the \acs{RMSE} uncertainties for the predictions of the two parameters $\sigma_{\rm k}$ ({\it blue}) and $h_{\rm c}$ ({\it red}) diminish with increasing number of neutron stars $n$ sampled from the simulations. We observe that both curves (with the appropriate rescaling) follow very similar trends. On the right, we show instead how the \acs{MRE}s evolve with $n$, indicating how the precision of the two-parameter prediction improves with the number of detected neutron stars. This plot shows that, under the assumptions that selection effects are unaltered and the underlying kick-velocity and birth-height distributions have a Maxwellian and exponential shape, respectively, our trained \acs{CNN} is able to predict $\sigma_{\rm k}$ and $h_{\rm c}$ with a relative error of $\sim 0.35$ for a sample of 2000 stars (see also lower panel of Figure~\ref{fig:ch5_icrs_proper_motion} as a reference). 

%-----------------------------------------------------------------
%-----------------------------------------------------------------

These results highlight that the number of neutron stars plays a crucial role for the level of accuracy that the \acs{CNN} can reach. As expected, a larger number of stars provides more information, which allows the \acs{CNN} to pinpoint differences between populations evolved from different initial conditions, also when selection effects are introduced. Future observational efforts aimed at detecting and characterising new pulsars will thus play an important role in constraining the birth properties of neutron stars. Specifically, the advent of the \acf{SKA} will represent an important step forward into this direction. Due to its large sensitivity (a factor of 10 better than other radio telescopes) and its long baseline (up to $\unit[3000]{km}$), \acs{SKA} has the potential to increase the number of discovered pulsars by a factor $10$ \citep{Smits2009, Smits2011}. This will allow more precise timing and astrometric measurements of pulsar positions as well as distances and proper motions. A larger and more precise dataset could also help to better constrain the shape of the kick-velocity distribution and differentiate between models that try to explain the origin of the natal kicks \citep{Tauris2015}.

%%%%%%%%%%%%%%%%%%%%%%%%%%%%%%%%%%%%%%%%%%%%%%%%%%
%%%%%%%%%%%%%%%%%%%%%%%%%%%%%%%%%%%%%%%%%%%%%%%%%%

\section{Summary}
\label{sec:ch5_summary}

In this work, we have analyzed the possibility of using machine-learning (ML) techniques to reconstruct the dynamical birth properties of an evolved population of isolated neutron stars. For this purpose, we developed a simplistic population-synthesis pipeline to simulate the dynamical evolution of Galactic neutron stars and subsequently used these simulations to train and validate two different neural networks. We specifically focused on their ability to predict two parameters that strongly impact on the phenomenology of the evolved population: the dispersion, $\sigma_{\rm k}$, of a Maxwell kick-velocity distribution and the scale height, $h_{\rm c}$, of an exponential distribution for the Galactic birth heights. This was achieved by providing the networks with two-dimensional stellar density and velocity maps in galactocentric and equatorial (ICRS) coordinate frames. We found that a \acf{CNN} is able to estimate the physical parameters with high accuracy when multiple input channels, i.e., position and velocity information, are provided. In particular, when simultaneously predicting $\sigma_{\rm k}$ and $h_{\rm c}$ from ICRS maps, the network is able to reach absolute uncertainties lower than $\unit[10]{km \, s^{-1}}$ and $\unit[0.05]{kpc}$, respectively, corresponding to a relative error of around $10^{-2}$ for both parameters. Albeit obtained under simplified assumptions, our feasibility study has thus demonstrated that \acs{ML} techniques are indeed suitable to infer information about the pulsar population. Our phenomenological analysis incorporating proper-motion measurements (an attempt at including observational biases in an agnostic way) has further highlighted that increasing the sample of known pulsars and accurately measuring their current characteristics with future telescopes is crucial to tightly constrain the birth properties of the neutron stars in the Milky Way. In particular, our trained \acs{CNN} is able to predict $\sigma_{\rm k}$ and $h_{\rm c}$ with a relative error of $\sim 0.35$ for a sample of 2000 pulsars with measured proper motions.

We also demonstrated that one of the main factors in limiting the accuracy of the \acs{CNN}'s predictions in our set-up is the degeneracy between $\sigma_{\rm k}$ and $h_{\rm c}$; as they both affect the evolved populations in a similar way, the network struggles to disentangle their effects. This limitation is a direct consequence of our simplified population-synthesis framework. In future works, we will go beyond modelling the dynamical evolution and focus on incorporating additional physics such as magneto-thermal and spin-period evolution. We will further model their emission in different electromagnetic bands and study corresponding detectability limits by addressing selection effects as well as observational survey biases. Such additional input information could potentially break the degeneracy between the kick-velocity and the Galactic height distributions and provide more accurate model constraints on $\sigma_{\rm k}$ and $h_{\rm c}$ as well as other input parameters. 

The ultimate goal will be to use multi-wavelength observations of the Galactic neutron star population and take advantage of \acs{ML}, combined with population synthesis, to recover their birth properties, such as the natal kick-velocity, spin-period or magnetic-field distribution. In the next chapter we will focus on inferring the magneto-rotational properties of the pulsar population by adopting a more sophisticated simulation based inference approach.

%%%%%%%%%%%%%%%%%%%%%%%%%%%%%%%%%%%%%%%%%%%%%%%%%%
%%%%%%%%%%%%%%%%%%%%%%%%%%%%%%%%%%%%%%%%%%%%%%%%%%

% Chapter 6

\chapter{Simulation-based inference for pulsar population synthesis} % Main chapter title

\label{Chapter6} % For referencing the chapter elsewhere, use \ref{Chapter1} 

%----------------------------------------------------------------------------------------

\section{Introduction}
\label{sec:ch6_intro}

As one of the end points of stellar evolution of massive stars, neutron stars are influenced by many extremes of physics including strong gravity, large densities, fast rotation and extreme magnetic fields (see Chapter~\ref{Chapter1}). Consequently, these compact objects have been connected with several of the most energetic transient phenomena in our Universe such as fast radio bursts, superluminous supernovae, ultra-luminous X-ray sources, long- and short-duration gamma-ray bursts, and gravitational-wave emission \citep[e.g.,][]{Bachetti2014, Metzger2014, Berger2014,  Abbott2017a, Margalit2018, Petroff2022}. Accurately modelling these processes requires a detailed understanding of neutron-star properties, which also set constraints on massive stellar evolution. Inferring the birth properties of neutron stars and the physics that govern their subsequent evolution is, thus, crucial for other fields of astrophysics.

Detecting and accurately characterizing individual objects within the entire neutron-star population is, hence, critical. As a result, the number of known pulsars (those neutron stars that emit regular electromagnetic pulses) has steadily increased since the first detection in 1967 \citep{Hewish1968} and we currently know around 3,000 of these objects \citep{Manchester2005}.\footnote{\url{https://www.atnf.csiro.au/research/pulsar/psrcat/}} These are visible across the full electromagnetic spectrum and their emission is predominantly driven by their enormous rotational energy reservoirs. Roughly 300 of these sources are in binaries. They were strongly influenced by accretion from their companions and spun up to short rotation periods earlier in their lives. The remaining sources are isolated neutron stars. Due to observational limitations and diverse emission properties, we cannot detect these with a single telescope, but instead have to focus on certain subpopulations. With around 1,100 members, a subset of isolated radio pulsars constitutes the largest fraction of neutron stars detected in a single survey \citep{Posselt2023}. However, these numbers only cover a tiny portion of the approximately one billion neutron stars expected in our Milky Way alone.

To bridge the gap between expected and observed neutron stars, we take advantage of population synthesis (see also Section~\ref{sec:ch1_popsyn} and Chapter~\ref{Chapter5}). This approach relies on producing a large catalogue of synthetic pulsar populations which are passed through a set of filters to mimic observational constraints. The resulting populations are then contrasted with the true observed sample to find those parameter regions that best explain the data. Although different versions of this methodology have been applied to pulsar data for several decades \citep[e.g.,][]{Narayan1990, Lorimer2004, Faucher2006, Gonthier2007, Bates2014, Gullon2014, Gullon2015, Cieslar2020}, the complexity of models that capture the properties of observed Galactic neutron stars significantly complicates the comparison between the simulated populations and the observed one. This is especially true if we are interested in quantifying uncertainties for our neutron-star parameters, because Bayesian \ac{MCMC} methods \citep[the standard tool for this kind of question, see, e.g.,]{Feroz2009, Foreman-Mackey2013, Sharma2017, Ashton2019, Speagle2020} become infeasible for pulsar population synthesis unless significant simplifications for simulation models and the likelihood function are made \citep{Cieslar2020}. The main reason for this is that we can no longer write down an explicit likelihood for realistic neutron-star simulation frameworks. In this chapter, we, thus, focus on \ac{SBI} \citep[also known as likelihood-free inference; see Chapter~\ref{Chapter2} and][for a recent review]{Cranmer2020} in the context of pulsar population synthesis for the first time.

In the past few years, \ac{SBI} has successfully challenged traditional approaches such as approximate Bayesian computation \citep[e.g.,][]{Rubin1984, Beaumont2002, Dean2011, Frazier2017} in those areas of science that rely on complex simulators which lead to intractable likelihoods. The existence of such a simulator, essentially acting as a forward model, is the only requirement for \ac{SBI}. As such, the approach is ideal for astrophysics and has been recently applied to parameter estimation in, e.g., cosmology \citep{Alsing2019, Lemos2023, Lin2023, Hahn2023}, high-energy astrophysics \citep{Mishra-Sharma2022, Huppenkothen2022}, gravitational-wave astronomy \citep{Dax2021, Cheung2022, Bhardwaj2023} and exoplanet research \citep{Vasist2023}. \ac{SBI} is particularly powerful in combination with neural networks, whose benefits for pulsar population synthesis studies was outlined in \citet{Ronchi2021} (see Chapter~\ref{Chapter5}) by inferring point estimates for the dynamical properties of radio pulsars in the Milky Way.

In this study, we take a Bayesian perspective to infer posteriors of neutron-star parameters using \ac{SBI}. For this purpose, we model the Galactic neutron-star dynamics, the magneto-rotational evolution and the radio emission properties. We then run snapshots of the total pulsar population at the current time through a set of filters to mimic observational limitations. The resulting simulation output are synthetic $P$-$\dot{P}$ diagrams (where $P$ and $\dot{P}$ denote the pulsar spin period and its time derivative, respectively) of the observed pulsar population. We then construct an \ac{SBI} pipeline, which we train, validate and test on a large database of these synthetic $P$-$\dot{P}$ diagrams to infer posterior distributions of our input parameters. We specifically focus on five parameters related to the initial period distribution of pulsars and their magnetic-field properties that crucially affect the positions of stars in the $P$-$\dot{P}$ plane. We then apply our optimized deep-learning framework, for the first time, to the radio pulsars detected in the \ac{PMPS} \citep{Manchester2001, Lorimer2006}, the \ac{SMPS} \citep{Edwards2001, Jacoby2009} and the low- and mid-latitude \ac{HTRU} survey \citep{Keith2010} (all recorded with Murriyang, the Parkes radio telescope).

This Chapter based on the work \citet*{Graber2023} is structured as follows: Sec.~\ref{sec:ch6_popsyn} summarizes our population synthesis framework. We then provide a general overview of \ac{SBI} as well as our choice of set-up in Secs.~\ref{sec:ch6_sbi_overview} and \ref{sec:ch6_architecture}, respectively, whereas Sec.~\ref{sec:ch6_experiments} summarizes the machine-learning experiments conducted for this study. We next address network training and inference results plus corresponding validation approaches in Sec.~\ref{sec:ch6_results}, specifically benchmarking our pipeline on test simulations before applying it to the observed pulsar population. Finally, we provide a detailed discussion of our approach and results as well as an outlook into the future in Sec.~\ref{sec:ch6_conclusions}.

%%%%%%%%%%%%%%%%%%%%%%%%%%%%%%%%%%%%%%%%%%%%%%%%%%%%%%%%%%%%%%%%%%%%%%%%%%%%%%%%%%%%%%%%%%%%%%%%%%%%%%%%%%%%%%%%%%%%%%%%%%%%%%

\section{Pulsar population synthesis}
\label{sec:ch6_popsyn}

%--------------------------------------------------------------
\begin{figure*}
	\centering
	\includegraphics[width=0.85\textwidth]{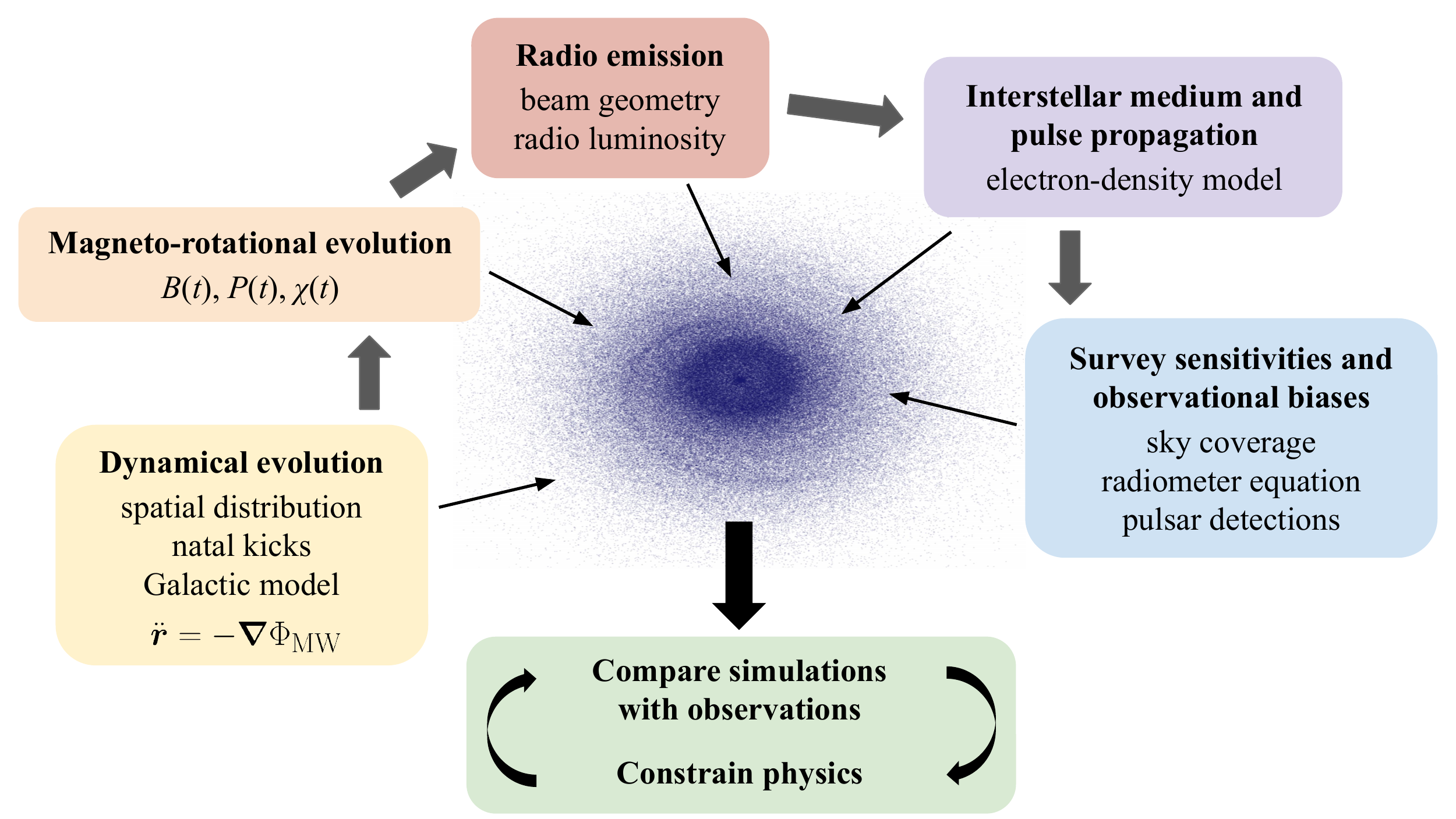}
	\caption[The key ingredients for pulsar population synthesis]{The key ingredients for pulsar population synthesis. Starting from the bottom left, this approach relies on modelling the neutron stars' dynamical evolution as well as their magneto-rotational properties. For a given beaming geometry and luminosity model, we then determine the pulsars' radio emission and its propagation across the Galaxy towards the Earth. For the neutron stars pointing towards us, we subsequently invoke survey limitations and sensitivity thresholds to determine those objects that are detectable. The resulting synthetic populations are compared to the observed ones to constrain input physics.}
	\label{fig:Pop_flowchart}
\end{figure*}
%--------------------------------------------------------------

Our population synthesis framework broadly follows earlier works (see, e.g., \citet{Faucher2006, Gullon2014, Cieslar2020, Ronchi2021}). In particular, our simulator employs a Monte-Carlo approach to sample the relevant neutron-star parameters at birth from corresponding probability density functions and subsequently evolves these parameters forward in time. However, to save computation time, we do not follow the procedure in Chapter~\ref{Chapter5} \citep[see also][]{Ronchi2021} and evolve the dynamical properties for each simulation run. As the dynamical and magneto-rotational properties are independent, we instead simulate a single dynamical database for a large number of current pulsar positions and velocities, and subsequently sample from these distributions before determining the magneto-rotational evolution. Prescriptions for both are summarised below in Sections~\ref{sec:ch6_dyn_evol} and \ref{sec:ch6_mr_evol}, respectively. We subsequently discuss the radio emission properties in Section~\ref{sec:ch6_emission_phys} and then run snapshots of the total pulsar population at the current time through a set of filters to mimic observational limitations (see Section~\ref{sec:ch6_obs_lims}). The resulting simulator output are synthetic $P$-$\dot{P}$ diagrams of the observed pulsar population (see Section~\ref{sec:ch6_sim_output}) that will serve as the starting point for our \ac{SBI} pipeline discussed in Section~\ref{sec:ch6_sbi}.

%%%%%%%%%%%%%%%%%%%%%%%%%%%%%%

\subsection{Dynamical evolution}
\label{sec:ch6_dyn_evol}

To create our dynamical database, we simulate $10^7$ neutron stars from birth to today. For each object, we randomly assign an age sampled from a uniform distribution up to a maximum age of $10^{8} \,$yr, which ensures a constant birth rate of $10$ stars per century. Note that this value, although somewhat higher than typically assumed \citep[e.g.,][]{Keane2008, Rozwadowska2021}, was chosen to populate our synthetic Milky Way with a sufficient number of neutron stars within reasonable computation time. As objects older than $10^{8} \,$yr are no longer detectable as radio pulsars (see below), the following approach provides a realistic description of the current positions of these sources.

We then define a cylindrical reference frame, $(r, \phi, z)$, whose origin is located at the Galactic centre. Here, $r$, $\phi$ and $z$ denote the distance from the origin in kpc, the azimuthal angle in radians and the distance from the Galactic plane in kpc, respectively. In particular, we position our Sun at $r=\unit[8.3]{kpc}$, $\phi= \pi/2$, and $z=\unit[0.02]{kpc}$ \citep[see][and references therein]{Pichardo2012}.

To determine the birth locations of individual neutron stars, we address the distributions of their massive progenitors in the $(r, \phi)$-plane and along $z$ separately. Considering the distribution of free electrons as a tracer of star formation in the Milky Way which correlates with the massive OB stars that evolve into neutron stars, we sample the initial positions in $r, \phi$ according to the Galactic electron density distribution of \cite{Yao2017} (see Figure~\ref{fig:ch1_e_density_model}). This will also allow consistency when relating pulsar distances with their dispersion measures \acp{DM} in Section~\ref{sec:ch6_obs_lims}. In addition, as the Galactic matter distribution is not static, we assume that the Milky Way rotates rigidly in clockwise direction with an angular velocity $\Omega = 2 \pi /T$, where $T \approx \unit[250]{Myr}$ \citep{Vallee2017, Skowron2019}. For a given stellar age, we can thus retrace the angular coordinate, $\phi$, at birth.

Moreover, we assume that pulsar birth positions along the $z$-direction follow an exponential disk model \citep{Wainscoat1992} and sample from a probability density function of the form
%----------------------------------------------------------------------------------------------
\begin{align}
\mathcal{P}(z) = \frac{1}{h_{\rm c}} \exp\left(-\frac{ \lvert z \rvert}{ h_{\rm c} } \right).
\end{align}
%----------------------------------------------------------------------------------------------
We follow the pulsar population studies of Chapter~\ref{Chapter5} \citep[see also][]{Gullon2014, Ronchi2021} and set the characteristic scale height, $h_{\rm c}$, to a fiducial value of $\unit[0.18]{kpc}$. Note that this is consistent with the distribution of young, massive stars in our Galaxy \citep{Li2019}. We then randomly assign each star's $z$-coordinate a positive or negative sign to distribute our population above and below the Galactic plane.

Next, we focus on the pulsars' birth velocities, which are a combination of the kick velocity, $\boldsymbol{v}_{\rm k}$, imparted during the supernova due to explosion asymmetries \citep[see][and references therein]{Janka2022, Coleman2022}, and the velocity, $\boldsymbol{v}_{\rm pr}$, inherited from the progenitors' orbital Galactic motion. Specifically, we sample the magnitude of the kick velocities, $v_{\rm k} \equiv |\boldsymbol{v}_{\rm k}|$, from a Maxwell distribution,
%-------------------------------------------------------------
\begin{equation}
\mathcal{P}(v_{\rm k}) = \sqrt{ \frac{2}{\pi} }  \frac{v_{\rm k}^2}{\sigma_{\rm k}^3}
\exp\left(-\frac{v_{\rm k}^2}{ \sigma_{\rm k}^2 } \right),
\label{eq:ch6_pdf_maxwell_kick}
\end{equation}
%-------------------------------------------------------------
and then assign a random direction to determine the kick along the $r$-, $\phi$- and $z$-directions. For the dispersion parameter, $\sigma_{\rm k}$, we take a fiducial value of $\sigma_{\rm k} \approx \unit[260]{km \, s^{-1}}$ \citep{Hobbs2005}, which is broadly consistent with observed proper motions of radio pulsars \citep{Hobbs2005, Faucher2006, Verbunt2017, Igoshev2020}.

The second velocity component due to the progenitors' motion depends on the Galactic gravitational potential, $\Phi_{\rm MW}$, and points along the azimuthal direction:
%-------------------------------------------------------------
\begin{align}
\boldsymbol{v}_{\rm pr} = \sqrt{ r \, \frac{\partial \Phi_{\rm MW} \left( r,z \right)}{\partial r} } \, \hat{\boldsymbol{\phi}},
\end{align}
%-------------------------------------------------------------
where $\hat{\boldsymbol{\phi}}$ is a unit vector in $\phi$-direction. For this study, we consider a Galactic potential that is given as the sum of four components, i.e., the nucleus, $\Phi_{\rm n}$, the bulge, $\Phi_{\rm b}$, the disk, $\Phi_{\rm d}$, and the halo, $\Phi_{\rm h}$, \citep{Marchetti2019}. The nucleus and bulge contributions are described by a spherical Hernquist potential \citep{Hernquist1990}:
%-------------------------------------------------------------
\begin{equation}
\Phi_{\rm n, b} = -\frac{ G M_{\rm n, b}}{ R_{\rm n, b} +  R},
\end{equation}
where $R = \sqrt{r^2 + z^2}$ is the spherical radial coordinate and $G$ the gravitational constant. The disk has a cylindrical Miyamoto-Nagai potential of the form \citep{Miyamoto1975}
%-------------------------------------------------------------
\begin{equation}
\Phi_{\rm d} = -\frac{ G M_{\rm d}}{ \sqrt{ \left(a_{\rm d} + \sqrt{z^2+b_{\rm d}^2}\right)^2 + r^2} },
\end{equation}
%-------------------------------------------------------------
where $a_{\rm d}$ and $b_{\rm d}$ represent the scale length and scale height of the disk, respectively. Finally, the halo is characterised by a spherical Navarro-Frenk-White potential \citep{Navarro1996}:
%-------------------------------------------------------------
\begin{equation}
\Phi_{\rm h} = -\frac{ G M_{\rm h}}{ R } \ln{ \left( 1 + \frac{R}{R_{\rm h}}\right) }.
\end{equation}
%-------------------------------------------------------------
The free parameters, $M_{\rm n, b, d, h}$, $R_{\rm n,b,h}$, $a_{\rm d}$ and $b_{\rm d}$, can be obtained through fits of the Milky Way's mass profile and are given in Table~\ref{tab:ch5_MW_pot_params_M19} (see also Tab.~1 of \citet{Marchetti2019} and \citet{Bovy2015}).

After determining the initial positions and velocities for each of our $10^7$ neutron stars, we perform the dynamical evolution by solving the Newtonian equation of motion in cylindrical coordinates, $\ddot{\boldsymbol{r}} = -\boldsymbol{\nabla} \Phi_{\rm MW}$, according to the stars' respective ages. In this way, we obtain a database of current pulsar positions and velocities in the Milky Way.

%%%%%%%%%%%%%%%%%%%%%%%%%%%%%%

\subsection{Magneto-rotational evolution}
\label{sec:ch6_mr_evol}

The primary diagnostic for the pulsar population is the $P$-$\dot{P}$ diagram. For our population synthesis study, we focus on rotation-powered radio pulsars, the largest class of neutron stars (see Section~\ref{sec:ch1_ns_zoo}). Corresponding period and period-derivative measurements for this population are enabled via radio timing. To first order, radio pulsars can be approximated as rotating magnetic dipoles, implying that their spin-down is driven by electromagnetic dipole radiation (see Section~\ref{sec:ch1_dip_spindown_evol}). The locations of individual neutron stars, and the shape of the population's distribution, in the $P$-$\dot{P}$ plane are, hence, determined by their dipolar magnetic fields and rotation periods at birth, and the subsequent magneto-rotational evolution. The latter couples the evolution of the pulsar period, $P$, the dipolar magnetic-field strength, $B$, at the pole and the inclination angle, $\chi$, between the magnetic and the rotation axis.

To capture these physics, we first sample the misalignment angle at birth, $\chi_0$, randomly in the range $[0, \pi/2]$ according to the probability density \citep{Gullon2014}
%--------------------------------------------------------------
\begin{equation}
\mathcal{P}(\chi_0) = \sin \chi_0.
\end{equation}
%--------------------------------------------------------------
We then sample the logarithm of the initial period, $P_0$, (measured in s) and the initial magnetic field, $B_0$, (measured in G) for each pulsar from normal distributions of the form \citep{Popov2010, Gullon2014, Igoshev2020, Igoshev2022, Xu2023}
%--------------------------------------------------------------
\begin{align}
\mathcal{P}(\log P_0) = \frac{1}{\sqrt{2 \pi} \sigma_{\log P}}
\, \exp\left(-\frac{\log P_0 - \mu_{\log P}}{2 \sigma_{\log P}^2} \right),
\label{eq:ch6_P_pdf} \\[1.8ex]
\mathcal{P}(\log B_0) = \frac{1}{\sqrt{2 \pi} \sigma_{\log B}}
\, \exp\left(-\frac{\log B_0 - \mu_{\log B}}{2 \sigma_{\log B}^2} \right).
\label{eq:ch6_B_pdf}
\end{align}
%--------------------------------------------------------------
The means, $\mu_{\log P}, \mu_{\log B}$, and the standard deviations, $\sigma_{\log P}, \sigma_{\log B}$, are free parameters of our model and four of those parameters, whose posteriors we set out to infer with our \ac{SBI} approach in Section~\ref{sec:ch6_sbi}. We will specifically explore the ranges $\mu_{\log P} \in [-1.5, -0.3]$, $\mu_{\log B} \in [12, 14]$, $\sigma_{\log P} \in [0.1, 1.0]$ and $\sigma_{\log B} \in [0.1, 1.0]$ to encompass results of earlier analyses \citep[e.g.,][]{Gullon2014}.

Assuming that pulsars spin down due to dipolar emission, we follow \citet{Philippov2014, Spitkovsky2006} and solve the following coupled differential equations (see Equations~\eqref{eq:ch1_rot_evol_forcefree_2}):
%--------------------------------------------------------------
\begin{align}
\dot{P} &= \frac{\pi^2}{c^3}\frac{B^2 R_{\rm NS}^6}{I_{\rm NS} P} \left( \kappa_0 + \kappa_1 \sin^2 \chi \right),
\label{eq:ch6_P_ode} \\[1.8ex]
\dot{\chi} &= -\frac{\pi^2}{c^3}\frac{B^2 R_{\rm NS}^6}{I_{\rm NS} P^2} \, \kappa_2 \sin\chi \cos\chi,
\label{eq:ch6_chi_ode}
\end{align}
%--------------------------------------------------------------
where $c$ is the speed of light, $R_{\rm NS} \approx \unit[11]{km}$ the neutron-star radius and $I_{\rm NS} \simeq 2 M_{\rm NS} R_{\rm NS}^2 / 5 \approx \unit[1.36 \times 10^{45}]{g \, cm^{2}}$ the stellar moment of inertia (for a fiducial mass $M_{\rm NS} \approx \unit[1.4]{M_{\odot}}$). For realistic pulsars surrounded by plasma-filled magnetospheres, we choose $\kappa_0 \simeq \kappa_1 \simeq \kappa_2 \simeq 1$, and note that Equation~\eqref{eq:ch6_chi_ode} implies that $\chi$ decreases with time, i.e., our pulsars move towards alignment.

%--------------------------------------------------------------
\begin{figure}
	\centering
	\includegraphics[width=0.7\columnwidth]{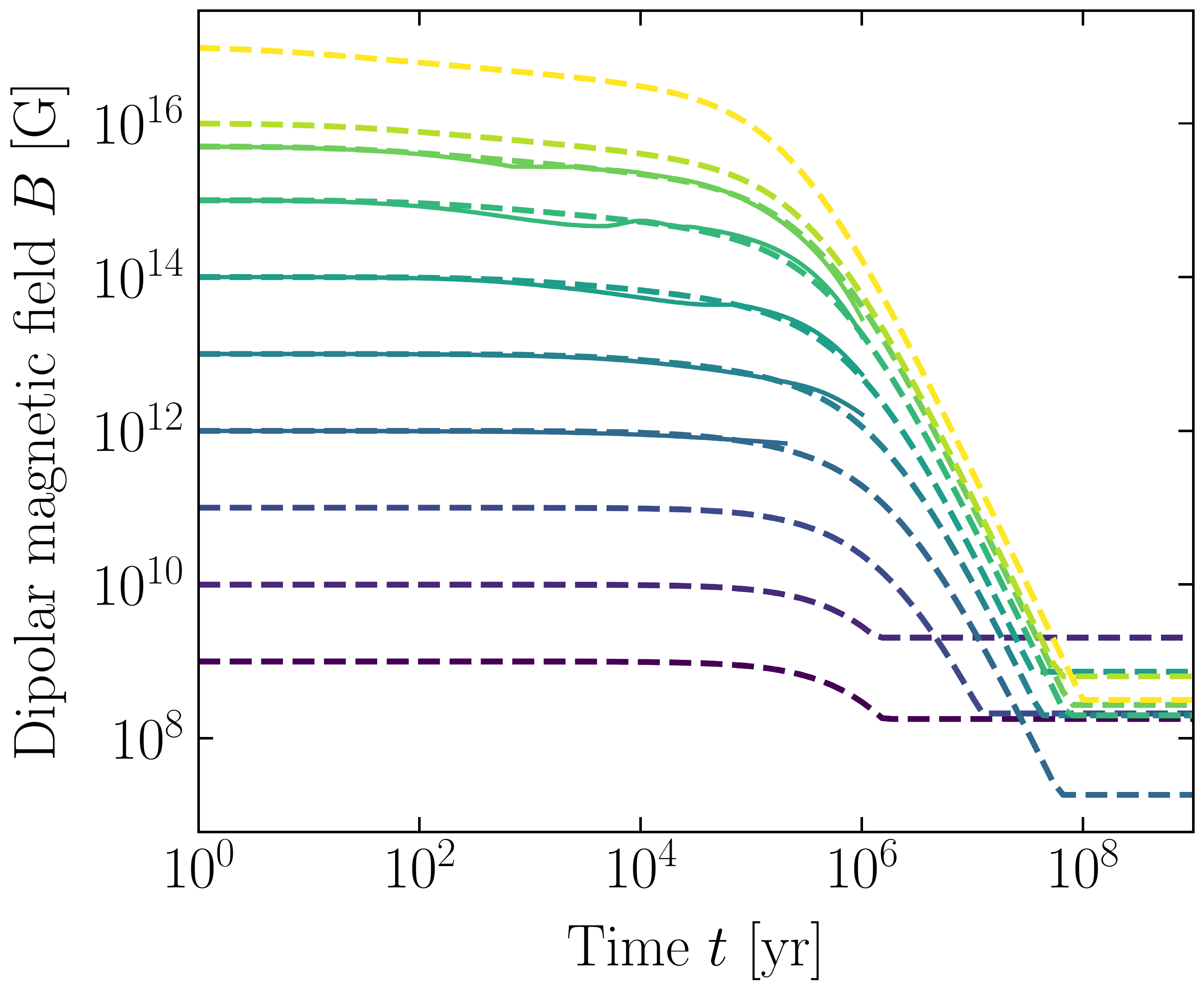}
	\caption[Illustration of the $B$-field parametrisation]{Illustration of the $B$-field parametrisation used for this study. The five solid curves represent realistic two-dimensional simulations of magneto-thermal evolution in the neutron-star crust \citep{Vigano2021}. We fit these together with the late-time power-law evolution of the magnetic field with several broken power laws. The dashed curves shown here are determined for $a_{\rm late} = -3.0$. The colours represent the initial magnetic-field strength, $B_0$. To avoid the field decaying to unrealistically small numbers at very late times, we sample the final fields from a Gaussian distribution. The procedure, which allows us to easily extract the dipolar field strength, $B$, at different times, $t$, to study the magneto-rotational evolution of our synthetic pulsars, is described in details in Appendix~\ref{app:B-field}.}
	\label{fig:ch6_B_fields}
\end{figure}
%--------------------------------------------------------------

The final ingredient is a suitable prescription for the evolution of the dipolar magnetic-field strength. While the $B$-field decay in the neutron-star crust is typically assumed to be driven by the combined action of the Hall effect and Ohmic dissipation \citep[e.g.,][]{Aguilera2008b}, changes in the magnetic field are strongly coupled to the thermal properties of the neutron-star interior \citep[e.g.,][]{Pons2019}. This is particularly important for strongly magnetised neutron stars with fields above $\sim \unit[10^{13}]{G}$ and, hence, relevant for a significant fraction of our simulated pulsar population. In the past decade, several theoretical and numerical efforts have begun to unveil the complex processes of magneto-thermal evolution \citep[e.g.,][see also Section~\ref{sec:ch1_B_evol}]{Vigano2013, Vigano2021, DeGrandis2021, Igoshev2021, Dehman2023}. As corresponding simulations are highly time-consuming, we instead develop a new approach, outlined in detail in Appendix \ref{app:B-field} and summarised in Figure~\ref{fig:ch6_B_fields}, that parametrises a range of magneto-thermal simulations for different magnetic-field strengths \citep{Vigano2021}. This prescription allows us to extract magnetic fields up to pulsar ages of around $\unit[10^6]{yr}$. Above this value, current numerical simulations become unreliable because they rely on implementations of complex microphysics that are unsuitable for cold, old stars. Moreover, they do not capture the highly uncertain physics of neutron-star cores, which become relevant at large ages. We instead incorporate the field evolution at late times by means of a power law of the form
%--------------------------------------------------------------
\begin{equation}
B(t) \propto \left(1+ \frac{t}{\tau_{\rm late}} \right)^{a_{\rm late}},
\label{eq:ch6_B_late}
\end{equation}
%--------------------------------------------------------------
where $\tau_{\rm late} \approx \unit[2 \times 10^6]{yr}$, $t$ is the time, and the power-law index, $a_{\rm late}$, is the fifth free parameter of our model. We note that although the details of core field evolution are not known, Equation~\eqref{eq:ch6_B_late} is physically motivated because several known mechanisms exhibit similar power-law behaviour (see Appendix \ref{app:B-field}). We will, hence, explore the parameter range $a_{\rm late} \in [-3.0, -0.5]$. Finally, to prevent the dipolar magnetic field from decaying to arbitrarily small values (in disagreement with observations of old, recycled millisecond pulsars; see, e.g., \citet{Lorimer2008}), we assume that the field eventually reaches a constant value. Therefore, we sample the logarithm of the field, $B_{\rm final}$, from a normal distribution with mean $\mu_{\log B, {\rm final}} = 8.5$ and a standard deviation $\sigma_{\log B, {\rm final}} = 0.5$ in line with observations of old pulsars.

Following this prescription allows us to determine the spin periods, dipolar field strengths and misalignment angles for our simulated pulsars at the current time.

%%%%%%%%%%%%%%%%%%%%%%%%%%%%%%

\subsection{Emission characteristics}
\label{sec:ch6_emission_phys}

We next implement a prescription for the radio emission geometry to determine those pulsars whose beams sweep over the Earth and are, in principle, detectable. In the \textit{canonical} model of radio pulsars, their emission is produced close to the stellar surface in the cone-shaped, open field-line region \citep[][see also Section~\ref{sec:ch1_radio_em_geometry}]{Lorimer2012, Johnston2020}. Assuming that this entire region is involved in the emission, geometric considerations allow us to estimate the half opening angle of the emission beam, $\rho_{\rm b}$, (in rad) via \citep[see Equation~\eqref{eq:ch1_beam_angular_aperture} and][]{Gangadhara2001}
%-----------------------------------------------------------------
\begin{equation}
\rho_{\rm b} \simeq \sqrt{\frac{9 \pi r_{\rm em}}{2 c P}},
\label{eq:ch6_rho_b}
\end{equation}
%-----------------------------------------------------------------
where $r_{\rm em}$ is the emission height. The latter is thought to be period independent and we set it to $\unit[300]{km}$ following \citet{Johnston2020} (see also references therein). Note that several studies of pulsars with stable emission profiles have recovered this $\rho_{\rm b} \propto P^{-1/2}$ behaviour \citep[e.g.,][and see also Figure~\ref{fig:ch1_w_P_relation}]{Kramer1994, Maciesiak2011b, Skrzypczak2018}. Knowledge of $\rho_{\rm b}$, then, allows us to obtain the solid angle, $\Omega_{\rm b}$, covered by a pulsar's two radio beams (see Equation~\eqref{eq:ch1_solid_angle}). More specifically,
%-----------------------------------------------------------------
\begin{equation}
\Omega_{\rm b} = 4 \pi (1 - \cos \rho_{\rm b}).
\label{eq:ch6_omega_b}
\end{equation}
%-----------------------------------------------------------------
As we do not expect biases in how we observe this conal emission for any given pulsar, we draw a random line-of-sight angle, $\alpha$, with respect to the rotation axis in the range $[0, \pi/2]$ using the probability density $\sin \alpha $. Combined with the half opening angle, $\rho_{\rm b}$, and the evolved inclination angle, $\chi$, we can then determine those pulsars whose radio beams are visible from Earth. We note that as a result of this purely geometric argument, between $\sim 60-95\%$ of our generated pulsars (depending on the specific choice of magneto-rotational parameters) are typically not detectable.

We proceed with determining the emission characteristics of those neutron stars that point towards the Earth. In particular, we follow \citet{Maciesiak2011a} and express the intrinsic pulse width (measured in s) of our simulated pulsars as follows (see Equation~\eqref{eq:ch1_pulse_width}):
%--------------------------------------------------------------
\begin{equation}
w_{\rm int} = \frac{2}{\pi} \arcsin{ \sqrt{ \frac{\sin^2\left( \frac{\rho_{\rm b}}{2} \right) 
			- \sin^2\left( \frac{\alpha - \chi}{2} \right)}
		{\sin \left( \alpha \right) \sin \left(\chi \right)}}} \, P,
\label{eq:ch6_instr_pulsewidth} 
\end{equation}
%--------------------------------------------------------------
where we replace $\beta = \alpha - \chi$. 
Finally, as the radio emission is ultimately powered by the stars' rotational energy reservoir, we follow the common procedure \citep[e.g.,][]{Faucher2006} and assume that the intrinsic radio luminosity, $L_{\rm int}$, (in ${\rm erg} \, {\rm s}^{-1}$) for each pulsar depends on the spin-down power, $\dot{E}_{\rm rot} \propto \dot{P} / P^3$ (see Section~\ref{sec:ch1_energetics}). In particular, we assume
%--------------------------------------------------------------
\begin{equation}
L_{\rm int} = L_0 \sqrt{ \frac{\dot{P}}{P^3}}. 
\end{equation}
%--------------------------------------------------------------
where $L_0$ is a normalisation factor whose logarithm we sample from a normal distribution with mean $\mu_{\log L} = 35.5$ and standard deviation $\sigma_{\log L} = 0.8$ to eventually recover observed luminosities \citep[see also][]{Faucher2006, Gullon2014}. 

%---------------------------------------------------------------
\begin{table}
	\centering
	\caption[Survey parameters for \acs{PMPS}, \acs{SMPS}, \acs{HTRU} surveys]{Survey parameters for the \acf{PMPS}, the \acf{SMPS}, the low- and mid-latitude \acf{HTRU} surveys taken from \citet{Manchester2001, Lorimer2006}, \citet{Edwards2001, Jacoby2009} and \citet{Keith2010}, respectively. We provide the survey region in Galactic longitude, $l$, and latitude, $b$, the central observing frequency $f$, the channel width $\Delta f$, the sampling time $\tau_{\rm samp}$, the telescope gain, $G$, the number of observed polarisations, $n_{\rm pol}$, the observing bandwidth $\Delta f_{\rm bw}$, the integration time $t_{\rm obs}$, the degradation factor $\beta$, the system temperature, $T_{\rm sys}$, and the $S/N$ threshold for each of the surveys. Corresponding units are given in brackets in the first column. \label{tab:ch6_SurveyParam}}
	\centering
	\small
	\begin{tabular}{c | c c c c}
		\toprule
		\tabhead{Survey} &
		\tabhead{PMPS} &
		\tabhead{SMPS} &
		\tabhead{HTRU mid} &
		\tabhead{HTRU low} \\
		\midrule
		sky region & $-100^{\circ} < l < 50^{\circ}$ & $-100^{\circ} < l < 50^{\circ}$ & 
		$-120^{\circ} < l < 30^{\circ}$ & $-80^{\circ} < l < 30^{\circ}$  \\
		& $|b| < 5^{\circ}$ & $5^{\circ}<|b| < 30^{\circ}$ & $|b| < 15^{\circ}$ &  $|b| < 3.5^{\circ}$  \\
		$f$ (GHz) & 1.374 & 1.374 & 1.352 & 1.352 \\
		$\Delta f$ (kHz) & 3000 & 3000 & 390.625 & 390.625 \\
		$\tau_{\rm samp}$ ($\mu$s) & 250 & 125 & 64 & 64 \\
		$G$ (K$\, {\rm Jy}^{-1}$) &  0.735 & 0.735 & 0.735 & 0.735 \\
		$n_{\rm pol}$ & 2 & 2 & 2 & 2 \\
		$\Delta f_{\rm bw}$ (MHz) & 288 & 288 & 340 & 340 \\
		$t_{\rm obs}$ (s) & 2100 & 265 & 540 & 4300 \\
		$\beta$ & 1.5 & 1.5 & 1.5 & 1.5 \\
		$T_{\rm sys}$ (K) & 21 & 21 & 23 & 23 \\
		$S/N$ threshold & 9 & 9 & 9 & 9 \\
		\bottomrule 
	\end{tabular}
\end{table}
%---------------------------------------------------------------

%%%%%%%%%%%%%%%%%%%%%%%%%%%%%%

\subsection{Simulating detections}
\label{sec:ch6_obs_lims}

Armed with the knowledge of intrinsic pulsar properties, we now turn to the possibility of detecting those objects whose emission beams cross our line of sight. Following Section~\ref{sec:ch1_propagation_ism}, first, the bolometric radio flux, $S$, that reaches us from any given simulated pulsar is equal to
%--------------------------------------------------------------
\begin{equation}
S = \frac{L_{\rm int}}{\Omega_{\rm b} d^2}, 
\end{equation}
%--------------------------------------------------------------
where $d$ is the distance known from the dynamical evolution outlined in Section~\ref{sec:ch6_dyn_evol}. To determine the corresponding radio flux density, $S_{f}$, (measured in Jy) at a specific observing frequency, $f$, we follow \citet{Lorimer2012} and assume that the radio emission spectrum follows a power law in $f$ (see Equation~\eqref{eq:ch1_flux_density}). In particular, we set the spectral index to $-1.6$ \citep{Jankowski2018}. We can, hence, approximate the total fluence of a pulse with width, $w_{\rm int}$, as $S_{f} w_{\rm int}$. Assuming that this fluence stays constant as the radio signal propagates from the pulsar to us, we estimate the flux density, $S_{f , {\rm obs}}$, that reaches Earth as
%--------------------------------------------------------------
\begin{equation}
S_{f , {\rm obs}} \simeq S_{f} \frac{ w_{\rm int}}{w_{\rm obs}}
\label{eq:ch6_obs_flux} 
\end{equation}
%--------------------------------------------------------------
where $w_{\rm obs}$ is the observed pulse width.

\begin{figure}
	\centering
	\includegraphics[width=0.8\textwidth]{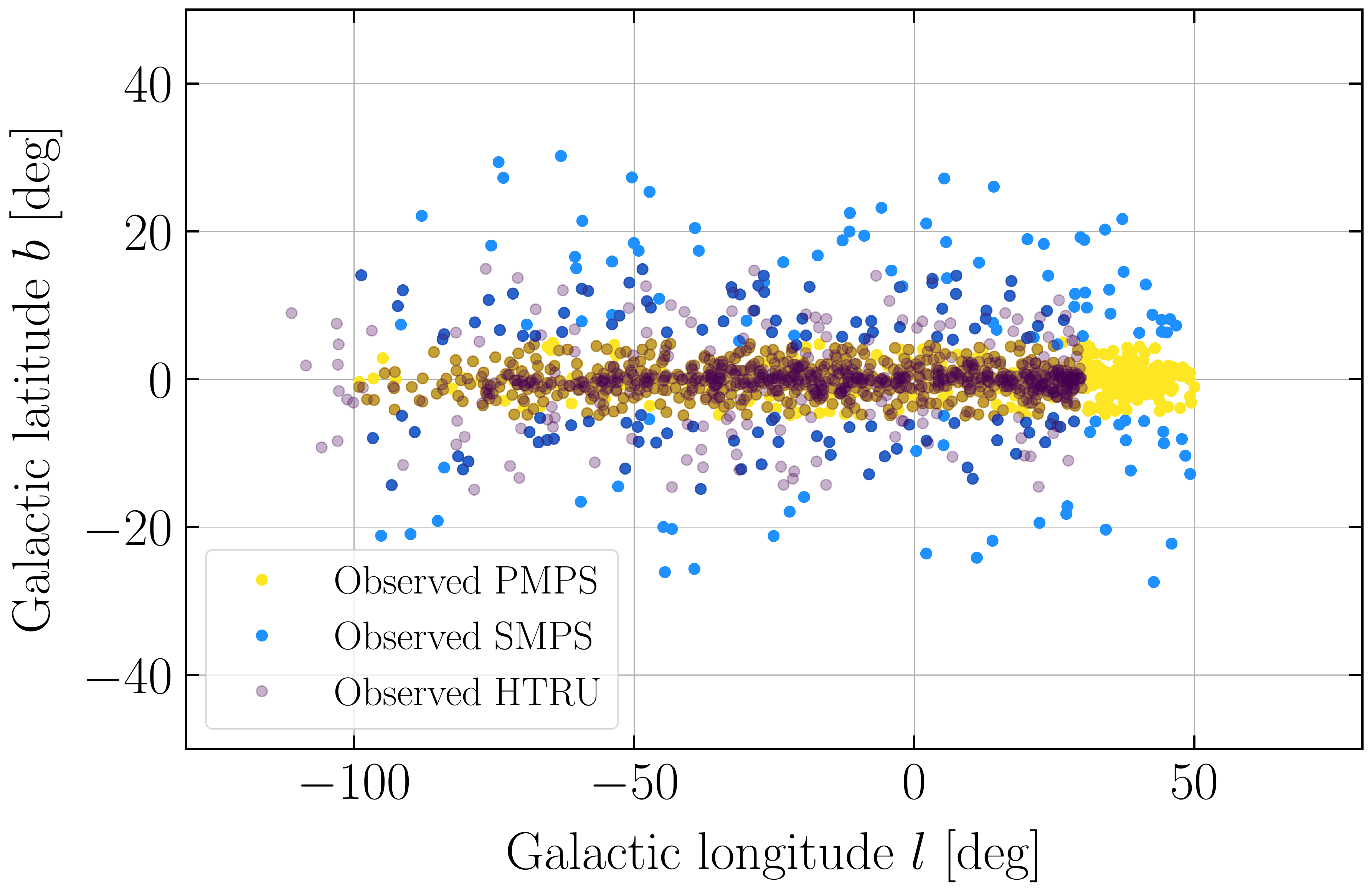}
	\hspace{0.2cm}
	\includegraphics[width=0.6\textwidth]{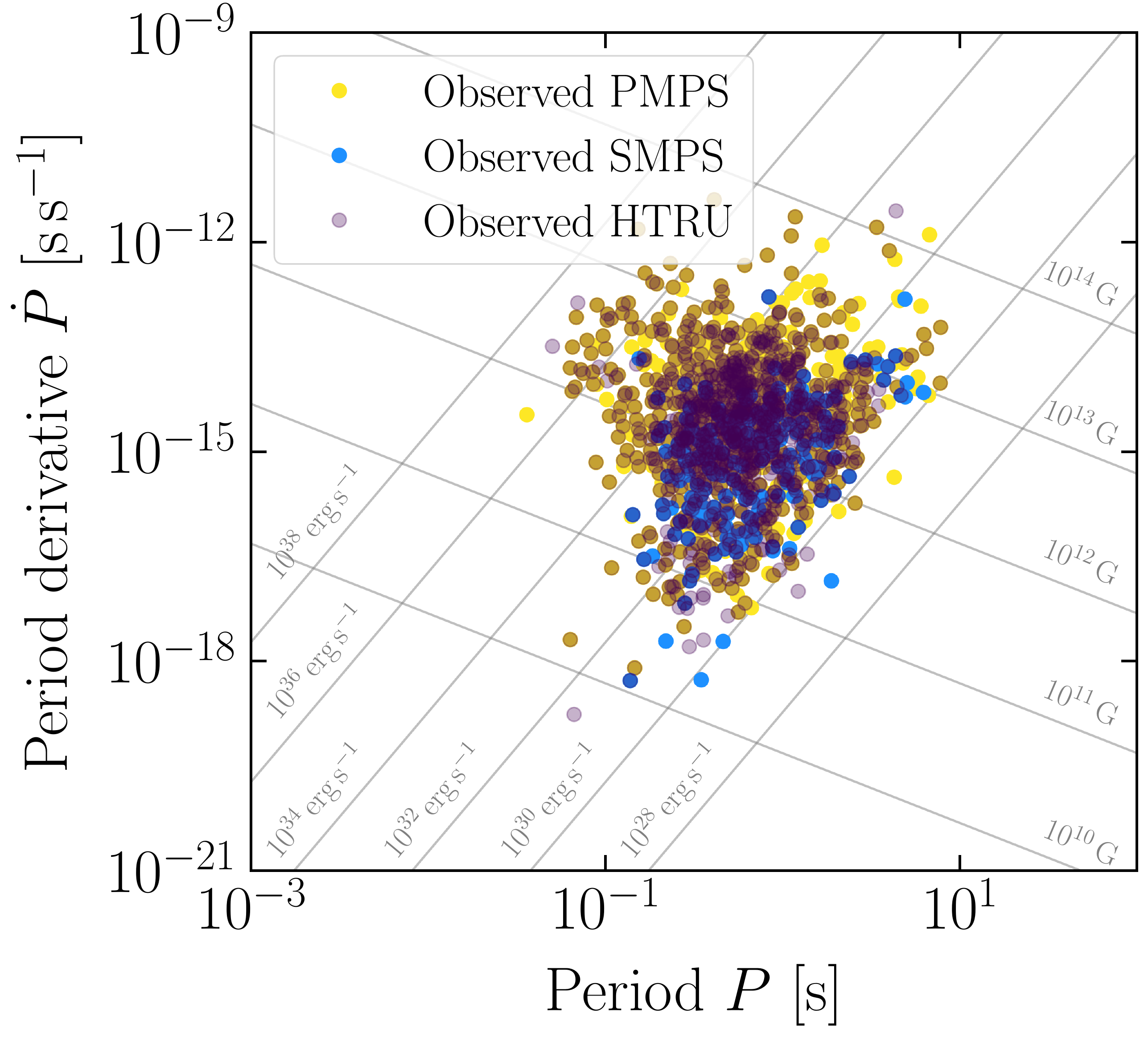}
	\caption[Observed populations of isolated Galactic radio pulsars detected with \acs{PMPS}, \acs{SMPS} and the \acs{HTRU} surveys]{Observed populations of isolated Galactic radio pulsars detected with the \acf{PMPS}, the \acf{SMPS} and the low- and mid-latitude \acf{HTRU} survey (highlighted in yellow, light blue and purple, respectively). The left panel shows the distribution of these three populations in Galactic latitude, $b$, and longitude, $l$, while the right panel depicts the detected pulsars in the period, $P$, and period derivative, $\dot{P}$, plane. In the latter, we also give lines of constant spin-down power, $\dot{E}_{\rm rot}$, and constant dipolar surface magnetic field, $B$, (estimated via Equation~\eqref{eq:ch6_P_ode} for an aligned rotator). Data taken from the ATNF Pulsar Catalogue \citep[][\url{https://www.atnf.csiro.au/research/pulsar/psrcat/}]{Manchester2005}. Observational filters are described in detail in the text.}
	\label{fig:ch6_pop_observed}
\end{figure}

Specifically, as a radio pulse propagates, it experiences dispersion and scattering caused by interactions with the free electrons and density fluctuations in the \ac{ISM}, respectively. Both mechanisms result in a broader pulse when compared with the intrinsic width, $w_{\rm int}$. Further broadening is caused by instrumental effects, which are dominated by the sampling time, $\tau_{\rm samp}$, of the hardware used to record radio observations (see Section~\ref{sec:ch1_propagation_ism}). Accounting for these processes, we can write the observed pulse width as \citep{Cordes2003}:
%--------------------------------------------------------------
\begin{equation}
w_{\rm obs} \simeq \sqrt{w_{\rm int}^2 + \tau_{\rm samp}^2 
	+ \tau_{DM}^2 + \tau_{\rm scat}^2}.
\label{eq:ch6_obs_pulsewidth} 
\end{equation}
%--------------------------------------------------------------
We follow \citet{Bates2014} to determine $\tau_{DM}$, encoding the pulse smearing due to dispersion for a single frequency channel of the telescope's receiver (see Equation~\ref{eq:ch1_smearing_DM}). Specifically,
%--------------------------------------------------------------
\begin{equation}
\tau_{DM} = \frac{e^2}{\pi m_e c} \, \frac{\Delta f_{\rm ch}}{f^3} \, DM,
\label{eq:ch6_tau_DM} 
\end{equation}
%--------------------------------------------------------------
where $e$ is the electronic charge, $m_{e}$ the corresponding mass, $\Delta f$ the hardware-specific width of a frequency channel at observing frequency, $f$, and $DM$ is the dispersion measure. We further use the empirical fit relationship from \citet{Krishnakumar2015} for $\tau_{\rm scat}$, the pulse smearing due to scattering of radio waves by an inhomogeneous and turbulent \ac{ISM}:
%--------------------------------------------------------------
\begin{align} \label{eq:ch6_tau_scat}
\tau_{\rm sc} = \unit[3.6 \times 10^{-9}]{s} \left(\frac{DM}{\unit[1]{pc \, cm^{-3}}} \right)^{2.2} \left[ 1.0 + 1.94 \times 10^{-3} \left(\frac{DM}{\unit[1]{pc \, cm^{-3}}} \right)^{2.0} \right].
\end{align}
%--------------------------------------------------------------
where $\tau_{\rm scat}$ is measured in s. We moreover account for a significant scatter in the underlying data \citep[see Fig.~3 in][]{Krishnakumar2015} by drawing $\log \tau_{\rm scat}$ values from a Gaussian distribution around the fit~\eqref{eq:ch6_tau_scat} with a standard deviation of 0.5. We also incorporate the fact that \citet{Krishnakumar2015} analysed observations at $\unit[327]{MHz}$ by rescaling to a given observing frequency, $f$, assuming a Kolmogorov spectrum, i.e., $\tau_{\rm scat} \propto f^{-4.4}$ \citep[see][for details]{Lorimer2012}. As both $\tau_{DM}$ and $\tau_{\rm scat}$ depend on the pulsars' respective dispersion measure, we again employ the Galactic electron density distribution of \cite{Yao2017} to convert our simulated neutron-star positions from Sec.~\ref{sec:ch6_dyn_evol} into $DM$ values. 

At this stage, we require information for the radio surveys we want to emulate. We specifically focus on three surveys recorded with Murriyang, the Parkes radio telescope: the \acf{PMPS} \citep{Manchester2001, Lorimer2006}, the \acf{SMPS} \citep{Edwards2001, Jacoby2009}, and the low- and mid-latitude \acf{HTRU} surveys \citep{Keith2010}. All relevant survey parameters (including the sampling time, $\tau_{\rm samp}$, the observing frequency, $f$, and the channel width, $\Delta f$, needed to calculate $w_{\rm obs}$) are summarised in Table~\ref{tab:ch6_SurveyParam}. 

To assess if those simulated stars that cross our line of sight are detectable with a given survey, we first determine if they are located in the surveys' fields of view. While \ac{PMPS} and \ac{HTRU} have a similar sky coverage, we highlight that \ac{SMPS} detected pulsars at higher Galactic latitude (see left panel of Figure~\ref{fig:ch6_pop_observed}). This survey is, thus, sensitive to older neutron stars which have had sufficient time to move away from their birth positions closer to the Galactic plane, providing complementary information on the pulsar population. For those objects that fall within our survey coverage, we subsequently establish if they are sufficiently bright to be detected. To do so, we calculate the pulsars' signal to noise ratio, $S/N$, using the radiometer equation \citep[see Section~\ref{sec:ch1_radiometer} and ][]{Lorimer2012}:
%--------------------------------------------------------------
\begin{equation}
S/N = \frac{ S_{f, \rm mean} G \sqrt{n_{\rm pol} \Delta f_{\rm bw} t_{\rm obs}} }
{ \beta \left[ T_{\rm sys} + T_{\rm sky }(l,b) \right] } 
\sqrt{\frac{P- w_{\rm obs}}{w_{\rm obs}}}.
\label{eq:ch6_radiometer}
\end{equation}
%--------------------------------------------------------------
Here, $S_{f, \rm mean} \simeq S_{f, {\rm obs}} w_{\rm obs} / P$ denotes the mean flux density averaged over a single rotation period $P$, $G$ is the receiver gain \citep[see][for details]{Lorimer1993, Bates2014}, $n_{\rm pol}$ is the number of detected polarisations, $\Delta f_{\rm bw}$ is the observing bandwidth, $t_{\rm obs}$ the integration time and $\beta > 1 $ a degradation factor that accounts for imperfections during the digitisation of the signal. Moreover, $T_{\rm sys}$ denotes the system temperature and $T_{\rm sky}(l,b)$ is the sky background temperature dominated by synchrotron emission of Galactic electrons which varies strongly with latitude, $l$, and longitude, $b$. To model the latter, we use results from \citet{Remazeilles2015}, who provided a refined version of the temperature map of \citet{Haslam1981, Haslam1982} (see Figure~\ref{fig:ch1_T_sky_distribution}. As the underlying data were obtained at 408 MHz, we rescale to the relevant observing frequencies by assuming a power-law dependence of the form $T_{\rm sky} \propto f^{-2.6}$ \citep{Lawson1987, Johnston1992}. 

A synthetic pulsar counts as detected, if the value obtained from Equation~\eqref{eq:ch6_radiometer} exceeds the surveys' sensitivity thresholds. We aim to recover the numbers of detected isolated Galactic radio pulsars for each survey, i.e. 
\begin{align}
&\text{\ac{PMPS}: $1009$ observed pulsars}, \nonumber \\
&\text{\ac{SMPS}: $218$ observed pulsars}, \label{eq:ch6_detected_objects} \\
&\text{\ac{HTRU}: $1023$ observed pulsars}. \nonumber
\end{align}
To obtain these values, we used the data from the ATNF Pulsar Catalogue \citep{Manchester2005}\footnote{\url{https://www.atnf.csiro.au/research/pulsar/psrcat/}} and removed extragalactic sources and those in globular clusters. We further applied a cut-off in period ($P > \unit[0.01]{s}$) and period derivative ($\dot{P} > \unit[10^{-19}]{s \, s^{-1}}$; for those pulsars with measured $\dot{P}$ values) to remove those objects that have (likely) undergone recycling from a companion star and cannot be modeled with the framework discussed so far. The locations of the observed objects in the $P$-$\dot{P}$ plane are shown in the right panel of Figure~\ref{fig:ch6_pop_observed}.

%%%%%%%%%%%%%%%%%%%%%%%%%%%%%%

%--------------------------------------------------------------
\begin{figure}
	\centering
	\includegraphics[height=0.44\columnwidth]{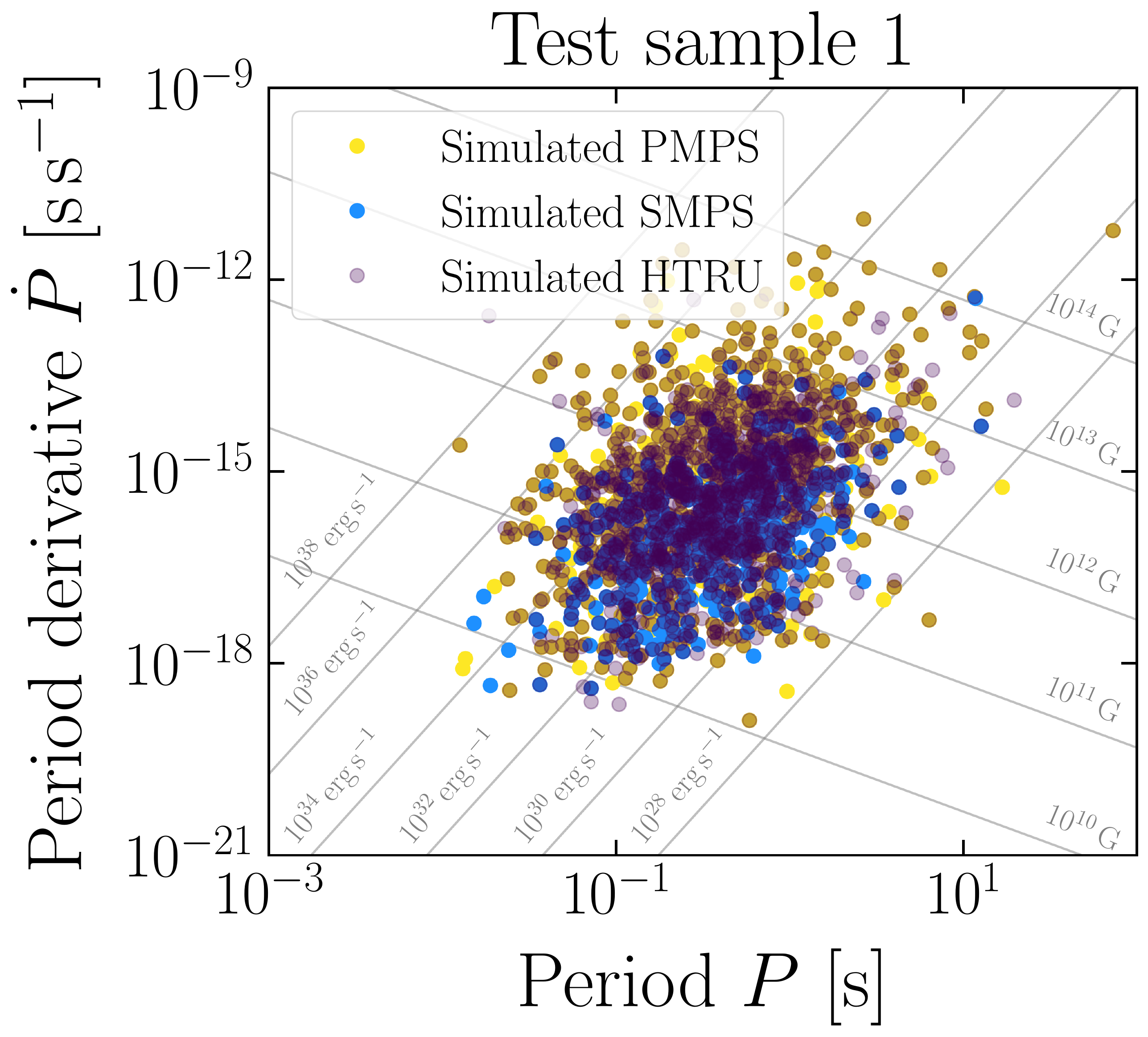}
	\hspace{0.1cm}
	\includegraphics[height=0.44\columnwidth]{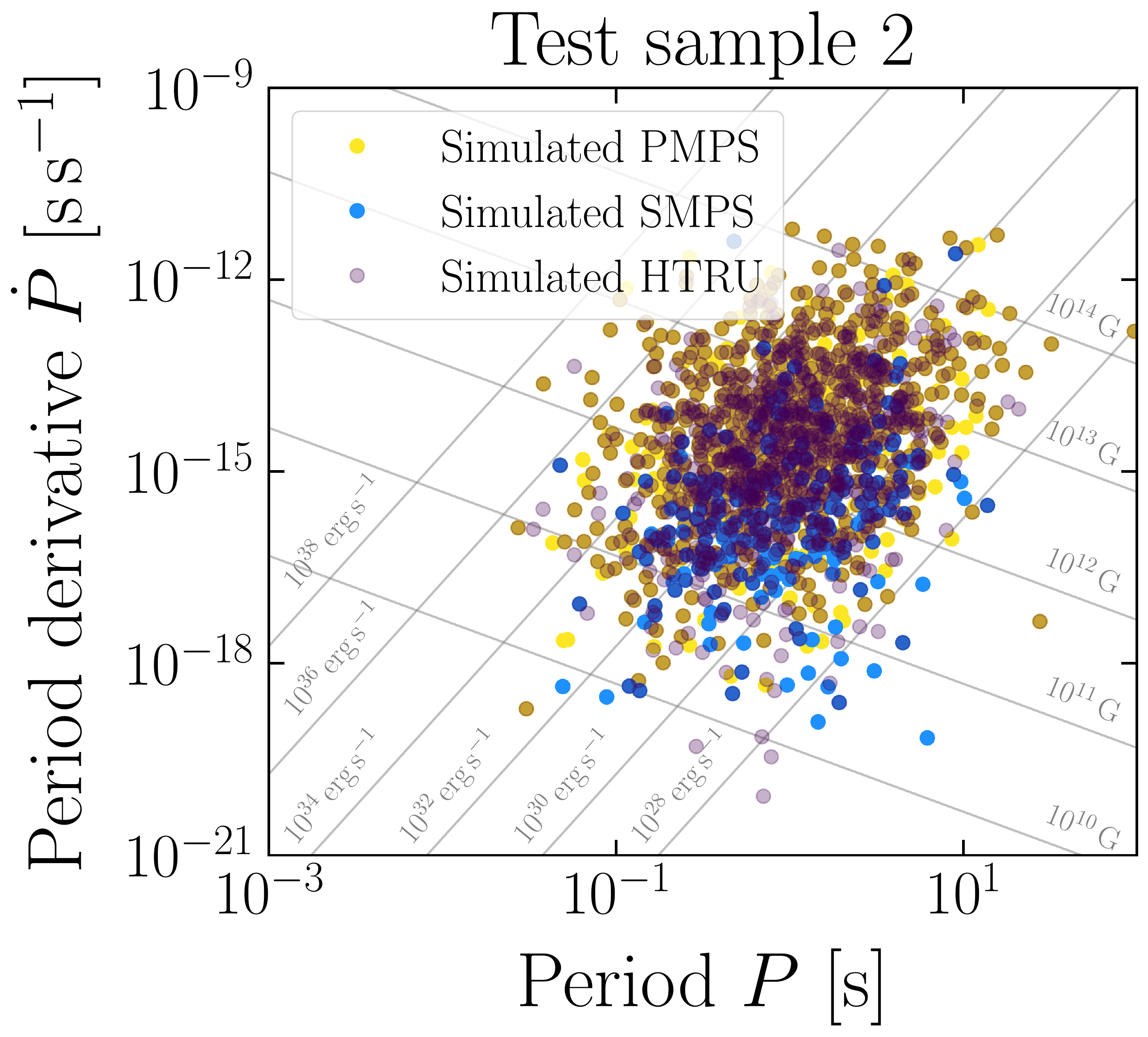}
	\hspace{0.1cm}
	\includegraphics[height=0.44\columnwidth]{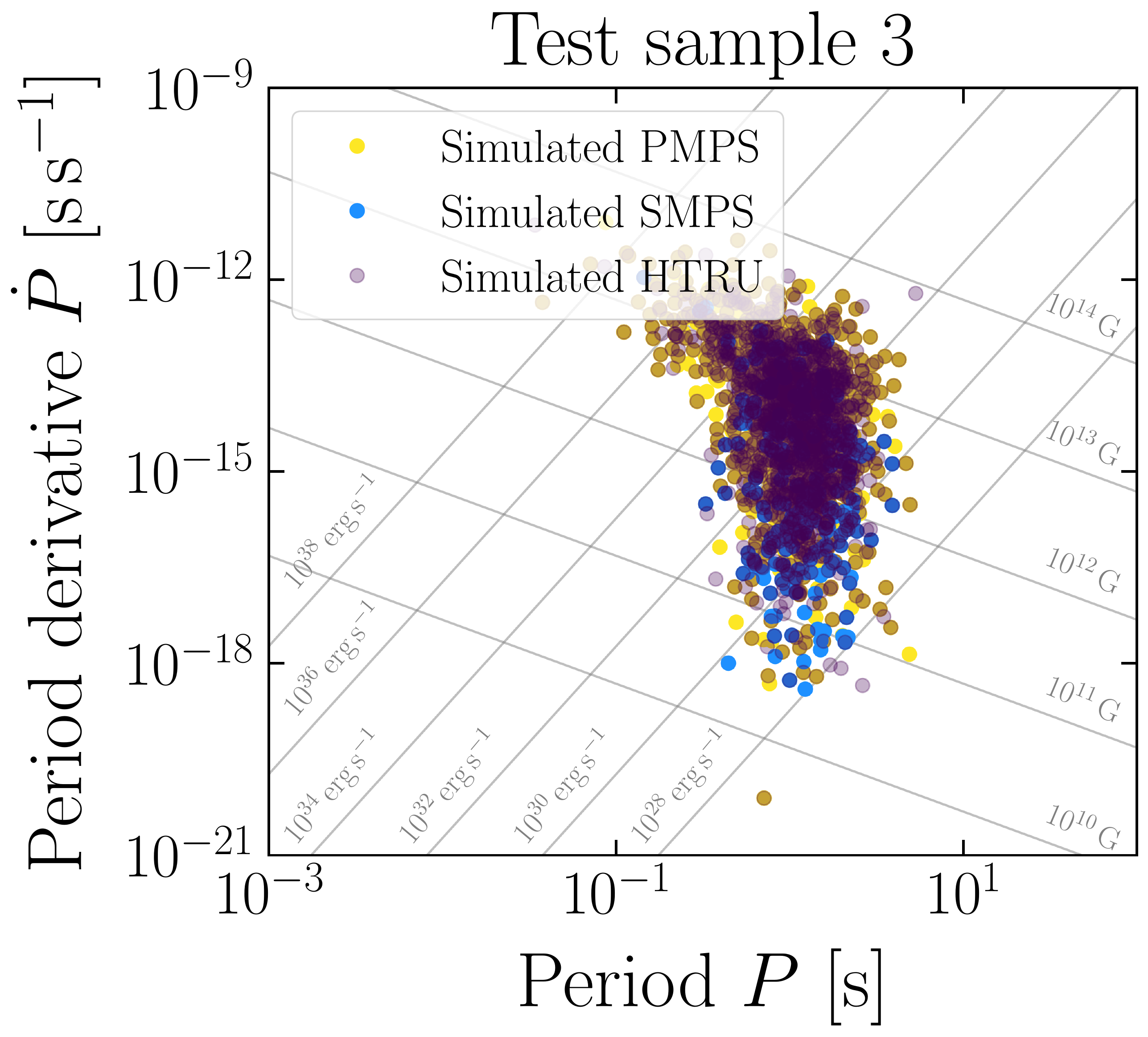} \vskip 0.2cm
	\includegraphics[height=0.30\columnwidth]{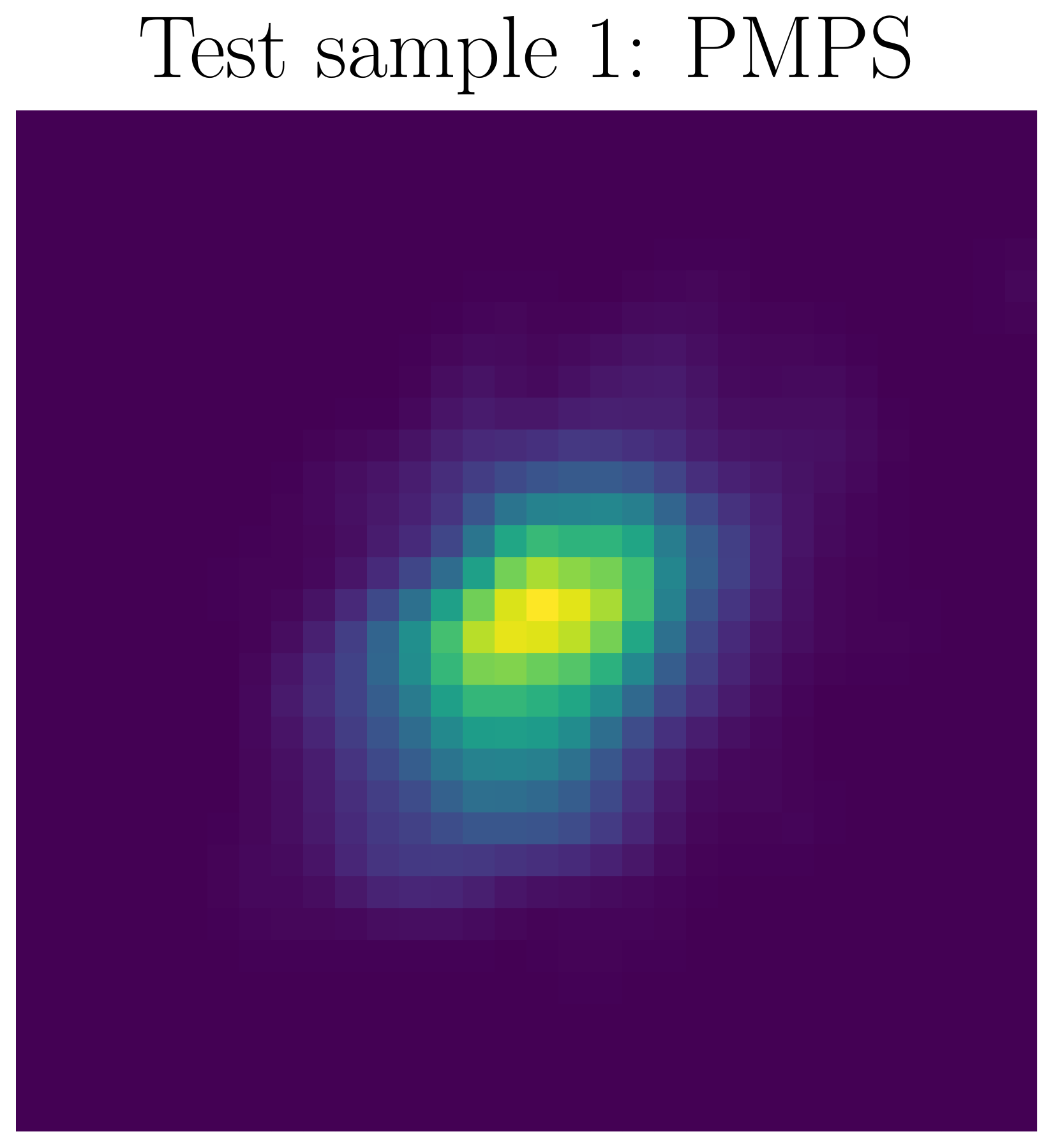}
	\hspace{1cm}
	\includegraphics[height=0.30\columnwidth]{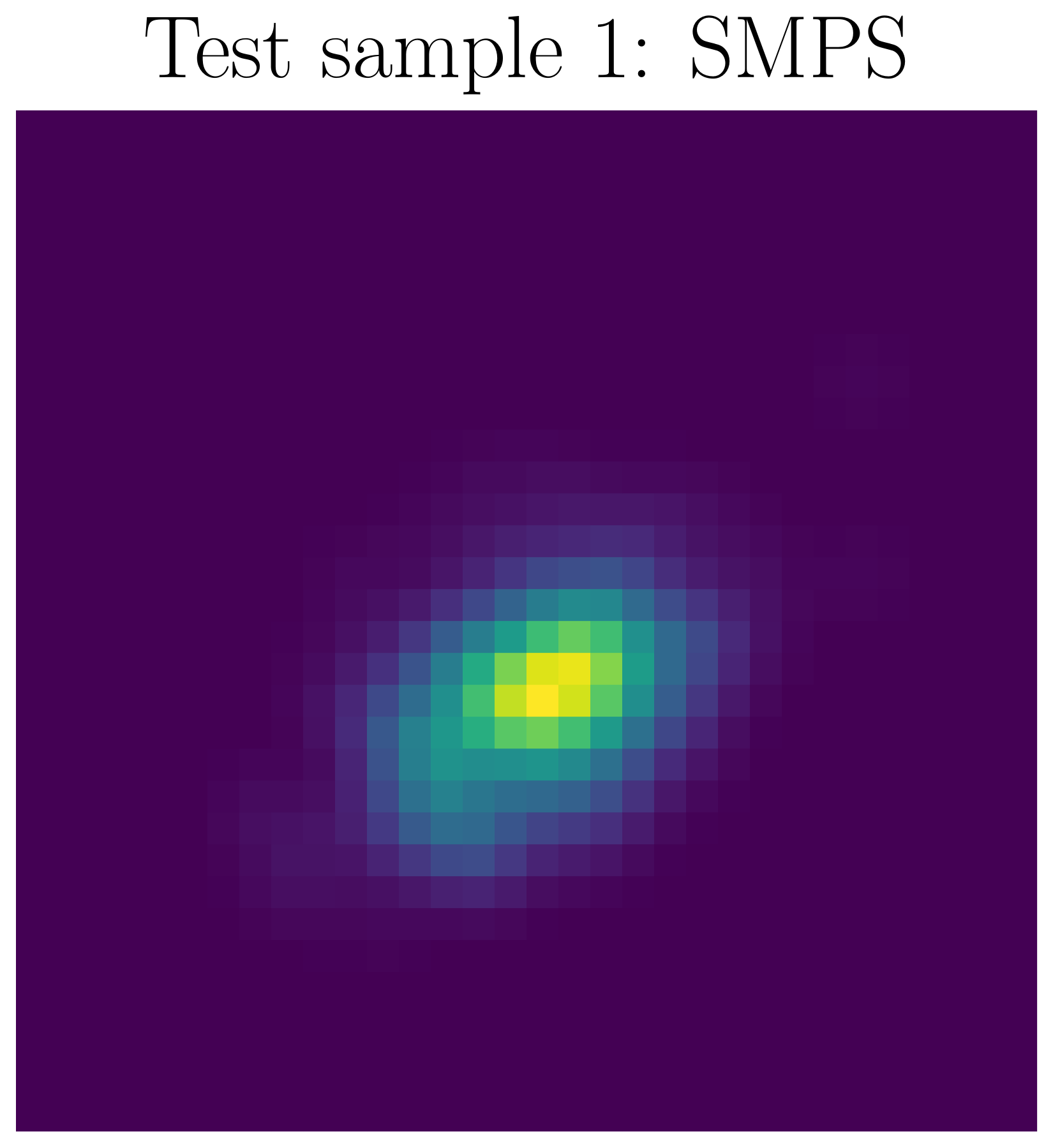}
	\hspace{1cm}
	\includegraphics[height=0.30\columnwidth]{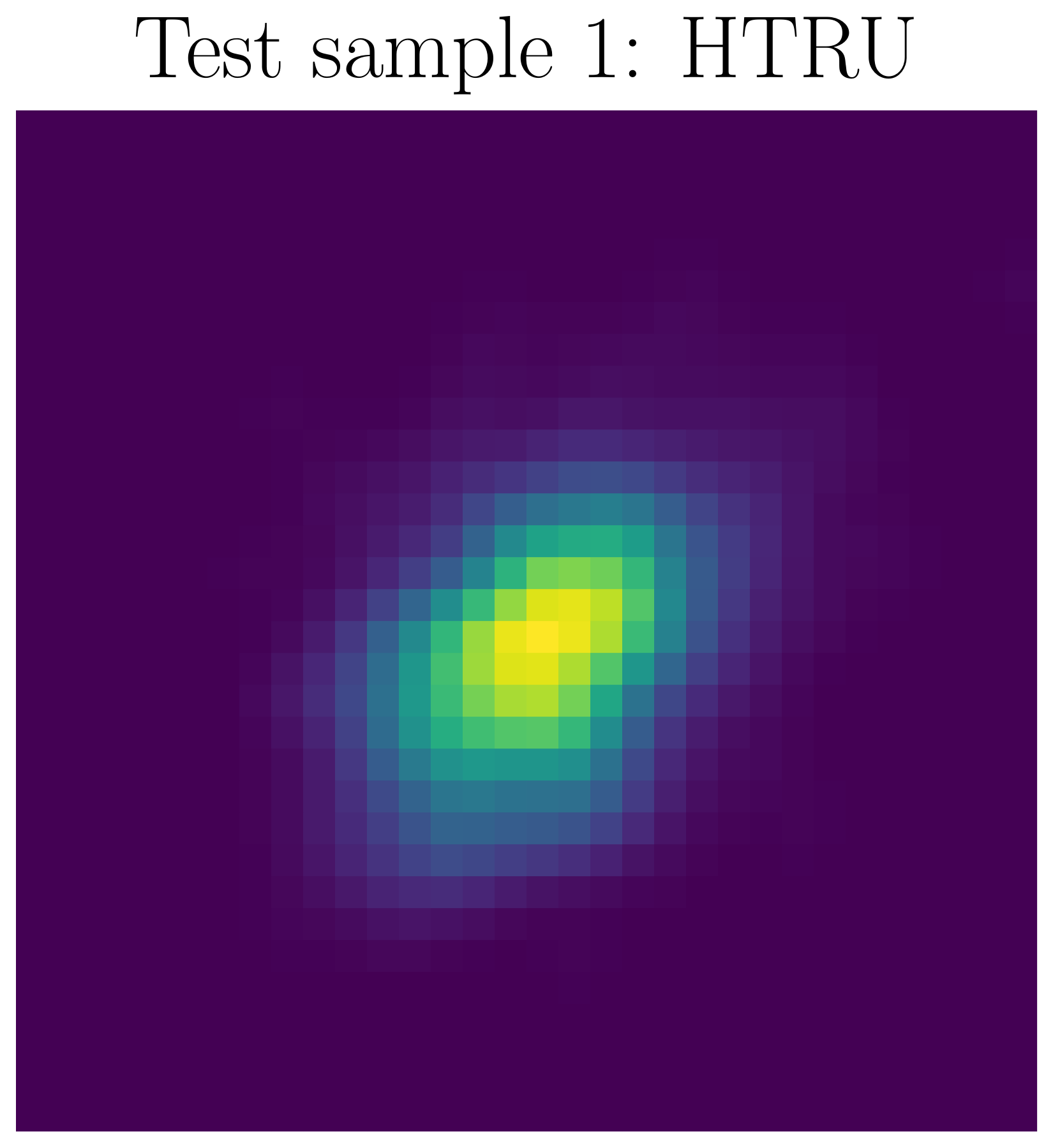}
	\caption[Examples of simulated pulsar populations and the corresponding density maps]{Examples of simulated pulsar populations and the corresponding density maps, which are fed into the simulation-based inference pipeline. The top three panels show synthetic $P$-$\dot{P}$ diagrams for the three surveys considered in this study generated from three random sets of magneto-rotation parameters. In particular, test sample 1 (\textit{top left}) is the result of a simulation with $\mu_{\log B} \approx 13.19, \sigma_{\log B} \approx 0.96, \mu_{\log P} \approx -0.85 , \sigma_{\log P} \approx 0.51$ and $a_{\rm late} \approx -0.86$, while test sample 2 (\textit{top right}) was generated with $\mu_{\log B} \approx 13.86, \sigma_{\log B} \approx 0.88, \mu_{\log P} \approx -0.42 , \sigma_{\log P} \approx 0.61$ and $a_{\rm late} \approx -1.71$. Finally, test sample 3 (\textit{top middle}) corresponds to $\mu_{\log B} \approx 13.35, \sigma_{\log B} \approx 0.24, \mu_{\log P} \approx -1.25 , \sigma_{\log P} \approx 0.60$ and $a_{\rm late} \approx -2.38$. The bottom row show the three density maps (one for each survey) generated with a resolution of 32 from the $P$-$\dot{P}$ diagram for test sample 1. Here, dark blue encodes regions where no neutron stars are present, while yellow bins represent the largest density for the binned pulsar distribution.}
	\label{fig:ch6_pop_simulated}
\end{figure}
%--------------------------------------------------------------

\subsection{Simulation output}
\label{sec:ch6_sim_output}

To simulate our \textit{mock} observed pulsar populations, we do not make any assumptions on the neutron-star birth rate at this stage. Instead, we randomly sample a subset of $10^5$ neutron stars from our dynamical database (see Section~\ref{sec:ch6_dyn_evol}). We subsequently evolve these stars magneto-rotationally as outlined in Section~\ref{sec:ch6_mr_evol} and assess how many of them are detected by each of the three surveys (see Sections~\ref{sec:ch6_emission_phys} and \ref{sec:ch6_obs_lims}), saving their respective properties. We iterate this process until the number of detected stars matches the number of observed stars in each survey. Note that we adaptively reduce the number of stars we draw from our dynamical database to $10^4$ and $5 \times 10^3$, once we have recovered 90\% and 95\% of the target values, respectively. The output of a single simulator run, which has a typical computation time of around $\unit[1-2]{hr}$, is a data frame containing the properties of those pulsars we can detect with \acs{PMPS}, \acs{SMPS} and \acs{HTRU}. 

The location of the resulting synthetic population and the shape of the stars' distribution in the $P$-$\dot{P}$ plane is directly controlled by the magneto-rotational parameters, $\mu_{\log B}, \sigma_{\log B}, \mu_{\log P}, \sigma_{\log P}$ and $a_{\rm late}$, the five parameters we set out to infer in the following. Three examples of synthetic $P$-$\dot{P}$ diagrams are shown in the top three panels of Figure~\ref{fig:ch6_pop_simulated}. 

We note that our prescription does not require a \textit{by-hand} implementation of a pulsar death line \citep[e.g.,][see also Section~\ref{sec:ch1_radio_em_deathlines}]{Chen1993, Rudak1994, Zhang2000}, beyond which radio emission ceases. Instead, pulsars become naturally undetectable if they approach the bottom right of the $P$-$\dot{P}$ plane in our framework. This is due the evolution towards ($i$) smaller misalignment angles, $\chi$, resulting in smaller beaming fractions, and ($ii$) smaller $\dot{P}$ and thus $E_{\rm rot}$, ultimately leading to sources that are too faint to be detected.

At this point, we also highlight that our approach provides information on the number of total stars generated over a time scale of $\unit[10^8]{yr}$ (the oldest possible age for stars in our dynamical database), implying that we can directly determine the birth rate required to reproduce observations for a given survey. Although not the primary focus of this work, we note two things here: First, the number of detectable neutron stars per iteration step described above and, thus, the birth rate (as well as the distribution of stars in the $P$-$\dot{P}$ plane) depends strongly on the five magneto-rotational parameters. For some parameter combinations, the birth rate can become unrealistically large and the computation time extensive. To overcome this issue, we stop our iterative simulation approach once the birth rate exceeds a conservative limit of $5$ neutron stars per century \citep[][see also Section~\ref{sec:ch1_birthrate}]{Keane2008, Rozwadowska2021}. While this implies that we do not reach the numbers of observed objects in these simulations, we still use them in the following to assess if our inference approach can identify those parameter combinations as unreasonable from the distribution of stars in the $P$-$\dot{P}$ plane alone. Second, for a single simulation run, we generally do not obtain the same birth rate for all three surveys and estimates can differ by a factor of $\sim 1-3$ neutron stars per century. In principle, we only expect the \textit{correct} physical simulator to produce the observed distributions of pulsars across different surveys. The correct simulation framework is, however, not known and constraining the relevant physics is the main goal of our analysis. To explore this behaviour, we therefore produce neutron stars until the target values in all three surveys are reached (or exceeded). While this implies that the number of detected objects in some simulations can be larger than the observed number of stars for a given survey (by up to a factor of $\sim 3$), our focus on the location and shape of the pulsar distribution in $P$ and $\dot{P}$ outlined below circumvents this issue. We will, however, return to the issue of the birth rate in the discussion in Section~\ref{sec:ch6_conclusions}, once we have explained our inference approach and provided results for our best estimates.

To provide a broad range of synthetic $P$-$\dot{P}$ diagrams for our inference pipeline, we explore the ranges outlined in Section~\ref{sec:ch6_mr_evol} and uniformly sample random combinations of the five parameters as follows:
\begin{align}
\mu_{\log B} &\in \mathcal{U}(12, 14), \nonumber \\
\sigma_{\log B} &\in \mathcal{U}(0.1, 1), \nonumber \\
\mu_{\log P} & \in \mathcal{U}(-1.5, -0.3), \label{eq:ch6_priors} \\
\sigma_{\log P} &\in \mathcal{U}(0.1, 1), \nonumber  \\
a_{\rm late} &\in \mathcal{U}(-3, -0.5) \nonumber.
\end{align}
We generate a total of 360,000 parameter combinations (which we refer to as our input parameters or labels in the following) and simulate the corresponding synthetic populations in parallel over the course of $\unit[6]{weeks}$.

To represent the discrete output of our simulator in a way that can be processed by a neural network, we follow our earlier study \citep[see Chapter~\ref{Chapter5} and][]{Ronchi2021} and convert a single $P$-$\dot{P}$ diagram for three surveys as seen in the top row of Figure~\ref{fig:ch6_pop_simulated} into three two-dimensional density maps (one for each survey) by counting the number of stars within a given bin. In particular, we set the limits $P \in [0.001, 100]\, {\rm s}$ and $\dot{P} \in [10^{-21}, 10^{-9}] \, {\rm s \, s^{-1}}$ and test the inference procedure for a resolution of 32 and 64 bins. To avoid sharp edges in our binned distributions, we apply a smoothing Gaussian filter (with radius $4 \sigma$ and $\sigma=1$), which will also improve the stability during the training of our machine-learning pipeline. An example of the resulting density maps is shown in the bottom row of Figure~\ref{fig:ch6_pop_simulated} for one of our simulations. 

The final preprocessing stage for our simulated data is either a normalisation or a standardisation step (depending on the choice of set-up discussed below) to provide the neural network with signals and labels of similar magnitude (see Section~\ref{sec:ch2_training_process}). In the former case, the bins in each individual density map are rescaled such that they contain continuous values between $0$ and $1$. The same holds for the corresponding labels, which are normalised over the entire parameter ranges given in Equation~\eqref{eq:ch6_priors}. On the other hand, standardisation is achieved by using $z$-scores, so that the resulting information in each map has a mean of $0$ and standard deviation of $1$. The same method is applied to the labels across our entire set of simulations.

%%%%%%%%%%%%%%%%%%%%%%%%%%%%%%%%%%%%%%%%%%%%%%%%%%%%%%%%%%%%%%%%%%%%%%%%%%%%%%%%%%%%%%%%%%%%%%%%%%%%%%%%%%%%%%%%%%%%%%%%%%%%%%

\section{Simulation-based inference}
\label{sec:ch6_sbi}

\subsection{Overview}
\label{sec:ch6_sbi_overview}

In Section~\ref{sec:ch2_sbi} we introduced the \acf{SBI} framework, here we will overview the main concepts. The pulsar population-synthesis pipeline summarised in Section~\ref{sec:ch6_popsyn} is a typical example of a stochastic forward model which aims to emulate real-world observations. We specifically introduced stochasticity by sampling relevant variables from underlying probability distributions using Monte-Carlo techniques. In particular, given the input parameter, $\bt = \{ \theta_1, \theta_2, \dots\}$, our simulator generates a synthetic realisation of the observed data, $\bx$. The key challenge is then to constrain our model parameters in such a way that they are consistent with true observations, $\bx_{\rm obs}$, and our prior knowledge, encoded in the prior distribution, $\pi(\bt)$. To this end, we want to compute the posterior distribution, $\pr(\bt | \bx)$, using Bayes' theorem
%------------------------------------------------------
\begin{equation}
\pr(\bt | \bx) = \frac{\pi(\bt) \pr(\bx |\bt)}{\pr(\bx)},
\label{eq:ch6_Bayes_theorem}
\end{equation}
%------------------------------------------------------
where $\pr(\bx |\bt)$ is the likelihood of our data, $\bx$, given the parameter, $\bt$, and $\pr(\bx)$ denotes the evidence obtained by marginalizing the likelihood over all $\bt$.
However, for complex simulators like ours, we typically cannot write down an explicit form of the likelihood function so $\pr(\bx |\bt)$ is essentially intractable. In addition, even if the likelihood were tractable, computing the evidence can become very costly as it involves
complicated integrals over a multidimensional parameter space.

Simulation based inference (SBI) circumvents these issues by taking advantage of the fact that our simulator encodes the likelihood function implicitly \citep[see Section~\ref{sec:ch2_sbi} and][for a recent review]{Cranmer2020}. These approaches have been particularly successful in combination with deep-learning techniques because neural networks can be used to learn a probabilistic association between a given simulation outcome, $\bx$, and the input parameters, $\bt$. This allows an approximation of the posterior distribution, $\pr(\bt |\bx)$, without the need to explicitly compute the likelihood.

For the following study, we choose the \acf{NPE} approach (see Section~\ref{sec:ch2_npe}) to directly learn the posterior conditional on our simulated data (avoiding the additional sampling step required for \acf{NLE} and \acf{NRE} as seen in Section~\ref{sec:ch2_sbi}) and take advantage of the corresponding implementation in the open-source Python package {\tt sbi} \citep{Tejero-Cantero2020}.\footnote{\url{https://github.com/sbi-dev/sbi}}

%%%%%%%%%%%%%%%%%%%%%%%%%%%%%%

%-----------------------------------------------------------------
\begin{figure}
	\centering
	\includegraphics[width = \textwidth]{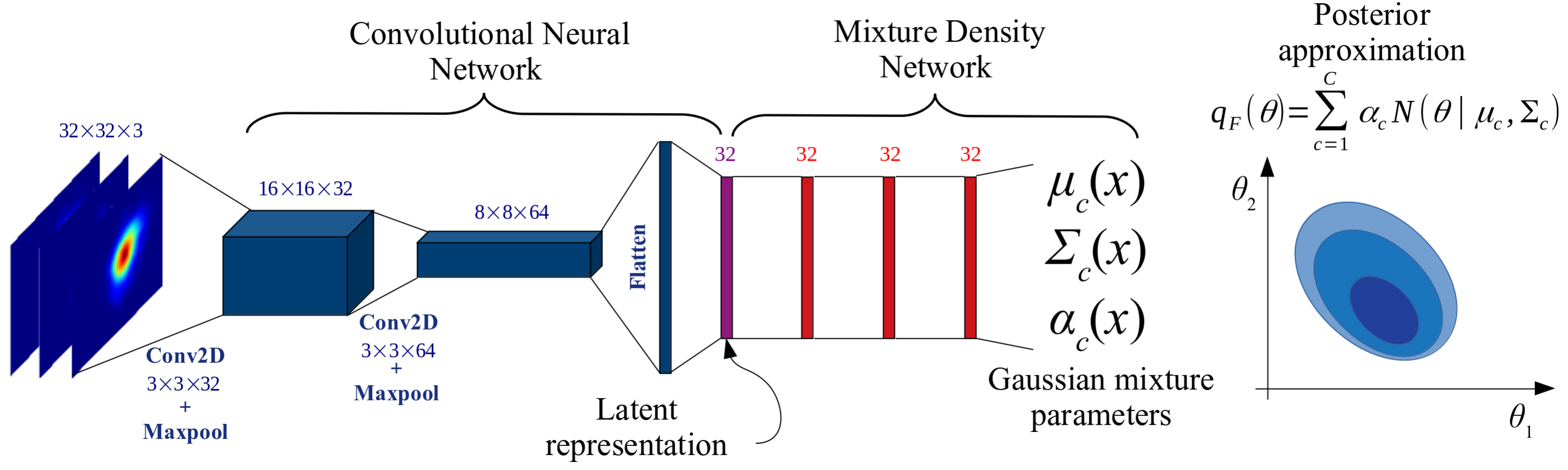}
	\caption[Schematic representation of our inference pipeline with NPE]{Schematic representation of our inference pipeline for three input $P-\dot{P}$ maps (one for each survey) with resolution $32 \times 32$. A \acf{CNN} is first used to extract features from our images and produce a compressed representation of our simulation output, $\bx$. We then train a Gaussian \acf{MDN}, a flexible neural density estimator, on this latent representation to approximate the posterior distribution of the simulation parameters, $\bt$.}
	\label{fig:sbi_architecture}
\end{figure}
%-----------------------------------------------------------------
\subsection{Deep-learning set-up}
\label{sec:ch6_architecture}

As described in Section~\ref{sec:ch2_npe}, for \ac{NPE}, we approximate the posterior using a family of densities, $q_{\bp}$, characterised by the distribution parameters, $\bp$. For our \ac{SBI} pipeline, we then use a neural network, $F$, to learn these $\bp$ for our simulator output, $\bx$, by adjusting the network weights, $\boldsymbol{w}$. In particular, we aim to optimise the neural density estimator such that $q_{F(\bx, \boldsymbol{w})}(\bt) \approx \pr (\bt | \bx)$. As described in Section~\ref{sec:ch2_npe} this can be achieved by minimizing the expectation value of the following loss function \citep{Papamakarios2016}
%------------------------------------------------------
\begin{equation}
\mathcal{L}(\boldsymbol{w}) = - \sum_{i=1}^N \log q_{F(\bx_i, \boldsymbol{w})} (\bt_i)
\label{eq:ch6_loss}
\end{equation}
%------------------------------------------------------
over a training data set $\{\bt_i, \bx_i\}$ of size $N$, provided that $N$ is large and the density estimator sufficiently flexible. In practice, we maximise the negative of $\mathcal{L}(\boldsymbol{w})$, i.e., the total log-posterior. A key advantage of the resulting posterior approximation is that the evaluation of $q_{F(\bx, \boldsymbol{w})}(\bt)$ corresponds to a simple forward pass through a neural network (without the need to simulate additional data), which is very fast. We will take advantage of this \textit{amortised nature} of the posterior to assess the quality of our inferences below. 

For our pulsar study, we have drawn the model parameters $\bt_i = \{\mu_{\log B},\, \sigma_{\log B},\,\\ \mu_{\log P},\, \sigma_{\log P},\, a_{\rm late}\}$ from uniform priors as defined previously in Equation~\eqref{eq:ch6_priors}. The corresponding output, $\bx_i$, of a single run through the simulator are the three $P$-$\dot{P}$ density maps (one for each survey) illustrated in the bottom row of Fig.~\ref{fig:ch6_pop_simulated}. In the following, we stack these maps together to form a three-channel input for our neural network. Of the 360,000 synthetic simulations produced, we use 90\% for training and validation reserving the remaining 10\% for testing purposes. The former data set is further split into 90\% for training ($291,600$ populations) and 10\% for validation ($32,400$ populations). We note that as each population is represented by three density maps, we train the following inference pipeline on roughly $875,000$ images. Performance results for the unseen test samples quoted in the following are computed for 10\% of the full test set ($3,600$ populations) for computational reasons. The full workflow and the network architecture are illustrated schematically in Figure~\ref{fig:sbi_architecture}. 

Due to the complexity of these data, we do not train a neural density estimator directly on the density maps. We instead first apply a \ac{CNN} to extract features from our images and embed the corresponding information in a lower-dimensional latent vector. Following a similar procedure as in Chapter~\ref{Chapter5} \citep[see also][]{Ronchi2021}, we choose the following baseline architecture for our embedding network:
\begin{itemize}
	\item 2D convolution layer with kernel size $3 \times 3$, 3 input channels, 32 output channels, stride 1, padding 1.
	\item 2D Max pooling layer with size $2 \times 2$, stride 2, no padding.
	\item 2D convolution layer with kernel size $3 \times 3$, 32 input channels, 64 output channels, stride 1, padding 1.
	\item 2D Max pooling layer with size $2 \times 2$, stride 2, no padding.
	\item Fully connected linear layer with the flattened output from the second pooling layer as input and 32 output neurons encoding the latent representation.
\end{itemize}
After each convolution and the fully connected layer, we apply a \acf{ReLU} activation function (see Section~\ref{sec:ch2_activation_function}). The weights for the \ac{CNN} are initialised using the Kaiming prescription \citep{Kaiming2015} to avoid exploding or vanishing gradients during the training process (see Section~\ref{sec:ch2_w_initialization}).

We subsequently pass the latent vector generated by the \acs{CNN} to a neural density estimator. We implement a \ac{MDN} and specifically opt for a \acf{GMM} in five dimensions to approximate the posterior, $q_{F(\bx, \boldsymbol{w})}(\bt)$, for our five free magneto-rotational parameters. This implies
%------------------------------------------------------
\begin{equation}
q_{F(\bx, \boldsymbol{w})}(\bt) = \sum_{c = 1}^{C} \alpha_c \, \mathcal{N} (\bt | \boldsymbol{\mu}_c, \boldsymbol{\Sigma}_c ),
\label{eq:ch6_posterior_gmm}
\end{equation}
%------------------------------------------------------
where, $C$ denotes the total number of Gaussian components used, $\alpha_{c}$ is the mixture weight and $\mathcal{N}(\bt | \boldsymbol{\mu}_c,\boldsymbol{\Sigma}_c)$ the multi-variate Gaussian distribution with mean vector $\boldsymbol{\mu}_c$ and covariance matrix $\boldsymbol{\Sigma}_c$ for the $c$-th component.

For our \ac{MDN}, we follow {\tt sbi}'s default implementation and use:
\begin{itemize}
	\item Three fully connected layers with 32 neurons each.
	\item Four fully connected output layers which encode the Gaussian mixture weights, $\alpha_{c}$, means, $\boldsymbol{\mu}_c$, diagonal and upper triangular components of the covariance matrices, $\boldsymbol{\Sigma}_c$, respectively. These contain $c$, $5 c$, $5 c$ and $10 c$ neurons, respectively.
\end{itemize}
We again apply the \ac{ReLU} activation function after each hidden layer, while weights for the \ac{MDN} are initialised with PyTorch's default initialisation \citep{Glorot2010}.

We subsequently train the entire pipeline using the gradient descent optimiser Adam \citep[][see also Section~\ref{sec:ch2_backprop}]{Kingma2014}. At each epoch the network undergoes a series of optimisation steps based on the information provided in the entire training data set before epoch-averaged training and validation metrics are computed based on the negative of Equation~\eqref{eq:ch6_loss}. In this way during optimisation the algorithm tend to maximise the training and validation metrics. Note that we also set an early stop of 20 to prevent overfitting, which implies that the training process is interrupted (and the weights of the best validation epoch recorded) once the validation metric has not improved for 20 epochs.
	
%%%%%%%%%%%%%%%%%%%%%%%%%%%%%%

%------------------------------------------------------
\begin{landscape}
	\begin{table}
		\centering
		\caption[Information for the 22 machine-learning experiments]{Information for the 22 machine-learning experiments summarising the specific training data and hyperparameters as well as the resulting metrics. The columns report the experiment number ($\#$), the resolution (res) for our $P$-$\dot{P}$ density maps, the included surveys and the fraction (frac; in $\%$) of the $291,600$ populations in the training set used for training, information on whether we standardised (std) or normalised (norm) the input, the number of Gaussian components (comp) in our \ac{MDN}, the batch size (BS), the learning rate (LR), the CNN architecture (we distinguish our baseline set-up and a deeper network; see Secs.~\ref{sec:ch6_architecture} and \ref{sec:ch6_experiments} for details), the best metric computed over the validation set (VM), the number of training epochs, the time it took to train the network in seconds and the average metric computed over our $3,600$ test samples (TM). In bold, we highlight those parameters that we have varied with respect to the baseline experiment $\#1$. Experiments with an asterisk ($\star$) are removed from following analysis due to training irregularities. \label{tab:ch6_experiments}}
		\small
		\begin{tabular}{c|cccc|cccc|cccc}
			\toprule
			\tabhead{$\#$} &
			\tabhead{res} &
			\tabhead{surveys} &
			\tabhead{frac ($\%$)} & 
			\tabhead{input} & 
			\tabhead{comp} &
			\tabhead{BS} &
			\tabhead{LR} &
			\tabhead{CNN} & 
			\tabhead{VM} & 
			\tabhead{epochs} & 
			\tabhead{time (s)} & 
			\tabhead{TM} \\
			\midrule
			\textbf{1}	& 32			&  PMPS, SMPS, HTRU	&	100		& std			& 10			& 8			&  0.0005			& baseline		& 3.65	& 38		& 9,373	& 3.64	\\
			\textbf{2}	& 32			&  PMPS, SMPS, HTRU	&	100		& std			& 10			& 8			&  0.0005			& \textbf{deep}	& 3.71	& 49		& 14,292	& 3.71	\\
			\textbf{3}	& \textbf{64}	&  PMPS, SMPS, HTRU	&	100		& std			& 10			& 8			&  0.0005			& baseline		& 3.55	& 55		& 78,837	& 3.54	\\
			\textbf{4}	& \textbf{64}	&  PMPS, SMPS, HTRU	&	100		& std			& 10			& 8			&  0.0005			& \textbf{deep}	& 3.64	& 89		& 128,119	& 3.64	\\
			\textbf{5}	& 32			&  PMPS, SMPS, HTRU	&	\textbf{75}	& std			& 10			& 8			&  0.0005			& baseline		& 3.74	& 71		& 13,232	& 3.78	\\
			\textbf{6}	& 32			&  PMPS, SMPS, HTRU	&	\textbf{50}	& std			& 10			& 8			&  0.0005			& baseline		& 3.56	& 58		& 7,000	& 3.55	\\
			\textbf{7}$\star$	& 32			&  PMPS, SMPS, HTRU	&	100		& \textbf{norm}	& 10			& 8			&  \textbf{0.01}		& baseline		& 3.47	& 30		& 7,445	& 3.73	\\
			\textbf{8}	& 32			&  PMPS, SMPS, HTRU	&	100		& \textbf{norm}	& 10			& 8			&  \textbf{0.001}	& baseline		& 9.66	& 54		& 13,015	& 9.60	\\
			\textbf{9}	& 32			&  PMPS, SMPS, HTRU	&	100		& std			& \textbf{8}	& 8			&  0.0005			& baseline		& 3.74	& 52		& 12,389	& 3.73	\\
			\textbf{10}	& 32			&  PMPS, SMPS, HTRU	&	100		& std			& \textbf{5}	& 8			&  0.0005			& baseline		& 3.83	& 118	& 27,973	& 3.86	\\
			\textbf{11}	& 32			&  PMPS, SMPS, HTRU	&	100		& std			& 10			& \textbf{16}	&  0.0005			& baseline		& 3.99	& 85		& 10,476	& 3.97	\\
			\textbf{12}	& 32			&  PMPS, SMPS, HTRU	&	100		& std			& 10			& \textbf{32}	&  0.0005			& baseline		& 4.11	& 79		& 5,346	& 4.06	\\
			\textbf{13}	& 32			&  PMPS, SMPS, HTRU	&	100		& std			& 10			& 8			&  \textbf{0.001}	& baseline		& 3.36	& 61		& 14,785	& 3.33	\\
			\textbf{14}	& 32			&  PMPS, SMPS, HTRU	&	100		& std			& 10			& 8			&  \textbf{0.0001}	& baseline		& 4.22	& 75		& 18,369	& 4.22	\\
			\textbf{15}	& 32			&  \textbf{HTRU}		&	100		& std			& 10			& 8			&  0.0005			& baseline		& 3.43	& 63 		& 15,568	& 3.42	\\
			\textbf{16}	& 32			&  \textbf{SMPS, HTRU}	&	100		& std			& 10			& 8			&  0.0005			& baseline		& 3.58	& 40		& 9,979	& 3.59	\\
			\textbf{17}	& 32			&  \textbf{PMPS, SMPS}	&	100		& std			& 10			& 8			&  0.0005			& baseline		& 3.41	& 69		& 16,937	& 3.41	\\
			\textbf{18}$\star$	& \textbf{64}	&  PMPS, SMPS, HTRU	&	\textbf{50}	& std			& 10			& 8			&  0.0005			& baseline		& 3.45	& 47		& 5,766	& 3.44	\\
			\textbf{19}	& 32			&  PMPS, SMPS, HTRU	&	100		& \textbf{norm}	& 10			& \textbf{32}	&  \textbf{0.001}	& baseline		& 10.05	& 44		& 2,864	& 10.20	\\
			\textbf{20}	& 32			&  PMPS, SMPS, HTRU	&	100		& \textbf{norm}	& 10			& \textbf{32}	&  \textbf{0.0001}	& baseline		& 10.31	& 90		& 5,815	& 10.49	\\\
			\textbf{21}	& 32			&  PMPS, SMPS, HTRU	&	100		& \textbf{norm}	& 10			& \textbf{16}	&  \textbf{0.001}	& baseline		& 9.82	& 77		& 9,901	& 9.98	\\
			\textbf{22}$\star$	& 32			&  PMPS, SMPS, HTRU	&	100		& \textbf{norm}	& 10			& \textbf{16}	&  \textbf{0.0001}	& baseline		& 10.45	& 124	& 15,603	& 10.55	\\
			\bottomrule
		\end{tabular}
	\end{table}
\end{landscape}
%------------------------------------------------------

\subsection{Experiments}
\label{sec:ch6_experiments}

Table~\ref{tab:ch6_experiments} summarises the 22 different experiments that we have conducted for this study to assess the performance of \ac{SBI} for pulsar population synthesis. For this purpose, we varied aspects of the training data as well as the hyperparameters of our deep-learning pipeline. In particular, for the input we explored two different resolutions for the $P$-$\dot{P}$ maps, 32 and 64, respectively, assessed the network performance when all three density maps or only two/one are provided, and whether normalisation or standardisation during preprocessing leads to different results. We further studied the impact of using the full training data set or smaller subsets. Moreover, for the network we varied the number of Gaussian mixture components in our neural density estimator, the batch size, and the learning rate, and we explored two different \acp{CNN} for our embedding net. In addition to the baseline architecture described in Section~\ref{sec:ch6_architecture}, we also conducted two experiments with a deeper network composed of four convolutional blocks. Here, the two convolutional layers introduced previously are followed by an additional layer with 32 and 64 input/output channels, respectively. Kernel size, stride, padding, subsequent pooling and fully-connected layers were kept as above. 

Due to the computational cost of each training experiment, a full grid search over all relevant configurations was beyond the scope of this work. We, therefore, opted to produce a representative set of experiments that provide sufficient information to study the variation of our inferred posteriors in Section~\ref{sec:ch6_results}. Finally note, that almost all of our optimisations are performed on a Tesla V100 SXM2 GPU with $\unit[32]{GB}$ memory. We only trained experiments $\#3$ and $\#4$, for which the full training data set with a resolution of 64 was too large to be optimised on the GPU, on a CPU with $\unit[32]{GB}$ RAM. In those two cases, training the network, thus, took markedly longer than for the other experiments (see below).

%%%%%%%%%%%%%%%%%%%%%%%%%%%%%%%%%%%%%%%%%%%%%%%%%%%%%%%%%%%%%%%%%%%%%%%%%%%%%%%%%%%%%%%%%%%%%%%%%%%%%%%%%%%%%%%%%%%%%%%%%%%%%%

\section{Results}
\label{sec:ch6_results}

%-----------------------------------------------------------------
\begin{figure}
	\centering
	\includegraphics[width=0.7\columnwidth]{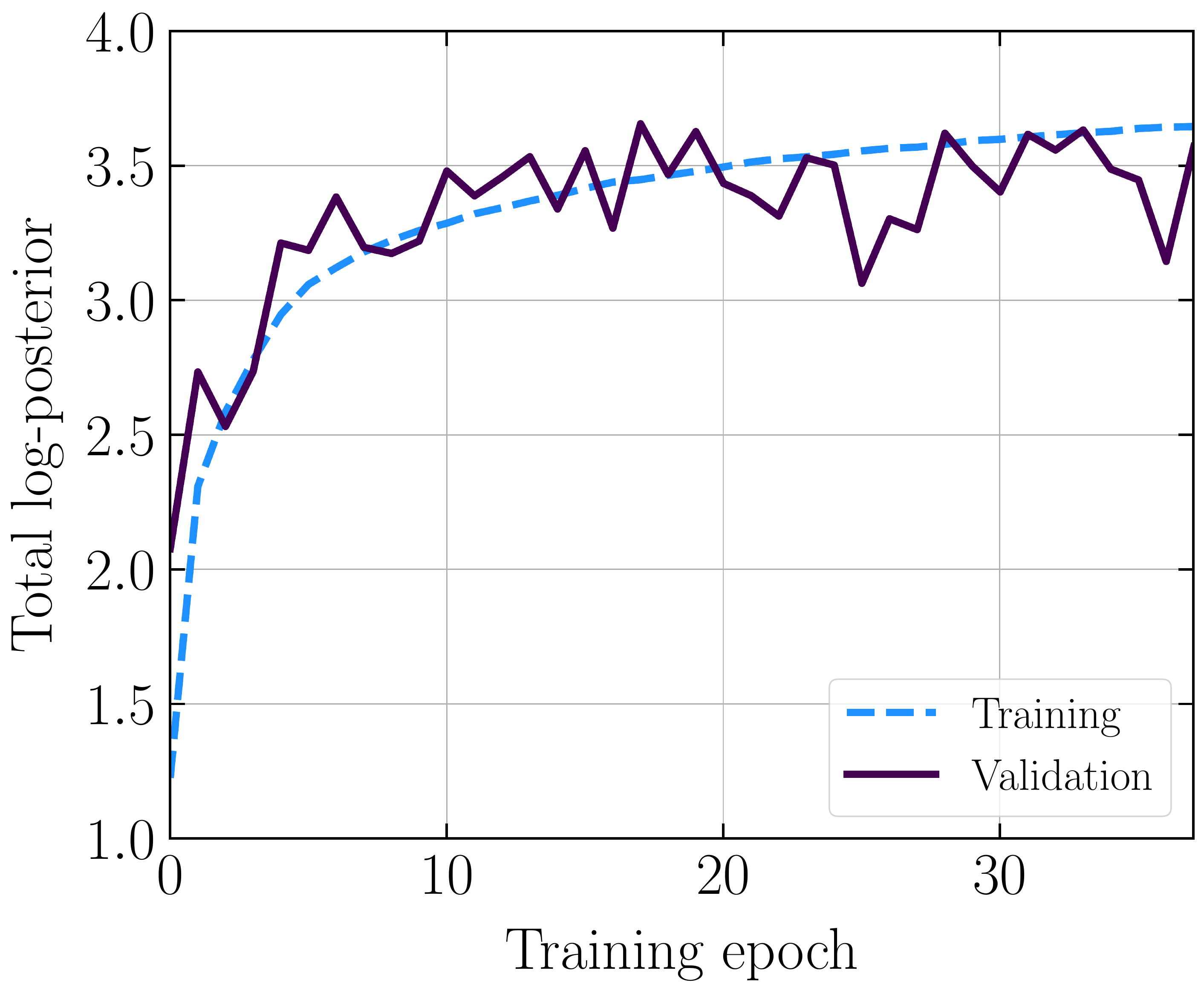}
	\caption[Training behaviour for baseline experiment $\#1$]{Training behaviour for baseline experiment $\#1$. We show training metric (dashed, light blue) and validation metric (solid, purple) as a function of the training epoch. We seek to maximise the total log-posterior, $\sum_{i=1}^N \log q_{F(\bx_i, \boldsymbol{w})} (\bt_i)$, over the training and validation data set, respectively, as the network learns. Both losses increase as expected and the validation curve closely tracks the training curve, i.e., we see little overfitting. The best validation loss is reached at epoch $17$ and the early stop criterion, thus, causes the training halt after $37$ training epochs.}
	\label{fig:ch6_training_curve}
\end{figure}
%-----------------------------------------------------------------

\subsection{Training}
\label{sec:ch6_training}

Several metrics for our experiments are summarised in the last four columns of Tab.~\ref{tab:ch6_experiments}. We observe that the optimisation of our neural networks take between $\sim \unit[1-8]{hr}$ on the GPU and on the order of a day on a CPU, completing $\sim 30- 124$ training epochs. In general, we find good training behaviour with the validation metric closely tracking the training metric and little or no overfitting. This is also evident in the network's generalisation ability illustrated by the average metrics computed over the unseen test set of $3,600$ simulations. The evolution of the training and validation metrics for experiment $\#1$ is shown in Fig.~\ref{fig:ch6_training_curve} as an example. We remind the reader that we aim to maximise the total log-posterior. After visual inspection of all training curves, we remove experiment $\#7$ due to irregularities in the training behaviour and experiments $\#18$ and $\#22$ due to a slight tendency to overfitting. Note that these shortcomings were not directly visible from the training metrics in Table~\ref{tab:ch6_experiments}. We also highlight that we find systematically larger training, validation and test metrics in those experiments where our input density maps were normalised. In the following, we however assess the quality of the corresponding posteriors and find that these do not result in better inferences. Beyond this difference, we cannot identify any significant variation in the metrics between the remaining configurations. We, hence, proceed with an analysis of all experiments apart from numbers $\#7, \#18$ and $\#22$.

%%%%%%%%%%%%%%%%%%%%%%%%%%%%%%

%---------------------------------------------------------------
\begin{table}
	\centering
	\caption[Magneto-rotational parameters for three random test samples and the observed pulsar population]{Magneto-rotational parameters for three random test samples and the observed pulsar population. The first five rows show the ground truths, $\bt$, used to simulate the test populations. The second block gives the median and $95\%$ \acp{CI} obtained from inferences with the neural network trained in experiment $\#1$. The final block contains the median and $95\%$ \acp{CI} determined from the ensemble posterior combining $19$ experiments. \label{tab:ch6_credible_intervals}}
	\small
	\begin{tabular}{ccc|cccc}
		\toprule
		\multicolumn3c{\tabhead{Parameters}} &
		\tabhead{Test sample $1$} &
		\tabhead{Test sample $2$} &
		\tabhead{Test sample $3$} &
		\tabhead{Observed population} \\
		\midrule
		\parbox[t]{0mm}{\multirow{5}{*}{\rotatebox[origin=c]{90}{Ground}}}	& \parbox[t]{2mm}{\multirow{5}{*}{\rotatebox[origin=c]{90}{truths, $\bt$}}}	& $\mu_{\log B}$	& $13.19$	& $13.86$	& $13.35$	& -	\\
		& 														& $\sigma_{\log B}$	& $0.96$	& $0.88$	& $0.24$	& -	\\
		& 														& $\mu_{\log P}$	& $-0.85$	& $-0.42$	& $-1.25$	& -	\\
		& 														& $\sigma_{\log P}$	& $0.51$	& $0.61$	& $0.60$	& -	\\
		&														& $a_{\rm late}$	& $-0.86$	& $-1.71$	& $-2.38$	& -	\\
		\hline
		\parbox[t]{0mm}{\multirow{5}{*}{\rotatebox[origin=c]{90}{$95\%$ \ac{CI}}}}	& \parbox[t]{2mm}{\multirow{5}{*}{\rotatebox[origin=c]{90}{experiment $\#1$}}}	& $\mu_{\log B}$	& $13.28^{+0.18}_{-0.18}$	& $13.73^{+0.15}_{-0.15}$	& $13.33^{+0.05}_{-0.04}$	& $13.07^{+0.07}_{-0.08}$	\\
		& 																	& $\sigma_{\log B}$	& $0.95^{+0.08}_{-0.08}$	& $0.79^{+0.07}_{-0.07}$	& $0.23^{+0.02}_{-0.02}$	& $0.43^{+0.03}_{-0.03}$	\\
		& 																	& $\mu_{\log P}$	& $-0.90^{+0.13}_{-0.13}$	& $-0.35^{+0.19}_{-0.18}$	& $-1.17^{+0.33}_{-0.34}$	& $-0.98^{+0.25}_{-0.29}$	\\
		& 																	& $\sigma_{\log P}$	& $0.49^{+0.10}_{-0.09}$	& $0.73^{+0.20}_{-0.15}$	& $0.73^{+0.25}_{-0.31}$	& $0.54^{+0.33}_{-0.25}$	\\
		&																	& $a_{\rm late}$	& $-0.83^{+0.06}_{-0.06}$	& $-1.88^{+0.35}_{-0.35}$	& $-2.47^{+0.43}_{-0.43}$	& $-1.77^{+0.35}_{-0.38}$	\\
		\hline
		\parbox[t]{0mm}{\multirow{5}{*}{\rotatebox[origin=c]{90}{$95\%$ \ac{CI}}}}	& \parbox[t]{2mm}{\multirow{5}{*}{\rotatebox[origin=c]{90}{ensemble}}}		& $\mu_{\log B}$	& $13.29^{+0.20}_{-0.20}$	& $13.74^{+0.19}_{-0.16}$	& $13.34^{+0.05}_{-0.05}$	& $13.10^{+0.08}_{-0.10}$	\\
		& 																	& $\sigma_{\log B}$	& $0.96^{+0.07}_{-0.08}$	& $0.78^{+0.09}_{-0.08}$	& $0.24^{+0.02}_{-0.02}$	& $0.45^{+0.05}_{-0.05}$	\\
		& 																	& $\mu_{\log P}$	& $-0.92^{+0.16}_{-0.15}$	& $-0.40^{+0.20}_{-0.27}$	& $-1.23^{+0.33}_{-0.34}$	& $-1.00^{+0.26}_{-0.21}$	\\
		& 																	& $\sigma_{\log P}$	& $0.49^{+0.10}_{-0.09}$	& $0.74^{+0.20}_{-0.17}$	& $0.67^{+0.30}_{-0.28}$	& $0.38^{+0.33}_{-0.18}$	\\
		&																	& $a_{\rm late}$	& $-0.84^{+0.06}_{-0.07}$	& $-1.76^{+0.39}_{-0.43}$	& $-2.34^{+0.43}_{-0.45}$	& $-1.80^{+0.65}_{-0.61}$	\\
		\bottomrule
	\end{tabular}
\end{table}
%---------------------------------------------------------------

%-----------------------------------------------------------------
\begin{figure}
	\centering
	\includegraphics[width=0.7\textwidth]{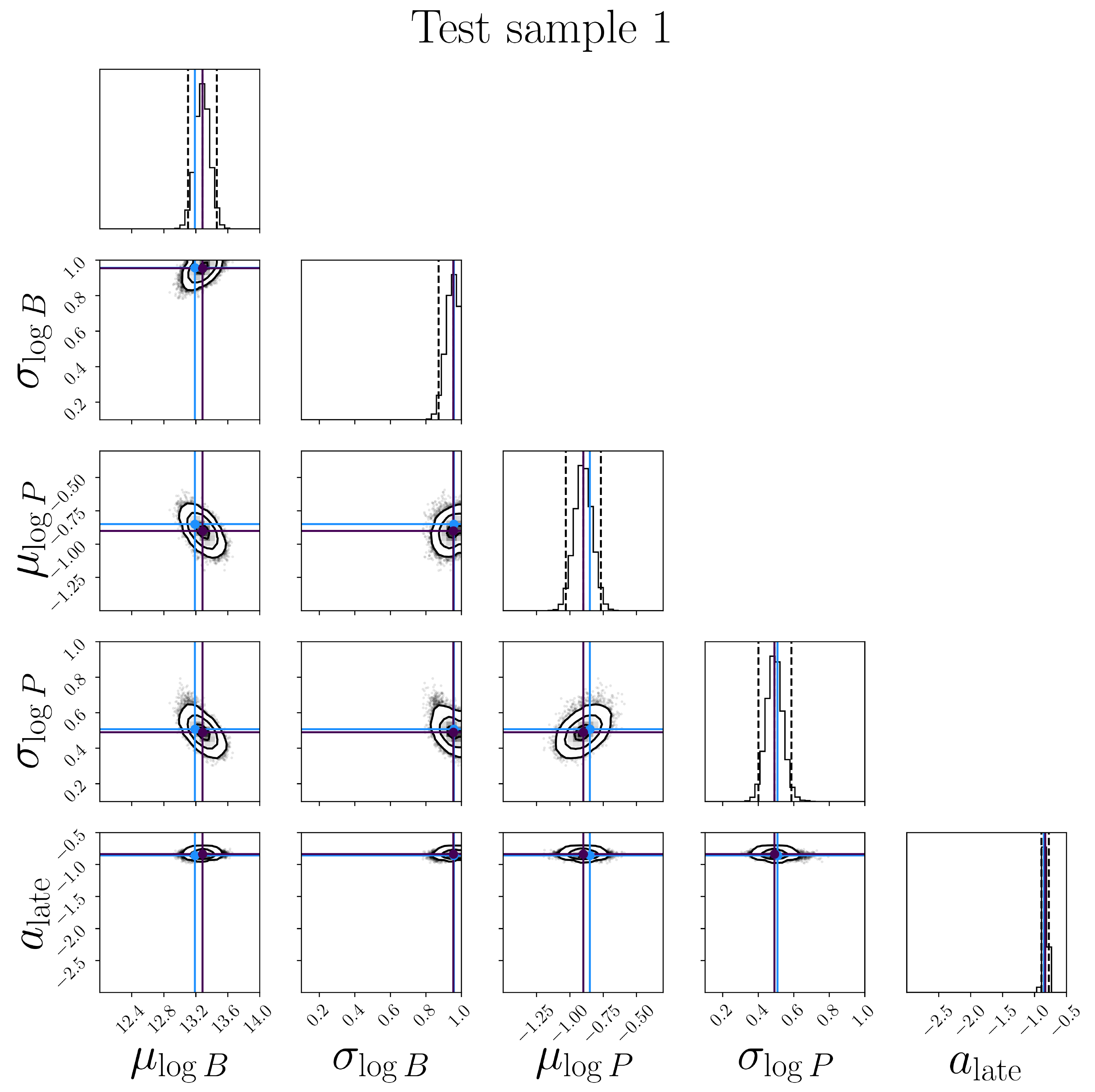}
	\includegraphics[width=0.7\columnwidth]{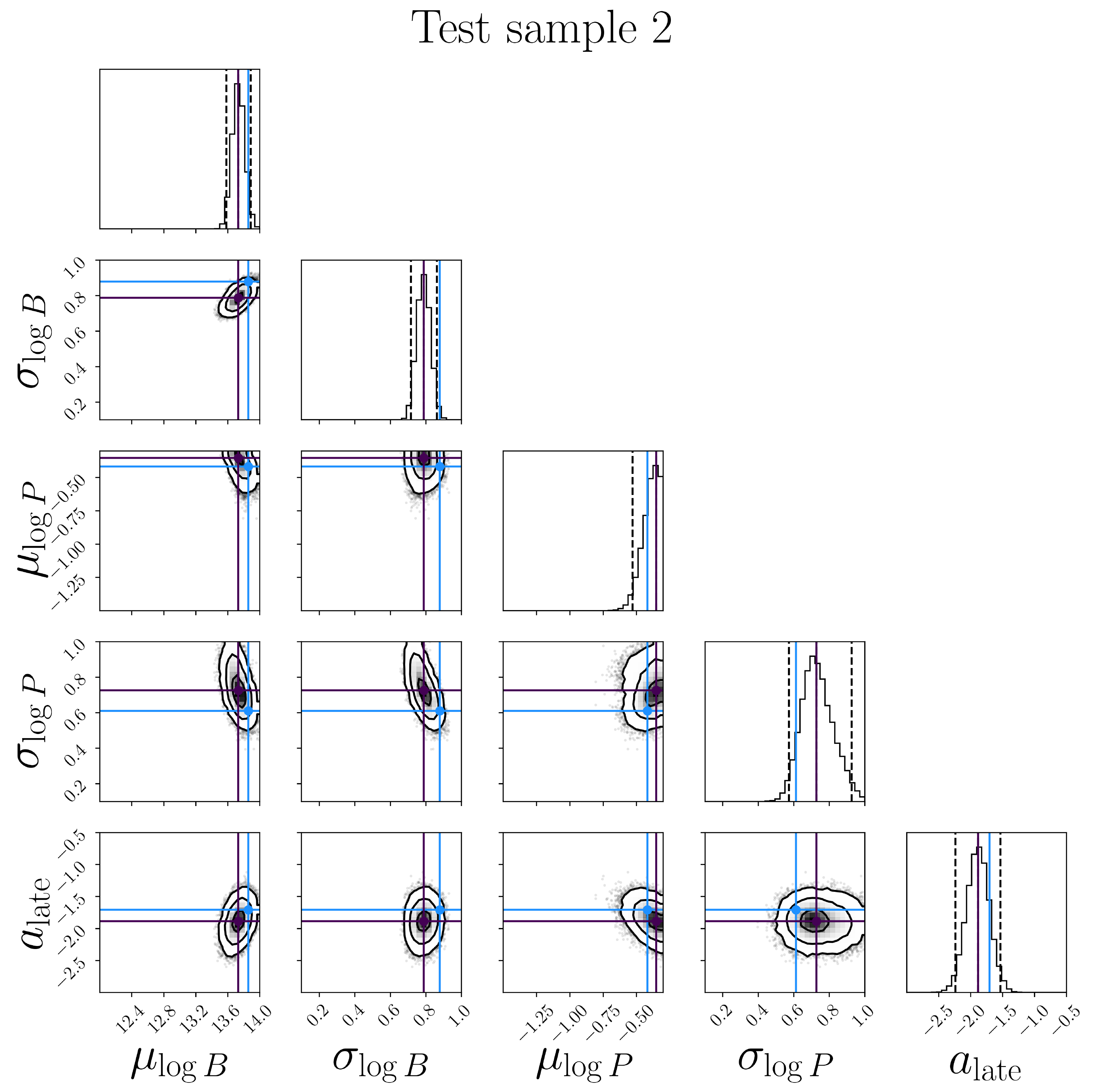}
	\caption[Benchmark inference for test simulation $1$ and 2 using the network from experiment $\#1$]{Benchmark inference for test simulation $1$ and 2 using the network from experiment $\#1$. The corner plot shows one- and two-dimensional marginal posterior distributions for the five magneto-rotational parameters. We also show corresponding ground truths, $\bt$, in light blue and the medians in purple. We observe that the posteriors cover the $\bt$ well. Corresponding $95\%$ \acp{CI} are summarised in Table~\ref{tab:ch6_credible_intervals}.}
	\label{fig:ch6_posterior_test12}
\end{figure}
%-----------------------------------------------------------------

%-----------------------------------------------------------------
\begin{figure}
	\centering
	\includegraphics[width=0.7\columnwidth]{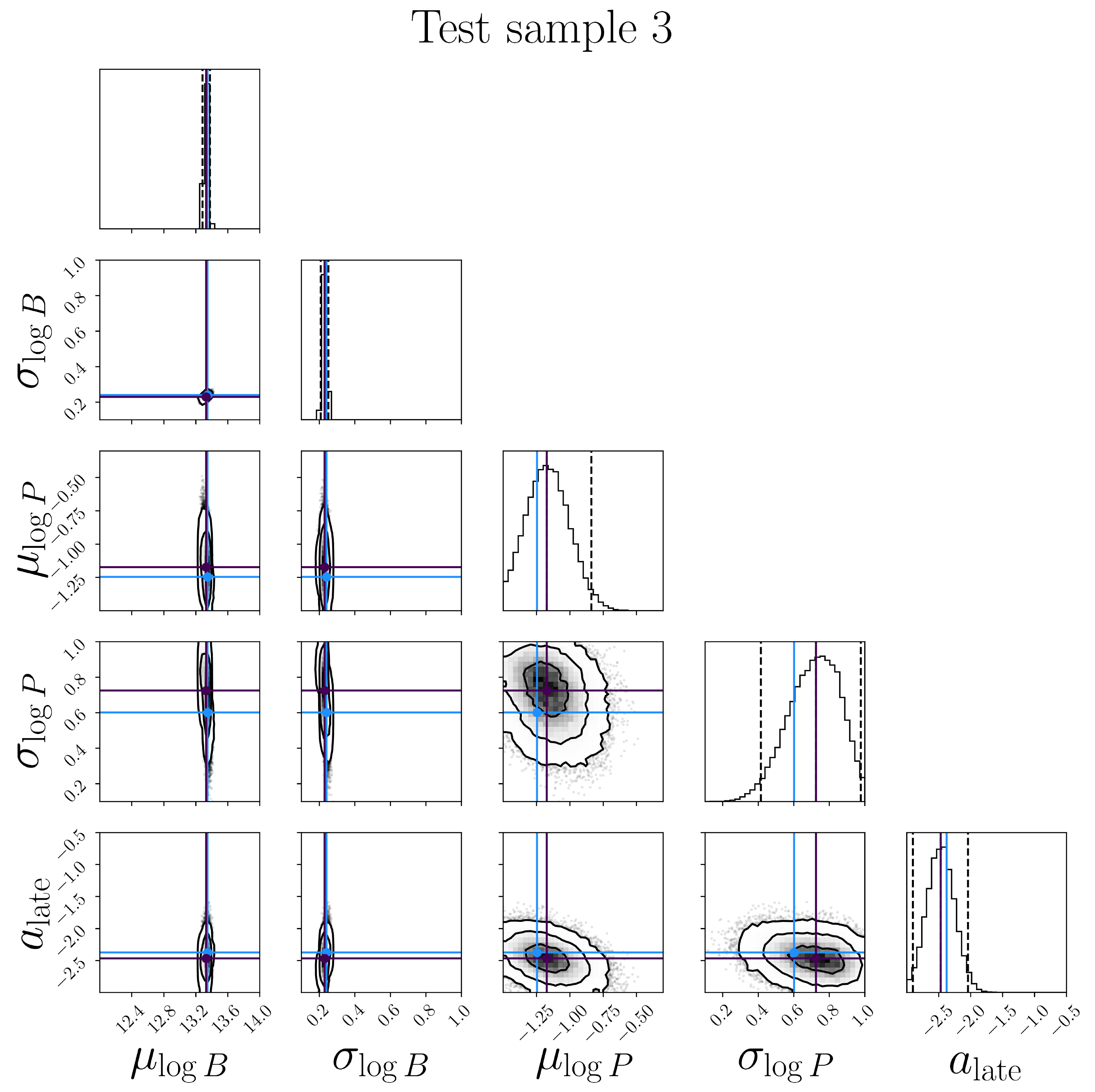}
	\caption[Benchmark inference for test simulation 3 using the network from experiment $\#1$]{Same as Fig.~\ref{fig:ch6_posterior_test12} but for test simulations $3$.}
	\label{fig:ch6_posterior_test3}
\end{figure}
%-----------------------------------------------------------------

%-----------------------------------------------------------------
\begin{figure}
	\centering
	\includegraphics[width=0.95\textwidth]{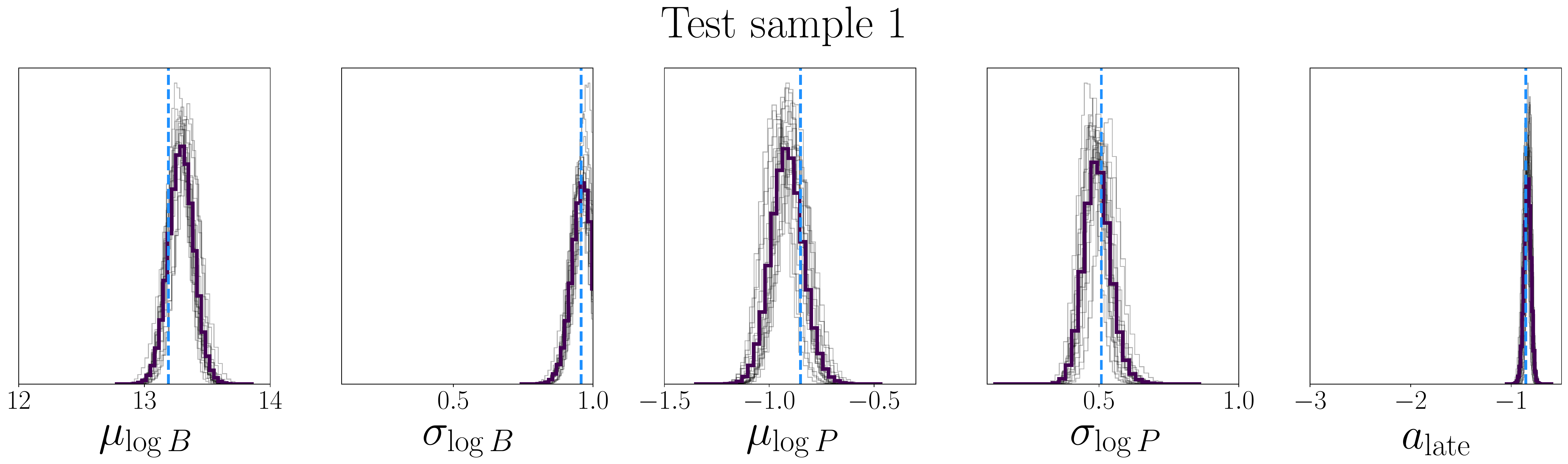}
	\vskip 0.2cm
	\includegraphics[width=0.95\textwidth]{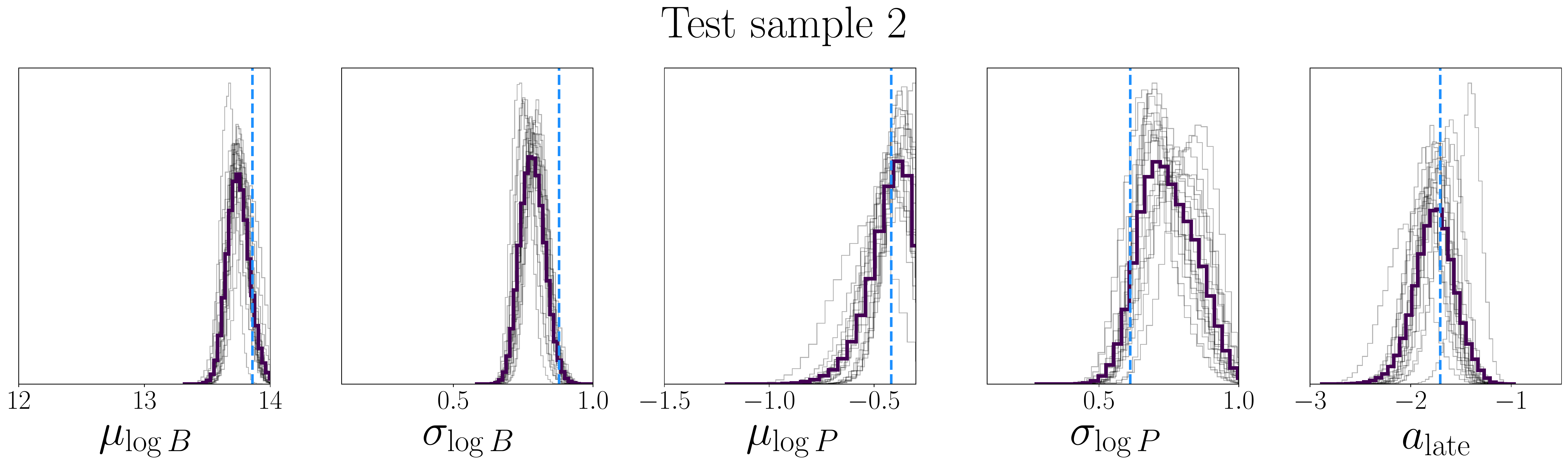}
	\vskip 0.2cm
	\includegraphics[width=0.95\textwidth]{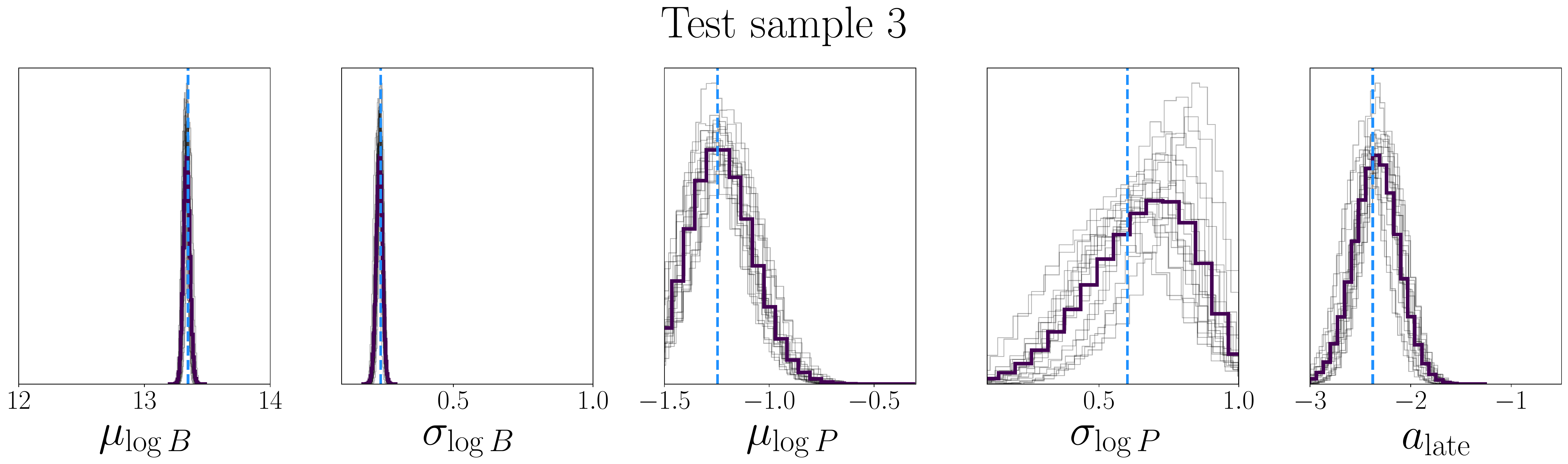}
	\caption[One-dimensional marginal posteriors for the five magneto-rotational parameters for the three test simulations]{One-dimensional marginal posteriors for the five magneto-rotational parameters for the three test simulations inferred using 19 different \ac{NPE} experiments shown in grey. The horizontal axes represent the parameters' prior ranges. The ground truths are shown as vertical dashed lines in light blue. We observe variation between the experiments, specifically for $\mu_{\log P}, \sigma_{\log P}$ and $a_{\rm late}$. We also plot the ensemble posteriors (purple) obtained as a weighted average of the individual posteriors.}
	\label{fig:ch6_posterior_comparison}
\end{figure}
%-----------------------------------------------------------------

%%%%%%%%%%%%%%%%%%%%%%%%%%%%%%

\subsection{Benchmark inferences}
\label{sec:ch6_inferences_test}

As a first assessment of our approximated posteriors, we focus on inferring the five magneto-rotational parameters, $\mu_{\log B}, \sigma_{\log B}, \mu_{\log P}, \sigma_{\log P}, a_{\rm late}$, for simulated populations where we know the input parameters, $\bt$. We specifically look at the three simulations, whose $P$-$\dot{P}$ diagrams were illustrated in the top row of Figure~\ref{fig:ch6_pop_simulated}. Corresponding ground truths, $\bt$, are summarised in the top five rows in Table~\ref{tab:ch6_credible_intervals}. In Figures~\ref{fig:ch6_posterior_test12} and~\ref{fig:ch6_posterior_test3}, we show the resulting one- and two-dimensional marginal posterior distributions obtained by repeatedly sampling from the neural network optimised during experiment $\#1$. For all three cases, the posteriors are well defined, significantly smaller than our prior ranges \eqref{eq:ch6_priors} shown along the axes, and centred around the ground truths, $\bt$, highlighted in light blue. To quantify this, we calculate the $1\sigma$, $2\sigma$ and $3\sigma$ credible regions, shown as contours in the two-dimensional posteriors. In the one-dimensional posterior panels, the corresponding $95\%$ credible intervals (\acp{CI}) are given as dashed, black lines, while medians are illustrated as solid, purple lines. Their numerical values are given in Table~\ref{tab:ch6_credible_intervals}. We observe that the ground truths, $\bt$, are typically contained within the $2\sigma$ credible regions, which we interpret as evidence that our \ac{NPE} approach is capable of producing reasonable posterior distributions. In general, the credible regions for the two parameters characterizing the initial magnetic-field distribution are narrower than those for the initial period distribution and the late-time magnetic-field decay. We confirm that the behaviour is qualitatively similar for the remaining $P$-$\dot{P}$ simulations in our test set.

We next compare the inferences for our various training experiments. To visualise corresponding differences, we plot the one-dimensional marginalised posteriors for all $19$ experiments for the three test samples in grey in Figure~\ref{fig:ch6_posterior_comparison}. Ground truths, $\bt$, are shown as dashed lines in light blue. We observe that the width of individual posterior approximations as well as their medians can vary somewhat between different test samples and magneto-rotational parameters. Compared across the full test set, this behaviour is again more dominant for the period and late-time magnetic-field parameters than for the initial $B$-field properties. However, no individual \acp{NPE} stand out by exhibiting either particularly good or poor posteriors. Further note that we also do not see any differences for those experiments with normalised input maps that showed systematically better metrics than those experiments trained on standardised data. This highlights that training behaviour alone does not provide sufficient information on the quality of the resulting inference.

In light of this, we also determine the combined posterior for all $19$ experiments \citep{Hermans2021}. We calculate the corresponding \textit{ensemble posterior}, $\overline{q} (\bt)$, as the weighted average of the individual posteriors:
\begin{equation}
\overline{q}(\bt) = \sum_{j=1}^{19} w_j q_{F_j} (\bt),
\end{equation}
where $w_j$ represents the weight of the $j$-th component. Giving equal importance to each experiment in the ensemble, we choose $w_j = 1/19$. The corresponding one-dimensional marginalised ensemble posteriors for $\mu_{\log B},$ $\sigma_{\log B}, \mu_{\log P}, \sigma_{\log P}$ and $a_{\rm late}$ for the three test simulations are illustrated as purple histograms in Figure~\ref{fig:ch6_posterior_comparison}. As expected, they fall within the individual posteriors. The corresponding $95\%$ \acp{CI} for the three test samples, which are typically comparable or slightly wider than those calculated for experiment $\#1$ posteriors alone, are summarised in the bottom five rows of Table~\ref{tab:ch6_credible_intervals}.

%%%%%%%%%%%%%%%%%%%%%%%%%%%%%%

%-----------------------------------------------------------------
\begin{figure}
	\centering
	\includegraphics[width=0.8\columnwidth]{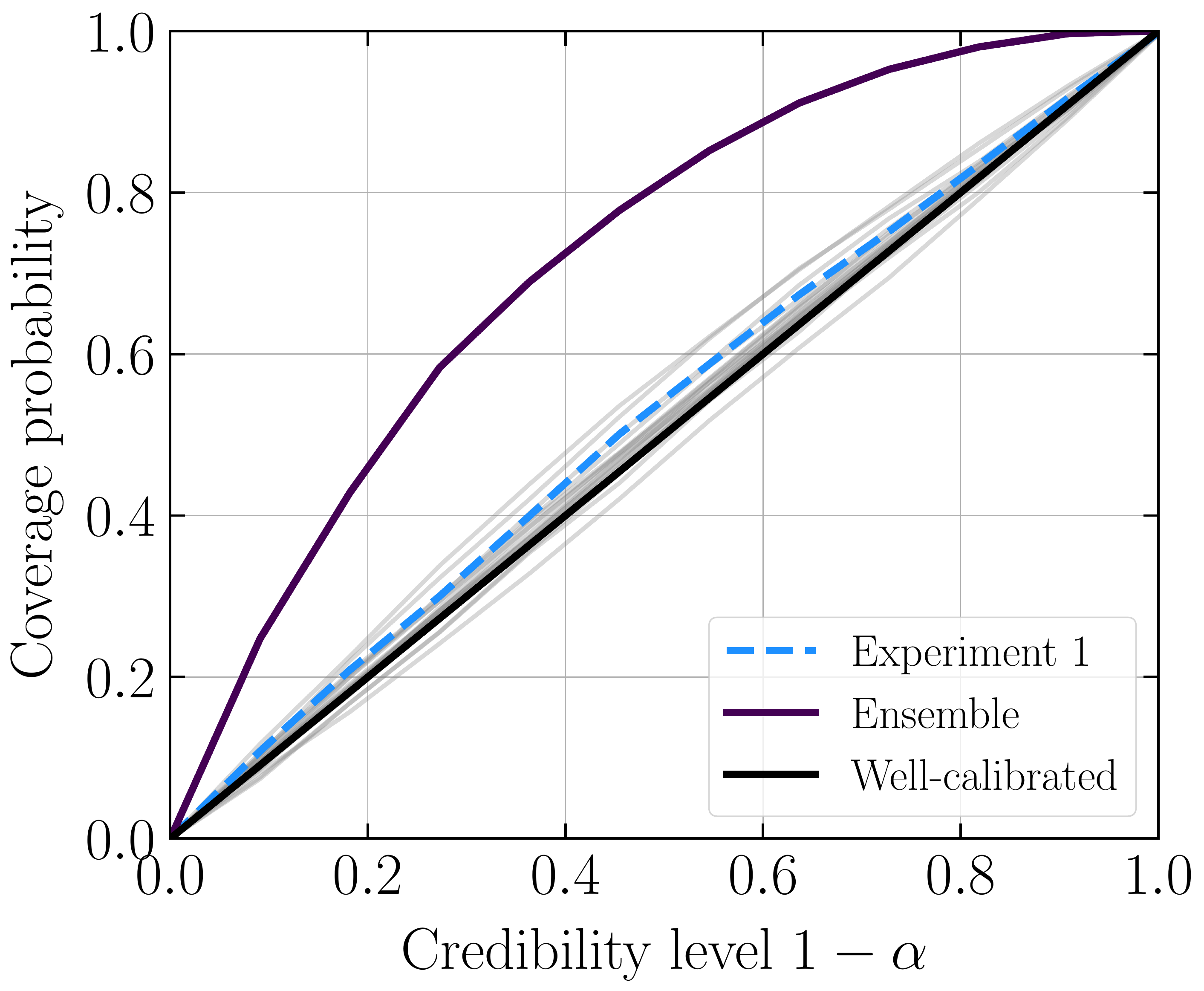}
	\caption[Coverage probability of the experiments]{Coverage probability as a function of the credibility level, $1 - \alpha$, for our approximate posteriors calculated for the $3,600$ test simulations. We specifically highlight the coverage for experiment $\#1$ as a dashed, light blue line and that for the ensemble as a solid, purple line. All remaining experiments are given in grey. For a well-calibrated posterior, the coverage follows the diagonal shown in black.}
	\label{fig:ch6_coverage}
\end{figure}
%-----------------------------------------------------------------

\subsection{Posterior validation}
\label{sec:ch6_valdation}

To further assess whether posterior estimates are well-calibrated, we determine their \textit{coverage} (see Section~\ref{sec:ch2_coverage}). As outlined in detail in Appendix~\ref{app:coverage}, the coverage probability measures the fraction of test samples for which (for a given credibility level $1 - \alpha$) the ground truths, $\bt$, fall within the corresponding $1 - \alpha$ region of their respective posteriors, $q_{F(\bx, \boldsymbol{\phi})} (\bt)$. For a well-calibrated posterior distribution and a sufficiently large number of test samples, this fraction should equal $1 - \alpha$. This implies that the coverage probability as a function of the credibility level is diagonal. In contrast, for a conservative posterior that is wider than the true posterior, we would recover a fraction larger than $1 - \alpha$. Conversely, for a narrower (overconfident) posterior, the corresponding fraction of test samples is less than $1 - \alpha$. In terms of the coverage, this corresponds to curves above and below the diagonal, respectively, and can, therefore, be used to assess the quality of approximate posteriors. 

We show the coverage probabilities for our different posterior estimates as a function of the credibility level, $1 - \alpha$, in Figure~\ref{fig:ch6_coverage}. We specifically highlight the coverage for the posterior from experiment $\#1$ (dashed, light blue) and the ensemble posterior (solid, purple). All remaining experiments are shown in grey. We observe that the approximate posteriors for individual experiments closely follow the diagonal, exhibiting either slightly conservative or overconfident behaviour. As expected, the most conservative estimate is given by our ensemble posterior, which incorporates variations in the inference for $19$ different machine-learning configurations across all $3,600$ test samples. These results provide additional support that our neural posteriors are trustworthy and have indeed learned to accurately infer magneto-rotational parameters from simulated $P$-$\dot{P}$ density maps.

%%%%%%%%%%%%%%%%%%%%%%%%%%%%%%

%-----------------------------------------------------------------
\begin{figure}
	\centering
	\includegraphics[width=\textwidth]{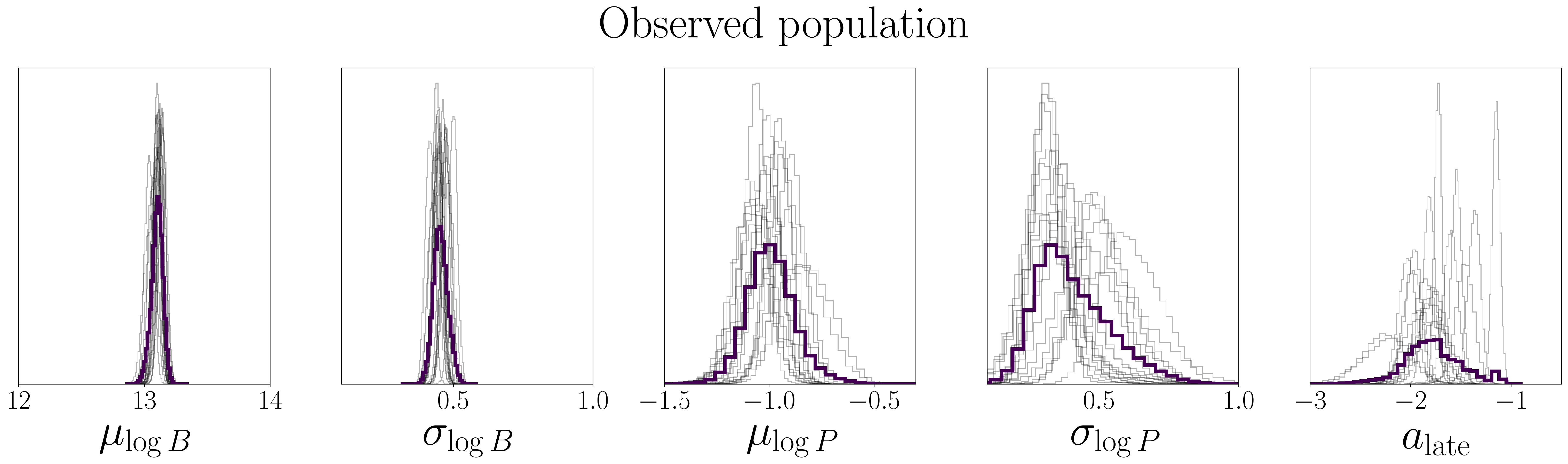}
	\caption[One-dimensional marginal posteriors for the five magneto-rotational parameters for the observed pulsar population]{One-dimensional marginal posteriors for the five magneto-rotational parameters for the observed pulsar population. We show inference results for 19 different \ac{NPE} experiments in grey and the ensemble posterior in purple.}
	\label{fig:ch6_posterior_comparison_atnf}
\end{figure}
%-----------------------------------------------------------------

\subsection{Inference on the observed population}
\label{sec:ch6_inference_observed}

Following the benchmark experiments and the coverage determination, we now turn our attention to the true pulsar populations observed with the \ac{PMPS}, the \ac{SMPS} and the low- and mid-latitude \ac{HTRU} survey. The corresponding $P$-$\dot{P}$ diagram was shown in the bottom panel of Fig.~\ref{fig:ch6_pop_observed}. We represent these populations as three density maps as outlined in Section~\ref{sec:ch6_sim_output} and subsequently feed them through our trained neural networks to infer the five parameters, $\mu_{\log B}, \sigma_{\log B}, \mu_{\log P}, \sigma_{\log P}$ and $a_{\rm late}$, assuming that our simulation framework provides a realistic description of the underlying physics.

\begin{figure}
	\centering
	\includegraphics[width=\textwidth]{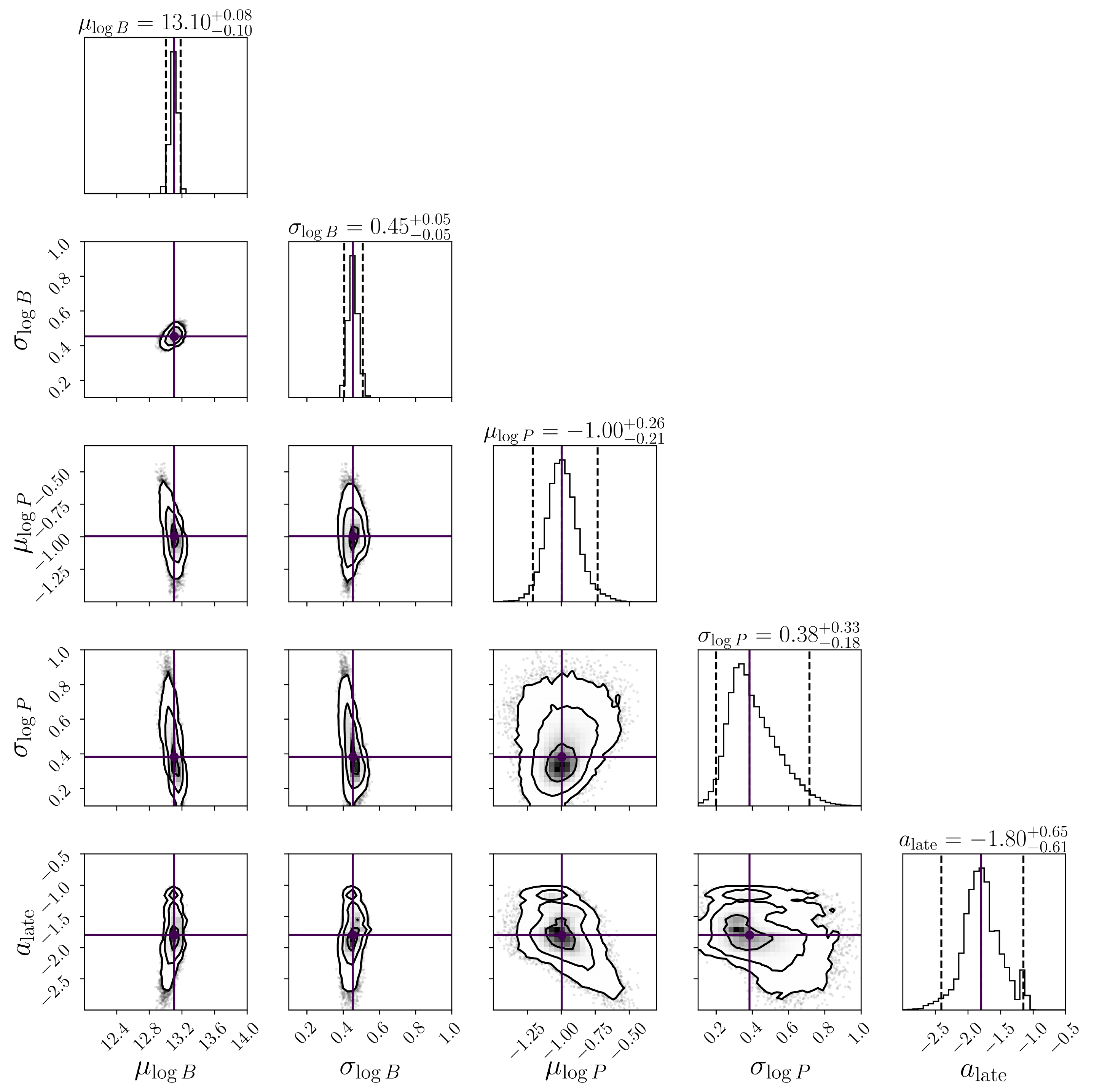}
	\caption[Inference results for the observed pulsar population using the ensemble posterior]{Inference results for the observed pulsar population using the ensemble posterior of $19$ different \acp{NPE}. The corner plot shows one- and two-dimensional marginal posterior distributions for the five magneto-rotational parameters. We highlight the medians in purple. Corresponding values and $95\%$ \acp{CI} are summarised above the panels and in Table~\ref{tab:ch6_credible_intervals}.}
	\label{fig:ch6_posterior_ensemble_atnf}
\end{figure}

We show the corresponding one-dimensional marginal posterior distributions for individual experiments (grey histograms) and the ensemble (purple histograms) in Figure~\ref{fig:ch6_posterior_comparison_atnf}. Additionally, a corner plot for the one- and two-dimensional ensemble posteriors is illustrated in Figure~\ref{fig:ch6_posterior_ensemble_atnf}. Corresponding medians (shown in purple in the corner plot) and $95\%$ \acp{CI} for experiment $\#1$ and the ensemble are also summarised in the last column of Table~\ref{tab:ch6_credible_intervals}.

The general trend (already observed for the simulated populations) that the initial magnetic-field parameters, $\mu_{\log B}$ and $\sigma_{\log B}$, are much better constrained by our $\ac{NPE}$ framework than the remaining three values also holds for the observed population. As seen in the first two panels of Figure~\ref{fig:ch6_posterior_comparison_atnf}, all $19$ experiments recover narrow posteriors around similar medians. For the initial period-distribution parameters, $\mu_{\log P}$ and $\sigma_{\log P}$, (see third and fourth panel, respectively), we obtain wider posteriors and a larger variety of median values between different experiments. These posteriors, however, cover similar regions within our prior ranges and are comparable to what we observed for the test samples. In contrast, the inferred posteriors for $a_{\rm late}$ (the final panel in Figure~\ref{fig:ch6_posterior_comparison_atnf}) exhibit different behaviour to our benchmark experiments. In particular, posteriors vary significantly in width between different experiments with those at the larger (smaller) end of the $a_{\rm late}$ range generally exhibiting narrower (larger) widths. Moreover, several distributions do not overlap at all. This is manifest as a relatively wide posterior in the ensemble which also shows a second peak, primarily driven by the right-most individual posterior resulting from experiment $\#2$. Note that this configuration did not cause irregularities during the network optimisation or unusual posteriors for our test samples. We, therefore, do not associate this behaviour with the network itself. The corresponding bi-modality is also visible in the final row of the corner plot in Figure~\ref{fig:ch6_posterior_ensemble_atnf}. We will discuss our interpretation of this below.

%-----------------------------------------------------------------
%-----------------------------------------------------------------

%%%%%%%%%%%%%%%%%%%%%%%%%%%%%%%%%%%%%%%%%%%%%%%%%%%%%%%%%%%%%%%%%%%%%%%%%%%%%%%%%%%%%%%%%%%%%%%%%%%%%%%%%%%%%%%%%%%%%%%%%%%%%%

\section{Discussion and conclusions}
\label{sec:ch6_conclusions}

In this study, we have successfully developed a new machine-learning pipeline that combines pulsar population synthesis with \acf{SBI} for the first time and tested the corresponding approach for inferring magneto-rotational properties of neutron stars. 

%%%%%%%%%%%%%%%%%%%%%%%%%%%%%%

\subsection{Simulation framework}
\label{sec:ch6_disc_simulator}

We first discussed our implementation of the forward model, i.e., the prescription for simulating the dynamical and magneto-rotational properties of the Galactic population of isolated radio pulsars, modelling their radio emission and subsequently mimicking observational limitations for the \acf{PMPS}, the \acf{SMPS} and the low- and mid-latitude \acf{HTRU} survey. We followed earlier frameworks \citep[e.g.,][]{Faucher2006, Bates2014, Gullon2014, Gullon2015, Cieslar2020} but implemented several key differences (see also Table~\ref{tab:ch1_pop_syn}). In particular, we sampled the birth positions of our pulsars from the Galactic electron distribution \citep{Yao2017} instead of following the typical approach of combining a spiral-arm model with a radial pulsar distribution of, e.g., \citet{Yusifov2004}. The latter is deduced for the observed, evolved pulsar sample and not the initial population. Moreover, we have included the (rigid) rotation of the Galaxy, which leads to a more consistent treatment of the pulsar birth positions compared to earlier analyses \citep[see also Chapter~\ref{Chapter5} and][]{Ronchi2021}. For the magnetic-field evolution, we used a similar approach to \citet{Gullon2014, Gullon2015} taking advantage of the newest two-dimensional magneto-thermal simulations \citep{Vigano2021} and solved for the coupled evolution of the spin period, $P$, and the misalignment angle, $\chi$, for a plasma-filled magnetosphere. To capture the field changes at late times, we developed a new physically motivated prescription in which the magnetic field, $B$, decays according to a power law captured by the index, $a_{\rm late}$. Together with the means, $\mu_{\log B}, \mu_{\log P}$, and standard deviations, $\sigma_{\log B}, \sigma_{\log P}$, which characterise the normally distributed logarithms of the initial periods and the initial fields, we hence obtained five parameters that control the neutron stars' magneto-rotational evolution. 

To simulate the detection of our synthetic pulsars, we make two main changes compared to earlier studies. First, we do not model the pulsars' pseudo luminosity (see Section~\ref{sec:ch1_pseudo_luminosity}) defined as $L_{f,\rm pseudo} \propto S_{f, {\rm mean}} d^2$ (where $S_{f, {\rm mean}}$ is the period-averaged detected flux at frequency $f$, and $d$ the pulsar distance) but instead assume that the intrinsic neutron-star luminosity, $L_{\rm int}$, is proportional to the spin-down power, $\dot{E}_{\rm rot}$ (see Section~\ref{sec:ch1_energetics}). In particular, we considered $L_{\rm int} \propto \dot{E}_{\rm rot}^{1/2}$ to determine the bolometric radio flux and subsequently propagate the corresponding pulsed emission towards the Earth. We also used a geometry-based description to determine the pulsars that are beamed towards us, which earlier works typically treat in an empirical manner. Finally, we not only looked at \ac{PMPS} and \ac{SMPS} but also incorporated the \ac{HTRU} survey for the first time. Using the resulting simulation framework, we then produced $360,000$ synthetic $P$-$\dot{P}$ diagrams which we converted to one density map per survey in preparation for the neural network. $90\%$ of these simulations were used for training and validation, and the remaining $10\%$ reserved for testing. 

%%%%%%%%%%%%%%%%%%%%%%%%%%%%%%

\subsection{Inference procedure}
\label{sec:ch6_disc_inference}

The second part of this study is centred on the implementation of the \ac{SBI} approach, specifically focusing on \acf{NPE}, to learn a probabilistic association between our simulator output and the input parameters, $\bt = \{\mu_{\log B}, \sigma_{\log B},\\ \mu_{\log P}, \sigma_{\log P}, a_{\rm late}\}$. To do so, we first used a \acf{CNN} to extract features from our high-dimensional $P$-$\dot{P}$ maps and obtain a compressed representation, which was then transferred into a flexible neural density estimator. By taking advantage of the open-source Python package {\tt sbi} \citep{Tejero-Cantero2020},\footnote{\url{https://github.com/sbi-dev/sbi}} we specifically implemented a Gaussian-mixture density model in five dimensions to approximate our posterior. To study the sensitivity of the \ac{NPE} results on the representation of our input data and the network hyperparameters, we conducted $22$ distinct experiments. An inspection of the corresponding training metrics led us to discard three experiments due to irregular training behaviour or overfitting. The remaining $19$ trained neural networks were analyzed further and we found no significant differences in the resulting inferences when benchmarked on three random test simulations. The same was observed when validating the posteriors through a coverage calculation over the test set with $3,600$ samples, highlighting that all $19$ posterior estimates are well-calibrated. From this we concluded, in particular, that the training behaviour is a poor identifier of subsequent inference quality, because normalisation of input maps led to systematically better training, test and validation metrics compared to standardizing the input but comparable inferences. Learning rate and batch size played a negligible role in both set-ups. 

We also point out that the use of smaller training data sets did not affect the inference quality either. While we expect that training sets of $\lesssim 10\%$ (i.e., $30,000$ simulations) will eventually have an effect on this, databases of $50 \%$ (i.e., $150,000$ simulations) are sufficient when inferring five parameters. For comparable studies, this would imply a significant reduction in simulation time, the most costly part of these analyses. Similar performances further justify optimizing our networks for density maps with a resolution of $32 \times 32$ bins instead of $64\times 64$ and the shallower baseline \ac{CNN} to speed up the training process. Additionally, we highlight that the use of different numbers of Gaussian mixture components also led to comparable optimisation metrics and inference results. Extracting the corresponding mixture weights, $\alpha_{\rm c}$, after the optimisation, we find that we only require two or three Gaussians to approximate our posteriors across the entire test data set. We do, however, point out that training with a larger number of components was faster. Finally, note that the use of fewer surveys (i.e., one or two density maps only) did not change the inference results for our five magneto-rotational parameters. Naively, one might think that complementary information on the pulsar population as, e.g., provided by \ac{SMPS}, which is sensitive to older stars at higher Galactic latitudes, would help the network learn better posteriors. We do, however, not observe such behaviour in our experiments. Although this might suggest that using single surveys in the future could be sufficient to constrain neutron-star parameters through population synthesis, we caution that different surveys, in principle, provide additional information on the neutron-star birth rate (see below) which was not supplied to our neural networks, i.e., we focused on the location and shape of the pulsar population in the $P$-$\dot{P}$ plane only.

Due to the variations in our inference results, and because we could not identify a single neural network as the best posterior estimator, we also determined the ensemble posterior through an equally weighted average of the individual experiments. The resulting posterior behaved as expected and showed more conservative behaviour than the ensemble members. For the next section, we, will, hence, follow the recommendation by \citet{Hermans2021} and use our (most conservative) ensemble posterior to analyze the observed pulsar population.

%%%%%%%%%%%%%%%%%%%%%%%%%%%%%%

\subsection{Inference results on the observed population}
\label{sec:ch6_disc_inference_obs}

Following the validation of our \ac{NPE} approach, we subsequently used the ensemble posterior estimator to infer the five magneto-rotational parameters for the true population of isolated Galactic radio pulsars observed with our three surveys. In particular, we found the following best estimates at $95\%$ credible level:
\begin{align}
\mu_{\log B} &= 13.10^{+0.08}_{-0.10}, \nonumber \\
\sigma_{\log B} &= 0.45^{+0.05}_{-0.05}, \nonumber \\
\mu_{\log P} &= -1.00^{+0.26}_{-0.21}, \label{eq:ch6_CI_best} \\
\sigma_{\log P} &= 0.38^{+0.33}_{-0.18}, \nonumber \\
a_{\rm late} &= -1.80^{+0.65}_{-0.61}. \nonumber
\end{align}
The corresponding corner plot was shown in Figure~\ref{fig:ch6_posterior_ensemble_atnf}. 
%show the corresponding PDFs in Fig. ...

As noted during the benchmarking experiments, we generally obtain narrower posterior distributions for the initial magnetic-field parameters when compared to the initial period parameters. Difficulties in constraining rotational birth properties are, however, not a shortcoming of our inference approach itself as this was also noted by earlier population-synthesis analyses \citep[e.g.,][]{Gullon2014, Gullon2015, Cieslar2020}. Instead, this has a physical reason that lies in the coupled evolution of the stars' misalignment angle, rotation period and magnetic field. While the $B$-field initially stays constant (see Figure~\ref{fig:ch6_B_fields}), pulsars move from the top left in the $P$-$\dot{P}$ plane diagonally towards the bottom right, following lines of constant magnetic field (see, e.g., right panel in Figure~\ref{fig:ch6_pop_observed}). As they do, stars with comparable field strengths but different initial periods evolve towards similar $P$ values. This is a consequence of the fact that the spin period evolution tends to lose memory of the initial spin periods for $t \gg t_{\rm em}$ where $t_{\rm em} \propto P_0^2 / B_0^2$, as seen in Section~\ref{sec:ch3_dipolar_losses}.In addition, the misalignment-angle evolution introduces further degeneracies because all $\chi$ decrease with time. However, as the field decays, spin-down and misalignment evolution slow down and pulsars begin to evolve almost vertically towards smaller $\dot{P}$ values. These processes depend further on $B_0$ and $P_0$ as stronger initial fields and smaller initial periods result in faster spin-down and faster evolution towards alignment. This is especially visible for test sample 3 (top right panel of Figure~\ref{fig:ch6_pop_simulated}), which is characterized by the smallest period mean, $\mu_{\log P}$, of all three test cases. The combined action of these effects is that stars born with different rotational properties attain similar $P$ at current times. This information loss on the initial period makes it harder to infer the corresponding parameters. As expected, test sample 3, therefore, shows the largest $95 \%$ \acp{CI} for $\mu_{\log P}$ and $\sigma_{\log P}$ out of our three test samples (see third column in Table~\ref{tab:ch6_credible_intervals} and last row in Figure~\ref{fig:ch6_posterior_comparison}).

%-----------------------------------------------------------------
\begin{figure}
	\centering
	\includegraphics[width=0.8\textwidth]{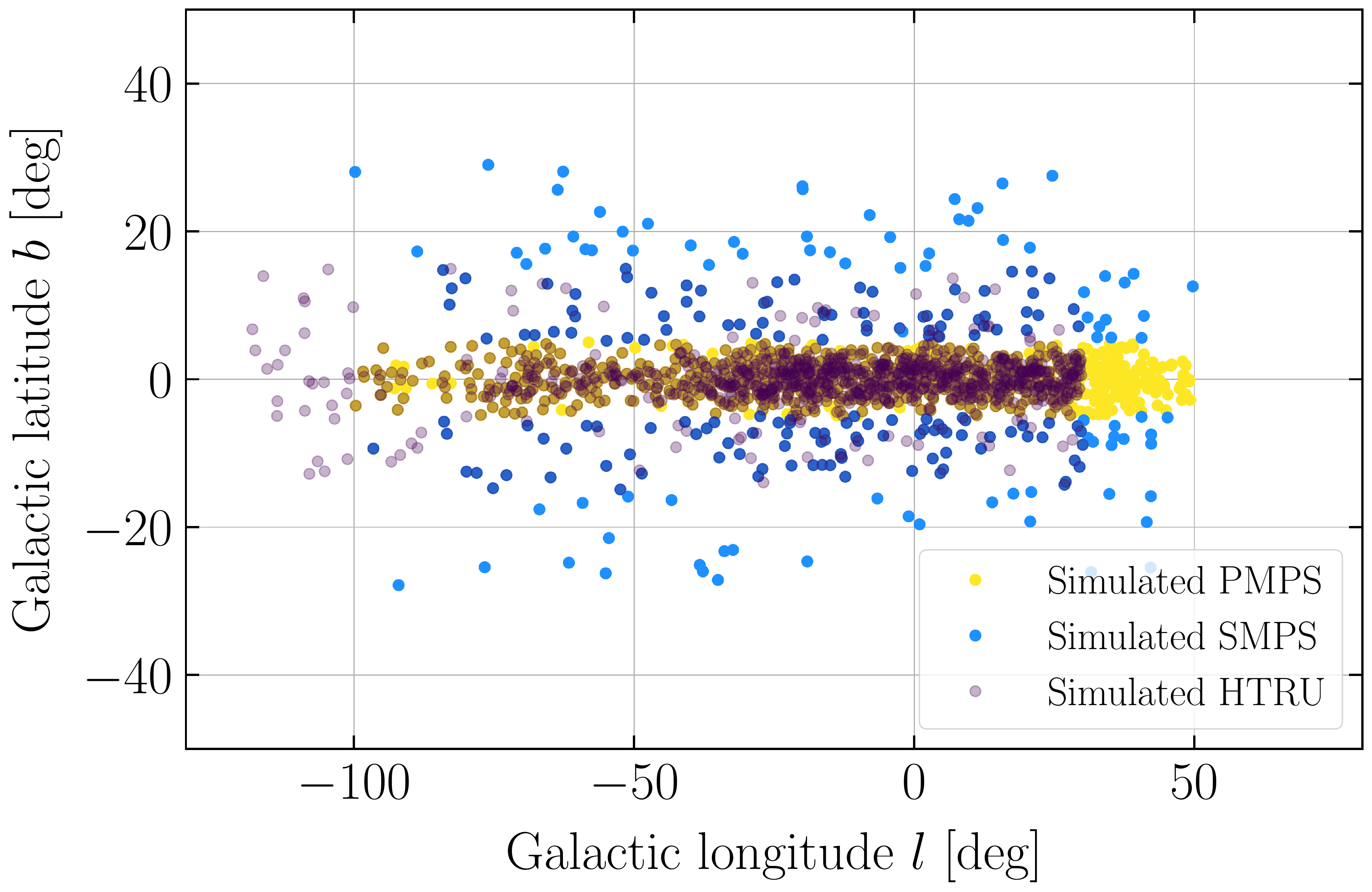}
	\hspace{0.2cm}
	\includegraphics[width=0.6\textwidth]{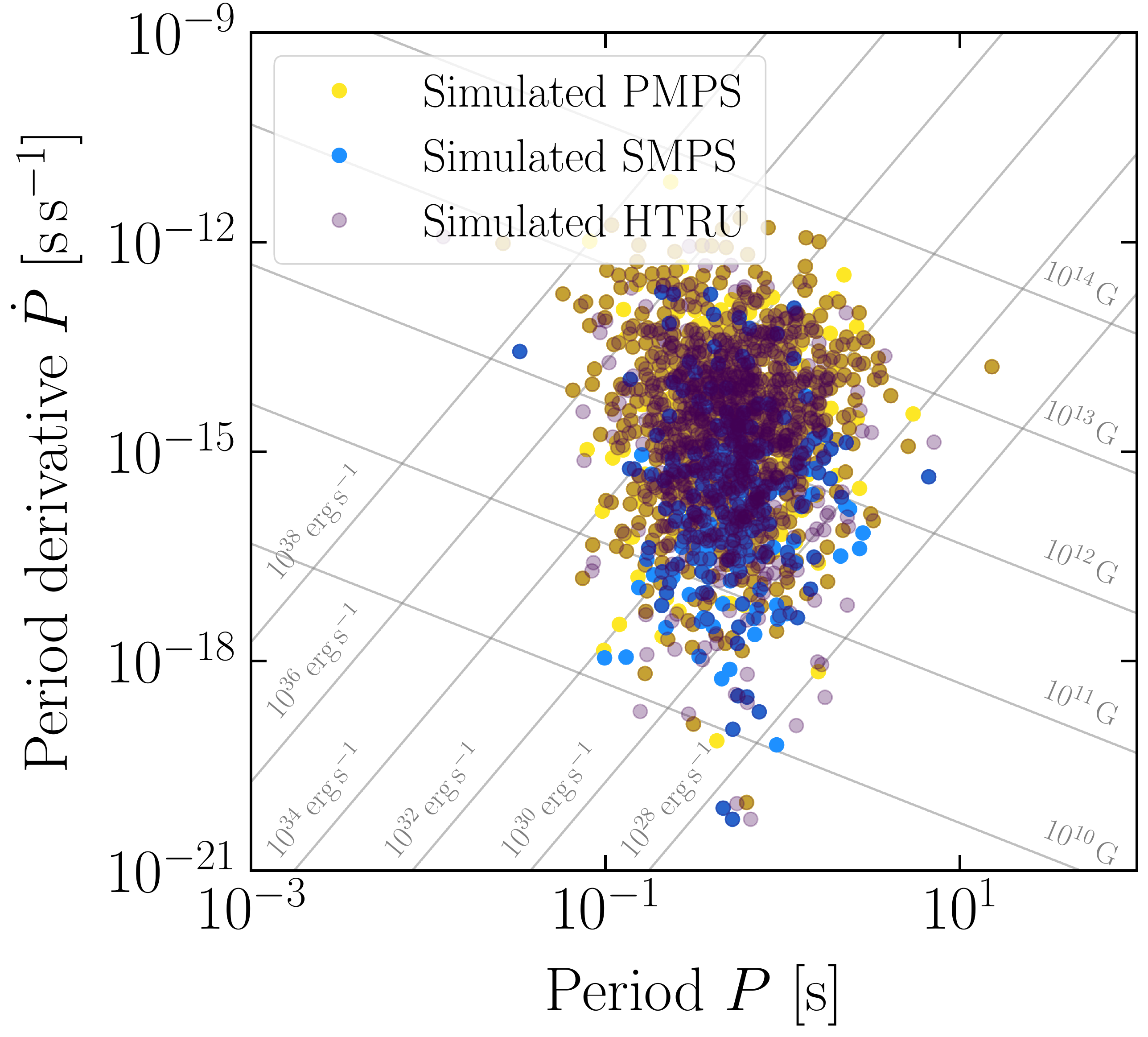}
	\caption[Simulated populations of isolated Galactic radio pulsars]{Simulated populations of isolated Galactic radio pulsars detected with the \ac{PMPS}, the \ac{SMPS} and the low- and mid-latitude \ac{HTRU} survey (highlighted in yellow, light blue and purple, respectively) for the parameters inferred via \ac{SBI} from the observed radio pulsar population (see Equation~\eqref{eq:ch6_CI_best}). The left panel shows the distribution of the simulated population in Galactic latitude, $b$, and longitude, $l$, while the right panel depicts the pulsars in the period, $P$, and period derivative, $\dot{P}$, plane. In the latter, we also give lines of constant spin-down power, $\dot{E}_{\rm rot}$, and constant dipolar surface magnetic field, $B$, (estimated via Equation~\eqref{eq:ch6_P_ode} for an aligned rotator). Both plots directly compare to the true (observed) population shown in Figure~\ref{fig:ch6_pop_observed}.}
	\label{fig:ch6_pop_best_simulated}
\end{figure}
%-----------------------------------------------------------------

Contrasting the posterior medians from Equation~\eqref{eq:ch6_CI_best} to other population-synthesis studies, we first note that our $\mu_{\log B}$ estimate is roughly consistent with \citet{Gullon2014, Gullon2015} but somewhat larger than those of \citet{Faucher2006} and \citet{Cieslar2020}. Moreover, we obtain a smaller $\sigma_{\log B}$ than \citet{Gullon2014, Gullon2015} and \citet{Faucher2006} but a slightly larger estimate than \citet{Cieslar2020}. Although all four studies determine optimal parameter ranges different to us, we expect the differences in the initial $B$ constraints to be mainly due to our more realistic prescription for the magnetic-field and coupled $P$-$\chi$ evolution. A direct comparison between our initial period parameters and earlier literature values is not possible at this stage, because (following recent results by \citet{Igoshev2022}; see also \citet{Xu2023}) we have considered the periods' logarithm and not the periods themselves to be normally distributed. We, however, highlight that our inferred initial period and magnetic-field parameters are comparable with those of \citet{Igoshev2022} obtained through a simplified analysis of $56$ young neutron stars in supernova remnants. As this study looked at magneto-rotational properties only, the authors were able to define an explicit likelihood function and perform statistical inference. In this context, we also point out that although \citet{Cieslar2020} derive (relatively narrow) posteriors for a range of pulsar properties using an \ac{MCMC} analysis, their underlying simulation framework is significantly reduced compared to ours invoking, e.g., (unrealistic) exponential field decay, vacuum magnetospheres, no coupling between periods and misalignment angles, and a simplified prescription of the beamed emission. In addition, they make an explicit assumption on the likelihood that might not accurately capture the complexity of the pulsar population synthesis even for their simplified model. We reiterate the robustness of our \ac{SBI} approach which eliminates the need for an explicit expression for the likelihood and is, therefore, also suitable for more complex simulators like ours. Moreover, as outlined above the use of a neural density estimator results in amortised posterior distributions that allow fast evaluation and sampling. We used this fact to determine the coverage and validate our posteriors, a procedure that is infeasible in \ac{MCMC} approaches due to the time-consuming need for repeated sampling.

%%%%%%%%%%%%%%%%%%%%%%%%%%%%%%

\subsection{Comparing results with earlier works}
\label{sec:lit_comparison}

%---------------------------------------------------------------
\begin{table}
	\centering
	\caption[Comparison between best parameters for the log-normal initial magnetic-field and initial period distributions in the literature.]{Comparison between best parameters for the log-normal initial magnetic-field and initial period distributions in the literature. We provide the references and the four relevant parameters. Note that the first three studies use a different prescription for the initial period, which prevents a direct comparison with our study. For \citet{Gullon2015} and \citet{Cieslar2020}, we compare with \textit{model D} for the radio-pulsar population and the \textit{rotational model}, respectively. The corresponding distributions are illustrated in Fig.~\ref{fig:log_normal_distributions}. \label{tab:best_param}}
	\begin{tabular}{c|cccc}
		\toprule
		\tabhead{References} &
		\tabhead{$\mu_{\log B}$} &
		\tabhead{$\sigma_{\log B}$} &
		\tabhead{$\mu_{\log P}$} &
		\tabhead{$\sigma_{\log P}$} \\
		\midrule
		\citeauthor{Faucher2006} & 12.65 & 0.55 & - & -  \\
		\citeauthor{Gullon2015} & 12.99 & 0.56 & - & - \\
		\citeauthor{Cieslar2020} & 12.67 & 0.34 & - & - \\
		\citeauthor{Igoshev2022} & 12.44 & 0.44 & $-1.04$ & 0.53 \\
		This work & 13.10 & 0.45 & $-1.00$  & 0.38 \\
		\bottomrule
	\end{tabular}
\end{table}
%---------------------------------------------------------------

%-----------------------------------------------------------------
\begin{figure}
	\centering
	\includegraphics[width=0.7\textwidth]{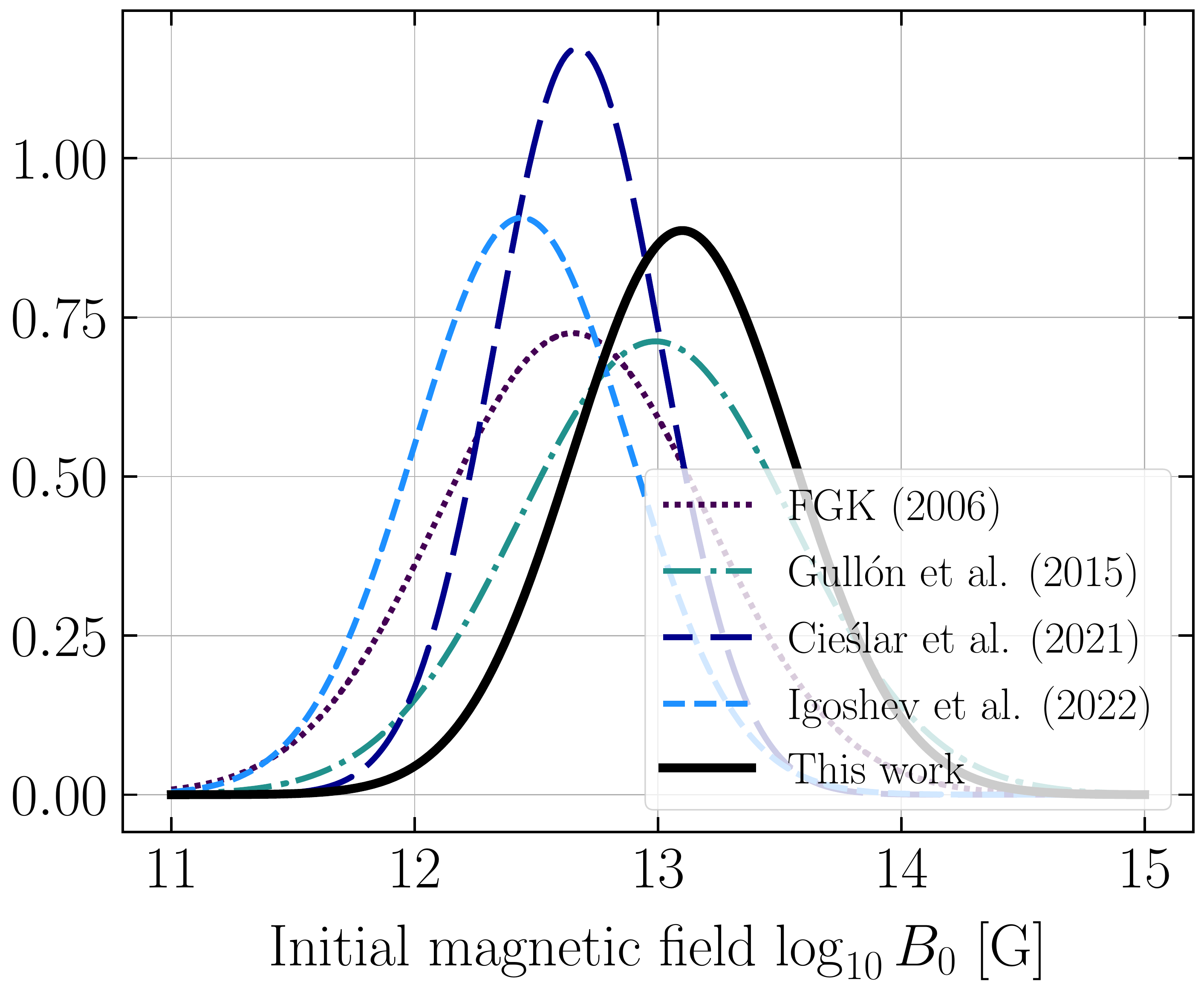}
	\vskip 0.3cm
	\includegraphics[width=0.7\textwidth]{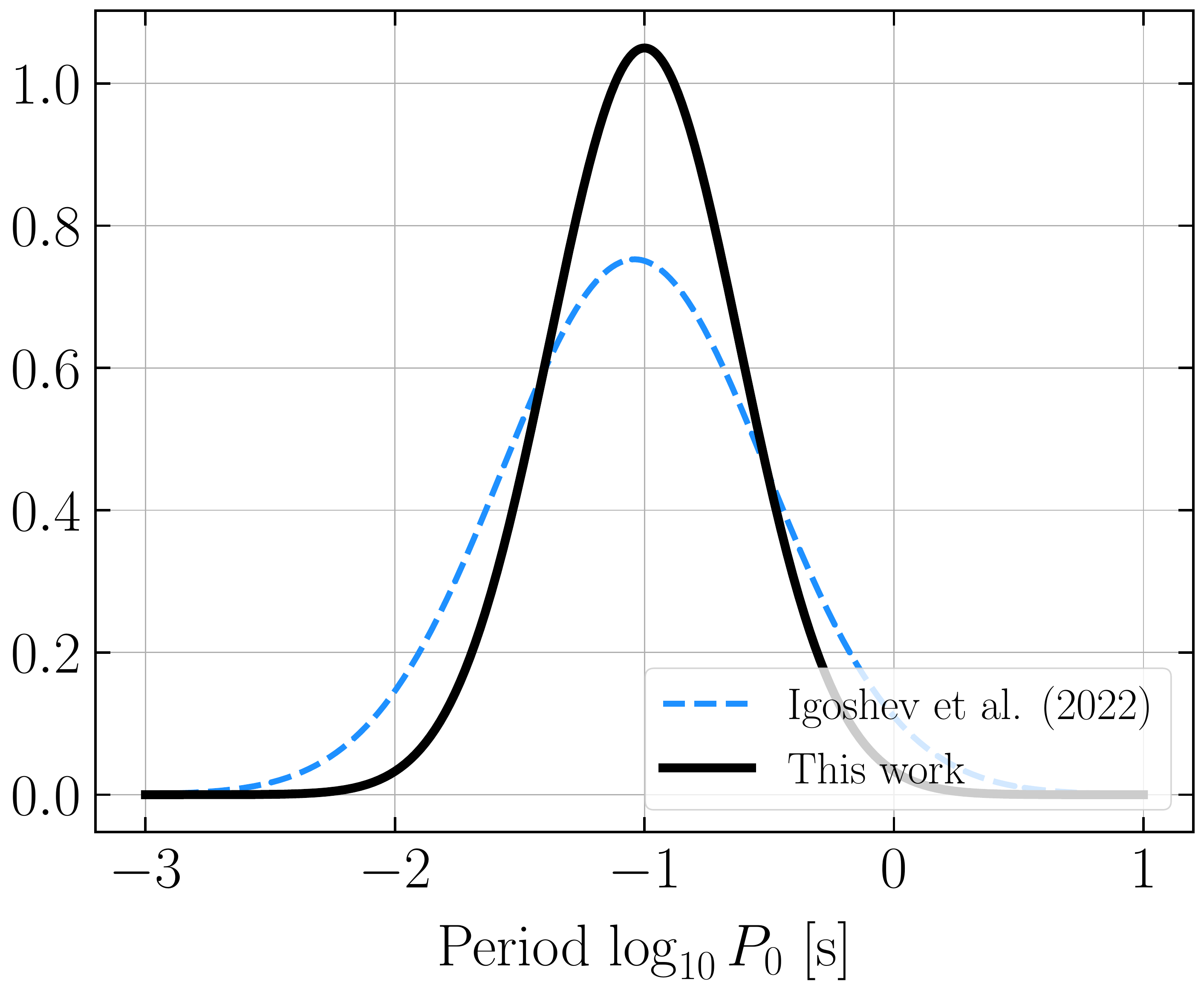}
	\caption[A comparison of initial magnetic-field and spin-period distributions for the radio pulsar population]{A comparison of initial magnetic-field, $B_0$, (\textit{top}) and period, $P_0$, (\textit{bottom}) distributions for the radio pulsar population. The logarithms of $B_0$ and $P_0$ are normally distributed (see Equations~\eqref{eq:ch6_B_pdf} and \eqref{eq:ch6_P_pdf}) and characterized by means, $\mu_{\log B, P}$, and standard deviations, $\sigma_{\log B, P}$, respectively. Corresponding numerical values are summarized in Table~\ref{tab:best_param}. The results of this work are illustrated as black, solid lines. Additional studies are shown as detailed in the legends.}
	\label{fig:log_normal_distributions}
\end{figure}
%-----------------------------------------------------------------

Contrasting the posterior medians from Equation~\eqref{eq:ch6_CI_best} with the results of earlier population-synthesis studies summarized in Table~\ref{tab:best_param} and Figure~\ref{fig:log_normal_distributions}, we first note that our $\mu_{\log B}$ estimate is roughly consistent with \citet{Gullon2014, Gullon2015} but somewhat larger than those of \citet{Faucher2006}, \citet{Cieslar2020} and \citet{Igoshev2022}. Moreover, while very close to \citet{Igoshev2022}, we obtain a smaller $\sigma_{\log B}$ than \citet{Gullon2014, Gullon2015} and \citet{Faucher2006} and a slightly larger estimate than \citet{Cieslar2020}. Although these works determine optimal parameter ranges different to us, we expect the variation in the $B_0$ constraints to be mainly due to our more realistic prescription for the field and the coupled $P$-$\chi$ evolution (see Table~\ref{tab:ch1_pop_syn}). 

A direct comparison of our initial period parameters and earlier population-synthesis literature is not possible, because (following recent results by \citet{Igoshev2022}; see also \citet{Xu2023}) we considered the periods' logarithm and not the periods themselves to be normally distributed. We, however, highlight that our inferred $\mu_{\log P}$ is comparable with that of \citet{Igoshev2022}, whereas our $\sigma_{\log P}$ is somewhat smaller (see bottom panel of Figure~\ref{fig:log_normal_distributions}). \citet{Igoshev2022} focused on a simplified analysis of $56$ young neutron stars in supernova remnants and looked at magneto-rotational properties only. The authors were, thus, able to define an explicit likelihood function and perform statistical inference. In this context, we also point out that although \citet{Cieslar2020} derive (relatively narrow) posteriors for a range of pulsar properties using an \ac{MCMC} analysis, their underlying simulation framework is significantly reduced compared to ours invoking, e.g., (unrealistic) exponential field decay, vacuum magnetospheres, no coupling between periods and misalignment angles, and a simplified prescription for the beamed emission. In addition, they make an explicit assumption on the likelihood that might not accurately capture the complexity of the pulsar population synthesis even for their simplified model. We reiterate the robustness of our \ac{SBI} approach which eliminates the need for an explicit expression for the likelihood and is, therefore, also suitable for more complex simulators like ours. Moreover, as outlined above, the use of a neural density estimator results in amortized posterior distributions that allow fast evaluation and sampling. We used this fact to determine the coverage and validate our posteriors, a procedure that is infeasible in \ac{MCMC} approaches due to the time-consuming need for repeated sampling.

%%%%%%%%%%%%%%%%%%%%%%%%%%%%%%

\subsection{Late-time magnetic-field decay}
\label{sec:late-time_decay}

We now turn our attention to the parameter, $a_{\rm late}$, the power-law index for the late-time magnetic-field decay. We newly introduced $a_{\rm late}$ in pulsar population synthesis to account for the highly uncertain, core-dominated field evolution above $\unit[10^6]{yr}$ in a phenomenological way. While corresponding inferences were satisfactory for our benchmark experiments, we found that posteriors for $a_{\rm late}$ inferred from the observed population differed significantly between our $19$ experiments, systematically resulting in larger $95\%$ \acp{CI} for smaller $a_{\rm late}$ medians and vice versa (see right most panel in Figure~\ref{fig:ch6_posterior_comparison_atnf}). In addition, several posteriors did not overlap at all across our prior range, leading to a bi-modality in the ensemble posterior. As we do not see anything similar for our synthetic simulations, we do not associate this behaviour with the networks' performance or the \ac{SBI} approach itself. Instead, we hypothesise that this is due to shortcomings in our simulation framework. Put differently, our statistical inferences are only as good as the simulation model used to train our density estimator. Consequently, we see the complications in inferring $a_{\rm late}$ as an indication that our treatment of the late-time field evolution via a power law (albeit physically motivated by the behaviour of known magnetic-field evolution mechanisms) is insufficient to model the observed pulsar population. 

Although further work is needed to better understand this aspect of neutron-star evolution, we can assure ourselves that our current power-law prescription is not too far away from the truth. To do so, we rerun our simulator with the best estimates summarised in Equation~\eqref{eq:ch6_CI_best} and show the distribution of the resulting population in Galactic longitude and latitude and $P$ and $\dot{P}$ in Figure~\ref{fig:ch6_pop_best_simulated}. These are analogous to the respective plots in Figure~\ref{fig:ch6_pop_observed}. While a detailed comparison between this simulated and the observed population and a study of implications for the neutron-star birth rate is beyond the scope of this work, we will highlight a few main aspects. The distributions looked markedly similar, giving a reasonable level of confidence in our underlying simulation framework. However, we do see a slight shift in the \ac{SMPS} population in the $P$-$\dot{P}$ diagram towards lower $\dot{P}$ values. This might again hint at missing physics at late times because \ac{SMPS} is sensitive to somewhat older pulsars compared to the other two surveys.

%%%%%%%%%%%%%%%%%%%%%%%%%%%%%%

\subsection{Neutron-star birth rate}
\label{sec:ch6_NS_birthrate}

We can further count the numbers of detected pulsars in all three synthetic surveys for our best-estimate simulation. Running our simulator ten times to account for its stochastic nature, we obtain average pulsar counts of $1013$, $242$ and $1298$ for the \ac{PMPS}, the \ac{SMPS}, and \ac{HTRU} survey, respectively. Comparing these to the true observed counts in Equation~\eqref{eq:ch6_detected_objects}, we find an equivalent number of objects in \ac{PMPS} (within the sensitivity limits of our iterative approach of generating and detecting pulsars as summarized in Section~\ref{sec:ch6_sim_output}), while we overestimate the \ac{SMPS} population by $\sim 11\%$ and the \ac{HTRU} population by $\sim 27\%$ on average.

To understand these small discrepancies, we return to our earlier discussion of the neutron-star birth rate in Section~\ref{sec:ch6_sim_output}. In particular, for our best estimates, we reach the observed target counts~\eqref{eq:ch6_detected_objects} for each survey for the following birth rates:
\begin{align}
&\text{\ac{PMPS}: $\sim 2.02 \pm 0.02$ neutron stars per century}, \nonumber \\
&\text{\ac{SMPS}: $\sim 1.84 \pm 0.03$ neutron stars per century}, \label{eq:ch6_BR_estimated} \\
&\text{\ac{HTRU}: $\sim 1.66 \pm 0.02$ neutron stars per century}, \nonumber
\end{align}
where we quote means and standard errors for the ten runs. These estimates are somewhat smaller than those obtained in earlier population-synthesis studies \citep{Gullon2014, Faucher2006} and very close to the recent core-collapse supernova estimate from \citet{Rozwadowska2021} ($1.63 \pm 0.46$ per century). The differences in Equation~\eqref{eq:ch6_BR_estimated} are sufficient to result in the slight overproduction of objects noted above. We remind that this is because we continue producing neutron stars until we hit the number of observed pulsars in all three surveys. In our specific case, \ac{PMPS} detections require a slightly larger birth rate than the other two surveys. As mentioned previously, the main reason for this is that we only expect the \textit{correct} physical model to produce the same birth rate across all surveys, again hinting that our simulator is missing some physics. Nonetheless, besides successfully constraining magneto-rotational parameters for pulsar population synthesis using \ac{SBI} for the first time, we do recover birth-rate results in Equation~\eqref{eq:ch6_BR_estimated} that are very similar across all surveys.
%%%%%%%%%%%%%%%%%%%%%%%%%%%%%%

\subsection{Future directions}
\label{sec:future}

In light of the previous conclusions, we intend to further develop our current approach in a number of ways. 

On the simulation side, we will investigate additional luminosity prescriptions that go beyond our assumption, $L_{\rm int} \propto \dot{E}_{\rm rot}^{1/2}$, as this is another quantity that can significantly affect the pulsar distribution. Varying the exponent in our simulations, which was beyond the scope of this study due to computational limitations, but using \ac{SBI} to constrain corresponding parameter ranges would be a first step in that direction. Moreover, while we followed \citet{Gullon2014, Gullon2015} and took a significant step forward in incorporating a realistic description of the neutron-star magnetic field, we already noted above that further investigations into the field evolution of the neutron-star core at late times will be important for future population-synthesis frameworks. Finally, new pulsar surveys (in the radio band as well as in other wavelengths) might hold the key to further constraining the neutron-star population. While we did not see a significant improvement in our inferences using information from one, two or three radio surveys, future studies will benefit from larger numbers of detected pulsars and accurate classification of telescope and detection biases. Furthermore, other wavebands, specifically X-rays or gamma-rays, provide complementary information on the neutron-star population. Our focus on realistic magnetic-field evolution and the expansion of our approach to new three-dimensional magneto-thermal simulations \citep[e.g.,][]{DeGrandis2021, Dehman2023} will be particularly crucial to determine realistic X-ray luminosities of the most strongly magnetized neutron stars. As highlighted by \citet{Gullon2015}, modeling these so-called magnetars and the isolated radio pulsar population consistently will be crucial to break degeneracies and constrain neutron-star physics further.

The increase in simulator complexity associated with these improvements will not only result in more free parameters but also inevitable lead to larger computation times for our forward model. The approach taken here, i.e., simulating a large database for input parameter combinations that cover the entire space sufficiently, will become infeasible. To overcome these hurdles, we will also have to explore new \ac{SBI} approaches. Sequential methods \citep[e.g.][]{Papamakarios2018, Deistler2022, Bhardwaj2023} that reduce the need for simulations by starting from a relatively small database and adaptively providing additional simulations (generated for those parts of the parameter space that are most useful for a neural density estimator to learn a posterior approximation) seem particularly suited to these tasks.

%%%%%%%%%%%%%%%%%%%%%%%%%%%%%%%%%%%%%%%%%%%%%%%%%%%%%%%%%%%%%%%%%%%%%%%%%%%%%%%%%%%%%%%%%%%%%%%%%%%%%%%%%%%%%%%%%%%%%%%%%%%%%% 
% Chapter 7

\chapter{Summary and conclusions} % Main chapter title

\label{Chapter7} % For referencing the chapter elsewhere, use \ref{Chapter1} 

%----------------------------------------------------------------------------------------

In this last chapter, I summarise the main results of this work, separately addressing the two main areas of this thesis, and outline the future directions.

\section{Towards a unified evolutionary scenario for the neutron-star population}

For this thesis we developed a flexible population-synthesis framework that is able to model the neutron-star birth properties and evolution in the Galaxy, starting from different prescriptions for the initial distributions in Galactic height, kick velocity, spin period, magnetic-field strength and magnetic-field evolution. Moreover, we simulated the neutron stars' electromagnetic emission in the radio band and modelled observational biases to emulate the detection with current radio facilities (see Chapters~\ref{Chapter1},~\ref{Chapter5} and~\ref{Chapter6}).
By varying the underlying model parameters, our simulation framework is able to generate different mock neutron-star populations that can be compared with the real observed population. 
To do this, we combine for the first time pulsar population synthesis with a machine learning framework with the aim of performing parameter inference on the observed data and constraining our physical models.  
Neural networks represent a powerful tool to perform this inference task due to their potential to process multi-dimensional data and extract relevant features in an automated way (see Chapter~\ref{Chapter2}). 

In Chapter~\ref{Chapter5} we focused on analysing the dynamics of Galactic neutron stars via simulations. In particular, by varying the natal kick-velocity distribution and the distribution of birth distances from the Galactic plane we evolved the pulsar trajectories in time, and generated a series of simulated populations. We then studied the feasibility of using convolutional neural networks to infer the dynamical properties at birth of a simulated population of neutron stars from density maps storing the information on their sky positions and proper motions. This proof-of-concept work showed that neural networks have the potential to predict the parameters governing the kick-velocity and Galactic height distributions with very good accuracy. The current limitation of this approach is the lack of observational data as we only know around 200 neutron stars with measured sky positions and proper motions. We demonstrated that a good inference accuracy would require about 10 times the current number of neutron stars with measured proper motions. Our work highlights the crucial need for detecting more neutron stars and accurately classifying them, a goal that can be reached with future radio telescopes such as the Square Kilometer Array. 
A future extension of this analysis should focus on the possibility of testing different kick-velocity distributions as recent studies have hinted at a possible bi-modality in the kick-velocity distribution due to different mechanisms producing the kicks in the supernova explosion. 

%Another limitation is the fact that a simple convolutional neural network only gives a point estimate of the values of the scale height and kick velocity parameters. An estimate of the uncertainty of the prediction in these parameters is required for meaningful inference results.

In Chapter~\ref{Chapter6}, we further improved our framework and focused on modelling the magneto-rotational evolution, the radio emission and the detection of neutron stars with current radio surveys. This study aimed to constrain the birth distributions of spin periods and magnetic fields and how these properties evolve in time for the radio pulsar population.
We primarily focused on the radio pulsar population as we have a wealth of data from large-scale surveys performed with Murriyang (the Parkes radio telescope) in Australia in the past decades. As we have a fairly good understanding of the biases and systematics that affect the observed sample from these surveys, they represent a safe testing ground for our deep-learning approach.

We specifically explored an approach known as simulation-based inference or likelihood-free inference which represents a step forward from the approach explored in Chapter~\ref{Chapter5}. There our analysis was limited to point estimates of dynamical neutron-star properties. With the approach presented in Chapter~\ref{Chapter6} we can instead evaluate a posterior distribution for the parameters. As introduced in Chapter~\ref{Chapter2}, in simulation-based inference a deep neural network is employed as a probability density estimator that directly maps a given simulation outcome to the posterior probability distribution of the underlying model parameters without the need to compute the likelihood. 
This is particularly useful for complex simulators with multi-dimensional parameter inputs like ours, where defining and computing a likelihood probability distribution can be complicated or even intractable.
To train the network for our specific problem, we employed density maps containing information on the spin period and spin period derivative of mock observed pulsar populations simulated with different distributions of initial magnetic fields and spin periods and different power laws for the magnetic-field evolution at late times. The trained neural network is then applied to the real empirical data to derive the posterior distribution of the parameters describing the real observed population of radio pulsars. We found that the observed radio pulsars can be best described by belonging to an underlying population born with the logarithm of the initial magnetic field following a normal distribution with best parameters $\mu_{\log B} = 13.10^{+0.08}_{-0.10}$, $\sigma_{\log B} = 0.45^{+0.05}_{-0.05}$ and the logarithm of initial spin period following a normal distribution with best parameters $\mu_{\log P} = -1.00^{+0.26}_{-0.21}$, $\sigma_{\log P} = 0.38^{+0.33}_{-0.18}$. We also put the first constraint on the magnetic-field evolution at late times. Our result suggests that assuming a late-time power law evolution, the magnetic field should decay with a power law-index $a_{\rm late} = -1.80^{+0.65}_{-0.61}$.

A further benefit of the simulation-based inference approach is that the robustness of the results and their uncertainty can be checked using the coverage probability test (see Chapter~\ref{Chapter2} and Appendix~\ref{app:coverage}). This allows us to determine if the inferred posterior is well-calibrated or shows signs of over-confidence or under-confidence. By applying this test to our trained posterior density estimator we showed that the predicted posteriors are well-calibrated and our results using a posterior-ensemble are conservative.

The work completed for this thesis will form the basis to further develop a simulation framework able to include the high-energy emission of neutron stars. To this end, we plan to consider the results of new magneto-thermal 3D simulations \citep{Gourgouliatos2016, DeGrandis2020, Dehman2023} to explain the thermal X-ray radiation from the neutron-star surface and a synchro-curvature model \citep{Vigano2015a, Vigano2015b} to describe the high-energy gamma-ray radiation originating in the neutron-star magnetospheres. We also seek to emulate the observational biases and detection with current X-ray and gamma-ray facilities in order to compare our models with the observed data. 
This multi-wavelength approach to studying the neutron-star population will allow us to consider and analyse the properties of different neutron star classes in a unified scenario. This will be particularly helpful to remove degeneracies and better constrain the parameters describing the neutron-star population as whole.  

Furthermore, we will further improve our simulation-based inference framework by including an active learning approach. After training the density estimator over an initial set of simulations (much smaller than what we outlined in Chapter~\ref{Chapter6}), the resulting initial estimate of the posterior distribution obtained for the observed data can be used to generate additional simulations that more closely reproduce the observed data. This new dataset of simulations can then be adopted as a new training set to refine the estimate of the posterior sequentially. This will allow a more effective inference procedure by concentrating the computational resources on the regions of the parameter space that is more compatible with the observed data.

%=======================================================================================================

\section{The mystery of long-period pulsars}

The second topic that I have explored in this thesis concerns the recent discovery of three intriguing sources: \mtp, the slowest radio neutron-star pulsar ever detected with a spin period of 76 s and two mysterious periodic radio sources, \gleam\ and \gpm, with periods of 1091 s and 1318 s, respectively. The emission properties of these sources are similar to those of magnetars but the nature of these last two objects is still uncertain. In Chapters~\ref{Chapter3} and~\ref{Chapter4} we studied possible scenarios to produce such long-period sources.

In particular in Chapter~\ref{Chapter3}, by assuming a neutron-star nature for these sources, I showed that the standard dipolar losses invoked to explain the currently observed radio pulsar population are insufficient to reach such long spin periods unless extreme (constant) magnetic fields are taken into account. 
We therefore explored the possibility that such slow pulsars might be neutron stars spun down during an initial phase of fallback accretion from the supernova debris. 
Soon after the supernova, the matter that remains gravitationally bound to the central compact object can start to fallback and circularise to form a disk. This disk can interact with the central neutron star and exchange angular momentum depending primarily on the mass accretion rate and the magnetic-field strength of the central neutron star. 
I showed that newly born neutron stars with birth magnetic fields of $B_0 \sim \unit[10^{14-15}]{G}$, experiencing a propeller accretion phase from a fallback disk with moderate initial accretion rates of $\unit[10^{22-27}]{g \, s^{-1}}$, can reach spin periods of thousands of seconds at relatively young ages ($\sim \unit[10^{3-5}]{yr}$). On the other hand, neutron stars with lower magnetic fields are unaffected by the presence of a fallback disk and do not experience a significant spin-down. Hence, while predicting the existence of a population of neutron stars with very long periods, this model also explains the normal radio pulsar population.  
%As the disk evolves and cools down it can undergo a thermal ionization instability and become unable to the transfer of angular momentum. As the accretion stops and the central neutron star's magnetosphere recovers an unperturbed configuration, its radio emission is reactivated. This could explain the observed coherent radio emission from these sources.
However deep X-ray observations have imposed upper limits on the X-ray luminosity of these two objects that challenge the magnetar interpretation. If they are neutron stars, they appear to be too cold to possess such strong magnetic fields according to our current understanding of magneto-thermal evolution in neutron stars.

In Chapter~\ref{Chapter4}, we considered both the neutron-star and magnetic white-dwarf origin for these periodic sources and further studied the implications for the two scenarios.
First, we showed that for both scenarios, these objects are difficult to reconcile with the standard model for radio emission due to particle acceleration in magnetospheric gaps. Moreover, we performed population synthesis simulations of both neutron stars and magnetic white dwarfs to assess the possibility of having long-period sources with a sufficient amount of rotational energy to power the observed radio emission.
Accounting for even the most extreme scenario of fallback accretion and constant magnetic fields, we showed that it is very unlikely to have a population of long-period neutron stars that still have the rotational energy budget to explain the observations of \gleam\ and \gpm.
On the other hand, since magnetic white dwarfs are born with longer spin periods and possess larger moments of inertia, a fraction of their population may possess the rotational energies and the magnetic fields necessary to power the observed emission.

More than 50 years after the first discovery of radio pulsations from neutron stars, the physical mechanism that produces coherent radio emission from rotating magnetosphere is still poorly understood. The discovery of \mtp, \gleam\ and \gpm\ poses further challenges to our current understanding of neutron-star and white-dwarf emission and evolution. The long-period domain in the radio band has been poorly explored in the past and the fact that we serendipitously detected these sources indicates that they may be very common. Future dedicated Galactic plane imaging surveys focusing on long-period sources will likely discover many more objects of this kind. The theoretical ground work presented in this thesis will therefore provide useful insights into future discoveries and help to shed light on the nature of this sources.

%----------------------------------------------------------------------------------------
%	THESIS CONTENT - APPENDICES
%----------------------------------------------------------------------------------------

\appendix % Cue to tell LaTeX that the following "chapters" are Appendices

% Include the appendices of the thesis as separate files from the Appendices folder
% Uncomment the lines as you write the Appendices

% Appendix A

\chapter{Timing tests} % Main appendix title

\label{app:timing_tests} % For referencing this appendix elsewhere, use \ref{AppendixA}

\section{Hardware and Software}
\label{app:hardware}

Our test machine features an Intel(R) Xeon(R) Gold 6230R CPU at 2.10GHz with a single NVIDIA GTX 2080 Ti GPU, 16 GiB RAM, and SSD drives. The system is running CentOS Linux release 7.8.2003 (Core) with PyTorch 1.2.0, CUDA toolkit 10.0.130 and GPU driver 455.32.00.

%%%%%%%%%%%%%%%%%%%%%%%%%%%%%%%%%%%%%%%%%%%%%%%%%%

%-----------------------------------------------------------------
\begin{figure*}
\centering
\includegraphics[width = 0.245\textwidth]{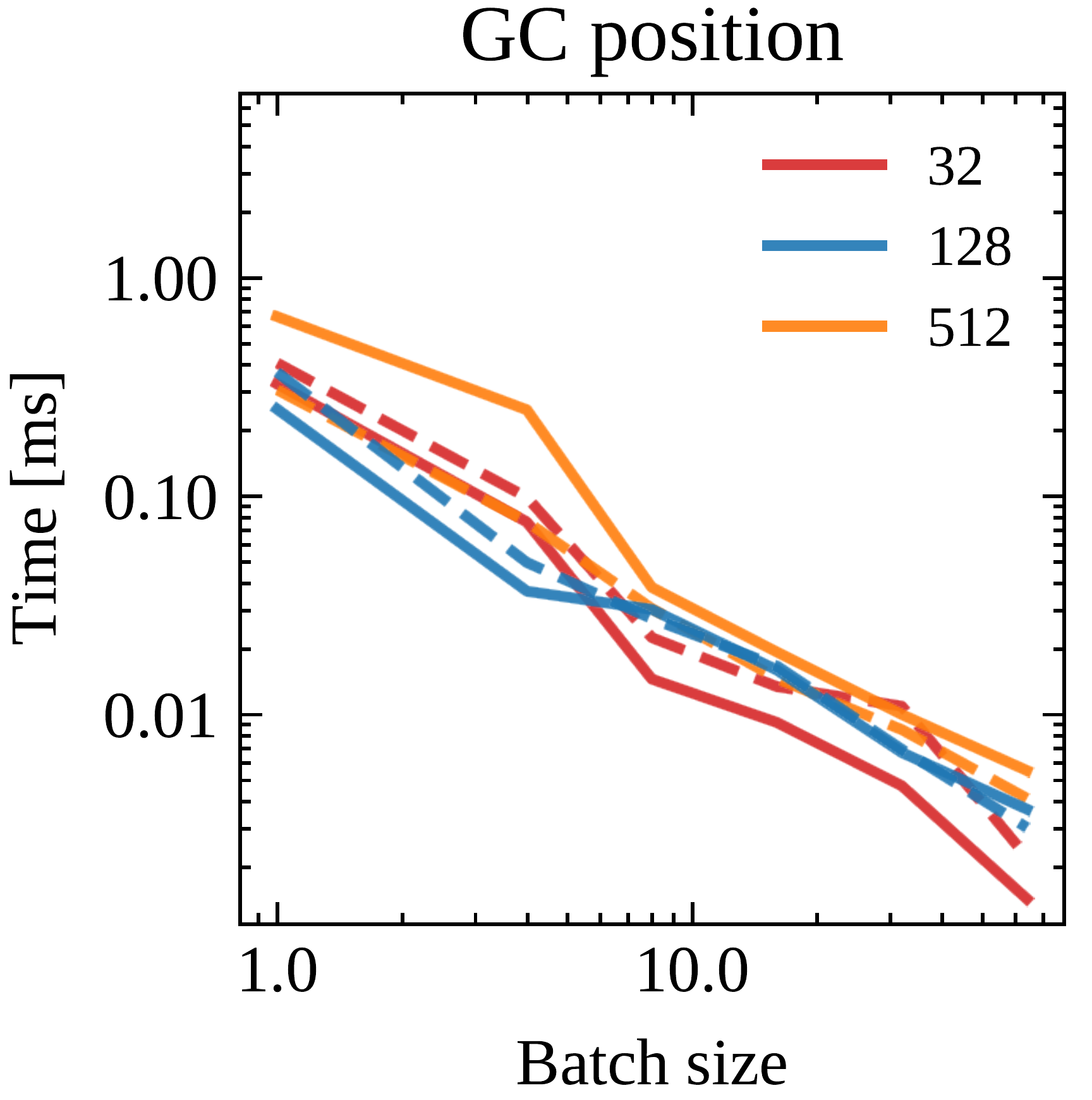}
\includegraphics[width = 0.245\textwidth]{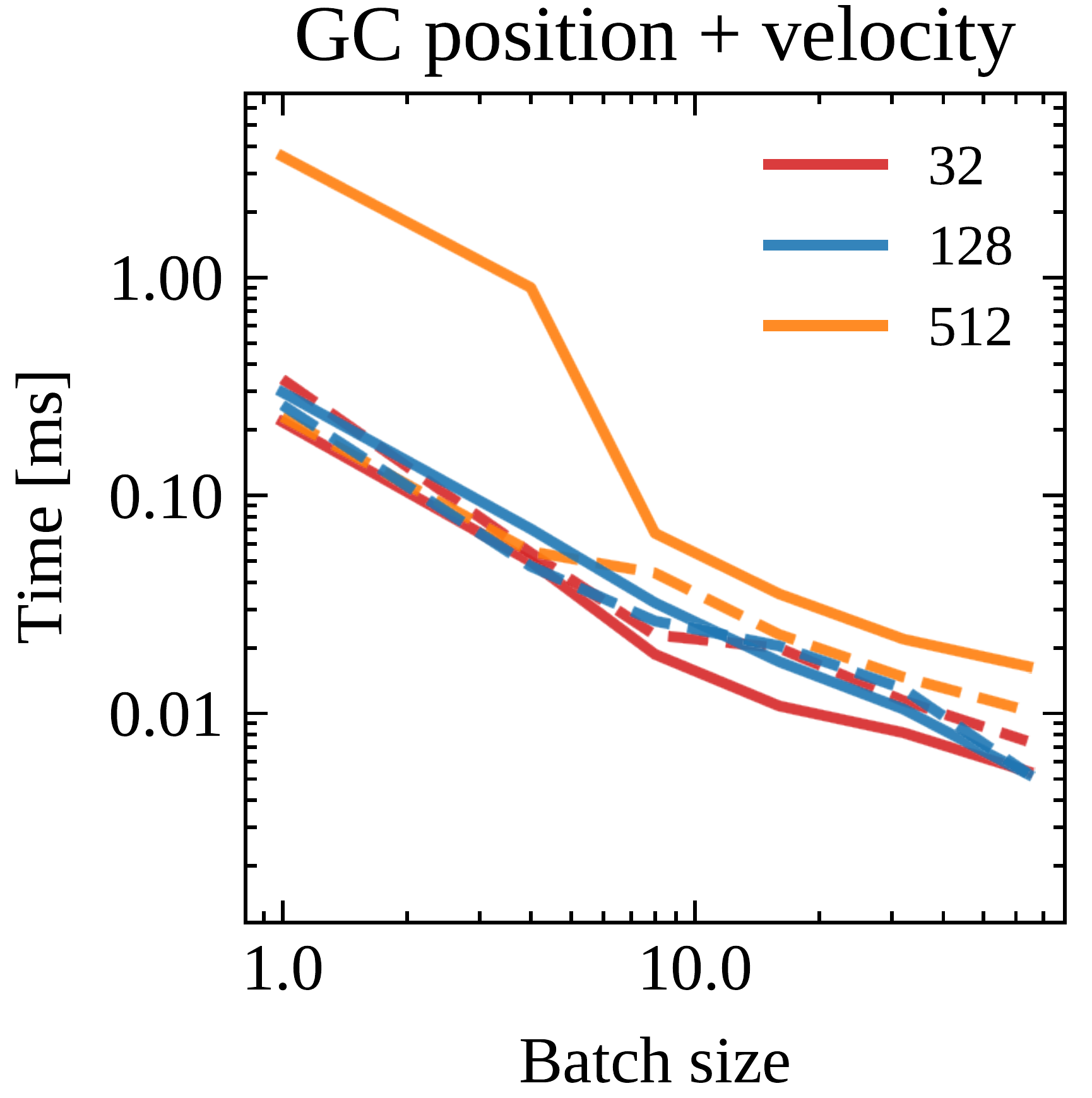}
\includegraphics[width = 0.245\textwidth]{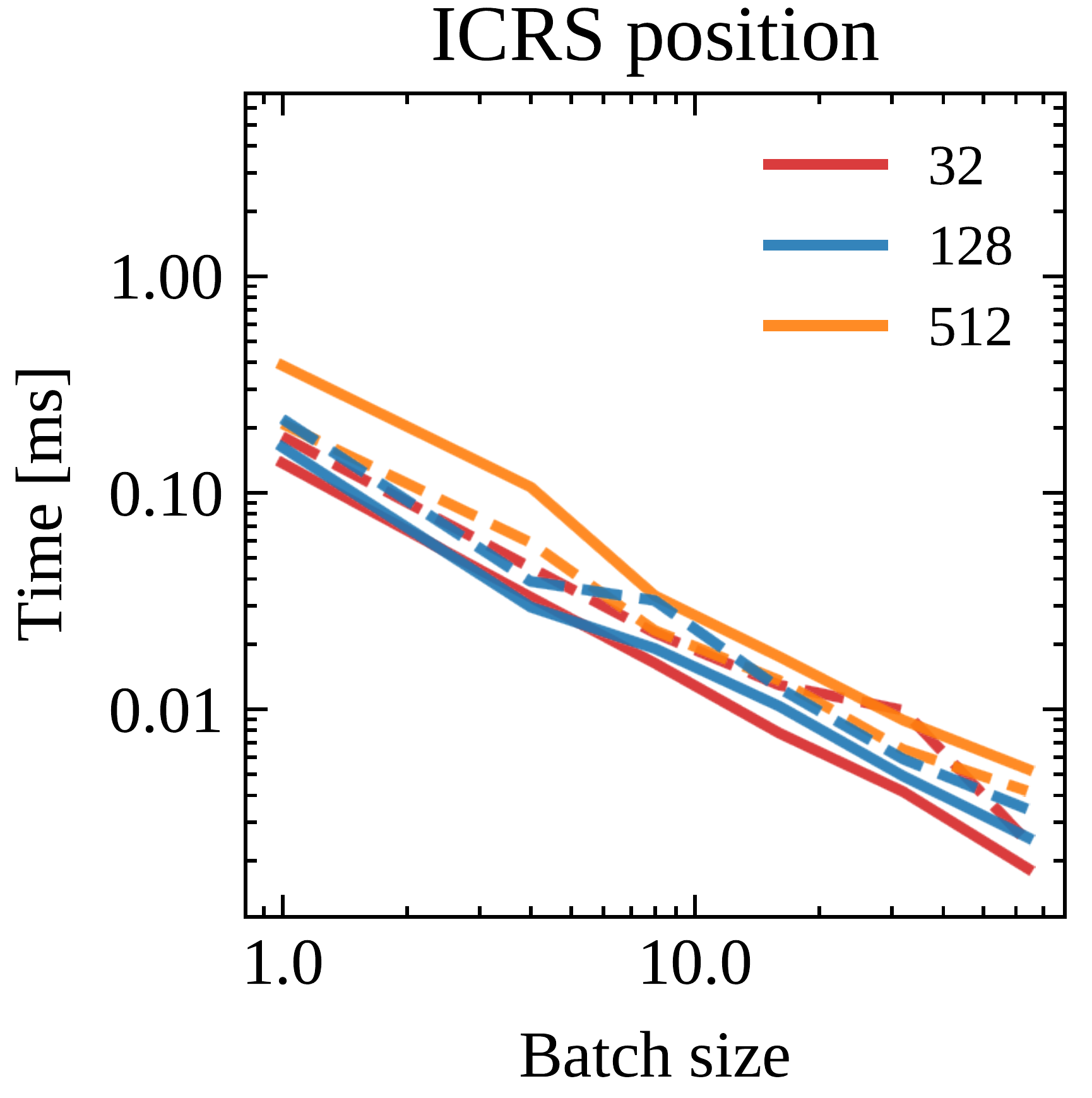}
\includegraphics[width = 0.245\textwidth]{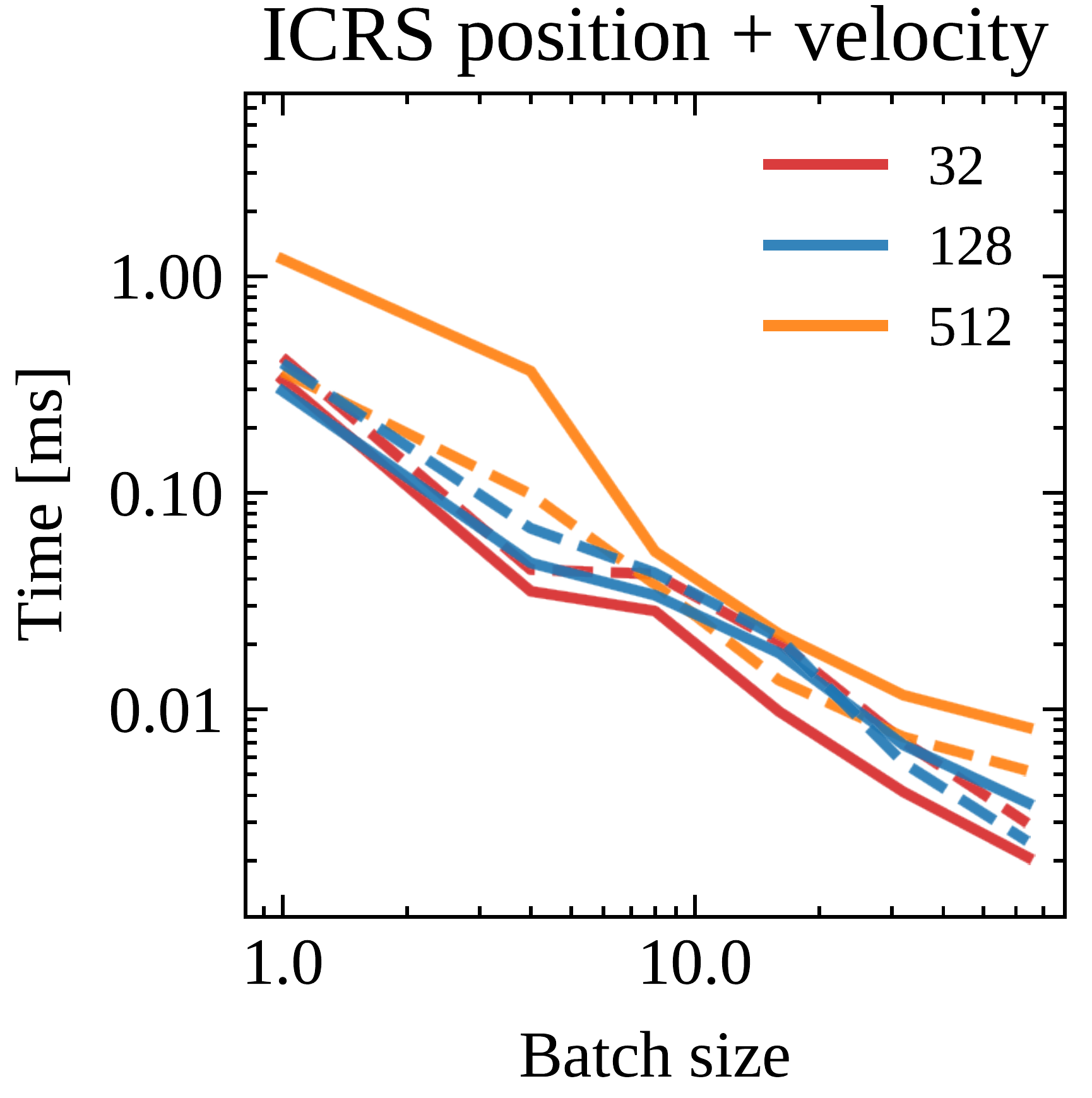}
\caption[\acs{MLP} forward and backward pass times per sample for the single parameter experiment]{{\acs{MLP} forward ({\it solid lines}) and backward ({\it dashed lines}) pass times per sample in ms for the training process on the single parameter $\sigma_{\rm k}$ of the Maxwell kick-velocity distribution, as a function of the batch size and the resolution ({\it red}, {\it blue}, and {\it orange} curves for 32, 128 and 512 respectively) using the four different input configurations T1 (GC position), T2 (GC position + velocity), T3 (ICRS position) and T4 (ICRS position + velocity).}}
\label{fig:app_lnn_time_experiment}
\end{figure*}  
%-----------------------------------------------------------------
%-----------------------------------------------------------------
\begin{figure*}
\centering
\includegraphics[width = 0.245\textwidth]{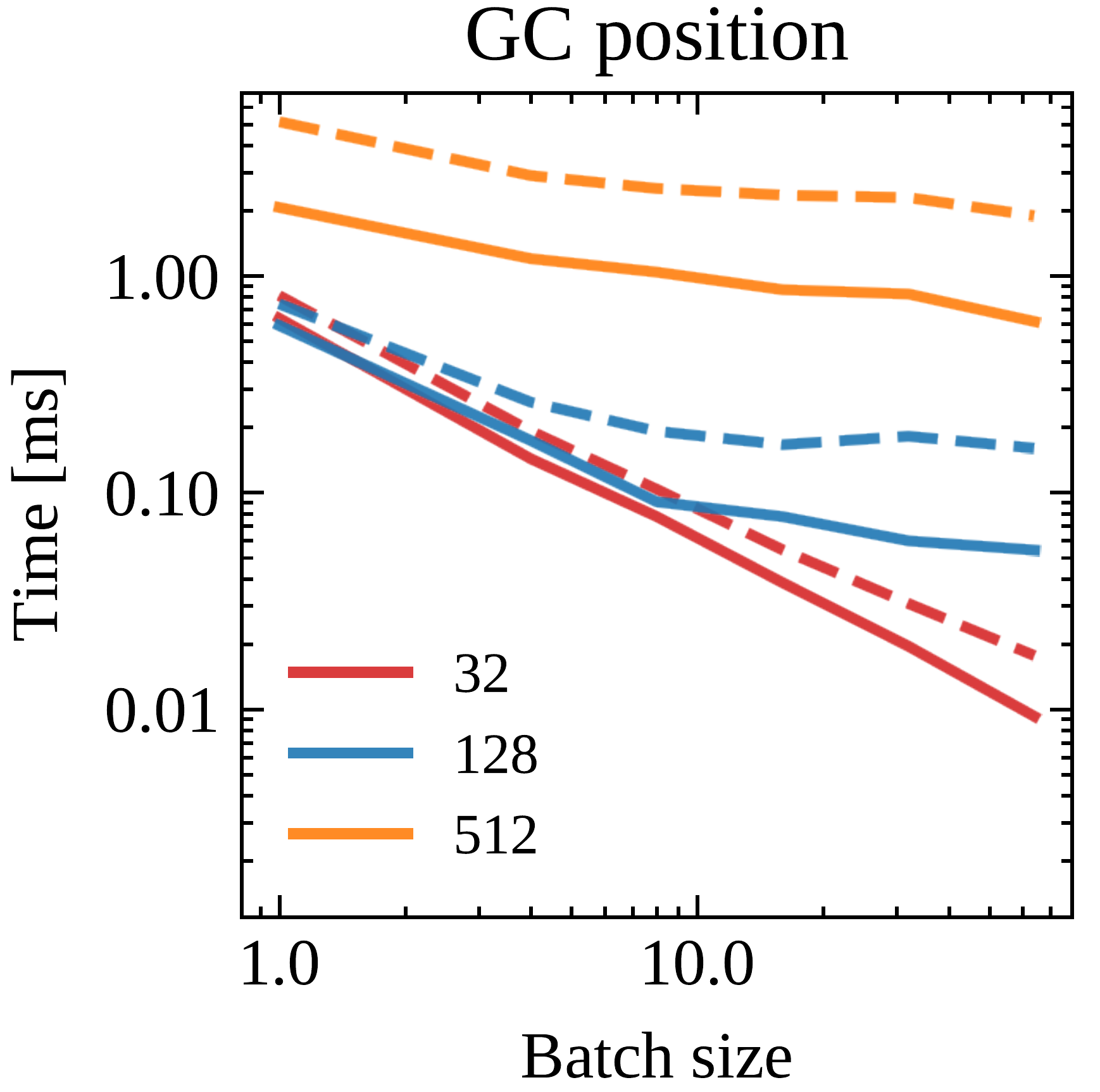}
\includegraphics[width = 0.245\textwidth]{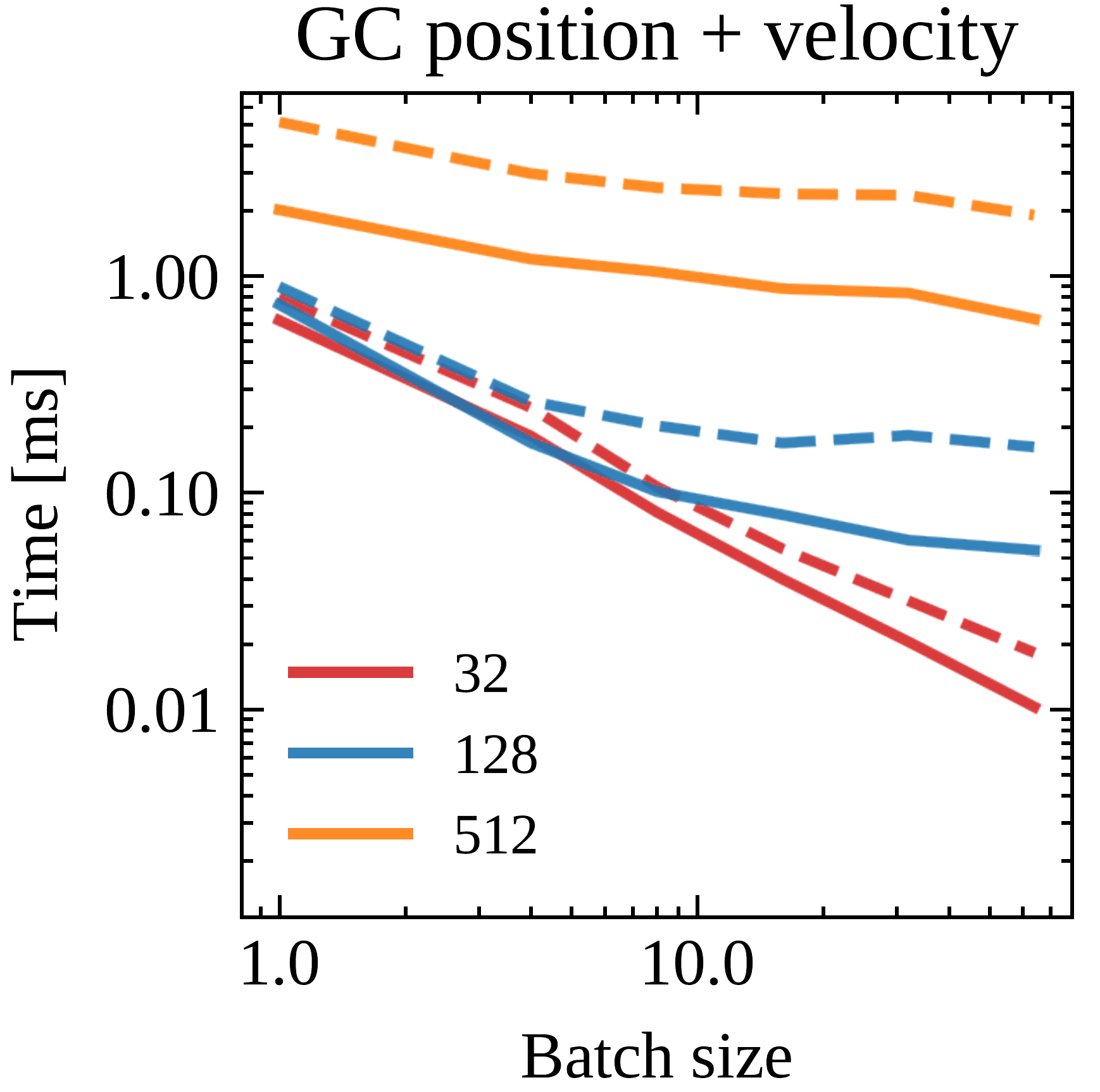}
\includegraphics[width = 0.245\textwidth]{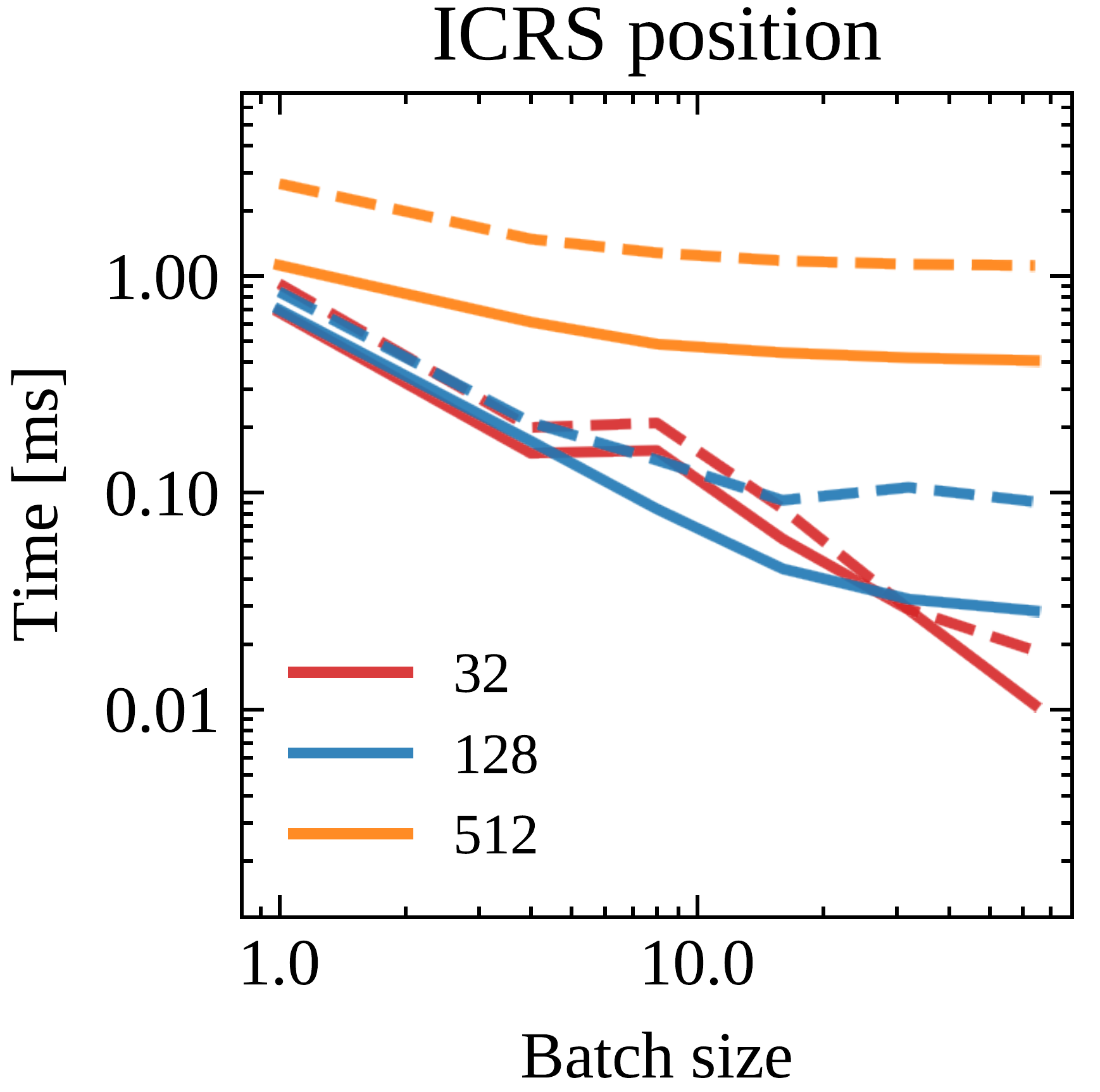}
\includegraphics[width = 0.245\textwidth]{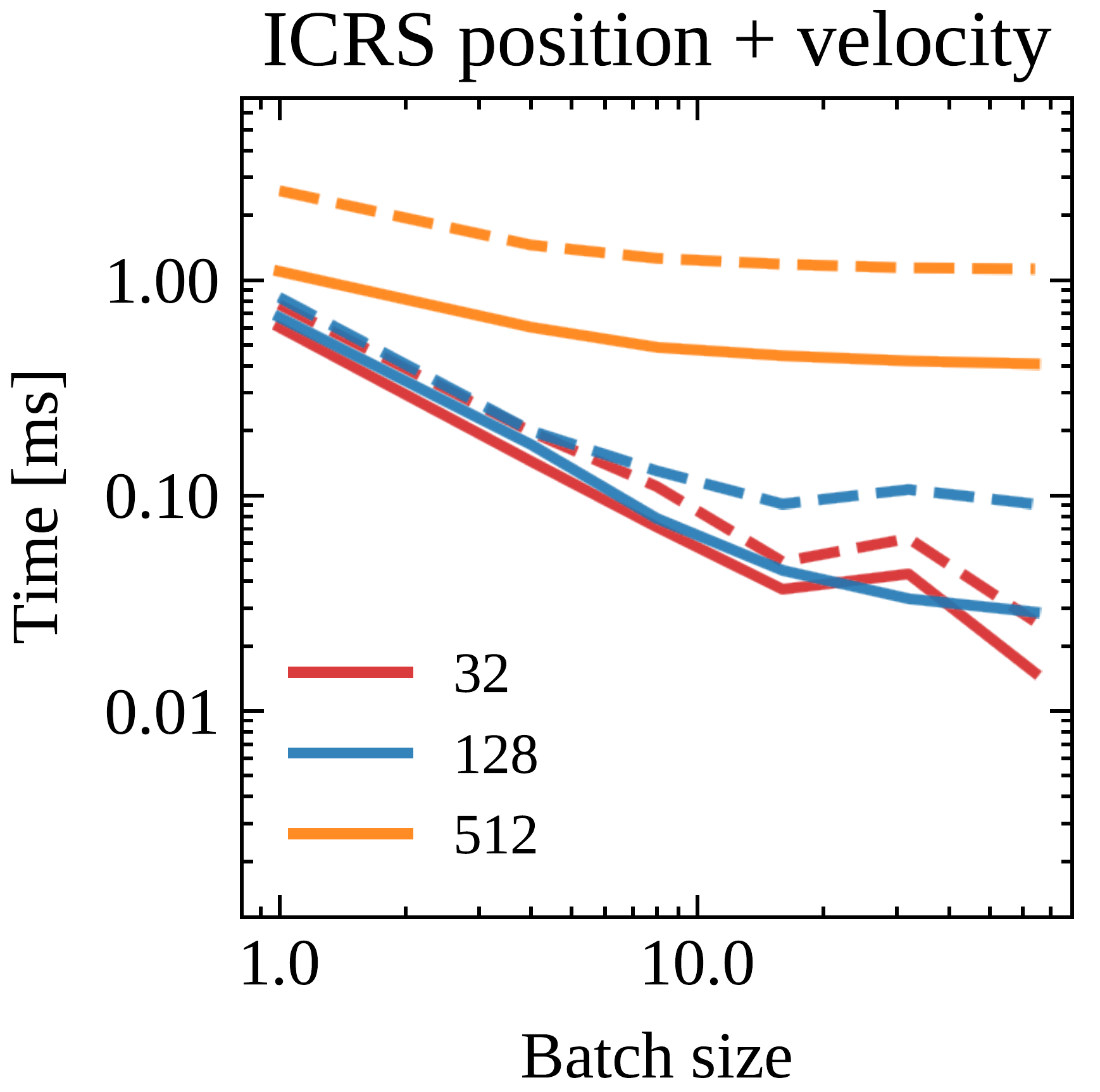}
\caption[\acs{CNN} forward and backward pass times per sample for the single parameter experiment]{{\acs{CNN} forward ({\it solid lines}) and backward ({\it dashed lines}) pass times per sample in ms for the training process on the single parameter $\sigma_{\rm k}$ of the Maxwell kick-velocity distribution, as a function of the batch size and the resolution ({\it red}, {\it blue}, and {\it orange} curves for 32, 128 and 512 respectively) using the four different input configurations T1 (GC position), T2 (GC position + velocity), T3 (ICRS position) and T4 (ICRS position + velocity).}}
\label{fig:app_cnn_time_experiment}
\end{figure*}  
%-----------------------------------------------------------------

\section{Timing Tests}

\subsection{Timing for Single-parameter Predictions}

We report here the timing benchmarks for the \acs{MLP} and \acs{CNN} during the single-parameter training experiments discussed in Section~\ref{sec:ch5_1par_tests}. We run our experiments on the test machine and individually record the forward pass time (the time needed to go through the samples in a batch and compute a prediction) and the backward pass time (the time to compute all the gradients and perform a single optimisation step) as a function of the batch size and resolution. Our benchmarks for the training data-sets from simulation run S1 using the four training configurations T1, T2, T3 and T4 are shown in Figures~\ref{fig:app_lnn_time_experiment} and~\ref{fig:app_cnn_time_experiment}, respectively.

The timing benchmark shows that the \acs{MLP} is slightly faster in performing an optimisation step. This is expected due to fewer trainable parameters when compared to the more complex \acs{CNN}. We can also see that the forward and backward pass times per sample decrease with increasing batch size for both the \acs{MLP} and \acs{CNN}. For a larger batch size, several input samples are transferred from the CPU to the GPU in one step, reducing the overall number of calls between the two. Thus, on average, the processing time for an individual sample reduces when the batch size is increased. Moreover, a higher resolution generally implies an increase in computational cost, albeit being more pronounced in the case of the \acs{CNN} than the \acs{MLP}. The number of input channels itself has very little effect on our timings. Finally, we note that ICRS maps are slightly faster to process (in particular for the higher resolutions), due to the fact that their size is smaller compared to the galactocentric maps (they have half the height in bins).

\subsection{Timing for Two-parameter Predictions}

\begin{figure}
\centering
\includegraphics[width = 0.245\textwidth]{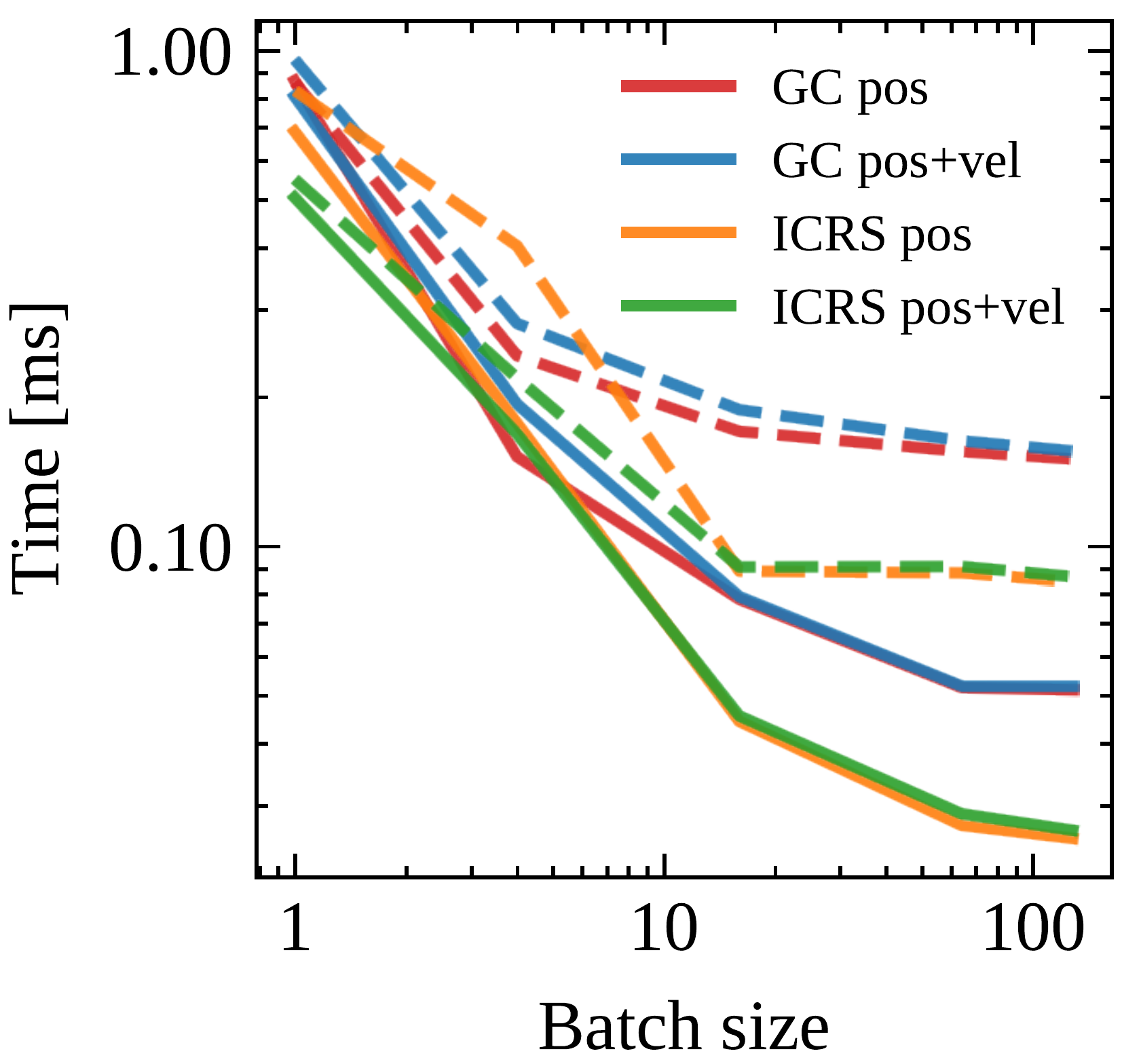}
\caption[\acs{CNN} forward pass and backward pass time per sample for the two-parameter experiments]{\label{fig:app_2par_cnn_time}{\acs{CNN} forward pass ({\it solid line}) and backward pass time ({\it dashed line}) per sample for the two-parameter experiments as a function of the batch size for the four different input configurations T1 (GC position), T2 (GC position + velocity), T3 (ICRS position) and T4 (ICRS position + velocity).}}
\end{figure}

Following the results for the single-parameter experiments, we restrict our two-parameter predictions to the \acs{CNN} model only and fix the resolution of the input maps to 128. The results of our timing benchmarks using the training data-sets from simulation run S3 with the four configurations T1, T2, T3 and T4 are shown in Figure~\ref{fig:app_2par_cnn_time}. We again report the timings for the forward and backward passes per sample as a function of the batch size and the type of input channels provided. As for the single-parameter case, we conclude that using ICRS maps ensures the lowest forward and backward pass times.
%-----------------------------------------------------------------
%-----------------------------------------------------------------

%%%%%%%%%%%%%%%%%%%%%%%%%%%%%%%%%%%%%%%%%%%%%%%%%%

% Appendix B

\chapter{Neutron stars with measured proper motion} % Main appendix title

\label{app:proper_motion} % For referencing this appendix elsewhere, use \ref{AppendixB}

In Table~\ref{tab:app_proper_motion}, we report the properties of 417 neutron stars with measured proper motions in RA and DEC. Data for these neutron stars are primarily collected from the ATNF catalogue\footnote{\url{https://www.atnf.csiro.au/research/pulsar/psrcat/}} \citep{Manchester2005}. In some cases, updated estimates are available and those values quoted and the corresponding references specified. Note that in those cases, where multiple proper-motion estimates are available, we choose the ones with the lowest absolute error. The columns report in order: (i) the object name based on J2000 coordinates; (ii) the right ascension (RA) in hour angle and (iii) declination (DEC) in degrees with the last digit uncertainty given in parentheses; the proper motion in (iv) RA and (v) DEC in milliarcseconds per year with corresponding uncertainties; (vi) the parallax measured in milliarcseconds with uncertainty where available; (vii) the position epoch in modified Julian days; (viii) the spin period in seconds; (ix) the spin-period derivative in seconds over seconds; (x) the dispersion measure in $\unit[]{[pc \, cm^{-3}}]$ with the last digit uncertainty given in parentheses; (xi) the heliocentric distance derived from the \acs{DM} using the YMW16 free-electron density model (for some objects the \acs{DM} exceeds the maximum Galactic \acs{DM} allowed by the YMW16 model, which assigns a default value of $\unit[25]{kpc}$; when available, we quote other distance estimates; * indicates a distance derived from other techniques especially for X-rays and gamma-ray sources, which have no measured \acs{DM}); (xii) the classification of the object, i.e., radio pulsar (PSR), binary pulsar (binary PSR), gamma-/X-ray pulsar (Gamma-/X-ray PSR), magnetar (MAG), X-ray dim isolated neutron star (XDINS); if the object is associated with a globular cluster (GC) or the Small Magellanic Cloud (SMC) this is reported in brackets; and (xiii) the reference for the proper motion measurements, indicated only if different from the ATNF catalogue, i.e., [1] \citet{Motch2009}, [2] \citet{Eisenbeiss2010}, [3] \citet{Walter2010}, [4] \citet{Stovall2014}, [5] \citet{Jennings2018}, [6] \citet{Perera2019}, [7] \citet{Dang2020}, [8] \citet{Danilenko2020}. 

%%%%%%%%%%%%%%%%%%%%%%%%%%%%%%%%%%%%%%%%%%%%%%%%%%

\begin{landscape}
	\centering
	{\setlength{\tabcolsep}{2pt}
		\scriptsize  % Switch to smaller text size otherwise table won't fit
\begin{longtable}{l l l l l l l l l l l l l}
			\caption{Up-to-date list of 417 neutron stars with measured proper motions in RA and DEC.
				\label{tab:app_proper_motion}} \\ 
			\toprule
			\tabhead{\multirow{2}{*}{JName}} &
			\tabhead{RA} &
			\tabhead{DEC} &
			\tabhead{$\mathbold{\mu_{\rm RA}}$} & 
			\tabhead{$\mathbold{\mu_{\rm DEC}}$} & 
			\tabhead{parallax} &
			\tabhead{Pos. Epoch} &
			\tabhead{$\mathbold{P}$} &
			\tabhead{$\mathbold{\dot{P}}$} & 
			\tabhead{\acs{DM}} & 
			\tabhead{$\mathbold{d_{\odot}}$} & 
			\tabhead{\multirow{2}{*}{Class(Assoc.)}} & 
			\tabhead{\multirow{2}{*}{Ref.}} \\
			& 
			\tabhead{[hms]} & 
			\tabhead{[$^{\circ}$ ' '']} & 
			\tabhead{[mas yr$^{-1}$]} & 
			\tabhead{[mas yr$^{-1}$]} & 
			\tabhead{[mas]} & 
			\tabhead{[MJD]} & 
			\tabhead{[s]} & 
			\tabhead{[s s$^{-1}$]} & 
			\tabhead{[pc cm$^{-3}$]} & 
			\tabhead{[kpc]} & 
			&  \\
			\midrule
			\endfirsthead
			% Define the table header for the other pages without the caption.
			\toprule
			\tabhead{\multirow{2}{*}{JName}} &
			\tabhead{RA} &
			\tabhead{DEC} &
			\tabhead{$\mathbold{\mu_{\rm RA}}$} & 
			\tabhead{$\mathbold{\mu_{\rm DEC}}$} & 
			\tabhead{parallax} &
			\tabhead{Pos. Epoch} &
			\tabhead{$\mathbold{P}$} &
			\tabhead{$\mathbold{\dot{P}}$} & 
			\tabhead{\acs{DM}} & 
			\tabhead{$\mathbold{d_{\odot}}$} & 
			\tabhead{\multirow{2}{*}{Class(Assoc.)}} & 
			\tabhead{\multirow{2}{*}{Ref.}} \\
			& 
			\tabhead{[hms]} & 
			\tabhead{[$^{\circ}$ ' '']} & 
			\tabhead{[mas yr$^{-1}$]} & 
			\tabhead{[mas yr$^{-1}$]} & 
			\tabhead{[mas]} & 
			\tabhead{[MJD]} & 
			\tabhead{[s]} & 
			\tabhead{[s s$^{-1}$]} & 
			\tabhead{[pc cm$^{-3}$]} & 
			\tabhead{[kpc]} & 
			&  \\
			\midrule
			\endhead
			
			J0014+4746   & 00:14:17.75(4)	     &  +47:46:33.4(3)      &	  $   19.3     \pm  1.8   $   &   $   -19.7    \pm   1.5   $   &   $   \_		     $   &  49664      &   $	1.24070     $	&    $    5.6446 \times 10^{-16	   } 	    $	&     $     30.405     (13)    $   &	 1.776       &  PSR		  &	    \\  		 					   
			J0023+0923   & 00:23:16.87910(2)     &  +09:23:23.8689(8)   &	  $   -12.4    \pm  0.5   $   &   $   -6.1     \pm   0.1   $   &   $   0.4	\pm   0.3    $   &  56567      &   $	0.00305     $	&    $    1.14234 \times 10^{-20   } 	    $	&     $     14.32810  (4)      $   &	 1.248       &  PSR		  &   [6]   \\  		 					   
			J0024-7204ab & 00:24:08.1615(5)      &  -72:04:47.602(2)    &	  $   4.2      \pm  0.6   $   &   $   -2.9     \pm   0.5   $   &   $   \_		     $   &  51600      &   $	0.00370     $	&    $    9.820 \times 10^{-21	   } 	    $	&     $     24.37     (2)      $   &	 2.54	     &  PSR(GC)	  	  &	    \\       		 					   
			J0024-7204C  & 00:23:50.3546(1)      &  -72:04:31.5048(4)   &	  $   5.2      \pm  0.1   $   &   $   -3.1     \pm   0.1   $   &   $   \_		     $   &  51600      &   $	0.00576     $	&    $    -4.98503 \times 10^{-20  } 	    $	&     $     24.5955   (8)      $   &	 2.594       &  PSR(GC)	  	  &	    \\  		 					   
			J0024-7204D  & 00:24:13.88092(6)     &  -72:04:43.8524(2)   &	  $   4.24     \pm  0.07  $   &   $   -2.24    \pm   0.05  $   &   $   \_		     $   &  51600      &   $	0.00536     $	&    $    -3.4220 \times 10^{-21   } 	    $	&     $     24.7432   (9)      $   &	 2.631       &  PSR(GC)	  	  &	    \\  		 					   
			J0024-7204E  & 00:24:11.10528(5)     &  -72:05:20.1492(2)   &	  $   6.15     \pm  0.03  $   &   $   -2.35    \pm   0.06  $   &   $   \_		     $   &  51600      &   $	0.00354     $	&    $    9.85103 \times 10^{-20   } 	    $	&     $     24.236    (2)      $   &	 2.509       &  PSR(GC)	  	  &	    \\
\end{longtable}
	}
\end{landscape}

%---------------------------------------------------------------
% Appendix C

\chapter{Parametrisation of the magnetic-field evolution} % Main appendix title

\label{app:B-field} % For referencing this appendix elsewhere, use \ref{AppendixC}

As outlined in Section~\ref{sec:ch6_mr_evol}, a key ingredient for the magneto-rotational evolution of radio pulsars is a realistic prescription for the evolution of the dipolar magnetic-field strength, $B$, up to neutron-star ages of $\unit[10^8]{yr}$. While earlier population-synthesis studies have typically either neglected magnetic-field decay entirely, or relied on simplified descriptions invoking decaying exponentials or power laws, we choose a different approach and take advantage of recent progress in modelling the magneto-thermal evolution of neutron-star crusts. In particular, we use a set of five magneto-thermal simulations \citep{Vigano2021} to fit the early-time magnetic-field evolution which is driven by the combined action of the Hall effect and Ohmic dissipation \citep[see, e.g.,][and Section~\ref{sec:ch1_B_evol} for details on these mechanisms]{Pons2019}. 

All five curves, shown as solid lines in Figure~\ref{fig:ch6_B_fields}, were simulated with realistic assumption on relevant physics. In particular, the stellar structure and composition are based on the equation of state SLy4 \citep{Douchin2001} for a neutron star of mass $\unit[1.4]{M_{\odot}}$, resulting in a radius of $\unit[11.74]{km}$. The impurity parameter at the highest densities in the inner crust is set to $100$ \citep{Pons2013}, representing the presence of resistive nuclear pasta phases \citep[see, e.g.,][]{Chamel2008}, whereas the impurity profile for other crustal densities matches the results of \citet{Carreau2020} (see their Fig.~5). Furthermore, the model for the neutron-star envelope is taken from \citet{Potekhin2015}, while specific parametrisation for the superfluid and superconducting energy gaps (SFB for the crustal neutrons, TToa for the core neutrons and CCDKp for the core protons) were adopted from \citet{Ho2015}. 

What varies between the different simulations is the initial poloidal magnetic-field strength, $B$, taking the values $10^{12}, 10^{13}, 10^{14}, 10^{15}, 5 \times 10^{15} \, {\rm G}$, respectively. This also implies different toroidal field strengths, which are typically a factor $10$ larger than the poloidal $B$s. We observe in Figure~\ref{fig:ch6_B_fields} that those runs with larger magnetic fields decay faster. This is a direct result of the Hall effect which depends on $B$ and acts to redistribute the magnetic-field energy to smaller scales, where it subsequently decays due to Ohmic dissipation. For sources with $B \lesssim \unit[10^{12}]{G}$ and coupled thermal evolution, this Hall cascade does not take place and magnetic fields remain pretty much constant on timescales of the order of $\unit[10^6]{yr}$. 

Above this timescale, however, current magneto-thermal simulations become unreliable because the implementation of relevant microphysics \citep{Potekhin2015} is unsuited to old neutron stars with temperatures $\lesssim \unit[10^6]{K}$. In addition, these simulations focus primarily on the crust and do not include a realistic treatment of the highly uncertain dynamics of the neutron-star core, which should become relevant above $\sim \unit[10^6]{yr}$. As we require a prescription for the field above $\unit[10^6]{yr}$ for our population synthesis, we develop a simplified parametrisation for the late-time magnetic-field evolution that encodes the unknown evolution of the stellar core. As highlighted in Equation~\eqref{eq:ch6_B_late}, we assume that field changes at late times can be captured by a power law characterised by the index, $a_{\rm late}$. This choice is physically motivated because several known magnetic-field evolution mechanisms exhibit the same functional form. For example, Hall-like physics are encoded by $a_{\rm late} = -1$ \citep{Aguilera2008b}, while ambipolar diffusion follows a power law with $a_{\rm late} = -0.5$ \citep{Goldreich1992}. 

To directly parametrise the behaviour of the magnetic fields across all relevant $B$ ranges and times $t$, we describe the field evolution with the following broken power laws:
\begin{align}
B(t) &= B_0 \left(1 + \frac{t}{\tau_1} \right)^{a_1} \left(1 + \frac{t}{\tau_2} \right)^{a_2 - a_1} 
\left(1 + \frac{t}{\tau_{\rm late}} \right)^{a_{\rm late} - a_2} \quad \text{for} \quad \tau_1 < \tau_2 < \tau_{\rm late}, 
\\[1.4ex]
B(t) &= B_0 \left(1 + \frac{t}{\tau_1} \right)^{a_1} \left(1 + \frac{t}{\tau_{\rm late}}\right)^{a_{\rm late} - a_1}
\quad \text{for} \quad \tau_1 < \tau_{\rm late} < \tau_2, 
\\[1.4ex]
B(t) &= B_0\left(1 + \frac{t}{\tau_{\rm late}} \right)^{a_{\rm late}}
\quad \text{for} \quad \tau_{\rm late} < \tau_1 < \tau_2. 
\end{align}
Here, the two timescales $\tau_1 \equiv A_1  B_0^{b_1}$ and $\tau_2 \equiv A_2 B_0^{b_2}$ depend on the initial magnetic field, $B_0$, while $\tau_{\rm late}$ is a constant. The latter together with the free parameters $A_{1,2}, b_{1,2}$ and the power-law indices $a_{1,2}$ can be adjusted to closely fit the numerical simulations. In particular, we choose $\tau_{\rm late} \approx \unit[2 \times 10^6]{yr}$, $A_{1} = 10^{14}$, $b_1 = -0.8$, $A_{2} = 6 \times 10^{8}$, $b_{2} = -0.2$, $a_{1} = -0.13$, and $a_{2} = -3.0$. 

For particularly steep power-law indices, $a_{\rm late}$, the current prescription, in principle, allows the magnetic field to decay to unrealistically small values in contrast with observations of old millisecond pulsars \citep{Lorimer2008}. To prevent this, we assume that the magnetic field eventually settles at a constant value, $B_{\rm late}$, for very late times. In line with detected old neutron stars, we randomly sample the logarithm of $B_{\rm late}$ from a normal distribution with a mean $\mu_{\log B, {\rm final}} = 8.5$ and a standard deviation $\sigma_{\log B, {\rm final}} = 0.5$ as already outlined previously. The result of this magnetic-field prescription for $a_{\rm late} = -3.0$ is shown as the dashed lines in Figure~\ref{fig:ch6_B_fields}.
% Appendix D

\chapter{Coverage calculation} % Main appendix title

\label{app:coverage} % For referencing this appendix elsewhere, use \ref{AppendixD}

To validate our neural posterior estimates, we follow \citet{Cook2006} who demonstrated that for a well-calibrated posterior distribution, the smallest volume that contains the ground truth, $\bt$, for a given sample in a test data set follows a uniform distribution. This, in turn, implies that the cumulative distribution function of these quantiles across the entire test set forms a diagonal line.
The graphical representation of this cumulative distribution function is commonly referred to as the \textit{coverage plot} (see Section~\ref{sec:ch2_coverage} and Figure~\ref{fig:ch6_coverage}). Put differently, if we consider a credibility level $1-\alpha$, we expect the ground truth, $\bt$, to fall into this region for a fraction $1 - \alpha$ of test samples if the coverage is diagonal.

To calculate the corresponding coverage for our posteriors and assess how well they are calibrated, we take advantage of the amortised nature of our approximate posterior. In particular, for each of our $3,600$ test samples, we have access to the ground truth, $\bt$, and the corresponding posterior approximation, $q_{F(\bx, \boldsymbol{\phi})}(\bt)$, where $F(\bx, \boldsymbol{w})$ represents a trained neural network. To determine the coverage, we need to calculate the quantiles for each $\bt$. In our case, where we infer five magneto-rotational parameters and the posterior, $q_{F(\bx, \boldsymbol{w})}(\bt)$, is a five-dimensional probability density function (see Equation~\eqref{eq:ch6_posterior_gmm}), we obtain corresponding quantiles by determining the so-called \acp{HDR}, i.e., those regions covering our sample space for a given probability $1-\alpha$ that have the smallest possible volume \citep{Hyndman1996}. To obtain these \acp{HDR} for each of our test samples, we first compute the total log-posterior at the ground truth, $\bt$, i.e., $\log q_{F(\bx, \boldsymbol{w})} (\bt)$. From each posterior, we subsequently draw samples, $\bt_s$, with $s\in \{1, \dots, S\}$, for which we also individually compute the log-posterior, i.e., $\log q_{F(\bx, \boldsymbol{w})} (\bt_s)$. The \ac{HDR} for a given test sample with ground truth, $\bt$, is now the percentage of samples, $\bt_s$, which satisfy the condition $\log q_{F(\bx, \boldsymbol{w})} (\bt_s) > \log q_{F(\bx, \boldsymbol{w})} (\bt)$. To compute the cumulative distribution function (coverage) across our test set, we repeat this process iteratively for all $3,600$ test samples to determine, for a given credibility level $1 - \alpha$, the fraction of test samples where the \ac{HDR} is smaller or equal to $1 - \alpha$. 
%For a well-calibrated posterior, we expect, e.g., in the case of a $95\%$ credibility interval, that the ground truths of $95\%$ of the test samples fall within the $0.95$ \ac{HDR}.

Deviations from the diagonal are present when posterior estimates are either too wide (conservative) or too narrow (over-confident). In the former case, ground truths would be enclosed within a given \acp{HDR} more often than expected for the true posterior, while in the latter scenario the opposite applies. The resulting coverage curves would, thus, lie above and below the diagonal, respectively, highlighting the benefit of the coverage plot in validating our posteriors.

Finally note that for our ensemble approach, we calculate the \ac{HDR} with the ensemble posterior, $\overline{q}(\bt)$, using the condition $\log \overline{q} (\bt_s) > \log \overline{q}(\bt)$. The remaining steps are identical to those outlined above.

%----------------------------------------------------------------------------------------
%	BIBLIOGRAPHY
%----------------------------------------------------------------------------------------

\printbibliography[heading=bibintoc]

%----------------------------------------------------------------------------------------

\end{document}